\documentclass[10pt,twoside,openright]{memoir}

\setlrmargins{53mm}{*}{*}
\checkandfixthelayout

\setsecnumdepth{subsection}
\usepackage[spanish,catalan,italian,english]{babel}
\usepackage[latin1]{inputenc}
\usepackage{amsthm}
\usepackage{amssymb}
\usepackage{amsmath}
\usepackage{bbold}
\usepackage{bbm}
\usepackage{nonfloat}
\usepackage{color}
\usepackage[pdftex]{hyperref}
\usepackage{braket}
\usepackage{dsfont}
\usepackage{mathdots}
\usepackage{mathtools}
\usepackage{enumerate}
\usepackage{csquotes}
\usepackage{stmaryrd}
\usepackage[cal=boondox]{mathalfa}
\usepackage{graphicx}
\usepackage{stackengine}
\usepackage{scalerel}
\usepackage[numbers]{natbib}
\usepackage{array}
\usepackage{makecell}
\newcolumntype{x}[1]{>{\centering\arraybackslash}p{#1}}
\usepackage{tikz}
\usepackage{tcolorbox}
\usepackage{etoolbox}
\usepackage{epigraph}

\DeclareGraphicsExtensions{.jpeg, .png, .pdf}

\newtheorem{thm}{Theorem}
\newtheorem*{thm*}{Theorem}
\newtheorem{prop}[thm]{Proposition}
\newtheorem*{prop*}{Proposition}
\newtheorem{lemma}[thm]{Lemma}
\newtheorem*{lemma*}{Lemma}
\newtheorem{cor}[thm]{Corollary}
\newtheorem*{cor*}{Corollary}
\newtheorem{cj}{Conjecture}
\newtheorem*{cj*}{Conjecture}
\newtheorem{Def}[thm]{Definition}
\newtheorem*{Def*}{Definition}
\newtheorem{question}[cj]{Question}
\counterwithin{thm}{chapter}
\counterwithin{cj}{chapter}

\theoremstyle{definition}
\newtheorem*{rem}{Remark}
\newtheorem*{note}{Note}
\newtheorem{ex}[thm]{Example}
\newtheorem{axiom}{Axiom}

\newcommand{\bb}{\begin{equation}}
\newcommand{\bbb}{\begin{equation*}}
\newcommand{\ee}{\end{equation}}
\newcommand{\eee}{\end{equation*}}
\newcommand{\Tr}{\text{Tr}\,}
\newcommand{\tr}{\text{Tr}}
\newcommand{\rk}{\text{rk}\,}
\newcommand*{\coloneqq}{\mathrel{\vcenter{\baselineskip0.5ex \lineskiplimit0pt \hbox{\scriptsize.}\hbox{\scriptsize.}}} =}

\newcommand{\texteq}[1]{\stackrel{\mathclap{\scriptsize \mbox{#1}}}{=}}
\newcommand{\textleq}[1]{\stackrel{\mathclap{\scriptsize \mbox{#1}}}{\leq}}
\newcommand{\textl}[1]{\stackrel{\mathclap{\scriptsize \mbox{#1}}}{<}}

\newcommand{\textgeq}[1]{\stackrel{\mathclap{\scriptsize \mbox{#1}}}{\geq}}
\newcommand{\cl}{\text{cl}}
\newcommand{\clit}{\text{\emph{cl}}}
\newcommand{\co}{\text{co}}
\newcommand{\coit}{\text{\emph{co}}}
\newcommand{\inter}{\text{int}}
\newcommand{\interit}{\text{\emph{int}}}
\newcommand{\Span}{\text{span}}
\newcommand{\Spanit}{\text{\emph{span}}}
\newcommand{\wstarlim}{\mathcal{w}^*\!\text{-lim}}

\newcommand{\aff}{\text{aff}}
\newcommand{\affit}{\text{\emph{aff}}}
\newcommand{\id}{{\mathds{1}}}
\newcommand{\tmin}{\ensuremath \!\! \raisebox{2.5pt}{$\underset{\begin{array}{c} \vspace{-3.9ex} \\ \text{\scriptsize min} \end{array}}{\otimes}$}\! \!}
\newcommand{\tmax}{\ensuremath \!\! \raisebox{2.5pt}{$\underset{\begin{array}{c} \vspace{-3.9ex} \\ \text{\scriptsize max} \end{array}}{\otimes}$}\!\! }
\newcommand{\tminit}{\ensuremath \!\! \raisebox{2.5pt}{$\underset{\begin{array}{c} \vspace{-3.9ex} \\ \text{\scriptsize \emph{min}} \end{array}}{\otimes}$}\!\! }
\newcommand{\tmaxit}{\ensuremath \!\! \raisebox{2.5pt}{$\underset{\begin{array}{c} \vspace{-3.9ex} \\ \text{\scriptsize \emph{max}} \end{array}}{\otimes}$}\!\! }
\newcommand{\tminfoot}{\ensuremath \!\!\! \raisebox{2.1pt}{$\scriptstyle\underset{\begin{array}{c} \vspace{-4.1ex} \\ \text{\fontsize{2}{4}\selectfont min} \end{array}}{\otimes}$}\!\!\!}
\newcommand{\tminitfoot}{\ensuremath \!\!\! \raisebox{2.1pt}{$\scriptstyle\underset{\begin{array}{c} \vspace{-4.1ex} \\ \text{\fontsize{2}{4}\selectfont \emph{min}} \end{array}}{\otimes}$}\!\!\!}

\newcommand{\lmatrix}{\left(\begin{smallmatrix}}
\newcommand{\rmatrix}{\end{smallmatrix}\right)}

\tcbset{colback=green!5,colframe=orange!50!red,
fonttitle=\bfseries, float=htb}

\newtcolorbox[auto counter]{example}[3][]
{float*=ht,title=Observation ~\thetcbcounter: #2,label= ex:#3 ,#1}

\renewenvironment{thebibliography}[1]{
  \begin{oldthebibliography}{#1}
    \setlength{\itemsep}{1ex}
    \setlength{\parskip}{0em}
}
{
  \end{oldthebibliography}
}

\stackMath
\def\dyhat{-0.15ex}
\newcommand\myhat[1]{\ThisStyle{%
              \stackon[\dyhat]{\SavedStyle#1}
                              {\SavedStyle\widehat{\phantom{#1}}}}}

\makeatletter
\newcommand*\rel@kern[1]{\kern#1\dimexpr\macc@kerna}
\newcommand*\widebar[1]{%
  \begingroup
  \def\mathaccent##1##2{%
    \rel@kern{0.8}%
    \overline{\rel@kern{-0.8}\macc@nucleus\rel@kern{0.2}}%
    \rel@kern{-0.2}%
  }%
  \macc@depth\@ne
  \let\math@bgroup\@empty \let\math@egroup\macc@set@skewchar
  \mathsurround\z@ \frozen@everymath{\mathgroup\macc@group\relax}%
  \macc@set@skewchar\relax
  \let\mathaccentV\macc@nested@a
  \macc@nested@a\relax111{#1}%
  \endgroup
}

\newcommand\diag[4]{%
  \multicolumn{1}{p{#2}|}{\hskip-\tabcolsep
  $\vcenter{\begin{tikzpicture}[baseline=0,anchor=south west,inner sep=#1]
  \path[use as bounding box] (0,0) rectangle (#2+2\tabcolsep,\baselineskip);
  \node[minimum width={#2+2\tabcolsep-\pgflinewidth},
        minimum  height=\baselineskip+\extrarowheight-\pgflinewidth] (box) {};
  \draw[line cap=round] (box.north west) -- (box.south east);
  \node[anchor=south west] at (box.south west) {#3};
  \node[anchor=north east] at (box.north east) {#4};
 \end{tikzpicture}}$\hskip-\tabcolsep}}

\begin{document}

\thispagestyle{empty}
$\,$

\begin{center}
{\large\textbf{Universitat Aut\`onoma de Barcelona}\par}
\vfill
{\huge \bfseries Non-classical correlations in quantum mechanics and beyond \par}
\vfill
{\LARGE by \par}
\smallskip
{\LARGE Ludovico Lami \par}
\smallskip
{\large under supervision of \par}
\smallskip
{\large Prof. Andreas Johannes Winter \par}
\vfill
{\large A thesis submitted in partial fulfillment for the \par}
{\large degree of Doctor of Philosophy \par}
\bigskip
\bigskip
{\large in \par}
{\large Unitat de F\'isica Te\`orica: Informaci\'o i Fen\`omens Qu\`antics \par}
{\large Departament de F\'isica \par}
{\large Facultat de Ci\`encies \par}
\bigskip
\bigskip
\bigskip
{\Large Publicly defended in Bellaterra on 31st October 2017 \par}
\begin{figure}[ht]
  \centering
  \mbox{\hspace{1.2ex} \includegraphics[height=4cm, width=4cm, keepaspectratio]{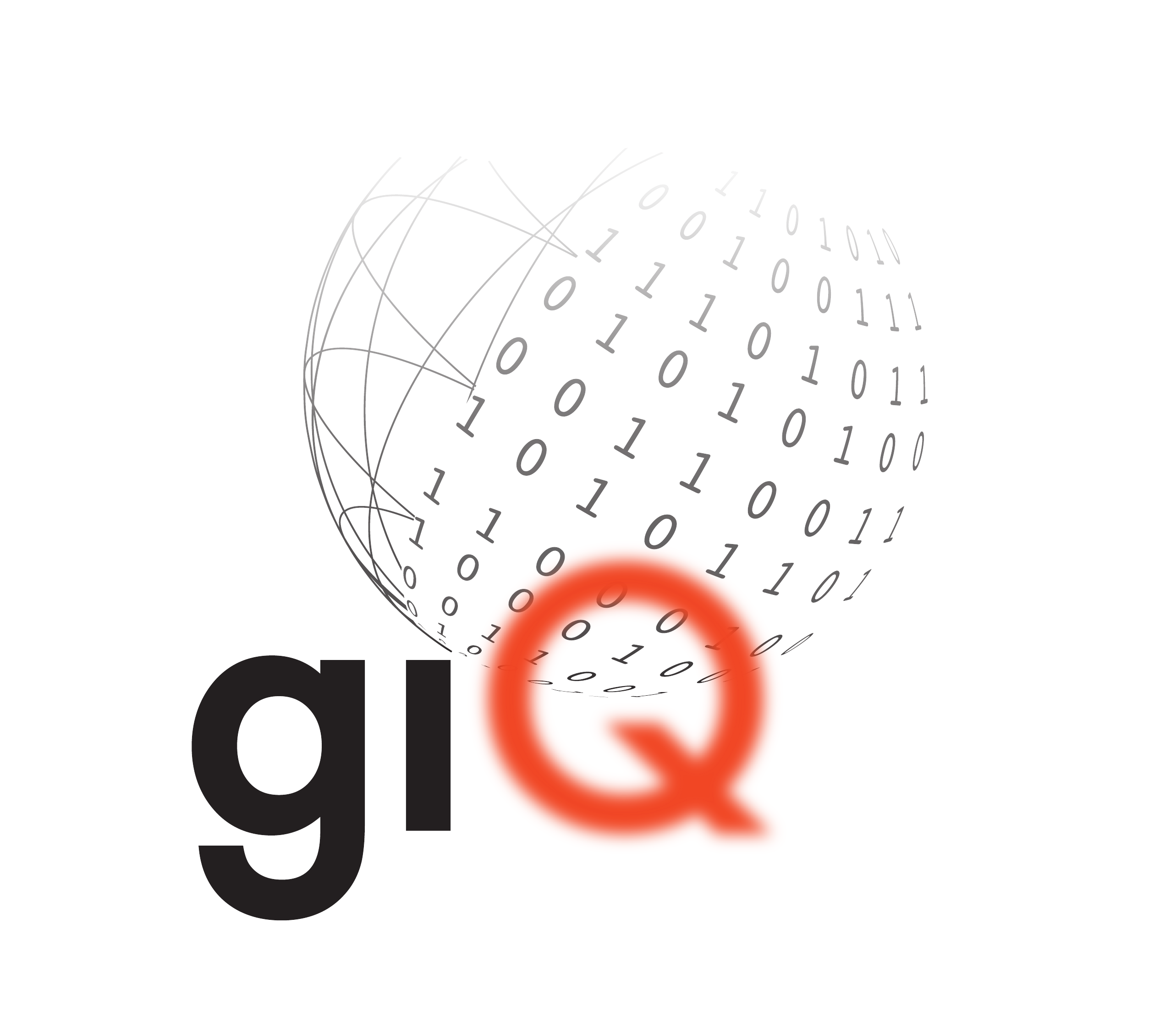}}
\end{figure}
\end{center}

\clearpage
\thispagestyle{empty}
\null
\newpage

\thispagestyle{empty}
\begin{otherlanguage}{italian}
\begin{vplace}[0.7]
\setlength{\epigraphwidth}{180pt}
\epigraph{``Noi ci allegrammo, e tosto torn\`o in pianto, \\
ch\'e della nova terra un turbo nacque \\
e percosse del legno il primo canto. \\[1.2ex]
Tre volte il f\'e girar con tutte l'acque; \\
alla quarta levar la poppa in suso \\
e la prora ire in gi\'u, com' altrui piacque, \\[1.2ex]
infin che 'l mar fu sovra noi richiuso".}{Ulisse \\ La Divina Commedia, Dante Alighieri}
\end{vplace}
\end{otherlanguage}

\cleardoublepage

\begin{abstract}
This thesis is concerned with a seemingly naive question: what happens when you separate two physical systems that were previously together?
One of the greatest discovery of the last century is that systems that obey quantum mechanical instead of classical laws remain inextricably linked even after they are physically separated, a phenomenon known as \emph{entanglement}. 
This leads immediately to another deep question: is entanglement an exclusive feature of quantum systems, or is it common to all non-classical theories? And if this is the case, how strong is quantum mechanical entanglement as compared to that exhibited by other theories?

The first part of the thesis deals with these questions by considering quantum theory as part of a wider landscape of physical theories, collectively called \emph{general probabilistic theories} (GPTs). Chapter~\ref{chapter1} reviews the compelling motivations behind the GPT formalism, preparing the ground for Chapter~\ref{chapter2}, where we translate the above questions into precise conjectures, and present our progress toward a full solution. In Chapter~\ref{chapter3} we consider entanglement at the level of measurements instead of states, which leads us to the investigation of one of its main implications, \emph{data hiding}.
In this context, we determine the maximal data hiding strength that a quantum mechanical system can exhibit, and also the maximum value among all GPTs, finding that the former scales as the square root of the latter.

In the second part of this manuscript we explore some problems connected with quantum entanglement. In Chapter~\ref{chapter4} we discuss its resistance to white noise,
as modelled by channels acting either locally or globally. Due to the limited number of parameters on which these channels depend, we are able to answer all the basic questions concerning various entanglement transformation properties. 
The following Chapter~\ref{chapter5} presents our view on the topic of Gaussian entanglement, with particular emphasis on the role of the celebrated `positive partial transposition criterion' in this context. Extensively employing matrix analysis tools such as Schur complements and matrix means, we present unified proofs of classic results, further extending them and closing some open problems in the field along the way.

The third part of this thesis concerns more general forms of non-classical correlations in bipartite continuous variable systems. In Chapter~\ref{chapter6} we look into Gaussian steering and problems related to its quantification,
moreover devising a general scheme that allows to consistently classify correlations of bipartite Gaussian states into `classical' and `quantum' ones. Finally, Chapter~\ref{chapter7} explores some problems connected with a `strong subadditivity' matrix inequality that plays a crucial role in our analysis of correlations in bipartite Gaussian states. Among the other things, the theory we develop allows us to conclude that a R\'enyi-2 Gaussian version of the elusive \emph{squashed entanglement} coincides with the corresponding entanglement of formation when evaluated on Gaussian states.
\end{abstract}

\newpage

\begin{otherlanguage}{spanish}
\begin{abstract}
Esta tesis versa sobre una cuesti{\'o}n aparentemente na{\'i}f: {?`}qu{\'e} ocurre cuando se separan dos sistemas f{\'i}sicos que estaban juntos previamente? Uno de los mayores descubrimientos del siglo pasado es que los sistemas que obedecen leyes mecano-cu{\'a}nticas, en lugar de cl{\'a}sicas, permanecen ligados inextricablemente incluso tras haber sido separados f{\'i}sicamente, un fen{\'o}meno conocido como \emph{entrelazamiento}. Aqu{\'i} nos preguntamos algo m{\'a}s profundo si cabe: {?`}es el entrelazamiento una caracter{\'i}stica exclusiva de los sistemas cu{\'a}nticos o es com{\'u}n a todas las teor{\'i}as no-cl{\'a}sicas? Y, si es este el caso, {?`}cu{\'a}n fuerte es el entrelazamiento mecano-cu{\'a}ntico comparado con aquel exhibido por otras teor{\'i}as?

La primera parte de esta tesis trata estas cuestiones considerando la teor{\'i}a cu{\'a}ntica como parte de un conjunto m{\'a}s amplio de teor{\'i}as f{\'i}sicas, colectivamente llamadas \emph{teor{\'i}as probabil{\'i}sticas generales} (TPG). En el Cap{\'i}tulo~\ref{chapter1} revisamos la s{\'o}lida motivaci{\'o}n que subyace al formalismo TPG, preparando el terreno para el Cap{\'i}tulo~\ref{chapter2}, donde traducimos las anteriores cuestiones a conjeturas precisas, y donde presentamos nuestro progreso hacia una soluci{\'o}n completa. En el Cap{\'i}tulo~\ref{chapter3} consideramos el entrelazamiento a nivel de medidas en vez de estados, lo cual conduce a la investigaci{\'o}n de una de sus implicaciones principales, la \emph{ocultaci{\'o}n de informaci{\'o}n}. En este contexto, determinamos el m{\'a}ximo poder de ocultaci{\'o}n de informaci{\'o}n que puede exhibir un sistema mecano-cu{\'a}ntico, as{\'i} como el mayor valor entre todas las TPG, hallando que el primero crece como la ra{\'i}z cuadrada del segundo.

En la segunda parte de este manuscrito exploramos algunos de los problemas relacionados con el entrelazamiento cu{\'a}ntico. En el Cap{\'i}tulo~\ref{chapter4} discutimos su resistencia al ruido blanco modelado por canales que act{\'u}an bien local o bien globalmente. Debido al n{\'u}mero limitado de par{\'a}metros de los que dependen estos canales, somos capaces de responder todas las preguntas b{\'a}sicas que conciernen a diversas propiedades de la transformaci{\'o}n del entrelazamiento. En el siguiente Cap{\'i}tulo~\ref{chapter5} presentamos nuestra perspectiva sobre el tema del entrelazamiento gaussiano, con un {\'e}nfasis particular sobre el papel del c{\'e}lebre ``criterio de la transposici{\'o}n parcial positiva'' en este contexto. Empleando extensivamente herramientas del an{\'a}lisis matricial como los complementos de Schur y las medias matriciales, presentamos pruebas unificadas de resultados cl{\'a}sicos, extendi{\'e}ndolos y cerrando algunos de los problemas abiertos en el campo.  

La tercera parte de esta tesis se ocupa de formas m{\'a}s generales de correlaciones no-cl{\'a}sicas en sistemas bipartitos de variable continua. En el Cap{\'i}tulo~\ref{chapter6} estudiamos el ``steering'' gaussiano y problemas relacionados con su cuantificaci{\'o}n, y dise{\~n}amos un esquema general que permite clasificar consistentemente correlaciones de estados gaussianos bipartitos en ``cl{\'a}sicas'' y ``cu{\'a}nticas''. Finalmente, en el Cap{\'i}tulo~\ref{chapter7} exploramos algunos problemas vinculados a una desigualdad matricial de ``subaditividad fuerte'' que desempe{\~n}a un papel crucial en nuestro an{\'a}lisis de las correlaciones en los estados gaussianos bipartitos. Entre otras cosas, la teor{\'i}a que desarrollamos nos permite concluir que una versi{\'o}n R{\'e}nyi-2 gaussiana del escurridizo \emph{squashed entanglement} coincide en estados gaussianos con el correspondiente entrelazamiento de formaci{\'o}n.
\end{abstract}
\end{otherlanguage}

\newpage

\begin{otherlanguage}{catalan}
\begin{abstract}
Aquesta tesis parteix d'una pregunta aparentment ing{\`e}nua: Qu{\`e} passa si es separen dos sistemes f{\'i}sics que estaven en contacte? Un dels descobriments m{\'e}s rellevants del segle passat {\'e}s que els sistemes que obeeixen les lleis de la mec{\`a}nica qu{\`a}ntica, en comptes de les lleis cl{\`a}ssiques, romanen intr{\'i}nsecament connectats fins i tot quan estan separats f{\'i}sicament. Aquest fenomen {\'e}s conegut com entrella\c{c}ament o \emph{entanglement}. Aqu{\'i}, ens preguntem quelcom m{\'e}s profund: pertany l'entrella\c{c}ament exclusivament als sistemes qu{\`a}ntics o {\'e}s com{\'u} a totes les teories no cl{\`a}ssiques? I, donat el cas,  com es pot comparar l'entrella\c{c}ament qu{\`a}ntic amb l'entrella\c{c}ament que pertany a d'altres teories?

La primera part de la tesis tracta amb aquestes q{\"u}estions considerant la teoria qu{\`a}ntica com a part d'un grup m{\'e}s ampli de teories f{\'i}siques anomenat \emph{general probabilistic theories} (GPTs). El Cap{\'i}tol~\ref{chapter1} repassa les motivacions que hi ha darrera el formalisme GPT,  contextualitzant el Cap{\'i}tol~\ref{chapter2}, on plantegem les preguntes mencionades en conjectures formals adjuntant-ne la nostre contribuci{\'o} cap a una soluci{\'o} completa. Al Cap{\'i}tol~\ref{chapter3}, considerem l'entrella\c{c}ament a nivell de mesures i no d'estats, la qual cosa ens porta cap a la investigaci{\'o} d'una de les seves principals implicacions, \emph{data hiding}. En aquest marc, determinem la m{\`a}xima efici{\`e}ncia de el \emph{data hiding} que un sistema qu{\`a}ntic pot exhibir i tamb{\'e} el m{\`a}xim valor entre tots els GPTs, trobant que els primers escalen amb l'arrel quadrada dels darrers.

En la segona part d'aquest manuscrit estudiem alguns problemes relacionats amb l'entrella\c{c}ament qu{\`a}ntic. Al Cap{\'i}tol~\ref{chapter4}, discutim la seva resist{\`e}ncia al soroll blanc, modelitzat amb canals que actuen tant local com globalment. Aquests canals depenen d'un nombre limitat de par{\`a}metres, aix{\`o} fa que siguem capa\c{c}os de respondre totes les preguntes b{\`a}siques relacionades amb les propietats de transformaci{\'o} de l'entrella\c{c}ament.  El Cap{\'i}tol~\ref{chapter5} presenta la nostre visi{\'o} sobre l'entrella\c{c}ament gaussi{\`a}, amb especial focus en el rol del anomenat \emph{`positive partial transposition criterion'} en aquest context. Extensament, fent servir t{\`e}cniques d'an{\`a}lisis de matrius com ara \emph{Schur complements} i \emph{matrix means}, presentem demostracions de resultats cl{\`a}ssics generalitzant-los i resolent algun dels problemes oberts existents en la mat{\`e}ria.

La tercera part de la tesis es basa en formes m{\'e}s generals de correlacions no cl{\`a}ssiques en sistemes bipartits i de variable cont{\'i}nua.  Al Cap{\'i}tol~\ref{chapter6} investiguem el \emph{Gaussian steering} i problemes relacionats en la seva quantificaci{\'o}, aix{\'i} com presentem un esquema general que permeti consistentment classificar correlacions de sistemes bipartits gaussians  en `cl{\`a}ssiques' i `qu{\`a}ntiques'. Finalment, el Cap{\'i}tol~\ref{chapter7} explora alguns dels problemes relacionats amb \emph{strong subadditivity} en desigualtats de matrius que juga un paper clau en el nostre an{\`a}lisis de correlacions en estats gaussians bipartits. Entre d'altres coses, la teoria que desenvolupem ens serveix per concloure que una R{\'e}nyi-2 versi{\'o} gaussiana del dif{\'u}s \emph{squashed entanglement} coincideix amb el corresponent entrella\c{c}ament de formaci{\'o} quan s'avalua en estats gaussians.
\end{abstract}
\end{otherlanguage}

\cleardoublepage

\chapter*{Acknowledgements}

If in these three years in Barcelona I have learnt anything, then my advisor, Andreas Winter, deserves large part of the credit. He has been an example to me, both scientifically and personally, and these few lines will never do justice to his guide. I thank him for having taught me that our personal taste should not determine the science we do more than its significance for real-world situations, and that I should never give in to the temptation of judging any scientist on the basis of the number of papers published or talks given. I am indebted to him for all the opportunities he has given me, all the conferences he has enthusiastically sent me to, and in general for the unconditional freedom I enjoyed during my PhD.

Having Howard Barnum as a referee of this manuscript is an honour I can hardly express in words. Let me just say that the first paper in quantum information I read, studied and loved is his `Largest separable balls around the maximally mixed bipartite quantum state' (in collaboration with L. Gurvits).
I am grateful to Antonio Ac\'in and John Calsamiglia for taking on the thankless task of refereeing this overly long manuscript. In particular, I thank Toni for driving me to Setcases safe and sound, meanwhile holding a conversation in flawless Italian. The countless discussions I had with John in front of a blackboard have taught me a great many things, so I thank him wholeheartedly.

I am beholden to Vittorio Giovannetti, my master's thesis advisor, for educating me in quantum information and setting me on this path. Part of the present thesis was completed while I was in Pisa, and I thank him and his group for their kind hospitality. Let me take the chance to express my gratitude to the Scuola Normale Superiore, my alma mater. I owe to the professors and students I met there much of my love for physics and mathematics, so they deserve a big shout-out.
The quantum optics course held there by Alessio Serafini calls for a special mention, as some of the work I will present here has its roots in those lectures.

I thank Gerardo Adesso for all the support he has given to me since the first time we have met. I am glad and honoured to start a new adventure with him in Nottingham.

I collaborated with many colleagues and friends over the last years. Apart from the already mentioned Andreas, Alessio and Gerardo, let me thank Carlos, Christoph, Fabien, Gl\'aucia, Kaushik, Marco, Marcus, Mark, and Siddhartha. This thesis would not be the same without your contribution.
I am also grateful to Abel and Mar\'ia, who helped me with the translation of the abstract of this thesis into Catalan and Spanish.

Ultimately, the success of a PhD is largely decided by the people you meet. And in my case, I have been as lucky as a person can be. A big shout-out to all GIQers, past and present, for never letting me forget that we are first people and then scientists.
Abel, C\'ecilia, Gianni, Krishna, Marcus, Mar\'ia, Milan, Moha and Sara deserve a special thought for the care they put into making me feel at home during my stay in Barcelona. Abel, as someone said, you are the most Catalan thing in the whole GIQ. C\'eci, how many cakes you baked are physically part of my body right now? Gianni, your psychological terrorism was decisive in making me defeat the dreadful Spanish bureaucratic machine. Krishna, good try, but there is no way you will look like a `brown David' before a substantial amount of exercise. Marcus, I have learnt more on Italian cuisine from you than from anybody else in the world. Mar\'ia, we will manage to attend an opera together, I promise. Moha, I hope you will like being mentioned here, even though this thesis pretends to be written in British English. Milan, I am very sorry to tell you that my Hungarian vocabulary has not gone much beyond a laughable `k{\"o}sz{\"o}n{\"o}m sz{\'e}pen'. Last but not least, Sara, thanks for listening to all my problems and complaints and for hugging me whenever I needed it.
I specially acknowledge support from emblematic institutions such as the late lamented LIQUID, the still burgeoning MAFIA, and the invaluable Graciosos group. 

Among the many people I have met in Barcelona outside academia, let me thank Valeria and Alfredo for the countless vermouths we downed together. Valeria, you have been a precious help in my vain attempt to extend my Sicilian vocabulary. A special thanks goes also to Elisa for having dragged me very confidently into salsa classes despite my apparent incompetence.

Going back to your hometown is always refreshing, but much more when an adorable gang of friends is ready to welcome you. Let me thank Afro, France, Luca, Mola and Ska for their unshakable trust in my capability of completing the PhD. The often legendary discussions I had with Mola and Afro on a variety of absurd topics have been a constant source of inspiration.
The friendship with the much more charming Arianna, Camilla and Diletta has been invaluable to me during these years abroad. I thank you wholeheartedly for all the strolls we took together.

I am grateful to Alvise, Enza, Franceschina and Piro for never leaving me alone when I felt disheartened. Alvise, trying to keep up with your moral integrity has been all but easy, but I tried my best.
Finally, I hope to find the right words to thank Francesca and Benedetta for being so close to me while I was in Barcelona, despite the long distance. Francesca, I enjoyed all the debates we set up on classical music, even if I still firmly believe Mozart's music to be better than anything else. Benedetta, if I had to properly express my gratitude for all you have done for me during these years, our afternoon tea would last forever. 

Without the everlasting love and care of my family this thesis would have never been completed, so I thank my parents and my brother Guglielmo from the bottom of my heart. Gug, I hope this will encourage you to pursue your dreams in spite of all difficulties.
My grandparents nonno Manrico and Grazia believed in me since I was little and have been a constant source of inspiration for me.

My final thought goes to Margherita, who made this whole endeavour possible through her sincere love, which I cherish beyond words.

\cleardoublepage

\tableofcontents
 
\mainmatter

\cleardoublepage

\chapter*{Introduction}
\addcontentsline{toc}{chapter}{Introduction}

When I started the PhD in 2014, the discovery of nonlocality as an intrinsic feature of quantum mechanical experimental predictions was exactly 50 years old. Half a century, half an ideal human life~\cite{COMMEDIA}, and we are still adding footnotes to the solution of the Einstein-Podolsky-Rosen paradox~\cite{EPR} given by John Stewart Bell in 1964~\cite{Bell}. And for how obscure and counterintuitive the laws of Nature may look, Bell's reasoning continues to show us the right way to grasp their deepest meanings.

Before we elaborate more on the contents of this thesis and on the way they reflect our standpoint on the topic, let us go back a bit and start with a brief historical note. The famous 1935 paper by Einstein, Podolsky and Rosen (EPR)~\cite{EPR} triggered an animated debate in the burgeoning community of quantum physicists. Bohr~\cite{Bohr} tried to answer some of the questions raised there, however not fully appreciating their foundational status. Schr\"odinger~\cite{schr} was struck by some of the implications of the EPR paper, but found it hard to believe that these corresponded to real physics.
It took almost 30 years to focus the problem, but finally, in 1964, John Stewart Bell~\cite{Bell} published his stunning answer. This answer, nowadays known under the name of \emph{Bell's theorem}, is one of the most profound discoveries in the history of science, and reveals us a fundamental feature of Nature that had eluded our previous investigations: \emph{if the experimental predictions of quantum mechanics are correct, no local hidden variable model can account for the behaviour of bipartite systems.} The point we want to stress here is that this is not a statement about the \emph{theory} of quantum mechanics, but rather concerns only its \emph{predictions}, which have proved to be correct in all situations where we have tested them, and in particular in the setting described by EPR~\cite{Aspect}. In other words, Bell's theorem forces us to rethink not our mathematical theory of reality, but rather our ontological view of reality itself. 

This last observation motivates our point of view on the investigation of the phenomena connected with bipartite physical systems, and in particular of \emph{entanglement}. A question that we find worth asking is as follows: \emph{to what extent do said phenomena depend on the mathematical description of the systems under examination?} In order to formulate more precisely the question, we need a framework that categorises physical theories in a systematic way. Theories belonging to such framework -- an example of which must be, naturally, quantum mechanics -- will encompass only the most basic physical requirements and allow only for the most basic predictions. Ideally, our formalism should not include anything that is not strictly necessary to make sense of the questions we want to ask. What we just described is the realm of \emph{general probabilistic theories} (GPTs). As we shall see, even in its most basic form, this formalism is rich enough to allow for the investigation of a wide variety of phenomena. 

The title of the thesis should be a bit less obscure by now. Our main subject of study is in fact the appearance, particularly in bipartite systems, of correlations that evade any `classical' explanation. While quantum systems and quantum phenomena still play a prominent role in our work, particular emphasis is laid on the `post-quantum' case, implying that all physical theories are treated on (a priori) equal footing, and a relevant part of our efforts are devoted to identifying their \emph{universal} behaviours. The manuscript is organised as follows.

The first part deals with general probabilistic theories, introducing the formalism and discussing within this framework some specific problems connected with the question raised above. More in detail, in Chapter~\ref{chapter1} we lay the foundations of GPTs, reviewing some compelling derivations of their formalism from first principles. Our main aim there is to provide a fully self-contained proof of one of the main foundational results in the field, the embedding theorem by Ludwig~\cite{LUDWIG}. The purpose of this theorem is to start from elementary and intuitive assumptions concerning the physical system under examination and our ability to access it through measurements, and to translate them into a convenient mathematical picture, which will constitute the arena where our questions find an adequate formulation.

This mathematical picture features a so-called \emph{base norm Banach space}, a particular type of ordered Banach space whose norm is closely related to its order relation. The main difficulty in the proof is that in general there is no reason to assume said space to be finite-dimensional. We review Ludwig's successful attempt to overcome this technical hurdle, which requires us to discuss some aspects of the theory of Banach spaces, with increasing specialisation to ordered Banach spaces and eventually to base norm and order unit spaces. Along the way, many classic results in the field are rederived. 
We believe that our review of this topic constitutes a valuable contribution, especially as we strove to give a unified and coherent presentation of the material dispersed over several original papers that are not always easy to read. For certain points in the proof of Ludwig's theorem we give alternative arguments, which may be helpful to the non-expert on this topic, and may be of independent interest.

In Chapter~\ref{chapter2} we review the basics of the GPT formalism in the simplified finite-dimensional setting, extending it as to include bipartite systems, and exploit its full power to ask the following question: \emph{is the appearance of entangled states logically implied when one joins two non-classical systems?} Here, a GPT is called `non-classical' if there is a set of pure states (pure meaning that they can not be obtained by any nontrivial probabilistic mixture) that is linearly dependent, in the sense that the corresponding expectation values on any observable are linearly dependent. A state of a bipartite system is called `entangled' if it can not be prepared with local operations and classical communication starting from uncorrelated states.

In this context, the question we ask translates to a precise mathematical problem concerning convex cones: \emph{is the maximal tensor product of any two non-simplicial cones always strictly larger than their minimal tensor product?} We conjecture that this is indeed the case, and present substantial evidence that supports this view. In particular, we prove the conjecture in the particular case when both cones have a centrally symmetric section. On a different line, we also consider a reformulation of the above question, where one looks at nonlocality rather than entanglement. Mimicking a classic argument by Namioka and Phelps~\cite{NP}, we show that the combination of any non-classical theory with a `gbit' (a model whose state space is a square) gives rise to states that are not only entangled but that can even violate the simplest Bell inequality, i.e. the CHSH inequality. Along the way, we develop tools to tackle the separability problem in arbitrary GPTs.

In Chapter~\ref{chapter3} we shift the focus of our analysis from entangled states to entangled measurements. One of the most striking consequences of the existence of global measurements that can not be performed locally is the phenomenon of \emph{data hiding}, i.e. the possibility of hiding a bit of information in a state discrimination query that two distant parties are unable to resolve unless they have access to joint measurements, even when classical communication is available for free~\cite{dh-original-1, dh-original-2}. We introduce a natural figure of merit for this task, called efficiency, and then ask ourselves: \emph{how high can the efficiency of quantum data hiding be, given the two dimensions of the local subsystems?} And more generally, \emph{what is the maximum efficiency an arbitrary GPT can exhibit, for fixed local dimensions?}

The answers to these questions, given here in full generality and up to a small universal constant (multiplicative in the first case, additive in the second), constitute two main results of the present thesis. Perhaps surprisingly, it turns out that the maximal quantum data hiding efficiency scales as the square root of the absolute maximum among all GPTs, which is instead achieved for other kinds of highly symmetric theories, namely those whose state space is a sphere.
Before we move on to the next chapter, few words about the proof techniques.
The argument we devise for the quantum case is surprisingly elementary, as it rests only on the celebrated teleportation protocol, yet it is so powerful that we are able to deduce a number of interesting corollaries from it, mainly leveraging results of~\cite{faithful}. These include the best known lower bound on the squashed entanglement~\cite{squashed} of a state in terms of its trace distance from the separable set, as well as a new algorithm for the separability problem whose running time scales only polynomially in the size of the larger subsystem.
In facing the case of arbitrary GPTs, which is significantly more involved, we uncover an unexpected connection with the branch of functional analysis that studies \emph{tensor norms}. The solution of the problem in that setting rests on the determination of the largest ratio between projective and injective tensor norms that can be achieved by any pair of Banach spaces of fixed dimensions, a result that may be of independent interest.

The second part of the present thesis focuses on some selected aspects of quantum entanglement, both in finite-dimensional and continuous variable systems. In Chapter~\ref{chapter4} we introduce a class of maps that are natural extensions of the well-known depolarising channel to the case of a bipartite input system. The operational significance of these maps, parametrised by three variables, rests on their capability of modelling both local and global white noise acting on a shared quantum state. In this context, we analyse all basic properties connected with entanglement transformation, thus determining the parameter regions for which the maps are: (a) positive; (b) completely positive; (c) entanglement-breaking, meaning that they break any entanglement that was previously shared between the system and the rest of the world; and (d) entanglement-annihilating, in the sense that any input (bipartite) state is transformed into a separable state. This latter characterisation allows us to close some open problems that have been raised in recent literature~\cite{EA3, EA4}. Along the way, we develop new tools to tackle the quantum separability problem, harnessing the properties of the Hadamard product in this context.

Chapter~\ref{chapter5} is devoted to Gaussian entanglement, i.e. entanglement of Gaussian states in continuous variable quantum systems. Due to the well-known fact that these states are in one-to-one relationship (up to local unitaries) with special positive definite matrices that we dub \emph{quantum covariance matrices}, we advocate the general strategy of using matrix analysis tools to solve problems on Gaussian states. Two tools are found to be specially useful in our approach, namely \emph{Schur complements} and \emph{matrix means}.
Extensively exploiting these techniques, we present a unified proof of the equivalence between separability and positivity of the partial transpose (so-called PPT condition) for Gaussian states of $1$ vs $n$ modes, for arbitrary $n$~\cite{Simon00, Werner01}. This constitutes a notable improvement over previous works.
Besides rederiving all classic results establishing said equivalence in various other special cases, we extend their validity considerably. For instance, we prove that bipartite Gaussian states that stay PPT under the application of any passive operations are necessarily separable, which closes an old open problem~\cite{passive}.

The third and last part of the manuscript is devoted to the study of other forms of non-classical correlations, especially in the continuous variable setting. In Chapter~\ref{chapter6} we look into \emph{steering}, a notion which lies between entanglement and nonlocality, laying emphasis on the case of Gaussian steering, i.e. steering of Gaussian states by Gaussian measurements. 
We start by showing that the Schur complements of quantum covariance matrices obey a matrix inequality whose structure resembles that of strong subadditivity of quantum entropy. Leveraging this result, we prove that a certain quantifier of Gaussian steerability based on the covariance matrix of the state~\cite{steerability} is an actual monotone of the corresponding resource theory, and thus a fully-fledged Gaussian steerability measure. The focus is then shifted towards the problem of `classifying' correlations of bipartite Gaussian states into classical and quantum. In the approach of~\cite{LiLuo}, this can be done in a consistent way if the quantifier of quantum correlations does not exceed \emph{half} the quantifier of total correlations. Using the theory of matrix means, we prove that the R\'enyi-2 Gaussian entanglement of formation and the corresponding R\'enyi-2 mutual information do satisfy this constraint when evaluated on Gaussian states. Monogamy and additivity of the former measure are immediately deduced from this result.

The investigation of these problems is pursued in the final Chapter~\ref{chapter7}. There, we examine some questions related to the strong subadditivity matrix inequality of Chapter~\ref{chapter6}. Exploring some connections with probability theory on the one hand and with matrix analysis on the other hand, we provide several improvements of the inequality, determining in particular several remainder terms from which the saturation conditions are easily read. Subsequently, we consider an entanglement measure for Gaussian states that is very similar in spirit to the \emph{squashed entanglement}~\cite{squashed}, with the important differences that: (i) the R\'enyi-2 entropy displaces the standard von Neumann entropy; and (ii) the extensions of the state are required to be Gaussian as well. Surprisingly, it can be shown that -- unlike in the finite-dimensional case -- this R\'enyi-2 Gaussian squashed entanglement coincides with the corresponding R\'enyi-2 Gaussian entanglement of formation on all Gaussian states. Among its many advantages, this remarkable result makes the numerical computation of the former measure possible, since it restricts the size of the extension from unbounded to bounded.

The various chapters of this thesis are based on the following publications or preprints.
\begin{itemize}

\item \textbf{Chapter~\ref{chapter3}.} L. Lami, C. Palazuelos, and A. Winter. Ultimate data hiding in quantum mechanics and beyond. \emph{Preprint arXiv:1703.03392}, 2017. To appear in \emph{Commun. Math. Phys.}

\item \textbf{Chapter~\ref{chapter4}.} L. Lami and M. Huber. Bipartite depolarizing maps. \emph{J. Math. Phys.}, 57(9):092201, 2016.

\item \textbf{Chapter~\ref{chapter5}.} L. Lami, A. Serafini, and G. Adesso. Gaussian entanglement revisited. \emph{New J. Phys.}, 20(2):023030, 2018.

\item \textbf{Chapter~\ref{chapter6}.} L. Lami, C. Hirche, G. Adesso, and A. Winter. Schur complement inequalities for covariance matrices and monogamy of quantum correlations. \emph{Phys. Rev. Lett.}, 117(22):220502, 2016.

\item \textbf{Chapter~\ref{chapter7}.} L. Lami, C. Hirche, G. Adesso, and A. Winter. From log-determinant inequalities to Gaussian entanglement via recoverability theory. \emph{IEEE Trans. Inf. Theory}, 63(11):7553--7568, 2017.

\end{itemize}

The first two chapters contain substantially new material, which we intend to publish soon. Finally, the papers that I have contributed to write during my PhD but that are not quite covered in the present thesis are the following.

\begin{itemize}

\item M.G. Genoni, L. Lami, and A. Serafini. Conditional and unconditional Gaussian quantum dynamics. \emph{Contemp. Phys.} 57(3):331--349, 2016.

\item F. Clivaz, M. Huber, L. Lami, and G. Murta. Genuine-multipartite entanglement criteria based on positive maps. \emph{J. Math. Phys.} 58(8):082201, 2017.

\item L. Lami, S. Das, and M.M. Wilde. Approximate reversal of quantum Gaussian dynamics. \emph{J. Phys. A}, 51(12):125301, 2018.

\item K.P. Seshadreesan, L. Lami, and M.M. Wilde. R{\'e}nyi relative entropies of quantum Gaussian states. \emph{Preprint arXiv:1706:09885}, 2017.

\end{itemize}

We conclude with a few remarks on the organisation of the manuscript. The first section of every chapter starts with few sentences that introduce the topic, followed by a quick overview of the structure of the chapter. Then there are two subsections: the first one introduces the problem in broad terms, while the second points the reader to our main original contributions. Chapter~\ref{chapter1} presents an additional subsection that aims at sketching the history of general probabilistic theories and giving a concise list of pertinent references.

We are now ready to set out, hoping that the reader will follow us on this journey.

\part{Non-classicality in general probabilistic theories}

\chapter{Foundations of general probabilistic theories} \label{chapter1}

\section{Introduction} \label{sec intro}

The glory of quantum mechanics penetrates the whole universe, and shines more brightly when one looks at it from certain perspectives, from others less. Throughout this chapter, we are going to take one of those points of view, from which we hope its brilliance to be more apparent. Our journey to this standpoint will lead us to seeing quantum mechanics as a part of a rich landscape whose landmarks are physical theories on their own, generically called General Probabilistic Theories (GPTs). This is profoundly instructive for at least two reasons. First, it will help us discerning the various peculiarities of quantum mechanics that distinguish it from a classical theory, and establishing a hierarchy among them. In fact, from this point of view it turns out that not all the surprising quantum effects we have been studying and understanding with difficulty throughout the past century should bewilder us to the same extent. Secondly, once this firm foundation has been established, we can turn our attention to the problem of singling out quantum mechanics by means of certain additional assumptions that are believed to be crucial to the construction of a complete physical theory. This latter approach seems to be one of the main reasons that has encouraged many scholars to investigate GPTs, and while it is certainly commendable and conceptually sound, the point of view we are going to adopt here is somewhat closer to the former.

The content of the present chapter is organised as follows. The rest of this section is devoted to sketching the history of the field (Subsection~\ref{subsec history GPTs}), to introducing the basic framework (Subsection~\ref{subsec prep measur}) and to discussing what distinguishes our approach from that taken by others and highlight our modest original contribution (Subsection~\ref{subsec path}). Sections~\ref{sec ord}--\ref{sec ord unit base norm} are intended to acquaint the reader with the mathematical tools that will be used later on. More in detail, we will first present the notions of ordered vector space (Section~\ref{sec ord}), of topological vector space (Section~\ref{sec topo}), and of Banach space (Section~\ref{sec Banach}), then combine them to study ordered Banach spaces in general (Section~\ref{sec ord Banach}) and two special cases known as order unit and base norm spaces (Section~\ref{sec ord unit base norm}). The reader who is already familiar with the above notions can jump straight away to Section~\ref{sec Ludwig}, where we apply all this mathematics to present the proof of Ludwig's embedding theorem, which lays the foundation of the so-called abstract state space formalism for GPTs.

\subsection{Brief history of General Probabilistic Theories} \label{subsec history GPTs}

As part of our introduction to GPTs, here we intend to give a brief account of the development of the field. The list of references we provide is by no means complete, since especially in recent years there has been much activity in trying to extend the framework to study all sort of problems. However, we hope that the reader will benefit from our glance into the history of GPTs.

As usual, drawing the starting line is a bit arbitrary, since in modern times every scientist is influenced by many others. That being said, the first author to develop an axiomatic approach to quantum mechanics that ultimately led to the formalisation of GPTs was perhaps Mackey~\cite{MACKEY} at the beginning of the Sixties. In his book, the prepare-and-measure scheme, the associated probability function and the axioms to be imposed on the latter are introduced and discussed. Seemingly independently, Ludwig~\cite{Ludwig-1} worked on a similar framework, further developed in a series of papers published in Communications in Mathematical Physics between 1967 and 1972 by Ludwig, D\"ahn and Stolz~\cite{Ludwig-2,Ludwig-3,Daehn-4,Stolz-5,Stolz-6,Ludwig-7}.

The authorship of the so-called `abstract state space formalism' that has nowadays become standard for discussing GPTs is hard to attribute. On the one hand, the first proof that the experimentally accessible probability function can be reconstructed by evaluating positive functionals on a subset of some Banach space seems to go back to Ludwig. The content of this crucial statement, which we report here as Theorem~\ref{Ludwig emb thm} together with a hopefully more accessible and simplified proof, reached its final form in~\cite{LUDWIG-DEUTUNG} (see also~\cite{LUDWIG}) after being developed throughout several papers~\cite{Ludwig-1,Ludwig-2,Ludwig-3}. 
On the other hand, in the very same year 1970 Davies and Lewis published a paper~\cite{Davies-1970} where they devise such a Banach space formalism in its modern form and use it as a starting point.

This latter paper triggered an animated debate during the next few years, and quick advances were made in improving the basics of the framework~\cite{Edwards-operational,Edwards-monotone,Edwards-simple-observables,Gudder-convex-structure}, expanding it as to encompass state transformations~\cite{Edwards-operations,Edwards-pure-operations}, and connecting it to previous approaches~\cite{Mielnik-filters,Daehn-algebra-filters,Daehn-symm-modularity}. More or less in the same years, Mielnik showed that this class of models is general enough as to support non-linear extensions of quantum mechanics, as could result for instance from a hypothetical inclusion of gravity in the picture~\cite{Mielnik-general-quantum}. From our point of view, a particularly important progress that was made several years later is the discussion of composite systems~\cite{tensor-rule-1, tensor-rule-2} (see also~\cite{telep-in-GPT} for a generalisation to the multipartite case). 

All the aforementioned authors made ample use of the mathematical theory of ordered topological vector spaces, whose development in the previous years is summarised in~\cite{SCHAEFER, PERESSINI, JAMESON}. Two particular classes of ordered topological (in fact, Banach) spaces quickly became central to the formalism, namely order unit spaces and base norm spaces, introduced by Ellis~\cite{Ellis-dual-base,Ellis-66} and Edwards~\cite{Edwards-base-norm}, respectively. 

In more recent times, the framework of general probabilistic theories has been used for instance to investigate non-locality phenomena~\cite{PR-boxes, nonloc-resource, Barrett-original, nonloc-polygon}, communication complexity questions~\cite{implausible, nonloc-comm-complex, PVV, Barrett-original, Lee-computation}, the complications arising from introducing post-measurement collapses in the picture~\cite{Pfister-no-disturbance,Pfister-Master}, and the status of the purification postulate~\cite{Chiribella-pur}. Several concepts that played a central role in the development of quantum information have been revisited within the GPT realm, such as teleportation~\cite{telep-in-GPT} and broadcasting~\cite{Barnum-no-broad} protocols (see also~\cite{Howard-info-processing}), steering~\cite{Barnum-steering}, entropy~\cite{Howard-entropy}, and entanglement~\cite{Jencova2017}, especially in connection with thermodynamics~\cite{ent-therm-GPT}.

On a different line, we already mentioned how the problem of deriving quantum theory from few reasonable axioms has attracted much attention since the very early days of the field. In fact, already von Neumann and co-workers looked into it from the algebraic point of view, and reconstructing quantum theory was one of the main motivations behind the works of Mackey~\cite{MACKEY} and Ludwig~\cite{LUDWIG-DEUTUNG,LUDWIG,LUDWIG-FOUNDATIONS-1} themselves. We will not attempt to sketch a history of the problem here, and for details on early attempts to solve it we refer the reader to~\cite{Gunson1967}. The advent of general probabilistic theories has seen a widespread renewal of interest in this kind of questions, and several new approaches were put forward in recent years. For the sake of the presentation, let us stick to the simplistic idea that these attempts differ from each other in the somewhat decisive axiom that singles out quantum theory at last, separating it from classical probability theory. In~\cite{Hardy2001, Schack2003, Masanes2011, quantum-info-unit} this is the existence of (continuous) reversible transformations among pure states, while in~\cite{Chiribella-info-der} an analogous role is played by the assumption that every state can be purified (in an almost unique way). The existence of a `symmetric faithful state' lies at the heart of the treatment given in~\cite{DAriano-1, DAriano-2}, while an algebraic postulate whose interpretation is not completely transparent does the job in~\cite{Wilce-4-1/2-axioms}. A somewhat different approach has been pursued in~\cite{no-3-order}, where more exotic theories are excluded based on the non-existence of correlations of `order' higher than two.

\subsection{Preparing and measuring procedures} \label{subsec prep measur}

The starting point of the reasoning that leads us to the study of general probabilistic theories is a lesson we learnt in a very early stage of the quantum revolution: \emph{our basic requests from a physical theory should not go much beyond a set of rules that allow us to deduce a probabilistic prediction of the outcome of an experiment from the detailed description of its preparation}. We do not claim that this is necessarily all we should expect from a physical theory, but rather that these are the minimal requirements for something to be considered an acceptable physical theory. 

Here we do not intend to delve into a complete discussion of the many subtleties incident to such an approach, for which we rather refer the reader to Ludwig's analysis~\cite{LUDWIG-FOUNDATIONS-1}. Let us just stress a couple of points. First, the probabilities themselves are not directly accessible experimentally. Under the hypothesis of repeatability of independent trials, what is accessible are the relative frequencies, that will converge to the abstract probabilities only in the limit. This aspect must be taken into account in the mathematical formulation of the theory. In Ludwig's approach, this is done by means of an additional `uniform structure of physical imprecisions'. Secondly, any axiomatisation of a physical theory that follows the above scheme requires an additional pre-theory to describe the experimental apparatus. Describing a quantum experiment of Stern-Gerlach type, for instance, requires a prior knowledge of electromagnetism. We now turn to detailing our pragmatic view of a physical experiment and of the associated theory.

In a somewhat simplistic view, an experimental setting consists in being given some physical system, say a laser, and some apparatus to measure it, say a set of lenses, mirrors and photon counters. An experiment will then consist in choosing a way of preparing the system and a configuration of the measuring apparatus. The set of all possible preparation procedures, also called \textbf{states}, will be denoted by $\Omega$. As for the measuring apparatus, in the era of digital devices it is clear that we can assume with basically no loss of generality that its output consists in a sequence of binary numbers (yes/no). Measuring the system means recording the yes/no pattern of a particular experiment. Therefore, it makes sense to define an \textbf{effect} as a configuration of the measuring apparatus together with a yes/no pattern that belongs to it. We will call $\Lambda$ the set of all possible effects.
From this point of view, and when restricted to this particular setting, a physical theory is nothing but a probability function $\mu:\Lambda \times \Omega\rightarrow [0,1]$, that associates a probability $\mu(\lambda, \omega)$ to the event of recording the effect $\lambda$ when measuring the system that has been prepared according to $\omega$.

Naturally, this function $\mu$ must be subjected to some nontrivial constraints if we want it to represent a physically realistic scenario. First of all, we can assume that we already made the preliminary step of identifying all preparations that were (even probabilistically!) indistinguishable from each other by all measurements, i.e. such that they always yielded the same probabilities when the same effect was measured. The same is true for the effects: we preliminarily chose to identify all effects that yielded the same probabilities when evaluated on the same states. This allows us to assume the following.

\begin{axiom} \label{ax separated}
States that are not distinguishable by any effect are to all intents and purposes the same state, and vice versa: mathematically, one says that $\mu$ separates points of both $\Omega$ and $\Lambda$, i.e. that for all distinct $\omega_1,\omega_2\in \Omega$ there is $\lambda\in \Lambda$ such that $\mu(\lambda,\omega_1)\neq \mu(\lambda,\omega_2)$, and vice versa that for all distinct $\lambda_1,\lambda_2\in \Lambda$ there is $\omega\in \Omega$ such that $\mu(\lambda_1,\omega)\neq \mu(\lambda_2,\omega)$.
\end{axiom}

Next, we can posit that the effects with certain outcomes `always yes' and `always no' are among those that we are allowed to consider, and that negating a valid effect yields another valid effect. This is the content of the following axiom.

\begin{axiom} \label{ax extreme effects}
$\Lambda$ contains the effects `always accept' and `never accept': mathematically, there are $u\in \Lambda$ (also called the \textbf{unit effect}) and $0\in \Lambda$ such that $\mu(u,\omega)=1$ and $\mu(0,\omega)=0$ hold for all states $\omega\in \Omega$. Furthermore, for all $\lambda\in \Lambda$ there is $\lambda'\in\Lambda$ such that $\mu(\lambda',\omega)=1-\mu(\lambda,\omega)$ for all $\omega\in\Omega$.
\end{axiom}

Finally, we want to be able to incorporate all pre- and post-randomisations of the preparations and of the outcomes in the picture. In other words, any probabilistic mixture of accessible preparations must be also accessible, and the same for the effects. We state this as follows.

\begin{axiom} \label{ax mixtures}
One can perform probabilistic mixtures of states (or effects) and obtain a valid state (or effect); this means that for all $\omega_1,\omega_2\in \Omega$ and $p\in [0,1]$ there exists $\tau\in \Omega$ such that $\mu(\lambda, \tau)=p\mu(\lambda,\omega_1)+(1-p)\mu(\lambda,\omega_2)$ for all $\lambda\in \Lambda$, and vice versa that for all $\lambda_1,\lambda_2\in \Lambda$ and $q\in [0,1]$ there exists $\eta\in \Lambda$ such that $\mu(\eta,\omega)=q\mu(\lambda_1, \omega)+(1-q)\mu(\lambda_2, \omega)$ for all $\omega\in \Omega$.
\end{axiom}

The axiomatic picture we just sketched is certainly convincing from the foundational perspective, but it has at least two annoying deficiencies. First, it lacks any geometrical interpretation, and this makes the discussion of even basics concepts a lot more cumbersome. As an example, we invite the reader to formulate the notion of pure state in the above language. Secondly, and perhaps more importantly, the seemingly poor structure with which the sets $\Omega$ and $\Lambda$ are endowed makes the picture totally inconvenient for computations. From a more pragmatic point of view, we could also add to this list the fact that since to this point there is no apparent connection between our framework and any well-studied mathematical theory, we will have a hard time employing standard mathematical tools to tackle our problems.

Fortunately, these deficiencies are really just superficial, and there is an equivalent way of formulating the above framework that fixes all of them at once. This alternative formulation comes in the form of an embedding theorem that allows us to see $\Omega$ and $\Lambda$ as convex subsets of a Banach space and of its dual (respectively), with the probability function $\mu$ being given by the canonical bilinear form (for details, see Theorem~\ref{Ludwig emb thm}). Ultimately, the purpose of this operation is to translate simple and easily understandable physical axioms into a convenient mathematical picture that constitutes the arena of GPTs. To the best of our knowledge, the first to present a complete proof of this remarkable fact -- i.e. a proof that does not rely on any further assumption such as the finite dimensionality of the involved linear spaces -- was Ludwig~\cite[IV \S\S 1, 3 and 4]{LUDWIG} (see also~\cite{LUDWIG-DEUTUNG}), and for this reason we name it \emph{Ludwig's embedding theorem}.

\subsection{The path to Ludwig's embedding theorem} \label{subsec path}

Ludwig's original proof~\cite[IV \S\S 3 and 4]{LUDWIG} of Theorem~\ref{Ludwig emb thm} is deep and elegant, but arguably a bit hard to read from a modern perspective, and even obscure in certain passages. In my personal view, this difficulty comes primarily from the heterogeneity of the mathematics he employs, and secondly from the sometimes bizarre notation he adopts. In particular, there the reader is assumed to have a strong background in basic topology, and to a lesser extent also in the theory of ordered Banach spaces. The result is a proof that is almost 17 pages long, the last 6 pages being an application (with proofs) of the duality theory of base norm and order unit spaces as developed by Edwards~\cite{Edwards-base-norm} and Ellis~\cite{Ellis-dual-base}. Since Ludwig's theorem is perhaps the foundational result of the field of general probabilistic theories, and its full proof is rather unknown, we believe it important to discuss it here in its entirety. We want to make our small contribution to Ludwig's cause by rewriting his proof the way we understood it, hoping that this will make it more accessible to a broader audience.

In our view, the strength of our presentation lies primarily in the fact that we will not assume any prior knowledge of anything but basic linear algebra. The relevant concepts and results in topology and Banach space theory are discussed and proved in the preliminary Sections~\ref{sec ord}--\ref{sec Banach}. We will state without proofs only few standard results that the reader can find in any textbook (Lemmas~\ref{Bourbaki lemma},~\ref{neigh 0 Banach} and Theorems~\ref{Hahn-Banach},~\ref{UBP},~\ref{Banach-Alaoglu}), and anyway appropriate references will be provided. Then, we move on to discussing the theory of ordered Banach spaces (Section~\ref{sec ord Banach}) and in particular of order unit and base norm spaces (Section~\ref{sec ord unit base norm}). We will spend some time proving from first principle important results like Theorem~\ref{GKE thm} on the duality of cones and Theorem~\ref{EE thm} on the duality between base norm and order unit spaces. Here we make no claim of originality, since the theory is well-known since the Sixties. Rather, we thought of improving the presentation by collecting all the results that are relevant to the proof of Ludwig's theorem in a single place. A similar operation had already been partially attempted in the series of lecture notes~\cite{FOUNDATIONS}, which was of inspiration to us.

With all the mathematical tools already developed in a systematic way in the preceding sections, we can finally delve into the proof of Theorem~\ref{Ludwig emb thm} in Section~\ref{sec Ludwig}. Most of the proof is a sequence of ingenious applications of elementary concepts, and we limit ourselves to spelling out some elementary steps that are missing in Ludwig's original presentation, for instance the proof of Proposition~\ref{R Banach prop} or the reasoning in Subsection~\ref{subsec uniqueness}, and to prove more directly some facts like Proposition~\ref{intersection symmetric prop}. However, there are two places where more advanced mathematical tools come into play, the first one being Subsection~\ref{subsec order structure}, where Theorem~\ref{EE thm} is applied to shorten Ludwig's reasoning significantly, and the second one being Subsection~\ref{subsec weak*-denseness}, where we are forced to invoke several nontrivial results in topology and Banach space theory. We tried to simplify this latter part by proposing a different and perhaps (depending on the personal taste) more elementary proof of Proposition~\ref{weak* closed are symmetric prop}. Finally, even if not strictly necessary for completing the proof, we provided an alternative and completely elementary reasoning to show the uniqueness part of the claim of Theorem~\ref{Ludwig emb thm} in Appendix~\ref{app Ludwig}.

As the reader might have guessed, the present chapter presents a series of results that are highly interconnected and ultimately work towards proving Theorem~\ref{Ludwig emb thm}. We tried to summarise the network of interdependencies by means of the following graph, whose vertices are lemmas, propositions, theorems and corollaries of this chapter, and whose directed edges represent logical connections. The reader can use it to choose what part of the preliminary mathematics he/she wants to gain a deeper knowledge of, if needed.

\begin{figure}[ht]
  \centering
  \includegraphics[height=12cm, width=12cm, keepaspectratio]{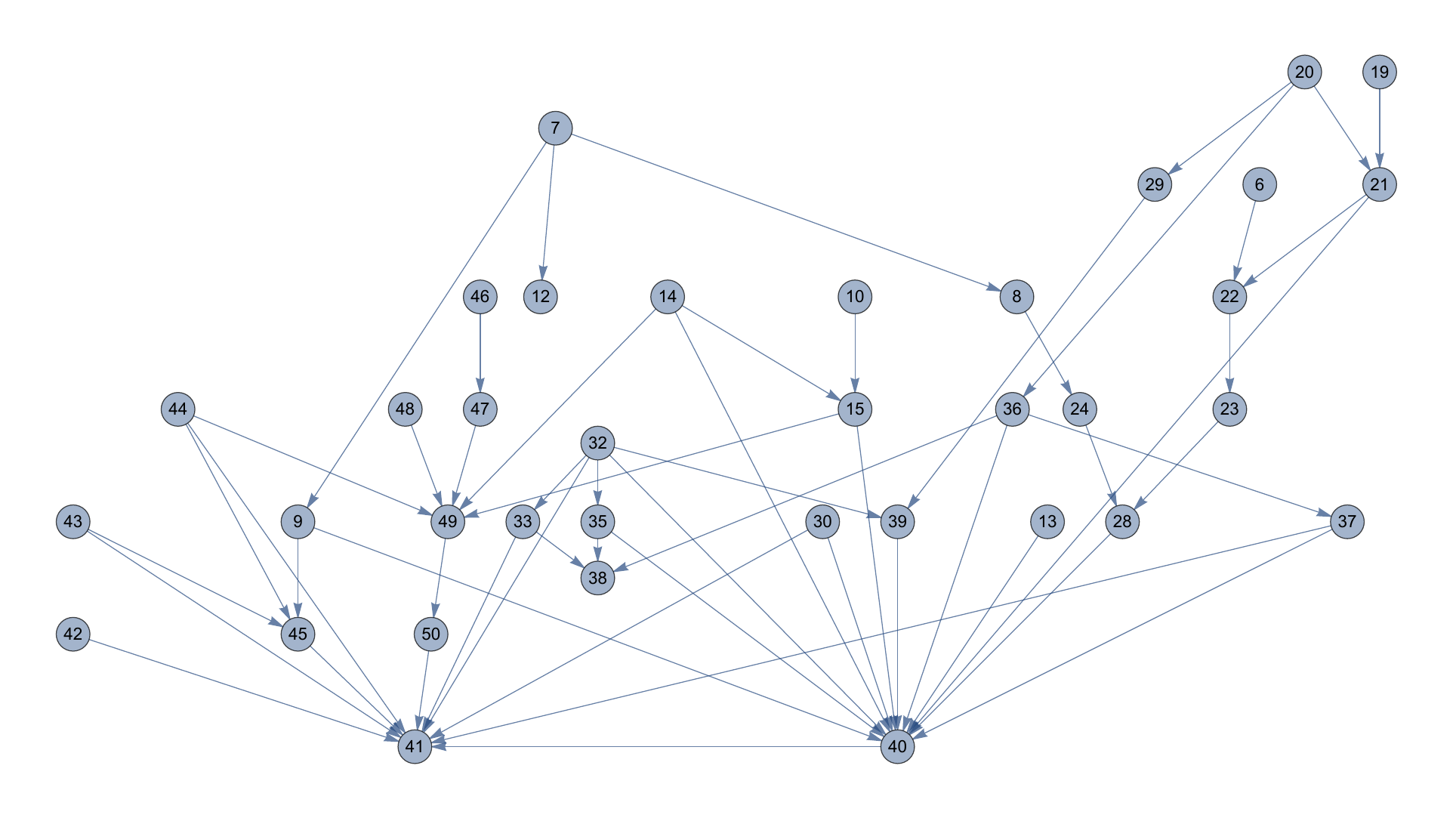}
  \caption{Conceptual links among all Lemmas, Propositions, Theorems and Corollaries of the present chapter. An arrow from vertex $x$ to vertex $y$ means that $x$ is needed to prove $y$.}
  \label{Ch1 scheme}
\end{figure}

\section{Ordered vector spaces} \label{sec ord}

Linear algebra has always been a useful tool in the hand of a physicist and in particular of a mathematical physicist. However, with the advent of quantum mechanics, it has acquired a somewhat different status, becoming rather a conceptually central field for the mathematical formulation of physics itself. Nowadays, the concept of vector space is so ubiquitous in mathematical physics, that it becomes not only interesting but almost unavoidable to study its interplay with some other major mathematical objects. For instance, in particle physics the concept of linear representation of a group becomes relevant, and representations are nothing but vector spaces endowed with some additional structure, in this case a subgroup of the corresponding linear group.

Analogously, a notion that plays a central role in this thesis is that of ordered vector space. We assume that the reader is familiar with vector spaces and their basic properties, but no more knowledge than this is required, in that any other concept we will need throughout the chapter will be introduced and discussed when relevant. Perhaps I should say at this point that in the field of quantum information vector spaces are often surreptitiously assumed to be finite-dimensional. It goes without saying, this technical assumption makes our life a lot easier, and for the purpose of asking the kind of questions we will be investigating in this thesis is even fundamental. Thus, starting from the next chapter we will always assume all vector spaces to be finite-dimensional. However, since now we want to discuss the foundations of general probabilistic theories, we will \emph{not} make this assumption in this chapter unless explicitly stated. 

The material of this section is covered by many excellent textbooks, most notably~\cite{SCHAEFER, PERESSINI, JAMESON} as well as the excellent set of lecture notes~\cite{FOUNDATIONS}. Let us start by reminding the reader that an \textbf{ordering} on a set $X$ is a binary relation $\leq$ on $X$ that is reflexive ($x\leq x$ for all $x\in X$), transitive ($x\leq y,\, y\leq z$ implies $x\leq z$ for all $x,y,z\in X$) and anti-symmetric ($x\leq y$ and $y\leq x$ is only possible when $x=y$). A crucial fact to keep in mind is that in general \emph{$\leq$ is not required to be total}, i.e. it might happen that for $x,y\in X$ neither $x\leq y$ nor $y\leq x$. We are now ready to introduce one of our main subjects of study throughout this chapter.

\begin{Def}
An \textbf{ordered vector space} is a real vector space $E$ equipped with an ordering $\leq$ which is: 
\begin{enumerate}[(i)]
\item translationally invariant, meaning that $x\leq y\,\Rightarrow\, x+z\leq y+z,\ \forall\, x,y,z\in E$;
\item positively homogeneous, i.e. $x\leq y\,\Rightarrow\, \lambda x\leq \lambda y,\ \forall\, x,y\in E,\, \lambda\geq 0$.
\end{enumerate}
\end{Def}

Let us now discuss some related concepts and constructions. By the end of this discussion, we hope that the reader who is less interested in the case of $E$ having infinite dimension will anyway be aware of the main differences with the finite-dimensional case. To help identifying these subtleties, we will point them out as soon as they arise. For further details, we refer the reader to~\cite[`Orderings of vector spaces']{FOUNDATIONS}. Let us start by fixing some terminology. 

\begin{Def} \label{def cone}
A subset $C\subseteq E$ of a real vector space $E$ is called a \textbf{cone} if:
\begin{enumerate}[(i)]
\item it is closed under sums, i.e. $C+C=C$;
\item it is closed under multiplication by positive scalars, i.e. $\lambda C\subseteq C$ for all $\lambda\geq 0$; and
\item does not contain nontrivial vector subspaces, i.e. $C\cap -C=\{0\}$.
\end{enumerate}
\end{Def}

\begin{note}
It is understood that for sets $X,Y\subseteq E$ one defines $X+Y\coloneqq \{x+y:\, x\in X,\, y\in Y\}$, and analogously $\lambda X = \{\lambda x:\, x\in X\}$ for scalars $\lambda\in\mathds{R}$.
\end{note}

\begin{note}
Some authors prefer to call a set that satisfies requirements (i)-(iii) of Definition~\ref{def cone} a `salient convex cone' or also a `proper cone'. Usually, when this nomenclature is adopted a cone is simply required to be closed under multiplication by positive scalars. The convention we adopt here follows that in~\cite{Ellis-dual-base}.
\end{note}

In every ordered vector space the set $E_+=\{x\in E:\, x\geq 0\}$ is easily seen to be a cone. In fact, since the order relation can be rewritten in an equivalent way as $x\leq y\Leftrightarrow y-x\in E_+$, one sees that giving an ordering on a vector space is the same as choosing a cone. As an easy consequence of this reformulation, observe that $x\leq y\Rightarrow -y \leq -x$. With a rather obvious terminology, a cone $C$ is said to be \textbf{spanning} if $\Span(C)=E$, which can equivalently be written as $C-C=E$. For $x,y\in E$ with $x\leq y$, the corresponding \textbf{interval} is defined as $[x,y]\coloneqq \{z:\, x\leq z\leq y\}$.

Given any real vector space $E$, we can consider the associated vector space of linear functionals $\varphi:E\rightarrow\mathds{R}$, called the \textbf{algebraic dual} of $E$ and denoted by $E'$. For $x\in E$ and $\varphi\in E'$, we will often write $\braket{\varphi, x}$ instead of $\varphi(x)$. This defines a bilinear form $\braket{\cdot,\cdot}: E'\times E\rightarrow \mathds{R}$, also called the \textbf{canonical bilinear form}.

\begin{note}
Most standard textbooks (like~\cite{RUDIN}) define the canonical bilinear form with the two arguments exchanged, writing $\braket{x,\varphi}$ instead of $\braket{\varphi,x}$. We prefer instead this latter convention since it is reminiscent of the Dirac notation in quantum mechanics, with dual vectors (bras) on the left and primal vectors (kets) on the right.
\end{note}

When $E$ is also an ordered vector space, also its dual $E'$ is naturally ordered by the \textbf{dual cone} $E_+^{\varoast}$ formed by all functionals that are positive on $E_+$. The subspace $E^\varoast\coloneqq E_+^\varoast - E_+^\varoast \subseteq E'$ is again an ordered vector space, called the \textbf{order dual} of $E$. It is not difficult to realise that if $E$ is finite dimensional and $E_+$ is in addition closed, then the order dual coincides with the dual vector space, i.e. $E^\varoast=E'$.

In this case, remember that we can identify $E'\simeq E$ upon the introduction of a nondegenerate scalar product. If a Euclidean scalar product is used, then the comparison between the positive cone $E_+$ and its dual $E_+^\varoast$ has a very intuitive geometrical meaning. In fact, it is not difficult to realise that the `narrower' the cone $E_+$ is, the `wider' its dual $E_+^\varoast$ will become, with `narrow' and `wide' referring to the opening angles.

\section{Topological vector spaces} \label{sec topo}

As we said, throughout this chapter we want to stay as general as possible, without making any assumption on the dimension of the vector spaces we will encounter along the way. The difficulties in dealing with infinite-dimensional spaces emerge especially when one comes to the somewhat risky business of taking duals.\footnote{For instance, it is known that the dual of an infinite-dimensional vector space has always a larger dimension -- in the sense of cardinalities -- than the original space.} For this reason, many vector spaces come with a topological structure that identifies a special class of functionals within the algebraic dual $E'$, namely the continuous functionals. 

In order to identify a concept of continuity of functions, we need some notion of neighbourhood. While finite-dimensional (real) spaces inherit such concepts from the isomorphism with $\mathds{R}^n$ identified by a basis, in the infinite-dimensional case we must enforce an analogous notion from the outside, or mathematically speaking we must equip the space with a topology. We remind the reader that a \textbf{topology} on a set $X$ is a collection $\tau\subseteq \mathcal{P}(X)$ of subsets of $X$ (called \textbf{open sets}) that~\cite[I]{KELLEY}:
\begin{enumerate}[(i)]
\item includes $\emptyset$ and $X$;
\item is closed under unions; and
\item is closed under finite intersections.
\end{enumerate}
A set $X$ endowed with a topology is called a \textbf{topological space}. Our archetypical example of a topological space is the real line $\mathds{R}$ equipped with the standard notion of open sets. Let us summarise quickly how some elementary constructions we learnt in that context generalise to arbitrary topologies. For more thorough discussions we refer the reader to~\cite{KELLEY}, while a concise yet comprehensive introduction to the basic concepts can also be found in~\cite[II]{MEGGINSON}.

\begin{itemize}

\item Topologies $\tau_1,\tau_2$ on the same set $X$ can be compared. If $\tau_1\subseteq \tau_2$ we say that $\tau_1$ is \textbf{coarser} (or \textbf{weaker}) than $\tau_2$.

\item A \textbf{neighbourhood} of $x\in X$ is an open set $U\in \tau$ that contains $x$. An extremely weak requirement to impose on a topology is that it is \textbf{Hausdorff} or \textbf{separating}, i.e. such that every two disjoint points admit disjoint neighbourhoods. All the topology we are going to discuss here will be Hausdorff.

For an arbitrary subset $S\subseteq X$ of a topological space $X$, those points $x\in S$ admitting a neighbourhood $U\subseteq S$ form the \textbf{interior} of $S$, denoted with $\inter(S)$ or with $\inter_\tau (S)$ when we want to specify the topology $\tau$. Observe that if $\tau_1$ is coarser than $\tau_2$, then $\tau_1$-neighbourhoods are a fortiori $\tau_2$-neighbourhoods, hence $\inter_{\tau_1}(S)\subseteq \inter_{\tau_2}(S)$.

\item A subset $K$ of a topological space $(X,\tau)$ is said to be \textbf{closed} if its complement $K^c$ in $X$ is open, i.e. if $K^c\in \tau$. Intersections and finite unions of closed sets are again closed. 

For a fixed set $S\subseteq X$, the smallest closed superset of $S$ in $X$ (i.e. the intersection of all closed supersets of $S$) is called the \textbf{closure} of $S$, and denoted with $\cl(S)$ or with $\cl_\tau (S)$ when there is need to specify the topology $\tau$. If $\tau_1$ is coarser than $\tau_2$, then $\tau_1$-closed sets are a fortiori $\tau_2$-closed, and hence $\cl_{\tau_1}(S)\supseteq \cl_{\tau_2}(S)$. A set whose closure is the entire space $X$ is called \textbf{dense} in $X$.

\item Another important concept in topology is that of compactness. A set $S\subseteq X$ in a topological space $(X,\tau)$ is said to be \textbf{compact} if every open cover of $S$ has a finite subcover, i.e. if for all families $\{U_i\}_{i\in I}\subseteq \tau$ such that $\bigcup_{i\in I} U_i\supseteq S$ one can find $n\in\mathds{N}$ and $i_1,\ldots,i_n\in I$ such that $\bigcup_{k=1}^n U_{i_k} \supseteq S$. Evidently, every finite collection of points is compact. It is not difficult to realise that closed subsets of compact sets are again compact. Moreover, if the underlying topology is Hausdorff, compact subsets are always closed. 

In a Euclidean space $\mathds{R}^n$ equipped the standard topology, $S$ is compact iff it is closed and bounded, but no such simple characterisations are available for more general topological spaces. An intuitive rule of thumb it is useful to remember is that in many circumstances compact sets behave like single points. Now, let us present a lemma that is not so commonly found but will be needed in order to prove Ludwig's embedding theorem (Theorem~\ref{Ludwig emb thm}).

\begin{lemma} \emph{\cite[I \S 9.4, Corollary 3]{BOURBAKI}.} \label{Bourbaki lemma}
Let $(X,\tau)$ be a compact topological space. A Hausdorff topology on $X$ that is coarser than $\tau$ must coincide with the latter.
\end{lemma}

\item Topological spaces can be multiplied, in the sense that the Cartesian product $X_1\times X_2$ of two topological spaces $X_1,X_2$ can be endowed with the \textbf{product topology} whose open sets are unions of products of local open sets.

\item A function $f:X_1\rightarrow X_2$ between two topological spaces is said to be \textbf{continuous} when the pre-images of open sets are again open, in symbols $f^{-1}(U_2)\in \tau_1$ for all $U_2\in \tau_2$. Alternatively, one can cast this condition in terms of closed sets by saying that $f$ is continuous when $f^{-1}(K_2)$ is closed for all $\tau_2$-closed sets $K_2$. It does not take much to realise that for real functions $f:\mathds{R}\rightarrow\mathds{R}$ this is just a rephrasing of the notion of continuity we learn in a standard course on calculus. 

Reverse engineering the process, one can use functions to define a topology. Namely, if $X$ is a set and $Y$ is a topological space, for a family $\mathcal{F}=\{ f_i \}_{i\in I}$ of functions $f_i:X\rightarrow Y$ there is a coarsest topology on $X$ such that all functions $f_i$ are continuous. The open sets of this topology are precisely the unions of finite intersections of sets of the form $f_i^{-1}(U)$, where $U\in \tau_Y$. In many cases we will have $Y=\mathds{R}$ with the standard topology, and we will refer to the above construction as the \textbf{initial topology} generated by $\mathcal{F}$, usually denoted by $\sigma (X,\mathcal{F})$.

\item In the case of $\mathds{R}$ with the standard topology, many useful notions can be conveniently rephrased in terms of sequences. Perhaps surprisingly, in the general case the notion of sequence is not quite sufficient, and we must resort to that of a \textbf{net}~\cite[II]{KELLEY}. A net on a set $X$ is a function $x:A\rightarrow X$, usually denoted by $(x_\alpha)_{\alpha\in A}$, whose domain $A$ is a \textbf{directed set}, i.e. an ordered set in which every two elements $a,b\in A$ admit a common upper bound $a,b\leq c\in A$. When $A=\mathds{N}$ is the set of natural numbers, a net becomes just a standard sequence. Nets on topological spaces are useful because one can give a definition of limit. Namely, $\tilde{x}\in X$ is said to be a \textbf{limit of a net} $(x_\alpha)_\alpha\subseteq X$ if for all neighbourhoods $V$ of $\tilde{x}$ there is $\alpha\in A$ such that $x_\beta\in V$ holds for all $\beta\in A$ such that $\beta\geq \alpha$. When the underlying topology is Hausdorff, which will be always the case in this thesis, limits of nets (when they exist) are unique. In this case, we write $\lim_\alpha x_\alpha = \tilde{x}$.

How to define subnets, in analogy with subsequences? We say that a net $(y_\beta)_{\beta\in B}$ is a \textbf{subnet} of a net $(x_\alpha)_{\alpha\in A}$ when there is a function $\phi:B\rightarrow A$ such that: (i) $y=x\circ \phi$; and (ii) for all $\alpha\in A$ there is $\beta\in B$ such that $\beta'\geq \beta\Rightarrow \phi(\beta')\geq \alpha$, for all $\beta'\in B$. Intuitively, this latter condition ensures that $\phi(\beta)$ becomes large when so does $\beta$. It is easy to show that if $\tilde{x}$ is a limit of the net $(x_\alpha)_\alpha$, then it is also a limit of all subnets $(y_\beta)_\beta$ of $(x_\alpha)_\alpha$.

The concept of net is useful because it allows to recast several topological concepts in a form that lends itself to convenient manipulations.
For instance, it can be shown that a subset $K\subseteq X$ is closed iff it contains all the limits of its nets, i.e. iff for all nets $(x_\alpha)_\alpha\subseteq K$ converging to $x\in X$ one has $x\in K$. Moreover, $K$ is compact iff every net $(x_\alpha)_\alpha\subseteq K$ admits a subnet that converges to some $x\in K$. Continuity of functions can be characterised in a similar way: a function $f:X\rightarrow Y$ between two topological spaces $X,Y$ is continuous iff for all nets $(x_\alpha)_{\alpha\in A}\subseteq X$ converging to $\tilde{x}\in X$ the corresponding net $(f(x_\alpha))_{\alpha\in A}\subseteq Y$ converges to $f(\tilde{x})\in Y$. These results are totally analogous to the usual reformulations of closedness and compactness in a Euclidean space.

\end{itemize} 

The following definition is an example of how the concept of topology can be reconciled with that of an underlying algebraic structure. In our case, the most interesting algebraic structure is that of a vector space. For details, we refer the reader to the monographs~\cite{SCHAEFER, KELLEY-NAMIOKA}.

\begin{Def} \label{def topo vector space}
A \textbf{topological vector space} is a (real) vector space $E$ equipped with a topology $\tau$ such that the sum $+:E\times E\rightarrow E$ and the scalar multiplication $\cdot: \mathds{R}\times E\rightarrow E$ are continuous functions with respect to the product topologies on $E\times E$ and $\mathds{R}\times E$.
\end{Def}

\section{Banach spaces} \label{sec Banach}

There is a very natural way to endow a vector space $E$ with a topology, namely through a \textbf{norm}, i.e. a function $\|\cdot\|:E\rightarrow \mathds{R}_+$ with codomain the set of non-negative real numbers and such that~\cite[\S 1.2]{RUDIN}:
\begin{enumerate}[(i)]
\item $\|x\|=0\Leftrightarrow x=0$;
\item $\|\lambda x\|=|\lambda| \|x\|$ for all $\lambda\in\mathds{R}$ and $x\in E$;
\item $\|x+y\|\leq \|x\|+\|y\|$ for all $x,y\in E$.
\end{enumerate}
Incidentally, if $\|\cdot\|$ satisfies only (ii) and (iii) then it is called a \textbf{semi-norm}. When a vector space is equipped with a norm, we can give the usual definition of an open set as a subset $U\subseteq E$ such that for all $x\in U$ there is $\epsilon>0$ with the property that $B_\epsilon (x)\subseteq U$, where $B_\epsilon (x)\coloneqq \{y\in E:\, \|x-y\|\leq \epsilon\}$ is the \textbf{ball} centered in $x$ with radius $\epsilon$. Open sets defined in this way form a topology on $E$, usually called \textbf{norm topology}. When dealing with normed spaces, if we refer to some topological concepts without specifying the underlying topology, this is understood to be the norm topology.

For our purposes, the most interesting case is when $E$ is equipped with a norm $\|\cdot\|$ that makes it \textbf{complete}. In this context, completeness means that every \textbf{Cauchy sequence} $(x_n)_n$ admits a limit, i.e. there is $x \in E$ satisfying $\lim_n \|x-x_n\|=0$. As in standard analysis, a sequence of vectors $(x_n)_{n\in \mathds{N}}\subset E$ is a Cauchy sequence if for all $\epsilon>0$ there exists $N\in\mathds{N}$ such that for all $n,m\geq N$ one has $\|x_n-x_m\|\leq \epsilon$. One gives the following fundamental definition~\cite[\S 1.2]{RUDIN}.

\begin{Def} \label{def Banach}
A \textbf{Banach space} is a real vector space $E$ endowed with a norm $\|\cdot\|$ that makes it complete.
\end{Def}

Useful references for the theory of Banach spaces are~\cite{RUDIN, MEGGINSON}, while a concise introduction can also be found in~\cite[IX]{SIMMONS}.

\subsection{Some immediate consequences of completeness}

The requirement of completeness is in some sense just technical. In fact, it turns out that every normed vector space $E$ can be thought of as a norm-dense subspace of a Banach space $\bar{E}$ called its \textbf{completion}.\footnote{Furthermore, the completion is unique up to isometric isomorphisms, and can be taken as the space of all equivalence classes of Cauchy sequences in $E$ obtained by the relation $(x_n)_n\sim (y_n)_n\Leftrightarrow \lim_n \|x_n-y_n\|=0$.} Many elementary results in Banach space theory hold even if one drops the completeness axiom. However, this assumption plays a crucial role in proving some deeper theorems that are widely regarded as cornerstones of the field such as the so-called uniform boundedness principle (Theorem~\ref{UBP}). Ultimately, many of these results stem from \emph{Baire's category theorem}~\cite[Theorem 2.2]{RUDIN}, which holds only in complete metric spaces. We will not state Baire's theorem, but will content ourselves with presenting one of its Banach space consequences, which we will make use of in Section~\ref{sec ord Banach}. For a direct proof that does not make use of Baire's theorem, we refer the reader to~\cite[pp.22-23]{MEGGINSON}.
Recall that a subset $K\subseteq E$ of a real vector space $E$ is said to be \textbf{convex} if $x,y\in K$ implies $px+(1-p)y\in K$, for all $p\in [0,1]$.

\begin{lemma} \emph{\cite[Theorem 1.3.14]{MEGGINSON}.} \label{neigh 0 Banach}
Let $K\subseteq E$ be a (norm-)closed, convex subset of a Banach space $E$. If $E=\bigcup_{n\in \mathds{N}} nK$, then $K$ contains a (norm-)neighbourhood of the origin, i.e. there is $\delta>0$ such that $\|x\|\leq \delta\Rightarrow x\in K$ for all $x\in E$.
\end{lemma}

\subsection{Hahn-Banach separation theorem}

If $E$ is a Banach space when equipped with the norm $\|\cdot\|$, its \textbf{Banach dual} $E^*$ is by definition the space of all linear functionals $\varphi:E\rightarrow \mathds{R}$ that are \textbf{continuous} with respect to $\|\cdot\|$, i.e. such that $\lim_n \braket{\varphi, x_n}=\braket{\varphi, x}$ whenever $(x_n)_n\subset E$ is a sequence that converges to $x$ in the norm topology, i.e. $\lim_n \|x-x_n\| = 0$. Because of linearity, $\varphi$ is continuous everywhere iff it is continuous in $0$, which is in turn equivalent to $\varphi$ being \textbf{bounded}, i.e. such that
\bb
\|\varphi\|_* \coloneqq \sup_{0\neq x\in E} \frac{|\braket{\varphi, x}|}{\|x\|}
\label{dual norm Banach}
\ee
is finite. Another useful fact to keep in mind is that $\varphi$ is continuous iff its kernel $\ker\varphi = \varphi^{-1}(0)$ is closed. The vector space $E^*$ becomes a Banach space itself when equipped with the norm $\|\cdot\|_*$ defined by~\eqref{dual norm Banach}. A useful fact we will make use of is the following. If a functional $\varphi:V\rightarrow \mathds{R}$ is defined only on a dense subspace $V\subseteq E$ of a Banach space $E$ and there is bounded (in the sense that $\sup_{0\neq x\in V}\frac{|\braket{\varphi,x}|}{\|x\|}<\infty$), then it can always be extended to a continuous functional on the whole space, i.e. with a slight abuse of notation we can write $\varphi\in E^*$~\cite[Theorem 3.6]{RUDIN}. Moreover, this process of \textbf{continuous linear extension}\footnote{Although we stated the continuous linear extension lemma for functionals, a similar result holds for general linear operators $\phi: V\rightarrow F$, where $F$ is any Banach space. In this case, the norm $\|\phi(x)\|$ displaces $|\braket{\varphi,x}|$ in expressions like~\eqref{dual norm Banach}.} preserves the norm, in the sense that
\bbb
\sup_{0\neq x\in V}\frac{|\braket{\varphi,x}|}{\|x\|} = \sup_{0\neq x\in E} \frac{|\braket{\varphi, x}|}{\|x\|}\, .
\eee

\begin{note}
Throughout the rest of this section, the Banach dual of a Banach space $E$ will be denoted with $E^*$. Instead, we reserve the notation $E^\varoast$ for the order dual of an ordered vector space $E$, as discussed in Section~\ref{sec ord}. This distinction is necessary, since these two subspaces of the algebraic dual $E'$ may happen to have nothing to do with each other. In fact, they can even exhibit a trivial intersection~\cite[Example 2.15(b)]{PERESSINI}.
\end{note}

We can of course iterate the construction of the dual and consider the space $E^{**}$ (the dual of the dual). As in the case of finite-dimensional spaces, there is a canonical linear embedding of $E$ into $E^{**}$, since every vector $x\in E$ can be seen as acting on functionals $\varphi\in E^*$ in a rather obvious way via the formula $x(\varphi)=\braket{\varphi,x}$. Since this action is continuous with respect to the norm~\eqref{dual norm Banach} on $E^*$, $x$ belongs to $(E^*)^*$. When $E$ is finite-dimensional we have $\dim E^{**}=\dim E^* =\dim E$, thus this embedding becomes an (isometric) isomorphism, and we can identify $E^{**}$ with $E$. However, when $E$ is infinite-dimensional it can happen that the image of $E$ through the canonical embedding is a proper subspace of $E^{**}$. When this is the case we say that $E$ is \textbf{non-reflexive}. While Hilbert spaces are well-known to be reflexive, it is important to bear in mind that non-reflexivity is not to be regarded as a pathological behaviour for Banach spaces. For instance, the standard sequence spaces we will examine in Examples~\ref{ex l infty} and~\ref{ex l1} turn out to be non-reflexive.

With the concept of dual space at hand, we can state an intuitive yet fundamental result in the theory of Banach spaces.

\begin{thm}[Hahn-Banach separation theorem] \label{Hahn-Banach} \emph{\cite[Theorem 3.4(b)]{RUDIN}.}
Let $M,N$ be disjoint, convex subsets of a Banach space $E$. Assume $M$ is compact and $N$ is closed. Then there are $\varphi\in E^*,\, \alpha\in\mathds{R},\, \epsilon>0$ such that
\bb
\braket{\varphi,x} \leq \alpha < \alpha+\epsilon \leq \braket{\varphi,y} \qquad  \forall\ x\in M,\ y\in N\, .
\label{Hahn-Banach eq}
\ee
\end{thm}

The interpretation of the above result is quite intuitive: any point which does not belong to a closed convex set can be strictly separated from it by means of a hyperplane. As one realises quickly, there is really nothing special about the numbers $\alpha,\epsilon$ we use to separate the two sets. In fact, in many cases one can rescale them as pleased by `zooming' in or out with the probe functional $\varphi$. For the application we have in mind, we will need a simplified version of the Hahn-Banach separation theorem that deals with the case of $N$ containing the origin or being invariant under multiplication by positive scalars. Let us state it and prove it now.

\begin{cor} \label{Hahn-Banach cor}
Let $M,N\subseteq E$ satisfy the hypotheses of Theorem~\ref{Hahn-Banach}.
\begin{enumerate}[(a)]
\item If additionally $0\in N$, then in~\eqref{Hahn-Banach eq} we can choose $\alpha=-1$ and $\epsilon<1$.
\item If $N$ is even invariant under multiplication by positive scalars, i.e. $N=t N$ for all $t>0$, then in~\eqref{Hahn-Banach eq} we can take $\alpha=-1$ and $\epsilon=1$, so that
\bb
\braket{\varphi,x}\leq -1\quad\text{and}\quad \braket{\varphi,y}\geq 0\qquad \forall\ x\in M,\ y\in N\, . 
\label{Hahn-Banach cor eq}
\ee
\end{enumerate}
\end{cor}

\begin{proof}
Let us start with claim (a). By taking $y=0$ in~\eqref{Hahn-Banach eq}, we see that $\alpha+\epsilon\leq 0$ and hence $\alpha+\frac{2}{3}\epsilon<0$. Defining $\varphi'\coloneqq \frac{\varphi}{|\alpha + \epsilon/3|}$ we get
\bbb
\braket{\varphi', x} \leq \frac{\alpha}{\left|\alpha + \frac{\epsilon}{3}\right|} \leq -1 < \frac{\alpha +\frac23 \epsilon}{\left|\alpha + \frac{\epsilon}{3}\right|} \eqqcolon -1+\epsilon' < 0 \leq \braket{\varphi', y}\quad \forall\ x\in M,\ y\in N\, .
\eee

As for claim (b), if $N=tN$ for all $t>0$ then $0\in N$, because $N$ is closed. Thus, we already know that we can take $\alpha=-1$. From~\eqref{Hahn-Banach eq} we see that $\braket{\varphi,y}\geq -1+\epsilon$ for all $y\in N$. Since $N$ is now closed under multiplication by positive scalars, for a fixed $y\in N$ we have also $t \braket{\varphi,y}\geq -1+\epsilon$ for all $t>0$. We conclude that $-1+\epsilon\leq 0$ and that $\braket{\varphi,y}\geq 0$. This amounts to taking $\epsilon=1$ from the start.
\end{proof}

Another useful consequence of the Hahn-Banach separation theorem is a dual formula for the norm of a vector. We present it as a separated statement since we will make repeated use of it later.

\begin{cor} \emph{\cite[\S 4.3]{RUDIN}.} \label{dual formula norm Banach}
Let $E$ be a Banach space and $x\in E$ a vector. Then
\bb
\|x\| = \sup_{\varphi\in E^*,\, \|\varphi\|_*\leq 1} |\braket{\varphi,x}|\, .
\label{dual formula norm Banach eq}
\ee
\end{cor}

\begin{proof}
On the one hand, by~\eqref{dual norm Banach} we obtain $|\braket{\varphi,x}|\leq \|\varphi\|_* \|x\|\leq \|x\|$ for all $\varphi\in E^*$ such that $\|\varphi\|_*\leq 1$. On the other hand, pick $\epsilon>0$ and apply Theorem~\ref{Hahn-Banach} with $M=\{x\}$ and $N=\{y\in E:\, \|y\|\leq (1-\epsilon)\|x\|\}$, which is evidently closed and convex. We obtain a functional $\varphi_1\in E^*$ such that $\braket{\varphi_1,x} < \braket{\varphi_1,y}$ for all $y\in N$. As a matter of fact, we can consider instead $\varphi_2\coloneqq -\varphi_1$, which is such that $\braket{\varphi_2,x} > \braket{\varphi_2,y}$. By definition of dual norm~\eqref{dual norm Banach}, we can find $y\in N$ such that $\braket{\varphi_2,y}\geq (1-\epsilon)\|\varphi_2\|_* (1-\epsilon)\|x\|\geq (1-2\epsilon)\|\varphi_2\|_* \|x\|$. Putting all together, we see that $\varphi_3\coloneqq \frac{\varphi_2}{\|\varphi_2\|_*}$ satisfies $\|\varphi_3\|_*=1$ and $\braket{\varphi_3,x} > (1 -2\epsilon)\|x\|$. Since the construction can be repeated for all $\epsilon>0$, we conclude that $\sup_{\varphi\in E^*,\, \|\varphi\|_*\leq 1} |\braket{\varphi,x}|\geq \|x\|$, yielding the claim.
\end{proof}

\subsection{Some classic results in Banach space theory}

As we saw before, the assumption of completeness is crucial in order to establish some more profound facts that lie at the heart of the theory of Banach spaces. One such fact is the \emph{uniform boundedness principle} (also called Banach-Steinhaus theorem), which we are now set to state in its simplest form. The most common way to prove it is via Baire's category theorem, but also direct proofs~\cite{TerryTao-UBP} or arguments based on non-standard analysis~\cite{non-standard-UBP} are available. 

\begin{thm}[Uniform boundedness principle] \emph{\cite[\S 2.5]{RUDIN} or~\cite[Theorem 1.6.9]{MEGGINSON}.} \label{UBP}
Let $E$ be a Banach space. If a family $\{\varphi_i\}_{i\in I}$ of continuous functionals $\varphi_i\in E^*$ satisfies
\bbb
\sup_{i\in I} |\braket{\varphi_i,x}| <\infty
\eee
for all $x\in E$, then
\bbb
\sup_{i \in I} \|\varphi_i\|_* < \infty\, .
\eee
\end{thm}

At this point, the reader might wonder, whether there is a real need to worry about the general concept of topological vector space if what we really care about here is Banach spaces, which carry a natural topology induced by their norms. As a matter of fact, although the norm topology really is the most intuitive topology one can endow a Banach space with, it is by no means the only one. In fact, the following alternative definitions will play an important role in what follows. For details, we refer the reader to~\cite[\S\S 3.11 and 3.14]{RUDIN} or~\cite[\S\S 2.5 and 2.6]{MEGGINSON}.

\begin{Def} \label{def weak weak*}
Let $E$ be a Banach space with dual $E^*$. The initial topology on $E$ generated by all continuous functionals $\varphi\in E^*$ is called \textbf{weak topology} and denoted by $\mathcal{w}$ or by $\sigma(E,E^*)$. The initial topology on $E^*$ generated by all vectors $x\in E$ (seen as functionals on $E^*$) is called \textbf{weak* topology} and denoted by $\mathcal{w}^*$ or by $\sigma(E^*,E)$.
\end{Def}

As is easy to see, $E$ and $E^*$ equipped with these topologies are topological vector spaces in the sense of Definition~\ref{def topo vector space}. Moreover, it turns out that the above definitions are nontrivial as soon as we step out of the finite-dimensional realm. In fact, norm topology and weak topology coincide on a Banach space $E$ iff $E$ is finite-dimensional~\cite[Proposition 2.5.13]{MEGGINSON}. The same is true for the norm and weak* topologies on $E^*$~\cite[Corollary 2.6.3]{MEGGINSON}.
Topological notions pertaining to either of the above two topologies are referred to by juxtaposing weak- (or weakly) and weak*- in front of the relevant term. For instance, the concept of weak*-denseness will play a relevant role in what follows. Weak- and weak*-limits will be denoted by $\text{$\mathcal{w}$-lim}$ and $\text{$\mathcal{w}^*\!$-lim}$, respectively.

Despite being different, weak topology and norm topology stem from the same underlying structure, i.e. the norm. Thanks to this common origin, there are indeed surprising connections between the two. An example of such a connection is provided by the following corollary to the Hahn-Banach separation theorem (Theorem~\ref{Hahn-Banach}).

\begin{cor} \emph{\cite[\S 3.12]{RUDIN} or~\cite[Theorem 2.5.16]{MEGGINSON}.} \label{norm cl = weak cl conv cor}
Norm closure and weak closure coincide for convex subsets of a Banach space $E$. In particular, a convex subset of $E$ is norm-closed iff it is weakly closed.
\end{cor}

\begin{proof}
Since the weak topology is coarser than the norm topology, the norm closure $\cl (K)$ of any set $K\subseteq E$ is clearly contained inside its weak closure $\cl_{\mathcal{w}}(K)$. However, by Theorem~\ref{Hahn-Banach} any $x\in E$ with $x\notin \cl(K)$ can be separated strictly from the latter set (and in particular from $K$) by means of a functional $\varphi\in E^*$. This shows that there is a weak neighbourhood of $x$ that does not contain points of $K$. Consequently, $x\notin \cl_{\mathcal{w}}(K)$, concluding the proof.
\end{proof}

Let us now explore some basic facts about the weak* topology. Naturally, all functionals on the dual $E^*$ that are induced by a vector $x\in E$ are considered to be continuous (this is in fact the definition of weak* topology). One could then wonder, whether these exhaust all the weak*-continuous functionals $E^*\rightarrow \mathds{R}$. This is in fact the case, as the following lemma establishes.

\begin{lemma} \emph{\cite[p.66]{RUDIN} or~\cite[Proposition 2.6.4]{MEGGINSON}.} \label{weak*-continuous are vectors}
Let $E$ be a Banach space. Then for a linear functional $f:E^*\rightarrow \mathds{R}$ the following are equivalent:
\begin{enumerate}[(a)]
\item $f$ is weak*-continuous;
\item its kernel $\ker (f)$ is weak*-closed;
\item $f(\cdot)=\braket{\cdot,x}$ for some $x\in E$.
\end{enumerate}
\end{lemma}

\begin{proof}
The equivalence between (a) and (b) is standard, and stems from the linearity of $f$. If $f$ is weak*-continuous then $f^{-1}(K)$ is weak*-closed for all closed sets $K\subseteq \mathds{R}$, and in particular $f^{-1}(0)=\ker (f)$ is weak*-closed. Conversely, pick a net $(\varphi_\alpha)_\alpha\subseteq E^*$ such that $\wstarlim_\alpha\, \varphi_\alpha =0$, and let us show that $\lim_\alpha f(\varphi_\alpha)=0$. Up to considering subnets, we can assume that $f(\varphi_\alpha)$ converges to some real number $r$. Pick $\lambda\notin\ker (f)$, and construct the weak*-converging net formed by the functionals $\varphi'_\alpha\coloneqq \varphi_\alpha - f(\varphi_\alpha) \frac{\lambda}{f(\lambda)}$. Since $\varphi'_\alpha\in \ker(f)$ for all $\alpha$ and $\ker(f)$ is weak*-closed,
\bbb
\ker (f) \ni \wstarlim_\alpha\,\varphi'_\alpha = \wstarlim_\alpha\, \varphi_\alpha - \frac{\lambda}{f(\lambda)} \lim_\alpha f(\varphi_\alpha) = - \frac{\lambda}{f(\lambda)} r\, ,
\eee
which is possible iff $r=0$.

Now, since (c) clearly implies (a), we have just to show that if $f$ is weak*-continuous then it acts as $f(\varphi)=\braket{\varphi,x}$ for some $x\in E$ and all $\varphi\in E^*$. By definition of continuity in topology, for all $\epsilon>0$ the set $f^{-1}\left( (-\epsilon,\epsilon)\right)\subseteq E^*$ is weak*-open. But weak*-open sets are unions of finite intersections of the form $\bigcap_{i=1}^n x_i^{-1}(U_i)$, where $x_i\in E$ is seen as a functional $x_i:E^*\rightarrow \mathds{R}$, and $U_i\subseteq \mathds{R}$ are open sets of real numbers (see the discussion in Section~\ref{sec topo}). Thus, since $0\in f^{-1}\left( (-\epsilon,\epsilon)\right)$, we can pick a finite number of vectors $x_i\in E$ such that $0\in \bigcap_{i=1}^n x_i^{-1}(U_i) \subseteq f^{-1}\left( (-\epsilon,\epsilon)\right)$. From $0\in \bigcap_{i=1}^n x_i^{-1}(U_i)$ we deduce immediately $0\in \bigcap_{i=1}^n U_i$. Take $\delta>0$ such that $(-\delta,\delta)\subseteq \bigcap_{i=1}^n U_i$, which is possible since all sets $U_i$ are open. Then for all $\varphi\in E^*$ one has
\begin{align*}
&|\braket{\varphi,x_i}|< \delta\ \ \forall\ i=1,\ldots,n \\
&\qquad\quad \Longrightarrow\quad \varphi\in \bigcap_{i=1}^n x_i^{-1} \left( (-\delta, \delta)\right) \subseteq \bigcap_{i=1}^n x_i^{-1}(U_i) \subseteq f^{-1}\left( (-\epsilon,\epsilon)\right) \\
&\qquad\quad \Longrightarrow\quad |f(\varphi)| < \epsilon\, .
\end{align*}
As a consequence, when $\varphi\in \bigcap_{i=1}^n \ker x_i$ we have $|t \braket{\varphi,x_i}|<\delta$ for all $i$ and $t\in\mathds{R}$, and thus $|t f(\varphi)|<\epsilon$ for all $t\in\mathds{R}$. This can happen only if $f(\varphi)=0$. We have just shown that
\bbb
\bigcap_{i=1}^n \ker x_i \subseteq \ker \varphi\, ,
\eee
a condition that is well-known to be equivalent to $\varphi\in \Span\{x_1,\ldots,x_n\}$ from elementary linear algebra.
\end{proof}

Many strange features of infinite-dimensional Banach spaces stem from the fact that the unit ball $B=\{x\in E:\, \|x\|\leq 1\}$ is never compact in the norm topology.\footnote{In fact, with the help of the so-called \emph{Riesz' lemma} it is not difficult to construct a sequence $(x_n)_n\subset B$ such that $\|x_n-x_m\|\geq 1/2$ for all $n\neq m$. Such a sequence does not admit any Cauchy (hence, convergent) subsequence, therefore $B$ can not be compact.} However, since the weak topology is coarser than the norm topology, it can happen that $B$ is weakly compact. In fact, an important theorem says that this is the case iff the space is reflexive~\cite[Theorem 2.8.2]{MEGGINSON}. One could expect similar exotic characterisations of the unit ball compactness to hold for the weak* topology. What comes as surprise now is that \emph{the dual unit ball is always weak*-compact.} This statement is the content of the classic Banach-Alaoglu theorem. 

\begin{thm}[Banach-Alaoglu] \emph{\cite[\S 3.15]{RUDIN} or~\cite[Theorem 2.6.18]{MEGGINSON}.} \label{Banach-Alaoglu}
Let $E$ be a Banach space. The dual unit ball $B_* \coloneqq \{\varphi\in E^*:\, \|\varphi\|_*\leq 1\}$ is weak*-compact.
\end{thm}

We can put in practise what we learnt until now and establish the following well-known result in Banach space theory. Since to the best of our knowledge the available demonstrations require some more advanced techniques, we include an elementary proof that makes instructive use of both the Banach-Alaoglu theorem and the uniform boundedness principle.

\begin{prop} \emph{\cite[Corollary 2.7.12]{MEGGINSON}.} \label{checking weak* closedness}
Let $W\subseteq E^*$ be a subspace of the dual $E^*$ of a Banach space $E$. Denote by $B_*=\{\varphi\in E^*:\, \|\varphi\|_*\leq 1\}$ the dual unit ball. Then $W$ is weak*-closed iff so is $W\cap B_*$.
\end{prop}

\begin{proof}
Since $B_*$ is weak*-compact by Theorem~\ref{Banach-Alaoglu} and hence also weak*-closed, if $W$ is weak*-closed then so is $W\cap B_*$. Conversely, assume that $W\cap B_*$ is weak*-closed, and let us show that the same holds for the whole $W$. First of all, since closed subsets of compact sets are again compact, we see that $W\cap B_*$ must in fact be weak*-compact (again, a consequence of Theorem~\ref{Banach-Alaoglu}). Then, take a net $(\varphi_\alpha)_\alpha\subseteq W$ that converges to $\varphi\in E^*$ in the weak* topology, i.e. $\wstarlim_\alpha \,\varphi_\alpha= \varphi$. If we prove that $\varphi\in W$, we are done. Rescale the functionals $\varphi_\alpha$ by defining $\lambda_\alpha\coloneqq \frac{\varphi_\alpha}{\|\varphi_\alpha\|_*}$. Now, since $(\lambda_\alpha)_\alpha \subseteq W\cap B_*$ and the latter set has been shown to be weak*-compact, the characterisation of compactness in terms of nets (see Section~\ref{sec topo}) guarantees that there is a subnet $(\lambda_\beta)_\beta$ of $(\lambda_\alpha)_\alpha$ such that $\wstarlim_\beta \,\lambda_\beta = \lambda$ for some $\lambda\in W\cap B_*$. Taking subnets of converging nets does not change the limit, so we still have $\wstarlim_\beta\, \varphi_\beta = \varphi$.

Now, since $\lim_\beta \braket{\varphi_\beta,x} = \braket{\varphi,x}$ and in particular $\sup_\beta |\braket{\varphi_\beta, x}| <\infty$ for all $x\in E$, the uniform boundedness principle (Theorem~\ref{UBP}) implies that there is $k\in\mathds{R}$ such that $\|\varphi_\beta\|_*\leq k$ for all $\beta$, or in other words that the net of real numbers $\left( \|\varphi_\beta\|_*\right)_\beta\subseteq \mathds{R}$ is bounded. We can then extract from $(\varphi_\beta)_\beta$ a subnet $(\varphi_\gamma)_\gamma$ such that $\lim_\gamma \|\varphi_\gamma\|_* = r\in\mathds{R}$. Of course, $(\lambda_\gamma)_\gamma$ converges again to $\lambda$ in the weak* topology, or $\wstarlim_\gamma \,\lambda_\gamma = \lambda$. Going back to the original net $(\varphi_\alpha)_\alpha$, which still satisfies $\wstarlim_\gamma\, \varphi_\gamma = \varphi$, we see that we must have 
\begin{align*}
\varphi &= \wstarlim_\gamma\, \varphi_\gamma \\
&= \wstarlim_\gamma\, \|\varphi_\gamma\|_* \frac{\varphi_\gamma}{\|\varphi_\gamma\|_*} \\
&= \wstarlim_\gamma\, \|\varphi_\gamma\|_* \lambda_\gamma \\
&= \left( \lim\nolimits_\gamma \|\varphi_\gamma\|_*\right) \left( \wstarlim_\gamma\, \lambda_\gamma \right) \\
&= r \lambda\, ,
\end{align*}
and since $\lambda\in W$ this implies that also $\varphi\in W$.
\end{proof}

\section{Ordered Banach spaces} \label{sec ord Banach}

Throughout Section~\ref{sec ord}, we discussed some natural ways to equip a vector space with an ordering. Instead, Section~\ref{sec topo} was devoted to the study of vector spaces endowed with topological structures, and in Section~\ref{sec Banach} we further specialised to Banach spaces. Now, we will see how these notions interplay, giving rise to the beautiful theory of ordered topological vector spaces. For a complete exposition of the subject, we refer the reader to~\cite[V]{SCHAEFER} or to the monographs~\cite{PERESSINI, JAMESON}. A more concise introduction can be found also in~\cite[Appendix A]{KELLEY-NAMIOKA}. Part of our presentation is taken from the lecture notes~\cite[`Duality of cones in locally convex spaces']{FOUNDATIONS}, but we try to be as self-contained as possible and to report all the proofs of the results we need.

In what follows, we are interested in ordered vector spaces that are also Banach spaces. Going back to the definitions, we see that we enforced no connection whatsoever between the order and the Banach structure. Hence, the reader will not be surprised by the fact that we need to do so before starting to prove nontrivial results. 
For an ordered Banach space $E$, an example of an outstanding problem in this context concerns the relation between its order dual $E^\varoast$ and its Banach dual $E^*$, as defined in Sections~\ref{sec ord} and~\ref{sec Banach}. Up to now, $E^\varoast$ and $E^*$ are two a priori different subspaces of the algebraic dual $E'$, and we already mentioned that in fact they can happen to have trivial intersection~\cite[Example 2.15(b)]{PERESSINI}. In light of this, the following question arises: \emph{under what hypotheses can we show a relation between order and Banach dual? For instance, when is it the case that every positive functional is automatically continuous, i.e. $E^\varoast\subseteq E^*$?}

We can of course take a pragmatic approach and restrict ourselves to work inside the Banach dual from the start. Observe that the cone $E_+^*$ formed by all positive \emph{and continuous} functionals turns $E^*$ into an ordered Banach space. We observe immediately that $E_+^*$ is always weak*-closed, for any net $(\varphi_\alpha)_\alpha\subseteq E_+^*$ such that $\wstarlim_\alpha\, \varphi_\alpha = \varphi\in E^*$ satisfies also $\braket{\varphi,x} = \lim_\alpha \braket{\varphi_\alpha,x}\geq 0$ for any given $x\in E_+$, implying that $\varphi\in E_+^*$.

\begin{note}
We collected a few different ways to take duals of spaces and cones, so perhaps a quick recap is helpful now.
\begin{itemize}

\item $E$ vector space: $E'$ is the algebraic dual, i.e. the space formed by all possible linear functionals $E\rightarrow\mathds{R}$.

\item $E, E_+$ ordered vector space: $E_+^\varoast$ denotes the set of all positive functionals inside the algebraic dual $E'$.

\item $E, E_+$ ordered vector space: $E^\varoast = E_+^\varoast - E_+^\varoast$ is the order dual.

\item $E$ Banach space: $E^*$ is the Banach dual, i.e. the space formed by all continuous (bounded) linear functionals.

\item $E,E_+$ ordered Banach space: $E_+^*$ denotes the cone of all positive and continuous linear functionals on $E$. This in turn defines an ordering on $E^*$.

\end{itemize}
\end{note}

\subsection{Duality of cones} \label{subsec GKE}

We start by asking ourselves: what are the possible requirements to impose on an ordered Banach space $E$ with the purpose of connecting Banach and order structure? A very natural choice is the closedness of the positive cone $E_+$ with respect to the norm topology (for more general topologies, this leads to the definition of \emph{ordered topological vector spaces}). Other central notions in the theory of ordered Banach spaces are that of \emph{normal cone} and \emph{strict $\mathcal{B}$-cone}.

Before delving into the technical details involved in the definition and investigation of these notions, let us try to explain the geometrical intuition behind them. Remember that in Definition~\ref{def cone} we chose not to call `cones' those geometrical cones that have a 180 degrees `opening angle' along some direction, in the sense that they contain a whole straight line. This would be good enough if we were to confine our investigation to just finite-dimensional spaces. Unfortunately, if the space is infinite dimensional there is another spooky possibility. In that case, in fact, there can be so many directions that it is possible for the cone to have \emph{increasingly large} (tending to 180 degrees) opening angles in some sequence of directions, while retaining the property of not being `completely' open in \emph{any} direction. A cone with this property is said to be not normal. 

Of course, there is also the opposite possibility. Instead of becoming wider and wider along different directions, the cone could become narrower and narrower, while still retaining the property of being spanning. Cones that do not exhibit such pathological behaviour are called strict $\mathcal{B}$-cones. 
At an informal level, one could say that the duality correspondence between `wide' and `narrow' cones is ultimately responsible for the Grosberg-Krein-Ellis theorem (Theorem~\ref{GKE thm}), which establishes the duality of the notions of normal and strict $\mathcal{B}$-cone in a rigorous sense.
Before stating the relevant definitions in precise terms, we need to fix some preliminary terminology and notation.

\begin{Def}
Let $M\subseteq E$ be a subset of an ordered vector space $E$. Its \textbf{saturated hull} $[M]$ and \textbf{convex kernel} $\,]M[$ are defined as
\begin{align}
[M] &\coloneqq (M+E_+)\cap (M-E_+)\, , \label{saturated hull} \\
]M[ &\coloneqq \coit\left( (M\cap E_+) \cup -(M\cap E_+) \right) . \label{convex kernel}
\end{align}
\end{Def}

Alternatively, one can think of the saturated hull as the union of all the intervals whose extreme points lie in $M$, in formula
\bb
[M] = \bigcup_{x,y\in M,\, x\leq y} [x,y]\, .
\label{saturated hull 2}
\ee
Clearly, convex kernels are always convex sets, and the saturated hull of a convex set is again convex. Moreover, if $M$ contains the origin then the same can be said of both $]M[$ and $[M]$. While $M\subseteq [M]$ holds for all sets $M$, note that the inclusion $]M[\subseteq M$ is valid provided that $M$ is convex. Despite all the desirable properties we have listed so far, neither of the two sets $[M]$ or $]M[$ is a priori guaranteed to be closed, not even when both $M$ and $E_+$ are such. 

\begin{Def} \label{def normal cone}
Let $\alpha,\beta\geq 1$ be real numbers. The positive cone $E_+$ of an ordered Banach space $E$ with unit ball $B\coloneqq \{x\in E:\, \|x\|\leq 1\}$ is said to be \textbf{$\alpha$-normal} if $[B] \subseteq \alpha B$, and \textbf{$\beta$-generating} if $B\subseteq \beta\, ]B[$. 
\end{Def}


The above definition can be cast into the following alternative form: $E_+$ is $\alpha$-normal if whenever $\|x\|,\|y\|\leq 1$ and $x\leq z\leq y$, one has $\|z\|\leq \alpha$, while it is $\beta$-generating if whenever $\|x\|\leq 1$ there are $x_\pm\geq 0$ such that $x=x_+-x_-$ and $\|x_\pm\|\leq \beta$. In particular, any $\beta$-generating cone is spanning. In the existing literature, a cone is usually called \textbf{normal} if it is $\alpha$-normal for some $\alpha\in\mathds{R}$, and a \textbf{strict $\mathcal{B}$-cone} if it is $\beta$-generating for some $\beta\in\mathds{R}$. Please note that our definitions match -- to some extent -- the intuition we gave at the beginning of this subsection.
Now, our immediate goal is to gain a better understanding the interplay between the two concepts of $\alpha$-normal and $\beta$-generating cones, which turn out to be dual to each other in a precise sense. Let us first give another definition.

\begin{Def} \label{polar def}
Let $E$ be a Banach space with dual $E^*$. The \textbf{polars} of subsets $M\subseteq E$ and $N\subseteq E^*$ are defined as
\begin{align}
M^\circ &\coloneqq \{\varphi\in E^*:\ \braket{\varphi,x}\leq 1\ \forall\, x\in M \}\, , \label{polar M} \\
N^\circ &\coloneqq \{x\in E:\ \braket{\varphi,x}\leq 1\ \forall\, \varphi\in N\}\, . \label{polar N}
\end{align}
\end{Def}

Naturally, starting from $M\subseteq E$ one can apply two times the polar operation and consider $M^{\circ\circ} \coloneqq (M^\circ)^\circ$. We show in Lemma~\ref{polar 1 lemma} of Appendix~\ref{app polars} that if $M\subseteq E$ and $N\subseteq E^*$ are convex and contain the origin then 
\begin{align}
M^{\circ\circ} &= \cl ( M )\, , \label{double polar 1} \\
N^{\circ \circ} &= \cl_{\mathcal{w}*} (N)\, . \label{double polar 2}
\end{align}
Polar sets behave well when it comes to taking saturated hulls or convex kernels. Namely, let $E$ be an ordered Banach space with closed positive cone $E_+$, and let its Banach dual $E^*$ be ordered by the cone $E_+^*$ of positive continuous functionals.
We denote by $B$ and $B^\circ$ the unit balls in $E$ and $E^*$, respectively (with the notation of Theorem~\ref{Banach-Alaoglu}, $B^\circ=B_*$). Then we have the following computation rules: 
\begin{align}
[B]^\circ &=\ ]B^\circ[ \, ,\label{computation rule 1} \\
[B^\circ]^\circ &= \cl\big( \, ]B[\, \big)\, , \label{computation rule 2}
\end{align}
where $\cl$ denotes norm closure. In Appendix~\ref{app polars} we report the proof of these formulae, adapted from the existing literature, see for instance~\cite[IV \S 1.5]{SCHAEFER} and~\cite{computation}. Here, we limit ourselves to the observation that if $[B^\circ]$ is always weak*-compact (in particular, closed), which can be seen as follows: (1) $B^\circ \cap E_+^*$ is the intersection of a weak*-closed set (i.e. $E_+^*$) with a weak*-compact set (i.e. $B^\circ$, see Theorem~\ref{Banach-Alaoglu}), hence it is weak*-compact; (2) convex hulls of the form $\co (M\cup N)$, where both $M$ and $N$ are compact, are themselves compact, because for all nets $(x_\alpha)_\alpha = (p_\alpha y_\alpha + (1-p_\alpha) z_\alpha)_\alpha \subseteq \co (M\cup N)$ we can successively extract a convergent subnet of $(y_\alpha)_\alpha\subseteq M$, of $(z_\alpha)_\alpha\subseteq N$, and finally of $(p_\alpha)_\alpha\subseteq [0,1]$.


With these tools in our hand, we can prove a duality result that will turn out to be very useful to develop the theory further. We need a preliminary lemma, which was stated for the first time in~\cite{Tukey-lemma} in a slightly different form. In~\cite{Klee-lemma} it is reported without proof, while Ellis~\cite[Lemma 7]{Ellis-dual-base} gives a complete argument, presented in a slightly simplified fashion in~\cite[p.18]{FOUNDATIONS}.

\begin{lemma}[Tukey] \label{interior closure lemma}
Let $M,N\subseteq E$ be closed, convex, and bounded subsets of $E$. Then
\bb
\interit\left( \clit\left( \coit (M\cup N)\right) \right)\subseteq \coit (M\cup N)\, .
\ee
Here interior and closure are taken with respect to the norm topology.
\end{lemma}

\begin{proof}
Let $x\in \inter\left( \cl\left( \co(M\cup N)\right) \right)$, i.e. assume that there is $\delta>0$ such that $\|x-y\|\leq 2\delta \Rightarrow y\in \inter \left( \cl\left( \co (M\cup N) \right)\right)$ for all $y\in E$. Pick $\tilde{y}_1\in E$ such that $\|x-\tilde{y}_1\|\leq \delta$. Clearly, $\tilde{y}_1\in \inter \left( \cl \left( \co (M\cup N) \right)\right)$. Since in particular $\tilde{y}_1\in \cl \left( \co (M\cup N) \right)$, we can choose $y_1\in \co(M\cup N)$ such that $\|y_1-\tilde{y}_1\|\leq \delta$. Now, define $z_1\coloneqq 2x - y_1$. One has $\|x-z_1\| = \|x-y_1\| \leq \|x-\tilde{y}_1\| + \|\tilde{y}_1 - y_1\|\leq \delta + \delta = 2\delta$, from which we deduce that $z_1\in \inter \left( \cl \left( \co(M\cup N) \right) \right)$. 

Until now, we have shown how to rewrite $x \in \inter \left( \cl \left( \co(M\cup N) \right) \right)$ as an average $x=\frac12 ( y_1 + z_1 )$, where $y_1\in \co (M\cup N)$ and $z_1\in \inter \left( \cl \left( \co(M\cup N)\right) \right)$. We can repeat this procedure with $z_1$ in place of $x$, and write $z_1=\frac12 (y_2+z_2)$ with $y_2\in \co (M\cup N)$ and $z_2\in \inter \left( \cl \left( \co(M\cup N)\right)\right)$. Going back to $x$, this gives us the representation $x=\frac12 y_1 + \frac14 y_2 + \frac14 z_2$. Iterating the process, we end up with an identity of the form $x=\sum_{n=1}^\infty 2^{-n} y_n$, where the series converges in norm. This latter fact can be seen as follows: since $M$ is bounded, the same is true for $\inter \left( \cl \left( \co(M\cup N)\right)\right)$, thus the remainder term must vanish in norm asymptotically.

Now, since $y_n\in \co (M\cup N)$ for all $n$ and both $M$ and $N$ are convex, we can write $y_n = p_n a_n + (1-p_n) b_n$ with $p_n\in [0,1]$, $a_n\in M$ and $b_n\in N$. Define $p\coloneqq \sum_{n=1}^\infty 2^{-n} p_n$, which is legitimate because the series on the right-hand side converges. We find
\begin{align*}
x &= \lim_n \sum_{k=1}^n 2^{-k} y_k \\
&= \lim_n \left( \sum_{k=1}^n 2^{-k} p_k a_k + \sum_{k=1}^n 2^{-k} (1-p_k) b_k \right) \\
&= \lim_n \sum_{k=1}^n 2^{-k} p_k a_k + \lim_n \sum_{k=1}^n 2^{-k} (1-p_k) b_k \\
&= p a + (1-p) b\, ,
\end{align*}
where $a\coloneqq \lim_n \sum_{k=1}^n \frac{p_k}{2^k p} a_k\in M$, since the sequence of partial sums is clearly Cauchy (because $M$ is bounded), $M$ is closed (hence complete) and convex. This proves that $x\in \co (M\cup N)$, as claimed.
\end{proof}

Let us state as a separate lemma the following elementary fact, which will turn out to be useful multiple times in what follows.

\begin{lemma} \label{1+epsilon closure lemma}
Let $M\subseteq E$ be a norm-bounded subset of a Banach space $E$. Then
\bb
\bigcap_{\epsilon>0} (1+\epsilon) M \subseteq \clit(M)\, .
\label{1+epsilon closure eq}
\ee
\end{lemma}

\begin{proof}
Let $x\in \bigcap_{\epsilon>0} (1+\epsilon) M$. Then for all $n\in\mathds{N}$ with $n\geq 1$ there is some $x_n\in M$ such that $x= \left( 1+\frac1n\right) x_n$. The norm-boundedness of $M$ implies that the sequence $(x_n)_n\subseteq M$ converges to $x$ in norm, since $\lim_n \|x-x_n\| = \lim_n \frac1n \|x_n\| =0$.
\end{proof}

Now we are ready to present the following result, arguably a cornerstone of the theory of ordered Banach spaces. The two claims it is made of are due to Grosberg and Krein~\cite{Krein-duality-cones} and to Ellis~\cite{Ellis-dual-base}. Here, we report the simplified proof found in~\cite{computation} and thoroughly discussed in~\cite[p.18]{FOUNDATIONS}.

\begin{thm}[Grosberg-Krein-Ellis] \label{GKE thm}
Let $E$ be an ordered Banach space with closed positive cone. Then:
\begin{enumerate}[(a)]
\item $E_+$ is $\alpha$-normal iff $E_+^*$ is $\alpha$-generating; and
\item $E_+^*$ is $\alpha$-normal iff $E_+$ is $(\alpha+\epsilon)$-generating for all $\epsilon>0$.
\end{enumerate}
\end{thm}

\begin{proof}
Let us start with statement (a). By definition, $E_+$ is $\alpha$-normal iff $[B]\subseteq \alpha B$. We claim that this condition is in turn equivalent to $[B]^\circ \supseteq \frac{1}{\alpha} B^\circ$. That the former implies the latter follows from the general rules $M\subseteq N\Rightarrow N^\circ \subseteq M^\circ$ and $(\alpha M)^\circ = \frac{1}{\alpha} M^\circ$ (where $\alpha>0$). Conversely, if $\frac{1}{\alpha}B^\circ\subseteq [B]^\circ$ by applying the same two rules we obtain $\alpha B^{\circ\circ} \supseteq [B]^{\circ\circ}$ and hence
\bbb
[B] \subseteq \cl\, [B] = [B]^{\circ\circ} \subseteq \alpha B^{\circ\circ} = \alpha B\, ,
\eee
where we applied twice~\eqref{double polar 1}, the first time with $M=[B]$, and the second with $M=B$. Now, we can use the computation rule~\eqref{computation rule 1} to deduce that $E_+$ is $\alpha$-normal iff $\frac{1}{\alpha} B^\circ \subseteq [B^\circ] =\ ]B^\circ[\,$, i.e. iff $E_+^*$ is $\alpha$-generating.

Proving claim (b) requires a bit more care. By definition, $E_+^*$ is $\alpha$-normal iff $[B^\circ]\subseteq \alpha B^\circ$. Using the fact that $[B^\circ]$ is weak*-closed (see the discussion after~\eqref{computation rule 2}) and hence $[B^\circ]^{\circ\circ} = [B^\circ]$ by~\eqref{double polar 2}, plus the usual identity $B^{\circ\circ}=B$, one sees that this is in turn equivalent to $[B^\circ]^\circ\supseteq \frac{1}{\alpha} B$. Applying~\eqref{computation rule 2}, this yields the equivalent condition $B\subseteq \alpha\, \cl \big(\, ]B[\,\big)$. Applying~\eqref{1+epsilon closure eq} to the norm-bounded set $]B[\,$, we see immediately that if $E_+$ is $(\alpha+\epsilon)$-generating for all $\epsilon>0$ then $E_+^*$ is $\alpha$-normal.

To show the converse, assume that $B\subseteq \alpha\, \cl\big( \, ]B[\, \big)$, and let us prove that $B\subseteq (\alpha+\epsilon)\,]B[$ for all $\epsilon>0$. By hypothesis,
\bbb
\inter\left( \alpha^{-1} B\right) \subseteq \inter \left( \cl\big( \, ]B[\, \big) \right) \subseteq \ ]B[\, ,
\eee
where the last step follows from Lemma~\ref{interior closure lemma}, which is applicable since $]B[\ = \co \left( (B\cap E_+) \cup -(B\cap E_+) \right)$ is the convex hull of the union of two closed, convex and bounded sets. Then, for all $\epsilon>0$ observe that $(\alpha+\epsilon)^{-1} B\subseteq \inter ( \alpha^{-1} B )\subseteq\ ]B[\,$, which amounts to saying that $E_+$ is $(\alpha+\epsilon)$-generating.
\end{proof}

\subsection{Order dual vs Banach dual} \label{subsec ord Banach dual}

We started this section by posing the question of the interplay between order and Banach duals of an ordered Banach space. The question is partially settled by the following results.

\begin{prop} \emph{\cite[V \S 3.5]{SCHAEFER} or~\cite[Corollary 1.28]{PERESSINI}.} \label{closed and spanning implies generating}
Let $E$ be an ordered Banach space with closed and spanning positive cone $E_+$. Then $E_+$ is a strict $\mathcal{B}$-cone, i.e. it is $\beta$-generating for some $\beta\in\mathds{R}$.
\end{prop}

\begin{proof}
Define $K\coloneqq \cl \big( \, ]B[ \, \big)$. We see immediately that $K$ is closed and convex, Furthermore, for any given $x\in E$ we can find a decomposition $x=x_+ - x_-$ with $x_\pm\geq 0$ by hypothesis. From this we deduce $x\in \left( \|x_+\| + \|x_-\| \right) \, ]B[$ and thus $E=\bigcup_{n\in\mathds{N}} n\, ]B[\ =\, \bigcup_{n\in\mathds{N}} n K$. Then, Lemma~\ref{neigh 0 Banach} implies that $K$ contains a neighbourhood of the origin, i.e. there is $\beta>1$ such that $\frac{1}{\beta-1} B\subseteq K$. As in the proof of Theorem~\ref{GKE thm}, thanks to~\eqref{computation rule 2} we obtain
\bbb
(\beta-1) B^\circ = \left( \frac{1}{\beta-1} B\right)^\circ \supseteq \left( \cl \big( \, ]B[ \, \big) \right)^\circ = \left( [B^\circ]^\circ \right)^\circ = [B^\circ]\, ,
\eee
implying that $E_+^*$ is $(\beta-1)$-normal. By Theorem~\ref{GKE thm}(b), we conclude that $E_+$ is $\beta$-generating, as claimed. 
\end{proof}

\begin{prop} \emph{\cite[V \S 5.5]{SCHAEFER} or~\cite[Corollary 2.17(b)]{PERESSINI}.} \label{positive continuous prop}
Let $E$ be an ordered Banach space with closed and spanning positive cone. Then every positive linear functional on $E$ is also continuous, in formula $E^\varoast \subseteq E^*$ (i.e. the order dual is contained inside the Banach dual).
\end{prop}

\begin{proof}
Assume by contradiction that there is a positive functional $\varphi\in E^*_+$ that is not continuous. Continuity and boundedness are equivalent concepts for linear functionals, hence $\varphi$ will be unbounded on $E = E_+ - E_+$. This is possible only if $\varphi$ is unbounded on the positive cone $E_+$ itself. In fact, thanks to Proposition~\ref{closed and spanning implies generating}, we know that $E_+$ is $\beta$-generating for some $\beta\geq 1$, i.e. that for all $x\in E$ there exists a decomposition $x=x_+ - x_-$ such that $x_\pm\geq 0$ and $\|x_\pm \|\leq \beta \|x\|$. If $\varphi$ were bounded on $E_+$, i.e. $|\braket{\varphi,x_\pm}|\leq k \|x_\pm\|$ for all $x_\pm \geq 0$, we would find $|\braket{\varphi,x}| \leq |\braket{\varphi,x_+}| +|\braket{\varphi,x_-}| \leq k \left( \|x_+\| + \|x_-\| \right) \leq 2 k \beta \|x\|$, contrary to the assumption that $\varphi$ is unbounded.

Therefore, we can construct a sequence $(x_n)_n\subset E_+$ with $\|x_n\|\leq 2^{-n}$ and $\braket{\varphi,x_n} \geq 1$ for all $n\in\mathds{N}$. The partial sums $y_n\coloneqq \sum_{k=0}^n x_k$ then form an increasing Cauchy sequence, which will converge to some $y = \lim_n y_n \in E_+$ (since $E_+$ is closed). From $y\geq y_n$ we get $\braket{\varphi,y} \geq \braket{\varphi,y_n} = \sum_{k=0}^n \braket{\varphi,x_k} \geq n+1$, which yields a contradiction once we take the limit on $n$.
\end{proof}

\begin{prop} \emph{\cite[V \S 3.2]{SCHAEFER}.} \label{order Banach dual coincide prop}
Let $E$ be a Banach space with normal positive cone. Then the cone formed by all bounded positive functionals generates the whole Banach dual $E^*$. In particular, if $E_+$ is normal, closed, and spanning, then we have $E^\varoast = E^*$.
\end{prop}

\begin{proof}
We have only to show that if $E_+$ is normal then every bounded functional $\varphi\in E^*$ admits a representation of the form $\varphi = \lambda - \mu$, where $\lambda,\mu$ are positive and bounded. To this purpose, let us augment the Banach space $E$ by taking the direct sum with a single copy of the real line. On the resulting space $\mathds{R}\oplus E$ one can impose several different norms, but the choice of the norm is not going to affect the proof we are presenting. For the sake of the argument, one can consider $\mathds{R}\oplus E$ to be normed by $\|(t,x)\|\coloneqq |t| + \|x\|$.

Now, for all $x\geq 0$ define the function
\bbb
p(x) \coloneqq \sup\{ \braket{\varphi,y}:\, 0\leq y\leq x \}\, .
\eee
We see immediately that: (a) $p(\lambda x) = \lambda p(x)$ for all $\lambda\geq 0$; (b) $p(x+y)\geq p(x) + p(y)$. This in turn shows that the set
\bbb
C \coloneqq \{(t,x):\ t\in\mathds{R},\, x\in E_+,\, 0\leq t\leq p(x)\} \subseteq \mathds{R}\oplus E
\eee
is in fact a cone. Now, since $E_+$ is normal we have that $\bigcup_{z_1,z_2\in B,\, z_1\leq z_2} [z_1,z_2] = [B] \subseteq \alpha B$ for some $\alpha\in \mathds{R}$, where $B$ is the unit ball of $E$. Therefore, any $y\in [0,x]$ satisfies $\frac{y}{\|x\|} \in \big[0,\frac{x}{\|x\|}\big]\subseteq [B]\subseteq \alpha B$, i.e. $\|y\|\leq \alpha \|x\|$. Consequently, we see that $p(x)\leq \alpha \|x\|$ holds for all $x\in E_+$. In turn, this implies that $(1,0)\notin \cl(C)$ (the closure of $C$ in the norm topology), as otherwise we could construct a sequence $(x_n)_n\subset E_+$ such that $\lim_n \|x_n\|=0$ but $\liminf_n p(x_n)\geq 1$, which is in contradiction with $p(x)\leq \alpha \|x\|$ for all $x\in E_+$.

Since the closed convex set $\cl(C)$ is easily seen to satisfy $\cl(C) = t\, \cl(C)$ for all $t>0$, and moreover $(1,0)\notin \cl(C)$, Corollary~\ref{Hahn-Banach cor} (b) allows us to construct a functional $h\in \left(\mathds{R}\oplus E\right)^*$ that is positive on $\cl(C)$ and such that $h(1,0)=-1$. Any bounded functional $h$ on $\mathds{R}\oplus E$ is of the form $h(s,y)\coloneqq a s + \braket{\lambda,y}$ for some $a\in \mathds{R}$ and $\lambda\in E^*$. In our case, we see immediately that $a=-1$ and that $-p(x)+\braket{\lambda,x}\geq 0$ holds for all $x\in E_+$. In particular, $\lambda$ is a positive functional and $\braket{\lambda,x}\geq \braket{\varphi,x}$ for all $x\in E_+$, from which the representation $\varphi = \lambda - (\lambda - \varphi)$ of $\varphi$ as the difference of two positive bounded functionals follows.
\end{proof}

\section{Order unit and base norm spaces} \label{sec ord unit base norm}

Throughout this section, we make another step toward increasing specialisation, and study two particular kinds of ordered Banach spaces, namely so-called \emph{order unit} and \emph{base norm} spaces. For details, we refer the reader to~\cite[III]{JAMESON} and to~\cite{FOUNDATIONS}. For the sake of completeness, we will also provide more specific references to the original papers when appropriate.
Unlike most of the content of Sections~\ref{sec topo}-\ref{sec ord Banach}, which becomes almost trivial when we deal with finite-dimensional vector spaces, the assumptions we are going to make here concern the shape of the unit ball and therefore restrict significantly the analysis even in finite dimension.

\begin{figure}[ht]
  \centering
  \includegraphics[height=7cm, width=7cm, keepaspectratio]{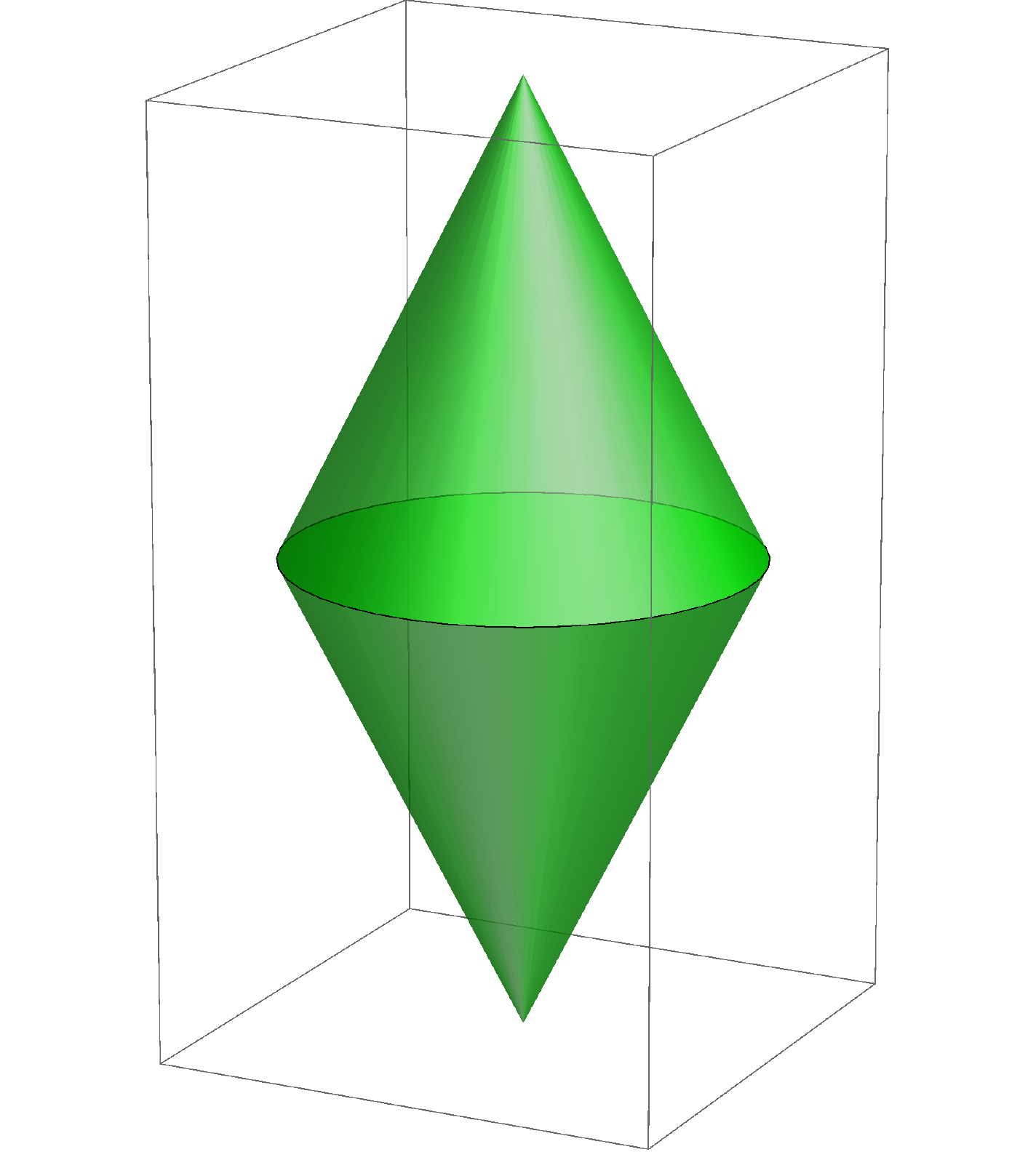}
  \caption{The unit ball of an order unit space, namely $\mathds{R}^3$ with the norm $\|(x,y,z)\|=2\sqrt{x^2+y^2} + |z|$.}
  \label{order unit ball fig}
\end{figure}

\begin{figure}[h]
  \centering
  \includegraphics[height=8.3cm, width=8.3cm, keepaspectratio]{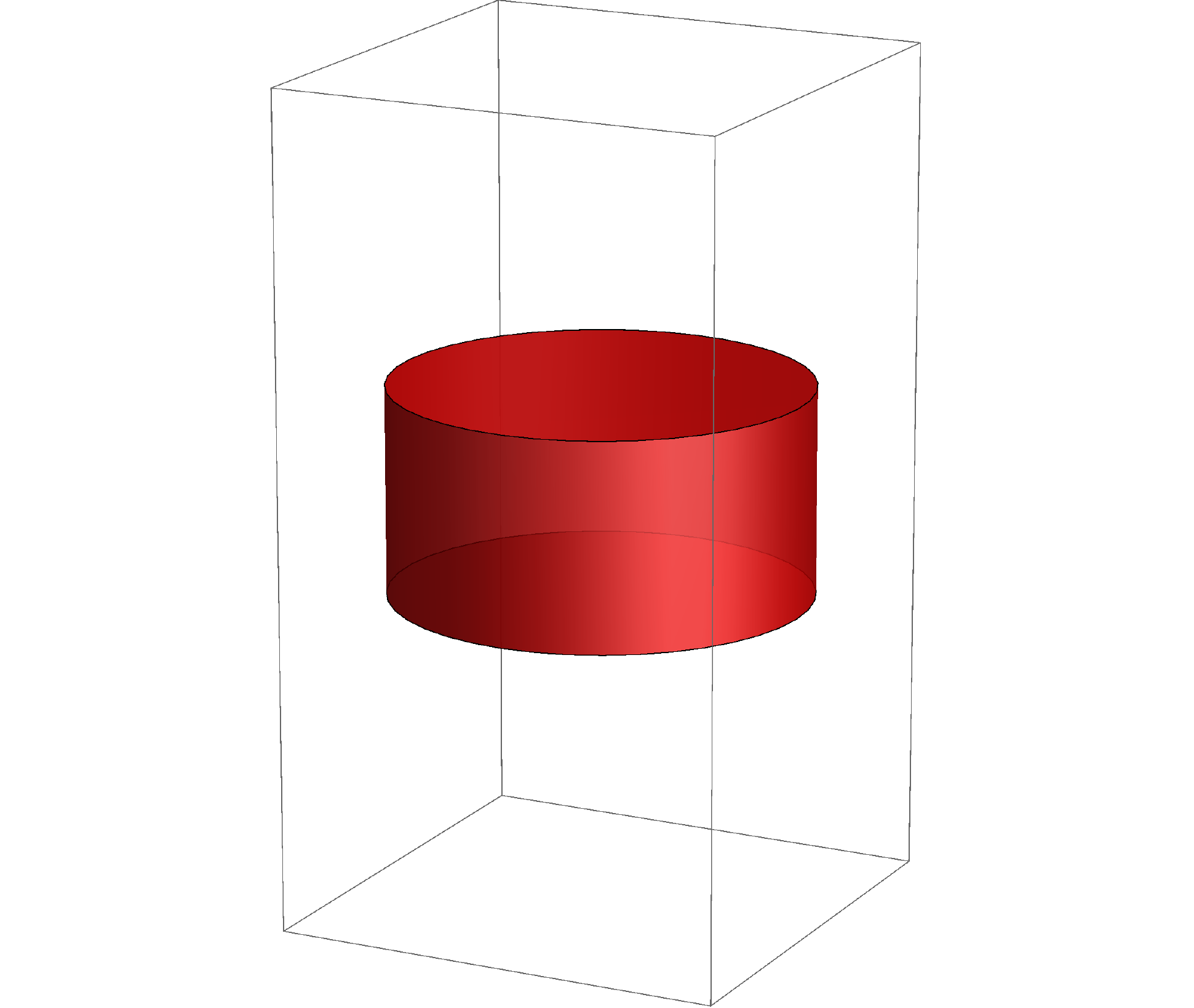}
  \caption{The unit ball of a base norm space, namely $\mathds{R}^3$ with the norm $\|(x,y,z)\|=\max\left\{\sqrt{x^2+y^2}, 2|z| \right\}$.}
  \label{base norm ball fig}
\end{figure}

Roughly speaking, an order unit space is a Banach space whose unit ball is made of two copies of the same cone, facing in opposite directions and glued together by their bases (Figure~\ref{order unit ball fig}). The original definition of order unit spaces in the context of ordered Banach spaces seems to go back to Ellis~\cite{Ellis-dual-base}.
The unit ball of a base norm space, instead, is generated via convex hull by two planar faces that are opposite to each other (Figure~\ref{base norm ball fig}). These kind of spaces seems to have been considered for the first time by Edwards~\cite{Edwards-base-norm}.
A remarkable feature of these kinds of spaces is that they are dual to each other. In fact, this is the case for the spaces depicted in Figures~\ref{order unit ball fig} and~\ref{base norm ball fig}. The purpose of this section is to define rigorously these concepts in the context of ordered Banach spaces, and present a duality theorem whose proof has been the subject of works by Krein~\cite{Krein-dual-base}, Edwards~\cite{Edwards-base-norm} and Ellis~\cite{Ellis-dual-base}.

\subsection{Order unit spaces} \label{subsec order unit}

Throughout this subsection we examine and discuss the notion of \emph{order unit space}~\cite{Ellis-dual-base}. We start by introducing an important yet very natural requirement on vector orderings.

\begin{Def}
An ordered vector space $E$ is called \textbf{Archimedean} if $x\leq 0$ whenever there is $y\in E$ such that $n x\leq y$ for all $n\in \mathds{N}$.
\end{Def}

Not all ordered vector spaces are Archimedean, not even in finite dimension. An example will help to understand why. Consider $E=\mathds{R}^2$ ordered by the cone $E_+=\{(x,y):\, \text{$x>0$ and $y>0$}\} \cup \{ (0,0)\}$. The vectors $x=(-1,0),\, y=(0,1)$ satisfy $nx\leq y$ for all $n\in \mathds{N}$, nevertheless $x\nleq 0$. Clearly, the construction of this counterexample was possible because the $E_+$ we chose was not closed. In fact, a finite-dimensional ordered vector space is Archimedean iff its positive cone is closed~\cite[V \S 4]{SCHAEFER}.

For the applications we have in mind, many ordered vector spaces come with a special element that is singled out by some physical axiom. Then, the following definition becomes relevant.

\begin{Def}
An \textbf{order unit} of an ordered vector space $E$ is a positive element $u\geq 0$ such that $E=\bigcup_{n\in \mathds{N}} n [-u,u]$, where $[-u,u]\coloneqq \{ x\in E:\, -u\leq x\leq u\}$.
\end{Def}

Again, the finite-dimensional case helps us to understand better the idea behind the definition. In fact, when $\dim E<\infty$, order units are all those elements that belong to the interior of $E_+$. An order unit is useful because it allows us to construct a norm on $E$. We are ready to give the following definition.

\begin{Def} \label{def order unit space}
An Archimedean ordered vector space $E$ with an order unit $u$ is called an \textbf{order unit space} if it is complete when equipped with the norm
\bb
p_u(x)\coloneqq \inf\left\{ t>0:\ x\in t [-u,u]\right\} .
\label{ord u norm}
\ee
\end{Def}

Note that the Archimedean axiom ensures that~\eqref{ord u norm} is a norm and not only a semi-norm. In fact, $p_u(x)=0$ implies that $x\in \frac1n [-u,u]$ for all $n\in\mathds{N}$. If $E$ is Archimedean, this is possible only if $x=0$.

\begin{ex} \label{ex l infty}
A paradigmatic example of an order unit space is the sequence space $\ell_\infty$~\cite[\S 1.6]{PERESSINI}, defined by
\bb
\ell_\infty \coloneqq \left\{ (a_n)_{n\in\mathds{N}}\subset \mathds{R}:\ \sup_{n\in\mathds{N}} |a_n|<\infty \right\} .
\label{l infty}
\ee
It is easy to verify that $\ell_\infty$ becomes complete when equipped with the norm
\bb
\|a\|_\infty \coloneqq \sup_{n\in\mathds{N}} |a_n|\, .
\label{norm infty}
\ee
Furthermore, when ordered by the (closed) cone $C \coloneqq \{ (a_n)_n: a_n\geq 0\ \forall\, n\}$, the norm~\eqref{norm infty} turns out to be induced by the order unit $u\coloneqq (1)_n$ (the constant sequence). Therefore, $\ell_\infty$ is an order unit space.
\end{ex}

Since order unit spaces are very special examples of ordered Banach spaces, they enjoy many of the properties we discussed in Section~\ref{sec ord Banach} in relation to the interplay between norm and order structure. An example is given by the following result, whose proof and corollaries can be deduced straight from the observations in~\cite[`Order unit and base norm spaces']{FOUNDATIONS}.

\begin{prop} \label{1-normal prop}
If $E$ is an order unit space, then the positive cone $E_+$ is spanning, closed in the norm topology, $1$-normal and $2$-generating. In particular, order dual and Banach dual coincide, i.e. $E^\varoast = E^*$.
\end{prop}

\begin{proof}
Clearly, the existence of an order unit $u$ already implies that the positive cone is spanning. Let us prove that $E_+$ is closed. If $(x_n)_n\subset E_+$ is a sequence of positive vectors that converges in norm to $x\in E$, then for all $n\in \mathds{N}$ we will have $\|x-x_m\|\leq \frac1n$ eventually in $m\in \mathds{N}$. Thus, $x-x_m\in \frac1n [-u,u]$ and in particular $-x\leq \frac1n u -x_m \leq \frac1n u$, i.e. $-nx\leq u$. Since this happens for all $n\in \mathds{N}$ and $E$ is Archimedean, we conclude that $-x\leq 0$ and hence that $x\geq 0$, as claimed. By Proposition~\ref{positive continuous prop}, this implies also $E^\varoast\subseteq E^*$.

To show that $E_+$ is $1$-normal, we employ the formula~\eqref{saturated hull 2} for the saturated hull $[B]$ of the unit ball $B=[-u,u]$:
\bbb
[B] = \bigcup_{x,y\in [-u,u],\, x\leq y} [x,y] = [-u,u] = B\, .
\eee
Checking that $E_+$ is $2$-generating is also easy, for any $x\in [-u,u]$ can be decomposed as $x= y_+ + y_- = 2\left( \frac12 y_+ + \frac12 y_-\right)$ with $y_\pm \coloneqq \frac{x\pm u}{2}$. Since $y_+\in [0,u] = [-u,u]\cap E_+ = B\cap E_+$ and $y_-\in [-u,0] = - B\cap E_+$, we get $x\in 2\, \co\left( (B\cap E_+) \cup -(B\cap E_+) \right)=\ ]B[$, and since $x$ was generic $B\subseteq\ ]B[$.
Finally, the equality $E^\varoast = E^*$ between order dual and Banach dual follows from Proposition~\ref{order Banach dual coincide prop}.
\end{proof}

\begin{cor} \label{unit ball order u space cor}
In an order unit space with order unit space $u$, the unit ball coincides with $[-u,u]$.
\end{cor}

\begin{proof}
From the definition~\eqref{ord u norm} it is clear that $[-u,u]\subseteq B$, where $B$ is the unit ball. On the other hand, if $x\in B$ then for all $\epsilon>0$ one has $x\in (1+\epsilon) [-u,u]$, and consequently $x\in \bigcap_{\epsilon>0} (1+\epsilon) [-u,u]\subseteq \cl \left( [-u,u] \right) = [-u,u]$, where we applied~\eqref{1+epsilon closure eq}, and the last inclusion holds because the positive cone is closed hence so are all intervals.
\end{proof}

For the sake of further developments of the theory, it is useful to have a criterion that allows us to establish that a space is an order unit space while checking as few conditions as possible. The following simple observation that partially reverses one of the claims of Proposition~\ref{1-normal prop} will come in handy.

\begin{lemma} \label{check ord u space}
Let $E$ be an ordered Banach space whose norm is given by~\eqref{ord u norm} for some $u\geq 0$. If the positive cone $E_+$ is closed, then $E$ is an order unit space.
\end{lemma}

\begin{proof}
We have just to check that $E$ is Archimedean. This is readily verified, since $-nx\leq y$ for all $n\in\mathds{N}$ implies that $\frac1n y - x\in E_+$, which in turn yields $-x\in \cl(E_+) = E_+$ upon taking the limit $n\rightarrow\infty$. From $-x\geq 0$ we deduce $x\leq 0$, as claimed.
\end{proof}

\subsection{Base norm spaces} \label{subsec base norm}

There is another way to induce a norm on an ordered vector space. In order to discuss this second alternative, we need a definition. 

\begin{Def} \label{def base}
Let $E$ be an ordered vector space. A non-empty convex subset $K$ is called a \textbf{base} of the positive cone $E_+$ if for all $x\geq 0$ there is a unique $t\geq 0$ such that $x\in t K$.
\end{Def}

Before we delve into the definition of base norm space, it is useful to have in hand a lemma characterising the concept of base in a way that lends itself better to computations. Let us start by introducing some terminology. As before, let $E$ be an ordered vector space. A positive linear functional $\varphi\in E_+^* \subseteq E^\varoast$ in the order dual is called \textbf{strictly positive} if $\braket{\varphi,x}>0$ whenever $x\in E_+$ with $x\neq 0$. In finite dimension, $\varphi$ is strictly positive iff it belongs to the relative interior of $E_+^*$. The following elementary result was originally found in~\cite{Ellis-dual-base, Edwards-base-norm}.

\begin{lemma} \label{strictly pos base lemma}
Let $E$ be an ordered vector space with positive cone $E_+$. A base $K$ for $E_+$ (if it exists) is always of the form
\bb
K = \left\{ x\geq 0:\ \braket{u,x}=1 \right\} ,
\label{base eq}
\ee
where $u \in E_+^*\subset E^\varoast$ is strictly positive.
\end{lemma}

\begin{proof}
Sets defined by~\eqref{base eq} are clearly bases, because for all $x\in E_+$, $x\in t K$ happens iff $t=\braket{u,x}$. Conversely, let $K$ be a base for $E_+$. Then let us show that the affine hull $\aff (K)$ of $K$ does not contain the origin. In fact, if $\sum_{i=1}^n \lambda_i a_i=0$ for some $a_i\in K$ and coefficients $\lambda_i$ summing to $1$, then constructing $a\coloneqq \frac1n \sum_{i=1}^n a_i \in K$ we obtain
\bbb
(1-t) a = (1-t)\frac1n \sum_{i=1}^n a_i = (1-t)\frac1n \sum_{i=1}^n a_i + t \sum_{i=1}^n \lambda_i a_i = \sum_{i=1}^n \left( \frac{1-t}{n} + t\lambda_i \right) a_i \, .
\eee
Since this is a convex combination of elements of $K$ for sufficiently small $t$, say $0<t<\epsilon$, we conclude that $(1-t) a\in K$ for all $0<t<\epsilon$, absurd since $K$ is a base. Now that we have shown that $0\notin \aff(K)$, we can construct a functional $u\in E'$ with the property that $\braket{u,x} = 1$ for all $x\in \aff(K)$. From the fact that $\braket{u,a}=1$ for all $a\in K$ we see that $u$ must be strictly positive on $E_+$.
\end{proof}

The following easily established result, which is concerned with cones and bases in ordered Banach spaces, will be needed later.

\begin{lemma} \label{take closures cones lemma}
Let $E$ be an ordered Banach space whose positive cone $E_+$ admits a base $K$. If the functional $u$ determined via Lemma~\ref{strictly pos base lemma} is continuous and satisfies $\|x\|\leq k \braket{u,x}$ for some $k\in\mathds{R}$ and all $x\in E_+$, then $\clit (E_+)$ is again a cone, it has a base $\clit(K)$, and moreover $\clit(K) = \clit(E_+) \cap u^{-1}(1)$.
\end{lemma}

\begin{proof}
We start by showing that under the aforementioned hypotheses $u$ is strictly positive on $\cl (E_+)$. Indeed, if $(x_n)_n\subset E_+$ is a sequence converging in norm to $x\in \cl(E_+)$, i.e. $\lim_n \|x-x_n\|=0$, we have $k \braket{u,x} = k \lim_n \braket{u,x_n} \geq \lim_n \|x_n\| = \|x\|$, where we used the continuity of $u$ and the fact that $\lim_n \|x_n\|=\|x\|$, easily seen to be a consequence of the triangle inequality $\left| \|x\| - \|x_n\| \right| \leq \|x-x_n\|$. Thus, $\braket{u,x}=0$ is possible only if $x=0$.

An immediate consequence of $u$ being strictly positive on $\cl(E_+)$ is that $\cl(E_+)$ is a legitimate cone, i.e. that $\cl(E_+)\cap -\cl(E_+) = \{0\}$. The set $\bar{K}\coloneqq \cl(E_+) \cap u^{-1}(1)$ is then a base by Lemma~\ref{strictly pos base lemma}, and it is clear that $\cl(K)\subseteq \bar{K}$ since the latter set is closed and contains $K$. On the other hand, if $(x_n)_n\subset E_+$ converges in norm to $x\in \cl(E_+)\cap u^{-1}(1)$, then $\lim_n\braket{u,x_n} = \braket{u,x}=1$ and thus the sequence $a_n\coloneqq \frac{x_n}{\braket{u,x_n}} \in K$ satisfies $\lim_n \|x-a_n\|=0$, in turn implying that $x\in \cl(K)$ and finally that $\cl(K) = \bar{K} = \cl(E_+)\cap u^{-1}(1)$.
\end{proof}

Given an ordered vector space whose positive cone $E_+$ is spanning and equipped with a base $K$, there is a natural way to turn it into a (semi-)normed space. In order to do this, it is useful (although not necessary) to employ the strictly positive functional $u$ that is uniquely associated with $K$ via~\eqref{base eq}.

\begin{Def} \label{def base norm space}
An ordered vector space $E$ whose positive cone $E_+$ is spanning and admits a base $K$ is called a \textbf{base norm space} if
\bb
q_K(x)\coloneqq \inf \left\{\braket{u, x_++x_-}:\ x=x_+-x_-,\ x_\pm\geq 0 \right\}
\label{base norm}
\ee
defines a norm on $E$ with respect to which $E$ is complete. Here, $u$ is associated with $K$ via~\eqref{base eq}.
\end{Def}

Some observations on the above definition are in order.
\begin{enumerate}[(a)]

\item The fact that $K$ is a base of a spanning cone ensures that the infimum appearing on the right-hand side of~\eqref{base norm} is always finite. However, in general~\eqref{base norm} will only define a semi-norm, unless the convex set $\co\left(K\cup (-K)\right)$ is \textbf{linearly bounded}, meaning that its intersections with one-dimensional subspaces are always bounded.

\item In practice, the reader will not be surprised by the fact that an example of the failure of $q_K$ to be a norm can already be constructed in finite dimension as long as we use non-closed cones. For instance, taking $E=\mathds{R}^2$ and $E_+=\{(x,y):\, y>0\} \cup \{ (0,0)\}$ yields $q_K(x,y)=|y|$, which is only a semi-norm. The non-closed nature of $E_+$ was a necessary ingredient in this example. In fact, in finite dimension $q_K$ is always a norm if $E_+$ is closed, or equivalently if $K$ is compact.\footnote{In finite dimension, a base of a closed cone is necessarily compact, and in fact every closed cone admits a compact base (see~\cite{Klee-extremal} or Lemma~\ref{finite dim cones have compact bases}).}

\item As anticipated, it is useful but not mandatory to have the functional $u$ in hand for the sake of the above definition. We could have defined the base norm directly as
\bb
\|x\| = \inf\left\{ |\alpha |+|\beta |:\ x=\alpha a + \beta b,\ a,b\in K,\ \alpha,\beta\in\mathds{R} \right\} .
\label{base norm K}
\ee
To see why~\eqref{base norm} translates to~\eqref{base norm K}, it suffices to realise that the optimal representation $x=\alpha a+ \beta b$ in~\eqref{base norm K} can always be chosen to have $\alpha\geq 0$ and $\beta\leq 0$. For instance, if $\alpha,\beta>0$ then $x=(\alpha+\beta)\frac{\alpha a + \beta b}{\alpha +\beta}$ is an alternative representation with $\alpha'=\alpha+\beta$ and $\beta'=0$, which then satisfies $|\alpha'|+|\beta'|\leq |\alpha|+|\beta|$.

\item The functional $u$ of~\eqref{base eq} is easily seen to be continuous with respect to the base norm. In fact, whenever $x=x_+ - x_-$ with $x_\pm\geq 0$ we find $|\braket{u, x}| = |\braket{u, x_+} - \braket{u,x_-}| \leq \braket{u,x_+} + \braket{u,x_-} = \braket{u,x_+ + x_-}$, and taking the infimum as in~\eqref{base norm} we see that $\|u\|_*\leq 1$. 

\end{enumerate}

Before we proceed further, let us discuss an explicit example of a base norm space.

\begin{ex} \label{ex l1}
The simplest example of a base norm space is perhaps the sequence space $\ell_1$~\cite[\S 1.6]{PERESSINI}, defined by
\bb
\ell_1 \coloneqq \left\{ (a_n)_{n\in\mathds{N}}\subset \mathds{R}:\ \sum_n |a_n|<\infty \right\} .
\label{l1}
\ee
Such a space is complete when equipped with the norm
\bb
\|a\|_1 \coloneqq \sum_n |a_n|\, ,
\label{norm 1}
\ee
and once ordered by the (closed) cone $E_+ \coloneqq \{ (a_n)_n: a_n\geq 0\ \forall\, n\}$ becomes a base norm space with base given by
\bbb
K \coloneqq \left\{ (a_n)_n\in E_+ \subset \ell_1 :\ \sum_n a_n=1 \right\} .
\eee
\end{ex}

Now, let us study a bit more in detail certain aspects of the theory of base norm spaces. Such spaces enjoy a lot of useful properties, one of the most important of which is undoubtedly the additivity of the norm on positive vectors. This simple observation is the content of the following lemma, for which we refer the reader again to~\cite[`Order unit and base norm spaces']{FOUNDATIONS}.

\begin{lemma} \label{base norm on positive}
Let $E$ be a base norm space with positive cone $E_+$, and let $u$ be the functional identified by the base $K$ via Lemma~\ref{strictly pos base lemma}. Then $\|x\|=\braket{u,x}$ for all $x\geq 0$, and thus $\|a\|=1$ for all $a\in K$.
\end{lemma}

\begin{proof}
If $x\geq 0$ then~\eqref{base norm} implies immediately that $\|x\|\leq \braket{u,x}$, but the converse is also true since any representation of the form $x=x_+ - x_-$ with $x_\pm\geq 0$ (and so $\braket{u,x_\pm}\geq 0$) satisfies $\braket{u,x_+} + \braket{u,x_-}\geq \braket{u,x_+} - \braket{u,x_-} = \braket{u,x}$.
\end{proof}

We can also wonder, how the unit ball of a base norm space can be characterised in terms of its base $K$. The following result will answer the question and help us to develop the theory further. For its proof and that of its corollary, see~\cite[`Order unit and base norm spaces']{FOUNDATIONS}.

\begin{prop} \label{unit ball base norm prop}
Let $E$ be a base norm space with associated base $K$. Then its unit ball is
\bbb
B = \clit \left( \coit (K\cup -K) \right) .
\eee
\end{prop}

\begin{proof}
From~\eqref{base norm} together with the fact that $\braket{u,a}=1$ for all $a\in K$ we see immediately that $\co (K\cup -K)\subseteq B$. Since the latter set is closed, we deduce $\cl \left( \co (K\cup -K) \right) \subseteq B$. To show the converse, we resort once more to~\eqref{base norm}, which tells us that $B\subseteq (1+\epsilon) \co(K\cup -K)$ holds for all $\epsilon>0$. The we find
\bbb
B \subseteq \bigcap_{\epsilon>0} (1+\epsilon) \co (K\cup -K) \subseteq \cl \left( \co (K\cup -K) \right) ,
\eee
where for the last step we used~\eqref{1+epsilon closure eq}.
\end{proof}

As an easy corollary of the above lemma, we find a slightly simplified formula for the norm associated with the dual of a base norm space. 

\begin{cor} \label{dual base norm cor}
Let $E$ be a base norm space with base $K$. Then the dual norm $\|\cdot\|_*$ on $E^*$ admits the expression
\bb
\|\varphi\|_* = \sup_{a\in K} |\braket{\varphi,a}|\, .
\label{dual base norm cor eq}
\ee
\end{cor}

\begin{proof}
By~\eqref{dual norm Banach} and using Proposition~\ref{unit ball base norm prop}, we have
\begin{align*}
\|\varphi\|_* &= \sup \left\{ |\braket{\varphi,x}|:\ x\in B\right\} \\
&= \sup \left\{ |\braket{\varphi,x}|:\ x\in \cl \left(\co\left( K\cup -K \right)\right) \right\} \\
&= \sup \left\{ |\braket{\varphi,x}|:\ x\in \co\left( K\cup -K \right)\right\} \\
&= \sup \left\{ |\braket{\varphi,a}|\ a\in K\right\} \, .
\end{align*}
Observe that for the third equality we used the continuity of $\varphi$.
\end{proof}

An obvious consequence of the expression~\eqref{dual base norm cor eq} is that $\|\cdot\|_*$ is monotone with respect to the dual order, i.e. that $\lambda\geq \varphi$ implies $\|\lambda\|_*\geq \|\varphi\|_*$.

We saw that the positive cone in an order unit space is necessarily closed (Proposition~\ref{1-normal prop}). A little thought should suffice to convince ourselves that the same needs not be true for base norm spaces, not even in finite dimension. One can try to amend this by taking the closure of the positive cone. A result that goes back to Ellis states that doing this does not affect the base norm~\cite{Ellis-66}.

\begin{prop} \label{(1+epsilon)-generating prop}
Let $E$ be a base norm space with positive cone $E_+$ and base $K$. Then:
\begin{enumerate}[(a)]

\item $E_+$ is spanning, $2$-normal and $(1+\epsilon)$-generating for all $\epsilon>0$;

\item $\clit (E_+)$ is again a cone with base $\clit (K)$, and $\clit (K)$ induces a base norm that coincides with that induced by $K$.

\end{enumerate}
In particular, Banach dual and order dual of a base norm space with closed positive cone coincide, in formula $E^\varoast = E^*$.
\end{prop}

\begin{proof}
Let us start with claim (a). The positive cone of a base norm space is spanning by definition. To show that it is $2$-normal, we pick $x\in [B] = (B+E_+)\cap (B-E_+)$ (where $B$ is the unit ball) and proceed to show that $\|x\|\leq 2$, i.e. that $x\in 2 B$. By hypothesis we can write $x=y_+ + z_+ = y_- - z_-$, where $\|y_\pm\|\leq 1$ and $z_\pm \geq 0$. Using Lemma~\ref{base norm on positive}, we see that $\|z_+\| + \|z_-\| = \|z_+ + z_-\| = \| y_- - y_+\|\leq \|y_-\| + \|y_+\|\leq 2$. Then by the triangle inequality
\bbb
\| 2 x\| = \left\| y_+ + z_+ + y_- - z_- \right\| \leq \| z_+\| + \| z_-\| + \|y_+\| + \|y_-\| \leq 4\, , 
\eee
i.e. $\|x\|\leq 2$. Checking that $E_+$ is $(1+\epsilon)$-generating for all $\epsilon>0$ is even easier. For a fixed $\epsilon>0$ and $x\in B$, from~\eqref{base norm} we see that we can find $x_\pm\geq 0$ such that $x=x_+ - x_-$ and $\braket{u,x_+ + x_-}\leq 1+\epsilon$. The representation $\frac{x}{1+\epsilon}=\frac{1}{1+\epsilon} (x_+ - x_-)$ then shows that $\frac{x}{1+\epsilon}\in\ ]B[$, yielding the claim.

Now we move on to proving claim (b). We already saw that the functional $u$ identified by~\eqref{base eq} is necessarily continuous with respect to the base norm (observation (d) above). Since $\braket{u,x}=\|x\|$ holds for all $x\in E_+$ by Lemma~\ref{base norm on positive}, we can apply Lemma~\ref{take closures cones lemma} and conclude that $\cl(E_+)$ is a legitimate cone with base $\cl(K) =  \cl(E_+) \cap u^{-1}(1)$. It remains to show that $\cl(K)$ induces the same base norm. This follows almost immediately from Proposition~\ref{unit ball base norm prop}, because it is easy to verify that
\bbb
\cl \left( \co (K\cup -K) \right) = \cl \left( \co \left( \cl\,K\cup - \cl\,K \right) \right)
\eee
and hence the two unit balls coincide.

Finally, the last claim is a consequence of (a) together with Proposition~\ref{order Banach dual coincide prop}.


\end{proof}

\begin{rem}
In light of Proposition~\ref{(1+epsilon)-generating prop}(b), it could seem at first sight that we can always assume without loss of generality that base norm spaces have closed positive cones. In particular, this could lead us into thinking that perhaps Banach and order dual coincide for \emph{all} base norm spaces, a stronger claim that that in Proposition~\ref{(1+epsilon)-generating prop}. However, this would not be fully justified, for the technical reason that although taking the closure of the positive cone does not affect the \emph{Banach} dual thanks to Proposition~\ref{(1+epsilon)-generating prop}(b), it can in principle affect the \emph{order} dual. In other words, there could be (unbounded) linear functionals that are positive on $E_+$ but not on its closure $\cl (E_+)$. While we do not have an explicit example of this behaviour, we wish to stress that its existence is conceivable. However, observe that Proposition~\ref{order Banach dual coincide prop} at least ensures that in any base norm space the \emph{bounded} positive functionals generate the whole Banach dual.
\end{rem}




\begin{rem}
The reader might have noticed a certain degree of asymmetry between the concept of order unit space (Subsection~\ref{subsec order unit}) and that of base norm space. That this is the case is manifest if one compares, for instance, Propositions~\ref{1-normal prop} and~\ref{(1+epsilon)-generating prop}: the positive cone is automatically closed for order unit spaces, but does not need to be such for base norm spaces (as examples in finite dimension already show). Ultimately, this asymmetry is due to the fact that we imposed the Archimedean axiom on order unit spaces from the start. If in Definition~\ref{def order unit space} we had only required $p_u$ to be a norm for some $u\geq 0$, then symmetry would have been somewhat restored. However, in most of the literature on the subject it is customary to refer to order unit spaces as complying with the requirements in Definition~\ref{def order unit space} as it is.
\end{rem}

\subsection{Duality of order unit and base norm spaces} \label{subsec duality}

As the reader might have guessed from our Examples~\ref{ex l infty} and~\ref{ex l1}, deliberately chosen to be closely related, the concepts of order unit and base norm spaces are in a certain sense dual to each other. This subsection is concerned with giving a precise sense to this intuition. Let us start by establishing an easy corollary to the characterisation of bases given in Lemma~\ref{strictly pos base lemma}.

\begin{cor} \label{base dual order unit cor}
Let $E$ be an order unit space with order unit $u$. Then
\bb
K_*\coloneqq \left\{\varphi\in E_+^*:\ \braket{\varphi,u}=1 \right\}
\label{base dual order unit}
\ee
is a base for the dual cone $E_+^*\subseteq E^*$.
\end{cor}

\begin{proof}
Once viewed as a functional on $E^*$, $u$ is clearly strictly positive on the dual cone $E_+^*$, for any $\varphi\in E_+^*$ with $\braket{\varphi,u}=0$ satisfies also $\braket{\varphi,x}=0$ for all $x\in [-u,u]$ and hence $\varphi=0$, because $[-u,u]$ is the unit ball of $E$ by Corollary~\ref{unit ball order u space cor}. By Lemma~\ref{strictly pos base lemma}, we conclude that $K_*$ defined via~\eqref{base dual order unit} is a base of $E_+^*$.
\end{proof}

Now, we are ready to present the duality theorem for base norm and order unit spaces, as was established by Edwards~\cite{Edwards-base-norm}  and Ellis~\cite{Ellis-dual-base} building upon previous progresses made by Krein~\cite{Krein-dual-base}. Our exposition of the proof is based on~\cite{FOUNDATIONS}.

\begin{thm}[Edwards-Ellis] \emph{\cite[Theorem 3.1]{FOUNDATIONS}.} \label{EE thm}
Let $E$ be an ordered Banach space whose positive cone $E_+$ is closed. Let the Banach dual $E^*$ be ordered by the cone $E_+^*$ of positive and continuous functionals. Then:
\begin{enumerate}[(a)]
\item $E$ is a base norm space iff $E^*$ is an order unit space; in this case, base and order unit are linked by~\eqref{base eq}.
\item dually, $E$ is an order unit space iff $E^*$ is a base norm space with weak*-compact base; in this case, order unit and base are linked by~\eqref{base dual order unit}.
\end{enumerate}
\end{thm}

\begin{proof}
Let us break down the proof into several steps.
\begin{itemize}

\item We start by showing that the dual of a base norm space is an order unit space with order unit given by~\eqref{base eq}. Thanks to Lemma~\ref{strictly pos base lemma}, given a base $K$ of the positive cone $E_+$, we can identify a strictly positive functional $u\in E'$ such that $K=\{x\in E_+:\, \braket{u,x}=1\}$. We know that $u$ is indeed bounded with respect to the base norm (see observation (d) in Subsection~\ref{subsec base norm}), so that $u\in E^*$ and we have a candidate for the order unit in $E^*$. Since $E_+$ is closed by hypothesis, Lemma~\ref{check ord u space} tells us that what is left to show is only that $\|\varphi\|= p_u(\varphi)$ for all $\varphi\in E^*$, where $p_u$ is given by~\eqref{ord u norm}. Applying Corollary~\ref{dual base norm cor}, we obtain
\begin{align*}
\|\varphi\|_* &= \sup \left\{ |\braket{\varphi,a}|:\ a\in K \right\} \\
&= \inf \left\{ t>0:\ -t\leq \braket{\varphi,a}\leq t \ \ \forall\ a\in K \right\} \\
&= \inf \left\{ t>0:\ -t \braket{u,a} \leq \braket{\varphi,a}\leq t \braket{u,a} \ \ \forall\ a\in K \right\} \\
&= \inf \left\{ t>0:\ -t \braket{u,x} \leq \braket{\varphi,x}\leq t \braket{u,x} \ \ \forall\ x\in E_+ \right\} \\
&= \inf \left\{ t>0:\ \varphi\in t [-u,u] \right\} \\
&= p_u(\varphi) ,
\end{align*}
as claimed.

\item Now we demonstrate that if $E^*$ is an order unit space with unit $u$ then $E$ is a base norm space with base $K\coloneqq \{x\geq 0:\, \braket{u,x}=1\}$, completing the proof of claim (a). We have to show that $K$ is really a base, and that the norm $\|\cdot\|$ on $E$ coincides with the function $q_K(x)$ in~\eqref{base norm}. First of all, as in the proof of Lemma~\ref{base norm on positive} we know that $q_K(x)=\braket{u,x}$ for all $x\geq 0$. We now employ formula~\eqref{dual formula norm Banach eq} to show that $\|x\|\leq q_K(x)$ whenever $x\in E_+-E_+$. Indeed, take a representation $x=x_+-x_-$ such that $x_\pm\geq 0$ and some $\varphi\in E^*$ such that $\|\varphi\|_*\leq 1$, i.e. $\varphi\in [-u,u]$. Then we have 
\bbb
|\braket{\varphi,x}| = |\braket{\varphi,x_+} -\braket{\varphi,x_-}| \leq |\braket{\varphi,x_+}| + |\braket{\varphi,x_-}| \leq \braket{u,x_+} + \braket{u,x_-}\, .
\eee
Taking the supremum over $\varphi\in [-u,u]$ and the infimum over $x_\pm$ this yields $\|x\|\leq q_K(x)$, as claimed. An immediate consequence is that $u$ is strictly positive on $E_+$, because any $x\geq 0$ such that $\braket{u,x}=0$ satisfies also $\|x\|\leq q_K(x)=\braket{u,x}=0$ and hence $x=0$. Then, Lemma~\ref{strictly pos base lemma} ensures that $K$ is really a base of $E_+$.

It remains to show that $q_K(x)\leq \|x\|$ for all $x\in E$ (implying in particular that $E_+$ is spanning). If $B$ is the unit ball of $\|\cdot\|$ we have clearly
\bbb
\|x\| = \inf\{ t>0:\ x\in tB\}\, .
\eee
Since $E_+^*$ is $1$-normal by Proposition~\ref{1-normal prop}, using Theorem~\ref{GKE thm} we see that $E_+$ must be $(1+\epsilon)$-generating for all $\epsilon>0$, and in particular spanning.
Then, picking $t>0$ such that $x\in tB$, since $B \subseteq (1+\epsilon)\, ]B[$ we obtain also
\bbb
x\in (1+\epsilon)t\ ]B[\ = (1+\epsilon)t\ \co\left( (B\cap E_+) \cup -(B\cap E_+) \right) ,
\eee
implying that there is a representation of the form $x= (1+\epsilon)t\, (p y_+ - (1-p)y_-)$, where $y_\pm\in B\cap E_+$ and $p\in [0,1]$. By definition of $q_K(x)$ we get
\bbb
q_K(x) \leq (1+\epsilon) t \left( p \braket{u,y_+} + (1-p) \braket{u,y_-}\right) \leq (1+\epsilon) t \, ,
\eee
where the last step follows because $\|u\|_*\leq 1$ and $y_\pm\in B$ together imply that $\braket{u,y_\pm}\leq \|u\|_* \|y_\pm\|\leq 1$. The above inequality holds for all $\epsilon>0$, so we conclude that $q_K(x)\leq t$, and after taking the infimum over $t$ this yields $q_K(x)\leq \|x\|$.

\item Let us move on to claim (b), and prove that if $E$ is an order unit space then $E^*$ is a base norm space with a weak*-compact base given by~\eqref{base dual order unit}. According to Definition~\ref{def base norm space}, we have to check that: (i) the cone $E_+^*$ is spanning; (ii) $K_*$ defined via~\eqref{base dual order unit} is a base of $E_+^*$; (iii) $K_*$ is weak*-compact; and (iv) the norm on $E^*$ is given by~\eqref{base norm}.

Now, (i) follows from Proposition~\ref{1-normal prop}, while (ii) is established by Corollary~\ref{base dual order unit cor}. As for (iii), note first that $K_*=E_+^*\cap u^{-1}(1)$ is the intersection of two weak*-closed sets and hence itself weak*-closed. For the argument that justifies why $E_+^*$ is weak*-closed, we refer the reader to the discussion at the beginning of Section~\ref{sec ord Banach}.
Next, we see that $\varphi\in K_*$ implies $\varphi\in E_+^*$ and hence $|\braket{\varphi,x}|\leq \braket{\varphi,u}=1$ for all $x\in B=[-u,u]$, with equality when $x=u$. By~\eqref{dual norm Banach}, we conclude that $\|\varphi\|_* = 1$, and in particular that $\varphi\in B^\circ$ with $B^\circ$ being the dual unit ball. Since $B^\circ$ is weak*-compact by the Banach-Alaoglu theorem (Theorem~\ref{Banach-Alaoglu}), and $K_*$ is weak*-closed, $K_*$ is a closed subset of a compact set and hence itself compact (in our case, weak*-compact). 

We now show (iv). Call $q_{K_*}(\varphi)$ the function defined as in~\eqref{base norm}, i.e.
\begin{align*}
q_{K_*} (\varphi) &= \inf \left\{ \braket{\varphi_+ + \varphi_-, u}:\ \varphi = \varphi_+ - \varphi_-,\ \varphi_\pm\in E_+^*\right\} \\
&= \inf\left\{ t>0 :\ \varphi\in t\, \co \left( K_* \cup -K_* \right) \right\} .
\end{align*}
Now, while proving claim (iii) we saw that $\|\varphi\|_*=\braket{\varphi, u}$ whenever $\varphi\in E_+^*$ (rescale and use the fact that $\varphi\in K_*\Rightarrow \|\varphi\|_*=1$). As a consequence, one gets $K_* = B^\circ \cap E_+^*$ and hence $\co \left( K_* \cup -K_* \right) =\ ]B^\circ[$ with the notation of~\eqref{convex kernel}. Since $E_+$ is $1$-normal by Proposition~\ref{1-normal prop} and hence $E_+^*$ is $1$-generating by Theorem~\ref{GKE thm}, we see that
\bbb
]B^\circ[\ = B^\circ\, .
\eee
Therefore,
\bbb
q_{K_*}(\varphi) = \inf \big\{ t>0:\, \varphi\in t \ ]B^\circ[\, \big\}  = \inf \left\{ t>0:\, \varphi\in t B^\circ \right\} = \|\varphi\|_*\, ,
\eee
concluding the proof of (iv).

\item It remains to prove that if $E^*$ is a base norm space with a weak*-compact base $K_*$ then $E$ must be an order unit space with order unit $u$ satisfying~\eqref{base dual order unit}. By Lemma~\ref{strictly pos base lemma}, we see that $K_*$ must be of the form $K_* = \{ \varphi\in E_+^*:\, f(\varphi)=1 \}$, where $f:E^*\rightarrow\mathds{R}$ is strictly positive on $E_+^*$. Then, we have to show that indeed $f$ acts as $f(\varphi)=\braket{\varphi,u}$, for some $u\in E$. Thanks to Lemma~\ref{weak*-continuous are vectors}, it is enough to show that $f$ is weak*-continuous, i.e. that $\ker(f)$ is weak*-closed. Proposition~\ref{checking weak* closedness} tells us that we have just to check that $\ker(f) \cap B^\circ$ is weak*-closed.

To this purpose, let us show that $\ker(f) \cap B^\circ = \frac12 \left( K_* - K_*\right)$. We start by proving that $\frac12 \left( K_* - K_*\right)\subseteq \ker(f) \cap B^\circ $. Pick $\varphi\in \frac12\left( K_*-K_*\right)$, i.e. assume that there is a decomposition $\varphi=\frac12 (\lambda-\mu)$ with $\lambda,\mu\in K_*$. Then, we see that $f(x) = \frac12 \left(f(\lambda)-f(\mu)\right) = 0$ and $\|\varphi\|_*\leq \frac12 \left( f(\lambda) + f(\mu) \right)=1$, i.e. $\varphi\in \ker(f)\cap B^\circ$. Note that we used the fact that $E^*$ is a base norm space to estimate $\|\varphi\|_*$ via~\eqref{base norm}. Conversely, take $\varphi\in \ker(f) \cap B^\circ$ and let us show that $\varphi\in \frac12 \left( K_*-K_* \right)$. Since $\|\varphi\|_*\leq 1$, by hypothesis for all $n>0$ one can find a decomposition $\varphi = \varphi_n^{+} - \varphi_n^{-} $ such that $f\left(\varphi_n^{+}\right) + f\left(\varphi_n^-\right) \leq 1+\frac1n$. From $\varphi\in \ker(f)$ we deduce $f\left(\varphi_n^+\right)= f\left(\varphi_n^-\right)$, so that $f\left(\varphi_n^\pm\right) \leq \frac{1+1/n}{2}$. Consequently, $\frac{2}{1+1/n} \varphi_n^{\pm}\in K_*$ for all $n\in\mathds{N}$. Since $K_*$ is weak*-compact, up to extracting subsequences we can assume that $2\varphi^\pm \coloneqq \wstarlim_n\, \frac{2}{1+1/n} \varphi_n^{\pm}\in K_*$ exist. We obtain $\varphi^+ - \varphi^- = \varphi$ and hence $\varphi\in \frac12 \left( K_* - K_* \right)$, as claimed.

With the above argument we have shown that $\ker(f) \cap B^\circ = \frac12 \left( K_* - K_*\right)$ (and indeed that $B^\circ=\co(K_* \cup -K_*)$, for that matter). Since $K_*$ is weak*-compact, the same is true for $\frac12 \left( K_* - K_* \right)$. Thus, also $\ker(f) \cap B^\circ$ is weak*-compact and in particular weak*-closed. As we said, via Proposition~\ref{checking weak* closedness} and Lemma~\ref{weak*-continuous are vectors} this implies that $f(\cdot) = \braket{\cdot, u}$ for some $u\in E$. Because of Lemma~\ref{check ord u space}, we have only to verify that the norm $\|\cdot\|$ on $E$ is given by~\eqref{ord u norm}. Thanks to the general formula~\eqref{dual formula norm Banach eq} and to Proposition~\ref{unit ball base norm prop}, we can basically repeat the reasoning used to show the first part of claim (a):
\begin{align*}
\|x\| &= \sup \left\{ |\braket{\varphi,x}|:\ \|\varphi\|_*\leq 1 \right\} \\
&= \sup \left\{ |\braket{\varphi,x}|:\ \varphi \in \cl \left( \co \left( K_* \cup -K_* \right) \right) \right\} \\
&= \sup \left\{ |\braket{\varphi,x}|:\ \varphi \in \co \left( K_* \cup -K_* \right) \right\} \\
&= \sup \left\{ |\braket{\varphi,x}|:\ \varphi\in K_* \right\} \\
&= \inf \left\{ t>0:\ -t\leq \braket{\varphi,x}\leq t \ \ \forall\ \varphi\in K_* \right\} \\
&= \inf \left\{ t>0:\ -t \braket{\varphi,u} \leq \braket{\varphi,x}\leq t \braket{\varphi,u} \ \ \forall\ \varphi\in K_* \right\} \\
&= \inf \left\{ t>0:\ -t \braket{\varphi,u} \leq \braket{\varphi,x}\leq t \braket{\varphi,u} \ \ \forall\ \varphi\in E_+^* \right\} \\
&= \inf \left\{ t>0:\ x\in t [-u,u] \right\} \\
&= p_u(x) ,
\end{align*}
where for the second last step we observed that $\braket{\varphi,x}\leq \braket{\varphi,y}$ for all $\varphi\in E_+^* = - E_+^\circ$ implies that $x-y\in \left(E_+^*\right)^\circ = - E_+^{\circ\circ} = -E_+$, this latter equality being a consequence of~\eqref{double polar 1} and of the identity $E_+ = \cl(E_+)$.

\end{itemize}
\end{proof}

\begin{rem}
In the statement of the above Theorem~\ref{EE thm}, one can freely change the expression `weak*-compact' to `weak*-closed'. In fact, a base of a base norm space is always a subset of the dual unit ball, which is in any case weak*-compact by Theorem~\ref{Banach-Alaoglu}. Thus, in this situation weak*-closedness and weak*-compactness are equivalent concepts.
\end{rem}

\section{Ludwig's embedding theorem} \label{sec Ludwig}

We are finally ready to state and prove Ludwig's embedding theorem, a cornerstone of general probabilistic theories. Referring to the discussion at the end of Subsection~\ref{subsec prep measur}, this result accomplishes two goals at once. First, it provides a geometrically intuitive representation of the set of states and effects as described in Subsection~\ref{subsec prep measur}. Secondly, it gives us a computationally convenient way to evaluate the probability functional $\mu$ subjected to Axioms~\ref{ax separated},~\ref{ax extreme effects},~\ref{ax mixtures}. In doing so, we arrive at the so-called `state space formalism' of general probabilistic theories, to be described in the next chapter, building it on firm and rigorous mathematical foundations. The theorem can be found in~\cite[IV]{LUDWIG} and~\cite{LUDWIG-DEUTUNG} (with proof) and also quoted in~\cite[III, Theorem 3.1]{LUDWIG-FOUNDATIONS-1} and~\cite[`The structure of ordered Banach spaces in axiomatic quantum mechanics']{FOUNDATIONS} (without proof).

\begin{thm}[Ludwig's embedding theorem] \emph{\cite[IV, Theorem 3.7]{LUDWIG} or~\cite[III, Theorem 3.1]{LUDWIG-FOUNDATIONS-1}.} \label{Ludwig emb thm}
Let $\Omega,\Lambda$ be sets equipped with a function $\mu: \Lambda \times \Omega \rightarrow [0,1]$ that satisfies Axioms~\ref{ax separated},~\ref{ax extreme effects},~\ref{ax mixtures}. Then, there is a base norm space $E$ with closed positive cone and an embedding of $\Omega$ into $E$ and of $\Lambda$ into the dual order unit space $E^*$ such that:
\begin{enumerate}[(a)]

\item (the images of) $\Omega$ and $\Lambda$ are convex, $0,u\in \Lambda$ (where $0\in E^*$ is the zero functional, and $u$ is the unit effect), and $\Lambda=u-\Lambda$;

\item $\mu$ coincides with the canonical bilinear form on $E^* \times E$, i.e. $\mu(\lambda, \omega)=\braket{\lambda,\omega}$ for all $\omega\in \Omega$ and $\lambda\in \Lambda$;

\item $E$ is normed by a base $K=\text{\emph{cl}}(\Omega)$ which is the norm closure of $\Omega$, and $u$ is the order unit that norms $E^*$;

\item the linear span of $\Lambda$ is weak*-dense in $E^*$.

\end{enumerate}
Moreover, the embedding satisfying all the above requirements is substantially unique, in the sense that any two embeddings are connected by an isometric order isomorphism (see the discussion at the beginning of the forthcoming Subsection~\ref{subsec uniqueness}).
\end{thm}

The proof we present through Subsections~\ref{subsec constructing} --~\ref{subsec uniqueness} is a re-elaborated version of the one in~\cite[pp.104-119]{LUDWIG}. 
The argument is quite long and especially in certain passages it relies heavily on many technical tools we have been developing through Sections~\ref{sec topo} --~\ref{sec ord Banach}, so for the sake of the presentation we break it down into simpler steps, and devote a subsection of the present section to discussing each of them.

\subsection{Constructing the basic vector spaces} \label{subsec constructing}

Let us start by defining the real vector space $V$ consisting of all the functions $x:\Lambda \rightarrow \mathds{R}$ that admit a representation of the form \bb
x(\cdot) =\sum_{i=1}^p \alpha_i \mu(\cdot, \omega_i),
\label{x emb}
\ee
for $p\in\mathds{N}$, $\alpha_i\in\mathds{R}$, and $\omega_i\in \Omega$. Clearly, $V$ is a vector space once equipped with the pointwise sum between functions. Moreover, there is an obvious embedding $\Omega \subset V$ given by $\Omega \ni \omega\mapsto \omega(\cdot)\coloneqq \mu(\omega,\cdot)$. Clearly, one can repeat the same construction exchanging $\Omega$ and $\Lambda$, and define a vector space $W$ as composed of all functions $y:\Omega\rightarrow \mathds{R}$ of the form
\bb
y(\cdot) =\sum_{j=1}^q \beta_j \mu(\lambda_j, \cdot) ,
\label{y emb}
\ee
with $q\in\mathds{N}$, $\beta_j\in\mathds{R}$, and $\lambda_j\in \Lambda$. Again, there is a natural embedding $\Lambda\subset W$ given by $\Lambda\ni \lambda\mapsto \lambda(\cdot) \coloneqq \mu(\cdot, \lambda)$. Let us stress here that the fact that these embeddings are indeed injective is a consequence of Axiom~\ref{ax separated}. We can also use Axiom~\ref{ax extreme effects} to deduce that: (1) the zero of $W$ as a vector space belongs to $\Lambda$; (2) $\Lambda$ contains also a special element $u$, i.e. (the image of) the unit effect, such that $\braket{u,\omega}=1$ for all $\omega \in \Omega$.

One can endow the two vector spaces we just constructed with a natural bilinear form $\braket{\cdot, \cdot}: W\times V \rightarrow \mathds{R}$. For $x\in V$ and $y\in W$ given by~\eqref{x emb} and~\eqref{y emb}, respectively, we define
\bb
\braket{y,x} \coloneqq \sum_{i=1}^p\sum_{j=1}^q \alpha_i \beta_j \mu(\lambda_j, \omega_i) .
\label{<> emb}
\ee
It is very easy to verify that once one thinks of $\Omega,\Lambda$ as subsets of $V,W$, one has $\braket{\cdot,\cdot}|_{\Lambda \times \Omega}= \mu(\cdot,\cdot)$. This bilinear form allows us to think of the elements of $V$ as linear functionals on $W$. This can be done by defining $W \ni y: V\rightarrow \mathds{R}$ as given by $y(x)\coloneqq \braket{y,x}$. In symbols, we write $W\subseteq V'$, where $V'$ is the algebraic dual of $V$.

An immediate advantage of having embedded $\Omega$ and $\Lambda$ into vector spaces is that Axiom~\ref{ax mixtures} now implies that $\Omega\subseteq V$ and $\Lambda\subseteq W$ are convex sets. Let us verify this claim by taking $\omega_1,\omega_2\in \Omega$, $p\in [0,1]$, and by showing that $p\omega_1 + (1-p) \omega_2\in \Omega$ (the proof for $\Lambda$ is totally analogous). Axiom~\ref{ax mixtures} ensures the existence of some $\tau\in \Omega$ such that $\mu(\lambda, \tau)=p\mu(\lambda,\omega_1)+(1-p)\mu(\lambda,\omega_2)$ for all $\lambda\in \Lambda$. This can be rephrased by saying that the function $\mu(\cdot,\tau)$ defined on $\Lambda$ coincides with the convex combination of the two functions $\mu(\cdot,\omega_1)$ and $\mu(\cdot,\omega_2)$. By definition of the embedding $\Omega\subseteq V$, we see that inside $V$ we have really $\tau=p\omega_1 +(1-p)\omega_2$, i.e. $\Omega$ is convex.

It is also clear from Axiom~\ref{ax extreme effects} that the equation $\Lambda=u-\Lambda$ holds. In fact, for $\lambda\in \Lambda$ there must exist $\lambda'\in\Lambda$ such that $\mu(\lambda',\cdot)=\mu(u,\cdot)-\mu(\lambda,\cdot)$ as functions on $\Omega$. Consequently, $\lambda'=u-\lambda$ and therefore $u-\lambda\in \Lambda$. As we shall see, this completes the verification of claim (a).

\subsection{Introducing a Banach space structure} \label{subsec introducing Banach}

Now, let us define
\bb
R \coloneqq \left\{ y\in V':\ \sup_{\omega\in \Omega} |\braket{y,\omega}| <\infty \right\} ,
\label{R emb}
\ee
which is readily verified to be a vector subspace of $V'$. We can of course turn $R$ into a normed vector space by equipping it with the norm
\bb
\|y\|^R \coloneqq \sup_{\omega\in \Omega} |\braket{y,\omega}| .
\label{norm R emb}
\ee
To check that this gives a true norm and not a semi-norm, observe that $0=\braket{y,\omega}=y(\omega)$ for all $\omega\in \Omega$ implies that $y:\Omega\rightarrow \mathds{R}$ is the zero function.

\begin{note}
We chose the notation $\|\cdot\|^R$ instead of the simpler $\|\cdot\|$ because we want to reserve the latter for the norm on $E$, the space we are aiming for.
\end{note}

Now, the following holds.

\begin{prop} \label{R Banach prop}
The space $R$ defined in~\eqref{R emb} becomes a Banach space once it is endowed with the norm~\eqref{norm R emb}.
\end{prop}

\begin{proof}
We have only to verify that $R$ is complete with the norm~\eqref{norm R emb}. Since showing this requires a bit of care, we break down the proof into elementary steps.
\begin{enumerate}[(1)]

\item Take a Cauchy sequence $(y_n)_{n\in\mathds{N}}\subset R$. A pointwise limit $y$ can be constructed via the formula $y(x)\coloneqq \lim_{n\rightarrow \infty} \braket{y_n, x}$, where the right-hand side is well defined because for a fixed $x$ as in~\eqref{x emb} we see that $(\braket{y_n,x})_{n\in\mathds{N}}\subset \mathds{R}$ is a Cauchy sequence, as follows from the inequalities
\begin{align*}
|\braket{y_n,x} - \braket{y_m,x}| &= \left| \sum_{i=1}^p \alpha_i \left( \mu(y_n,\omega_i) - \mu(y_m,\omega_i) \right) \right| \\
&\leq \sum_{i=1}^p |\alpha_i | \left| \mu(y_n, \omega_i) - \mu(y_m, \omega_i) \right| \\
&\leq \sum_{i=1}^p |\alpha_i | \left( \|y_n\|^R - \|y_m\|^R \right) \\
&\leq \left(\sum_{i=1}^p |\alpha_i |\right) \|y_n - y_m\|^R
\end{align*}
together with the definition of Cauchy sequence in a Banach space.

\item Now that we have constructed a plausible limit $y$, let us examine its properties. The bilinearity of $\braket{\cdot,\cdot}$ and the linearity of limits together ensure that $y(\cdot)$ is a linear functionals, so that $y\in V'$.

\item We have to verify that $y\in R$, i.e. that $\sup_{\omega\in \Omega} |\braket{y,\omega}| < \infty$. This can be done by writing
\begin{align*}
\sup_{\omega\in \Omega} |\braket{y,\omega}| &= \sup_{\omega\in \Omega} \lim_{n\rightarrow\infty} |\braket{y_n,\omega}| \\
&\leq \sup_{\omega\in \Omega} \sup_{n\in\mathds{N}} |\braket{y_n, \omega}| \\
&= \sup_{n\in\mathds{N}} \sup_{\omega\in \Omega} |\braket{y_n, \omega}| \\
&= \sup_{n\in\mathds{N}} \|y_n\|^R \\
&<\infty\, ,
\end{align*}
where the last inequality follows from the fact that $\left(\|y_n\|^R\right)_{n\in\mathds{N}}\subset \mathds{R}$ is itself a Cauchy sequence, which is in turn a consequence of the triangle inequality $\left|\|y_n\|^R-\|y_m\|^R\right|\leq \|y_n-y_m\|^R$. Cauchy sequences are convergent and in particular bounded, which yields the last step.

\item Finally, we have to check that the convergence $y_n \rightarrow y$ happens in the Banach space norm and not only pointwise. In other words, we have to prove that $\lim_{n\rightarrow \infty} \|y-y_n\|^R=0$. Assume by contradiction that there exists $\epsilon_0>0$ such that $\|y-y_n\|^R> 2\epsilon_0$ holds for infinitely many values of $n$. Then we will show that $(y_n)_{n\in\mathds{N}}\subset R$ can not be a Cauchy sequence, i.e. for all $N\in\mathds{N}$ there are $n,m\geq N$ such that $\|y_n-y_m\|^R\geq \epsilon_0>0$. To this purpose, let $N\in\mathds{N}$ be given. By hypothesis, we can pick $n\geq N$ such that $\sup_{\omega\in \Omega} |\braket{y-y_n,\omega}| = \|y-y_n\|^R > 2 \epsilon_0$, and consequently $\omega\in \Omega$ such that $|\braket{y-y_n, \omega}|\geq 2\epsilon_0$. Since $\braket{y-y_n, \omega}=\lim_{m\rightarrow\infty}\braket{y_m-y_n, \omega}$, this shows the existence of $m\geq N$ such that $|\braket{y_m-y_n, \omega}|\geq \epsilon_0$, finally implying that $\|y_n-y_m\|^R\geq \epsilon_0$.

\end{enumerate}
\end{proof}

Now, let us examine few further properties of $R$ that we will need in the following. First, $W$ and $R$ are both vector subspaces of $V'$, and in fact it turns out that $W\subseteq R$. In fact, for all $y\in W$ represented as in~\eqref{y emb}, we obtain
\begin{align*}
\sup_{\omega\in \Omega} | \braket{y,\omega} | &= \sup_{\omega\in \Omega} \Bigg| \sum_{j=1}^q \beta_j \mu(\lambda_j, \omega) \Bigg| \\
&\leq \sum_{j=1}^q |\beta_j |\, |\mu(\lambda_j,\omega)| \\
&\leq \sum_{j=1}^q |\beta_j | \\
&< \infty\, ,
\end{align*}
which implies $y\in R$, by definition. This fact has an analogous at the level of primal spaces. Namely, one can show in pretty much the same way that elements of $V$ are continuous (equivalently, bounded) when seen as linear functionals on $R$, or in other words that $V\subseteq R^*$, where $R^*$ is the Banach dual of $R$. 
This is because an element $x\in V$ represented as in~\eqref{x emb}, once it is seen in a natural way as a functional on $R$ via the formula $y\mapsto x(y)\coloneqq \braket{y,x}$, is clearly bounded, since
\begin{align*}
|x(y)| &= |\braket{y,x}| \\
&= \left| \sum_{i=1}^p \alpha_i \mu(y,\omega_i) \right| \\
&\leq \left( \sum_{i=1}^p |\alpha_i| \right) \sup_{\omega\in \Omega} |\braket{y,\omega}| \\
&= \left( \sum_{i=1}^p |\alpha_i| \right) \|y\|^R . 
\end{align*}
This shows that the norm of $x\in V$ as a functional is bounded from above, or more explicitly that $\|x\|^R_* \leq \sum_{i=1}^p |\alpha_i|<\infty$. In turn, this proves that $x\in R^*$.

\subsection{Inducing new norms on $V$} \label{subsec new norms}

Now that we have a norm on $R$, we can use it to construct new norms on $V$. For instance, for any given subspace $Z$ such that $W\subseteq Z\subseteq R$, we can construct a norm $\|\cdot\|_{Z}$ on $V$ given by
\bb
\|x\|_{Z} \coloneqq \sup_{z\in Z,\, \|z\|^R\leq 1} |\braket{z,x}|\, .
\label{norm tilde emb}
\ee
Because of the constraint $\|z\|^R\leq 1$, one sees that $\|x\|_{Z}\leq \|x\|^R_*<\infty$, which implies in particular that~\eqref{norm tilde emb} is well defined for all $x\in V$. Clearly, we obtain also the identity $\|\cdot\|_{R}=\|\cdot\|^R_*$. Observe that~\eqref{norm tilde emb} defines a norm, since $\braket{\lambda,x}=0$ for all $\lambda\in \Lambda\subseteq W\subseteq Z$ shows that $x$ is the zero function on $\Lambda$.

Observe that by its very definition~\eqref{norm R emb} the norm $\|\cdot\|^R$ satisfies
\bb
\|\lambda\|^R\leq 1 \qquad \forall\ \lambda\in \Lambda\, .
\label{norm R on Lambda}
\ee
This implies that all effects can be plugged into~\eqref{norm tilde emb}, yielding $\|x\|_{Z}\geq \sup_{\lambda \in \Lambda} |\braket{\lambda,x}|$. In particular, when one considers $x=\omega\in \Omega$, the unit effect $u\in \Lambda$ gives us $\|\omega\|_{Z}\geq |\braket{u,\omega}|=1$. On the other hand, since $|\braket{y,\omega}|\leq 1$ whenever $\|y\|^R\leq 1$ (by the very definition of $\|\cdot\|^R$), we know that $\|\omega\|_{Z}\leq 1$. Putting these two observations together we see that
\bb
\|\omega\|_{Z} = \braket{u,\omega} = 1\quad \forall\ \omega\in \Omega\, .
\label{norm tilde states emb}
\ee

\subsection{Constructing new Banach spaces} \label{subsec new Banach}

Now that we have endowed $V$ with the family of norms $\|\cdot\|_{Z}$, we can consider the associated completions $V_{Z}$ of $V$. Observe that even if we know that $V\subseteq R^*$, we can not a priori claim that $V_{Z}\subseteq R^*$, because $R^*$ is complete with respect to the norm $\|\cdot\|^R_*$, while the completion $V_{Z}$ is taken with respect to another (a priori incomparable) norm $\|\cdot\|_{Z}$. In other words, it will not be the case in general that all elements $x\in V_{Z}$ are bounded when viewed as functionals on $R$ with respect to $\|\cdot\|^R$. They are however bounded on $Z$ with the same norm $\|\cdot\|^R$, since for $y\in Z$ one has $|\braket{y,x}|\leq \|y\|^R \|x\|_{Z}$. This shows that $V_{Z}\subseteq Z^*$ holds true, a weaker inclusion than the one discussed above because the set of linear functionals on $R$ that are bounded only on $Z$ is in general a proper superset of $R^*$ (inside the algebraic dual $R'$).

Dually, with the same reasoning we see that every $y\in Z$ is also bounded as a functional on $V$ with respect to the norm $\|\cdot\|_{Z}$, which tells us that it can be extended to a bounded functional on $V_{Z}$. In other words, we can regard $Z$ as a subset of $V_{Z}^*$. In turn, this latter subspace of $V'$ happens to be be contained inside $R$. In fact, consider a functional $y\in V_{Z}^*\subseteq V'$, and let us show that it satisfies~\eqref{R emb}. Write
\begin{align*}
\sup_{\omega\in \Omega} |\braket{y,\omega}| &\texteq{(1)} \sup_{\omega\in \Omega\subseteq V} \frac{|\braket{y,\omega}|}{\|\omega\|_{Z}} \\
&\textleq{(2)} \sup_{0\neq x\in V} \frac{|\braket{y,x}|}{\|x\|_{Z}} \\
&\textl{(3)} \infty
\end{align*}
The justification of these steps is as follows: (1) we employed~\eqref{norm tilde states emb} to divide by $\|\omega\|_{Z}$; (2) we extended the supremum from $\Omega$ to the whole $V\supset \Omega$; (3) we used the fact that $y$ is bounded as a functional on the normed space $\left(V,\|\cdot\|_{Z}\right)$ by hypothesis. The above derivation shows also that  
\bb
\|y\|_{Z*} \leq \|y\|^R \qquad \forall\ y\in V_Z^*\, .
\label{Z* vs standard norm emb}
\ee
Let us summarise the above discussion by stating that
\bb
W\subseteq Z \subseteq V_{Z}^*\subseteq R\, .
\label{incl emb}
\ee

At this point, let us ask ourselves whether the expression~\eqref{norm tilde emb} is still valid for all $x\in V_Z$, while it was initially conceived only for $x\in V$. The answer is affirmative, as we proceed to show.

\begin{lemma} \label{expression for norm valid on all VZ}
Equation~\eqref{norm tilde emb} is valid on the whole completion $V_Z$ of $V$.
\end{lemma}

\begin{proof}
First of all, pick $x\in V_Z$ and $z\in Z\subseteq V_Z^*$ such that $\|z\|\leq 1$. Using~\eqref{Z* vs standard norm emb}, we immediately deduce that $|\braket{z,x}| \leq \|x\|_Z \|z\|_{Z*}\leq \|x\|_Z \|z\| \leq \|x\|_Z$, so that
\bbb
\|x\|_Z \geq \sup_{z\in Z,\, \|z\|^R\leq 1} | \braket{z,x} |\, .
\eee
To show the opposite inequality, consider a sequence $(v_n)_n\subset V$ such that $\lim_n \|x-v_n\|_Z = 0$. This sequence satisfies $\braket{z,x}\coloneqq \lim_n \braket{z,v_n}$ (remember that we are allowed to consider $z$ as a functional in $V_Z^*$, thus acting also on $x$). In fact, such a sequence can be used to \emph{define} the scalar product $\braket{z,x}$. But there is more: the above limit is also uniform on the set of $z\in Z$ such that $\|z\|^R\leq 1$. This is because once we fix an $\epsilon>0$ and construct $N\in \mathds{N}$ such that $\|x-v_n\|_Z\leq \epsilon$ for all $n\geq N$, using~\eqref{Z* vs standard norm emb} we will also have $|\braket{z,x}-\braket{z,v_n}|=|\braket{z,x-v_n}|\leq \|z\|_{Z*} \|x-v_n\|_Z\leq \|z\|^R \|x-v_n\|_Z \leq \epsilon$ independently of $z\in Z$ as long as $\|z\|^R\leq 1$. In turn, this allows us to conclude that
\bbb
\sup_{z\in Z,\, \|z\|^R\leq 1} |\braket{z,x}| \geq \sup_{z\in Z,\, \|z\|^R\leq 1} |\braket{z,v_n}| - \epsilon = \|v_n\|_Z -\epsilon\, .
\eee
Taking the limit on $n$ yields
\bbb
\sup_{z\in Z,\, \|z\|^R\leq 1} |\braket{z,x}| \geq \|x\|_Z -\epsilon
\eee
for all $\epsilon> 0$, i.e.
\bbb
\sup_{z\in Z,\, \|z\|^R\leq 1} |\braket{z,x}| \geq \|x\|_Z\, ,
\eee
as claimed.
In the above derivation, we used the fact that also the norm of $x$ can be written as a limit, namely $\|x\|_Z=\lim_n \|v_n\|_Z$. This can be seen easily by applying the triangle inequality to show that $\lim_n \left|\|x\|_Z - \|v_n\|_Z\right|\leq \limsup_n \|x-v_n\|_Z = 0$.
\end{proof}

\subsection{Identifying symmetric subspaces} \label{subsec symmetric}

The Banach spaces $V_Z$ are our candidates for the $E$ we want to construct. Observe that the expression~\eqref{norm R emb} of the norm on $R$ coincides with the formula for the dual of a base norm given by Corollary~\ref{dual base norm cor}. Since we want to retain this feature, we would like to consider subspaces $Z$ such that the norm on $V_Z^*$ as a Banach space coincides with the restriction of the norm on $R$. We proceed now to identify a class of $Z$s such that this happens. We need the following elementary result.

\begin{lemma} \label{norms of VZ* and R coincide on Z lemma} \emph{\cite[IV, Theorem 3.4]{LUDWIG}.}
For all $Z$ such that $W\subseteq Z\subseteq R$, the norm associated with $V_Z^*$ and the restriction of that of $R$ coincide at least on $Z$ itself.
\end{lemma}

\begin{proof}
Since~\eqref{Z* vs standard norm emb} holds, we have only to show that $\|z\|^R\leq \|z\|_{Z*}$ for all $z\in Z$. We proved that the formula~\eqref{norm tilde emb} is satisfied not only for $x\in V$ but more generally for $x\in V_Z$, hence we see immediately that $\frac{|\braket{z,x}|}{\|z\|^R}\leq \|x\|_Z$ for all $z\in Z$, $z\neq 0$. Therefore,
\bbb
\|z\|_{Z*} = \sup_{x\in V_Z,\, \|x\|_Z\leq 1} |\braket{z,x}| \leq \sup_{x\in V_Z,\, \|x\|_Z\leq 1} \|z\|^R \|x\|_Z \leq \|z\|\, ,
\eee
concluding the proof of the claim.
\end{proof} 

The above calculation shows that there is a simple way to ensure that the norm of $R$ coincides with that of $V_Z^*$ on the whole $V_Z^*$, i.e. requiring that $Z=V_Z^*$. Subspaces $Z$ such that $W\subseteq Z\subseteq R$ and $Z=V_Z^*$ will play a special role in what follows, and will be called \emph{symmetric}. The simplest example of a symmetric subspace is $R$ itself, as~\eqref{incl emb} with $Z=R$ implies that $R=V_R^*$.

\subsection{Constructing the smallest symmetric subspace} \label{subsec smallest symmetric}

Symmetric subspaces have the remarkable properties of being closed under intersections, as we proceed to show.

\begin{prop} \label{intersection symmetric prop}
If $\left\{Z_i \right\}_{i\in I}$ is a family (finite or infinite) of symmetric subspaces, their intersection $Z_\star \coloneqq \bigcap_{i\in I} Z_i$ is again a symmetric subspace.
\end{prop}

\begin{proof}
We need only to show that $V_{Z_\star}^*\subseteq Z_\star$, since the converse is generally true~\eqref{incl emb}. We start by noticing that because of the very definition of $\|\cdot\|_{Z}$~\eqref{norm tilde emb}, from the inclusions $Z_\star\subseteq Z_i$ the inequality $\|\cdot\|_{Z_\star}\leq \|\cdot\|_{Z_i}$ follows immediately. In turn, such inequalities show that functionals that are continuous with respect to $\|\cdot\|_{Z_\star}$ are also continuous with respect to $\|\cdot\|_{Z_i}$, in formula $V_{Z_\star}^*\subseteq V_{Z_i}^*$. Taking the intersection over $i\in I$ and using the identities $V_{Z_i}^* = Z_i$, one obtains 
\bbb
V_{Z_\star}^* \subseteq \bigcap_{i\in I} V_{Z_i}^* = \bigcap_{i\in I} Z_i = Z_\star\, ,
\eee
yielding the claim.
\end{proof}

Now, we can meaningfully consider the smallest symmetric subspace, call it $E_*$. Evidently, $E_*$ can also be written as the intersection of all symmetric subspaces, in formula
\bb
E_* \coloneqq \bigcap_{\scriptsize \begin{array}{c} W\subseteq Z\subseteq R\, , \\[-0.2ex] \text{$Z$ symmetric} \end{array}} Z\, .
\label{E* emb}
\ee
Now, define
\bb
E \coloneqq V_{E_*}\, .
\label{E emb}
\ee
Since $E_*$ is symmetric, we have $E_*=V_{E_*}^*=E^*$, i.e. $E^*$ is the dual Banach space of $E$. Thus, from now on we will forget about $E_*$ and rather call it $E^*$. The norm on $E$ will be denoted by $\|\cdot\|$, while $\|\cdot\|_*$ will be the corresponding dual norm on $E^*$.

We have $V\subseteq E$ and $W\subseteq E^*$, and clearly the canonical bilinear form $\braket{\cdot,\cdot}$ on $E^*\times E$ coincides with $\mu(\cdot, \cdot)$ on $\Lambda\times\Omega$, meeting requirement (b). Applying Lemma~\ref{norms of VZ* and R coincide on Z lemma} to the symmetric subspace $E^*$, and using the expression~\eqref{norm tilde emb} generalised to all $x\in V_{E_*}=E$ by Lemma~\ref{expression for norm valid on all VZ}, we establish that
\bb
\|\varphi\|_* = \|\varphi\|^R = \sup_{\omega\in \Omega} |\braket{\varphi,\omega}|\qquad \forall\ \varphi\in E^*\, .
\label{norm E* emb}
\ee
The norm on $E$ is correspondingly given by~\eqref{dual formula norm Banach eq}:
\bb
\|x\| = \sup_{\varphi\in E^*,\, \|\varphi\|_*\leq 1} |\braket{\varphi,x}| = \|x\|_{E^*} \qquad \forall\ x\in E\, .
\label{norm E emb}
\ee

\subsection{Order structure: meeting requirement (c)} \label{subsec order structure}

We now show that $E$ defined by~\eqref{E* emb} and~\eqref{E emb} can be given a structure of base norm space. In order to do this, we first define a cone on $E$ given by $C\coloneqq \mathds{R}_+\cdot \Omega = \{ t\omega:\, t\geq 0,\, \omega\in \Omega\}$. Obviously, $\Omega$ is a base for $C$, the strictly positive functional determined by Lemma~\ref{strictly pos base lemma} coinciding with the unit effect $u\in \Lambda$. By~\eqref{norm R on Lambda} and~\eqref{norm E* emb}, we see that $u$ satisfies $\|u\|_*\leq 1$, in particular it is continuous. Moreover,~\eqref{norm tilde states emb} guarantees that $\|x\|=\braket{u,x}$ holds for all $x\in C$. Then, we can apply Lemma~\ref{take closures cones lemma} and conclude that $E_+\coloneqq \cl(C)$ is a legitimate cone and thus defines an ordering on $E$. Furthermore, we also know that $\cl(\Omega) = \cl(C) \cap u^{-1}(1)$ is a base of $E_+$.

It remains to show that: (1) $E$ equipped with the closed cone $E_+$ is a base norm space with corresponding base $\cl(\Omega)$; and (2) $E$ is an order unit space with order unit $u$. Thanks to Theorem~\ref{EE thm}, we see that (2) implies (1), so we restrict ourselves to proving (2). We can follow the same style of reasoning as the one adopted in the proof of Theorem~\ref{EE thm}, and write
\begin{align*}
\|\varphi\|_* &\texteq{(1)} \sup \left\{ | \braket{\varphi,\omega}| : \, \omega\in \Omega \right\} \\
&\texteq{(2)} \sup \left\{ | \braket{\varphi,\omega}| : \, \omega\in \cl(\Omega) \right\} \\
&= \inf \left\{ t>0:\ -t \leq \braket{\varphi,\omega} \leq t\ \ \forall\ \omega\in \cl(\Omega) \right\} \\
&\texteq{(3)} \inf \left\{ t>0:\ -t \braket{u,\omega} \leq \braket{\varphi,\omega} \leq t \braket{u,\omega}\ \ \forall\ \omega\in \cl(\Omega) \right\} \\
&\texteq{(4)} \inf \left\{ t>0:\ -t \braket{u,x} \leq \braket{\varphi,x} \leq t \braket{u,x}\ \ \forall\ x\in E_+ \right\} \\
&= \inf \left\{ t>0:\ \varphi \pm t u\geq E_+^* \right\} \\
&= \inf \left\{ t>0:\ \varphi \in t [-u,u] \right\} \\
&= p_u(\varphi)\, ,
\end{align*}
where $p_u$ is given by~\eqref{ord u norm}. The justification of the above steps is as follows: (1) we applied~\eqref{norm E* emb}; (2) we recalled that $\braket{\varphi,\cdot}$ is continuous as a functional acting on $E$; (3) we employed the identity $\cl(\Omega) = \cl(C) \cap u^{-1}(1)$; (4) we used the fact that $\cl(\Omega)$ is a base of the cone $E_+$.
Now, $E_+^*$ is weak*-closed and hence norm-closed because of the discussion at the beginning of Section~\ref{sec ord Banach}. Then, Lemma~\ref{check ord u space} ensures that $\|\varphi\|_* = p_u(\varphi)$ for all $\varphi\in E^*$ is sufficient to guarantee that $E^*$ is an order unit space. As we mentioned above, Theorem~\ref{EE thm} tells us that $E$ is a base norm space with corresponding base $\cl(\Omega)$.

\subsection{Weak*-denseness: meeting requirement (d)} \label{subsec weak*-denseness}

This is perhaps the most delicate part of the proof of Ludwig's theorem, as it relies crucially on many results in Banach space theory as well as in general topology. Since many different topologies will play a significant role in what follows, throughout this subsection we suspend the usage of expressions like weak and weak* topology on (say) $E$ and $E^*$, and rather more pedantically we spell them out as $\sigma(E,E^*)$ and $\sigma(E^*,E)$, respectively.  

Let us start with some notation. Given any Banach space $F$ with dual $F^*$, and subsets $M\subseteq F$ and $N\subseteq F^*$, the \textbf{annihilators} $M^\perp$ and $^\perp N$ are given by
\begin{align}
M^\perp &\coloneqq \left\{ \varphi\in F^*:\ \braket{\varphi,x}=0\ \forall\ x\in M \right\} , \label{annihilator M} \\
^\perp \! N &\coloneqq \left\{ x\in F:\ \braket{\varphi,x}=0\ \forall\ \varphi\in N \right\} . \label{annihilator N}
\end{align}
By confronting these definitions with those of polars,~\eqref{polar M} and~\eqref{annihilator N}, we see that $M^\perp = M^\circ$ and $^\perp\! N=N^\circ$ whenever $M$ or $N$ are subspaces. Using Lemma~\ref{polar 1 lemma}, we arrive at the formulae
\begin{align}
^\perp (M^\perp) = \cl (M)\, , \label{double annihilator M} \\
(^\perp \! N)^\perp = \cl_{\mathcal{w}*}(N)\, , \label{double annihilator N}
\end{align}
valid when $M,N$ are subspaces. The interested reader can look up this standard observation in many textbooks in functional analysis (see for instance~\cite[Theorem 4.7(b)]{RUDIN} or~\cite[Proposition 2.6.6(c)]{MEGGINSON}).

We now turn to the first results of this subsection. Let us remind the reader that because of~\eqref{incl emb} one can identify $R$ with the dual subspace $V_R^*$.

\begin{lemma} \label{lemma limit norm}
For $W\subseteq Z \subseteq R=V_R^*$ and $V\subseteq V_R$, pick $x\in\, \!\! ^\perp\! Z\subseteq V_R$ and construct a sequence $(v_n)_n\subset V$ such that $\lim_{n} \|x-v_n\|_R=0$. Then
\bbb
\lim_n \|v_n\|_Z = 0\, .
\eee
\end{lemma}

\begin{proof}
In fact, if this were not the case we could find $\epsilon_0>0$ such that $\|v_n\|_Z \geq 2\epsilon_0$ happens frequently in $n$. We could then construct a sequence $(z_n)_n\subset Z$ such that $\|z_n\|^R\leq 1$ and $|\braket{z_n,v_n}|\geq \epsilon_0$ frequently in $n$. Since $x\in\, \!\! ^\perp\! Z$, we would have
\bbb
\|x-v_n\|_R = \sup_{y\in R,\, \|y\|^R\leq 1} |\braket{y,x-v_n}| \geq |\braket{z_n,x-v_n}| = |\braket{z_n,v_n}| \geq \epsilon_0\, ,
\eee
in contradiction with the assumption that $\lim_n \|x-v_n\|_R=0$.
\end{proof}

\begin{prop} \emph{\cite[IV, Theorem 3.6]{LUDWIG}.} \label{weak* closed are symmetric prop}
Let $Z$ be a subspace such that $W\subseteq Z\subseteq R=V_R^*$. If $Z$ is $\sigma(V_R^*,V_R)$-closed, then it is symmetric.
\end{prop}

\begin{proof}
We have to show that a $\sigma(V_R^*,V_R)$-closed subspace $Z$ satisfies also $V_Z^*\subseteq Z$. Naturally, the topology $\sigma(V_R^*, V_R)$ is nothing by the weak* topology induced by the Banach space $V_R$ on its dual $V_R^*$. By~\eqref{double annihilator N} we can write $Z=(^\perp \! Z)^\perp$, where the annihilators have to be taken with respect to $V_R$ and $V_R^*$. Then, pick $y\in V_Z^*$ and $x\in\, \!\! ^\perp\! Z \subseteq V_R$, and let us show that $\braket{y,x}=0$, so that we will have $y\in (^\perp\! Z)^\perp$ and the claim will follow. 

Take a sequence $(v_n)_n\subset V$ converging to $x$ in the norm $\|\cdot\|_R$. Since $y\in V_Z^*\subseteq V_R^*$, on the one hand we have $\braket{y,x} = \lim_n \braket{y,v_n}$. On the other hand, since $y$ is bounded on $V$ with respect to $\|\cdot\|_Z$, one has $|\braket{y,v_n}|\leq \|y\|_{Z*} \|v_n\|_Z$ and consequently $\lim_{n} |\braket{y,v_n}|\leq \|y\|_{Z*} \lim_n \|v_n\|_Z=0$, where the last step is an application of Lemma~\ref{lemma limit norm}.
\end{proof}

Until now, we only looked at topologies induced by the global Banach spaces $V_R,V_R^*$. However, claim (d) of Theorem~\ref{Ludwig emb thm} concerns an `intrinsic' topology, induced by the smallest symmetric subspace $E^*$ on itself. Therefore, to complete the proof we need some tools to relate the two concepts. This is the purpose of the forthcoming Proposition~\ref{inferno prop}.

\begin{lemma} \label{inferno preliminary lemma}
Let $Z$ be a subspace such that $W\subseteq Z\subseteq V_Z^*\subseteq R=V_R^*$. Then the topologies $\sigma(V_R^*,V_R)\big|_{V_Z^*}$ and $\sigma(V_Z^*,V_Z)$ satisfy $\sigma(V_R^*,V_R)\big|_{V_Z^*}\subseteq \sigma(V_Z^*,V_Z)$. In other words, the restriction of the global weak* topology to $V_Z^*$ is coarser than its own weak* topology.
\end{lemma}

\begin{proof}
First of all, observe that if you have a family $\mathcal{F}$ of functions $f: X\rightarrow \mathds{R}$ on a set $X$, and a subset $Y\subseteq X$, the restriction to $Y$ of the initial topology generated by $\mathcal{F}$ on $X$ coincides with the initial topology on $Y$, i.e. $\sigma \left( X, \mathcal{F} \right) \big|_{Y} = \sigma\left(Y,\mathcal{F}\right)$. This is a trivial consequence of the identities $\left(f|_Y\right)^{-1}(U) = f^{-1}(U)\cap Y$, which are valid for all (open) subsets $U\subseteq \mathds{R}$.

Therefore, the restriction of $\sigma(V_R^*,V_R)$ to $V_Z^*$ is identical to $\sigma(V_Z^*, V_R)$, the initial topology generated directly on $V_Z^*$ by the family of functionals of the form $x: V_Z^*\rightarrow \mathds{R}$ (acting as $x(\cdot)=\braket{\cdot, x}$), where $x\in V_R$.
By definition of initial topology (Section~\ref{sec topo}), in order to show that $\sigma(V_Z^*, V_R)\subseteq \sigma(V_Z^*, V_Z)$, we have to prove that every function $x: V_Z^*\rightarrow \mathds{R}$ with $x\in V_R$ is continuous with respect to $\sigma(V_Z^*, V_Z)$.

Using the characterisation of continuity with nets, our claim is that every net $(\varphi_\alpha)_\alpha\subseteq V_Z^*$ that converges to $\varphi\in V_Z^*$ in the $\sigma(V_Z^*, V_Z)$ topology satisfies also $\lim_\alpha \braket{\varphi_\alpha,x} = \braket{\varphi,x}$ for all $x\in V_R$ (i.e. $x:V_Z^*\rightarrow \mathds{R}$ is continuous). Now, pick $x\in V_R$, so that there exists a sequence $(v_n)_n\subset V$ with $\lim_n \|x-v_n\|_R = 0$. As a consequence, we have also $\braket{\psi, x}=\lim_n \braket{\psi,v_n}$ for all $\psi\in V_R^*$, and in particular for all $\psi\in V_Z^*$. Then, our claim becomes
\bb
\lim_\alpha \lim_n \braket{\varphi_\alpha,v_n} = \lim_n \braket{\varphi,v_n}\, .
\label{inferno preliminary eq}
\ee

Now, $(v_n)_n$ is Cauchy with respect to the norm $\|\cdot\|_R$, hence also with respect to $\|\cdot\|_Z\leq \|\cdot\|_R$. Thus, there is $x'\in V_Z$ such that $\lim_n \|x'-v_n\|_Z=0$. Since $\varphi_\alpha, \varphi \in V_Z^*$, this implies that $\lim_n \braket{\varphi_\alpha,v_n} = \braket{\varphi_\alpha, x'}$ and $\lim_n \braket{\varphi,v_n} =\braket{\varphi,x'}$. Thus,~\eqref{inferno preliminary eq} becomes
\bbb
\lim_\alpha \braket{\varphi_\alpha,x'} = \braket{\varphi,x'}\, ,
\eee
which is satisfied by hypothesis since $(\varphi_\alpha)_\alpha\subseteq V_Z^*$ converges to $\varphi\in V_Z^*$ in the $\sigma(V_Z^*, V_Z)$ topology, and $x'\in V_Z$.
\end{proof}

\begin{prop} \emph{\cite[IV, Theorem 3.5]{LUDWIG}.} \label{inferno prop}
Let $Z$ be a symmetric subspace, and let $T\subseteq Z$ be a subspace of $Z=V_Z^*$ that is $\sigma(V_Z^*,V_Z)$-closed. Then $T$ is also $\sigma(V_R^*, V_R)$-closed, in particular symmetric.
\end{prop}

\begin{proof}
In what follows, we will denote by $B_F$ the unit ball of a generic Banach space $F$. Let us start by noting that a subspace $T$ is $\sigma(V_R^*, V_R)$-closed (i.e. weak*-closed with respect to the Banach spaces $V_R$ and $V_R^*$) iff $T\cap B_R$ is $\sigma(V_R^*,V_R)$-closed, by Proposition~\ref{checking weak* closedness}. From now on, we will thus focus on this latter set. Since by Lemma~\ref{norms of VZ* and R coincide on Z lemma} the restriction of the norm of $R=V_R^*$ to $Z=V_Z^*$ coincides with the norm naturally carried by the space $V_Z^*$, we have $B_Z = B_R \cap Z = B_{V_Z^*}$. Then, from $T\subseteq Z$ we find
\bbb
T\cap B_R = (T\cap Z)\cap B_R = T\cap (Z \cap B_R) = T\cap B_Z\, .
\eee

Now, using again Proposition~\ref{checking weak* closedness} (the `easy' direction) we see that since $T$ is $\sigma(V_Z^*,V_Z)$-closed, the same happens to $T\cap B_{V_Z^*}$, i.e. to $T\cap B_Z$ (remember that $B_Z=B_{V_Z^*}$). In fact, $T\cap B_{V_Z^*}$ is a $\sigma(V_Z^*,V_Z)$-closed subset of the unit ball $B_{V_Z^*}$, which is in turn $\sigma(V_Z^*,V_Z)$-compact by the Banach-Alaoglu theorem (Theorem~\ref{Banach-Alaoglu}). Then, we conclude that $T\cap B_Z=T\cap B_{V_Z^*}$ is in fact $\sigma(V_Z^*,V_Z)$-compact. On $T\cap B_Z\subseteq V_Z^*$ the $\sigma(V_Z^*,V_Z)$ topology is weaker than (the restriction of) the $\sigma(V_R^*, V_R)$ topology because of Lemma~\ref{inferno preliminary lemma}, and clearly Hausdorff because any weak* topology is such. But by Lemma~\ref{Bourbaki lemma} a coarser Hausdorff topology on a compact set must coincide with the original topology, hence $\sigma(V_Z^*,V_Z)$ and $\sigma(V_R^*, V_R)\big|_{V_Z^*}$ are indeed the same. In particular, since $T\cap B_Z=T\cap B_R$ was $\sigma(V_Z^*,V_Z)$-compact, it is also $\sigma(V_R^*,V_R)$-compact, and in particular $\sigma(V_R^*,V_R)$-closed. As discussed above, this implies that $T$ is $\sigma(V_R^*,V_R)$-closed, and by Proposition~\ref{weak* closed are symmetric prop} that $T$ is symmetric.
\end{proof}

Now, we are ready to state the result that completes the proof of claim (d) of Theorem~\ref{Ludwig emb thm}.

\begin{prop}
Let $E,E^*$ be defined by~\eqref{E emb},~\eqref{E* emb}. Then $W$ is $\sigma(E^*,E)$-dense in $E^*$.
\end{prop}

\begin{proof}
We remind the reader that $E^*$ is nothing but the smallest symmetric subspace. Assume by contradiction that $W$ is not $\sigma(E^*,E)$-dense in $E^*$, and consider its $\sigma(E^*,E)$-closure in $E^*$, call it $\widebar{W}\subsetneq E^*$. Since $E^*$ is symmetric and $\widebar{W}$ is $\sigma(E^*,E)$-closed, we can apply Proposition~\ref{inferno prop} and conclude that also $\widebar{W}$ is symmetric. This is a contradiction because $E^*$ was the smallest symmetric subspace.
\end{proof}

\begin{rem}
We now have another way to characterise the smallest symmetric subspace $E^*$: it is the $\sigma(V_R^*, V_R)$-closure of $W$ inside $V_R^*$.
\end{rem}

\subsection{Verifying uniqueness up to isomorphisms} \label{subsec uniqueness}

Before we start proving this last claim, let us clarify the concept of order and isometric isomorphisms in the context of ordered Banach spaces. We remind the reader that a bijective linear map $\Phi:E_1\rightarrow E_2$ between two ordered vector spaces $E_1,E_2$ is called an \textbf{order isomorphism} if for all $x,y\in E_1$ one has $x\leq y \Leftrightarrow \Phi(x)\leq \Phi(y)$, or equivalently if $x\geq 0 \Leftrightarrow \Phi(x)\geq 0$. If instead $E_1,E_2$ are Banach spaces, the relevant concept is that of \textbf{isometric isomorphism}, which is a surjective linear map $\Phi:E_1\rightarrow E_2$ such that $\|\Phi(x)\|_2=\|x\|_1$ for all $x\in E_1$. Isometric isomorphisms are automatically injective and hence bijective. 
We are mainly interested in the case of $E_1,E_2$ being ordered Banach spaces. To preserve these two structures simultaneously, a surjective linear map $\Phi:E_1\rightarrow E_2$ must be at the same time an order isomorphism and an isometric isomorphism, i.e. an \textbf{isometric order isomorphism}.

Let us come to the proof of the uniqueness claim. As it turns out, we can rely on Ludwig's construction of $E$ (see~\eqref{E* emb} and~\eqref{E emb}) to show that such a subspace is isometrically and order isomorphic to any other Banach space $F$ satisfying requirements (b), (c) and (d) of Theorem~\ref{Ludwig emb thm}. Here we are going to follow this route, which is also Ludwig's route, to conclude the proof of the theorem. However, it is also instructive to prove the uniqueness of $E$ up to isometric order isomorphisms from scratch, using only the constraints expressed by the statement of Theorem~\ref{Ludwig emb thm}. For the interested readers, we report such an argument in Appendix~\ref{app Ludwig}. Here, let us confine ourselves to arguing as follows.
\begin{enumerate}[(1)]

\item Since $V,W$ are obtained via finite linear combinations from $\Omega,\Lambda$, we can think of them as linear subspaces of $F,F^*$, i.e. $V\subseteq F$ and $W\subseteq F^*$.

\item $F$ is a base norm space, hence by Corollary~\ref{dual base norm cor} the norm on $F^*$ is given by
\bbb
\|\varphi\|_* = \sup_{\omega'\in \cl(\Omega)} |\braket{\varphi,\omega'}| = \sup_{\omega\in \Omega} |\braket{\varphi,\omega}|\, .
\eee
This means that one can embed $F^*$ in $R$, i.e. $F^*\subseteq R$, and that on $F^*$ the norm of $R$ coincides with that of $F^*$.

\item Moreover, because of the above reasoning $F^*$ turns out to be a symmetric subspace, which implies that $E^*\subseteq F^*$.

\item On the other hand, we saw that $E^*$ is nothing but the closure of $W$ inside $V_R^*=R$ with respect to the $\sigma(V_R^*,V_R)$ topology. By Lemma~\ref{inferno preliminary lemma}, one has $\sigma(V_R^*,V_R)\big|_{F^*}\subseteq \sigma (F^*,F)$. Using the fact that $W$ is $\sigma(F^*,F)$-dense in $F^*$, we have
\bbb
F^* = \cl_{\sigma(F^*,F)}(W) \subseteq \cl_{\sigma(V_R^*,V_R)}(W) = E^*\, .
\eee
This proves that $E^*=F^*$.

\item Now, $E=V_{E^*}$ by definition~\eqref{E* emb}, while $F=V_{F^*}$ because by requirement (c) $V$ is norm-dense in $F$ (as is easy to verify). Therefore, one gets also $E=F$ (as Banach spaces).

\item Finally, the construction of the order structure in $E$ discussed in Subsection~\ref{subsec order structure} ensures that $E$ and $F$ are also order isomorphic with the same isomorphism that makes $E=F$ as Banach spaces. 

\end{enumerate}

\chapter{Composite systems in general probabilistic theories} \label{chapter2}

\section{Introduction} \label{sec2 intro}

Throughout the previous chapter we laid the foundation of the so-called abstract state space formalism of general probabilistic theories (GPTs). Now, we will rather take this as a starting point, and turn our attention to the study of more advanced constructions. A wise approach is to let ourselves guide by the history of quantum mechanics, and a fundamental lesson we learnt in the past century is that many interesting phenomena appear when we look at composite systems, i.e. systems made of different parties, each one with its share of a physical system.
In quantum mechanics this approach ultimately led to the discovery of nonlocality~\cite{Bell, Aspect}, which has revealed us a fundamental feature of Nature that had escaped our previous investigations. Within the context of GPTs, we will therefore study how the state space of a bipartite system can be related to those of its components.

The present chapter is organised as follows. The rest of this section is devoted to presenting the fundamental questions we want to address in a non-technical language (Subsection~\ref{subsec2 fundamental}) and to discussing our original contributions (Subsection~\ref{subsec2 contributions}). For the sake of completeness, throughout Section~\ref{sec2 basic} we provide an overview of the mathematical machinery of GPTs for single systems and introduce the formalism to deal with bipartite systems. In trying to make this chapter as self-contained as possible, we will quickly review some of the concepts already introduced in Chapter~\ref{chapter1}. In Section~\ref{sec2 examples}, we list some examples of GPTs or classes of GPTs. Besides that of acquainting the reader with the formalism, this has the purpose of providing us with an essential sample of models on which we can test conjectures and hypotheses, something that will turn out useful also in Chapter~\ref{chapter3}. In Section~\ref{sec2 sep problem}, which constitutes in some respects a mathematical detour, we are concerned with examining the separability problem, well-known from quantum mechanics, in the broader GPT setting. Despite spurring primarily from mathematical curiosity, many tools we will develop there will be needed in the rest of the thesis. The last two sections constitute the heart of the chapter and contain the most significant original material. In Section~\ref{sec2 min vs max} we introduce a problem that is in our view central to our understanding of composite systems in general probabilistic theories, and review what is already known about it. Section~\ref{sec2 beyond} is then devoted to the presentation of our progress towards a full solution. Along the way, we provide novel insights into some mathematical problems in functional analysis.

\subsection{Some fundamental questions} \label{subsec2 fundamental}

Among all the scientific revolutions the twentieth century has brought to us, we are mainly concerned here with the dramatic shift of paradigm that led us to change our picture of the physical world forever. I am naturally speaking of the discovery that our naive understanding of Nature as a classical reality is inconsistent at a fundamental level. As part of a standard undergraduate course, a student is usually taught that the reason for this is the \emph{superposition principle}, which dictates that microscopic objects have access to an infinite number of `incomparable' states, incomparable meaning that none of them can be prepared as a probabilistic mixture of the others (in mathematical language, they are extreme points of the set of states). This phenomenon corresponds to what is usually called the \emph{non-classicality} of quantum theory. As we will see, this type of non-classicality is indeed common to almost all GPTs, since classical theories have a very peculiar geometric structure (Subsection~\ref{subsec2 ex class}).

There is however another situation where the incompatibility with a classical theory emerges clearly, namely when one considers bipartite quantum states and the correlations they can exhibit. Can we analyse this kind of questions in the more general GPT setting? As a start, given two GPTs $A$ and $B$ describing two physical systems, what can we say about the new system obtained by joining them? As we saw in Chapter~\ref{chapter1}, general probabilistic theories are in a sense nothing more than a way to embed reasonable physical axioms into a coherent mathematical picture, and for this mathematical picture to have some reason to be, we should be able to describe the joint system by means of another GPT, call it $AB$.

Naturally, the GPT $AB$ is subjected to certain constraints that are the direct mathematical translations of compelling physical assumptions. A detailed explanation of these constraints will be the subject of Subsection~\ref{subsec2 bipartite}. Leaving aside the technical details, we will see that there is no unique way to construct a meaningful $AB$ that is compatible with the local models $A$ and $B$. And in fact we should not expect to find such a uniqueness, for there can be many ways one can (or can not) have access to global states that were not available when the systems were physically disjoint. As is well known, this is exactly what happens for quantum mechanical systems: states that are accessible with local operations assisted by classical communication between the two parties are commonly called \emph{separable} (originally `classically correlated'~\cite{Werner}), while global states that can not be prepared in such a way are dubbed \emph{entangled}. As the reader may know, entanglement is necessary (although not sufficient~\cite{Werner}) to generate nonlocal correlations out of a physical state via local measurements. States that are sufficiently entangled to be able to produce such correlations are called \emph{nonlocal}.

It would be hard to overestimate the importance of entanglement from the purely conceptual and -- we are tempted to say -- ontological level. Already from the early days of quantum mechanics, the centrality of the concept was very clearly perceived, although not all of the consequences were immediately appreciated. The memorable debate on the status of this phenomenon had as protagonists scientists of the calibre of Einstein~\cite{EPR}, Schr\"odinger~\cite{schr} and Bohr~\cite{Bohr}, and ultimately led to the discovery of nonlocality by Bell~\cite{Bell}. Besides, and on a different line, over the last 30 years quantum information science has proved, among the other things, how crucial entanglement is to overcome classical constraints at the operational level~\cite{Horodecki-review}.

As it turns out (Subsection~\ref{subsec2 generalities}), the same definitions of separable, entangled and nonlocal state make sense in the GPT setting as well. In view of the above discussion, we want to gain a deeper understanding of what happens when one combines two different non-classical theories. This goal can be seen as the common thread that binds together the whole first part of the present thesis.
Before detailing our programme further, let us stress an important point. In order to make the questions independent of the particular way of joining the systems, we should focus not on a specific composition rule but rather on all possible composition rules. In fact, in Subsection~\ref{subsec2 bipartite} it will emerge that we are always free to prescribe \emph{de iure} that the only allowed states of the bipartite system are separable. That being said, we move on to describing how our programme is articulated.
\begin{enumerate}[(i)]

\item First, and most importantly, we want to address the following question: \emph{is the existence of entangled states always contemplated when one combines two non-classical theories?} This question, which will be adequately formalised and treated in Section~\ref{sec2 beyond}, plays a major role in our investigation and in our conception of the problem. One could complement it with an analogous yet stronger query, by replacing the `entanglement' with `Bell nonlocality'. This latter version of the question is very well motivated from the foundational and operational point of view, as it is formulated using concretely measurable probabilities instead of rather abstract algebraic properties.

\item An even more ambitious (and much more difficult to tackle) question is then: \emph{is it possible to lower bound the maximal entanglement that can be exhibited by a composite theory, once it is known how non-classical the local theories are?} The problem here is intentionally ill-defined, and we refer the reader to Section~\ref{sec2 conclusions} for more details. Again, the same question with `Bell nonlocality' in place of `entanglement' is of interest.

\item In general, \emph{out of the many non-classical phenomena quantum information helped us to understand, which ones are generic features of non-classical theories?} This is the defining question of a very commendable programme~\cite{Howard-info-processing}, which has been pursued by many authors and already led to important results~\cite{Howard2003, Barnum-no-broad, telep-in-GPT}. Here, we want to lay emphasis on the quantitative version the the above general query: \emph{how strong is the non-classicality exhibited by quantum mechanics, in all its various forms?} Is it the strongest that is possibly conceivable, or are there more extreme examples? We will devote Chapter~\ref{chapter3} to answering this question for the particular non-classical phenomenon known as \emph{data hiding}~\cite{dh-original-1, dh-original-2}. 

\item Even though it is of perhaps secondary importance, we want to \emph{develop tools to tackle the separability problem in general probabilistic theories}, in analogy with what has been done in quantum mechanics. This means finding necessary and sufficient conditions for a state of a bipartite theory to be separable, a problem that can naturally be tackled both in general and for specific classes of models. Section~\ref{sec2 sep problem} is devoted to presenting some results in this direction.

\end{enumerate}

To the careful reader it will not escape that we are putting in practice the approach we outlined in the discussion at the beginning of Section~\ref{sec intro}, which in turn reflects our personal view on the field: general probabilistic theories should not be seen merely as a framework within which one tries to \emph{derive} quantum theory\footnote{Obviously, I do not think that the axiomatic approach to quantum theory boils down to this, which would be both ungenerous and simplistic. Instead, the many excellent works on the subject are praiseworthy in their effort to condense the peculiarity of quantum mechanics in a short list of requirements.}, but rather -- or complementarily -- as a chance to put some order into the plethora of surprising phenomena we have discovered after the advent of the quantum revolution.\footnote{Let me add a purely personal note. I would find it very disappointing if someone were to come up with a derivation of quantum mechanics from first and indisputable principles, that is, with some kind of \emph{proof} that Nature \emph{must be} quantum mechanical. To me, it would mean that there is -- at a fundamental level -- nothing else we have to understand about why things are the way they are, which is, by definition, disheartening.} Arguably, not all of these phenomena are at the same level, since some are common to all non-classical theories, some are not, and the latter must be regarded as features that are specifically quantum and thus tell us something even deeper about the structure of reality.

\subsection{Original contributions} \label{subsec2 contributions}

The purpose of this subsection is to give the reader a brief overview of the original contributions contained in this chapter, mostly in Sections~\ref{sec2 sep problem} and~\ref{sec2 beyond}. Our interest in the problems we discuss here was spurred mainly by the study of state discrimination subjected to locality constraints, as conducted in Chapter~\ref{chapter3}. Contrary to the order of appearance in the present thesis, the forthcoming Chapter~\ref{chapter3} contains what came out of our my original PhD project, while we arrived at formulating many of the problems presented in the present chapter at a later stage. In fact, how it is often the case in science, more foundational and abstract questions arise only after one has gained a deeper understanding of some of the subtleties of the theory. 

In view of all this, the reader will not be surprised by our lack of a full understanding of the many questions we formulate here. This is reflected in the abundance of conjectures and questions we put forward here (Conjectures~\ref{ultimate cj} and~\ref{ultimate cj weaker}, Questions~\ref{question Bell} and~\ref{question tensor norms}), which contribute to make this chapter more an exposition of a research project than a complete work. And in fact, almost all the results we obtain here are still unpublished, although we are confident that they will be given some publishable form soon. 
 However, as for every research project worthy of the name, what we \emph{do} have clear in mind are the questions themselves, and motivating and presenting those questions is our main goal here.

That being said, let us try to point the reader to some of the original contributions of this chapter. Section~\ref{sec2 sep problem} is devoted to the study of the separability problem in the context of GPTs, something that will come in handy later, and that to the extend of our knowledge was not the subject of previous investigation. Besides providing an elementary extension of the Woronowicz criterion~\cite{Woronowicz} to the GPT realm (Proposition~\ref{Woronowicz prop}), we also show that a low enough tensor rank is sufficient to ensure separability of a state in an arbitrary GPT (Theorem~\ref{t-rk 2 sep}). This extends an analogous quantum mechanical result by Cariello~\cite{Cariello13,Cariello14}. There are at least two good reasons why we include this generalisation among our original results: first, taking this abstract approach ends up telling us something we did not know even in the quantum mechanical case (Corollary~\ref{tensor rank 2 QM cor}); and secondly, after a few common initial steps, the proof departs significantly from the one in~\cite{Cariello13}, and becomes significantly more involved. We then move on to a discussion of the separability problem for special examples of GPTs called \emph{centrally symmetric models}. In Proposition~\ref{sep centr prop} we establish an unexpected connection between the separability problem for these models and the beautiful theory of \emph{tensor norms} as originally put forward by Grothendieck~\cite{DEFANT, RYAN}.

Section~\ref{sec2 min vs max} is devoted to turning some of the problems discussed in the above Subsection~\ref{subsec2 fundamental} into mathematically precise conjectures. Particularly central to our investigation will be Conjecture~\ref{ultimate cj}, the very formulation of which constitutes, to the extent of our knowledge, an original contribution. Despite the fact that we deem it of prime importance for our understanding of the intrinsic features of the phenomenon of entanglement in physical theories, the problem does not seem to have drawn enough attention in the existing literature. The rest of the section is dedicated a detailed exposition of what is currently known about the question, which is mostly due to Namioka and Phelps~\cite{NP}. We report their findings as Theorem~\ref{NP thm}. 

Next, Section~\ref{sec2 beyond} presents our progress towards answering Conjecture~\ref{ultimate cj} and some related questions. We first investigate Question~\ref{question Bell}, which asks for a strengthening of Conjecture~\ref{ultimate cj} with the notion of nonlocality displacing that of entanglement. In Theorem~\ref{NP nonlocal}, we extend the results of Namioka and Phelps (Theorem~\ref{NP thm}) to show that such a strengthened conjecture holds true at least for the special case considered in said Theorem~\ref{NP thm}. We then go back to Conjecture~\ref{ultimate cj} (or more precisely its slightly weaker form, Conjecture~\ref{ultimate cj weaker}), and examine it within the context of centrally symmetric models. We will explain in detail why we deem this restriction of the problem worth investigation. While looking at the specific features of this restricted problem, we will stumble upon some universal Banach space constants, defined in~\eqref{R} and~\eqref{S}, whose investigation may be of independent interest in functional analysis. Here we limit ourselves to showing a nontrivial lower bound on the second such constant (Theorem~\ref{thm c ratio}), which can be regarded as the main original achievement of the chapter, since it leads us to answering Conjecture~\ref{ultimate cj weaker} in the affirmative for the special case of centrally symmetric models (Corollary~\ref{thm c ratio cor}). 
Although we will try to give the reader some intuition why Theorem~\ref{thm c ratio} holds, the fact that its claim concerns all Banach spaces makes its proof somewhat long and laborious. We discuss some limitations on the constant appearing in Theorem~\ref{thm c ratio} through a somewhat illuminating example in Appendix~\ref{app S(2)}.

Finally, let us mention some minor contributions that are believed to be partly original in spirit, either for the approach we adopt or for the proof technique we use. Lemma~\ref{tensor product lemma}, which tells us how to join two physical systems that are locally described by GPTs, is very well known. However, we hope that the readers will find the simple proof we present in Appendix~\ref{app tensor} useful and worth the space it takes. The definition of locally constrained sets of measurements (Definition~\ref{locally constr}) comes straightforwardly from the quantum mechanical case. In spite of this, some amount of work is required to formalise what we mean by LOCC protocols in the GPT setting.\footnote{I am grateful to Howard Barnum and Matthias Christandl for enlightening discussions on this topic.} Next, centrally symmetric models as given in Definition~\ref{centr} seem to have been considered here (and in our paper~\cite{ultimate}) for the first time. Finally, although it does not represent a substantial progress in any respect, we are particularly proud of the approach to Bell inequalities and Bell nonlocality outlined in Subsection~\ref{subsec2 other ex}, for we found it conceptually clarifying and notationally convenient (although this is also a matter of personal taste).

\section{Basic theory} \label{sec2 basic}

\subsection{Some preliminaries on Banach spaces and tensor norms} \label{subsec2 tensor norms}

We start this section with some mathematical notions on finite-dimensional Banach spaces and tensor norms. The assumption of finite dimension simplifies heavily the technical tools needed in analysing many of the problems we will investigate, but this is not the only reason why we make it. In fact, some of those problems themselves make only sense in the finite-dimensional setting. This is the case, for instance, for the questions raised in Chapter~\ref{chapter3}.

\begin{note}
From now on, all vector spaces we will deal with are surreptitiously assumed to be finite-dimensional.
\end{note}

One of the immediate advantages of the finite-dimensional setting is that we have no longer to worry about specifying the topology carried by our vector space $V$. In fact, there is a natural topology that comes from the isomorphism with $\mathds{R}^d$ ($d=\dim V$). When we make use of topological concepts like closedness, this is the topology we are referring to. A well-known property of this kind of topologies is that a subset $K\subseteq V$ is compact iff it is closed and bounded. Another consequence of the isomorphism with $\mathds{R}^d$ is that the assumption of completeness in Definition~\ref{def Banach} can be dropped with no repercussions. We will still speak of Banach spaces, but in finite dimension this notion coincides with that of normed space.
Furthermore, there are some results that hold specifically for finite-dimensional Banach spaces. An example that is worth mentioning since we will make use of it is the following.

\begin{lemma}[Auerbach] \emph{\cite[Appendix A.4]{DEFANT}.} \label{Auerbach lemma} Let $\left(V, \|\cdot\|\right)$ be a Banach space. If $d=\dim V<\infty$, then there is a basis $\{v_1, \ldots, v_d\}$ of $V$ (called the \textbf{Auerbach basis}) such that $\|v_i\|=1=\|v_i^*\|_*$ for all $i=1,\ldots,d$, where $\{v_i^*\}_i$ denotes the corresponding dual basis, and $\|\cdot\|_*$ the dual norm as in~\eqref{dual norm Banach}.
\end{lemma}

\begin{rem}
The above result is nontrivial because we are requiring a basis and its dual basis to be \emph{simultaneously} normalised.
\end{rem}

We now move on to discussing possible ways to turn a tensor product of two Banach spaces into a Banach space itself. In the finite-dimensional setting, the only problem we have to worry about is how to construct a norm on the tensor product of the two spaces. As we shall see, there are some natural requirements to be imposed on such norm, most notably that we would like the norm of a simple tensor $x\otimes y$ to equal the product of the local norms of $x$ and $y$, with an analogous equality holding at the dual level. Norms meeting these requirements are called \emph{tensor norms}~\cite{DEFANT, RYAN}.

The reason why we want to devote a full subsection to a discussion of these mathematical objects may not be apparent yet, but tensor norms will turn out to play a major role throughout the present chapter (particularly in Subsection~\ref{subsec2 sep centr} and~\ref{subsec2 univ ent centr}) and to a larger extent in the forthcoming Chapter~\ref{chapter3}. There, we will even sharpen some of our technical tools and use them to tackle problems related to the ultimate effectiveness of state discrimination subjected to locality constraints.

We start by reminding the reader~\cite{DEFANT, RYAN} that given two finite-dimensional Banach spaces $V_{A},V_{B}$ (whose norms will be equally denoted by $\|\cdot\|$ for simplicity), there are two notable ways in which one can construct a norm on the tensor product $V_{A}\otimes V_{B}$. The first construction yields the so-called \textbf{injective norm}, which can be expressed as
\bb
\|Z\|_{\varepsilon} \coloneqq \max\left\{ \braket{\varphi\otimes \lambda, Z}:\ \, \varphi\in V_{A}^{*},\ \|\varphi\|_{*}\leq 1,\ \lambda\in V_{B}^{*},\ \|\lambda\|_{*}\leq 1 \right\} .
\label{inj}
\ee
The second norm we are interested in goes under the name of \textbf{projective norm}, and is defined as
\bb
\|Z\|_{\pi} \coloneqq \min\left\{ \sum_{i=1}^{n} \|x_{i}\|\,\|y_{i}\| :\ \, n\in\mathds{N},\ Z=\sum_{i=1}^{n} x_{i}\otimes y_{i} \right\} .
\label{proj}
\ee
These two norms are dual to each other in the following sense. Thinking of $V_{A}^{*}, V_{B}^{*}$ as Banach spaces equipped with the dual norms $\|\cdot\|_{*}$, we can construct the associated injective and projective tensor norms on $V_{A}^{*} \otimes V_{B}^{*}$, denoted by $\|\cdot\|_{*\varepsilon}$ and $\|\cdot\|_{*\pi}$, respectively. One could wonder how these norms compare to the duals to~\eqref{inj} and~\eqref{proj}, denoted by $\|\cdot\|_{\varepsilon*}, \|\cdot\|_{\pi*}$, respectively. As it turns out, one has
\bb
\|\cdot\|_{*\varepsilon} = \|\cdot\|_{\pi*}\, ,\qquad \|\cdot\|_{*\pi} = \|\cdot\|_{\varepsilon*}\, .
\label{dual inj proj}
\ee

\begin{note}
The notation is intended to help the reader via simple graphic rules. For instance, the symbol $\|\cdot\|_{\varepsilon*}$ stands for `first construct the injective norm, then take the dual', and conversely $\|\cdot\|_{*\pi}$ means `first take the dual norms, then construct the projective norm out of them'. Then, the above identities can be recovered easily by remembering that taking a $*$ from the external to the internal position (or vice versa) causes an exchange $\varepsilon\leftrightarrow \pi$.
\end{note}

Concerning the comparison between injective and projective norms, the inequality $\|Z\|_{\varepsilon} \leq \|Z\|_{\pi}$ is easily seen to hold for all $Z\in V_{A}\otimes V_{B}$. 
To see why this is the case, take a tensor $Z\in V_A\otimes V_B$, and consider a decomposition $Z=\sum_{i=1}^n x_i\otimes y_i$ as in~\eqref{proj}. For any two functionals $\varphi\in V_A^*$ and $\lambda\in V_B^*$ such that $\|\varphi\|_*, \|\lambda\|_*\leq 1$, one has $|\braket{\varphi, x_i}|\leq \|x_i\|$ and $|\braket{\lambda,y_i}|\leq \|y_i\|$, so that
\begin{align*}
|\braket{\varphi \otimes \lambda, Z}| &= \bigg| \sum_{i=1}^n \braket{\varphi, x_i} \braket{\lambda, y_i} \bigg| \\
&\leq \sum_{i=1}^n |\braket{\varphi, x_i}|\, |\braket{\lambda, y_i}| \\
&\leq \sum_{i=1}^n \|x_i\|\, \|y_i\| \\
&\leq \|Z\|_\pi\, ,
\end{align*}
where in the last step we employed~\eqref{proj}. Maximising over $\varphi,\lambda$ as in~\eqref{inj} yields the claim.
We will write symbolically
\bb
\|\cdot\|_{\varepsilon} \leq \|\cdot\|_{\pi}\, .
\label{inj=<proj}
\ee
As it turns out, both $\|\cdot\|_\varepsilon$ and $\|\cdot\|_\pi$ yield the product of the local norms when evaluated on simple tensors, i.e. the identity $\|x\otimes y\|_{\varepsilon}=\|x\otimes y\|_{\pi}=\|x\|\, \|y\|$ holds. Because of~\eqref{dual inj proj}, the same equality holds at the level of the dual spaces, which makes injective and projective norm `tensor norms' in the sense outlined above. Establishing an inequality that complements~\eqref{inj=<proj} will be one of the main steps for the solution of the problems presented in Chapter~\ref{chapter3} (Theorem~\ref{prop proj=<ninj}).

Let us now present an example of the construction of injective and projective norms we sketched above. The reader will soon recognise in our case of study more familiar notions of matrix norms.

\begin{ex} \label{ex inj proj}
Consider two Euclidean spaces $V_{A}$ and $V_{B}$, whose norms we denote by $|\cdot|_{2}$. We can identify $V_{A}\otimes V_{B}$ with the set of $d_{A}\times d_{B}$ real matrices, denoted by $\mathds{R}^{d_{A}\times d_{B}}$. With this convention, for a given vector $Z\in V_{A}\otimes V_{B}$ and two functionals $v \in \mathds{R}^{d_{A}} \simeq V_{A}^{*},\, w\in \mathds{R}^{d_{B}}\simeq V_{B}^*$, we can write $\braket{v\otimes w, Z}=v^{T} Z w$. Then, remembering that Euclidean norms are self-dual, from~\eqref{inj} we obtain in this special case
\begin{align*}
\|Z\|_{\varepsilon} &= \max\{ v^{T}Zw:\ |v|_{2*},|w|_{2*}\leq 1 \} \\
&= \max\{ v^{T}Zw:\ |v|_{2},|w|_{2}\leq 1 \} \\
&= \|Z\|_{\infty}\, ,
\end{align*}
where $\|Z\|_\infty$ denotes the operator norm of the matrix $Z$, i.e. its largest singular value. The fact that $|\cdot|_{2*}=|\cdot|_{2}$ also implies that $\|\cdot\|_{*\varepsilon}=\|\cdot\|_{\varepsilon}$. This together with~\eqref{dual inj proj} shows that
\bbb
\|\cdot\|_{\pi} = \|\cdot\|_{*\varepsilon*} = \|\cdot\|_{\varepsilon*} = \|\cdot\|_{\infty *} = \|\cdot\|_{1}\, ,
\eee
where $\|\cdot\|_1$ is the trace norm, i.e. the sum of all singular values.
\end{ex}

\subsection{Single systems} \label{subsec2 single}

As we already mentioned, the theory we develop in the present chapter builds upon Ludwig's embedding theorem (Theorem~\ref{Ludwig emb thm}). In fact, the definition of GPT we are going to adopt makes use of the concepts appearing in the claim of Theorem~\ref{Ludwig emb thm}.



Let us start by reminding the reader that a subset $K$ of a real vector space $V$ is called \textbf{convex} if $px+(1-p)y\in K$ whenever $x,y\in K$ and $p\in [0,1]$. For an introduction to the basics of convex geometry, we refer the reader to~\cite{SIMON}. Let us present here a brief review of the few properties we will need in the following.

A point $x$ of a convex set $K$ is called \textbf{extreme} if it can not be written as a nontrivial convex combination of other points in the set, i.e. if $x=p y + (1-p) z$ for $y,z\in K$ and $p\in (0,1)$ implies that $y=z=x$. The extreme points of a compact convex set $K$, collectively denoted as $E(K)$, have the remarkable property that their convex combinations cover the whole $K$, as expressed by the celebrated Minkowski-Carath\'eorody theorem~\cite{MINKOWSKI, Caratheodory}.

\begin{thm}[Minkowski-Carath\'eorody] \emph{\cite[Theorem 8.11]{SIMON}.} \label{MC thm}
Let $K\subseteq V$ be a compact convex subset of a finite-dimensional real vector space $V$ of dimension $d$. Then every point of $K$ can be written as a convex combination of at most $d+1$ extreme points of $K$. 
\end{thm}

We now move on to reviewing the basics of general probabilistic theories, or GPTs for short. Before we define rigorously the notion of GPT, let us give a quick summary of the concepts discussed in Chapter~\ref{chapter1} and of their specialisations to the finite dimensional case. The reader who went through Sections~\ref{sec ord} and~\ref{sec ord unit base norm} of Chapter~\ref{chapter1} can jump directly to Definition~\ref{GPT def}. 

An \textbf{ordered vector space} (Section~\ref{sec ord}) is a real vector space $V$ equipped with a \textbf{cone} $C$ (also denoted by $V_+$) that defines the positive elements. A cone (Definition~\ref{def cone}) is any subset $C\subset V$ that is: (i) closed under sums; (ii) closed under multiplications by positive scalars; and (iii) does not contain nontrivial vector subspaces, or equivalently is such that $C\cap -C=\{0\}$. If $\Span (C)=C-C=V$, then $C$ is called \textbf{spanning}. Most cones we will encounter in the rest of this thesis are spanning. The ordering on $V$ is such that $x\leq y$ iff $y-x\in C$. A positive vector $x\in C$ that is also internal to $C$, in symbols $x\in\inter(C)$, is called \textbf{strictly positive}, and we will write $x>0$.

When $V$ is an ordered vector space with spanning positive cone $C$, its \textbf{dual} $V^*$, which consists of all linear functionals $V\rightarrow \mathds{R}$, can also be given the structure of an ordered vector space. The positive cone $C^*=V_+^*$, also called \textbf{dual cone}, is defined as to include all functionals $\varphi\in V^*$ that are non-negative on $C$, i.e. such that $\braket{\varphi,x}\geq 0$ for all $x\geq 0$ (which means $x \in C$). As usual, we use the notation $\braket{\varphi,x}\coloneqq \varphi(x)$, which defines the canonical bilinear form $\braket{\cdot,\cdot}:V^*\times V\rightarrow\mathds{R}$.
In analogy with the nomenclature adopted for vectors, a functional $\varphi\in C^*$ is said to be \textbf{strictly positive} if it belongs to the interior of $C^*$, or equivalently if $\braket{\varphi,x}=0$ for some $x\in C$ is possible only if $x=0$.

As is easy to realise, dual cones are always closed and spanning. Furthermore, if a spanning cone is closed then it coincides with its double dual, i.e. $C=C^{**}$.

Let us take the chance to fix some more nomenclature. Since we are dealing with finite-dimensional spaces, $V$ and $V^*$ are always isomorphic. If $V$ is equipped with a cone $C\subset V$, it makes sense to ask whether there is a linear isomorphism $T: V\rightarrow V^*$ such that $T(C)=C^*$. If this is the case, the cone $C$ is said to be \textbf{weakly self-dual}. If $C$ is weakly self-dual with the corresponding isomorphism $T$ being induced by a nondegenerate scalar product on $V$, then $C$ is called \textbf{strongly self-dual}.


A \textbf{base} of a cone $C\subset V$ is a convex subset $\Omega\subset C$ such that for all $x\in C$ there is a unique real number $t\geq 0$ satisfying $x\in t\Omega$ (Definition~\ref{def base}). It turns out that a base of the positive cone $C$ of an ordered vector space is always of the form
\bb
\Omega = \left\{x\geq 0:\ \braket{u,x}=1\right\} ,
\label{strictly pos base}
\ee
where $u\in V^*$ is a strictly positive functional (Lemma~\ref{strictly pos base lemma}). Using the above characterisation, it is not difficult to show the following.

\begin{lemma} \emph{\cite{Klee-extremal}.} \label{finite dim cones have compact bases}
Let $\Omega$ be a base of a cone $C\subset V$, with $V$ finite-dimensional. Then $\Omega$ is bounded. If $C$ is in addition closed, then $\Omega$ is even compact. Furthermore, every finite-dimensional cone admits a bounded base, and, if closed, a compact one. 
\end{lemma} 

\begin{proof}
Throughout this proof, we assume without loss of generality that the cone is spanning, i.e. that $V=C-C$.
Let us start with the first claim. Let $\Omega\subset C \subset V$ be a base of a cone in dimension $d=\dim V$, and let $u \in \inter (C^*)$ be the strictly positive functional associated with $\Omega$ via~\eqref{strictly pos base}. Then, it is not difficult to verify that $u$ can be written as a strictly positive linear combination of positive functionals forming a basis, i.e. $u=\sum_{i=1}^d \alpha_i \varphi_i$ with $\varphi\in \inter(C^*)$ and $\alpha_i>0$. This can seen as follows. Take a closed ball centered on $u$ and entirely contained inside $\inter (C^*)$. Pick functionals $\varphi_1, \ldots, \varphi_d$ that: (i) lie on the boundary of said ball; and (ii) are such that $u\in\inter \left( \co \{\varphi_1, \ldots, \varphi_d\}\right)$. In particular, they must form a basis of the dual space. Now, by construction $u$ can be written as a strictly positive linear combination of $\varphi_1, \ldots, \varphi_d$, say $u = \sum_{i=1}^d \alpha_i \varphi_i$ with $\alpha_i>0$ for all $i$, as we claimed.
Now, for $x\in\Omega$ we obtain $1=\braket{u, x}=\sum_{i=1}^d \alpha_i \braket{\varphi_i,x}$, that together with the positivity of each addend leads to $0\leq \braket{\varphi_i, x}\leq 1/\alpha_i$ for all $i=1,\dots, d$. Since all functionals $\varphi_i$ (for $i=1,\ldots, d$) are bounded on $\Omega$ and together they form a basis of the dual space, $\Omega$ is bounded as a set.

Observe that since by~\eqref{strictly pos base} the base $\Omega$ is the intersection of $C$ with a hyperplane, if $C$ is closed the same is true for $\Omega$. Since in finite dimension closed and bounded sets are compact, we obtain that bases of closed cones are necessarily compact.

As for the second part of the claim, it suffices to show that all finite-dimensional cones admit a base, i.e. by~\eqref{strictly pos base} a strictly positive functional. This boils down to the observation that in finite dimension all spanning cones have internal points. In our case, we are naturally concerned with the dual $C^*$ of a cone $C\subset V$. Since $C^*$ is spanning, it will contain a basis $\varphi_1,\ldots, \varphi_d$ of $V^*$. Then it is easy to verify that $\sum_{i=1}^d \varphi_i$ belongs to the interior of $C^*$, thus is a strictly positive functional and as such induces a basis of $C$ via~\eqref{strictly pos base}. 
\end{proof}

Thanks to the above result, we can translate properties of compact sets into properties of cones (in finite dimension). See for instance the discussion in~\cite[XI]{SIMON}. A simple example of this strategy in action concerns the Minkowski-Carath\'eodory theorem (Theorem~\ref{MC thm}). Let us start with a simple definition. A nonzero element $x\in V_+$ of the positive cone of an ordered vector space $V$ is called an \textbf{extremal vector} if for all $y\in V$ the inequalities $0\leq y\leq x$ together imply that $y=\alpha x$ for some scalar $\alpha$. In this case the set $\{\alpha x:\, \alpha\in\mathds{R}\}$ is called an \textbf{extremal ray} of $V_+$. Once a base $\Omega$ of $V_+$ has been fixed, and $u$ is the corresponding strictly positive functional given by~\eqref{strictly pos base}, it is not difficult to see that a nonzero $x\in V_+$ is an extremal vector of $V_+$ iff $\frac{x}{\braket{u,x}}\in \Omega$ is an extreme point of $\Omega$ as a convex set.
For a more thorough discussion of these and related concepts, we refer the reader to~\cite[\S 1.6]{ALIPRANTIS}. We can now deduce the following corollary of Theorem~\ref{MC thm}.

\begin{cor} \label{conic MC cor}
Let $C\subset V$ be a closed cone in dimension $d=\dim V<\infty$. Then every point of $C$ can be written as a sum of at most $d$ extremal vectors of $C$. 
\end{cor}

\begin{proof}
The claim follows immediately by putting together Theorem~\ref{MC thm} and the above outlined correspondence between extremal vectors of $V_+$ and extreme points of $\Omega$.
\end{proof}

An ordered vector space whose positive cone is spanning and has a base $\Omega$ is called a \textbf{base norm space} (Definition~\ref{def base norm space}) if its norm is given by
\bb
\|x\| = \inf\left\{ \braket{u, x_+ + x_-}:\ x=x_+ -x_-,\, x_\pm\geq 0 \right\}
\label{dual base eq}
\ee
for all $x\in V$, where the functional $u$ is given by~\eqref{strictly pos base}, as usual. It is easy to verify that in a base norm space $\|x\|=\braket{u,x}$ whenever $x\geq 0$, and that in particular $\|\omega\|=1$ for all $\omega\in \Omega$ (Lemma~\ref{base norm on positive}). The base norm is then seen to be additive on the positive cone.

The dual of a base norm space is a so-called \textbf{order unit space} (Definition~\ref{def order unit space}), and its norm is given by
\bb
\|\varphi\|_* = \max\left\{ t>0:\ \varphi\in t [-u,u]\right\} ,
\label{ord unit norm eq}
\ee
where $[-u,u]\coloneqq \{\varphi\in V^*:\, -u\leq \varphi\leq u\}$, the ordering in the dual space being defined as usual by means of the dual cone $C^*$. Observe that the above equation~\eqref{ord unit norm eq} defines a norm iff $u$ is a strictly positive functional, i.e. if it is internal to the dual positive cone. It is clear from~\eqref{ord unit norm eq} that $[-u,u]$ is the dual unit ball. Observe that the maximum on the right-hand side of~\eqref{ord unit norm eq} exists because $C^*$ is closed and thus so are all intervals. Let us now come to the rigorous mathematical definition of a general probabilistic theory.

\begin{Def}[General probabilistic theories~\cite{Davies-1970}] \label{GPT def}
A \emph{general probabilistic theory} (GPT) is a finite-dimensional base norm space whose positive cone is closed.
\end{Def}

\begin{note}
Let us note in passing that in~\cite{Davies-1970} GPTs are rather called \emph{state spaces} (and no assumption of finite dimension is made). However, here we will reserve this latter term for the set $\Omega$ of normalised states.
\end{note}

In what follows, we will specify a GPT by giving the triple $(V,C,u)$, where: (i) $V$ is the underlying vector space; (ii) $C$ is its positive cone; and (iii) $u$ is a strictly positive functional that identifies a base via~\eqref{strictly pos base}. We will call $u$ the \textbf{unit effect}. Usually, capital letters like $A,B$ are reserved for GPTs. The base norm of a GPT $A$ will be occasionally denoted by $\|\cdot\|_A$ when there is some form of ambiguity, for instance if there are multiple GPTs defined on the same vector space. 

Now, Definition~\ref{GPT def} does not tell us anything about the physics behind a GPT model. To be considered a valid physical theory, a mathematical object must be endowed with a set of rules that specify how to translate physical experiments into its language and how to extract predictions out of the mathematical machinery. These rules can be called collectively `phyisical interpretation', since they connect the physical world with the mathematics we use to model it. We already discussed the most basic of those rules in Subsection~\ref{subsec prep measur}, where we translated them to postulates on the $\mu$ probability function (Axioms~\ref{ax separated},~\ref{ax extreme effects},~\ref{ax mixtures}). Let us briefly recap what we have established until now, and recast it into the language provided by Theorem~\ref{Ludwig emb thm}.
In the following, let $(V,C,u)$ be a GPT.
\begin{enumerate}[(I)]

\item The base $\Omega$ identified by~\eqref{strictly pos base} is the \textbf{state space} of the theory. States of the physical system under examination are represented by elements of $\Omega$.

\item An \textbf{effect}, i.e. a configuration of the measuring apparatus together with a pattern of outputs of the counters that constitute it, is represented by a functional $e\in [0,u]$. Accordingly, a \textbf{measurement} is represented by a (finite) collection of effects $(e_{i})_{i\in I}\subset [0,u]$ such that $\sum_{i\in I} e_{i}=u$. If $|I|=2$ then the measurement is called \textbf{binary}.

\item Given a state $\omega \in \Omega$ and an effect $e\in [0,u]$, the real number $\braket{e,\omega}\in [0,1]$ is the probability of registering the outcome specified by $e$ when the system is measured according to the procedure identified by $e$.

\item If the preparation of the system in a state $\omega_i$ is conditioned on the outcome of a (discrete) random variable $i$, and after the preparation all the information on particular realisation of $i$ is deleted, the resulting state of the system is represented by the convex combination $\widebar{\omega} = \sum_i p_i \omega_i$. Analogously, if our apparatus measures the effect $e_i$ conditioned on $i$ the net measured effect is $\widebar{e}=\sum_i p_i e_i$.

\end{enumerate}

Some comments are in order. First, and most importantly, in the simple picture we just sketched only \emph{destructive measurements} are taken into account. In other words, there is no rule in the above list that specifies how a system transforms \emph{after} a measurement has been performed. We will see how this can be amended when we will discuss the definition of LOCC protocols, before Definition~\ref{locally constr}. However, including the post-measurement collapses in the picture can come at a cost, especially if one wants to impose further restrictions. An example of such a restriction is that the outcomes coming with probability one should not cause collapses. In~\cite{Pfister-no-disturbance, Pfister-Master} it is shown that this rules out all GPTs whose state space is a polytope.

Second, in (ii) we considered only complete measurements, i.e. collections $(e_i)_{i\in I}\subset [0,u]$ such that the probabilities of detection add up to $1$ for all input states, i.e. $\sum_{i\in I} \braket{e_i,\omega}=1$ for all $\omega\in \Omega$, or equivalently $\sum_{i\in I} e_i =u$. This is no loss of generality, as any incomplete measurement can be completed by adding to the collection of effects the functional $u-\sum_{i\in I} e_i$, corresponding to the `error symbol'. If the reader is familiar with the basics of quantum information, the measurements as defined here correspond
exactly to the so-called \emph{positive operator-valued measurements} (POVMs).
On a different line, let us remark that for the sake of the presentation we restricted to finite alphabets when dealing with random variables or measurements. This poses no hurdle in principle, as all definitions make perfect sense when one uses instead integrals over measurable spaces. 

There is at least another question that this very minimal scheme of a physical theory has left unanswered. We know that effects are associated with functionals $e\in [0,u]$, but \emph{conversely, do all the functionals $e\in [0,u]$ correspond to physical effects?} And even more generally, \emph{do all collections $(e_i)_{i\in I}\subset [0,u]$ correspond to physical measurements?} Here, a physical measurement is understood to be a configuration of the counters, and correspondingly an effect $e_i$ that belongs to it represents one of the possible outputs of the counters. In the context of GPTs, this is usually called the \emph{no-restriction hypothesis}~\cite{no-restriction}. Here, we are not going to discuss the status of this hypothesis, and we will limit ourselves to adding it as a separate assumption to the above list. 
\begin{enumerate}[(V)]
\item All collections $(e_i)_{i\in I}\subset [0,u]$ represent legitimate physical measurements. We shall denote the set of all such collections as $\mathbf{M}$.
\end{enumerate}
In particular, the above assumption (V) implies that every effect $e\in [0,u]$ is actually physically measurable. In the notation of Theorem~\ref{Ludwig emb thm}, this amounts to saying that $\Lambda = [0,u]$. 

For the applications we have in mind, we will often deal with restricted sets of measurements, i.e. subsets $\mathcal{M}\subseteq\mathbf{M}$. It is then useful to fix some nomenclature here. For a restricted set of measurements $\mathcal{M}\subseteq\mathbf{M}$, we will denote by $\langle \mathcal{M} \rangle$ the set generated by $\mathcal{M}$ via \textbf{coarse graining}, i.e. by a posteriori declaring some of the outcomes of a measurement in $\mathcal{M}$ as the same. In formula,
\begin{equation}
\begin{split}
\langle \mathcal{M} \rangle \coloneqq \Big\{ (e_{j})_{j\in J}:\ \exists\ I\ \text{finite},\ \{I_{j}\}_{j\in J}\ \text{partition of}\ I,\\
(e'_{i})_{i\in I}\in\mathcal{M}:\ e_{j}=\sum_{i\in I_{j}} e'_{i}\ \, \forall\, j\in J \Big\} .
\end{split}
\label{coarse}
\end{equation} 

We conclude this section by listing some alternative formulae to express the base norm~\eqref{dual base eq} and its dual norm~\eqref{ord unit norm eq} associated with a GPT $(V,C,u)$. First, since the positive cone $C$ is assumed to be closed, it is easy to see that the infimum in~\eqref{dual base eq} can be replaced with a minimum. Moreover, we can apply the dual formula~\eqref{dual formula norm Banach eq} for the norm of a Banach space to our special case and conclude that
\begin{align} 
\|x\| &= \max_{\varphi \in [-u,u]} \braket{\varphi,x} \label{base norm 1} \\ 
&= \max_{e\in [0,u]} \{|\braket{e,x}| + |\braket{u-e,x}|\}\, . \label{base norm 2}
\end{align}
The dual to a base norm can be found not only via the expression~\eqref{dual base eq}, but also by performing the maximisation
\bb
\|\varphi\|_* = \sup_{\omega\in \Omega} |\braket{\varphi,\omega}|\, ,
\label{ord unit norm sup eq}
\ee
as detailed in Corollary~\ref{dual base norm cor}. In particular, we see that $\|\cdot\|_*$ is necessarily order monotone, i.e. $\varphi\leq \lambda$ implies $\|\varphi\|_*\leq \|\lambda\|_*$. In the finite-dimensional case, it is a simple and instructive exercise to translate minimisations like~\eqref{dual base eq} to maximisations like~\eqref{base norm 1} using the duality theory of convex programming~\cite{BOYD}.

\subsection{Bipartite systems} \label{subsec2 bipartite}

Until now we only cared about modelling single systems. However, a lot of interesting physics emerges when one considers instead composite systems. Here we will focus only on bipartite systems, while for the treatment of larger composites we refer the reader to~\cite{telep-in-GPT}.
Clearly, if we want it to represent real physics, the GPT machinery should encompass some kind of way to build a bipartite system $AB=(V_{AB},C_{AB},u_{AB})$ out of two single systems $A=(V_A,C_A,u_A)$ and $B=(V_B,C_B,u_B)$. Throughout this subsection, we will review the basic physical requirements to be imposed on such a construction.

First, preparing two independent local states $\omega_{A}$ and $\tau_{B}$ should be possible for all $\omega_{A}\in \Omega_{A}$ and $\tau_{B}\in \Omega_{B}$. The resulting `composition map', which sends the pair $(\omega_{A},\tau_{B})$ to the state of the composite system that corresponds to the physical procedure of preparing $\omega_A$ on $A$ and $\tau_B$ on $B$ in an uncorrelated way, must be convex-linear in both arguments if the probabilistic interpretation (assumption (IV) in Subsection~\ref{subsec2 bipartite}) has to be respected. The same is naturally true for the effects, as all pairs of local measurements $(e_{A},f_{B})$ must give rise to a legitimate global measurement, with the corresponding mapping being convex-linear in both arguments. Moreover, the probability of measuring a combined effect $(e_{A},f_{B})$ on a combined state $(\omega_{A},\tau_{B})$ has to be given by $\braket{e_A,\omega_A}\braket{f_B,\tau_B}$. Finally, it is reasonable to postulate that `local statistics' are sufficient to determine any state of the joint system, i.e. that from the set of numbers $\braket{(e_A, f_B),\, \eta_{AB}}$ (for all $e_A\in [0,u_A]$ and $f_B\in [0,u_B]$) one can perfectly reconstruct the state $\eta_{AB}\in \Omega_{AB}$. The same for the effects: measuring an effect on all uncorrelated states $(\omega_A, \tau_B)$ should suffice to determine it completely.\footnote{For how reasonable this assumption may appear, there are relevant examples of GPTs that do \emph{not} satisfy it. This happens for instance when one considers real quantum mechanics, where the allowed density matrices are forced to be real symmetric instead of Hermitian.} Summarising, we state the following additional axioms.

\begin{axiom} \label{ax composition maps}
There are convex-bilinear `composition maps' $\mathcal{j} :\Omega_A\times \Omega_B\rightarrow \Omega_{AB}$ and $\mathcal{j}_*: [0,u_A]\times [0,u_B]\rightarrow \left[0,u_{AB}\right]$.
\end{axiom}

\begin{axiom} \label{ax factorisation probabilities}
For all states $\omega_A,\tau_B$ and all effects $e_A, f_B$, it holds 
\bbb
\braket{\mathcal{j}_* (e_A, f_B),\, \mathcal{j}(\omega_A , \tau_B)} = \braket{e_A,\omega_A} \braket{f_B,\tau_B}\, .
\eee
\end{axiom}

\begin{axiom}[Local tomography principle] \label{ax local tomography}
Bipartite states are fully determined by the statistics resulting from local measurements, and conversely bipartite effects are fully determined by the statistics resulting from measuring uncorrelated states.
\end{axiom}

Under these assumptions, the following can be shown~\cite{tensor-rule-1, tensor-rule-2}.

\begin{lemma} \label{tensor product lemma}
Let the finite-dimensional GPTs $A,B$ together with their composite $AB$ satisfy Axioms~\ref{ax composition maps},~\ref{ax factorisation probabilities}, and~\ref{ax local tomography}. Then there are isomorphisms $\mathcal{J}:V_A\otimes V_B\rightarrow V_{AB}$ and $\mathcal{J}_*:V_A^*\otimes V_B^*\rightarrow V_{AB}^*$ such that:
\begin{enumerate}[(a)]
\item $\mathcal{J}(\omega_A\otimes \tau_B) = \mathcal{j}(\omega_A, \tau_B)$ for all states $\omega_A\in \Omega_A$, $\tau_B\in \Omega_B$;
\item $\mathcal{J}_*(e_A\otimes f_B) = \mathcal{j}_*(e_A, f_B)$ for all effects $e_A\in [0,u_A]$, $f_B\in [0,u_B]$;
\item $\mathcal{J}_*(u_A\otimes u_B) = u_{AB}$; and
\item $\mathcal{J}_*^{-1} = \mathcal{J}^*$, where $\mathcal{J}^*$ is the adjoint of $\mathcal{J}$.
\end{enumerate}
\end{lemma}

Since we are interested only in finite-dimensional GPTs, the proof of the above result is really elementary. We report it in Appendix~\ref{app tensor}. Because of Lemma~\ref{tensor product lemma}, we will always identify the joint vector space $V_{AB}$ with the tensor product of the two local spaces, i.e. 
\begin{equation}
V_{AB} = V_{A}\otimes V_{B}\, .
\label{tensor spaces}
\end{equation}
Similarly, we shall write $u_{AB}=u_{A}\otimes u_{B}$, and claims (a), (b) and (d) in Lemma~\ref{tensor product lemma} tell us that we must add the following interpretation rule to the list in Subsection~\ref{subsec2 single}.
\begin{enumerate}[(VI)]

\item If two uncorrelated physical systems are prepared in local states $\omega_A$ and $\tau_B$, the resulting state of the system $AB$ is represented by $\omega_A\otimes \tau_B$. Similarly, if measurements $\big(e_A^{(i)}\big)_{i\in I}\subset [0,u_A]$ and $\big(f_B^{(j)}\big)_{j\in J}\subset [0,u_B]$ are performed on any state of the global system, the resulting measurement is given by $\big(e_A^{(i)}\otimes f_B^{(j)}\big)_{(i,j)\in I\times J}$.

\end{enumerate}

As usual, particular care must be taken in ensuring that the model we are constructing is \emph{non-signalling}, i.e. it does not allow instantaneous transmission of information between the two parties. As is easy to see, \emph{within the GPT formalism the non-signalling principle is automatically implemented}: if $AB$ is in a state $\omega_{AB}$, whatever operation is carried out on the local party $B$, the reduced state of the system $A$ will be given by the `partially evaluated' vector $\braket{u_B,\eta_{AB}}\in \Omega_A$, implicitly defined by the equations
\bb
\braket{\varphi_A, \braket{u_B,\eta_{AB}}}\coloneqq \braket{\varphi_A\otimes u_B, \eta_{AB}}\qquad \forall\ \varphi_A\in V_A^*\, .
\ee

Until now we have only talked about the host space $V_{AB}$, saying nothing about the set of \emph{states} we can legitimately prepare there. In other words, in order to specify the GPT $AB = (V_{AB}, C_{AB}, u_{AB})$ fully, we have yet to tell what the positive $C_{AB}$ is.
Remarkably, $C_{AB}$ is not fully determined by the above axioms, and moreover there seems to be no reasonable a priori assumption we can posit to specify it completely. The reason why this is the case will be apparent from the zoo of examples we will examine in Section~\ref{sec2 examples}.

However, there are indeed two minimal requirements we have to impose on $C_{AB}$, which descend directly from the above interpretation rule (VI). On the one hand, products of local states have to be legitimate global states, and since $C_{AB}$ has to be convex we end up requiring that $C_{A}\tmin C_{B} \subseteq C_{AB}$, where the \textbf{minimal tensor product} is defined by
\begin{equation}
C_{A}\tmin C_{B} \coloneqq \co\left( C_{A}\otimes C_{B} \right) ,
\label{minimal}
\end{equation}
with $C_A\otimes C_B = \left\{ x\otimes y:\, x\in C_A,\, y\in C_B \right\}$. Since $C_A, C_B$ are both closed cones, $C_A \tmin C_B$ can be shown to be closed as well. States in the minimal tensor product are usually called \textbf{separable}. This terminology comes from quantum information, in the context of which the notion of separable states was originally put forward (see~\cite{Werner}, where those states are called `classically correlated').

On the other hand, a similar reasoning holds at the level of effects, enforcing $C_{A}^* \tmin C_{B}^* \subseteq C_{AB}^*$. We can turn this into a statement on the primal cone $C_{AB}$ by taking the dual of both sides, which yields the inclusion $C_{AB}\subseteq C_A \tmax C_B$, where
\begin{align}
C_{A}\tmax  C_{B} &\coloneqq \left\{ Z\in V_{A}\otimes V_{B}:\ \braket{\varphi \otimes \lambda, Z}\geq 0\ \ \forall\ \varphi\in C_{A}^{*},\, \lambda\in C_{B}^{*} \right\} \label{maximal} \\
&= \Big( C_{A}^{*}\tmin C_{B}^{*}\Big)^{*} ,
\label{maximal2}
\end{align}
and the equivalence of the above two definitions of \textbf{maximal tensor product} is elementarily established. Accordingly, one has also
\bb
\Big( C_A \tmax C_B \Big)^* = C_A^* \tmin C_B^*\, .
\label{minimal2}
\ee
In what follows, borrowing the terminology from quantum information, functionals in the maximal tensor product are usually called \textbf{entanglement witnesses} or simply \textbf{witnesses}.
The original definitions of minimal and maximal tensor products seem to go back to Peressini and Sherbert~\cite{Peressini-minmax} and independently to Hulanicki and Phelps~\cite{Hulanicki-minmax}. Both works were written toward the end of the sixties, when the theory of ordered Banach spaces was reaching a certain degree of maturity. Substantial progress on problems related with these concepts is due to Namioka and Phelps~\cite{NP}, as we shall see in Section~\ref{sec2 min vs max}. Summarising, we have established that the positive cone of the composite system must obey the two-sided bound
\begin{equation}
C_{A}\tmin C_{B} \subseteq C_{AB} \subseteq C_{A}\tmax  C_{B}\, .
\label{CAB bound}
\end{equation}
Minimal and maximal tensor products of cones will be the main subject of study in Section~\ref{sec2 min vs max}. 

With a slight abuse of notation, given two GPTs $A=(V_{A},C_{A},u_{A})$ and $B=(V_{B},C_{B},u_{B})$, we will refer to the composites
\begin{align}
A\tmin B &\coloneqq \Big( V_{A}\otimes V_{B},\, C_{A}\tmin C_{B},\, u_{A}\otimes u_{B}\Big)\, , \label{minimal GPTs} \\
A\tmax  B &\coloneqq \Big( V_{A}\otimes V_{B},\, C_{A}\tmax C_{B},\, u_{A}\otimes u_{B}\Big) \label{maximal GPTs}
\end{align}
as the minimal and maximal tensor products of the GPTs $A$ and $B$, respectively.

\subsection{Classes of measurements in bipartite GPTs} \label{subsec2 measur bip}

Now that we have the construction of bipartite GPTs at our disposal, we can define some interesting classes of measurements on a composite (bipartite) system. As usual, the intuition we have developed in studying quantum mechanics (assuming any such intuition exists at all) can guide us in giving operationally meaningful definitions.
In fact, when dealing with bipartite quantum systems, some restricted sets of protocols come into play quite naturally as deriving from operational constraints. Examples of such sets include:
\begin{itemize}
\item local operations (LO), i.e. protocols that can be realised when both parties have full control over their local systems but are forbidden to communicate in any way, at least while they can still access the systems (communication is allowed afterwards);
\item local operations assisted by one-way classical communication ($\text{LOCC}_{\rightarrow}$), when besides local operations one gives in addition the possibility of using a one-way classical communication device; and finally
\item general LOCC protocols, when local operations are complemented by two-way classical communication (we give the two parties, say, a phone).
\end{itemize}
For the following comment, we are going to assume that the reader is familiar with basic concepts of quantum information such as the formalisation of general measurements as POVMs and the notion of separability. We will anyway provide rigorous definitions in the forthcoming Section~\ref{sec2 examples}.
Once the above classes of protocols have been introduced, it is also convenient to define suitable mathematical relaxations that can be much easier to handle. For instance, consider the set formed by all quantum POVMs $(E_{i})_{i\in I}$ on a bipartite system $AB$ for which $E_{i}$ is a separable operator for all $i\in I$. We call these \emph{separable measurements}, and denote them collectively as SEP. It is easy to see that $\text{LOCC}\subseteq\text{SEP}$, and -- less trivially -- the inclusion can be shown to be strict. For a comprehensive review of these and related problems, we refer the reader to~\cite{LOCC}.

Until now we have discussed only quantum theory. Perhaps surprisingly, it turns out that all these restricted classes of measurements can be defined in any bipartite GPT $AB=(V_A\otimes V_B,\, C_{AB},\, u_A\otimes u_B)$ that is constructed out of two local theories $A$ and $B$ in such a way that the constraints~\eqref{CAB bound} are met. Before we provide the general definitions below, let us briefly discuss how to add \emph{dynamical prescriptions} to our mathematical formalism. The purpose of these rules is to specify how states transform after a measurement, extending the picture we have been describing so far, which is rather oriented towards the outcomes and their probabilities. In practice, we will not make use of these prescriptions, and in fact our results are totally independent of any assumption concerning them beyond the mere consistency with the operational interpretation of the theory. However, this apparatus is needed to define a generic LOCC protocol, which requires multiple, interactive rounds of operations on the same systems.\footnote{I thank Howard Barnum and Matthias Christandl for an enlightening discussion on this point, from which the forthcoming definition of LOCC protocols in an arbitrary GPT ultimately came out.}

Following~\cite{Davies-1970}, we can define \textbf{instruments} on one of the two systems, say $A=(V_A, C_A, u_A)$, as collections $(\phi_i)_{i\in I}$ of linear maps $\phi_i: V_A\rightarrow V_A$ that are \textbf{completely positive}, i.e. satisfy $\left((\phi_i)_A \otimes I_B\right)(C_{AB})\subseteq C_{AB}$, and sum up to a normalisation-preserving map, i.e. $\sum_{i\in I} \phi^*(u_A) = u_A$, with $\phi_i^*: V_A^*\rightarrow V_A^*$ being the adjoint of $\phi_i$ \eqref{adjoint}. A totally analogous definition can be given for instruments on the $B$ system. In the operational interpretation of the theory, an instrument describes a non-destructive measurement, with $\phi_i(\omega)$ representing the unnormalised post-measurement state when the outcome $i$ has been recorded on the initial state $\omega$, and the normalisation coefficient $\braket{u, \phi_i(\omega)} = \braket{ \phi_i^*(u), \omega}$ being the probability that the process yields the outcome $i$ (accordingly, observe that $\left(\phi_i^*(u)\right)_{i\in I}$ is a valid measurement in the GPT sense thanks to the interpretation rule (V) in Subsection~\ref{subsec2 single}). 

With the concept of instrument at hand, in order to define LOCC protocols we can follow the steps described in~\cite[\S 2.2]{LOCC}. We will not repeat the construction here since it is totally analogous to the quantum mechanical one once the concept of instrument in GPTs has been clarified, and it is not indispensable for the applications we have in mind.

\begin{Def} \label{locally constr}
Let $A=(V_{A}, C_{A}, u_{A})$ and $B=(V_{B}, C_{B}, u_{B})$ be two GPTs, and let the composite system \mbox{$AB=(V_A\otimes V_B,\, C_{AB},\, u_A\otimes u_B)$} satisfy~\eqref{CAB bound}. Then local operations (LO), local operations assisted by one-way classical communication ($\text{LOCC}_{\rightarrow}$) or two-way classical communication ($\text{LOCC}$), and separable measurements ($\text{SEP}$) are subsets of the set $\mathbf{M}_{AB}$ of all measurements on $AB$ given by:
\begin{align}
\text{\emph{LO}} &\coloneqq \left\langle \left\{ (e_{i}\otimes f_{j})_{(i,j)\in I\times J}:\ (e_{i})_{i\in I}\in\mathbf{M}_{A},\ (f_{j})_{j\in J}\in\mathbf{M}_{B} \right\}\right\rangle , \label{LO} \\[1ex]
\text{\emph{LOCC}}_{\rightarrow} &\coloneqq \left\langle\left\{ \big(e_{i}\otimes f_{j}^{(i)}\big)_{(i,j)\in I\!\times\! J}:\, (e_{i})_{i\in I}\!\in\!\mathbf{M}_{A},\, (f_{j}^{(i)})_{j\in J}\!\in\!\mathbf{M}_{B}\ \forall\, i\in I \right\}\right\rangle , \label{1-way LOCC} \\[1ex]
\text{\emph{LOCC}} &\coloneqq \left\{ \left( \Phi_i^*(u_A\otimes u_B) \right)_{i\in I}:\ \text{$\left( \Phi_i \right)_{i\in I}$ LOCC instrument on $AB$} \right\} \label{LOCC} \\[1ex]
\text{\emph{SEP}} &\coloneqq \left\{ (E_i)_{i\in I} \in \mathbf{M}_{AB}:\ E_i\in C_A^* \tminit C_B^*\ \forall\ i \right\} . \label{SEP}
\end{align}
Here, $\langle\cdot\rangle$ denotes coarse graining as defined by~\eqref{coarse}. The above sets will be collectively called \textbf{locally constrained sets of measurements}.
\end{Def}

It is easy to verify that
\begin{equation}
\text{LO}\, \subseteq\, \text{LOCC}_{\rightarrow}\, \subseteq\, \text{LOCC}\, \subseteq\, \text{SEP}\, .
\label{chain M}
\end{equation}
The last inclusion is slightly less trivial than the others, but its proof is a straightforward generalisation of the argument one gives for the quantum mechanical case. Namely, referring for details and nomenclature to~\cite{LOCC}, one can observe that: (1) one-way local instruments are separable, in the sense each component is a sum of tensor products of completely positive maps; (2) coarse-graining preserves separability; (3) an instrument that is LOCC-linked to a separable one is again separable; (4) separability is preserved under limits; and finally (5) if $(\Phi_i)_{i\in I}$ (acting on $AB$) is separable as an instrument, $\left( \Phi_i(u_A\otimes u_B) \right)_{i\in I}$ is separable as a measurement.

\begin{rem}
Let us stress here that \emph{LOCC is the only locally constrained set of measurements that depends explicitly on the choice of the positive cone $C_{AB}$ of the bipartite system.} In fact, it is easy to realise that only the structure of the local GPTs appears instead in~\eqref{LO},~\eqref{1-way LOCC}, and~\eqref{SEP}. Following the discussion before Definition~\ref{locally constr}, we see that this dependence is hidden inside the concept of completely positive map, in turn necessary to define local instruments.
\end{rem}

\section{Examples} \label{sec2 examples}

Until now, we have been developing the mathematical machinery of the GPT framework in full generality. However, time has come to see it in action with specific examples. Apart from the obvious goal of familiarising the reader with the formalism, the purpose of doing this is twofold. First, we will convince ourselves that the realm of GPTs is sufficiently rich, as to encompass not only all physically relevant theories, but also nearly all theories that are physically \emph{conceivable}. This is to be expected, as the assumptions we made so far are minimal. Secondly, we will also set the stage for more specialised constructions that refer only to particular examples of GPTs or classes of GPTs. These constructions obviously come at the price of a loss of generality, but on the one hand they can still give valuable insights into general behaviours, and on the other hand they can tell us something about models we care about particularly because of their physical relevance.

\subsection{Classical probability theory} \label{subsec2 ex class}

The simplest instance of a GPT is undoubtedly classical probability theory on a finite alphabet. Although it is trivial in some respects, its understanding is crucial in appreciating more complicated examples to be treated in the rest of this section.

The state space of classical probability theory is the set of probability distributions over a finite alphabet. The corresponding GPT can be defined as the triple
\begin{equation}
\text{Cl}_{d}\coloneqq \left(\mathds{R}^{d},\, \mathds{R}^{d}_{+},\, u\right)\, ,
\label{classical}
\end{equation}
where $\mathds{R}^{d}_{+}\coloneqq\{ x\in \mathds{R}^{d}:\, x_{i}\geq 0\ \forall\, i=1,\ldots, d\}$ is the positive octant, and the unit effect acts as $\braket{u,y}=\sum_{i=1}^{d}y_{i}$ for all $y\in \left( \mathds{R}^{d}\right)^{*}\simeq\mathds{R}^{d}$. The base norm associated with classical probability theory is easily seen to be the $l_{1}$-norm $|x|_{1}=\sum_{i=1}^{d}|x_{i}|$. 

Of course, what we have described so far is the standard form of a classical theory. Other GPTs can be obtained from it via the application of linear isomorphisms. Since we will be interested in these models in the following, it is worth it to give the following definition.

\begin{Def} \label{def class}
A \textbf{simplicial cone} in a real vector space $V$ of finite dimension $d$ is the set of non-negative linear combinations of $d$ linearly independent vectors of $V$. A GPT $(V,C,u)$ is called \textbf{classical} if the cone of states $C$ is simplicial, i.e. if the ordered vector spaces $(V,C)$ and $(\mathds{R}^d,\, \mathds{R}_+^d)$ (where $d=\dim V$) are linear- and order-isomorphic (see Subsection~\ref{subsec uniqueness} for a definition of order isomorphism).
\end{Def}

Let us comment on some elementary properties of simplicial cones. First of all, simplicial cones are closed and spanning by definition. Moreover, the dual cone $C^*$ of a closed and spanning cone $C$ is simplicial iff $C$ is itself simplicial.
Finally, if the dimension of the host space is low enough, namely $d\leq 2$, then every closed cone is simplicial~\cite[\S 2.1, Exercise 12]{ALIPRANTIS}.

We now come back to the standard probability theory as given by~\eqref{classical}. Until now, we have dealt with a single system. As for bipartite systems, let us start by observing that if $A=\text{Cl}_{d_A}$ and $B=\text{Cl}_{d_B}$ the vector space pertaining to the joint system $AB$, given by the rule~\eqref{tensor spaces}, is easily seen to be
\bb
\mathds{R}^{d_A} \otimes \mathds{R}^{d_B} = \mathds{R}^{d_A d_B}\, .
\ee
Moreover, it can be seen by direct inspection that the lower and upper bound in~\eqref{CAB bound} coincide, hence the composition rule is trivial: if $A=\text{Cl}_{d_{A}}$ or $B=\text{Cl}_{d_{B}}$, then necessarily
\bb
C_{AB} = C_A\tmin C_B = C_A\tmax  C_B = \mathds{R}_+^{d_A d_B}\, ,
\ee
or in other words
\bb
AB = A\tmin B = A \tmax B = \text{Cl}_{d_A d_B}\, .
\ee
In Section~\ref{sec2 min vs max} we will see that this is a particular example of a more general fact: \emph{whenever one of the two local GPTs is a classical probability theory, the minimal and maximal tensor products necessarily coincide} (Lemma~\ref{simplicial min=max lemma}).

\subsection{Quantum mechanics} \label{subsec2 ex QM}

We now turn our attention to the description of finite-dimensional quantum systems within the formalism of GPTs. It goes without saying, quantum mechanics is the most important example of GPT and the archetypical example of a non-classical theory.
In what follows, we describe a quantum system with $n$ levels. Vectors in $\mathds{C}^n$ will be represented in Dirac notation as $\ket{\alpha}, \ket{\beta}$ and so on. The canonical Hermitian product is given by $\braket{\alpha|\beta}=\sum_{i=1}^n \alpha_i^* \beta_i$.\footnote{According to the physicists' convention, we take it to be antilinear in the first entry and linear in the second.}

As is well known, the quantum mechanical cone of unnormalised states consists of the positive semidefinite $n\times n$ matrices (collectively denoted by $\text{PSD}_{n}$), embedded in the real space of Hermitian matrices (called $\mathcal{H}_{n}$) whose real dimension is $d=n^2$. Normalised quantum states, also called \textbf{density matrices} (or `mixed states') are in addition required to have unit trace.
Then, we see immediately that the unit effect coincides with the trace. We write symbolically
\begin{equation}
\text{QM}_{n} \coloneqq \left( \mathcal{H}_{n},\, \text{PSD}_{n},\, \Tr \right) ,
\label{quantum}
\end{equation}
remembering that
\begin{equation}
\dim \text{QM}_{n} = n^{2}\, .
\label{dim quantum}
\end{equation}
The extremal points in the set of density matrices are exactly the (normalised) rank-one orthogonal projectors $\ket{\psi}\!\!\bra{\psi}$, called \textbf{pure states}.

Observe that the positive semidefinite cone is strongly self-dual, i.e. $\text{PSD}_{n}^{*}=\text{PSD}_{n}$, when one endows $\mathcal{H}_n$ with the natural \textbf{Hilbert-Schmidt scalar product} given by $\braket{X,Y}\coloneqq \Tr X Y$. The base norm in quantum mechanics can be proved to coincide with the \textbf{trace norm} $\|X\|_{1} = \Tr |X| =\sum_{i=1}^{n}|\lambda_{i}(X)|$, where $\lambda_{i}(X)$ are the eigenvalues of $X\in\mathcal{H}_{n}$, and the last equality holds because $X$ is Hermitian.

Now, let us discuss the composition rules for bipartite systems. First of all, it is easy to see that the real vector spaces of Hermitian matrices combine according to a very simple rule under tensor products (as prescribed by~\eqref{tensor spaces}):
\bb
\mathcal{H}_{n_A} \otimes \mathcal{H}_{n_B} = \mathcal{H}_{n_A n_B}\, .
\ee
The fact that there is equality above is really a special feature of Hermitian matrices, and follows straightforwardly from dimensional considerations. An analogous statement does not hold, for instance, for real symmetric matrices, in which case one gets only the strict inclusion $\mathcal{S}_{n_A} \otimes \mathcal{S}_{n_B}\subsetneq \mathcal{S}_{n_A n_B}$. To see this latter fact, it suffices again to evaluate the dimensions of both sides: one obtains $\frac{n_A(n_A+1)}{2}\frac{n_B (n_B+1)}{2}$ on the left and $\frac{n_A n_B (n_A n_B +1)}{2}$ (strictly larger as soon as $n_A, n_B>1$) on the right.
This is the reason why real quantum mechanics, where all density matrices are prescribed to be real symmetric rather than complex Hermitian, does not respect the local tomography principle (Axiom~\ref{ax local tomography}) and thus does not satisfy~\eqref{tensor spaces}.

As for the positive cone of a bipartite quantum system, using the definitions~\eqref{minimal} and~\eqref{maximal}, we see that
\begin{align}
\begin{split}
\text{PSD}_{n_{A}}\tmin \text{PSD}_{n_{B}} &= \Big\{ \sum\nolimits_{i\in I} P_{i}\otimes Q_{i} :\ I\ \text{finite}, \\[-0.4ex]
&\qquad\quad P_{i}\in\text{PSD}_{n_{A}},\, Q_{i}\in \text{PSD}_{n_{B}}\ \forall\ i\in I \Big\}\, ,
\end{split}
\label{separable} \\[1ex]
\begin{split}
\text{PSD}_{n_{A}}\tmax  \text{PSD}_{n_{B}} &= \Big\{ W\in \mathcal{H}_{n_{A}n_{B}}:\ \Tr[ P\otimes Q\, W ]\geq 0\\[-0.4ex]
&\qquad\quad\forall\ P\in\text{PSD}_{n_{A}},\, Q\in \text{PSD}_{n_{B}} \Big\}\, .
\end{split}
\label{witnesses}
\end{align}
As we mentioned, in the context of quantum information states belonging to~\eqref{separable} are usually called \emph{separable} or \emph{classically correlated}~\cite{Werner}, while elements of~\eqref{witnesses} are variously called \textbf{entanglement witnesses}, \textbf{separability witnesses} or \textbf{block-positive operators}. This latter name comes from the fact that since in~\eqref{witnesses} we can restrict $P$ and $Q$ to be pure states (i.e. rank-one projectors), the defining condition for belonging to the set amounts to imposing the positivity of the diagonal blocks in all product bases. Interestingly enough, Nature has a preferred choice for the cones of bipartite systems, which is neither the maximal nor the minimal tensor product. Instead, if $A=\text{QM}_{n_{A}}$ and $B=\text{QM}_{n_{B}}$ then $AB=\text{QM}_{n_{A}n_{B}}$, i.e.
\begin{equation}
\text{PSD}_{n_{A}}\tmin \text{PSD}_{n_{B}} \subsetneq C_{AB} = \text{PSD}_{n_{A}n_{B}} \subsetneq \text{PSD}_{n_{A}}\tmax  \text{PSD}_{n_{B}}\, .
\label{cone bipartite quantum}
\end{equation}

The fact that all three inclusions are strict is an elementary result in entanglement theory, whose proof we do not report here. We refer the interested reader to~\cite{NC} or to the monograph~\cite{GeometryQuantum}.
Let us stress here that quantum mechanics is our first example of a GPT whose composition rules prescribe the bipartite cone to be neither the minimal nor the maximal tensor product of the local cones.

In fact, we can let ourselves inspire by the above observation and use it to define a new composition rule that allows only separable bipartite states. Such a construction is mentioned in~\cite[\S IV.E]{Barrett-original} and considered there at a purely hypothetical level. We take the trouble to write out the definition because we will find this example useful and instructive later.

\begin{Def} \label{W theory}
We call \textbf{W-theory} the class of GPT models that can obtained from the family $\left\{ \text{\emph{QM}}_n \right\}_{n\in\mathds{N}}$ of single quantum systems by taking minimal tensor products according to the rule~\eqref{minimal GPTs}.
\end{Def}

In $W$-theory, the only allowed states of a multipartite system are fully separable, while the set of possible effects contains all entanglement witnesses (equivalently, all elements in the cone~\eqref{witnesses}). Thus, the base norm of a Hermitian operator $X_{AB}\in\mathcal{H}_{n_A n_B}$ will be given by
\bb
\begin{split}
\|X\|_W \coloneqq &\ \|X\|_{\text{QM}_{n_A}\tminfoot \text{QM}_{n_B}} \\
=&\ \max\left\{\Tr XY:\ \big|\! \braket{\alpha\beta|Y|\alpha\beta} \!\big|\leq \braket{\alpha|\alpha}\! \braket{\beta|\beta}\ \, \forall\, \ket{\alpha}\in\mathds{C}^{n_A},\, \ket{\beta}\in\mathds{C}^{n_B} \right\} ,
\end{split}
\label{W norm}
\ee
where the last line corresponds to the formula~\eqref{base norm 1}.

\subsection{Centrally symmetric models} \label{subsec2 centr symm}

Until now, we have examined standard examples of GPTs that have been previously considered in the literature. Throughout this subsection, we want to introduce a novel class of so-called `centrally symmetric' models, whose state space $\Omega$ is invariant under `inversion' with respect to a suitably chosen state. We will see what this means precisely in a moment.

Restricting the analysis to centrally symmetric models, as we will do under some circumstances in Section~\ref{sec2 min vs max}, has undoubtedly some drawbacks, the most serious one being that we are excluding by default physically relevant examples such as classical probability theory and quantum mechanics (Subsections~\ref{subsec2 ex class} and~\ref{subsec2 ex QM}, respectively). However, there are also lots of reasons why this is both instructive and meaningful. First, we still have at our disposal an ample family of theories that can be used to test conjectures, for they are theoretically simpler, and optimality of certain bounds, for they are computationally convenient. We will see instances of both approaches in action in Section~\ref{sec2 min vs max} and later in Chapter~\ref{chapter3}. Secondly, centrally symmetric models reveal a surprising connection between GPTs and a certain branch of functional analysis that deals with tensor products of Banach spaces. In some cases this connection can be exploited in full generality (beyond the centrally symmetric case), and this idea will be instrumental in finding the main results of Chapter~\ref{chapter3}. Let us start with the following definition.

\begin{Def}[Centrally symmetric models] \label{centr}
A GPT of the form $(\mathds{R}^{d}, C, u)$, with $u=(1,0,\ldots,0)^T$, is said to be \textbf{centrally symmetric} if there exists a norm $|\cdot|$ on $\mathds{R}^{d-1}$ such that 
\bb
C=\left\{(x_{0},\widebar{x})\in \mathds{R}\oplus\mathds{R}^{d-1}:\ x_{0}\geq |\widebar{x}|\right\} .
\label{centr eq}
\ee
\end{Def}

\begin{note}
It is somewhat unusual to denote a vector norm by $|\cdot|$. Here we make this choice in order to distinguish this latter object, which is a norm on $\mathds{R}^{d-1}$, from the base norm $\|\cdot\|$, which instead has the global space $\mathds{R}^d$ as its domain.\footnote{Some mathematicians would rather introduce a third bar, preferring the notation $|||\cdot|||$, which however looks to us too outrageous to be considered.}
\end{note}

As can be easily verified, the dual cone to a `centrally symmetric' cone as in~\eqref{centr eq} shares the same structure, being given by
\bb
C ^* = \left\{(y_{0},\widebar{y})\in \mathds{R}\oplus\mathds{R}^{d-1}:\ y_{0}\geq |\widebar{y}|_*\right\} ,
\label{centr dual eq}
\ee
where $|\cdot|_*$ is the dual to the norm $|\cdot|$. With a straightforward calculation via~\eqref{base norm 1}, the base norm of a centrally symmetric GPT can be easily seen to be given by
\bb
\|x\|=\|(x_0,\widebar{x})\| = \max\left\{|x_0|, |\widebar{x}|\right\} .
\label{centr base norm}
\ee
Another peculiarity of centrally symmetric models is the existence of a privileged state, denoted by $u_{*}\coloneqq (1,0,\ldots,0)^T$. The vector space can then be written as 
\bb
\mathds{R}^{d} = \mathds{R} u_{*}\oplus \mathds{R}^{d-1}\, ,
\label{V decomp}
\ee
the second component being sometimes referred to as the \textbf{section} of the host vector space $\mathds{R}^d$. We will keep calling $x_0, \widebar{x}$ the two components of $x\in \mathds{R}^d$ according to the above decomposition. Along the same lines, for $v\in \mathds{R}^{d-1}$ we will write $\myhat{v}\coloneqq 0\oplus v$. A remarkable feature of centrally symmetric models is the existence of a simple yet nontrivial order isomorphism, given by the linear map $T:\mathds{R}^{d}\rightarrow \mathds{R}^{d}$ defined as
\bb
T \coloneqq 1 \oplus (-I_{d-1})\, .
\label{T map}
\ee
Here, the direct sum is with respect to the decomposition~\eqref{V decomp}, and $I_{d-1}$ denotes the identity on $\mathds{R}^{d-1}$. The fact that $T$ is an order isomorphism follows from the readily established equality $T(C)=C$. Accordingly, for all $x\in \mathds{R}^d$ one has $\|x\|=\|T(x)\|$, which is also obvious from the explicit formula for the base norm given above. Now, let us turn to bipartite systems. From~\eqref{V decomp} we infer the natural decomposition
\bb
\begin{split}
V_{AB} &= \mathds{R}^{d_{A}}\otimes \mathds{R}^{d_{B}}\\
&= \left(\mathds{R}\, u_{A*}\!\otimes u_{B*}\right) \oplus \left(u_{A*}\!\otimes \mathds{R}^{d_{B}-1}\right) \oplus \left(\mathds{R}^{d_{A}-1}\!\otimes u_{B*}\right) \oplus \left(\mathds{R}^{d_{A}-1}\!\otimes  \mathds{R}^{d_{B}-1}\right) .
\end{split}
\label{VAB decomp}
\ee

Tensors belonging to the bipartite vector space $\mathds{R}^{d_{A}}\otimes\mathds{R}^{d_{B}}$ (or to its dual) can be thought of as $d_{A}\times d_{B}$ matrices $Z\in\mathds{R}^{d_{A}\times d_{B}}$. We shall find useful to denote by $\widebar{Z}$ the $(d_{A}-1)\times (d_{B}-1)$ submatrix of $Z\in\mathds{R}^{d_{A}\times d_{B}}$ which is obtained by cutting off the zeroth components $Z_{i0},\, Z_{0j}$ of the latter. Complementarily, given $M\in\mathds{R}^{(d_{A}-1)\times (d_{B}-1)}$ we call $\widehat{M}$ the $d_{A}\times d_{B}$ `lifted' matrix whose entries are
\bbb
\widehat{M}_{ij} \coloneqq \left\{ \begin{array}{cl} 0 & \text{ if $i=0$ or $j=0$,} \\[0.5ex] M_{ij} & \text{ if $i,j\geq 1$.} \end{array}  \right.
\eee
In what follows, the tensor product of the two unit functionals will be occasionally denoted by $U\coloneqq u_A\otimes u_B$.
Also the composite system hosts a `privileged state' inherited by the two local spaces, namely $U_* \coloneqq u_{A*}\otimes u_{B*}$. 

As for the cone $C_{AB}$ of bipartite states, there is no a priori preferred rule to compose two centrally symmetric models, so we will use different ones depending on the circumstances. 

Now, there is a natural example of centrally symmetric model to which we want to devote special attention. Our starting point is the following elementary observation: as is well known, the state space of a two-level quantum system is identifiable with a $3$-dimensional ball (Bloch ball). We can let ourselves inspire by this observation to define a hypothetical class of physical models whose state space is a Euclidean ball of \emph{arbitrary} dimension. These GPTs have been considered recently in connections to attempts of reconstructing quantum mechanics starting from few physically motivated axioms~\cite{Hardy2001, Wilce-4-1/2-axioms, quantum-info-unit}. This is to be expected in light of a famous classification theorem by Koecher and Vinberg~\cite{Koecher-hom-cones, Vinberg-hom-cones}, implying among the other things that spherical models are one of the few classes of models that are strongly self-dual, the most notable of which is precisely quantum mechanics~\cite{telep-in-GPT}.
In light of the above discussion, we define the \textbf{spherical model} as the GPT
\begin{equation}
\text{Sph}_{d} \coloneqq \left( \mathds{R}^{d},\, C_d,\, u \right) .
\label{spherical}
\end{equation}
Here, $C_{d}$ is the Lorentz or `ice-cream' cone
\bb
C_{d} \coloneqq \left\{ x \in \mathds{R}^{d}:\ |\widebar{x}|_{2} \leq x_{0} \right\} , \label{ice cream}
\ee
where $|\cdot|_{2}$ is the standard Euclidean norm in $\mathds{R}^{d-1}$, and $u$ acts as $\braket{u,y}\coloneqq y_{0}$ for all $y\in \left( \mathds{R}^{d} \right)^{*}$. Naturally, the spherical model is an example of a centrally symmetric GPT, as \eqref{ice cream} is a special case of \eqref{centr eq}. 
Observe that with the canonical identification $\left( \mathds{R}^{d} \right)^{*} \simeq \mathds{R}^d$ one has $C_d^*=C_d$, hence the spherical model is also strongly self-dual, as we mentioned.
As for the base norm in $\text{Sph}_{d}$, we can specialise~\eqref{centr base norm} to our particular case and obtain
\bb
\|x\|_{\text{Sph}_d}=\|(x_0,\widebar{x})\|_{\text{Sph}_d}=\max\left\{|x_{0}|,|\widebar{x}|_{2}\right\} .
\label{sph base norm}
\ee

\subsection{Generalised non-signalling theories} \label{subsec2 other ex}

In this subsection, we will discuss standard examples of GPTs known as generalised non-signalling theories~\cite{Barrett-original}. The usefulness of these models is that every experimental setting can be used to map an arbitrary physical state into a state of such a theory. Therefore, in some sense, every experimentally accessible feature of a physical theory should be analysable via these mappings. And in fact the class of generalised non-signalling theories is sufficiently rich so that this is typically what happens.

There is another good reason to care about these models. Namely, concepts like Bell inequalities and nonlocality admit a very natural formulation within the formalism of GPTs applied to the specific case of generalised non-signalling theories. We especially like this formulation because of the crystal clear mathematical representation of the phenomenon of nonlocality that it gives, and not least because of the compactness of the notation, free from complicated sums running over many indices. We hope that this work will contribute to spreading the word and making the usage of this framework more common among quantum information theorists.\footnote{Almost all the time I read a paper on nonlocality, I personally really feel the need of a more compact notation. The situation somewhat reminds me of the times when linear algebra was a great deal a matter of writing down and manipulating linear systems with spelled-out coefficients. The gradual introduction, throughout the 19th century, of objects called matrices, not only simplified the notation considerably, but shed also new light on the conceptual meaning of linear systems and linear substitutions. My favourite instance of this phenomenon that one may call of notation compactification occurred however in electromagnetism. At the dawn of the modern theory, the equations of the electromagnetic field were written \emph{in components}, which makes them significantly harder to read, visualise, and ultimately manipulate. Maxwell's paper \emph{`A dynamical theory of the electromagnetic field'}, for instance, features long lists of scalar equations instead of the compact vector equations we all know and love. Of course, these secondary issues did not stop Maxwell from gaining a full command of the subtleties behind said equations, but this is another story.}


A generalised bit, or \textbf{gbit} for short~\cite{Barrett-original}, is one of the simplest example of a non-classical system. As a GPT, we can define it as $G_{2} \coloneqq \left( \mathds{R}^{3},\, C_{\square},\, u \right)$, where
\bb
C_{\square} \coloneqq \left\{ (x,y,z)\in\mathds{R}^{3}:\ |x|+|y|\leq z\right\}
\label{C square}
\ee
and $\braket{u,(x,y,z)}\coloneqq z$, or equivalently $u=(0,0,1)^T$. As is easily seen, the state space of $G_{2}$ is a two-dimensional square. Thus, a straightforward generalisation of the above notion takes as state space a Cartesian product of $n$ identical $k$-simplices, in symbols $\Omega_{n}^k\coloneqq \Delta_{k}^{\times n}$~\cite{Barrett-original}. Here, a $k$-simplex is just the convex space of all probability distributions on an alphabet of $k$ symbols (its linear dimension is thus $k-1$, the missing degree of freedom coming from the normalisation condition). The corresponding GPT will be denoted by $G_{n}^k$, and the associated cone by $C_n^k$. The underlying vector space is naturally identified with a subspace of $\mathds{R}^{n \times k}$ of dimension $\dim G_{n}^k = n (k-1) + 1$. States of $G_{n}^k$ can thus be identified with $n \times k$ matrices of probabilities $P=(p_{ij})_{i=1,\ldots, n}^{j=1,\ldots, k}\in \mathds{R}^{n \times k}$. With this notation, $(p_{i1}, \ldots, p_{ik})^T$ is nothing but the component of $P\in \Delta_{k}^{\times n}$ corresponding to the $i$-th copy of the simplex $\Delta_k$.

This family of GPTs is particularly useful because any state of an arbitrary GPT $A=(V, C, u)$ can be used to generate a state of $G_{n}^k$ with the help of a \emph{measurement apparatus}. We think of any such apparatus as a measuring device with a selectable setting, indexed by $i=1,\ldots, n$. Any particular value $i$ of the setting index performs a given $n$-outcome measurement $(e_{i1},\ldots, e_{ik})$ on an input state $\omega\in \Omega$ of $A$. We can then give the following definition.

\begin{Def}[Apparatuses] \label{def apparatus}
Given a GPT $A=(V,C,u)$ and two positive integers $n,k$, an $(n,k)$-\textbf{apparatus} on $A$ is a linear map of the form
\bb
\begin{array}{llll}
\Gamma : & V & \longrightarrow & \mathds{R}^{n \times k} \\
                 & x & \longmapsto      & \left(\braket{e_{ij}, x} \right)_{i=1,\ldots, n}^{j=1,\ldots, k} ,
\end{array}
\ee
where $(e_{ij})_j\in \mathbf{M}$ is a measurement on $A$ for all $i=1,\ldots, n$. Observe that $\Gamma$ maps states of $A$ to states of $G_{n}^k$.
\end{Def}

Particularly appealing for its simplicity is the case $k=2$, with the corresponding state space being an $n$-dimensional hypercube. We will call this latter example $G_{n}^2\eqqcolon G_n$ \textbf{cubic model}. We remark here that \emph{cubic models are centrally symmetric in the sense of Subsection~\ref{subsec2 centr symm}}, hence all the theory we developed and notation we fixed there applies also in this case. In particular, the base norm in $G_n$ can be easily deduced from~\eqref{centr base norm}, and becomes
\bb
\|x\|_{G_n} = \|(x_0, \widebar{x})\|_{G_n} = \max\left\{|x_0|, |\widebar{x}|_\infty \right\} ,
\label{cubic base norm}
\ee
where for $v\in\mathds{R}^n$ one has $|v|_\infty\coloneqq \max_{1 \leq i \leq n} |v_i|$. 

The main reason for introducing these toy models is that they provide a useful tool in understanding nonlocal correlations. In fact, the celebrated Popescu-Rohrlich (PR) box~\cite{PR-boxes} can be seen as states of the maximal tensor product $G_{2}\tmax G_{2}$. In general, GPTs of the class $G_{n}^k$ equipped with the maximal tensor product composition rule are called `generalized non-signalling theories' (GNST) in~\cite{Barrett-original}. Within this framework, nonlocality can be formulated somewhat more easily. Let us give the following definition.

\begin{Def}[Bell inequalities] \label{def Bell}
A \textbf{Bell inequality} with $n$ measurements having $k$ outcomes each is a linear inequality that is obeyed by all states in $G_n^k \tminit G_n^k$ but violated by some states in $G_n^k \tmaxit G_n^k$.
\end{Def}

The above definition makes sense because probability distributions in the minimal tensor product are exactly those that are reproducible via a so-called \emph{local hidden variable model}. Thus a Bell inequality separates local hidden variable models from probability patterns that are signatures of genuine nonlocal phenomena. Providing a summary of the extensive literature on the subject is almost impossible, so we refer the interested reader to the excellent review in~\cite{Brunner-review}.

Naturally, in a realistic scenario correlations are produced by means of physical states. Therefore, we should have a way to refer the concept of violation of a Bell inequality to physical states rather than to abstract state spaces such as $G_n^k$. This goal is accomplished by the following definition, with which we close this subsection.

\begin{Def} \label{def Bell violation}
Given two GPTs $A=(V_A, C_A, u_A)$, $B=(V_B, C_B, u_B)$, and a way to combine them into a bipartite GPT $AB = (V_{AB}, C_{AB}, u_{AB})$, where $C_{AB}$ obeys~\eqref{CAB bound}, we say that $AB$ \textbf{violates a Bell inequality} if for some integers $n,k,n',k'\geq 2$ there are an $(n,k)$-apparatus $\Gamma_A$ on $A$ and an $(n',k')$-apparatus $\Gamma_B$ on $B$ such that
\bbb
\left(\Gamma_A \otimes \Gamma_B\right) (C_{AB}) \not\subseteq G_n^k \tminit G_{n'}^{k'}\, .
\eee
\end{Def}

It is not difficult to observe that if $AB = A \tmin B$ (i.e. only separable bipartite states are allowed) then no violation of a Bell inequality can occur. In particular, the forthcoming Lemma~\ref{simplicial min=max lemma} ensures that if either $A$ or $B$ is a classical theory according to Definition~\ref{def class}, then $AB$ does not violate any Bell inequality.

\section{Separability problem} \label{sec2 sep problem}

Throughout this section, we will explore some features of the separability problem in general probabilistic theories. This corresponds to point (iv) of the discussion in Subsection~\ref{subsec2 fundamental}. While the core of the present chapter is mainly formed by the forthcoming Sections~\ref{sec2 min vs max} and~\ref{sec2 beyond}, and in some respects this section could be seen as a mathematical divertissement, we will see that some of the tools we develop here will turn out useful several times throughout the thesis. In fact, we chose to present this material before delving into the more fundamental problems discussed in the forthcoming sections exactly because some of the questions that we shall raise later in the chapter have a lot to do with the separability problem. Furthermore, some of the results we present here will be instrumental in Chapter~\ref{chapter3}.

We remind the reader that in quantum information the separability problem~\cite{GeometryQuantum} consists in establishing computationally convenient criteria to decide whether a given state is separable or entangled. Analogously, in GPTs we will have to decide whether a given vector in the maximal tensor product of two cones belongs also to the minimal tensor product or not.

For quantum states there is a wealth of available criteria, even if the problem is known to be NP-hard in general (under certain assumptions on the required error)~\cite{GurvitsNPhard, GharibianNPhard}. However, many of those can not expected to carry over to the GPT realm, because they depend on features of the problem that are specific to the quantum case. For instance, the positive partial transpose (PPT) criterion~\cite{PeresPPT, HorodeckiPPT}, as well as the reduction criterion~\cite{Horodecki1999} depend on the existence of a special map that preserves the positive cone, while the reshuffling criterion~\cite{Resh1, Resh2, Resh3} makes use of the fact that each local vector space is a space of linear operators (Hermitian matrices), and thus itself a tensor product.
The same is true at the global level for the range criterion~\cite{Horodecki-PPT-entangled}, which indeed involves the notion of support and thus refers to the density matrices as operators and not merely as vectors in a real space. This is common to all the conditions that are related to the spectra, like the majorization criterion~\cite{Nielsen-majorization}, the separable ball construction~\cite{GurvitsBarnum}, and the minimal eigenvalue sufficient condition~\cite{VidalTarrach}.



 
Of all the mainstream entanglement criteria, we are left with three: what we could call (following the terminology of~\cite{GurvitsBarnum}) the `Woronowicz condition'~\cite{Woronowicz}, and the complete extendibility criterion~\cite{complete-extendibility}. 
Furthermore, there is a less known sufficient condition based on the tensor rank of the density matrix itself~\cite{Cariello13,JBlog,Cariello14}. The rest of this section is dedicated to investigating these conditions in the GPT framework.

\subsection{Woronowicz and complete extendibility criteria} \label{subsec2 Woronowicz}

We start with the Woronowicz criterion, since it is the simplest. For the quantum separability problem, it states roughly that a state is separable if and only if it has a positive expectation value on all entanglement witnesses, where an entanglement witness is an observable that in turn has positive expectation value on all separable states. As one sees immediately, this has nothing to do with specific features of quantum theory, and is rather an instance of the Hahn-Banach separation theorem (Theorem~\ref{Hahn-Banach}), which allows us to identify a closed convex set with the intersection of all the closed half-spaces containing it. An equivalent way of recasting the Woronowicz condition is via positive maps: a bipartite quantum state $\omega_{AB}$ is separable iff $\left(T_{A\rightarrow A'}\otimes I\right)(\omega_{A B})$ is still a positive semidefinite operator for all positive maps $T_{A\rightarrow A'}:\mathcal{H}_{n_{A}} \rightarrow \mathcal{H}_{n_{A'}}$, where positive means that $T_{A \rightarrow A'}(\text{PSD}_{n_A})\subseteq \text{PSD}_{n_{A'}}$.

This latter notion admits an obvious generalisation to GPTs. Given two ordered vector spaces $V_1,V_2$, a map $T:V_1\rightarrow V_2$ is said to be \textbf{positive} if it sends positive elements to positive elements, i.e. if $T(x)\geq 0$ whenever $x\geq 0$. This is a different notion than that of `complete positivity' we investigated in Subsection~\ref{subsec2 measur bip}. Every completely positive map must be positive, but the converse is generally not true. In quantum mechanics, the transposition provides a classical example of a positive yet not completely positive map~\cite{PeresPPT}.

We can rephrase the positivity requirement for maps within the language we have been developing so far if we remember that one can equivalently see a map $T:V_1\rightarrow V_2$ as a tensor $W_T\in V_1^* \otimes V_2$. This identification is made possible by the canonical isomorphism $T\leftrightarrow W_T$ defined by the identity
\bb
\braket{W_T, Z} = \text{tr}\left[ (T\otimes I)(Z) \right]\qquad \forall\ Z\in V_1\otimes V_2^*\, ,
\label{canonical isomorphism}
\ee
where for an arbitrary vector space $V$ one defines the \textbf{trace function} 
\bb
\begin{split}
\text{tr}: V\otimes V^* &\longrightarrow \mathds{R} \\
 x \otimes \varphi\ \; &\longmapsto \braket{\varphi, x}\, .
\end{split}
\ee
It is very simple to verify that $T$ is a positive map iff $W_T\in C_1^* \tmax C_2$, where $C_1$ and $C_2$ are the positive cones of $V_1$ and $V_2$, respectively.

Equipped with these tools, we can formulate a generalisation of Woronowicz condition to the GPT setting.

\begin{prop}[Woronowicz condition for GPTs] \label{Woronowicz prop}
Let $V_A, V_B$ be finite-dimensional ordered vector spaces with positive cones $C_A, C_B$. For a tensor $Z\in C_A \tmaxit C_B$, the following are equivalent:
\begin{enumerate}[(a)]
\item $Z\in C_A \tminit C_B$;
\item for all $W\in C_A^* \tmaxit C_B^*$ one has $\braket{W,Z}\geq 0$;
\item for all positive maps $T:V_A\rightarrow V_{B}^*$, one has $\text{\emph{tr}}\left[ (T\otimes I)(Z)\right] \geq 0$.
\end{enumerate}
\end{prop}

\begin{proof}
The equivalence between (a) and (b) is a consequence of a duality relation analogous to \eqref{minimal2}, which reads
\bb
\Big(C_A^* \tmax C_B^* \Big)^* = C_A \tmin C_B\, .
\label{minimal3}
\ee
Moreover, the existence of the canonical isomorphism in~\eqref{canonical isomorphism} implies that (c) is just a reformulation of (b).
\end{proof}

Until now we have discussed only the Woronowicz criterion, which depends only on the convex structure of the set of separable states and therefore admits an immediate generalisation to the GPT setting. Let us turn our attention to another separability criterion that has found extensive application in quantum information theory, namely the complete extendibility criterion~\cite{complete-extendibility}, which states that every separable density matrix $\rho_{AB}$ admits a symmetric extension of arbitrarily high order, i.e. for all $k=1,2,\ldots$ there is a legitimate (i.e. positive semidefinite) quantum state that: (i) extends $\rho_{AB_1}$ on $AB_1\ldots B_k$, where the systems $B_i$ are identical copies of $B$, and (ii) is symmetric under any permutation of the $B_i$ systems.
More than a single condition, this is instead a whole hierarchy of increasingly stringent conditions. Each level of the hierarchy can be decided in a computationally efficient fashion, since it is nothing but a semidefinite program, and furthermore the hierarchy is complete, meaning that if a symmetric extension is found for all $k=1,2,\ldots $, then the original state was separable.

In the GPT setting, we do not have a unique notion of positivity at our disposal, but we can still observe that if an arbitrary tensor $Z\in C_A \tmax C_B$ is separable, i.e. $Z\in C_A\tmin C_B$, then as before for all positive integers $k$ there is a symmetric extension $Z_{AB_1\ldots B_k}$ that belongs to $Z\in C_A \tmax C_{B_1} \ldots \tmax C_{B_k}$. In fact, if $Z=\sum_i x_i \otimes y_i$ with $x_i, y_i\geq 0$, and we assume without loss of generality that $\braket{u,y_i}=1$, a symmetric extension can be constructed as $Z_{AB_1\ldots B_k} = \sum_i x_i \otimes y_i^{\otimes k}$. It is clear that $Z_{AB_1\ldots B_k}\in C_A \tmin C_{B_1} \ldots \tmin C_{B_k}\subseteq C_A\tmax C_{B_1} \ldots \tmax C_{B_k}$.

Whether the complete extendibility condition is also sufficient in general, it does not seem transparent. We limit ourselves to illustrating why the standard way of proving the completeness of the hierarchy in quantum mechanics fails already in $W$-theory, as given by Definition~\ref{W theory}. We remind the reader that states in $W$-theory are represented by entanglement witnesses, which are in general non-positive. This lack of positivity deprives us of an important tool, i.e. the existence of purifications. Most standard proofs of the quantum de Finetti theorem (see for instance~\cite{1-1/2-de-Finetti} or~\cite[\S 9]{math-ent}) seem to rest crucially on the possibility of purifying a symmetric system while retaining the symmetry. Thus, we are forced to leave this problem open, hoping that further investigations will provide either a proof of the universality of the complete extendibility criterion or an appropriate counterexample.

\subsection{Separability under symmetry} \label{subsec2 sep symm}

Our main focus in this subsection is on how to simplify the separability problem when the states under examination possess further symmetries. As it turns out, these simplifications can be crucial in making the problem analytically solvable for specific classes of states, as we shall see in Section \ref{sec EB}. In the case of quantum theory, what follows has become part of the common knowledge in the entanglement theory community \cite{Werner-symmetry, EggelingPhD}.

First of all, let us fix some notation. Consider two GPTs $A=(V_A, C_A, u_A)$ and $B=(V_B, C_B, u_B)$, and take a vector subspace $\mathcal{V}\subseteq V_{AB}=V_A\otimes V_B$ of the global vector space. Naturally, we will want to consider subspaces that contain some nontrivial states, i.e. such that $\mathcal{V}\cap \big( C_A \tmax C_B \big) \neq \{0\}$.

The problem we address here consists in deciding separability of states that belong to $\mathcal{V}$. In particular, we want to identify what properties of $\mathcal{V}$ can simplify the analysis significantly. In what follows, we denote by $\mathcal{C}_{\min}$ and $\mathcal{C}_{\max}$ the cones on $\mathcal{V}$ obtained by intersecting it with $C_A \tmin C_B$ and $C_A \tmax C_B$, respectively. In formulae,
\begin{align}
\mathcal{C}_{\min} &\coloneqq \mathcal{V} \cap \Big( C_A \tmin C_B \Big)\, , \label{section min cone} \\
\mathcal{C}_{\max} &\coloneqq \mathcal{V} \cap \Big( C_A \tmax C_B \Big)\, . \label{section max cone}
\end{align}

Naturally, one can think of $\mathcal{V}^*$ as ordered by any of the two dual cones $\mathcal{C}^*_{\min},\, \mathcal{C}^*_{\max}\subset \mathcal{V}$.
For a totally generic subspace $\mathcal{V}$, a very basic statement concerning the characterisation of separability is obtained by `projecting' the Hahn-Banach separation theorem onto $\mathcal{V}$, as expressed by the following result.

\begin{prop} \label{proj HhBh} 
For $Z\in\mathcal{V}$, we have that $Z\in \mathcal{C}_{\min}$ iff
\begin{equation}
\braket{W, Z}\geq 0\quad \forall\ W\in \mathcal{C}_{\min}^*\, .
\label{proj HhBh eq}
\end{equation}
\end{prop}

The problem with the above statement is that $\mathcal{C}_{\min}^*$ is in general not easy to characterise. In fact, even though the extremal rays of $C_A \tmin C_B$ have a mathematically convenient form, once the intersection with $\mathcal{V}$ is carried out it is no longer clear how to identify the new extremal rays, i.e. those of $\mathcal{C}_{\min}$. Consequently, imposing $W \in \mathcal{C}_{\min}^*$, i.e. $\braket{W, Z}\geq 0$ for all vectors $Z$ that generate extremal rays of $\mathcal{C}_{\min}$, can easily be analytically unfeasible.

From the above discussion it is clear that the solution of the separability problem would be greatly simplified if one had in hand an efficient way of characterising $\mathcal{C}_{\min}^*$. This is possible for a relevant class of subspaces $\mathcal{V}$ that we are now set to define. We remind the reader that a \textbf{projection} onto a subspace $V_2$ of a vector space $V_1$ is a linear map $\Pi: V_1\rightarrow V_1$ such that: (i) $\Pi^2=\Pi$; and (ii) $\Pi(V_1)=V_2$.

\begin{Def} \label{cent sect def}
A subspace $\mathcal{V}\subseteq V_{AB}$ of a bipartite system is called a \textbf{central section} (of $C_A\tminit C_B$) if there exists a projection $\Pi$ of $V_{AB}$ onto $\mathcal{V}$ that is also separability-preserving, i.e. such that
\bb
\Pi \Big( C_A \tminit C_B \Big)\subseteq \mathcal{V}\cap \Big( C_A \tminit C_B \Big) = \mathcal{C}_{\min}\, .
\label{cent sect def eq}
\ee
\end{Def}

Before proceeding further, let us stop for a moment and look into a particular class of example of central sections that are of prominent importance in quantum information.

\begin{ex} \label{local groups ex}
An interesting particular case of separability-preserving projection is of the form 
\bb
\Pi_G \coloneqq \int_G dg\ \zeta_1(g)\otimes \zeta_2(g)\, ,
\label{local groups projection}
\ee
where $\zeta_i : G\rightarrow GL(V_i)$ are (real) representations of a compact group $G$ on $V_1, V_2$, and $\int_G dg$ is the Haar integral. In that case, $\mathcal{V}$ must be taken to be the set of fixed points of the representation $\zeta_1 \otimes \zeta_2$, that coincides (up to the Choi isomorphism in an orthogonal basis for a $G$-invariant product) with the set of $G$-commuting isomorphisms $V_1 \rightarrow V_2$, and therefore can be easily determined with the aid of representation theory. What we just described constitutes the typical application of this framework \cite{Werner, Werner-symmetry}. We will stumble upon it ourselves in Section \ref{sec EB}.   
\end{ex}

Returning to the general framework, we see that the existence of a (for now generic) projection $\Pi$ of $V_{AB}$ onto $\mathcal{V}$, which can be equivalently seen as a map $\Pi:V_{AB}\rightarrow \mathcal{V}$, enables us to give some more structure to the problem. In fact, we can construct also the adjoint projection $\Pi^*: \mathcal{V}^*\rightarrow V_{AB}^*$ and identify $\mathcal{V}^* \simeq \Pi^*(V_{AB}^*)$. This amounts to thinking of $\mathcal{V}^*$ as a subspace of $V_{AB}^*$ by saying that $\varphi \in\mathcal{V}^*$ can be extended to $\tilde{\varphi}\in V_{AB}^*$ through $\braket{\tilde{\varphi}, \cdot}\coloneqq \braket{\varphi, \Pi(\cdot)}$. Therefore, from now on we can safely let $\varphi \in\mathcal{V}^*$ act on global states.

Once we fix a projection onto $\mathcal{V}$, so that $\mathcal{V}^*\subseteq V_{AB}^*$, we can define two other cones in the dual space:
\begin{align}
\mathcal{C}_{*\min} &\coloneqq \mathcal{V}^* \cap \Big( C_A^* \tmin C_B^* \Big)\, , \label{section* min cone} \\
\mathcal{C}_{*\max} &\coloneqq \mathcal{V}^* \cap \Big( C_A^* \tmax C_B^* \Big)\, . \label{section* max cone}
\end{align}
Once we take into account also the duals of \eqref{section min cone} and \eqref{section max cone}, we now have four possible cones in $\mathcal{V}^*$. However, not all of these cones are independent of each other. For instance, one can easily show that
\begin{align}
\mathcal{C}_{*\min} &\subseteq \mathcal{C}_{\max}^*\, , \label{bound section primal min} \\
\mathcal{C}_{*\max} &\subseteq \mathcal{C}_{\min}^*\, , \label{bound section primal max}
\end{align}
and consequently that
\begin{align}
\mathcal{C}_{\max} &\subseteq \mathcal{C}_{*\min}^*\, , \label{bound section dual max}
\\
\mathcal{C}_{\min} &\subseteq \mathcal{C}_{*\max}^*\, . \label{bound section dual min}
\end{align}

Now, remember that we are looking for a convenient mathematical characterisation of the cone $\mathcal{C}_{\min}^*$. Indeed, such a characterisation would allow us to improve the efficiency of the criterion in Proposition \ref{proj HhBh}. The importance of central sections in this context rests primarily on the following observation.

\begin{lemma} \label{cent sect lemma}
Let $\mathcal{V}\subseteq V_{AB}$ be a central section. Then the cones $\mathcal{C}_{\min},\, \mathcal{C}_{*\max}$ associated with it through \eqref{section min cone} and \eqref{section* max cone} satisfy
\bb
\mathcal{C}_{*\max} = \mathcal{C}_{\min}^*\, ,
\label{cent sect eq}
\ee
i.e. the bound in \eqref{bound section primal max} (equivalently, that in \eqref{bound section dual min}) is saturated.
\end{lemma}

\begin{proof}
Take a $W\in \mathcal{C}_{\min}^*\subseteq \mathcal{V}^*\subseteq V_{AB}^*$, and let us show that $W \in\mathcal{C}_{*\max}$. Because of the definition \eqref{section* max cone}, all we have to prove is that $W\in C_A^* \tmax C_B^*$.
Thanks to the duality relation \eqref{minimal3}, this translates to
\bbb
\braket{W, Z}\geq 0\quad \forall\ Z\in C_A \tmin C_B\, .
\eee
Now, the way the inclusion of $\mathcal{V}^*\subseteq V_{AB}^*$ is designed implies indeed that
\bbb
\braket{W, Z} = \braket{W, \Pi(Z)} \geq 0\, ,
\eee
because $\Pi(Z)\in \Pi \big( C_A \tmin C_B \big)\subseteq \mathcal{C}_{\min}$, as follows from \eqref{cent sect def eq}. This concludes the proof.
\end{proof}

As an immediate consequence, we obtain an improved version of Proposition~\ref{proj HhBh}.

\begin{prop} \label{impr HhBh}
For $\mathcal{V}$ central section and $Z\in\mathcal{V}$, we have that $Z\in \mathcal{C}_{\min}$ iff
\begin{equation}
\braket{W, Z}\geq 0\quad \forall\ W \in \mathcal{C}_{*\max}\, .
\label{impr HhBh eq}
\end{equation}
\end{prop}

\begin{proof}
The necessity of \eqref{impr HhBh eq} is obvious, while the converse statement follows by putting together Proposition \ref{proj HhBh} and Lemma \ref{cent sect lemma}.
\end{proof}

Rephrasing the claim of Theorem \ref{impr HhBh}, we could say that whenever $\mathcal{V}$ is a central section, in order to decide whether $Z\in\mathcal{V}$ is separable or not it suffices to test the inequality $\braket{W,Z}\geq 0$ on all functionals $W\in \mathcal{V}^*\subseteq V_{AB}^*$ that happen to be global entanglement witnesses.

\subsection{Tensor rank and separability} \label{subsec2 tensor rank}

This subsection is concerned with exploring possible connections between tensor rank and entanglement properties. Our interest in this kind of questions was spurred by recent results of Cariello~\cite{Cariello13, Cariello14}, which show in particular that any bipartite quantum state whose tensor rank is $2$ or less is necessarily separable. For a concise presentation of this result, we refer the reader to~\cite{JBlog}. Here we aim to generalise this statement to the GPT setting, a goal achieved by Theorem~\ref{t-rk 2 sep}.

In order to arrive at the main result of this subsection we need a couple of preliminary lemmas. Let us start by reminding the reader that a vector $x$ of an ordered vector space $V$ is called strictly positive, and we write $x>0$, if it is internal to the positive cone, in symbols $x\in \inter(V_+)$.

\begin{lemma} \label{int sep}
Let $V_{A},V_{B}$ be finite-dimensional ordered vector spaces, whose positive cone we denote by $C_A,C_B$. If $a\in V_A$ and $b\in V_B$ are both strictly positive, i.e. $a,b>0$, then $a\otimes b \in \interit \big(C_{A} \tminit C_{B}\big)$, i.e. $a \otimes b >0$ with respect to the ordering determined on $V_A\otimes V_B$ by the minimal tensor product $C_A\tminit C_B$.
\end{lemma}

\begin{proof}
Fix two bases $\{v_{i}\}_{i=1}^{d_A}$ on $V_{A}$ and $\{w_{j}\}_{j=1}^{d_B}$ on $V_{B}$. For all vectors $\delta \in V_{A}$ expressed uniquely as $\delta =\sum_{i=1}^{d_A} \alpha_{i} v_{i}$, define the norms $|\delta|_{\infty}\coloneqq \max_{i} |\alpha_{i}|$, $|\delta|_1 \coloneqq \sum_i |\alpha_i|$, and analogously for $V_{B}$. The same constructions can be applied to the tensor product $V_{A}\otimes V_{B}$ equipped with the basis $\{v_{i}\otimes w_{j}\}_{i,j=1}^{d_A d_B}$. Now, take $\epsilon>0$ such that $|\delta|_{\infty}\leq \epsilon$ implies $a+\delta\geq 0$ for all $\delta\in V_A$ and $b+\delta\geq 0$ for all $\delta\in V_B$. Then we claim that for all $\Delta \in V_{A}\otimes V_{B}$ we have that $|\Delta|_{1}\leq \epsilon^2$ implies $a \otimes b + \Delta \in C_{A}\otimes_{\min}C_{B}$. In fact, taking $\Delta=\sum_{ij} \alpha_{ij} v_{i}\otimes w_{j}$ such that $|\Delta|_1\leq \epsilon^2$, we have
\begin{align*}
&a \otimes b +\Delta \\
&\quad= a\otimes b + \sum_{i,j} \alpha_{ij} v_{i}\otimes w_{j} \\
&\quad= \left(1 - \frac{|\Delta|_1}{\epsilon^2}\right) a\otimes b + \frac{1}{\epsilon^2} \sum_{ij} |\alpha_{ij}|\, \left( a\otimes b + s_{ij} \epsilon^2 v_{i} \otimes w_{j} \right) \\
&\quad= \left(1 - \frac{|\Delta|_1}{\epsilon^2}\right) a\otimes b \\
&\qquad + \frac{1}{2\epsilon^2} \sum_{ij} |\alpha_{ij}| \big( (a+ \epsilon\, v_{i})\!\otimes\! (b+s_{ij} \epsilon\, w_{j}) + (a - \epsilon\, v_{i})\!\otimes\! (b-s_{ij} \epsilon\, w_{j}) \big) \, ,
\end{align*}
where $s_{ij}=\text{sign}(\alpha_{ij})$ are the signs of the coefficients $\alpha_{ij}$. The above representation is explicitly separable, since the choice of $\epsilon>0$ makes
\bbb
a\pm \epsilon\,v_{i} \geq 0\, ,\qquad b\pm \epsilon\,w_{j}\geq 0
\eee
for all $i$ and $j$. We thus have $a\otimes b+\Delta\in C_{A}\tmin C_{B}$, as claimed.
\end{proof}

\begin{lemma} \label{boundary}
Let $V$ be a finite-dimensional ordered vector space $V$. For $a>0$ and $v\in V$ with $v\ngeq 0$, define $t_{0}\coloneqq \max\{t\geq 0\in\mathds{R}:\ a+tv\geq 0\}$. Then there exists a positive functional $\varphi_0$ that: (a) defines an extremal ray of the dual positive cone; (b) is such that $\braket{\varphi, a+t_{0}v}=0$; and (c) satisfies $\braket{\varphi,a}>0$.
\end{lemma}

\begin{proof}
We denote by $C$ the positive cone of $V$. A little thought reveals that by construction $a+t_0 v$ must lie on the boundary of $C$. This implies that there is a positive functional $\varphi_0\in C^*$ such that $\braket{\varphi_0, a+t_0 v}=0$, meeting (b). In fact, it this were not the case, we could take a norm on $V$, denoted by $\|\cdot\|$, then consider
\bbb
\epsilon\coloneqq \inf\left\{ \braket{\varphi, a+t_0 v}:\ \varphi\in C^*,\, \|\varphi\|_*=1 \right\} > 0\, ,
\eee
where the last inequality follows by compactness, and note that the vector $a+\left( t_0 + \frac{\epsilon}{\|v\|} \right) v$ belongs to $C=C^{**}$ since it has a positive expectation value on all functionals $\varphi\in C^*$. This would be in contradiction with the definition of $t_0$.

We can of course decompose $\varphi_0 \in C^*$ as a convex combination of functionals lying on extreme rays of $C^*$, and naturally all of those functionals vanish when evaluated on $a+t_0 v$. It follows that we can safely assume that $\varphi_0$ determines an extremal ray itself, meeting requirement (a).
Finally, since $a$ is internal to $C$ and $\varphi_0\in C^*$, we must have $\braket{\varphi, a}>0$.
\end{proof}

We are now ready to present the main result of this subsection.
The first part of its proof follows closely that given for the quantum case in~\cite{Cariello13} (see also~\cite{JBlog}). However, we will see that we are forced to depart from it at some point.

\begin{thm} \label{t-rk 2 sep}
Let $V_A, V_B$ be finite-dimensional ordered vector spaces with positive cones $C_A, C_B$. Whenever $Z\in V_{A}\otimes V_{B}$ has tensor rank $2$ or less, $Z \in C_{A} \tmaxit C_{B}$ if and only if  $Z\in C_{A} \tminit C_{B}$.
\end{thm}

\begin{proof}
Clearly, we have just to show that every vector $Z \in C_{A} \tmax C_{B}$ of tensor rank $2$ belongs also to $C_{A} \tmin C_{B}$, as the inclusion $C_A \tmin C_B\subseteq C_A\tmax C_B$ always holds. As usual, we consider $V_{A},V_{B}$ as endowed with the partial orders given by $C_{A},C_{B}$. We write
\begin{equation}
Z = a\otimes b + x\otimes y\, . \label{alpha}
\end{equation}
Let us break down the proof into several steps.

\begin{enumerate}

\item First of all, we want to show that we can suppose without loss of generality that $a,b\geq 0$ (each of them in its own space). To see this, pick strictly positive functionals $u_A\in \inter\left(C_A^*\right)$ and $u_B\in \inter\left(C_B^*\right)$. Since $u_A\otimes u_B$ is again internal to $C_A^{*}\tmin C_B^{*}$ thanks to Lemma~\ref{int sep}, applying~\eqref{minimal2} we conclude that $\braket{u_A\otimes u_B, Z}>0$. Therefore, without loss of generality we can assume that
\bbb
\braket{u_A\otimes u_B, Z} = \braket{u_A, a} \braket{u_B, b} + \braket{u_A, x} \braket{u_B, y}=1\, .
\eee
Now, let us write
\begin{align*}
Z &= \left( \braket{u_B,b} a + \braket{u_B,y} x \right) \otimes  \left( \braket{u_A,a} b + \braket{u_A,x} y \right) \\
&\qquad + \left( \braket{u_A,x} a - \braket{u_A, a} x \right) \otimes  \left( \braket{u_B,y} b - \braket{u_B,b} y \right)\, .
\end{align*}
In the above equation, which has the same form as~\eqref{alpha}, we see that $\braket{u_B,b} a + \braket{u_B,y} x = \braket{u_B,Z}\geq 0$, where $\braket{u_B,Z}$ corresponds to a partial evaluation of the functional $u_B$ on the tensor $Z$, and is therefore a positive vector since $Z\in C_A\tmax C_B$. Analogously, $\braket{u_A,a} b + \braket{u_A,x} y = \braket{u_A,Z} \geq 0$.

\item In fact, we can do better, and suppose in~\eqref{alpha} that $a,b>0$. This is because for all $a,b\geq 0$ there are always sequences $a_n, b_n$ such that $a_n >a,\, b_n>b$ for all $n\in\mathds{N}$, with equality in the limit $n\rightarrow \infty$. To obtain such sequences, it suffices to add to $a,b$ asymptotically small multiples of a strictly positive vector. The new tensor $Z_n \coloneqq a_n \otimes b_n + x\otimes y$ still belongs to $C_A\tmax C_B$, as it can be verified that the application of a nonzero product functional always yields a larger result than that one would obtain with $Z$. Since $\lim_{n \rightarrow \infty} Z_n = Z$ and the minimal tensor product is closed, it suffices to show that $Z_n\in C_A\tmin C_B$ for all $n$.

\item At this point Cariello~\cite{Cariello13} operates a local filtering operation to assume that $a,b$ are both the identity matrix. This is obviously not possible in the GPT setting, so we are forced to take a different route. The originality of our proof lies mainly in the rest of the argument.

We refer again to~\eqref{alpha} but now we take $a,b>0$. We claim that we can suppose without loss of generality that there are two positive functionals $\varphi\in C_A^{*}$, $\lambda\in C_B^{*}$ such that $\braket{\lambda,a} \braket{\eta, b}=-\braket{\lambda,x}\braket{\eta,y}>0$. In fact, defining
\bbb
k\coloneqq \max\left\{t\geq 0: a\otimes b + t\, x\otimes y\in C_A\tmax C_B\right\}\geq 1\, ,
\eee
we see that
\begin{equation}
Z = \left(1-\frac1k \right) a\otimes b + \frac1k \left( a\otimes b + k\, x\otimes y\right)\, .
\end{equation}
Clearly, proving that $a\otimes b + k\, x\otimes y\in C_A\tmin C_B$ would allow us to conclude. Observe that by Lemma~\ref{boundary} and identity~\eqref{minimal2} there is a functional $\Phi \in \big(C_A\tmax C_B\big)^{*}=C_A^{*}\tmin C_B^{*}$ corresponding to an extremal ray and such that $\braket{\Phi ,a\otimes b + k\, x\otimes y}=0$ and $\braket{\Phi, a\otimes b}>0$. It is elementary to verify that the extremal rays in the minimal tensor product are just tensor products of positive functionals, i.e. $\Phi = \lambda\otimes \eta$ with $\lambda \in C_A^{*},\, \eta\in C_B^{*}$. Renaming $k x\mapsto x$ yields the claim.

\item Consider the two positive functionals $\lambda,\eta $ we constructed above. According to step 3, we now assume that $Z$ written as in~\eqref{alpha} satisfies $\braket{\lambda,a} \braket{\eta, b}=-\braket{\lambda,x}\braket{\eta,y}>0$. Applying $\lambda,\eta$ on one side or the other of $Z$ we get
\begin{align*}
\braket{\eta, Z} &= \braket{\eta, b} a + \braket{\eta, y} x  \eqqcolon c \geq 0\, ,\\
\braket{\lambda, Z} &= \braket{\lambda,a} b + \braket{\lambda,x} y \eqqcolon d \geq 0\, . 
\end{align*}
Since $\braket{\lambda, a}\braket{\eta, b}>0$, proving that $\braket{\lambda, a}\braket{\eta, b} Z\in C_A\tmin C_B$ will suffice. We obtain
\bb
\begin{split}
\braket{\lambda, a}\braket{\eta, b} Z &= \braket{\eta,b} a\otimes \braket{\lambda,a} b + \braket{\lambda, a}\braket{\eta, b} x\otimes y \\
&= (c - \braket{\eta, y} x)\otimes (d-\braket{\lambda,x} y) + \braket{\lambda, a}\braket{\eta, b} x\otimes y \\
&= c\otimes d - \braket{\eta,y} x\otimes d - \braket{\lambda, x} c\otimes y \\
& \eqqcolon W\, .
\end{split}
\label{beta}
\ee

\item By the same reasoning as in step 2, we can suppose that $c,d>0$. Exactly as in step 3, we can again assume without loss of generality that there are two positive functionals $\mu\in C_A^*,\, \nu\in C_B^*$ satisfying $\braket{\mu\otimes\nu, W}=0$ and $\braket{\mu, c} \braket{\nu, d} >0$. Then, it suffices to prove that $\braket{\mu, c} \braket{\nu, d} W\in C_A\tmin C_B$. Applying $\mu$ and $\nu$ on $W$ written as in~\eqref{beta} we get
\begin{align*}
\braket{\nu, W} &= \left(\braket{\nu,d} - \braket{\lambda,x} \braket{\nu,y}\right) c - \braket{\eta,y} \braket{\nu,d} x \eqqcolon p \geq 0\, ,\\
\braket{\mu,W} &= \left(\braket{\mu,c} - \braket{\eta,y} \braket{\mu,x} \right) d - \braket{\lambda, x} \braket{\mu, c} y \eqqcolon q \geq 0\, .
\end{align*}
Finally,
\begin{align*}
&\braket{\mu, c} \braket{\nu, d} W \\[0.5ex]
&\quad = \braket{\mu, c} \braket{\nu, d} c\otimes d - \braket{\eta,y} \braket{ \nu, d} x\otimes \braket{\mu,c} d - \braket{\nu,d} c\otimes \braket{\lambda,x} \braket{\mu,c} y \\[0.5ex]
&\quad= \braket{\mu,c} \braket{\nu,d} c\otimes d + \left(p - \left(\braket{\nu,d} -\braket{\lambda,x}\braket{\nu,y}\right) c\right)\otimes \braket{\mu,c} d \\
&\quad\quad + \braket{\nu,d} c\otimes \left(q - \left(\braket{\mu,c} - \braket{\eta,y}\braket{\mu,x}\right) d\right) \\[0.5ex]
&\quad = \left( \braket{\mu,c} \braket{\nu,d} - \braket{\mu,c} \left(\braket{\nu,d} - \braket{\lambda,x} \braket{\nu,y}\right) \right. \\
&\quad\quad \left.- \braket{\nu,d} \left(\braket{\mu,c} - \braket{\eta,y}\braket{\mu,x}\right) \right)\, c\otimes d \\
&\quad\quad + \braket{\mu,c} p\otimes d + \braket{\nu,d} c\otimes q \\[0.5ex]
&\quad = \left( -\braket{\mu,c} \braket{\nu,d} + \braket{\mu,c} \braket{\lambda,x} \braket{\nu,y} + \braket{\nu,d} \braket{\eta,y}\braket{\mu,x} \right)\, c\otimes d \\
&\quad\quad + \braket{\mu,c} p\otimes d + \braket{\nu,d} c\otimes q \\[0.5ex]
&\quad = - \braket{\mu\otimes \nu, W} c\otimes d + \braket{\mu,c} p\otimes d + \braket{\nu,d} c\otimes q \\[0.5ex]
&\quad= \braket{\mu,c} p\otimes d + \braket{\nu,d} c\otimes q \\[0.5ex]
&\quad\in\, C_A\tmin C_B\, ,
\end{align*}
which allows us to conclude.
\end{enumerate}
\end{proof}

\begin{rem}
A little thought reveals that we have also proved that any $Z \in C_A\tmax C_B$ of tensor rank $2$ can be always written as $Z=\sum_{i=1}^{4} a_{i}\otimes b_{i}$, with $a_{i},b_{i}\geq 0$.
\end{rem}

When restricted to quantum mechanics, Theorem~\ref{t-rk 2 sep} tells us the following.

\begin{cor} \label{tensor rank 2 QM cor}
Every bipartite operator $W\in\mathcal{H}_{n_A}\otimes\mathcal{H}_{n_B}$ which is an entanglement witness, i.e. belongs to~\eqref{witnesses}, and whose tensor rank is $2$ or less is necessarily separable, i.e. it belongs to the cone~\eqref{separable}.
\end{cor}

\begin{proof}
Straightforward application of Theorem~\ref{t-rk 2 sep} to quantum mechanics as discussed in Subsection~\ref{subsec2 ex QM}. 
\end{proof}

The above Corollary~\ref{tensor rank 2 QM cor} generalises the result obtained by Cariello~\cite{Cariello13, JBlog}, which states that any \emph{positive} operator of tensor rank $2$ or less is separable. Things can not be pushed further, as in~\cite{Cariello14} it is proved that there exist two-qutrit entangled states whose tensor rank is $3$.

\subsection{Separability problem for centrally symmetric models} \label{subsec2 sep centr}

We now turn our attention to the class of centrally symmetric models (Definition~\ref{centr}). The reason why we should care about these models will become clear in Chapter~\ref{chapter3}. For now, as we said at the beginning of the section, let us see this as a mathematical divertissement. For notation and conventions, we refer the reader to Subsection~\ref{subsec2 centr symm}. 

Our focus here is on bipartite systems whose local parties are modelled by centrally symmetric theories. The main result of this subsection is the solution of the separability problem for states of the particular form $Z= U_* + \widehat{M}$, where $M\in \mathds{R}^{(d_A-1) \times (d_B-1)}$. As we shall see in a moment, the solution involves the theory of tensor norms as we discussed it in Subsection~\ref{subsec2 tensor norms}.

\begin{prop} \label{sep centr prop}
Let $Z$ be a state of a bipartite theory whose local systems are represented by centrally symmetric GPTs. Then
\begin{align}
Z \in C_A \tminit C_B\quad &\Longrightarrow\quad \big|\widebar{Z}\big|_\pi \leq Z_{00}\, , \label{sep centr eq1} \\
Z \in C_A \tmaxit C_B\quad &\Longrightarrow\quad \big|\widebar{Z}\big|_\varepsilon\leq Z_{00}\, , \label{sep centr eq2}
\end{align}
where $|\cdot|_\pi,\, |\cdot|_\varepsilon$ are the projective and injective tensor norms constructed out of the local norms $|\cdot|$. Furthermore, the above conditions are also sufficient when all cross terms $Z_{i0}, Z_{0j}$ ($i,j\geq 1$) vanish.
\end{prop}

\begin{proof}
We start by proving~\eqref{sep centr eq1}. If $Z=\sum_{k}x^{(k)} (y^{(k)})^{T}$ with $\big|\widebar{x}^{(k)}\big| \leq x_{0}^{(k)},\, |\widebar{y}^{(k)}|\leq y_{0}^{(k)}$ for all $k$, then
\begin{align*}
\big|\widebar{Z}\big|_\pi &= \bigg|\sum_{k} \widebar{x}^{(k)} (\widebar{y}^{(k)})^{T}\bigg|_\pi \\
&\leq \sum_{k} \big|\widebar{x}^{(k)}\big|\, \big|\widebar{y}^{(k)}\big| \\
&\leq \sum_{k} x_{0}^{(k)} y_{0}^{(k)} \\
&= Z_{00}\, .
\end{align*}

Now, let us turn our attention to the proof of the sufficiency claim. Suppose that $Z_{ij}=0$ whenever $i=0,\, j\geq 1$ or $i\geq 1,\, j=0$, and further assume that $\big|\widebar{Z}\big|_{\pi}\leq Z_{00}$. Then by definition of projective norm~\eqref{proj} there are vectors $v^{(k)}\in\mathds{R}^{d_{A}-1}$ and $w^{(k)}\in\mathds{R}^{d_{B}-1}$ such that $\widebar{Z}=\sum_{k} v^{(k)} (w^{(k)})^T$ and $\big|\widebar{Z}\big|_{\pi}=\sum_{k} \big|v^{(k)}\big|\, \big|w^{(k)}\big|$. Define vectors $x_{\pm}^{(k)}\coloneqq \big|v^{(k)}\big| u_{A} \pm \myhat{v}^{(k)}\in \mathds{R}^{d_{A}}$, i.e.
\bbb
\left(x_{\pm}^{(k)}\right)_{i} = \left\{ \begin{array}{cl} \big|v^{(k)}\big| & \text{ if $i=0$,} \\[0.5ex] \pm\, \left(v^{(k)}\right)_i & \text{ if $i\geq 1$,} \end{array} \right.
\eee
and analogously for $y_{\pm}^{(k)}\coloneqq \big|w^{(k)}\big| u_{B} \pm \myhat{w}^{(k)} \in \mathds{R}^{d_{B}}$. The definition of centrally symmetric models~\eqref{centr eq} tells us that $x_\pm^{(k)}\in C_A$ and $y_\pm^{(k)}\in C_{B}$. Then it is easy to see that
\begin{align*}
Z &= \left(Z_{00} - \big|\widebar{Z}\big|_\pi \right) u_A u_B^T + \frac12 \sum_{k} \left( x^{(k)}_{+} (y^{(k)}_{+})^{T} +  x^{(k)}_{-} (y^{(k)}_{-})^{T} \right) ,
\end{align*}
which shows that $Z$ is separable, as claimed. Alternatively, one can show separability by exploiting the `Woronowicz condition' (Proposition~\ref{Woronowicz prop}(b)).

We can perform an analogous reasoning at the level of the dual, which shares the same centrally symmetric structure, as apparent from~\eqref{centr dual eq}. We obtain immediately that $W\in C_A^*\tmin C_B^*$ implies that $\big|\widebar{W}\big|_{*\pi}\leq W_{00}$. An arbitrary $Z\in C_A \tmax C_B$ will satisfy $\braket{W,Z}\geq 0$ for all such $W$. Optimising this condition yields $Z_{00} - \big|\widebar{Z}\big|_{*\pi *}\geq 0$, which becomes $\big|\widebar{Z}\big|_{\varepsilon}\leq Z_{00}$ after employing the duality formulae~\eqref{dual inj proj}. This concludes the proof of~\eqref{sep centr eq2}. Exploiting again duality in the form of~\eqref{maximal2}, one shows that~\eqref{sep centr eq2} expresses a sufficient condition when all cross terms $Z_{i0}, Z_{0j}$ ($i,j\geq 1$) vanish.
\end{proof}

Of course, the above Proposition~\ref{sep centr prop} constitutes merely a rephrasing of the problem, if we do not know how to compute injective and projective norms in the specific case we want to analyse. And in fact in general the optimisations problems in~\eqref{inj} and~\eqref{proj} are hard to solve explicitly. However, as we saw in Subsection~\ref{subsec2 tensor norms}, there is an important special case where some sort of closed expression is indeed available, i.e. when one takes two Euclidean spaces. At the level of GPTs, this is naturally the same as considering two spherical models, as defined in~\eqref{spherical},~\eqref{ice cream}. We obtain the following.

\begin{cor} \label{sep sph cor}
Let $Z$ be a state of a bipartite theory whose local systems are modelled by spherical models as in~\eqref{spherical}. Then
\begin{align}
Z \in C_{d_A} \tminit C_{d_B} \quad &\Longrightarrow\quad \big\|\widebar{Z}\big\|_1 \leq Z_{00}\, , \label{sep sph eq1} \\
Z \in C_{d_A} \tmaxit C_{d_B} \quad &\Longrightarrow\quad \big\|\widebar{Z}\big\|_\infty \leq Z_{00}\, , \label{sep sph eq2}
\end{align}
where the cones are defined by~\eqref{ice cream}, and $\|\cdot\|_1,\, \|\cdot\|_\infty$ denote the trace and operator norm, respectively. Furthermore, the above conditions are also sufficient when all cross terms $Z_{i0}, Z_{0j}$ ($i,j\geq 1$) vanish.
\end{cor}

\begin{proof}
Follows by putting together Proposition~\ref{sep centr prop} and Example~\ref{ex inj proj}.
\end{proof}

\section{Minimal and maximal tensor product} \label{sec2 min vs max}

\subsection{Generalities} \label{subsec2 generalities}

In Subsection~\ref{subsec2 bipartite}, we have discussed two constructions that are fundamental in defining the composition rules of GPTs, i.e. minimal and maximal tensor products of cones. As we mentioned, these constructions were originally developed in~\cite{Peressini-minmax, Hulanicki-minmax} in connection with the theory of tensor products of ordered Banach spaces. Although certainly meaningful from the mathematical point of view, the reader might wonder why we should care about them here, since this thesis is primarily about mathematical \emph{physics}. The reason is simple: \emph{bipartite states that are in the maximal but not in the minimal tensor product are physically entangled}, meaning that they can not be prepared with LOCC operations~\eqref{LOCC}. Therefore, the question (i) discussed in Subsection~\ref{subsec2 fundamental} can be rephrased in the GPT setting as follows: \emph{given two non-classical GPTs $A$ and $B$, is it always the case that $C_A\tminit C_B\neq C_A \tmaxit C_B$?} This happens for quantum mechanical systems, as we saw in Subsection~\ref{subsec2 ex QM}, but we suspect that this is indeed the case in full generality. With the help of Definition~\ref{def class}, we formulate this conjecture as follows.

\begin{cj} \label{ultimate cj}
If $C_A$ and $C_B$ are two non-simplicial cones, then
\bb
C_A \tminit C_B \neq C_A \tmaxit C_B\, .
\label{ultimate cj eq}
\ee
\end{cj}

When the positive cones $C_A, C_B$ of two GPTs $A=(V_A, C_A, u_A)$, $B = (V_B, C_B, u_B)$ satisfy $C_A \tmin C_B \neq C_A \tmax C_B$, we say that $A$ and $B$ are \textbf{entangleable}.
Despite its outstanding physical and conceptual relevance, the question in Conjecture~\ref{ultimate cj} seems to have been considerably overlooked. In fact, we could not find any clear statement of it anywhere in the rich literature on the subject of nonlocality in GPTs~\cite{Oppenheim-Wehner,Banik-2013,Busch-2013,Stevens-Busch,Cavalcanti-2016,Plavala-2016,Jencova-Plavala,Jencova2017}. 
The first to have investigated and solved a closely related question have been Namioka and Phelps. In their paper~\cite{NP}, they prove a weaker form of Conjecture~\ref{ultimate cj} whose validity was apparently suggested to them by E. Effros. Namely, they show that there is a particularly simple choice of a non-simplicial cone $C_B$ such that for any other cone $C_A$, the equality in~\eqref{ultimate cj eq} is possible iff $C_A$ is simplicial. This `test' cone $C_B$ can be taken as that of the gbit, defined by~\eqref{C square}. For details, we refer the reader to Subsection~\ref{subsec2 NP thm}, where we will state and prove Namioka-Phelps' theorem.

Before we proceed further, we want to discuss a slightly less general scenario than that treated in Conjecture~\ref{ultimate cj}. Namely, we can consider the special case of two copies of the same model, and ask whether they can be entangled. We thus consider the following conjecture.

\begin{cj} \label{ultimate cj weaker}
If $C$ is a non-simplicial cone, then
\bb
C \tminit C \neq C \tmaxit C\, .
\label{ultimate cj weaker eq}
\ee
\end{cj}

This is clearly weaker than Conjecture~\ref{ultimate cj}. The reason to consider this variation is twofold. First, we believe it captures all the essential features of the general problem, and a complete solution of it could easily lead to solving Conjecture~\ref{ultimate cj} as well. Second, it is technically a bit easier to work with, for a variety of reasons. For instance, we will see in Subsection~\ref{subsec2 univ ent centr} that we can show an explicit construction of an entangled state in the case of two copies of the same centrally symmetric model (thus establishing the validity of Conjecture~\ref{ultimate cj weaker} in this setting), but that the same construction for two \emph{different} centrally symmetric models does not quite work.

Before we delve into the discussion of Conjecture~\ref{ultimate cj} and of its cousin Conjecture~\ref{ultimate cj weaker}, let us start with almost trivial sanity check. Namely, we want to verify that the assumption that both $C_A$ and $C_B$ are non-simplicial is essential for the claim of Conjecture~\ref{ultimate cj} to hold. The following is a well-known result that confirms this intuition.

\begin{lemma} \emph{\cite{NP}.} \label{simplicial min=max lemma}
If either $C_A$ or $C_B$ are simplicial, then
\bb
C_A \tminit C_B = C_A \tmaxit C_B\, .
\ee
\end{lemma}

\begin{proof}
Since we work in finite dimension, a direct proof of the result that differs slightly from the one presented in~\cite{NP} is available. Let us present this rather intuitive proof. We make no claim of originality, since the argument is extremely simple and part of the folklore surrounding the problem. Assume that $C_A$ is a simplicial cone. According to Definition~\ref{classical}, this amounts to saying that
\bbb
C_A = \co\left\{ v_1, \ldots, v_d\right\}
\eee
for some basis $\{ v_1, \ldots, v_d\}$ of $V_A$. Let $\left\{v_1^*, \ldots, v_d^*\right\}$ be the corresponding dual basis of $V_A^*$. It is not difficult to realise that $v_i^*\in C_A^*$ for all $i$, since writing any $x\in C_A$ as $x = \sum_j \alpha_j v_j\in C_A$ for some coefficients $\alpha_j\geq 0$ ensures that $\braket{v_i^*, x} = \alpha_i \geq 0$.

Since the minimal tensor product is always included in the maximal one, we have to show only the opposite inclusion. For an arbitrary tensor $Z\in C_A \tmax C_B$ expressed as $Z=\sum_{i=1}^d v_i \otimes y_i$ for some vectors $y_i\in V_B$, the defining property of the maximal tensor product~\eqref{maximal} guarantees that
\bbb
0\leq \braket{v_i^*\otimes \varphi, Z} = \braket{\varphi, y_i}
\eee
for all $\varphi\in C_B^*$. This implies that $y_i\in C_B^{**}=C_B$, thus ensuring that $Z\in \co \left( C_A \otimes C_B\right) = C_A \tmin C_B$.
\end{proof}

In order to make some progress toward establishing Conjecture~\ref{ultimate cj} at least in some special cases, we need to gain a better understanding of classical systems. The purpose of the following Subsections~\ref{subsec2 lattices},~\ref{subsec2 dec interpol} and~\ref{subsec2 dec lemma} is to introduce some basic notions of \emph{lattice theory}, which treats ordered sets (in our case, ordered vector spaces) enjoying the additional property that every pair of elements admits an \emph{infimum}, i.e. a minimal common upper bound, and a \emph{supremum}, i.e. a maximal common lower bound. As we will see, this seemingly innocent property is a strong requirement to impose on an ordered vector space, in fact so strong that in finite dimension vector lattices corresponds exactly to simplicial cones as given by Definition~\ref{def class}. This alternative characterisation is heavily exploited in the proof of Namioka-Phelps' theorem (Theorem~\ref{NP thm}) in Subsection~\ref{subsec2 NP thm}, which is the ultimate reason why we are about to devote some time reviewing the basics of lattice theory.

\subsection{Lattices and vector lattices} \label{subsec2 lattices}

Throughout this subsection, we discuss some properties of abstract mathematical structures called lattices, with special attention devoted to the case of vector lattices. The theory we will review here is covered by standard textbooks like~\cite{BIRKHOFF} (for lattice theory) and~\cite{ALIPRANTIS,SIMON} (for vector lattices). A brief introduction to vector lattices can also be found in~\cite[\S 10]{PHELPS}. Our exposition follows mostly~\cite[I]{ALIPRANTIS}. Let us start with the following general definition.

\begin{Def} \label{def lattice}
A \textbf{lattice} is a partially ordered set $(X,\leq)$ in which for every two elements $x,y\in X$ there exists the minimum simultaneous upper bound $x \vee y \coloneqq \min\left\{z\in X:\, z\geq x\ \text{and}\ z\geq y \right\}$ (supremum) and the maximum simultaneous lower bound $x\wedge y \coloneqq \min\left\{z\in X:\, z\leq x\ \text{and}\ z\leq y \right\}$ (infimum).
\end{Def}

The above notion of lattice is fully general, and it will be helpful to have it at hand. However, in our case we will be mostly concerned with lattices on vector spaces. The relevant definition is as follows. 

\begin{Def} \label{def vector lattice}
A \textbf{vector lattice} (also called a \textbf{Riesz space}) is an ordered vector space that is also a lattice (with respect to the same order relation). 
\end{Def}

It can be easily proved that if $V$ is a vector lattice then its positive cone $V_+$ is necessarily spanning. In fact, any $x\in V$ can be written as 
\bbb
x=(x-x\wedge 0) - (-\,x\wedge 0)\in V_+ - V_+\, .
\eee
Here we do not list all the elementary properties of vector lattices, for which we refer the reader to~\cite[I]{BIRKHOFF} or to~\cite[Theorem 1.17]{ALIPRANTIS}. However, there are few simple observations we want to make because they will turn out to be useful in what follows.

\begin{prop} \label{lower bounding lattice prop}
Let $V$ be a vector lattice. Then the infimum $\wedge$ satisfies the following identities:
\begin{enumerate}[(a)]
\item translational invariance: $(x+z)\wedge (y+z)=x\wedge y + z$ for all $x,y,z\in V$;
\item for all $x,y,z\geq 0$ one has $(x+y)\wedge z \leq x\wedge z + y\wedge z$.
\end{enumerate}
Analogous relations can be obtained for the supremum $\vee$ with the help of the identity $x \vee y= -\left((-x)\wedge (-y)\right)$.
\end{prop}

\begin{proof} $ \\[-4ex] $
\begin{enumerate}[(a)]
\item On the one hand, $x\wedge y\leq x,\,y$ implies $x\wedge y +z \leq x+z,\, y+z$, from which we deduce $x\wedge y +z \leq (x+z)\wedge (y+z)$. Applying this latter relation with the substitutions $x\mapsto x+z$, $y\mapsto y+z$ and $z\mapsto -z$ gives also the opposite inequality, so that equality is established.
\item Define $w\coloneqq (x+y)\wedge z$ and observe that: (1) $w=(x+y)\wedge(x+z-x)=x+y\wedge(z-x)\leq x+y\wedge z$, because $b\leq c\Rightarrow a\wedge b \leq a\wedge c$ ; (2) $w\leq z\leq z+y\wedge z$. Since $w$ is therefore a lower bound for both $x+y\wedge z$ and $z+y\wedge z$, we conclude that $w\leq (x+y\wedge z)\wedge (z+y\wedge z)=x\wedge z + y\wedge z$.
\end{enumerate}
\end{proof}


\subsection{Decomposition and interpolation properties} \label{subsec2 dec interpol}

Let us now turn our attention to the study of two important features that an ordered vector space can exhibit, namely the so-called decomposition and interpolation properties. Here we define them separately and then prove that they are equivalent, and in finite dimension even equivalent to the underlying ordered vector space being a vector lattice, provided that the positive cone is closed.

\begin{Def} \emph{\cite[Definition 1.50]{ALIPRANTIS}.} \label{def dec}
An ordered vector space is said to have the \textbf{decomposition property} if for any finite families of positive vectors $x_i, y_j\geq 0$ ($i\in I$, $j\in J$ finite sets) satisfying $\sum_i x_i = \sum_j y_j$, there exists a matrix of positive vectors $z_{ij}\geq 0$ over $I\times J$ such that
\begin{align}
x_i &=\sum_j z_{ij}\quad \forall\ i\in I\, , \label{dec eq1}\\
y_j &=\sum_i z_{ij}\quad \forall\ j\in J\, . \label{dec eq2}
\end{align}
We will say that the space has the binary decomposition property if said positive decomposition can always be found provided that $|I|,|J|\leq 2$.
\end{Def}

\begin{Def} \emph{\cite[Definition 1.52]{ALIPRANTIS}.} \label{def interpol}
An ordered vector space is said to have the interpolation property if for all $a_1,a_2 \leq b_1, b_2$ (all the four inequalities satisfied), there exists a vector $c$ such that (with the same notation) $a_1,a_2\leq c\leq b_1,b_2$.
\end{Def}

\begin{rem}
We did not distinguish a finite and a binary interpolation properties, because the two are elementarily seen to be equivalent.
\end{rem}

Interestingly enough, the three aforementioned properties turn out to be equivalent in a fully general setting, as the next lemma due to Riesz~\cite{Riesz} will show.

\begin{lemma} \emph{\cite[Theorem 1.54]{ALIPRANTIS}.} \label{dec=interpol lemma}
Let $V$ be an ordered vector space. The following are equivalent:
\begin{enumerate}[(a)]
\item $V$ has the decomposition property;
\item $V$ has the binary decomposition property;
\item $V$ has the interpolation property.
\end{enumerate}
\end{lemma}

\begin{proof} $ \\[-4ex] $
\begin{description}

\item[$(a)\!\Rightarrow\! (b).$] Obvious.

\item[$(b)\!\Rightarrow\! (a).$] We show how to reconstruct the generic case of finite $|I|,|J|$ by separate induction on the two cardinalities.
We start by looking at $|I|=2$ and performing an induction on $|J|=m>2$.

Consider $x_1,x_2,y_1,\ldots, y_m\geq 0$ such that $x_1+x_2=\sum_{j=1}^m y_j$; we want to find an appropriate set of $z$ vectors that decomposes this set. Since we assume that the claim has been established for $m-1$, we know we can find $z'_{ij}\geq 0$ ($i=1,2$, $j=1,\ldots,m-1$) that satisfy
\begin{align*}
\sum_{j=1}^{m-1} z'_{ij} &= x_i\, , \\
z'_{1j}+z'_{2j} &= \left\{ \begin{array}{lr} y_{m-1}+y_m & \text{if $j=m-1$,} \\ y_j & \text{otherwise.} \end{array} \right.
\end{align*}
The construction of the requested vectors $z_{ij}$ ($i=1,2$, $j=1,\ldots,m$) is in two steps. If $j\leq m-2$, we simply identify $z_{ij}\coloneqq z'_{ij}$. In order to construct the remaining four elements, we can exploit the equality $z'_{1,m-1}+z'_{2,m-1}=y_{m-1}+y_m$ and the $|I|=|J|=2$ case of the decomposition property in order to find $z_{1,m-1},z_{2,m-1},z_{1m},z_{2m}\geq 0$ such that
\begin{align*}
z'_{1,m-1} &= z_{1,m-1}+z_{1m},\\
z'_{2,m-1} &= z_{2,m-1}+z_{2m}, \\
y_{m-1} &= z_{1,m-1}+z_{2,m-1}, \\
y_m &= z_{1,m}+z_{2,m}\, .
\end{align*}
It is very easy to verify that one has $x_i=\sum_{j=1}^m z_{ij}$ and $y_j=z_{1j}+z_{2j}$ for all $i=1,2$ and $j=1,\ldots, m$, respectively.

Finally, the induction on $|I|=n$ can be performed in a totally analogous way by adding one column after the other to the basic case $n=2$. These additions are made possible by the above $2\times m$ decomposition property.

\item[$(b)\!\Rightarrow\! (c).$] Given $a_1,a_2\leq b_1,b_2$, define $x_1 \coloneqq b_1-a_1$, $x_2 \coloneqq b_2-a_2$, $y_1\coloneqq b_1-a_2$, $y_2 \coloneqq b_2-a_1$. By hypothesis $x_1,x_2,y_1,y_2\geq 0$, and moreover $x_1+x_2=b_1+b_2-a_1-a_2=y_1+y_2$. Applying the binary decomposition property yields four positive vectors $z_{11}, z_{12}, z_{21}, z_{22} \geq 0$ such that
\begin{align*}
x_1&=z_{11}+z_{12}\, ,\\
x_2&=z_{21}+z_{22}\, ,\\
y_1&=z_{11}+z_{21}\, ,\\
y_2 &=z_{12}+z_{22}\, .
\end{align*}
We now define $c\coloneqq z_{12}+a_1=z_{21}+a_2$, which clearly satisfies $c\geq a_1, a_2$. Furthermore, $c\leq z_{12}+z_{11}+a_1=b_1$ and analogously $c\leq b_2$, so the construction is complete.

\item[$(c)\!\Rightarrow\!(b).$] Consider arbitrary positive vectors $x_1,x_2,y_1,y_2\geq 0$ such that $x_1+x_2=y_1+y_2$. Since it is easy to verify that the four inequalities $0,x_1-y_1\leq x_1, y_2$ hold, by the interpolation property we can find $z_{12}\geq 0$ such that $0,x_1-y_1\leq z_{12}\leq x_1, y_2$. The other three elements of the $z$ matrix are now fixed by the equalities in Definition~\ref{def dec}:
\begin{align*}
z_{11} &\coloneqq x_1-z_{12}\geq 0\,,\\
z_{22}&\coloneqq y_2-z_{12}\geq 0\, ,\\
z_{21}&\coloneqq x_2-y_2+z_{12}=y_1-x_1+z_{12}\geq 0\, .
\end{align*}
By construction, these four $z$ vectors satisfy the sum constraints required by the decomposition property.
\end{description}
\end{proof}

\subsection{Decomposition lemma and converse thereof} \label{subsec2 dec lemma}

If we spent so much time discussing the above properties is because they lie at the heart of the concept of vector lattice. In fact, as we mentioned, they are even equivalent to it for closed cones in finite dimension. Let us start with a general result, that holds in full generality, not only for finite-dimensional spaces.

\begin{lemma}[Decomposition lemma] \emph{\cite[Corollary 1.55]{ALIPRANTIS}.} \label{dec lemma}
All vector lattices have the decomposition property.
\end{lemma}

\begin{proof}
Thanks to the above discussion, it suffices to observe that any vector lattice has the interpolation property, since $a_1, a_2 \leq b_1,b_2$ implies $a_1 \leq b_1\wedge b_2$ and analogously $a_2 \leq b_1\wedge b_2$, so that $c\coloneqq b_1\wedge b_2$ satisfies $a_1, a_2\leq c\leq b_1, b_2$.
\end{proof}

\begin{proof}[Alternative proof (direct)]
We just prove the binary version of the decomposition property. Given $x_1,x_2,y_1,y_2\geq 0$ such that $x_1+x_2=y_1+y_2$, the equations to be satisfied are
\begin{align*}
x_1 &= z_{11}+z_{12}\, ,\\
x_2 &= z_{21}+z_{22}\, ,\\
y_1 &= z_{11}+z_{21}\, ,\\
y_2 &=z_{21}+z_{22}\, .
\end{align*}
First of all, we define $z_{11}\coloneqq x_1\wedge y_1$, so that necessarily $z_{12}\coloneqq x_1-x_1\wedge y_1$, $z_{21}\coloneqq y_1-x_1\wedge y_1$ and finally $z_{22} \coloneqq y_2-x_1+x_1\wedge y_1 = x_2-y_1+x_1\wedge y_1$ (where for the last equality we used $x_1+x_2=y_1+y_2$). It is left to prove that the expression we gave for $z_{22}$ identifies a positive vector. This can be done by applying the lower bounding technique in Proposition~\ref{lower bounding lattice prop}(b):
\begin{align*}
z_{22} &= x_2-y_1+x_1\wedge y_1\\
& \geq x_2 - y_1 + (x_1+x_2)\wedge y_1 - x_2\wedge y_1\\
&= x_2 - y_1 + (y_1+y_2)\wedge y_1 - x_2\wedge y_1\\
&= x_2 - y_1 + y_1 - x_2\wedge y_1\\
&= x_2 - x_2\wedge y_1\\
&\geq 0\, .
\end{align*}
Since it is apparent from the definitions that also $z_{11},z_{12},z_{21}\geq 0$, and these $z$ vectors satisfy the required identities by construction, the proof is complete. 
\end{proof}

The connection between decomposition property and lattice axioms is so deep that in finite dimension the two are in fact equivalent under the only assumption of topological closeness of the positive cone. Moreover, finite-dimensional closed cones that define a lattice order are exactly the simplicial cones considered in Definition~\ref{def class}, a result that goes back to Yudin~\cite{Yudin}.

\begin{thm}[Yudin] \emph{\cite[Theorem 3.21]{ALIPRANTIS} or~\cite[Proposition 10.10]{PHELPS}.} \label{lattice=simplex thm}
Let $V$ be a finite-dimensional ordered vector space whose positive cone $V_+$ is closed and spanning. Then the following are equivalent:
\begin{enumerate}[(a)]
\item $V$ is a vector lattice;
\item $V$ has the decomposition property;
\item $V_+$ is simplicial.
\end{enumerate}
\end{thm}

\begin{proof} 
We give on purpose a redundant proof of this important result. In what follows, $d=\dim V< \infty$.
\begin{description}

\item[$(a)\!\Rightarrow\! (b).$] See Lemma~\ref{dec lemma}.

\item[$(b)\!\Rightarrow\! (a).$]
We have to prove that the decomposition property ensures that $V$ is a lattice. Thanks to Lemma~\ref{dec=interpol lemma}, we can instead assume that the interpolation property is satisfied. Consider $x,y\in V$ and let us prove that there exists a maximum simultaneous lower bound (the reasoning for the minimum simultaneous upper bound is totally analogous). First of all, denoting with $x=x_+-x_-$ and $y=y_+-y_-$ two decompositions such that $x_+,x_-,y_+,y_-\geq 0$ (which exist because $V_+$ is spanning), we note that $x\wedge y$ (if it exists) can be sought inside the interval
\bbb
K_0 \coloneqq [-x_--y_-,\ x_++y_+] = \{z\in V:\, -x_- - y_-\leq z\leq x_++y_+\}\, .
\eee
This is a consequence of the interpolation property: whenever $w\leq x,y$, taking into account that also $-x_- - y_- \leq x,y$, we can construct $w'$ such that $-x_- - y_-,w \leq w'\leq x,y\leq x_+ + y_+$.

We now observe that any interval $[a,b]$ in a finite-dimensional ordered vector space is necessarily bounded. To show this fact, it suffices to consider a basis $\{\varphi_1,\ldots, \varphi_d\}$ of the dual space $V^*$, and to prove that $\braket{\varphi_i ,z}$ is bounded when $z\in [a,b]$, for all $i=1,\ldots, d$. Since the dual cone $V_+^*$ is spanning, we can take a basis in which each $\varphi_i$ is a positive functional. Thanks to the positivity, from $a\leq z\leq b$ the relation $\braket{\varphi_i, a}\leq \braket{\varphi_i , z}\leq \braket{\varphi_i, b}$ follows immediately. In particular, $\braket{\varphi_i , z}$ is bounded for all $i=1,\ldots, d$ and for all $z\in [a,b]$.

Since $V_+$ is assumed to be closed, the intervals $[a,b]$ are also closed and thus compact (in finite-dimensional spaces a set is compact iff it is closed and bounded). The (non-empty) set $K\coloneqq \{z\in K_0:\, z\leq x,y\}$ is compact, too, because it is a closed subset of a compact set. It can be shown that compact sets have always a maximal element in an ordered topological space, and in particular in a vector space ordered by a closed cone. This intuitive fact is usually referred to as Wallace's lemma~\cite{Wallace}.

Therefore, there must exists a maximal element $z_0$ of $K$. Thanks to the interpolation property, $z_0$ has indeed to be the maximum, because for any other $z\in K$ in the set the inequalities $z,z_0\leq x,y$ guarantee the existence of $z'_0$ such that $z,z_0\leq z'_0\leq x,y$, which in turn implies $z'_0=z_0$ and therefore $z_0\geq z$.

\item[$(a)\!\Rightarrow\! (c).$] We will prove that there can not be more than $d=\dim V$ extremal rays in $V_+$ if $V$ has to be a vector lattice. Since $V_+$ is also spanning, this will suffice to prove that their number is in fact exactly $d$. Suppose by contradiction that $x_0,\ldots, x_d\geq 0$ are all extremal (and none of them is a multiple of another). There must be a linear dependence relation $\sum_{j=0}^d \alpha_j x_j=0$, so that for all $i=0,\ldots, d$ one obtains, thanks to Proposition~\ref{lower bounding lattice prop}(b), 
\begin{equation*}
0 = x_i \wedge \sum_{j=0}^d \alpha_j x_j \leq \sum_{j=0}^d \alpha_j (x_i\wedge x_j) = \alpha_i\, ,
\end{equation*}
because if $x_i$ and $x_j$ are both extremal then $x_i\wedge x_j$ satisfies $x_i\wedge x_j \leq x_i,x_j$, i.e. it must lie at the same time on the ray generated by $x_i$ and on that generated by $x_j$, and therefore has to be zero. We are left with $\sum_{i=0}^d \alpha_i x_i=0$ and $\alpha_i\geq 0$ for all $i=0,\ldots,d$, which is possible only if $\alpha_i \equiv 0$ identically. 

\item[$(c)\!\Rightarrow\! (a).$] The order induced by a simplicial cone $V_+=\{\alpha_1 v_1+\ldots + \alpha_n v_n: \alpha_i\geq 0\}$ (where the vectors $v_i$ form a basis of $V$) makes $V$ a lattice. For instance, if $x=\sum_{i=1}^n\alpha_i v_i$ and $y=\sum_{i=1}^n\beta_i v_i$, it is easy to verify that $x\wedge y\coloneqq \sum_{i=1}^n\min\{\alpha_i,\beta_i\}\, v_i$ is the least simultaneous upper bound of $x$ and $y$.

\item[$(b)\!\Rightarrow\! (c).$] The following classic argument is a modification of the one at the very end of~\cite[\S 10]{PHELPS}. Suppose by contradiction that $v_0,\ldots, v_d\geq 0$ are all extremal vectors (and none of them is a multiple of another). There will exist a linear dependence relation $\sum_{i=0}^n \alpha_i v_i=0$ among said vectors. Let us rewrite it as $\sum_{i\in I} x_i = \sum_{j\in J} y_j$, where $I=\{0\leq i\leq n: \alpha_i>0\}$, $J=\{0\leq j\leq n: \alpha_j<0\}$ are both non-empty, $x_i \coloneqq \alpha_i v_i\neq 0$ and $y_j \coloneqq -\alpha_j v_j\neq 0$. Apply now the decomposition property in order to construct $z_{ij}\geq 0$ such that $\sum_j z_{ij} = x_i$ for all $i$ and $\sum_i z_{ij} = y_j$ for all $j$. Since all the vectors $x_i$ are extremal, we deduce immediately $z_{ij}\propto x_i$ and analogously $z_{ij}\propto y_j$, so that the existence of a single $z_{ij}\neq 0$ leads to the absurd conclusion $x_i\propto y_j$.

\item[$(c)\!\Rightarrow\! (b).$] The binary decomposition property for simplicial cones reduces immediately to the same property for the ordered space $(\mathds{R},\mathds{R}_+)$, whose proof is trivial.

\end{description}
\end{proof}

The equivalence expressed by the above Theorem~\ref{lattice=simplex thm} breaks down in infinite dimension. Ultimately, this happens because there is no converse to the decomposition lemma (Lemma~\ref{dec lemma}) for generic ordered Banach spaces. For explicit examples showing this phenomenon in action, we refer the interested reader to~\cite[Examples 1.56--1.58]{ALIPRANTIS}. A great deal of the existent literature on Banach lattices is devoted to finding suitable replacements for the crucial result expressed in Theorem~\ref{lattice=simplex thm}. It is hopeless to make a complete list of references here, so we limit ourselves to mentioning the cornerstone result known as Ando's theorem. In brief, this states that under certain regularity assumptions on the cone (i.e. that it is spanning and \emph{normal}, see Definition~\ref{def normal cone}), the decomposition property at the level of the primal space is equivalent to that on the dual space. For details, see the original paper by Ando~\cite{Ando-cones} or the excellent presentation in~\cite[Theorem 2.47]{ALIPRANTIS}. 

Let us now discuss some consequences of Theorem~\ref{lattice=simplex thm}. As an immediate corollary, we find the following result.

\begin{cor} \emph{\cite[Theorem 3.9]{ALIPRANTIS}.} \label{dual simplex cor}
A finite-dimensional ordered vector space $V$ whose positive cone is closed and spanning is a vector lattice iff its dual $V^*$ is a vector lattice.
\end{cor}

\begin{proof}
Follows by putting together Theorem~\ref{lattice=simplex thm} with the observation that a closed spanning cone is simplicial iff its dual is such, as discussed in Subsection~\ref{subsec2 ex class}.
\end{proof}

The above consequence admits an appealing physical interpretation. Let us consider a GPT $(V,C,u)$. We remind the reader that two measurements $(e_i)_i, (f_j)_j\in \mathbf{M}$ are said to be \textbf{jointly measurable} (or simply \textbf{compatible}) if there is a third measurement $(g_{ij})_{ij}\in\mathbf{M}$ such that
\begin{align}
e_i &= \sum_j g_{ij}\quad \forall\ i\, , \label{joint measur 1}\\
f_j &= \sum_i g_{ij}\quad \forall\ j\, . \label{joint measur 2}
\end{align} 
This is probably not the most intuitive representation of what `compatibility' of two measurements means. One could formulate an alternative notion of \textbf{coexistent} measurements, by requiring the existence of a third measurement $(g_k)_{k\in K}$ such that for all $i\in I$ and $j\in J$ there are subsets $K_i, K^j\subseteq K$ with the property that $\sum_{k\in K_i} g_k = e_i$ and $\sum_{k\in K^j} g_k = f_j$. While it turns out that joint measurability and coexistence are equivalent concepts for binary measurements~\cite{KRAUS, joint-1, joint-2}, this equivalence does no longer hold already when $|I|=2$ and $|J|=3$. In what follows, we deal mostly with binary measurements. 

Equations~\eqref{joint measur 1} and~\eqref{joint measur 2} have the same form as those appearing in the definition of the decomposition property,~\eqref{dec eq1} and~\eqref{dec eq2}. There is however a small difference, namely, that the sum $\sum_i e_i = \sum_j f_j=u$ is fixed to a specific value (i.e. the order unit), and can not be changed. A consequence of this fact is that we can not use Theorem~\ref{lattice=simplex thm} directly to conclude that every non-classical GPT admits two binary measurements that are not jointly measurable. Anyway, this happens to be the case, as proved in~\cite{Plavala-2016}.
From Theorem~\ref{lattice=simplex thm}, however, it \emph{does} follow -- by applying a suitable rescaling -- that \emph{in every non-classical theory there are measurements with at most $3$ outcomes that are not jointly measurable} (which is exactly the claim in the title of~\cite{Plavala-2016}).

\subsection{Namioka-Phelps theorem} \label{subsec2 NP thm}

Now that we have at hand the powerful Theorem~\ref{lattice=simplex thm}, which provides an alternate characterisation of classical GPTs as defined in Definition~\ref{def class}, we can turn our attention to reviewing the proof of a result by Namioka and Phelps~\cite[Theorem 1.4]{NP} that confirms Conjecture~\ref{ultimate cj} at least in an important special case, namely when one of the two cones in question has a square base, i.e. is of the form~\eqref{C square}. 

\begin{thm}[Namioka-Phelps] \label{NP thm}
Let $C$ be a closed spanning cone in a finite-dimensional real space $V$. Then the following are equivalent:
\begin{enumerate}[(i)]
\item $C$ is simplicial;
\item $C \tminit \tilde{C} = C \tmaxit \tilde{C}$ for all closed spanning finite-dimensional cones $\tilde{C}$;
\item $C \tminit C_{\square} = C \tmaxit C_{\square}$, with $C_{\square}$ being given by~\eqref{C square}.
\end{enumerate}
\end{thm}

\begin{proof}
In light of Lemma~\ref{simplicial min=max lemma}, the only nontrivial implication is (iii) $\Rightarrow $ (i). Employing also Theorem~\ref{lattice=simplex thm}, it suffices to prove that if $C \tmin C_{\square} = C \tmax C_{\square}$ then $C$ has the binary decomposition property. For simplicity, we will consider $V$ as a vector space ordered by $C$.

Let us start with some notation. Denote the vectors generating the four extremal rays of $C_{\square}$ by
\bb
\begin{array}{ll}
v_1 = (1,1,1)\, ,\quad & v_2 = (1,-1,1)\, , \\
v_3 = (-1,1,1)\, ,\quad & v_4 = (-1,-1,1)\, .
\end{array}
\label{extremal C square}
\ee
Observe that $v_4=v_2+v_3-v_1$. The extremal rays of the dual cone $C_{\square}^*$ are generated by
\bb
\begin{array}{ll}
\varphi_1=(1,0,1)\, \quad & \varphi_2=(-1,0,1)\, ,\\
\varphi_3=(0,1,1)\, ,\quad & \varphi_4=(0,-1,1)\, .
\end{array}
\label{extremal C square*}
\ee

Now, take three arbitrary positive vectors $x_1,x_2,y_1\geq 0$ in $V$ such that $y_1\leq x_1+x_2$, and define the \textbf{Namioka-Phelps state}
\bb
Z_{NP} \coloneqq (x_1-y_1)\otimes v_1 + y_1\otimes v_2 + x_2\otimes v_3\, .
\label{NP state}
\ee
We want to check that $Z_{NP}\in C\tmax C_{\square}$. This can be easily done by applying systematically all the extremal functionals $\varphi_i$ ($i=1,2,3,4$) on the second subsystem, and by verifying that the resulting `partially evaluated' vectors on the first subsystem belong to $C$. A simple graphical picture will help us understanding why this is the case.

\begin{figure}[ht]
  \centering
  \includegraphics[height=6cm, width=6cm, keepaspectratio]{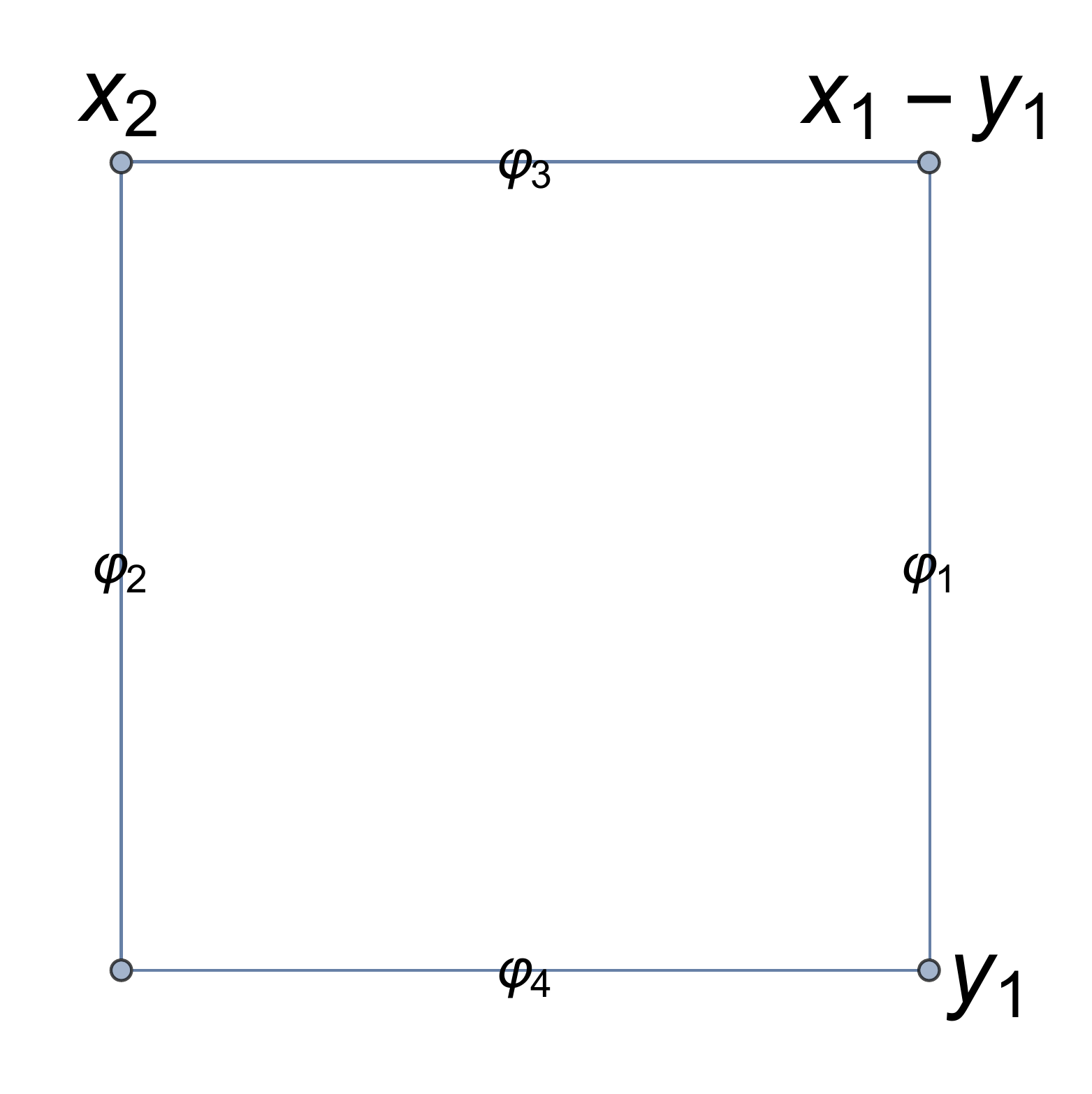}
  \caption{Schematic representation of the Namioka-Phelps state in~\eqref{NP state}. The vertices of the graph represent, in clockwise order from the upper right corner, $v_1, v_2, v_4, v_3$. The vertices are labelled by the $V$ vector they are associated with via tensor product. Edges are labelled by the functionals among those in~\eqref{extremal C square*} that take the maximal value on the corresponding straight lines.}
  \label{NP drawing}
\end{figure}

With reference to Figure~\ref{NP drawing}, we see that evaluating any of the functionals $\varphi_i$ on $Z_{NP}$ yields (up to a constant factor) a vector in $V$ that is the sum of the labels of the two vertices contained in the edge identified by $\varphi_i$. In formulae,
\begin{align*}
\frac12 \braket{\varphi_{1}^B, Z_{NP}^{AB}} &= x_1^A\geq 0\, ,\\
\frac12 \braket{\varphi_{2}^B, Z_{NP}^{AB}} &= x_2^A \geq 0\, ,\\
\frac12 \braket{\varphi_{3}^B, Z_{NP}^{AB}} &= x_1^A - y_1^A + x_2^A = y_2^A \geq 0\ ,\\
\frac12 \braket{\varphi_{4}^B, Z_{NP}^{AB}} &= y_1^A\geq 0\, .
\end{align*}
This is enough to show that $Z_{NP}\in C\tmax C_{\square}$. In fact, a consequence of the calculations we performed is that $\braket{\lambda^A\otimes \varphi_i^B, Z_{NP}^{AB}}\geq 0$ for all $i=1,2,3,4$ and for all $\lambda^A\in C^*$. Since every $\varphi^B\in C_{\square}^*$ is a non-negative combination of the extremal functionals $\varphi_i$ (by Corollary~\ref{conic MC cor}), this shows that 
\bbb
\braket{\lambda^A\otimes \varphi^B, Z_{NP}^{AB}}\geq 0\qquad \forall\ \lambda^A\in C^*,\ \varphi^B\in C_{\square}^*\,,
\eee
implying that $Z_{NP}\in C\tmax C_{\square}$ by the very definition of maximal tensor product~\eqref{maximal}.

We now proceed to show that if $Z_{NP}\in C\tmin C_{\square}$ then $x_1,x_2,y_1,y_2$ obey the binary decomposition property. To this purpose, we use Corollary~\ref{conic MC cor} to write any tensor $Z\in C\tmin C_{\square}$ in the minimal tensor product as $Z=\sum_{i=1}^4 z_i\otimes v_i$, where $z_i\geq 0$ for all $i=1,2,3,4$. In the case of $Z_{NP}$ we would obtain the equality
\begin{align*}
Z_{NP} &= (x_1-y_1)\otimes v_1 + y_1\otimes v_2 + x_2\otimes v_3 \\
&= z_1 \otimes v_1 + z_2 \otimes v_2 + z_3 \otimes v_3 + z_4 \otimes v_4 \\
&= z_1 \otimes v_1 + z_2 \otimes v_2 + z_3 \otimes v_3 + z_4 \otimes (v_2+v_3 - v_1) \\
&= (z_1-z_4) \otimes v_1 + (z_2+z_4) \otimes v_2 + (z_3+z_4) \otimes v_3\, .
\end{align*}
Since $v_1,v_2,v_3$ are linearly independent, this is possible only if
\begin{align*}
x_1 - y_1 &= z_1 - z_4\, ,\\
y_1 &= z_2 + z_4\, ,\\
x_2 &= z_3 + z_4\, .
\end{align*}
Adding the first two equations and writing out also the decomposition for $y_2=x_1+x_2-y_1$ yields
\begin{align*}
x_1 &= z_1 + z_2\, ,\\
x_2 &= z_3 + z_4\, , \\
y_1 &= z_2 + z_4\, ,\\
y_2 &= z_1 + z_3\, .
\end{align*}
Up to relabelling, these are the equations that define the binary decomposition property. We are then led to conclude that if $Z_{NP}\in C\tmin C_{\square}$ for all $x_1,x_2,y_1,y_2\geq 0$ such that $x_1+x_2=y_1+y_2$, then the cone $C$ satisfies the binary decomposition property, and thus (thanks to Theorem~\ref{lattice=simplex thm}) is simplicial.
\end{proof}

From our point of view, the above results is important primarily because it confirms Conjecture~\ref{ultimate cj} when one of the two cones corresponds to the simplest non-classical theory, i.e. the gbit discussed in Subsection~\ref{subsec2 other ex}. We state this observation as a corollary.

\begin{cor} \label{NP cor}
Conjecture~\ref{ultimate cj} holds when one of the two cones has a square base~\eqref{C square}.
\end{cor}

However, the importance of Theorem~\ref{NP thm} lies also in the proof technique itself. In fact, the Namioka-Phelps state~\eqref{NP state} is to the best of our knowledge the first and most general example of an entangled state. Its generality stems from the fact that the construction can be applied to all cones $C_A$, provided that $C_B$ is fixed to be the square-based cone in~\eqref{C square}. In this sense there is a lot of work to be done yet, in order to extend the construction (if at all possible) to the case of an arbitrary $C_B$. 

Let us stress here that~\eqref{NP state} anyway encompasses as special cases other examples of entangled states that have been discovered much later. For instance, the so-called \emph{PR-box}~\cite{PR-boxes} can be seen as the state of $G_2\tmax G_2$ obtained from~\eqref{NP state} by taking $C=C_\square$, $x_1=\frac12 v_2$, $x_2=\frac12 v_3$, $y_1=\frac12 v_1$ and $y_2=\frac12 v_4$.

\section{Beyond Namioka-Phelps} \label{sec2 beyond}

Throughout this section, we explore some possible generalisations and extensions of the Namioka-Phelps theorem, Theorem~\ref{NP thm}. In doing so, we will review and formalise many of the questions listed in Subsection~\ref{subsec2 fundamental}. To the best of my knowledge, most of what I will present here constitutes an original contribution.

\subsection{Nonlocality in maximal tensor products} \label{subsec2 nonlocal}

As we mentioned in Subsection~\ref{subsec2 fundamental}, point (i), there is a stronger version of Conjecture~\ref{ultimate cj} that is perhaps more operationally motivated. Namely, instead of asking whether two non-classical GPTs are entangleable, we could rather look at the presence of nonlocal correlations. With the language of Definition~\ref{def Bell violation}, we are thus asking the following question.

\begin{question} \label{question Bell}
Given two non-classical GPTs $A=(V_A, C_A, u_A)$ and $B=(V_B, C_B, u_B)$, is it always the case that $A\tmaxit B$ violates a Bell inequality according to Definition~\ref{def Bell violation}? 
\end{question}

We did not make this question a conjecture because even if Conjecture~\ref{ultimate cj} holds true, it may well be the case that the entanglement contained in certain exotic models can not be `extracted' via measurements and transformed into nonlocal correlations. This is, however, pure speculation, as to the present day we do not have evidence of such behaviour in any explicit example. 

As the reader may know, the simplest Bell inequality of all is the so-called \textbf{CHSH inequality}~\cite{CHSH}. In the terminology we fixed in Definition~\ref{def Bell}, CHSH separates point of $G_2^2\tmax G_2^2$ from local hidden variable models in $G_2^2\tmin G_2^2$, it deals with the case of two $(2,2)$-apparatuses acting on the two shares of a bipartite state. We change slightly the notation of Definition~\ref{def apparatus} to adapt it to the most common conventions.

Let us denote the physical state used to generate the correlations as $\omega\in C_A\tmax C_B$, with $\braket{u_A\otimes u_B, \omega_{AB}}=1$. In most literature on the subject, the settings of the apparatuses on the $A$ and $B$ sides are labelled by $\mathbf{x}=0,1$ and $\mathbf{y}=0,1$, respectively, while the corresponding outcomes are indexed by $\mathbf{a}=\pm 1$ and $\mathbf{b}=\pm 1$. The two measurements on $A$ will then be denoted by $(e_{0+}, e_{0-})$ and $(e_{1+}, e_{1-})$, while the measurements on $B$ will be $(f_{0+}, f_{0-})$ and $(f_{1+}, f_{1-})$.
With these conventions, the CHSH inequality takes the form
\bb
\braket{\mathds{B}, \omega} \coloneqq \sum_{\mathbf{a},\mathbf{b},\mathbf{x},\mathbf{y}} \mathbf{a}\mathbf{b}(1-2\mathbf{x}\mathbf{y}) \braket{e_{\mathbf{x}\mathbf{a}}\otimes f_{\mathbf{y}\mathbf{b}}, \omega} \leq 2\, .
\label{CHSH}
\ee
The functional acting on $\omega$ on the left-hand side of the above inequality is called \textbf{Bell functional}. A state $\omega$ of $A\tmax B$ for which~\eqref{CHSH} in not obeyed is said to \textbf{violate CHSH}. In particular, any such state violates a Bell inequality according to Definition~\ref{def Bell violation}, but the converse is not a priori true, since it is well-known that for higher values of $n,k,n',k'$ there are many more Bell inequalities coming into play~\cite{Brunner-review}.

We have seen that the state~\eqref{NP state} considered by Namioka and Phelps is basically the only explicit construction of an entangled state that works for all non-classical GPTs on the $A$ system, provided that $B=G_2$ is the gbit discussed in Subsection~\ref{subsec2 other ex}. It is then natural to wonder whether such a state violates any Bell inequality. The first original result we present here is the observation that \emph{the Namioka-Phelps state indeed violates CHSH, for all non-classical GPTs $A$.} 

Before we state and prove this curious result, let us discuss some consequences of the non-classicality of a GPT $A$. By the results in~\cite{Plavala-2016}, we know that any such theory admits two binary measurements that are not jointly measurable (see the discussion at the end of Subsection~\ref{subsec2 dec lemma}). The concept of incompatibility is important in this context because it is not difficult to realise that \emph{compatible measurements can not violate any Bell inequality}~\cite{Fine1982}.
There is an intuitive way of \emph{quantifying} how much two binary measurements fail to be jointly measurable via a convex optimisation program. This idea was introduced in~\cite{Wolf-incompatible}\footnote{In this case, the convex program becomes a semidefinite program, or SDP.}, and subsequently translated into the GPT framework~\cite{Banik-2013,Busch-2013,Stevens-Busch,Plavala-2016,Jencova2017}.

Given two binary measurements $(e,u-e)$ and $(f,u-f)$ on a GPT $A=(V,C,u)$, by direct inspection of~\eqref{joint measur 1} and~\eqref{joint measur 2} we see that they are jointly measurable iff there is $g\in V^*$ such that
\bb
0,e+f-u\leq g\leq e,f\, .
\ee
Therefore, one can consider the convex optimisation program~\cite{Wolf-incompatible, Stevens-Busch}
\bb
\begin{array}{llll}
t[e,f] & \coloneqq & \inf_{t,g}
& t \\[0.8ex]
& & \text{s.t.} & 0,e+f-u \leq g\leq e+tu ,f+tu\, .
\end{array}
\label{convex program inc}
\ee
In what follows, $t[e,f]$ will be called the \textbf{incompatibility degree} or simply \textbf{incompatibility}.\footnote{Other authors variously call closely related measures `unsharp parameter'~\cite{Banik-2013}, `degree of incompatibility'~\cite{Plavala-2016}, and so on. We follow the terminology of~\cite{Jencova2017}.} By construction, we see that the two measurements defined by $e$ and $f$ are jointly measurable iff $t[e,f]\leq 0$. In general, the range of $t[e,f]$ is the interval $[-1/2,1/2]$. It is indeed very easy to see that $(t,g)$ has a chance of being a feasible point for the program in~\eqref{convex program inc} only when $t\geq -1/2$, and that $t=1/2,\, g=\frac12 (e+f)$ is always feasible. The extremal value $-1/2$ is achieved for instance by $e=f=u/2$, while $1/2$ is achieved by the two measurements on a gbit identified, in the notation of~\eqref{extremal C square*}, by $e=\frac12 \varphi_1$ and $f=\frac12 \varphi_3$~\cite{Stevens-Busch}. 

As shown in~\cite{Wolf-incompatible, Stevens-Busch}, $t[e,f]$ has an interpretation in terms of the minimal amount of `noise' that is needed in order to make the two measurements $(e,u-e)$ and $(f,u-f)$ jointly measurable. Namely, the maximal $\lambda$ such that $\lambda e + \frac{1-\lambda}{2} u$ and $\lambda f + \frac{1-\lambda}{2} u$ identify two jointly measurable measurements is $\lambda[e,f]=\frac{1}{1+2 t[e,f]}$.

Exactly as their cousins SDPs, also convex optimisation programs come with a dual. In our case, the dual to~\eqref{convex program inc} can be taken in exactly the same way as in~\cite{Wolf-incompatible}. We obtain
\bb
\begin{array}{llll}
t[e,f] & = & \sup_{x,y,z}
& \braket{e+f-u,z} - \braket{e,x} - \braket{f,y} \\[0.8ex]
& & \text{s.t.} & x,y,z\geq 0\, , \\
& & & z\leq x+y\, , \\
& & & \braket{u,x+y} = 1\, .
\end{array}
\label{dual convex program inc}
\ee
Observe that since the dual program is strictly feasible, its value coincides with that of the primal~\cite{BOYD}.

Since every non-classical GPT $A$ admits two binary measurements that are not jointly measurable, it is clear that the \textbf{global incompatibility}
\bb
T[A] \coloneqq \sup_{e,f\in [0,u]} t[e,f]
\label{lambda[A]}
\ee 
satisfies $T[A]>0$ iff $A$ is non-classical. Using the dual program~\eqref{dual convex program inc}, we can rewrite
\begin{align*}
T[A] &= \sup_{e,f\in [0,u]} t[e,f] \\[1ex]
&= \sup_{e,f\in [0,u]} \sup_{\scriptsize \begin{array}{c} x,y,z \geq 0 \\ z\leq x+y \\ \braket{u,x+y}=1 \end{array}} \left\{ \braket{e,z-x} + \braket{f,z-y} - \braket{u,z} \right\} \\
&\texteq{(1)} \sup_{\scriptsize \begin{array}{c} x,y,z \geq 0 \\ z\leq x+y \\ \braket{u,x+y}=1 \end{array}} \sup_{e,f\in [0,u]} \left\{ \braket{e,z-x} + \braket{f,z-y} - \braket{u,z} \right\} \\
&\texteq{(2)} \sup_{\scriptsize \begin{array}{c} x,y,z \geq 0 \\ z\leq x+y \\ \braket{u,x+y}=1 \end{array}} \sup_{\varphi,\eta \in [-u,u]} \frac12 \left\{ \braket{\varphi, z-x} + \braket{\eta, z-y} - 1 \right\} \\
&\texteq{(3)} \sup_{\scriptsize \begin{array}{c} x,y,z \geq 0 \\ z\leq x+y \\ \braket{u,x+y}=1 \end{array}} \frac12 \left\{ \|z-x\| + \|z-y\| -1 \right\} .
\end{align*}
The justification of the above calculation is as follows: (1) exchanging two suprema is always allowed; (2) we substituted $\varphi \coloneqq 2e-u$ and $\eta\coloneqq 2f-u$; (3) we employed the expression~\eqref{base norm 1} for the base norm. In conclusion, we have found that the global incompatibility measure can be written in the form
\bb
\begin{array}{llll}
T[A] & = & \sup_{x,y,z}
& \frac12 \left\{ \|z-x\| + \|z-y\|  -1 \right\} \\[0.8ex]
& & \text{s.t.} & x,y,z\geq 0\, , \\
& & & z\leq x+y\, , \\
& & & \braket{u,x+y} = 1\, ,
\end{array}
\label{inc eq}
\ee
where we exchanged the supremum with a maximum by virtue of the compactness of the domain.
Upon completion of this chapter, we realised that this expression for $T[A]$ was independently found in~\cite[Proposition 6]{Jencova2017}.

We now proceed to show that for all GPTs $A$, the incompatibility $T [A]$ quantifies exactly the violation of the CHSH inequality~\eqref{CHSH} exhibited by the best Namioka-Phelps state (and indeed by any state in the composite system).

\begin{thm} \label{NP nonlocal}
For $A$ an arbitrary GPT, consider the composite system $A\tmaxit G_2$. Then the following holds.
\begin{enumerate}[(a)]
\item The value of the Bell functional in~\eqref{CHSH} does not exceed $2(1+2 T[A])$, for all bipartite states $\omega$~\cite{Banik-2013}.
\item There is a normalised state $Z_{NP}$ of Namioka-Phelps kind~\eqref{NP state} such that $\braket{\mathds{B}, Z_{NP}} = 2 (1+ 2 T[A])$.
\item In particular, all non-classical theories $A$ are such that $A\tmaxit G_2$ violates CHSH. The answer to Question~\ref{question Bell} is thus affirmative when $B=G_2$.
\end{enumerate}
\end{thm}

\begin{proof}
Let us prove the various claims one by one.
\begin{enumerate}[(a)]

\item In the GPT setting, this claim was originally proved in~\cite{Banik-2013}, the main idea coming once again from~\cite{Wolf-incompatible}. Let us repeat the argument here for the convenience of the reader. Denote by $\mathds{B}_{e,f}$ the Bell functional~\eqref{CHSH} corresponding to some measurements $(e,u-e)$ and $(f,u-f)$ on the first subsystem, and some unspecified but fixed measurements on the second subsystem. One observes that under the transformation rule $e'\coloneqq \lambda e+\frac{1-\lambda}{2} u$, $f'\coloneqq \lambda f+\frac{1-\lambda}{2} u$ the Bell functional behaves according to
\bbb
\mathds{B}_{e',f'} = \lambda\,\mathds{B}_{e,f}\, .
\eee
Since we saw that for $\lambda=\lambda[e,f] = \frac{1}{1+2 t [e,f]}$ the transformed effects $e',f'$ give rise to compatible measurements~\cite{Stevens-Busch}, and that compatible measurements can never lead to a violation of a Bell inequality, we deduce that
\bbb
\frac{B_{e,f}}{1+2 t[e,f]} \leq 2\, ,
\eee
which yields the claim upon maximisation over $e,f\in [0,u]$.

\item For arbitrary $x_1,x_2,y_1,y_2\geq 0$ such that $x_1+x_2=y_1+y_2$ is a normalised state, i.e. such that $\braket{u,x_1+x_2}=1$, construct the Namioka-Phelps state $Z_{NP}$ as in~\eqref{NP state}. We want to choose appropriate measurements on $A$ and $B=G_2$ such that the associated Bell functional $\mathds{B}$ of~\eqref{CHSH} evaluates to $\braket{\mathds{B}, Z_{NP}} = 2 \left( 1+ 2 T[A] \right)$. To this purpose, the measurements on the gbit that constitutes the second subsystem are chosen as follows. Adopting the notation of~\eqref{CHSH} and~\eqref{extremal C square*}, we take
\bb
\begin{array}{ccc}
f_{0+} = \frac12 \varphi_1\, , & & f_{0-} = \frac12 \varphi_2\, , \\[1ex]
f_{1+} = \frac12 \varphi_4\, , & & f_{1-} = \frac12 \varphi_3\, .
\end{array}
\label{choice B}
\ee
As in the proof of Theorem~\ref{NP thm}, one verifies that partial evaluation of these functionals on $Z_{NP}$ yields 
\bb
\begin{array}{ccc}
\braket{f_{0+}^B, Z_{NP}^{AB}} = x_1^A \, , & & \braket{f_{0-}^B, Z_{NP}^{AB}} = x_2^A \, , \\[1ex]
\braket{f_{1+}^B, Z_{NP}^{AB}} = y_1^A \, , & & \braket{f_{1-}^B, Z_{NP}^{AB}} = y_2^A\, .
\end{array}
\ee
Equipped with these relations, we can evaluate all the expectation values $\braket{e_{\mathbf{xa}} \otimes f_{\mathbf{yb}},\, Z_{NP}}$, for all $\mathbf{x},\mathbf{y}=0,1$ and $\mathbf{a},\mathbf{b}=\pm$. These calculations are reported in Table~\ref{Bell table NP}.

\begin{table}[hhh]
\setlength{\extrarowheight}{0.2cm}
\begin{tabular}{x{1.5cm} | x{1.5cm} x{1.5cm} x{1.5cm} x{1.5cm}} 
\diag{0.1em}{1.5cm}{$\mathbf{ab}$}{$\mathbf{xy}$} & $++$ & $+-$ & $-+$ & $--$ \\ \hline
00 & $\braket{e_{0+}, x_1}$ & $\braket{e_{0+}, x_2}$ & $\braket{e_{0-}, x_1}$ & $\braket{e_{0-}, x_2}$ \\ 
01 & $\braket{e_{0+}, y_1}$ & $\braket{e_{0+}, y_2}$ & $\braket{e_{0-}, y_1}$ & $\braket{e_{0-}, y_2}$ \\ 
10 & $\braket{e_{1+}, x_1}$ & $\braket{e_{1+}, x_2}$ & $\braket{e_{1-}, x_1}$ & $\braket{e_{1-}, x_2}$ \\ 
11 & $\braket{e_{1+}, y_1}$ & $\braket{e_{1+}, y_2}$ & $\braket{e_{1-}, y_1}$ & $\braket{e_{1-}, y_2}$ \\ 
\end{tabular}
\vspace{1ex}
\caption{Evaluation of all the terms $\braket{e_{\mathbf{xa}} \otimes f_{\mathbf{yb}},\, Z_{NP}}$ appearing in~\eqref{CHSH}. Here, $Z_{NP}$ is the Namioka-Phelps state in~\eqref{NP state}, while the measurements on the second subsystem are chosen according to~\eqref{choice B}.}
\label{Bell table NP}
\end{table}

With the notation of~\eqref{CHSH}, we deduce that
\bb
\begin{split}
\braket{\mathds{B}, Z_{NP}} &= \sum_{\mathbf{a},\mathbf{b},\mathbf{x},\mathbf{y}} \mathbf{a}\mathbf{b}(1-2\mathbf{x}\mathbf{y}) \braket{e_{\mathbf{x}\mathbf{a}}\otimes f_{\mathbf{y}\mathbf{b}}, Z_{NP}} \\
&= \braket{e_{0+}-e_{0-}, x_1-x_2 + y_1-y_2} \\
&\quad + \braket{e_{1+}-e_{1-}, x_1-x_2 - y_1 + y_2} \\[0.5ex]
&\leq \| x_1-x_2 + y_1 -y_2\| + \| x_1 - x_2 -y_1+y_2\|\, .
\end{split}
\label{NP nonlocal eq1}
\ee
Observe that the upper bound in the last line is achievable by a careful choice of $e_{0+}, e_{1+}\in [0,u]$, since for instance $e_{0+}-e_{0-}$ ranges over the whole dual unit ball $[-u,u]$, and~\eqref{base norm 1} holds.
Now, let $x,y,z$ be the vectors that achieve the maximum in~\eqref{dual convex program inc}. We have $x,y,z\geq 0$, $z\leq x+y$, and $\braket{u,x+y}=1$, i.e. $x+y$ is a legitimate state. Construct the Namioka-Phelps state in~\eqref{NP state} with $x_1 \coloneqq x$, $x_2\coloneqq y$, $y_1\coloneqq z$, and $y_2\coloneqq x+y-z$, so that $x_1+x_2=y_1+y_2$ is a legitimate state as required. The upper bound in~\eqref{NP nonlocal eq1} then reads
\begin{align*}
\| x_1-x_2 + y_1 -y_2\| + \| x_1 - x_2 -y_1+y_2\| &= 2\| z - y \| + 2 \| z - x \| \\[0.5ex]
&= 2 (1+2T[A])\, ,
\end{align*}
which concludes the proof of claim (b).

\item Thanks to the results of~\cite{Plavala-2016}, we know that any non-classical GPT admits two incompatible binary measurements. Therefore, $T[A]>0$ for all non-classical theories $A$. By claim (b), this implies that there is a Namioka-Phelps state $Z$ of $A\tmax G_2$ that violates CHSH~\eqref{CHSH}.

\end{enumerate}
\end{proof}

The above Theorem~\ref{NP nonlocal} provides good evidence that the answer to Question~\ref{question Bell} might be affirmative in general, after all. 
Further investigation into this problem will be conducted elsewhere.

\subsection{Universal entangleability of centrally symmetric models} \label{subsec2 univ ent centr}

This subsection is concerned with proving Conjecture~\ref{ultimate cj weaker} in the special case when both $A=B$ is a centrally symmetric model in the sense of Definition~\ref{centr}. After some time spent in introducing the necessary mathematical machinery, this goal is accomplished by Theorem~\ref{thm c ratio}. In our view, pursuing this programme is important for several good reasons. First, this will provide us with some evidence that Conjecture~\ref{ultimate cj} or at least its weaker version Conjecture~\ref{ultimate cj weaker} holds also for high-dimensional GPTs, besides the simplest case treated in Theorem~\ref{NP thm}, in which the dimension of the second subsystem is fixed to the smallest value that does not trivialise the problem, i.e. $\dim B = 3$. Second, while investigating this question we will develop the only other construction of an entangled state that we are aware of, besides that of Namioka and Phelps. Such construction is semi-universal, in the sense that it works for all centrally symmetric models, but it is not obvious how to extend it to the general setting.
Last, but not least, we will see in Chapter~\ref{chapter3} that the tools we employ here  -- for some a priori unclear reason -- will turn out to be extremely useful in tackling general problems that have nothing to do with the special class of centrally symmetric models.

As a by-product of the proof, we will obtain some novel mathematical insights into the theory of tensor norms. In fact, to the extent of our knowledge, the content of Theorem~\ref{thm c ratio} was not known to the functional analysis community. Building upon this, in Chapter~\ref{chapter3} we will go a bit more in depth and prove a complementary result (Theorem~\ref{prop proj=<ninj}) that might also be of independent interest in functional analysis.

For notation and conventions on centrally symmetric models, we refer the reader to Subsection~\ref{subsec2 centr symm}. 
Recall that if the local systems $A$ and $B$ are modelled by two centrally symmetric GPTs of dimensions $n+1$ and $m+1$, states of the bipartite system $AB$ can be represented via $(n+1) \times (m+1)$ matrices. A particularly important class of bipartite states are of those of the form $\omega = U_* + \widehat{Z}$, where $U=u_{A*}u_{B*}^T$ with $u_* = (1,0,\ldots, 0)^T$ (the length of the vector depending on the dimension of the system), and $Z\in\mathds{R}^{n\times m}$ occupies the bottom right corner of $\omega$.

The separability problem in centrally symmetric models has already been faced in Subsection~\ref{subsec2 sep centr}, and our starting point is Proposition~\ref{sep centr prop}. A remarkable consequence of this result is that a state of the form $\omega = U_* + \widehat{Z}$ satisfies $\omega\in C_A \tmax C_B$ iff $|Z|_\varepsilon\leq 1$, and is separable iff $|Z|_\pi \leq 1$. Therefore, to construct an entangled state, which is our ultimate goal if we want to show that Conjecture~\ref{ultimate cj} holds, it suffices to provide an example of a tensor $Z$ satisfying $|Z|_\varepsilon < |Z|_\pi$. Once such a $Z$ has been constructed, and we normalise it as to make $|Z|_\varepsilon \leq 1$, it is clear that $\omega = U_* + \widehat{Z}$ will be an entangled state. Of course, the problem is nontrivial only when $n,m \geq 2$ (that is, when $Z$ is a matrix of size at least $2\times 2$).

\begin{note}
Since the problem has been translated to a general statement on tensor norms, it is now completely independent from the framework of centrally symmetric GPTs. Then, for the sake of readability, we will denote the norms on the two sections $\mathds{R}^n$ and $\mathds{R}^m$ with the usual symbol $\|\cdot\|$, instead of $|\cdot|$.
\end{note}

It does not take long to convince ourselves that tensors $Z$ satisfying $\|Z\|_\varepsilon< \|Z\|_\pi$ should exist \emph{for all local norms $\|\cdot\|$}, as it seems really implausible that in~\eqref{inj} one can always find functionals $\varphi,\lambda$ achieving $\braket{\varphi,x_i} = \|x_i\|$ and $\braket{\lambda, y_i}=\|y_i\|$ for all vectors $x_i, y_i$ appearing in some decomposition of $Z$ as in~\eqref{proj}. A simple example of this phenomenon is provided by the paradigmatic example of two Euclidean spaces (Example~\ref{ex inj proj}). In that case, the injective norm coincides with the matrix operator norm (i.e. the largest singular value), while the projective norm is nothing but the trace norm (i.e. the sum of all singular values). Apart from those of rank-one, all other matrices exhibit a gap between the two, the maximal such gap being displayed by multiples of orthogonal matrices. Our intuition was thus correct in this case.

One might be tempted to try to generalise this reasoning straight away for the case of two arbitrary norms. However, a little thought shows that the reason why we were able to identify a gap between the two norms in the Euclidean case is that we knew what their expressions in terms of the list of singular values were. That is, that proof strategy rests crucially on the singular value decomposition theorem (and on the Euclidean nature of the local norms, of course). Despite belonging to the field of standard linear algebra, this is in turn a highly nontrivial result. This is, in a nutshell, the reason why the same statement $\|\cdot\|_\varepsilon \neq \|\cdot\|_\pi$ (which is the content of Theorem~\ref{thm c ratio}) requires a so much more careful (and involved) proof in the case of arbitrary Banach spaces.

In light of the above discussion, the main goal of this subsection is the following: \emph{given two arbitrary Banach spaces $V, W$ of dimensions $n,m\geq 2$, show that there exists a tensor $Z\in V \otimes W$ with the property that $\|Z\|_\varepsilon<\|Z\|_\pi$.}\footnote{Of course for specific tensors (e.g. product ones) it may happen that the two norms are the same.}
Following the discussion in Subsection~\ref{subsec2 fundamental}(ii), we can make this question quantitative and define the universal constants
\begin{align}
R(n,m) &\coloneqq \inf_{{\footnotesize \begin{array}{c} \dim V_A=n \\ \dim V_B=m \end{array}}} \sup_{0\neq Z \in V_A\otimes V_B} \frac{\|Z\|_\pi}{\|Z\|_\varepsilon}\, , \label{R} \\
S(n) &\coloneqq \inf_{\dim V = n}\, \sup_{0\neq Z \in V\otimes V} \frac{\|Z\|_\pi}{\|Z\|_\varepsilon}\, . \label{S}
\end{align}
Here, the infima are taken over all Banach spaces $V_A,V_B,V$ of fixed dimensions. We dub the above functions \textbf{inseparability ratios}.
These quantities are universal in the sense that they characterise the theory of Banach spaces itself rather than specific examples.

A more careful analysis of the inseparability ratios will be conducted elsewhere, as now we want to focus on the question of establishing the validity of Conjectures~\ref{ultimate cj} and~\ref{ultimate cj weaker} for centrally symmetric models. In the language we have been developing so far, our central question becomes the following.

\begin{question} \label{question tensor norms}
Do the universal constants defined in~\eqref{R} satisfy $R(n,m)>1$ for all $n,m\geq 2$?
\end{question}

As we mentioned earlier, we will only deal with the simplest case $A=B$ and $V_A = V_B \eqqcolon V$. This corresponds to focusing on Conjecture~\ref{ultimate cj weaker} instead of its stronger version Conjecture~\ref{ultimate cj}. At the level of inseparability ratios, we are thus looking at Question~\ref{question tensor norms} with $S(n)$ instead of $R(n,m)$. We believe it possible to extend the proof strategy we will discuss to this latter case, however doing so does not boil down to a simple change of notation, and instead requires a more careful manipulation of the technical tools we use. 
We are now ready to state and prove the main result of this section.

\begin{thm} \label{thm c ratio}
There exists a universal constant $c>1$ such that for all $n\geq 2$ one has $S(n)\geq c$. A possible choice is $c=1+\frac{1}{29+12\sqrt{6}}$.
\end{thm}

\begin{proof}
Let $V$ be two Banach space of dimension $n$. In what follows, we will denote with $\tau_{ij}$ ($i,j=1,2$) the four entries of the matrix
\bb
\tau \coloneqq \begin{pmatrix} 1 & 1 \\ 1 & -1 \end{pmatrix} .
\label{tau}
\ee
For arbitrary vectors $x_1, x_2 \in V$, consider the tensor
\begin{equation}
Z[x_1, x_2] \coloneqq \sum_{i,j=1,2} \tau_{ij}\, x_i \otimes x_j \in V\otimes V\, . \label{Z}
\end{equation}
An analogous construction can be performed at the level of the duals, in which case we adopt a different notation, that is, for $\varphi_1, \varphi_2 \in V^*$ we define
\begin{equation}
H[\varphi_1, \varphi_2] \coloneqq \sum_{i,j=1,2} \tau_{ij}\, \varphi_i\otimes \varphi_j \in V^*\otimes V^* \, . \label{H}
\end{equation}
It is easy to show that $\left\|Z[x_1,x_2]\right\|_{\varepsilon}\leq 2$ whenever $\|x_i\| \leq 1$ for $i=1,2$. This is an immediate consequence of the fact that $|\alpha_i|,|\beta_j|\leq 1$ ensures that $\sum_{ij} \tau_{ij} \alpha_i \beta_j\leq 2$. Analogously, $\|\varphi_i\|_* \leq 1$ implies that
\begin{equation}
\|H[\varphi_1, \varphi_2 ]\|_{\pi*} = \|H[\varphi_1, \varphi_2]\|_{*\varepsilon} \leq 2\, , \label{H vee'}
\end{equation}
where we employed also~\eqref{dual inj proj}.

Now, we use Auerbach's lemma (Lemma~\ref{Auerbach lemma}) to construct an Auerbach basis $\{v_1,\ldots, v_n\}$ on $V$. This means the following:
\begin{equation}
\braket{v_i^*, v_j}=\delta_{ij}\, ,\qquad \|v_i^*\|_*, \|v_j\|= 1\quad \forall\ i=1,\ldots, n,\ j=1,\ldots, m \, . \label{Auerbach}
\end{equation}
Denote by $P$ the Auerbach-orthogonal projector onto $\text{Span}\{v_1^*, v_2^*\}$, whose action is defined by
\bb
P \varphi \coloneqq v_1^* \braket{\varphi, v_1} + v_2^* \braket{\varphi, v_2}\, .
\label{P proof}
\ee
As is easy to see, for all $\varphi \in V^*$ and $x\in\Span \{v_1, v_2\}$ one has $\braket{\varphi, x} = \braket{P \varphi, x}$. Define
\begin{equation}
1+k \coloneqq \max\{ |\alpha_1|+|\alpha_2| :\ \exists\ \varphi\in V^* : \|\varphi\|_*\leq 1\, ,\ P \varphi = \alpha_1 v_1^* +\alpha_2 v_2^* \}\, , \label{k proof}
\end{equation}
so that $0\leq k\leq 1$. Intuitively, we are looking at the convex body given by the projection of the global unit ball of the dual norm $\|\cdot\|_*$ onto the subspace spanned by $v_1^*, v_2^*$, and determining the size of the smallest `tilted square' containing such a body. Up to changing sign to $v_1^*$ or $v_2^*$ (and to the corresponding primal vector) and up to swapping them, we can suppose that the $\alpha_1,\alpha_2$ achieving the maximum in~\eqref{k proof} are non-negative, and that $\alpha_1 \leq \alpha_{2}$.

We now construct the $Z$ tensor as in~\eqref{Z} by taking $x_1=v_1$, $x_2 = v_2$.
Lemma~\ref{lemma chsh quasi sat} below implies that
\begin{equation}
\|Z[v_1, v_2]\|_{\varepsilon}\leq 2-(1-k)^{2}\, . \label{Z vee}
\end{equation}

To see why this is the case, consider an arbitrary product functional $\eta \otimes \xi$ entering the maximisation that defines the injective norm as in~\eqref{inj}, so that $\|\eta\|_*, \|\xi\|_* \leq 1$. We can expand both $\eta$ and $\xi$ in the dual Auerbach basis, obtaining $\eta = \sum_{i=1}^n \alpha_i v_i^*$ and $\xi =\sum_{j=1}^n\beta_j  v_j^*$. Clearly, we have $|\alpha_i|,|\beta_j|\leq 1$ for all $i$ and $j$, because for instance $|\alpha_i| = |\braket{\eta, v_i}| \leq \|\eta\|_* \|v_i\| \leq 1$. Furthermore, if $P$ is defined as in~\eqref{P proof} one has $P \eta = \alpha_1 v_1^* + \alpha_2 v_2^*$. By definition of $k$, this implies that $|\alpha_1|+|\alpha_2|\leq 1+k$. The same is naturally true on the second space, so that $|\beta_1| + |\beta_2|\leq 1+k$ holds as well. Thanks to Lemma~\ref{lemma chsh quasi sat}, one obtains
\bbb
\braket{\varphi\otimes\lambda, Z[v_1, v_2]} = \alpha_1 (\beta_1+\beta_2) + \alpha_2 (\beta_1-\beta_2) \leq f(k) = 2 - (1-k)^{2}\, .
\eee
Since this holds for all functionals $\eta,\xi$ such that $\|\eta\|_*, \|\xi\|_*\leq 1$, we conclude that $\|Z[v_1, v_2]\|_{\varepsilon}\leq 2-(1-k)^{2}$, as claimed.

Next, consider the tensor $H[v_1^*, v_2^*]$ constructed out of the first two functionals of the dual Auerbach basis as in~\eqref{H}. 
Since~\eqref{H vee'} holds, we can estimate the projective norm of $Z[v_1, v_2]$ by plugging the ansatz $\frac12 H[v_1^*, v_2^*]$ into the generic formula~\eqref{dual norm Banach}. This yields
\begin{equation}
\begin{split}
\|Z[v_1, v_2]\|_{\pi} &\geq \frac12 \|H[v_1^*, v_2^*]\|_{\pi*} \|Z[v_1, v_2]\|_{\pi} \\
&\geq \frac12 \braket{H[v_1^*,v_2^*], Z[v_1, v_2]} \\
&= 2\, ,
\end{split}
\label{Z wedge 1}
\end{equation}
where we used the definitions~\eqref{Z} and~\eqref{H} of $Z$ and $H$ together with the properties~\eqref{Auerbach} of the Auerbach basis. 

Now that we have an upper bound for the injective norm~\eqref{Z vee} and a lower bound for the projective one~\eqref{Z wedge 1}, we can hope to prove the existence of a gap between the two.
Unfortunately, what we have in hand right now is not enough to conclude, because the right-hand side of~\eqref{Z vee} can coincide with that of~\eqref{Z wedge 1} when $k=1$. 
We then look for a new lower estimate of the projective norm $\|Z[v_1, v_2]\|_{\pi}$. 
In order to do so, we construct another witness of the form $ H[\lambda_1, \lambda_2] $, with $\lambda_1, \lambda_2\in V^*$. As long as $\|\lambda_i \|_* \leq 1$, we know from~\eqref{H vee'} that $H[\lambda_1, \lambda_2]_{*\varepsilon}=H[\lambda_1, \lambda_2]_{\pi*}\leq 2$.

The trick now is to focus on the values of $k$ corresponding to the `worst case scenario'. As we saw, when $k\approx 1$ the right-hand side of~\eqref{Z vee} approaches $2$, making~\eqref{Z wedge 1} useless for the purpose of establishing a gap between projective and injective norm.
However, we can exploit the very definition of $k$ to try to say something about this case, too.
Namely, we can invoke the existence of a new functional $\lambda_1$ that saturates the maximisation in~\eqref{k proof}. Such functional will be obtained by `tilting' the pair $(v_1^*, v_2^*)$.

More precisely, one can find $\lambda_1 \in V^*$ with $\|\lambda_1\|_*\leq 1$ and such that
\bb
P\lambda_1 = \alpha_1 v_1^* + \alpha_2 v_2^*\, ,
\ee
with $0\leq \alpha_1\leq \alpha_2\leq 1$ and $\alpha_1+ \alpha_2 =1+k$. To construct $\lambda_2$ we simply appeal to the triangle inequality. Namely, for $|s_1|+|s_2|\leq 1$ we notice that $\lambda_2 \coloneqq s_1 v_1^* + s_2 v_2^*$ obeys $\|\lambda_2\|_* \leq 1$. A quick computation yields
\begin{align*}
&\braket{H[\lambda_1, \lambda_2], Z[v_1, v_2]} \\
&\quad = \braket{H[\alpha_1 v_1^*+ \alpha_2 v_2^*,\, s_1 v_1^* + s_2 v_2^*], Z[v_1, v_2] } \\
&\quad= \alpha_1^{2} + 2\alpha_1\alpha_2 - \alpha_2^{2} + 2(\alpha_1+\alpha_2) s_1 + 2(\alpha_1-\alpha_2) s_2 - s_1^{2} - 2s_1s_2 + s_2^{2}\, . 
\end{align*}
We are free to maximise over $s_1, s_2$ subjected to the constraint $|s_1|+|s_2|\leq 1$. Using Lemma~\ref{lemma max} below, we obtain
\begin{align*}
\|Z[v_1, v_2]\|_{\pi} &\geq \max_{|s_1|+|s_2|\leq 1} \frac12 \braket{H[\lambda_1, \lambda_2] , Z[v_1, v_2] } \\
&= \frac32 \alpha_1^{2} + \alpha_1 (\alpha_2-1) - \frac12 (\alpha_2-1)^{2} + 1\, .
\end{align*}

Now, we do not know what the actual values of $\alpha_1, \alpha_2$ are, but we know that they are subjected to the constraints $0\leq \alpha_1\leq \alpha_2 \leq 1$ and $\alpha_1+\alpha_2 = 1+k\geq 1$. Therefore, we can estimate the above lower bound by taking the minimum over all such pairs $\alpha_1, \alpha_2$. This can be done thanks to Lemma~\ref{lemma min} below, and the result is
\begin{equation}
\left\| Z[v_1, v_2] \right\|_{\pi} \geq 1+\frac32 k^{2}\, . \label{Z wedge 2}
\end{equation}
Putting together~\eqref{Z wedge 1} and~\eqref{Z wedge 2} yields
\begin{equation}
\|Z[v_1, v_2]\|_{\pi} \geq \max\left\{ 2,\, 1+\frac32 k^{2} \right\} . \label{Z wedge final}
\end{equation}
Finally, we see that
\begin{equation}
\frac{\|Z[v_1, v_2]\|_{\pi}}{\|Z[v_1, v_2]\|_{\varepsilon}} \geq \frac{\max\left\{ 2,\, 1+\frac32 k^{2} \right\}}{2-(1-k)^{2}} \geq 1+\frac{1}{29+12\sqrt{6}}\, ,
\end{equation}
where for the last step we used Lemma~\ref{universal min} below.
\end{proof}

\begin{rem}
A by-product of the proof is that it is always possible to achieve $\|Z\|_{\pi}\big/\|Z\|_{\varepsilon}\geq c$ with a tensor $Z$ whose tensor rank is $2$.
\end{rem}

\begin{rem}
The constant we found is most likely prone to substantial improvement. We know we must have $c\leq 2$, though, because as we saw in Example~\ref{ex inj proj} the two-dimensional Euclidean norm satisfies $\|Z\|_{\pi}=\|Z\|_{1}\leq 2 \|Z\|_{\infty}= 2\|Z\|_\varepsilon$, where we thought of $Z$ as a $2\times 2$ real matrices, $\|\cdot\|_{1},\|\cdot\|_{\infty}$ denote the sum of the singular values and the largest singular value, respectively, and thus $2$ is the tightest possible constant in the above inequality.

By examining tensor products of simple two-dimensional Banach spaces like $l_2^2$ (whose unit ball is the unit circle) or  $l_1^2$ (whose unit ball is the unit square), one might be tempted to conjecture that the best constant $c$ satisfying $\|Z\|_\pi \leq c\|Z\|_\varepsilon$ can be taken as large as $2$. However, this turns out not to be the case, as there are examples of `highly local' Banach spaces such that the largest $c$ satisfying Theorem~\ref{thm c ratio} is $\sqrt{3}$, as we show in Appendix~\ref{app S(2)}.
\end{rem}

\begin{cor} \label{thm c ratio cor}
For all centrally symmetric cones $C$ defined as in~\eqref{centr eq}, one has
\bb
C \tminit C \neq C \tmaxit C\, .
\ee
\end{cor}

Let us repeat here that we think of the above Corollary~\ref{thm c ratio cor} as providing some strong evidence in favour of Conjecture~\ref{ultimate cj}. Indeed, to the extent of our knowledge, it constitutes the first attempt to prove such conjecture for models of arbitrarily high dimension (even if subjected to a further symmetry constraint). The other major partial result, Namioka-Phelps theorem (Theorem~\ref{NP thm}), in fact, covers only the case of one of the two local theories being a particular, low-dimensional GPT.
Furthermore, the proof technique is relatively original and might give some hints on how to prove Conjecture~\ref{ultimate cj} in general. Last, but not least, Theorem~\ref{thm c ratio} may be of independent interest in functional analysis, since it deals with universal behaviours of Banach spaces. 

Below we state and prove the four lemmas we used in the proof of Theorem~\ref{thm c ratio}.

\begin{lemma} \label{lemma chsh quasi sat}
Let $\alpha_1,\alpha_2, \beta_1,\beta_2\in [-1,1]$ be real numbers. Let $k\in[-1,1]$ be such that
\begin{equation}
1+k \geq \max\{ |\alpha_1|+|\alpha_2|,\, |\beta_1|+|\beta_2| \}\, . \label{k}
\end{equation}
For $t\in [-1,1]$, consider the monotonically increasing function
\begin{equation}
f(t) \coloneqq \left\{ \begin{array}{lr} (1+t)^{2} & -1\leq t\leq 0\, , \\[1ex] 2-(1-t)^{2} & 0\leq t\leq 1\, . \end{array}\right. \label{phi}
\end{equation}
Then
\begin{equation}
\alpha_1 (\beta_1+\beta_2) + \alpha_2 (\beta_1-\beta_2) \leq f(k)\, . \label{chsh quasi sat}
\end{equation}
\end{lemma}

\begin{proof}
We have to prove that the function $g:[0,1]\rightarrow\mathds{R}$ defined as
\begin{equation*}
g(k) \coloneqq \max\left\{ \sum_{i,j=1,2} \tau_{ij}\, \alpha_i \beta_j :\ |\alpha_i|,|\beta_j|\leq 1,\ \sum_{i=1,2} |\alpha_i|,\ \sum_{j=1,2} |\beta_j| \leq 1+k \right\}
\end{equation*}
coincides with that given in~\eqref{phi}. Let us start by assuming $0<k\leq 1$. The maximisation over $\alpha_1, \alpha_2$ can be carried out with the help of the general formula
\begin{align*}
&\max \left\{ \alpha_1 r_1 + \alpha_2 r_2:\ |\alpha_1|, |\alpha_2|\leq 1,\ |\alpha_1|+|\alpha_2|\leq 1+k \right\} \\
&\quad= \max\left\{k |r_1|+|r_2|,\, |r_1|+k |r_2| \right\} \\
&\quad= k (|r_1|+|r_2|) + (1-k) \max\{|r_1|, |r_2|\} \, ,
\end{align*}
which is an easy consequence of the linearity of the function inside the maximum on the left-hand side and is valid provided that $0\leq k\leq 1$. We are left with
\begin{align*}
g (k) &= \max \big\{ k (|\beta_1 + \beta_2|+|\beta_1- \beta_2|) + (1-k) \max\{|\beta_1+\beta_2|, |\beta_1 - \beta_2|\} : \\
&\qquad \qquad |\beta_1|,|\beta_2|\leq 1,\ |\beta_1|+|\beta_2|\leq 1+k \big\} \\
&= \max\big\{ 2k\max\{|\beta_1|, |\beta_2|\} + (1-k) (|\beta_1| + |\beta_2|):\\
&\qquad\qquad |\beta_1|,|\beta_2|\leq 1,\ |\beta_1|+|\beta_2|\leq 1+k \big\} \\
&= 2k + (1-k)(1-k) \\
&= 2- (1-k)^2 \\
&=f(k)\, .
\end{align*}
If $-1\leq k\leq 0$ we can perform exactly the same steps, which become even easier since the geometric shapes involved in the maximisations are squares and not octagons. For instance, one uses
\begin{equation*}
\max \{ \alpha_1 r_1 + \alpha_2 r_2:\ |\alpha_1|,|\alpha_2|\leq 1,\ |\alpha_1|+|\alpha_2|\leq 1+k \}\, =\, (1+k) \max\,\{|r_1|,|r_2|\}\, ,
\end{equation*}
to eliminate $\alpha_1,\alpha_2$, so that
\begin{align*}
g (k) &= \max\big\{ (1+k) \max\{ |\beta_1 + \beta_2|, |\beta_1-\beta_2| \}: \\
&\qquad\qquad |\beta_1|,|\beta_2|\leq 1,\ |\beta_1|+|\beta_2|\leq 1+k \big\} \\
&= (1+k)^{2} \\
&=f(k)\, ,
\end{align*}
concluding the proof.
\end{proof}

\begin{lemma} \label{lemma max}
For $\alpha_2 \geq \alpha_1\geq 0$ such that $\alpha_1+\alpha_2 \geq 1$ we have
\begin{align*}
&\max\left\{ 2(\alpha_1+\alpha_2) s_1 + 2(\alpha_1-\alpha_2)s_2 - s_1^{2} - 2s_1 s_2 + s_2^{2}:\ |s_1|+|s_2|\leq 1 \right\} \\
&\quad = 2\alpha_1^{2} - 2\alpha_1 + 2\alpha_2 +1\, .
\end{align*}
\end{lemma}

\begin{proof}
Call $w(s_1,s_2)\coloneqq 2(\alpha_1+\alpha_2)s_1 + 2(\alpha_1-\alpha_2)t - s_1^{2} - 2s_1s_2 + s_2^{2}$. It is easily seen that $\partial_{s_1} w=0=\partial_{s_2}w$ happens only for $s_1=\alpha_1,\, s_2=\alpha_2$. However, the constraint $\alpha_1+\alpha_2\geq 1$ imposes that this point is not inside the tilted square $|s_1|+|s_2|\leq 1$. Therefore, in order to find the maximum we have to look at the values of $w$ on the border.

Rewriting $-s_1^{2}-2s_1s_2+s_2^{2}=(s_1-s_2)^{2}-2s_1^{2}=-(s_1+s_2)^{2}+2s_2^{2}$ we see that $w$ is convex on the two sides of the square where $s_1+s_2=\pm1$ and concave on the other two where $s_1-s_2=\pm 1$. Therefore, these latter two sides (together with their vertices) are all that matters.

Now, on the side identified by $s_2=s_1-1$ ($0\leq s_1\leq 1$) we obtain
\begin{equation*}
w(s_1,s_1-1) = 2(\alpha_1+\alpha_2)s_1 + 2(\alpha_1-\alpha_2) (s_1-1) +1 -2s_1^{2}  \, ,
\end{equation*}
which achieves his maximum for $s_1=\alpha_1$:
\begin{equation}
\max_{0\leq s_1\leq 1} w(s_1,s_1-1) = 2\alpha_1^{2} - 2\alpha_1 + 2\alpha_2 +1\, . \label{max 1}
\end{equation}
On the other side $s_2=s_1+1$ ($-1\leq s_1\leq 0$) we have
\begin{equation*}
w(s_1,s_1+1) = 2(\alpha_1+\alpha_2)s_1 + 2(\alpha_1-\alpha_2) (s_1+1) +1 -2s_1^{2}\, ,
\end{equation*}
which is maximum on the extreme points since the derivative is never zero for $s_1<0$. Thus,
\begin{equation}
\begin{split}
\max_{-1\leq s_1\leq 0} w(s_1,s_1+1) &= \max\{ w(-1,0),\, w(0,1) \} \\
&= \max\{ -2(\alpha_1+\alpha_2)-1,\, 2(\alpha_1 - \alpha_2)+1 \} \\
&= 2(\alpha_1-\alpha_2)+1\, .
\end{split}
\label{max 2}
\end{equation}
Thanks to $\alpha_2 \geq \alpha_1$, the rightmost side of~\eqref{max 2} is clearly upper bounded by~\eqref{max 1}, which becomes the absolute maximum we are looking for.
\end{proof}

\begin{lemma} \label{lemma min}
For all $0\leq k\leq 1$, we have
\begin{equation}
\begin{split}
&\min \Big\{ \frac32 \alpha_1^{2} + \alpha_1 (\alpha_2 - 1) - \frac12 (\alpha_2 - 1)^{2} + 1: \\
&\qquad\quad 0\leq \alpha_1\leq \alpha_2\leq 1,\ \alpha_1+\alpha_2 \geq 1+k \Big\} \\
&\quad= 1 + \frac32 k^{2}\, .
\end{split}
\label{eq min}
\end{equation}
\end{lemma}

\begin{proof}
Call $\nu(\alpha_1,\alpha_2)$ the objective function and $\mu(k)$ the minimum on the left-hand side of~\eqref{eq min}. The region of the plane $\alpha_1,\alpha_2$ determined by the inequalities $0\leq \alpha_1\leq \alpha_2\leq 1$ and $\alpha_1+\alpha_2\geq 1+k$ is a triangle with vertices $(k,1),\, (1,1),\, \big(\frac{1+k}{2},\frac{1+k}{2}\big)$. For a fixed $\frac{1+k}{2}\leq \alpha_2 \leq 1$, the range of $\alpha_1$ is easily seen to be $[1+k-\alpha_2,\alpha_2]$. The function $\nu(\alpha_1,\alpha_2)$ is convex in $\alpha_1$ for fixed $\alpha_2$, but the point $\alpha_1=\frac{1-\alpha_2}{3}$ in which the derivative $\partial_{\alpha_1}\nu(\alpha_1,\alpha_2)$ is zero is never inside the range, because $\frac{1-\alpha_2}{3}+\alpha_2=\frac{1+2\alpha_2}{3}\leq 1\leq 1+k$. Instead, for the allowed values of $\alpha_1$ we always have $\partial_{\alpha_1}\nu(\alpha_1,\alpha_2)\geq 0$. Therefore, in order to find the minimum we have to look at the smallest possible $\alpha_1$, and we get
\begin{align*}
\mu(k) &= \min_{\frac{1+k}{2}\leq \alpha_2\leq 1} \nu(1+k-\alpha_2,\alpha_2) \\
&= \min_{\frac{1+k}{2}\leq \alpha_2\leq 1} \left( 1 -2k(\alpha_2-1) + \frac32 k^{2} \right) \\ &= 1+\frac32 k^{2}\, ,
\end{align*}
as claimed.
\end{proof}

\begin{lemma} \label{universal min}
The following holds:
\begin{equation}
\min_{0\leq k\leq 1} \frac{\max\left\{ 2,\, 1+\frac32 k^{2} \right\}}{2-(1-k)^{2}} = 1+\frac{1}{29+12\sqrt{6}}\, .
\end{equation}
\end{lemma}

\begin{proof}
For $0\leq k\leq \sqrt{\frac23}$ the function to be maximised is $2\big/\left( 2-(1-k)^{2} \right)$, which is decreasing for $0\leq k\leq 1$. For $\sqrt{\frac23}\leq k\leq 1$, we obtain instead $\left( 1+\frac32 k^{2} \right)\big/\left( 2-(1-k)^{2} \right)$, whose derivative is positive outside $[-2,\frac13]$. Therefore, the minimum is achieved for $k=\sqrt{\frac23}$, and we can conclude.
\end{proof}

\section{Conclusions and outlook} \label{sec2 conclusions}

Throughout this chapter, we reviewed the basic of the abstract state space formalism of GPTs, and we discussed the problem of the composition of local systems within this framework. We looked into some known examples of GPTs such as classical probability theory, quantum mechanics, or generalised non-signalling theories, and saw how the formalism of GPTs can be useful in rephrasing known concept such as that of Bell nonlocality in a more compact way. We also introduced new classes of theories like the centrally symmetric models, which will turn out to be instrumental in the forthcoming Chapter~\ref{chapter3}.

Inspired by its importance (and beauty) in quantum information, we spent some time studying some features of the separability problem in GPTs. We extended some recent results on separability of low-rank tensors to the GPT setting, and uncovered a connection between the theory of tensor norms in functional analysis and that of entanglement of centrally symmetric models.

Next, we formulated and motivated some conjectures about general composite systems, whose solution is in our view central to deepen our understanding of the connections between non-classicality of local theories and the possible existence of entangled states when they are composed to form a single bipartite system. We presented what is known about the problem and strengthened an old argument of Namioka and Phelps to include also a statement about nonlocality, which is a strictly stronger concept than entanglement.
Moreover, we answered said conjecture in the affirmative when the local subsystems are described by two copies of a centrally symmetric model. Along the way, we proved a general result on Banach spaces that might be of independent interest in functional analysis.

As the careful reader will have noticed, we (intentionally) evaded the problem of giving substance to point (ii) of the discussion in Subsection~\ref{subsec2 fundamental}. There, we were asking for a lower estimate of the entangleability of two local GPTs in terms of their `degree of non-classicality'. In order to pursue this goal, the following steps must be followed.
\begin{enumerate}[(i)]

\item  We have to find some appropriate measure of non-classicality of GPTs, call it $\mathcal{N}$, which vanishes for classical theories and somehow quantifies how far from this extremal case we are. An example of such measure could be the global incompatibility quantifier $T$ of \eqref{lambda[A]}, but many other choices are possible.

\item We have to be able to quantify how much entanglement is contained in a state of $A\tmax B$, for $A,B$ GPTs. The maximal entanglement achievable in the maximal tensor product will identify some `entangleability measure' $\mathcal{E}[A,B]$. Perhaps surprisingly, here there is a somewhat natural choice, for it is safe to say that the only standard quantum entanglement measure that carries over to the GPT case, as it rests only on general concepts such as convexity, is the entanglement robustness \cite{VidalTarrach, Jencova2017}.

\item The third, decisive step is then combining the above two quantifiers and prove an inequality of the form
\bb
\mathcal{E}[A,B] \geq \mathcal{N}[A]\, \mathcal{N}[B]\, ,
\label{dream eq}
\ee
to be obeyed by all GPTs $A,B$ of any dimension. 
\end{enumerate} 

An inequality like~\eqref{dream eq}, if established, would be incredibly suggestive and inspiring. It would tell us that there is a deep, quantitative link between a totally self-referential feature like non-classicality, which reveals itself when a GPT is considered on its own, and a global phenomenon such as the entanglement. Finally, as we saw in Subsection~\ref{subsec2 fundamental}, one could alternatively consider an analogue of~\eqref{dream eq} with the concept of Bell nonlocality displacing that of entanglement.

As the abundance of conjectures and questions we stated in the chapter suggests, the analysis carried out here leaves many open problems, and is prone to substantial improvement and deepening. This is partly due to the fact that my interest in these problems is relatively new, as it was spurred by the results of the forthcoming Chapter~\ref{chapter3}, which came chronologically earlier in my PhD. Anyway, I hope that the reader has enjoyed our tour of the vast field of study of nonlocality in general probabilistic theories, and appreciated my sincere enthusiasm for the deep problems we were faced with.

\chapter{Ultimate data hiding \\ in quantum mechanics and beyond} \label{chapter3}

\section{Introduction} \label{sec3 intro}

While discussing the scientific programme of Chapter~\ref{chapter2}, in Subsection~\ref{subsec2 fundamental}, we stressed the importance of identifying which ones among the many non-classical phenomena we discovered with the advent of the quantum revolution pertain to all non-classical theories, and which ones on the contrary have more to do with the particular nature of quantum mechanics. In point (iii) of that discussion, we pushed things a bit further, and suggested that this conceptually important question can be made \emph{quantitative} instead of merely \emph{qualitative}. Namely, for any particular non-classical phenomenon, we can ask how strong that specific type of non-classicality is in quantum mechanics compared to its ultimate limits within the broad realm of GPTs. One of the main purposes of the present chapter is to conduct this analysis thoroughly for the particular case of \emph{data hiding}.

The present chapter is organised as follows. The rest of this section is devoted to introducing in broad terms the main questions we will face and the answers we will obtain (Subsection~\ref{subsec3 dh}), and to pointing the reader to our original contributions (Subsection~\ref{subsec3 original}).
Section~\ref{sec dh GPTs} is devoted to giving precise definitions of what we mean by data hiding in the GPT setting. Then, throughout Section~\ref{sec3 ex} we complete the solution of the important special case constituted by quantum mechanical data hiding. In Section~\ref{sec3 univ} we present the main result of the chapter, i.e. the determination of the optimal universal upper bound on data hiding ratios against locally constrained sets of measurements (Theorem~\ref{thm univ}). Section~\ref{sec3 special} presents a body of techniques to compute data hiding ratios in specific classes of GPT models satisfying some further assumptions. Finally, Section~\ref{sec3 conclusions} is devoted to the conclusions.

\subsection{Data hiding} \label{subsec3 dh}

In quantum mechanics, data hiding is usually intended as the existence of pairs of states of a bipartite system that are perfectly distinguishable with global measurements yet almost indistinguishable when only protocols involving local operations and classical communication (LOCC) are allowed~\cite{dh-original-1, dh-original-2}. In view of the discussion in Subsection~\ref{subsec2 measur bip}, we can easily extend the relevant definitions as to encompass more general forms of data hiding in arbitrary general probabilistic theories (Definition~\ref{dh}). In this context, the effectiveness of discriminating protocols is measured by the minimal probability of error $P_e^{\mathcal{M}}(\rho,\sigma; p)$ in the task of distinguishing the two states $\rho,\sigma$ with a priori probabilities $p,1-p$ respectively, when only operations from the class $\mathcal{M}$ are available. The archetypical example of a pair of states exhibiting data hiding is given by the normalised projectors onto the symmetric and antisymmetric subspaces in $\mathds{C}^{n}\otimes\mathds{C}^{n}$, denoted by $\rho_{S}$ and $\rho_{A}$, respectively. While $P_e^{\text{ALL}}\left(\rho_{S}, \sigma_{\mathcal{A}}, \frac12\right) = 0$, because the two states have orthogonal support, it can be shown that $P_e^{\text{LOCC}}\left(\rho_{S}, \sigma_{\mathcal{A}}, \frac12\right) = \frac{2}{n+1}$~\cite{dh-original-1, dh-original-2}. 

For these discussions, a more convenient quantity is the \emph{distinguishability norm} associated to $\mathcal{M}$~\cite{VV-dh}, denoted by $\|\cdot\|_{\mathcal{M}}$ and defined by $P_{e}^{\mathcal{M}}(\rho,\sigma; p) = \frac12 \left( 1 - \| p\rho-(1-p)\sigma\|_{\mathcal{M}}  \right)$, as detailed in Definition~\ref{def d norm} and Lemma~\ref{discr GPT}; it quantifies the advantage of making an observation over pure guessing (i.e. the prior information). It is immediately obvious that higher values of $\| p\rho-(1-p)\sigma\|_{\mathcal{M}}$ correspond to an increased discriminating power of the set $\mathcal{M}$. Thus, the central object of our investigation is a quantity that we name \emph{data hiding ratio}, which depends on the GPT as well as on the restricted set of measurements we consider, and is given by
\begin{equation*}
R (\mathcal{M}) = \max \frac{\| p\rho-(1-p)\sigma\|_{\text{ALL}}}{\| p\rho-(1-p)\sigma\|_{\mathcal{M}}},
\end{equation*}
where $\|\cdot\|_{\text{ALL}}$ denotes the norm associated to the whole set of possible measurements and the maximisation ranges over all pairs of states $\rho,\sigma$ and a priori probabilities $p$ (Definition~\ref{dh} and Proposition~\ref{dh ratio}).

We will look at the quantity $R(\mathcal{M})$ in the case of physically or operationally significant restricted sets of protocols $\mathcal{M}$. Concretely, we focus on the case when the system under examination is bipartite, and $\mathcal{M}$ is some set of locally constrained measurements (Definition~\ref{locally constr}).
Our interest in this kind of questions has been spurred by a result in~\cite{VV-dh}, stating that on a finite-dimensional quantum mechanical system $\mathds{C}^{n_A}\otimes\mathds{C}^{n_B}$ the data hiding ratio against LOCC protocols (which constitute the most operationally relevant class) satisfies
\bb
\Omega\left(n\right) \leq R_{\text{QM}}(\text{LOCC}) \leq O \left( \sqrt{n_A n_B} \right),
\label{VV bound 0}
\ee
where $n \coloneqq \min\{n_A, n_B\}$. In particular, when $n_A=n_B=n$ one obtains 
\bbb
R_{\text{QM}}(\text{LOCC})=\Theta(n)\, .
\eee 
In this chapter, we will conduct a thorough analysis of $R_{\text{QM}} (\text{LOCC})$ and of analogous quantities constructed in arbitrary GPTs. For details, see the following Subsection~\ref{subsec3 original}.

\subsection{Our contributions} \label{subsec3 original}

The material presented in this chapter is part of the homonymous paper~\cite{ultimate}:

\begin{itemize}
\item L. Lami, C. Palazuelos, and A. Winter. Ultimate data hiding in quantum mechanics and beyond. \emph{Preprint arXiv:1703.03392}, 2017. To appear in \emph{Commun. Math. Phys.}
\end{itemize}

Let us now point the reader to our main original contributions. From our review of what is already known about data hiding in quantum mechanics, it is immediately clear that there is a gap between lower and upper bound in~\eqref{VV bound 0}, which can become arbitrarily large for very different values of the local dimensions. This gap can be reduced but not entirely filled with the help of~\cite[Lemma 20]{Brandao-area-law}, which entails the upper bound $R_{\text{QM}}(\text{LOCC}) \leq n^2$, where once again $n\coloneqq \min\{n_A, n_B\}$.

We are thus interested in the determination of the exact scaling of the data hiding ratio $R_{\text{QM}}(\text{LOCC})$ with respect to $n_A$ and $n_B$. Our first contribution is an intuitive argument that uses the quantum teleportation protocol to show that $R_{\text{QM}}(\text{LOCC})=\Theta(n)$ holds in fact for all $n_A, n_B$ (Theorem~\ref{dh QM}). The beauty of the argument lies not only in its crystal clear simplicity, but also in the fact that: (i) it works for all $n_A, n_B$, thus providing a general solution to the problem of computing $R_{\text{QM}}(\text{LOCC})$ up to multiplicative constants; (ii) it is constructive, in the sense that it provides an explicit one-way LOCC protocol to achieve the lower bound on the probability of error; (iii) it yields better constants that the one in~\cite{VV-dh}; in fact, our bounds will be so close to being tight that they will enable us to compute \emph{exactly} the `relaxed' ratio $R_{\text{QM}}(\text{SEP})$ at least when $n_A=n_B$; (iv) the argument is potentially generalisable to GPTs other than quantum mechanics, as long as they support teleportation~\cite{telep-in-GPT}.

We will spend some time examining some of the consequences of Theorem~\ref{dh QM} and of the teleportation argument in standard quantum information (Subsection \ref{subsec3 telep arg}). In particular, we will exhibit the best known lower bound on the squashed entanglement in terms of the trace norm (Corollary \ref{lower sq cor}) and an alternate quantum de Finetti theorem (Corollary \ref{de Finetti cor}).

Back to the data hiding problem, we already mentioned that Theorem~\ref{dh QM} implies that $R_{\text{QM}}(\text{LOCC})=\Theta\left(\min\{n_A, n_B\}\right)$. Note that the local real dimensions of the quantum cones of states (in other words, the dimensions of the local GPTs in the sense of Definition~\ref{GPT def}) are $d_A=n_A^2$ and $d_B=n_B^2$. Using these as parameters, one could write the value of the quantum data hiding ratio as $R_{\text{QM}}(\text{LOCC}) = \Theta \left( \min\{ \sqrt{d_A}, \sqrt{d_B} \}\right)$.

Along the same lines of thought, we compute the data hiding ratios against locally constrained sets of measurements for another significant GPT called `spherical model' (Subsection~\ref{subsec3 sph}). As the name suggests, the state space of such a model is a Euclidean ball. We prove that for fixed local real dimensions, the data hiding ratio displayed by the spherical model is quadratically larger than the quantum mechanical one, i.e. $R_{\text{Sph}}(\mathcal{M}) = \Theta \left( \min\{ d_A, d_B \} \right)$.

Then, the problem we are faced with is to establish an \emph{optimal, universal upper bound} for $R(\mathcal{M})$ that depends only on the local real dimensions $d_A, d_B$ and \emph{not} on the particular GPTs we choose.
Here, $\mathcal{M}$ is a locally constrained set of measurements, as usual. The answer to this central question is the content of the main result of the present chapter, Theorem~\ref{thm univ}, which states that $R(\mathcal{M})\leq \min\{d_A, d_B\}$ holds for all bipartite GPTs of local dimensions $d_A, d_B$. Since we have seen that such a scaling characterises spherical models, we deduce that \emph{$\min\{d_A, d_B\}$ is the optimal universal upper bound on the data hiding ratio against locally constrained sets of measurements.} This answers our fundamental question on the ultimate effectiveness of data hiding when the local systems have bounded size.

The originality of the proof of Theorem~\ref{thm univ} is twofold. On the one hand, it rests crucially on a deep connection between the theory of locally constrained distinguishability norms and that of tensor norms (as introduced in Subsection~\ref{subsec2 tensor norms}). This connection is uncovered here for the first time, and we believe it may lead to further advances in the understanding of both sides of the problem. On the other hand, in order to arrive at the proof of the central claim we have to deal with a question concerning tensor norms constructed out of arbitrary Banach spaces of fixed dimension. This problem is addressed in Proposition~\ref{prop proj=<ninj}, which constitutes -- to the extent of our knowledge -- an original mathematical contribution on its own.

Section~\ref{sec3 special} is less essential to the comprehension of the foundational questions posed in this chapter, and is mainly motivated by genuine curiosity. There, we analyse the problem of computing the data hiding ratios for special classes of GPTs. More in detail, we look at centrally symmetric models (Definition~\ref{centr}), and reduce the task of computing the data hiding ratio to that of evaluating a certain quantity constructed out of some injective and projective tensor norms (Theorem~\ref{centr ratio}). This approach constitutes a generalisation of the aforementioned solution of the spherical model, and represents our third instance (after those described in Theorems~\ref{thm c ratio} and~\ref{thm univ}) of applications of the theory of tensor norms to information theoretic problems in general probabilistic theories. Finally, we show how to generalise the concept of data hiding Werner states to arbitrary GPTs that possess an adequate symmetry. It turns out that the maximal effectiveness of data hiding that is achievable using those highly symmetric states depends only on few geometric parameters rather than on the whole structure of the involved GPTs, as detailed in Theorem~\ref{dh Werner}.

\section{Data hiding in general probabilistic theories} \label{sec dh GPTs}

Throughout this section, we are going to define the central object of our investigation, i.e. the phenomenon known as data hiding. Since we want to address the problem in full generality, our treatment encompasses the case of an arbitrary GPT. In the next Section~\ref{sec3 ex}, we will analyse in detail the important case of quantum mechanical data hiding.

\subsection{State discriminiation} \label{subsec3 state discr}

In the following, let $(V,C,u)$ be a given GPT in the sense of Definition~\ref{GPT def}. As customary, we will denote by $\Omega$ its state space. The binary distinguishability problem consists of choosing secretly one of the two states $\rho,\sigma\in \Omega$ with known a priori probabilities $p, 1-p$ and handing it over to an agent, whose task is to discriminate between the two alternatives. Naturally, the larger the set of measurements the agent has at his disposal, the lower the associated probability of error will be (for fixed states and a priori probabilities). In general, it will make sense to consider measurements that are at least \emph{informationally complete}, meaning that from their complete statistics the full state can be reconstructed unambiguously. A formal definition is below. Following the notation developed in Subsection~\ref{subsec2 single}, we remind the reader that a measurement $(e_{i})_{i\in I}\in \mathbf{M}$ is a finite collection of effects $e_i\in [0,u]$ such that $\sum_{i\in I} e_i = u$.

\begin{Def} \label{info complete}
Let $(V,C,u)$ be a GPT. Then a measurement $(e_{i})_{i\in I}\in \mathbf{M}$ is said to be \emph{informationally complete} if $\text{\emph{span}}\{e_{i}:\, i\in I\}=V^{*}$. A set $\{\mu_t\}_{t\in T}$ made of measurements $\mu_t=\big(e_i^{(t)}\big)_{i\in I_t}$ is deemed informationally complete if $\text{\emph{span}}\big\{e_{i}^{(t)}:\, t\in T, i\in I_t\big\}=V^{*}$.
\end{Def}

If $\mathcal{M}\subseteq \mathbf{M}$ is a set of measurements in an arbitrary GPT, we can define an associated norm by translating to GPTs the analogous definition in~\cite{VV-dh}.

\begin{Def} \label{def d norm}
Let $\mathcal{M}\subseteq\mathbf{M}$ be an informationally complete set of measurements in a GPT $(V,C,u)$. The associated \emph{distinguishability norm} $\|\cdot\|_{\mathcal{M}}$ is a norm on $V$ given by
\begin{equation}
\|x\|_{\mathcal{M}} \coloneqq \sup_{(e_{i})_{i\in I}\in \mathcal{M}} \sum_{i} |\braket{e_{i},x}|
\label{d norm}
\end{equation}
for all $X\in V$.
\end{Def}

As is easy to see, the function defined in~\eqref{d norm} is truly a norm on $V$ thanks to the informational completeness of $\mathcal{M}$. Among its elementary properties, we note three: (i) the identity $\|x\|_{\mathcal{M}}=u(x)$, valid on positive vectors $x\geq 0$ (independently of $\mathcal{M}$); (ii) the general bound $\|x\|_{\mathcal{M}} \geq |u(x)|$, an elementary consequence of the definition~\eqref{d norm}; (iii) the monotonicity of $\|\cdot\|_\mathcal{M}$ in $\mathcal{M}$, with the partial order defined by the inclusion; and (iv) the fact that $\|\cdot\|_{\mathbf{M}}$ coincides with the base norm $\|\cdot\|$.  This latter fact can be seen as follows: on the one hand, from~\eqref{base norm 2} it is clear that $\|\cdot\|_{\mathbf{M}}\geq \|\cdot\|$, because $(e,u-e)$ is a legitimate measurement for all $e\in [0,u]$; on the other hand, $\|\cdot\|\geq \|\cdot\|_{\mathbf{M}}$ because if $\mathcal{M}=\mathbf{M}$ then in~\eqref{d norm} we can restrict ourselves to binary measurements with no loss of generality. In fact, for an arbitrary $(e_i)_{i\in I}\in\mathbf{M}$ one has
\begin{align*}
\sum_{i\in I} |\braket{e_i, x}| &= \sum_{i\in I_+} \braket{e_i, x} - \sum_{i\in I_-} \braket{e_i, x} \\
&= \Big\langle\sum_{i\in I_+} e_i, x \Big\rangle - \Big\langle \sum_{i\in I_-} e_i, x \Big\rangle \\
&\leq \bigg| \Big\langle \sum_{i\in I_+} e_i, x\Big\rangle \bigg| + \bigg| \Big\langle \sum_{i\in I_-} e_i, x \Big\rangle \bigg|\, ,
\end{align*}
where we defined $I_+\coloneqq \{i\in I:\, \braket{e_i,x}\geq 0\}$ and $I_-\coloneqq I\backslash I_+$.

The above Definition~\ref{def d norm} makes sense by virtue of its link to the operational task of state discrimination as given by the following lemma, totally analogous to~\cite[Theorem 5]{VV-dh}.

\begin{lemma} \label{discr GPT}
Let $(V,C,u)$ be a GPT with state space $\Omega$, and consider an informationally complete set of measurements $\mathcal{M}\subseteq \mathbf{M}$. Then the lowest probability of error for discriminating between two states $\rho,\sigma\in \Omega$ with a priori probabilities $p,1-p$, respectively, is given by
\begin{equation}
P_{e}^{\mathcal{M}}(\rho,\sigma; p)\, =\, \frac12 \left( 1 - \| p\rho-(1-p)\sigma\|_{\mathcal{M}}  \right) , \label{pr error}
\end{equation}
where $\|\cdot\|_{\mathcal{M}}$ is the distinguishability norm given by~\eqref{d norm}.
\end{lemma}

\begin{proof}
This goes in complete analogy with the corresponding argument for quantum mechanics~\cite{HELSTROM, VV-dh, state-discr-GPT}, but we explain it here for the sake of completeness. Without loss of generality, we can assume that the protocol consists of measuring the state with a measurement $(e_{i})_{i\in I}$ and performing a (possibly probabilistic) post-processing of the classical outcome $i$. To encompass the possibility of such probabilistic post-processing, assume that the outcome $i$ yields $\rho$ or $\sigma$ as final answers with probabilities $q_{i}$ and $1-q_{i}$, respectively. Then the probability of error is given by
\begin{align*}
P_{e} &= p \sum_{i\in I} \braket{e_{i},\rho} (1-q_{i}) + (1-p) \sum_{i} \braket{e_{i},\sigma} q_{i} \\
&= p - \sum_{i} q_{i} \braket{e_{i},\, p\rho-(1-p)\sigma}\, ,
\end{align*}
where we employed the normalisation relation $\sum_{i\in I} e_{i}=u$. Minimising over all probabilities $q_{i}$ one obtains
\begin{equation*}
P_{e} = \frac12 \left( 1 - \sum_{i\in I} |\braket{e_{i},\, p\rho-(1-p)\sigma}| \right) ,
\end{equation*}
and finally~\eqref{pr error} after a minimisation over all measurements $(e_{i})_{i\in I}\in \mathcal{M}$.
\end{proof}

The analogy with quantum mechanics goes much beyond this. In fact, all the results of~\cite[\S 2]{VV-dh} (with the exception of Proposition 8 there) carry over to GPTs. In translating the statements one has just to remember that the quantum mechanical trace norm $\|\cdot\|_{1}$ becomes the base norm in the GPT framework, and that similarly the identity becomes the unit effect. For details on the interpretation of quantum mechanics as a GPT, we refer the reader to Subsection~\ref{subsec2 ex QM}. Here we limit ourselves to providing a formulation of of~\cite[Theorem 4]{VV-dh} for arbitrary GPTs. Remember from Subsection~\ref{subsec2 single} that for a set of measurements $\mathcal{M}\subseteq\mathbf{M}$ we will denote by $\langle \mathcal{M} \rangle$ the set generated by $\mathcal{M}$ via the procedure of coarse graining~\eqref{coarse}.

\begin{lemma} \label{unit ball dual d norm}
The unit ball of the dual to the distinguishability norm~\eqref{d norm} is given by
\begin{equation}
B_{\|\cdot\|_{\mathcal{M},*}} = \clit \left( \coit  \big\{ 2e-u:\ \{ e, u-e \}\in \langle \mathcal{M}\rangle \big\} \right) .
\label{unit ball dual d norm eq}
\end{equation}
Equivalently, $\|\cdot\|_{\mathcal{M}}$ can be computed as
\bb
\|x\|_{\mathcal{M}} = \sup\left\{ \braket{\varphi, x}:\ \left(\frac{u+\varphi}{2},\, \frac{u-\varphi}{2} \right) \in \langle \mathcal{M}\rangle \right\} .
\label{d norm altern}
\ee
Consequently, there is a one-to-one correspondence between distinguishability norms~\eqref{d norm} and closed symmetric convex bodies $K$ such that $\pm u\in K\subseteq [-u,u]$.
\end{lemma}

\begin{proof}
See~\cite{VV-dh}.
\end{proof}

\vspace{0ex}
\begin{rem}
From Lemma~\ref{unit ball dual d norm} it follows in particular that $\|\cdot\|_{\mathcal{M}}$ depends only on the set of measurements generated by $\mathcal{M}$ via coarse graining, and in fact only on the right-hand side of~\eqref{unit ball dual d norm eq}.
\end{rem}

At this point, the reader should be familiar enough with the body of techniques we have discussed so far, to be able to work out the translations of the other results in~\cite[\S 2]{VV-dh} by herself. As for us, we believe it more appropriate to devote the forthcoming Subsection~\ref{subsec3 dh statement} to discussing the phenomenon of data hiding, first in quantum mechanics (where it was originally discovered), and subsequently in an arbitrary GPT.

\subsection{Data hiding and statement of the problem} \label{subsec3 dh statement}

Throughout this section, a generalisation of the concept of data hiding against LOCC measurements in quantum mechanics as originally conceived in~\cite{dh-original-1, dh-original-2} is discussed. On the one hand, we will extend this notion to an arbitrary GPT, and on the other hand we will allow for data hiding against an arbitrary set of measurements, without any a priori assumption on its nature. We give the following definition.

\begin{Def} \label{dh}
Let $(V,C,u)$ be a GPT with state space $\Omega$. For an informationally complete set of measurements $\mathcal{M} \subseteq \mathbf{M}$, we say that there is \emph{data hiding against $\mathcal{M}$ with efficiency $R\geq 1$} if there are two normalised states $\rho,\sigma\in\Omega$ and a real number $p\in [0,1]$ such that the probability of error defined in~\eqref{pr error} satisfies
\begin{equation}
P_{e}^{\mathbf{M}}(\rho,\sigma; p) = 0\, ,\qquad P_{e}^{\mathcal{M}}(\rho,\sigma; p) = \frac12 \left( 1-\frac1R\right) .
\label{dh eq}
\end{equation}
The highest data hiding efficiency against $\mathcal{M}$ is called \emph{data hiding ratio against $\mathcal{M}$} and will be denoted by $R(\mathcal{M})$.
\end{Def}

\begin{rem}
A priori, the only meaningful way to define $R(\mathcal{M})$ is as the supremum of all achievable data hiding efficiencies. However, we will see in a moment that this supremum is actually a maximum (Proposition~\ref{dh ratio}).
\end{rem}

In the above Definition~\ref{dh}, we chose not to restrict ourselves to equiprobable pairs of states. There are several reasons that justify this choice. On the one hand, it can be shown that any pair that exhibits data hiding with high efficiency is approximately equiprobable, the approximation becoming better and better for higher efficiencies. On the other hand, even considering only the case of exact equality $p=1/2$ from the start, the obtained data hiding ratio does not differ by more than a factor of two (additive constants apart) from that we have defined here. We devote Appendix~\ref{app equi} to exploring the consequences of restricting the definition of data hiding to the equiprobable case.

We now go back to the investigation of data hiding in the sense of Definition~\ref{dh}. An elementary yet fruitful observation is that if a set of measurements exhibits data hiding with high efficiency, then the associated distinguishability norm -- given by~\eqref{d norm} -- has to be very different from the base norm. This is again a straightforward consequence of Lemma~\ref{discr GPT}. The following result establishes the converse, i.e. that if the two norms are very different on some vectors then there is highly efficient data hiding.

\begin{prop} \label{dh ratio}
For an informationally complete set of measurements $\mathcal{M}\subseteq\mathbf{M}$ in an arbitrary GPT $(V,C,u)$, the data hiding ratio $R(\mathcal{M})$ is given by
\begin{equation}
R(\mathcal{M}) = \max_{0\neq x\in V} \frac{\|x\|}{\|x\|_{\mathcal{M}}}\, .
\label{dh ratio eq}
\end{equation}
Adopting the terminology of~\cite{VV-dh}, we can rephrase~\eqref{dh ratio eq} by saying that $R(\mathcal{M})$ is the constant of domination of $\|\cdot\|_{\mathcal{M}}$ on $\|\cdot\|$, i.e. the smallest $k\in \mathds{R}$ such that $\|\cdot\|\leq k\|\cdot\|_{\mathcal{M}}$.
\end{prop}

\begin{proof}
Let the GPT $(V,C,u)$ have state space $\Omega$. Then, from Definition~\ref{dh} and from~\eqref{pr error} we see that
\begin{align*}
R(\mathcal{M}) = \max \big\{ &\|p\rho-(1-p)\sigma\|_{\mathcal{M}}^{-1}:\ \rho,\sigma\in\Omega,\ 0\leq p\leq 1, \\
&\|p\rho-(1-p)\sigma\|=1 \big\}\, .
\end{align*}
Now, the crucial observation is that the set $K$ of vectors $x\in V$ that can be represented as $x=p\rho-(1-p)\sigma$ for appropriate $\rho,\sigma\in\Omega$ and $p\in[0,1]$ coincides with the unit ball of the base norm, i.e. $K=B_{\|\cdot\|}$. This is an easy consequence of~\eqref{dual base eq} and of~\eqref{strictly pos base}, something that we exploited already in writing down~\eqref{base norm K}.
Thanks to this observation, we rewrite the above representation of $R(\mathcal{M})$ as
\begin{align*}
R(\mathcal{M}) &= \max\left\{\|x\|_{\mathcal{M}}^{-1}:\, \|x\|=1\right\} \\
&= \max_{x\neq 0}\, \frac{\|x\|}{\|x\|_{\mathcal{M}}}\, .
\end{align*}
For the last step, we used the positive homogeneity of the norms, and we converted the supremum over the (compact) unit ball of the base norm into a maximum.
\end{proof}

Some elementary properties of the data hiding ratio are as follows.

\begin{lemma} \label{elem dh}
Let $\mathcal{M}$ be an informationally complete set of measurements in an arbitrary GPT. Then:
\begin{itemize}
\item the data hiding ratio given in Definition~\ref{dh} satisfies $R(\mathcal{M})\geq 1$, with equality iff the set on the left hand side of~\eqref{unit ball dual d norm eq} coincides with the full interval $[-u,u]$;
\item $R(\mathcal{M})$ is monotonically non-increasing as a function of $\mathcal{M}$, where the partial order on sets of measurements is the one given by inclusion.
\end{itemize}
\end{lemma}

As expected, not much can be said about data hiding ratios for a single system and when the sets of measurements are completely arbitrary. In fact, it is easy to show that already in a classical GPT (Subsection~\ref{subsec2 ex class}) such as $(\mathds{R}^2, \mathds{R}^2_+, u)$ with $u=(1,1)$, so that $\braket{u,(x,y)} = x+y$, for the particular case when $\mathcal{M}$ is made of just one measurement $\big( (\varepsilon,0),\, (1-\varepsilon, 1)\big)$ we have $R(\mathcal{M})=(2-\varepsilon)/\varepsilon$, so that the data hiding ratio can even be unbounded in a system of fixed dimension.

The situation changes dramatically when we consider bipartite systems (Subsection~\ref{subsec2 bipartite}) and data hiding against locally constrained sets of measurements (Definition~\ref{locally constr}). This follows closely as originally done in the context of data hiding, which was initially defined against LOCC protocols in bipartite quantum systems~\cite{dh-original-1, dh-original-2}.
An important feature of locally constrained sets of measurements, which makes them suitable for the study of data hiding phenomena, is informational completeness. This is a consequence of the local tomography principle (Axiom~\ref{ax local tomography}), which we assumed to hold in general when we combine local GPTs to describe a bipartite system. We formulate this elementary observation as a follows.

\begin{lemma} \label{locally constr info complete}
All the locally constrained sets of measurements (Definition~\ref{locally constr}) are informationally complete according to Definition~\ref{info complete}.
\end{lemma}

\begin{proof}
Because of~\eqref{chain M}, we just need to show that $\text{LO}$ is informationally complete. In order to prove this elementary fact, consider a local measurement $(e_{i}\otimes f_{j})_{(i,j)\in I\times J}\in \text{LO}$ such that $\text{span}\{e_{i}\}_{i\in I}=V_{A}^{*}$ and $\text{span}\{f_{j}\}_{j\in J}=V_{B}^{*}$. Then, obviously, $\text{span}\{e_{i}\otimes f_{j}\}_{(i,j)\in I\times J}=V_{A}^{*}\otimes V_{B}^{*}=V_{AB}^*$, which yields the claim. Here, we made use of the dual of the identity~\eqref{tensor spaces}, which rests crucially on Axiom~\ref{ax local tomography}.
\end{proof}

Thanks to~\eqref{chain M} and Lemma~\ref{elem dh}, we also find
\bb
R(\text{SEP}) \leq R(\text{LOCC}) \leq R(\text{LOCC}_{\rightarrow}) \leq R(\text{LO})
\label{chain dh}
\ee
for all fixed GPTs. Our primary interest from now on is in data hiding against such operationally restricted sets of measurements. In particular, we want to understand how the best data hiding ratio \emph{scales} with the dimensions of the local GPTs. To be more precise, we give the following definition.

\begin{Def}[Ultimate data hiding ratio] \label{univ dh ratio}
For a locally constrained set of measurements $\mathcal{M}$, the \emph{ultimate data hiding ratio against $\mathcal{M}$} for fixed local dimensions, denoted by $R_{\mathcal{M}}(d_{A},d_{B})$, is the supremum over all data hiding ratios $R(\mathcal{M})$ achieved by composite GPTs that satisfy~\eqref{CAB bound} and have local dimensions $d_{A},d_{B}$.
\end{Def}

With this concept at hand, we are ready to formulate the question lying at the heart of our investigation in the present chapter, namely, \emph{what is the scaling of the ultimate ratio $R_\mathcal{M}(d_{A}, d_{B})$ with the local dimensions $d_{A},d_{B}$?} Clearly, thanks to the chain of inequalities~\eqref{chain dh}, we find immediately
\bb
R_{\text{SEP}} \leq R_{\text{LOCC}} \leq R_{\text{LOCC}_{\rightarrow}} \leq R_{\text{LO}}
\label{chain univ dh}
\ee
for all positive integers $d_A, d_B$ (the dependence on which of all the above ratios has been omitted for convenience).

We stress that the supremum in Definition~\ref{univ dh ratio} has to be taken over \emph{all} local GPTs of the given dimensions, and over all \emph{composition rules} adopted to join the system (i.e. among all the global cones respecting the bounds~\eqref{CAB bound}). Now, we will show that at least this latter maximisation can be carried out explicitly when $\mathcal{M}$ is a locally constrained set of measurements different from $\text{LOCC}$, the optimal composite being always given by the \emph{minimal} tensor product~\eqref{minimal}. To see why, notice that the exclusion of $\text{LOCC}$ implies that for a fixed $Z\in V_A\otimes V_B$ only the global base norm $\|Z\|$ depends on the composition rule we chose. On the contrary, the locally constrained norm $\|\cdot\|_{\mathcal{M}}$ ($\mathcal{M}\neq \text{LOCC}$) will depend on the local structure only. Then, maximising the ratio between the former and the latter amounts to maximise the global base norm. In order to do so, a large set of global effects and thus a small set of states are required, and according to~\eqref{CAB bound} the smallest possible positive cone in a bipartite system is given by the minimal tensor product.

The above reasoning is perhaps not obvious from Definition~\ref{dh} alone, because restricting the set of available bipartite states gives less freedom in choosing the data hiding pair. However, this restriction plays no role once Proposition~\ref{dh ratio} is available. This way around the problem is made possible by the fact that any difference of two normalised states can be thought of as a positive multiple of the difference of two separable states, the multiplication coefficient being given by the base norm induced by the minimal tensor product. We summarise this whole discussion stating the following.

\begin{prop} \label{min is optimal}
Given two local GPTs $A=(V_A,C_A,u_A)$, $B=(V_B,C_B,u_B)$, and a locally constrained set of measurements $\mathcal{M}\neq \text{\emph{LOCC}}$, the maximal data hiding ratio against $\mathcal{M}$ is achieved when the cone of bipartite states is given by the minimal tensor product, i.e. $C_{AB}=C_A\tminit C_B$.
\end{prop}

\begin{rem}
Some intuitive understanding of Proposition~\ref{min is optimal} can be gained by looking at the opposite case, i.e. when the composite system is formed via the maximal tensor product. When $AB=A \tmax  B$, in fact, the global base norm coincides with the separability norm. This is ensured by the fact that every allowed effect within this theory is automatically separable.
\end{rem}

Before proceeding further, let us reassure those readers that might be worried about the exclusion of the LOCC protocols from Proposition~\ref{min is optimal}. After all, there are operationally motivated reasons to consider LOCC as the most important among the locally constrained set of measurements. However, this exclusion will turn out to have no effect on our final result, Theorem~\ref{thm univ}. The reason why this is the case is that, for how difficult to characterise the LOCC set can be, we know that it obeys the two-sided inclusion bound~\eqref{chain M}, which translates to~\eqref{chain univ dh} in our case. Ultimately, this will allow us to capture some features of the behaviour of LOCCs without having to deal with them separately.

\section{Data hiding in quantum mechanics} \label{sec3 ex}

Throughout this section, we will apply the mathematical machinery we developed through the above Subsection~\ref{subsec3 dh statement} to the special case of finite-dimensional quantum mechanics, seen as a GPT as in Subsection~\ref{subsec2 ex QM}. The reason for doing this lies primarily in the outstanding physical importance of this example of GPT, which -- as far as we understand -- represents Nature as it happens to be.
However, from a purely mathematical point of view, one can see this also as an attempt to obtain a lower bound to the ultimate data hiding ratios by investigating a specific example.


\subsection{Quantum data hiding ratios} \label{subsec3 quantum dh ratios}

The purpose of this subsection is to determine the scaling of the data hiding ratios against locally constrained sets of measurements in the case of a composite quantum system $\text{QM}_{n_A n_B} = \text{QM}_{n_A} \otimes \text{QM}_{n_B}$, as defined in Subsection~\ref{subsec2 ex QM}. In particular, we will be interested in the quantity $R_{\text{QM}}(\text{LOCC})$ as a function of $n_A$ and $n_B$. Here, `determining the scaling' means `computing up to universal constants that do not depend on $n_A, n_B$'.

Let us start by recalling in brief how quantum mechanics fits the scheme of GPTs. For details, we refer the reader to Subsection~\ref{subsec2 ex QM}. The cone of states of an $n$-level quantum mechanical system, called $\text{PSD}_{n}$, comprises all positive semidefinite $n\times n$ matrices. The host vector space is that of real Hermitian matrices (denoted by $\mathcal{H}_{n}$), and its dimension is $d=n^2$. Density matrices are the positive matrices with trace one, and therefore the unit effect is nothing but the trace. Therefore, we write $\text{QM}_{n} = \left( \mathcal{H}_{n}, \text{PSD}_{n}, \Tr \right)$ as in~\eqref{quantum}.
The quantum mechanical base norm coincides with the trace norm $\|\cdot\|_{1}$.

The composition rule for bipartite systems is very simple. If $A=\text{QM}_{n_{A}}$ and $B=\text{QM}_{n_{B}}$, then $AB=\text{QM}_{n_{A}n_{B}}$. This corresponds to choosing a composite cone which is neither the minimal nor the maximal tensor product of the local cones~\eqref{cone bipartite quantum}. While this is clearly not optimal in the sense of data hiding because of Proposition~\ref{min is optimal}, it deserves special attention because of its prime importance in physics. In the forthcoming Subsection~\ref{subsec3 dh W}, we will also show that modifying the tensor product rule does not lead to any substantial improvement of the data hiding ratio of the theory.

The computation of data hiding ratios in quantum mechanics has been the subject of many papers, whose main results we are now about to summarise. The original example of a data hiding pair involves the normalised projectors onto the symmetric and antisymmetric subspace in $\mathds{C}^{n}\otimes\mathds{C}^{n}$, denoted by $\rho_{S}$ and $\rho_{A}$, respectively~\cite{dh-original-1, dh-original-2}. While $\|\rho_{S}-\rho_{A}\|_{1}=2$ because the two states have orthogonal support, it can be shown that $\|\rho_{S}-\rho_{A}\|_{\text{LOCC}}=2/(n+1)$~\cite{dh-original-1, dh-original-2}. The fact that the two states are mixed is crucial for this construction to work, as it can be shown that for pure states trace norm and LOCC norm always coincide~\cite{no-dh-pure-1, no-dh-pure-2}. In general, from~\cite{VV-dh-Chernoff, VV-dh} it is known that
\bb
\|\cdot\|_2\leq \sqrt{153} \|\cdot\|_{\text{LO}}\, ,
\label{VV bound HS}
\ee
where $\|Z\|_2\coloneqq \sqrt{\Tr Z^\dag Z}$ is the Hilbert-Schmidt norm. The above relation, whose proof is surprisingly intricate, is useful and interesting per se, and we will come to that later. Once translated in terms of the trace norm thanks to $\|\cdot\|_1\leq \sqrt{n_A n_B} \|\cdot\|_2$, it yields the upper bound $R_{\text{QM}}(\text{LO})\leq \sqrt{153\, n_A n_B}$. Combining this with the lower bound on $R_{\text{QM}}(\text{SEP})$ as described in~\cite{VV-dh}, one obtains 
\bb
\frac{\min\{n_A, n_B\}+1}{2} \leq R_{\text{QM}}(\text{SEP}) \leq R_{\text{QM}}(\text{LOCC}) \leq R_{\text{QM}}(\text{LO}) \leq \sqrt{153\, n_A n_B}\, ,
\label{VV bound 1}
\ee
while for $R_{\text{QM}}(\text{SEP})$ the tighter bound
\bb
\frac{\min\{n_A, n_B\}+1}{2} \leq R_{\text{QM}}(\text{SEP}) \leq \sqrt{n_A n_B}
\label{VV bound 2}
\ee
is available. As shown in~\cite[Corollary 17]{VV-dh}, the above relations solve the problem of determining the optimal scaling in $n$ of all the data hiding ratios against locally constrained measurements when the subsystems have equal dimensions $n_A=n_B=n$. On the contrary, a problem arises when $n_A$ and $n_B$ are very different and thus $\min\{n_A,n_B\}\ll \sqrt{n_A n_B}$. In this case, the leftmost and rightmost side of~\eqref{VV bound 1} are no longer of the same order of magnitude, and an alternative argument has to be designed.

This scaling problem is somehow mitigated by~\cite[Lemma 20]{Brandao-area-law}, which implies that
\bb
R_{\text{QM}}(\text{LOCC})\, \leq\, R_{\text{QM}} (\text{LO})\, \leq\, \min\{n_A^2,\, n_B^2\}\, .
\ee
Although this upper bound behaves better than that in~\eqref{VV bound 1} when $n_A$ and $n_B$ are very different from each other, its quadratic nature prevents us from determining -- for instance -- the exact scaling of the operationally relevant data hiding ratio against LOCC protocols.

Here we provide a simple reasoning that shows that in fact $O(\min\{n_A,n_B\})$ is still an upper bound for $R_{\text{QM}}(\text{LOCC})$ (and hence for $R_{\text{QM}}(\text{SEP})$, too). Furthermore, our reasoning yields much better constants for both the leftmost and the rightmost side of~\eqref{VV bound 1} (where $R_{\text{QM}}(\text{LO})$ is excluded, though). In fact, these constants are so close to being optimal that we are even able to compute $R_{\text{QM}}(\text{SEP})$ \emph{exactly} when $n_A=n_B=n$.

\begin{thm}[Teleportation argument] \label{dh QM}
For a bipartite quantum mechanical system with Hilbert space $\mathds{C}^{n_A}\otimes \mathds{C}^{n_B}$, define $n\coloneqq\min\{n_A,n_B\}$. Then the distinguishability norm against $\text{LOCC}_{\rightarrow}$ protocols satisfies
\bb
\|\cdot\|_1 \leq (2n-1) \|\cdot\|_{\text{\emph{LOCC}}_{\rightarrow}}\, .
\label{bound telep}
\ee
In particular, the data hiding ratios are such that
\bb
n \leq R_{\text{\emph{QM}}}(\text{\emph{SEP}}) \leq R_{\text{\emph{QM}}}(\text{\emph{LOCC}}) \leq R_{\text{\emph{QM}}}(\text{\emph{LOCC}}_{\rightarrow}) \leq 2n-1\, ,
\label{dh QM eq}
\ee
where the communication direction in $\text{\emph{LOCC}}_{\rightarrow}$ is from the smaller to the larger subsystem. Moreover, if $n_A=n_B=n$ then $R_{\text{\emph{QM}}}(\text{\emph{SEP}})=n$.
\end{thm}

\begin{proof}
Let us start by proving the upper bound on $R_{\text{QM}}(\text{LOCC}_\rightarrow)$ in~\eqref{dh QM eq}.
We can assume without loss of generality that $n=n_{A}\leq n_{B}$, and that the classical communication line goes from $A$ to $B$.
We now have to show that $\|Z_{AB}\|_1\leq (2n-1)\|\cdot\|_{\text{LOCC}_\rightarrow}$ holds for all Hermitian matrices $Z_{AB}$ of size $n_An_B\times n_A n_B$. We remind the reader that the maximally entangled state $\ket{\Phi}=\frac{1}{\sqrt{n}}\,\sum_{i=1}^{n} \ket{ii}\in\mathds{C}^{n}\otimes \mathds{C}^{n}$, whose corresponding rank-one projector we denote by $\Phi$, has the property that there is a separable state $\sigma$ such that $\frac1n \Phi+\frac{n-1}{n}\, \sigma$ is again separable (in the language of~\cite{VidalTarrach}, $\Phi$ has entanglement robustness $r(\Phi)=n-1$). For instance, using classic results on the entanglement of isotropic states~\cite{Horodecki1999}, it is not difficult to see that $\sigma = \frac{\mathds{1} - \Phi}{n^{2}-1}$ satisfies all the requirements.

Now, since we can always produce any separable state with $\text{LOCC}_{\rightarrow}$ operations, we are free to evaluate the $\text{LOCC}_{\rightarrow}$ norm on $Z_{AB} \otimes \left( \frac1n \Phi+\frac{n-1}{n}\, \sigma \right)_{A'B'}$ instead of $Z_{AB}$. Here, the systems $A',B'$ have dimension $n_{A'}=n_{B'}=n_A=n$, and the operations are $\text{LOCC}_{\rightarrow}$ with respect to the splitting $AA':BB'$.
We are ready to apply the quantum teleportation protocol from $A$ to $B$~\cite{teleportation}. This is an $\text{LOCC}_{\rightarrow}$ operation $\tau$ mapping states of the system $AA'BB'$ to states of $B'B$, which can be defined as follows. For $p,q=0,\ldots,n-1$, introduce the unitary matrices
\bb
\begin{split}
\mathbf{X}(p) &\coloneqq \sum_{k=1}^n \ket{k\oplus p}\!\!\bra{k}\, ,\\
\mathbf{Z}(q) &\coloneqq \sum_{k=1}^n e^{2 q k \pi i/n} \ket{k}\!\!\bra{k}\, ,\\
U(p,q) &\coloneqq \mathbf{X}(p) \mathbf{Z}(q)\, ,
\end{split}
\label{HW}
\ee
where $\oplus$ denotes sum modulo $n$. Incidentally, these are easily seen to be suitable generalisations of the Pauli matrices from qubit to qudit systems.
Then the teleportation $\tau$ is given by
\bb
\begin{split}
&\tau(Z_{AA'BB'}) \\
&\quad \coloneqq \sum_{p,q=0}^{n-1} U(p,q)_{B'}\ \text{Tr}_{AA'}\,\big[ Z_{AA'BB'}\, U(p,q)_A \Phi_{AA'} U(p,q)_A^\dag \big]\ U(p,q)_{B'}^{\dag}\, .
\end{split}
\label{telep}
\ee
Most notably, observe that $\tau\left(Z_{AB} \otimes \Phi_{A'B'}\right)=Z_{B'B}$ (meaning that the same operator $Z$ is written in the registers $B'\simeq A$ and $B$ instead of $A$ and $B$). Now, on the one hand, after the protocol has been performed, the local constraint plays no role any more, and any desired measurement can be applied to $B'B$, showing that $\|Z_{B'B}\|_{\text{LOCC}_{\rightarrow}} = \|Z_{AB}\|_{1}$. On the other hand, $\tau(Z_{AB}\otimes \sigma_{A'B'})$ is obtained from $Z_{AB}$ via an $\text{LOCC}_{\rightarrow}$ protocol, hence $\|\tau (Z_{AB} \otimes \sigma_{A'B'})\|_{\text{LOCC}_{\rightarrow}}\leq\|Z_{AB}\|_{\text{LOCC}_{\rightarrow}}$.
Putting all together, we obtain the following chain of inequalities:
\begin{align}
\|Z_{AB}\|_{\text{LOCC}_{\rightarrow}} &= \left\| Z_{AB} \otimes \left( \frac1n \Phi+\frac{n-1}{n}\, \sigma \right)_{A'B'} \right\|_{\text{LOCC}_{\rightarrow}}
\label{telep 1} \\[0.5ex]
&\geq \left\| \tau\left(Z_{AB} \otimes \left( \frac1n \Phi+\frac{n-1}{n}\, \sigma \right)_{A'B'}\right) \right\|_{\text{LOCC}_{\rightarrow}} \label{telep 2} 
\end{align}

\begin{align}
\quad \quad &= \left\| \frac1n\, Z_{B'B} +\frac{n-1}{n}\, \tau(Z_{AB}\otimes \sigma_{A'B'}) \right\|_{\text{LOCC}_{\rightarrow}} \label{telep 3} \\
&\geq \frac1n\, \|Z_{B'B}\|_{\text{LOCC}_{\rightarrow}} - \frac{n-1}{n}\, \|\tau(Z_{AB}\otimes \sigma_{A'B'}) \|_{\text{LOCC}_{\rightarrow}} \label{telep 4} \\[0.5ex]
&\geq \frac1n\, \|Z_{AB}\|_{1} - \frac{n-1}{n}\, \|Z_{AB} \|_{\text{LOCC}_{\rightarrow}}\, . \label{telep 5}
\end{align}
We conclude that
\bbb
\|Z_{AB}\|_{\text{LOCC}_{\rightarrow}}\, \geq\,  \frac{1}{2n-1}\, \left\| Z_{AB} \right\|_1\, , 
\eee
enforcing $R_{\text{QM}}(\text{LOCC}_{\rightarrow})\leq 2n-1$ in view of Proposition~\ref{dh ratio}.

In order to deduce the lower bound $R_{\text{QM}}(\text{SEP})$, we appeal to Werner states~\cite{Werner, Werner-symmetry}. These can be thought of as convex combinations of the normalised projectors onto the symmetric and antisymmetric subspace of $\mathds{C}^{n}\otimes \mathds{C}^{n}$, denoted by $\rho_{S}$ and $\rho_{A}$, respectively. In terms of the `flip operator' $F$ defined by $F\ket{\alpha\beta} = \ket{\beta\alpha}$ for all $\ket{\alpha},\ket{\beta}\in\mathds{C}^{n}$, we have
\begin{equation}
\rho_{S} = \frac{\mathds{1}+F}{n(n+1)}\, ,\qquad \rho_{A} = \frac{\mathds{1}-F}{n(n-1)}\, .
\label{symm antisymm proj}
\end{equation}
Since $n=\min\{n_{A},n_{B}\}$, we can safely imagine to give one share of this bipartite system to $A$ and the other to $B$. We already saw how the preparation with equal a priori probabilities of the two extremal states is well-known to produce data hiding, as shown by the fact that $\|\rho_{S}-\rho_{A}\|_{1}=2$ but $\|\rho_{S}-\rho_{A}\|_{\text{SEP}}=\|\rho_{S}-\rho_{A}\|_{\text{LOCC}}=2/(n+1)$~\cite{VV-dh, VV-dh-Chernoff}. Curiously, there is an optimised version of this construction with different weights that does not seem to have been considered before. Namely, via the same techniques it can be shown that
\bb
\begin{split}
&\left\| \frac{n+1}{n}\, \rho_{S} - \frac{n-1}{n}\, \rho_{A} \right\|_{1} = 2\, ,\\[0.5ex]
&\left\| \frac{n+1}{n}\, \rho_{S} - \frac{n-1}{n}\, \rho_{A} \right\|_{\text{SEP}} = \frac{2}{n}\, .
\end{split}
\label{bound Werner}
\ee
Observe that $\frac{n+1}{n}\, \rho_{S} - \frac{n-1}{n}\, \rho_{A}=\frac{2}{n^2} F$. Since the proof of~\eqref{bound Werner} is just a variation of a standard calculation, we relegate it to Appendix~\ref{app Werner}. Thanks to Proposition~\ref{dh ratio}, this yields the lower bound in the claim. Finally, combining $R_{\text{QM}}(\text{SEP})\geq n$ with the upper bound in~\eqref{VV bound 2}, we see that when $n_{A}=n_{B}=n$ we must have $R_{\text{QM}}(\text{SEP})=n$.
\end{proof}

\begin{rem}
The fact that the upper bound for $R_{\text{QM}}(\text{LOCC})$ in Theorem~\ref{dh QM} scales only linearly in $\min\{n_{A},n_{B}\}$ is crucial in solving the data hiding problem in quantum mechanics (up to constants) for all pairs $(n_{A}, n_{B})$. To our knowledge, this complete solution was not known before. We find the simplicity of the above proof quite instructive on its own, but it does not seem like the teleportation argument can encompass the case of purely local measurements, by its very nature. Therefore, we must leave open the problem of finding the optimal scaling of $R_{\text{QM}}(\text{LO})$ in the general case.
\end{rem}

\begin{rem}
We do not want to give the reader the wrong impression that the inequality~\eqref{VV bound HS} found in~\cite{VV-dh} is now useless, since what you deduce from it concerning data hiding ratios, i.e.~\eqref{VV bound 1}, is a looser bound than our~\eqref{dh QM eq}. On the contrary,~\eqref{VV bound HS} has to be thought as giving some kind of complementary information. First, it tells us that in order to achieve high data hiding efficiency we need to engineer high-rank states, something not apparent from~\eqref{bound telep}. Second, it involves the norm $\|\cdot\|_{\text{LO}}$ instead of the larger $\|\cdot\|_{\text{LOCC}_\rightarrow}$ that appears in~\eqref{bound telep}. Finally,~\eqref{VV bound HS} can be generalised to the multipartite case. While the same is true for our~\eqref{bound telep}, in this more general case these two approaches can be shown to lead to incomparable results concerning the ratio $\|\cdot\|_1 / \|\cdot\|_{\text{LOCC}}$.
\end{rem}

\begin{rem}
In terms of the real dimensions of the local spaces, which are given by $d_A=n_A^2,\, d_B=n_B^2$ according to~\eqref{dim quantum}, Theorem~\ref{dh QM} shows that the data hiding ratio against separable protocols scales as $\min\{\sqrt{d_{A}},\,\sqrt{d_{B}} \}$. Thus, we deduce a first estimate $R_{\text{SEP}}(d_{A},d_{B})\geq \min\{\sqrt{d_{A}},\,\sqrt{d_{B}}\}$ (valid when $\sqrt{d_{A}},\sqrt{d_{B}}$ are integers).
\end{rem}

\subsection{Data hiding in $W$-theory} \label{subsec3 dh W}

In view of Proposition~\ref{min is optimal}, the reader might wonder, whether considering a modified version of quantum mechanics in which composite systems are obtained via the minimal tensor product~\eqref{separable} exhibits better data hiding properties. We already encountered such a modified version of standard quantum mechanics, and dubbed it $W$-theory (Definition~\ref{W theory}). Remember that the base norm of such theory takes the form~\eqref{W norm}. Perhaps surprisingly, it turns out that the answer to the aforementioned question is substantially negative, meaning that the improvement in the highest data hiding efficiency is bounded by a constant. 

In order to arrive at this conclusion, we need to investigate a bit more the concept of LOCC protocols in $W$-theory. Here, by protocol we mean a state transformation that is achieved by combining elementary operations, possibly subjected to operational constraints, and possibly probabilistic. In the language of Subsection~\ref{subsec2 measur bip}, we could also call these `instruments'. As remarked there, this has not to be confused with the concept of measurement. The outcome of a measurement is a classical label, while the outcome of a protocol or instrument is a state (possibly together with a classical label).

Let us then consider the class of protocols that can be realised in the $W$-theory framework when two agents $A$ and $B$ are allowed to use just local operations and $A\rightarrow B$ classical communication. We denote this class of protocols by $\text{\textbf{LWCC }}_\rightarrow$. Despite the fact that $\text{\textbf{LWCC }}_\rightarrow$ constitutes a more general class of \emph{protocols} than $\text{LOCC}_\rightarrow$ in standard quantum mechanics, it is not difficult to see that they are not more powerful than the latter within the context of state discrimination. This is a consequence of the fact that the \emph{measurements} one can implement with $\text{\textbf{LWCC}}_\rightarrow$ operations are -- up to coarse graining -- necessarily of the form $\big(E_i\otimes F_j^{(i)}\big)_{(i,j)\in I\times J}$ for some local measurements $(E_i)_{i\in I}$ on $A$ and $\big( F^{(i)}_j\big)_{j\in J}$ on $B$. Since local measurements in $W$-theory are the same as in standard quantum mechanics, the same measurement is also obtainable via $\text{LOCC}_\rightarrow$ operations. More generally, we saw already that the locally constrained sets of measurements defined in~\eqref{LO},~\eqref{1-way LOCC},~\eqref{SEP} do not depend on the composition rule we choose for assembling multipartite systems.

The above discussion allows us to write the identity
\bb
\|\cdot\|_{\text{LOCC}_{\rightarrow}} = \|\cdot\|_{\text{\textbf{LWCC}}_{\rightarrow}}\, .
\label{confusing}
\ee
Here, the right-hand side can be thought of as defined via the minimal error probability according to~\eqref{pr error}, where said error probability has to be achieved by $A$ and $B$ running a $\text{\textbf{LWCC}}_\rightarrow$ protocol followed by a classical decision.
Of course, one can also look at it as defined by the usual formula~\eqref{d norm}, where it is understood that the corresponding set $\mathcal{M}$ includes in this case all those measurements that are implementable through an $\text{\textbf{LWCC}}_{\rightarrow}$ protocol.



As it turns out, data hiding in $W$-theory is not much better than in quantum theory, in the sense that the scaling with the local dimensions is exactly the same.
The proof of this latter result constitutes another example of how the techniques used in~\cite{VV-dh-Chernoff, VV-dh} seem not to be applicable in a more general scenario. In fact, the approach taken there, based on~\eqref{VV bound HS}, would start with the inequality $\|\cdot\|_{\text{LOCC}_{\rightarrow}}\geq \kappa \|\cdot\|_2$, for $\kappa$ universal constant and $\|\cdot\|_2$ the Hilbert-Schmidt norm. As we saw, when the system under examination is $\mathds{C}^n\otimes \mathds{C}^n$ the elementary relation $\|\cdot\|_2\leq \frac{1}{n} \|\cdot\|_1$ yields $\|\cdot\|_{\text{LOCC}_{\rightarrow}}\geq \frac{\kappa}{n} \|\cdot\|_1$, which is optimal up to a constant factor. However, it is not difficult to prove that the ratio between the norm $\|\cdot\|_W$ defined in~\eqref{W norm} and the Hilbert-Schmidt norm $\|\cdot\|_2$ can be asymptotically as large as $n^{3/2}$ (see Appendix~\ref{app W norm}). Therefore, the tighter inequality we can deduce by making use of the Hilbert-Schmidt norm in an intermediate step is $\|\cdot\|_{\text{LOCC}_\rightarrow}\geq \kappa \|\cdot\|_2 \geq \frac{\kappa}{n^{3/2}} \|\cdot\|_W$. As we will see in a moment, the scaling of the lower bound is not tight.

While a direct approach via the other techniques previously exploited in the literature does not lead to a satisfactory answer to the problem, the teleportation argument can be quickly adapted to compute \emph{exactly} the data hiding ratios against $\text{SEP}$ or $\text{LOCC}_{\rightarrow}$ in $W$-theory. 

\begin{prop} \label{dh W}
For a bipartite $W$-theory with local Hilbert spaces of dimensions $n_{A},n_{B}$, define $n\coloneqq\min\{n_A,n_B\}$. Then the data hiding ratios against separable and $\text{\emph{LOCC}}_{\rightarrow}$ measurements are given by
\bb
R_{W}(\text{\emph{SEP}})\, =\, R_{W}(\text{\emph{LOCC}}_{\rightarrow})\, =\, 2n-1\, ,
\ee
where the communication direction in $\text{\emph{LOCC}}_{\rightarrow}$ is from the smaller to the larger subsystem.
\end{prop}

\begin{proof}
We assume without loss of generality that $n_B\geq n_A=n$. Let us start by showing that the above data hiding ratios can be upper bounded by $2n-1$. The idea is that the argument in~\eqref{telep 1}-\eqref{telep 5} can be adapted to encompass also the case of $W$-theory, by replacing everywhere $\|\cdot\|_{\text{LOCC}_{\rightarrow}}$ with $\|\cdot\|_{\text{LWCC}_{\rightarrow}}$ and the trace norm $\|\cdot\|_{1}$ with the minimal tensor product base norm~\eqref{W norm}.
The first step consists in acknowledging the fact that we can compute the $\text{LOCC}_{\rightarrow}$ distinguishability norm by making use of more general $\text{\textbf{LWCC}}_{\rightarrow}$ protocols that are available in $W$-theory, as expressed in~\eqref{confusing}.

Now, we choose a particular $\text{\textbf{LWCC}}_\rightarrow$ protocol in order to lower bound the norm $\|Z_{AB}\|_{\text{\textbf{LWCC}}_{\rightarrow}}$. Such a protocol resembles the one we devised for the proof of Theorem~\ref{dh QM}, with one important difference.

Since separable states can be created with local operations and shared randomness, we can safely start by supplying $A$ and $B$ with a separable isotropic state $\left( \frac1n \Phi+\frac{n-1}{n}\, \sigma \right)_{A'B'}$, defined on an ancillary system $A'B'$ with local dimension $n_{A'}=n_{B'}=n$. As usual, $\sigma$ is an appropriate separable state. Then, we perform the teleportation $\tau$ defined in~\eqref{telep}, which is an $\text{\textbf{LWCC}}_\rightarrow$ (even $\text{LOCC}_\rightarrow$) operation with respect to the splitting $AA':BB'$, where classical communication goes from $AA'$ to $BB'$. After applying the triangle inequality, we are left with two terms, i.e. $\|Z_{B'B}\|_{\text{\textbf{LWCC}}_\rightarrow}$ and $\|\tau(Z_{AB}\otimes \sigma_{A'B'})\|_{\text{\textbf{LWCC}}_\rightarrow}$. The first one can be computed exactly, since the operator $Z_{B'B}$ obtained after teleportation belongs to the local subsystem $BB'$, and therefore measuring any witness $H_{B'B}$ is an allowed $\text{\textbf{LWCC}}_\rightarrow$ operation, leading to the equality $\|Z_{B'B}\|_{\text{\textbf{LWCC}}_\rightarrow}=\|Z\|_W$. As for the second term, we observe that $\|\tau(Z_{AB}\otimes \sigma_{A'B'})\|_{\text{\textbf{LWCC}}_\rightarrow}\leq \|Z\|_{\text{\textbf{LWCC}}_\rightarrow}$, since adding the ancillary system $A'B'$ in a separable state $\sigma_{A'B'}$ and subsequently applying $\tau$ is clearly an $\text{\textbf{LWCC}}_\rightarrow$ protocol (which is why this latter inequality is in fact an equality). The above reasoning can be summarised in the following chain of inequalities, totally analogous to~\eqref{telep 1}-\eqref{telep 5}:
\begin{align*}
\|Z_{AB}\|_{\text{LOCC}_\rightarrow} &= \|Z_{AB}\|_{\text{\textbf{LWCC}}_\rightarrow} \\[0.5ex]
&= \left\| Z_{AB} \otimes \left( \frac1n \Phi +\frac{n-1}{n}\, \sigma \right)_{A'B'} \right\|_{\text{\textbf{LWCC}}_\rightarrow} \\[0.5ex]
&\geq \left\| \tau\left(Z_{AB} \otimes \left( \frac1n \Phi +\frac{n-1}{n}\, \sigma \right)_{A'B'}\right) \right\|_{\text{\textbf{LWCC}}_\rightarrow} \\[0.5ex]
&= \left\| \frac1n\, Z_{B'B} +\frac{n-1}{n}\, \tau(Z_{AB}\otimes \sigma_{A'B'}) \right\|_{\text{\textbf{LWCC}}_\rightarrow} \\
&\geq \frac1n\, \|Z_{B'B}\|_{\text{\textbf{LWCC}}_\rightarrow} - \frac{n-1}{n}\, \|\tau(Z_{AB}\otimes \sigma_{A'B'}) \|_{\text{\textbf{LWCC}}_\rightarrow} \\[0.5ex]
&\geq \frac1n\, \|Z_{AB}\|_{W} - \frac{n-1}{n}\, \|Z_{AB} \|_{\text{\textbf{LWCC}}_\rightarrow} \\[0.5ex]
&= \frac1n\, \|Z_{AB}\|_{W} - \frac{n-1}{n}\, \|Z_{AB} \|_{\text{LOCC}_\rightarrow}\, .
\end{align*}
In conclusion, we find
\bb
\|Z_{AB}\|_{\text{LOCC}_{\rightarrow}}\, =\, \|Z_{AB}\|_{\text{LWCC}_{\rightarrow}}\, \geq\,  \frac{1}{2n-1}\, \left\| Z_{AB} \right\|_W\, , \label{bound telep W}
\ee
which implies $R_{W}(\text{LOCC}_{\rightarrow})\, \leq\, 2n-1$. Once more, to derive a lower bound on $R_{W}(\text{SEP})$ we use Werner states~\cite{Werner, Werner-symmetry}. With the same notation as in the proof of Theorem~\ref{dh QM}, it can be shown that
\bb
\begin{split}
&\left\| \frac{n+1}{2n-1}\, \rho_{S}\, -\, \frac{n-1}{2n-1}\, \rho_{A} \right\|_{W} = 2\, , \\[0.5ex]
&\left\| \frac{n+1}{2n-1}\, \rho_{S}\, -\, \frac{n-1}{2n-1}\, \rho_{A} \right\|_{\text{SEP}} = \frac{2}{2n-1}\, , 
\end{split}
\label{bound Werner W}
\ee
enforcing the complementary bound $R_{W}(\text{SEP}) \geq 2n-1$. The proof of~\eqref{bound Werner W} is provided in Appendix~\ref{app Werner}.
\end{proof}

\subsection{Further implications of the teleportation argument} \label{subsec3 telep arg}

This subsection is devoted to investigating some implications of the bound~\eqref{bound telep} we proved thanks to the teleportation argument (Theorem~\ref{dh QM}). As it turns out, the fact that~\eqref{bound telep} depends only linearly on the minimal local dimension has many applications, particularly in improving previously known results. It is our purpose here to examine some of these improvements in detail, although a more thorough analysis will be conducted elsewhere.\footnote{Most of what I will say here came out of a series of very instructive conversations with Matthias Christandl, whom I thank wholeheartedly.}

\subsubsection{Lower bounds on squashed entanglement}

Among all the many entanglement measures that have been explored in the vast literature on the subject, one of the most mathematically appealing is perhaps \textbf{squashed entanglement}~\cite{Tucci1999, squashed}. For a bipartite quantum state $\rho_{AB}$, it is defined via the equation
\bb
E_{\text{sq}} (A:B)_\rho \coloneqq \inf_{\rho_{ABC}} \frac12 I(A:B|C)_\rho\, , 
\label{squashed eq}
\ee
where: (i) the infimum runs over all possible \emph{extensions} $\rho_{ABC}$ of $\rho_{AB}$, i.e. over all possible ancillary quantum systems $C$ and over all possible density matrices $\rho_{ABC}$ such that $\text{Tr}_C\, \rho_{ABC} = \rho_{AB}$; (ii) the \textbf{quantum conditional mutual information} $I(A:B|C)_\rho$ is defined via
\bb
I(A:B|C)_\rho \coloneqq S(\rho_{AC}) + S(\rho_{BC}) - S(\rho_{ABC}) - S(\rho_C)\, ,
\label{cond mutual info}
\ee
and
\bb
S(\rho) \coloneqq - \Tr [\rho \log \rho]
\label{entropy}
\ee
is the celebrated \textbf{von Neumann entropy}. Here, $\log$ stands for the logarithms to base $2$. It is a classic result of quantum information theory that the conditional mutual information~\eqref{cond mutual info} is always non-negative in the quantum case as well, the corresponding inequality being known as \textbf{strong subadditivity}~\cite{lieb73a, lieb73b, lieb73c}. Consequently, the squashed entanglement of any state is also non-negative.
The fact that it is a legitimate entanglement measure, in the sense that it can not be increased by LOCC and it vanishes on separable states, was proved already in~\cite{squashed}. However, it took few years to show that the squashed entanglement is also \emph{faithful}, in the sense that it vanishes \emph{only} on separable states. This fact was finally proved in~\cite{faithful} by means of the inequality~\cite[Eq. (9) and (12)]{faithful}
\bb
I(A:B|C)_\rho \geq 2E_{\text{sq}}(A:B)_\rho \geq \frac{1}{8 \ln 2} \left\| \rho_{AB} - \mathcal{S}_{A:B} \right\|_{\text{LOCC}_\rightarrow}^2\, ,
\label{faithful bound}
\ee
where
\bbb
\left\| \rho_{AB} - \mathcal{S}_{A:B} \right\| \coloneqq \inf_{\sigma\in \mathcal{S}_{A:B}} \left\| \rho_{AB} - \sigma_{AB} \right\|
\eee
is the `distance' of $\rho_{AB}$ from the set $\mathcal{S}_{A:B}$ of normalised separable states of the bipartite system $AB$, as measured in an arbitrary norm $\|\cdot\|$. Observe that since the squashed entanglement is invariant under the exchange of $A$ and $B$, there is no need to specify the direction of the one-way classical communication (as long as it is fixed from the start). Clearly, since $\|\cdot\|_{\text{LOCC}_\rightarrow}$ is a norm and not only a seminorm,~\eqref{faithful bound} shows that a vanishing squashed entanglement ensures that the state under examination is separable.

Later on, in~\cite[p.4]{rel-ent-sq} the bound~\eqref{faithful bound} was improved by a factor $4$ to
\bb
I(A:B|C)_\rho \geq 2E_{\text{sq}}(A:B)_\rho \geq \frac{1}{2 \ln 2} \left\| \rho_{AB} - \mathcal{S}_{A:B} \right\|_{\text{LOCC}_\rightarrow}^2
\label{rel ent sq bound}
\ee
as a by-product of the approach put forward in~\cite{rel-ent-sq}.

In~\cite{faithful}, the lower bound~\eqref{faithful bound} is employed in conjunction with~\eqref{VV bound HS} to obtain relations like
\begin{align}
I(A:B|C)_\rho &\geq 2E_{\text{sq}}(A:B)_\rho \\
&\geq \frac{1}{306 \ln 2} \left\| \rho_{AB} - \mathcal{S}_{A:B} \right\|_{2}^2 \label{faithful bound 2} \\
&\geq \frac{1}{306 \ln 2\, n_A n_B} \left\| \rho_{AB} - \mathcal{S}_{A:B} \right\|_1^2\, , \label{faithful bound 3}
\end{align}
where we made use of the better constant in~\eqref{rel ent sq bound}.
A different strategy to tackle the problem of finding lower bounds on the squashed entanglement was pursued in~\cite{VV-Markov}, where results of~\cite{Fawzi-Renner} are exploited in order to obtain the inequality~\cite[Corollary 2]{VV-Markov}
\bb
\left\| \rho_{AB} - \mathcal{S}_{A:B} \right\|_{1} \leq 3.1\, n\, E_{\text{sq}}^{1/4}(A:B)_\rho\, ,
\label{VV schifo inequality}
\ee
where $n\coloneqq \min\{n_A, n_B\}$, as usual.
However, now that we have~\eqref{bound telep}, we can employ it to obtain the best lower bound on the squashed entanglement in terms of the trace norm that is currently available. 

\begin{cor} \label{lower sq cor}
For all states $\rho_{AB}$ of a bipartite $n_A\times n_B$ quantum system, one has
\bb
I(A:B|C)_\rho \geq 2E_{\text{sq}}(A:B)_\rho \geq \frac{1}{2 \ln 2 (2n-1)^2} \left\| \rho_{AB} - \mathcal{S}_{A:B} \right\|_{1}^2\, ,
\label{lower sq}
\ee
where $n\coloneqq \min\{n_A, n_B\}$, or equivalently
\bb
\begin{split}
\left\| \rho_{AB} - \mathcal{S}_{A:B} \right\|_{1} &\leq 4 \sqrt{\ln 2}\, ( n -1/2 ) \sqrt{E_{\text{sq}(A:B)_\rho}} \\
&\leq 3.34 \, ( n - 1/2) \sqrt{E_{\text{sq}(A:B)_\rho}} \, .
\end{split}
\label{lower sq bis}
\ee
\end{cor}

\begin{proof}
The bound~\eqref{lower sq} is found by chaining~\eqref{rel ent sq bound} with~\eqref{bound telep}, while~\eqref{lower sq bis} follows after straightforward algebraic manipulations.
\end{proof}

The lower bound on the squashed entanglement in Corollary~\ref{lower sq cor} combines the advantages of: (i) what you get from~\cite[Eq. (12)]{faithful} by using the results of~\cite{VV-dh}, here reported as~\eqref{faithful bound 3}; and (ii)~\cite[Corollary 2]{VV-Markov}, here reported as~\eqref{VV schifo inequality}. Namely, the trace-norm dependence is quadratic as in the former (and not quartic as in the latter), but the constant depends only on the minimal dimension as in the latter (and not on both local dimensions as in the former).

\subsubsection{A modified quantum de Finetti theorem}

We already encountered the complete extendibility criterion at the end of Subsection~\ref{subsec2 Woronowicz}. In brief, it states that any separable state $\rho_{AB}$ admits a \emph{symmetric $k$-extension} for all positive integers $k$. This is a state $\rho_{AB_1\ldots B_k}$ of a quantum system $AB_1\ldots B_k$ (here the $B_i$ are all isomorphic to $B$) which is: (i) symmetric under the exchange of any two $B$ systems; and (ii) an extension of $\rho_{AB}$, in the sense that $\text{Tr}_{B_2\ldots B_k}\, \rho_{AB_1\ldots B_k}$ is nothing but the original state $\rho$ (written in the registers $AB_1$). 
We also saw that a bipartite state that is $k$-extendible for all $k$, also called completely extendible, is necessarily separable~\cite{complete-extendibility}. Of course, checking infinitely many values of $k$ is not practically feasible, hence an approximate version of this latter statement is needed.
Greatly simplifying the matter, we could say that a \emph{quantum de Finetti theorem} serves this purpose, for it provides an estimate of how \emph{close} a $k$-extendible state must be from the separable set. Delving into the details of the rich theory of quantum de Finetti theorems is beyond the scope of this thesis. We therefore limit ourselves to pointing the interested reader to some of the many papers dedicated to the subject~\cite{deFinetti0, deFinetti1, deFinetti2, 1-1/2-de-Finetti, faithful, deFinetti4}.\footnote{The list is by no means complete nor representative.} A good introduction can also be found in~\cite[\S 9]{math-ent}. The theorem is named after Bruno de Finetti, who proved an earlier (classical) version of the statement in 1937~\cite{deFinetti}.

Our goal here is much less ambitious. We follow~\cite{faithful} closely and rely heavily on the results presented there to show the following. 

\begin{cor} \label{de Finetti cor}
Let $\rho_{AB}$ be a quantum state of a bipartite $n_A\times n_B$ quantum system that is $k$-extendible on the $B$ side. Calling $n\coloneqq \min\{n_A, n_B\}$, we have
\bb
\|\rho_{AB} -\mathcal{S}_{A:B}\|_1 \leq 4\sqrt{\ln 2}\, (n-1/2) \sqrt{\frac{\log n}{k}}\, .
\label{bound k ext}
\ee
\end{cor}

\begin{proof}
It suffices to combine~\cite[Corollary 2]{faithful} (with the better constant found in~\eqref{rel ent sq bound}) with our estimate~\eqref{bound telep}. Alternatively, one can resort directly to~\eqref{lower sq bis} and to the observation that the squashed entanglement of any $k$-extendible state is no larger than $\frac{\log n}{k}$~\cite{faithful}. 
\end{proof}

An interesting feature of the right hand side of~\eqref{bound k ext} is that it stays finite when the dimension of one of the two subsystems diverges.

\section{Ultimate bound on data hiding effectiveness} \label{sec3 univ}

Throughout this section, we will determine the ultimate data hiding ratios of Definition~\ref{univ dh ratio} up to an additive constant, thus proving the main result of this chapter, Theorem~\ref{thm univ}. In order to do this, we need to: (i) find an explicit example of two GPTs that combined achieve a high data hiding ratio (Subsection~\ref{subsec3 sph}); and (ii) show that this is the largest ratio that is compatible with our assumptions on the composition of local theories (Subsection~\ref{subsec3 ultimate}).
As discussed at the end of Section~\ref{subsec3 original}, this answers the central question of our investigation, that is, the determination of the general constraints that data hiding is forced to obey in any GPTs.

\subsection{Data hiding in spherical models} \label{subsec3 sph}

As it turns out, quantum mechanical data hiding, whose maximum efficiency scales as $\min \left\{\sqrt{d_A}, \sqrt{d_B}\right\}$, is not the strongest possible within the realm of GPTs. Instead, throughout this subsection we will show that the spherical model, as introduced at the end of Subsection~\ref{subsec2 centr symm} (see for instance~\eqref{spherical}), exhibits a data hiding ratio against separable operations that is as large as $\min \{d_A, d_B\}$, thus quadratically larger than that in quantum mechanics.

Let us start with a quick recap of the notation we adopt for spherical models. For more details, we refer the reader to Subsection~\ref{subsec2 centr symm}.
The arena is as usual the vector space $\mathds{R}^d$. Vectors are denoted by $x=(x_0,x_1,\ldots, x_{d-1})^T$, and since $u\coloneqq (1,0,\ldots,0)^T$, the $0$-th component represents the normalisation of $x$. The space can be thought of as $\mathds{R}^d = \mathds{R} u_* \oplus\mathds{R}^{d-1}$, where $u_* = (1,0,\ldots, 0)^T$ is a distinguished normalised state, and the vector component $\mathds{R}^{d-1}$ is equipped with the Euclidean norm $|\cdot |_2$.
A vector $v\in\mathds{R}^{d-1}$ can be turned into an element of the global space by simply adding a zero as the first component, which we denote by $\myhat{v} = (0,v) \in \mathds{R}^d$. Conversely, we can extract from any $x = (x_0, \widebar{x})\in\mathds{R}^d$ its vector component $\widebar{x}\in\mathds{R}^{d-1}$.
With this notation, the cone of unnormalised states takes the form $C_d=\{(x_0, \widebar{x}):\, x_0\geq |\widebar{x}|_2\}$. Remarkably, with the canonical identification $\mathds{R}^d \simeq \big(\mathds{R}^d\big)^*$, we have $C_d\simeq C_d^*$, i.e. the spherical model is (strongly) self-dual.

Consider now two spherical models of dimensions $d_A$ and $d_B$. States of the bipartite system $AB$ are naturally identifiable with matrices $Z$ of size $d_A \times d_B$. The tensor product of the two vector components of the local spaces, that is, $\mathds{R}^{d_B-1}\otimes \mathds{R}^{d_B-1}$, is represented by the $(d_A-1)\times (d_B-1)$ bottom right corner of $Z$. Retaining only this corner of $Z$ yields another matrix that we denote by $\widebar{Z} \in \mathds{R}^{(d_A-1) \times (d_B-1)}$. As usual, one can do the opposite as well, and build a $d_A\times d_B$ matrix out of a $(d_A-1)\times (d_B-1)$ matrix $M$ by adding zeros on the first row and column. Explicitly,
\bbb
\myhat{M}_{ij} = \left\{ \begin{array}{cl} 0 & \text{ if $i=0$ or $j=0$,} \\[0.5ex] M_{ij} & \text{ if $i,j\geq 1$.} \end{array}  \right.
\eee
In what follows, we will often use the notation $U=u_A \otimes u_B$ and $U_* = u_{A*} \otimes u_{B*}$.

Since our primary interest is in the exploration of the data hiding properties, according to Proposition~\ref{min is optimal} we construct a bipartite system $AB$ by joining two spherical models $A=\text{Sph}_{d_{A}}$ and $B=\text{Sph}_{d_{B}}$ via the minimal tensor product, i.e. taking $AB=\text{Sph}_{d_{A}}\tmin\text{Sph}_{d_{B}}$. This latter assumption will be made throughout the rest of this subsection.

According to Proposition~\ref{dh ratio}, the data hiding ratio against separable measurements can be computed once we know the expressions for both the separability norm and the base norm induced by the minimal tensor product. Instead of treating the general case, we show how to compute these norms for a restricted yet large class of matrices, that is, those having zero entries in the first row and column.

\begin{lemma} \label{norms sph}
Consider a bipartite system $AB=\text{\emph{Sph}}_{d_{A}}\tminit\text{\emph{Sph}}_{d_{B}}$. Then for all $M\in\mathds{R}^{(d_{A}-1)\times (d_{B}-1)}$ we have
\bb
\big\| \myhat{M} \big\| = \|M\|_1\, ,\qquad \big\| \myhat{M}\big\|_{\text{\emph{SEP}}} = \|M\|_\infty\, , \label{norms sph eq}
\ee
where it is understood that the base norm $\| \cdot \|$ is induced by the minimal tensor product, and $\|M\|_{1},\, \|M\|_{\infty}$ denote the trace and operator norm of $M$, i.e. the sum and the largest of its singular values, respectively.
\end{lemma}

\begin{proof}
Consider an arbitrary dual tensor $W\in\mathds{R}^{d_{A}\times d_{B}}$. Since the spherical model is self-dual, we can safely apply Corollary~\ref{sep sph cor} and conclude that: (a) if $W$ is separable then necessarily $\big\|\widebar{W}\big\|_{1}\leq W_{00}$; and (b) this condition is also sufficient when all cross terms $W_{i0}, W_{0j}$ ($i,j\geq 1$) vanish.

Now, let us compute the distinguishability norm against separable operations, denoted by $\|\cdot\|_{\text{SEP}}$. On the one hand, the above necessary condition for separability of effects shows that
\bb
\begin{split}
\big\|\myhat{M}\big\|_\text{SEP} &= \max \left\{ \braket{W, \myhat{M}}:\ U\pm W \in C_{d_A}^*\tmin C_{d_B}^* \right\} \\[0.5ex]
&\leq \max \left\{ \braket{W,\myhat{M}}:\ \big\|\widebar{W}\big\|_1 \leq 1\pm W_{00} \right\} \\[0.5ex]
&= \max \left\{ \Tr \big[\widebar{W}^{T} M\big] :\ \big\|\widebar{W}\big\|_1 +|W_{00}| \leq 1 \right\} \\[0.5ex]
&= \|M\|_\infty\, ,
\end{split}
\label{norms sph proof 4-5}
\ee
where we employed~\eqref{d norm altern} to find a more compact expression for the separability norm, and we exploited the fact that trace norm and operator norm are dual to each other. On the other hand, the fact that $\|N\|_1\leq 1$ is sufficient to guarantee the separability of $U \pm \myhat{N}$ leads us, again via~\eqref{d norm altern}, to the complementary bound
\bb
\begin{split}
\big\| \myhat{M}\big\|_\text{SEP} &\geq \max \left\{ \big\langle \myhat{N}, \myhat{M} \big\rangle:\ \|N\|_1\leq 1 \right\} \\[0.5ex]
&= \max \left\{ \Tr [N^T M]:\ \|N\|_1\leq 1 \right\} \\[0.5ex]
&= \|M\|_\infty\, .
\end{split}
\label{norms sph proof 6}
\ee

Our final task is the calculation of the base norm induced by the minimal tensor product. Thanks to the formula~\eqref{dual base eq}, we can write
\bb
\big\|\myhat{M}\big\| = \min\left\{ \braket{U, Z_+ + Z_-}:\ Z_\pm\in C_{d_A}\tmin C_{d_B},\ \myhat{M}=Z_+ - Z_- \right\}\, .
\label{norms sph proof 7}
\ee
Since $Z_\pm\in C_{d_A}\tmin C_{d_B}$ implies $\braket{U, Z_\pm}\geq \big\|\widebar{Z}_\pm\big\|_1$ and $\myhat{M} = Z_+ - Z_-$ implies $M=\widebar{Z}_+ - \widebar{Z}_-$, we see that
\bb
\begin{split}
\big\|\myhat{M}\big\| &\geq \min\left\{ \big\|\widebar{Z}_+\big\|_1+\big\|\widebar{Z}_-\big\|_1:\ M=\widebar{Z}_+ - \widebar{Z}_-,\ Z_\pm\in C_{d_A}\tmin C_{d_B} \right\} \\
&\geq \|M\|_1\, ,
\end{split}
\label{norms sph proof 8}
\ee
where the last lower bound follows from the triangle inequality. On the other hand, we construct an ansatz for $Z_\pm$ achieving the above lower bound. From the singular value decomposition theorem, it is immediately seen that for all real matrices $M$ there exists a decomposition $M=M_+ - M_-$ such that $\|M_+\|_1=\|M_-\|_1=\frac{\|M\|_1}{2}$, and consequently $\|M\|_1=\|M_+\|_1+\|M_-\|_1$. Then, consider $Z_\pm=\frac{\|M\|_1}{2}\, U + \myhat{M}_\pm$, so that $Z_+ - Z_- = \myhat{M}$. Since $(Z_\pm)_{i0} = 0 =(Z_\pm)_{0j}$ when $i,j\geq 1$, Corollary~\ref{sep sph cor} tells us that the condition $(Z_\pm)_{00}\geq \|\widebar{Z}_\pm\|_1$ (satisfied by construction) is sufficient to ensure the separability of $Z_\pm$, hence this is a valid ansatz. We find
\bb
\big\|\myhat{M}\big\| \leq \|M_+\|_1+\|M_-\|_1 = \|M\|_1\, ,
\label{norms sph proof 9}
\ee
thus showing that $\big\|\myhat{M}\big\| = \|M\|_1$.
\end{proof}

\begin{cor} \label{dh sph}
In the bipartite GPT $AB=\text{\emph{Sph}}_{d_{A}}\tminit\text{\emph{Sph}}_{d_{B}}$, the data hiding ratio against separable measurements can be lower bounded as
\bb
R_{\text{\emph{Sph}}}(\text{\emph{SEP}})\, \geq\, \min\{d_{A},d_{B}\} - 1\, .
\ee
Consequently, the ultimate data hiding ratios against locally constrained sets of measurements obey
\bb
\begin{split}
R_{\text{\emph{LO}}}(d_{A},d_{B}) &\geq R_{\text{\emph{LOCC}}_{\rightarrow}}(d_{A},d_{B}) \\
&\geq R_{\text{\emph{LOCC}}}(d_A,d_B) \\
&\geq R_{\text{\emph{SEP}}}(d_{A},d_{B}) \\
&\geq \min\{d_{A},d_{B}\} - 1\, .
\end{split}
\label{univ dh ratio lower}
\ee
\end{cor}

\begin{proof}
From Lemma~\ref{norms sph} we know that $R_\text{Sph}(\text{SEP})$ can be lower bounded by the maximal ratio between trace norm and operator norm of matrices in $\mathds{R}^{(d_{A}-1)\times (d_{B}-1)}$, which is well known to be $\min\{d_A-1,d_B-1\}=\min\{d_A,d_B\}-1$. Since we can provide an example of bipartite GPT with local dimensions $d_{A},d_{B}$ for which the data hiding ratio against separable measurements is no smaller than $\min\{d_A,d_B\}-1$, this constitutes a lower bound on the ultimate data hiding ratio $R_{\text{SEP}}(d_{A},d_{B})$.
\end{proof}

The problem of finding a complementary upper bound for the data hiding ratios in the spherical model is solved by the general result expressed in the forthcoming Theorem~\ref{thm univ}, which implies that $R_{\text{Sph}}(\text{LO})\leq \min\{d_A,d_B\}$. This yields the almost tight, two-sided bound
\bb
\begin{split}
\min\{d_{A},d_{B}\} - 1 &\leq R_{\text{Sph}}(\text{SEP}) \\
&\leq R_{\text{Sph}}(\text{LOCC}) \\
&\leq R_{\text{Sph}}(\text{LOCC}_\rightarrow) \\
&\leq R_{\text{Sph}}(\text{LO}) \\
&\leq \min\{d_A,d_B\}\, ,
\end{split}
\label{double bound sph}
\ee
which fully determines the scaling of all the data hiding ratios against locally constrained measurements up to an additive constant.

\begin{rem}
Corollary~\ref{dh sph} shows that quantum mechanics, even when it is modified to encompass the minimal tensor product composition rule according to Proposition~\ref{min is optimal}, is not optimal from the point of view of data hiding, in the sense that there exist GPTs with the same local dimensions but with a (quadratically) higher data hiding ratio against all locally constrained sets of measurements.
\end{rem}

\subsection{A result on tensor norms}

In this subsection, we will lay the foundations for the proof of the main result of the present chapter, Theorem~\ref{thm univ} in the forthcoming Subsection~\ref{subsec3 ultimate}. A decisive tool in our analysis will be the theory of tensor norms as discussed in Subsection~\ref{subsec2 tensor norms}. The reason why these objects play an important role here may not be a priori clear. However, consider that we are mainly interested in understanding the asymptotic behaviour of certain quantities constructed out of local Banach spaces as the dimension of those spaces goes to infinity. This study fits in the so called local theory of Banach spaces, which is concerned with the quantitative analysis of $d$-dimensional normed spaces (as $d\rightarrow \infty$). The investigation of tensor norms and their relations with operator ideals and factorizing operators is a crucial part of those studies~\cite{DEFANT}. Thus, in that respect it is not surprising that tensor norms play an important role in our approach. 

More in detail, a crucial problem for us will be the comparison of injective and projective norm constructed out of general local Banach spaces of fixed (finite) dimension. Let us mention in passing that the same type of problem in the setting of infinite dimensional Banach leads to very deep results in functional analysis~\cite{Pisier, PisierII}.
In our context, the problem we want to address asks for the smallest constant $\kappa = \kappa(d_{A},d_{B})$ such that $\|\cdot\|_{\pi}\leq \kappa \|\cdot\|_{\varepsilon}$ holds true for all Banach spaces $V_{A},V_{B}$ of dimensions $d_{A},d_{B}$, respectively. At this stage it is not even obvious that such a quantity will be finite. The answer to this question is provided by the following result, which could be known to experts in the topic, although we did not find any explicit reference.

\begin{prop} \label{prop proj=<ninj}
For all pairs of finite dimensional Banach spaces $V_{A},V_{B}$ with dimensions $d_{A}=\dim V_{A}$, $d_{B}=\dim V_{B}$, we have
\bb
\|\cdot\|_{\pi} \leq \min\{d_{A},d_{B}\}\, \|\cdot\|_{\varepsilon}\, .
\label{proj=<ninj}
\ee
Furthermore, the constant on the right-hand side of the above inequality is the best possible for all pairs of positive integers $d_A, d_B$.
\end{prop}

\begin{proof}
We start by recalling Auerbach's lemma, which we reported here as Lemma~\ref{Auerbach lemma}. This states that any finite-dimensional Banach space admits a basis $\{v_{i}\}_{i}$, whose associated dual basis we denote by $\{v_{j}^{*}\}_{j}$, such that~\eqref{Auerbach} holds.
Suppose without loss of generality that $d_{A}\leq d_{B}$. Expand any tensor $Z\in V_{A}\otimes V_{B}$ in the local Auerbach basis for $V_{A}$, that is, $Z=\sum_{j=1}^{d_{A}} v_{j}\otimes y_{j}$.\footnote{We thank Guillaume Aubrun for having brought Auerbach's lemma and its usefulness in the context of this problem to our attention.} Choose $d_{A}$ functionals $\lambda_{i}\in V_{B}^{*}$ such that $\|\lambda_{i}\|_{*}\leq 1$ and $\braket{\lambda_{i}, y_{i}}=\|y_{i}\|$. Since also $\|v_{i}^{*}\|_{*}\leq 1$, using~\eqref{inj} we can lower bound the injective norm as follows:
\begin{align*}
\|Z\|_{\varepsilon} &\geq \braket{v_{i}^{*}\otimes \lambda_{i}, Z} \\
&= \sum_{j=1}^{d_{A}} \braket{v_{i}^{*}, v_{j}} \braket{\lambda_{i}, y_{j}} \\
&= \sum_{j=1}^{d_{A}} \delta_{ij} \braket{\lambda_{i}, y_{j}} \\
&= \braket{\lambda_{i}, y_{i}} \\
&= \|y_{i}\|\, .
\end{align*}
Then, the definition of projective norm as given in~\eqref{proj} tells us that
\begin{align*}
\|Z\|_{\pi} &\leq \sum_{i=1}^{d_{A}} \|v_{i}\|\, \|y_{i}\| \\
&\leq \sum_{i=1}^{d_{A}} \|v_{i}\|\, \|Z\|_{\varepsilon} \\
&= d_{A}\, \|Z\|_{\varepsilon}\, ,
\end{align*}
where we employed also the other defining property of the Auerbach basis, i.e. $\|v_{i}\|\leq 1$ for all $i$.

To see that the constant $\min\{d_{A}, d_{B}\}$ is optimal for all $d_{A},d_{B}$, we resort to the results of Example~\ref{ex inj proj}. There, we saw that when $V_A, V_B$ are both Euclidean spaces, the injective norm $\|\cdot\|_\varepsilon$ coincides with the matrix operator norm $\|\cdot\|_\infty$, while the corresponding projective norm is nothing but the matrix trace norm $\|\cdot\|_1$. Since $\|Z\|_{1} = \min\{d_{A},d_{B}\}\, \|Z\|_{\infty}$ whenever all the singular values of $Z$ coincide, we see that the constant in~\eqref{proj=<ninj} is the best possible.
\end{proof}

\subsection{Ultimate data hiding ratios} \label{subsec3 ultimate}

This subsection is devoted to the proof of the main result of the present chapter, Theorem~\ref{thm univ}. Namely, we want to show that the lower bound~\eqref{univ dh ratio lower} on the ultimate data hiding ratios is substantially tight. In order to be able to do this, we have to address all GPTs at once, demonstrating that none of them can exhibit a larger ratio.

As we anticipated, the theory of tensor norms and in particular Proposition~\ref{proj=<ninj} are going to play a major role in our approach. In order to apply Proposition~\ref{prop proj=<ninj} to the problem, we need to relate the distinguishability norms on a composite system to the injective and projective norm constructed out of the local base norms. This translation between purely mathematical and physically motivated quantities is the subject of our next result.

\begin{prop} \label{d norms inj proj}
Let $A=(V_{A},C_{A},u_{A})$ and $B=(V_{B},C_{B},u_{B})$ be two GPTs. The local base norms turn $V_{A},V_{B}$ into Banach spaces, and we can construct injective and projective tensor norms on $V_{A}\otimes V_{B}$, denoted simply by $\|\cdot\|_{\varepsilon}$ and $\|\cdot\|_{\pi}$. Considering $V_{A}\otimes V_{B}$ as the vector space associated with the joint system $AB$, we can also define on is: (i) the base norm $\|\cdot\|_{A\tminitfoot B}$ associated to the minimal tensor product $A \tminit B$ taken as in~\eqref{minimal GPTs}; and (ii) the local distinguishability norm $\|\cdot\|_{\text{\emph{LO}}}$. Then the following holds true:
\bb
\|\cdot\|_{\pi} = \|\cdot\|_{A {\tminitfoot} B}\, ,\qquad \|\cdot\|_{\varepsilon} \leq \|\cdot\|_{\text{\emph{LO}}}\, . \label{d norms inj proj eq}
\ee
\end{prop}

\begin{proof}
We start by showing the first relation in~\eqref{d norms inj proj eq}. Consider $Z\in V_{A}\otimes V_{B}$, and decompose it as $Z=\sum_{i=1}^{n} x^{i}\otimes y^{i}$ in such a way that $\|Z\|_{\pi} =\sum_{i=1}^{n} \|x^{i}\|\,\|y^{i}\|$ (see the definition~\eqref{proj}). According to~\eqref{dual base eq}, we can construct $x^{i}_{\pm}\in C_{A}$ and $y^{i}_{\pm}\in C_{B}$ such that $x^{i}=x^{i}_+ - x^{i}_-$ and $\|x^{i}\|= \braket{u_A, x^{i}_{+}+x^{i}_{-}}$ and analogously for $y^{i}_\pm$. Then
\bbb
X = \sum_{i=1}^{n} (x^{i}_{+}\otimes y^{i}_{+} + x^{i}_{-} \otimes y^{i}_{-})\, - \sum_{i=1}^{n} (x^{i}_{+}\otimes y^{i}_{-} + x^{i}_{-}\otimes y^{i}_{+})\, ,
\eee
and since
\bbb
\sum_{i=1}^{n} (x^{i}_{+}\otimes y^{i}_{+} + x^{i}_{-} \otimes y^{i}_{-}),\ \sum_{i=1}^{n} (x^{i}_{+}\otimes y^{i}_{-} + x^{i}_{-} \otimes y^{i}_{+})\, \in\, C_{A} \tmin C_{B}\, ,
\eee
the dual formula~\eqref{dual base eq} yields
\begin{align*}
\|X\|_{A\tminfoot B} &\leq \Big\langle u_{A}\otimes u_{B},\, \sum_{i=1}^{n} (x^{i}_{+}\otimes y^{i}_{+} + x^{i}_{-} \otimes y^{i}_{-}) \Big\rangle \\
&\quad + \Big\langle u_{A}\otimes u_{B},\, \sum_{i=1}^{n} (x^{i}_{+}\otimes y^{i}_{-} + x^{i}_{-} \otimes y^{i}_{+}) \Big\rangle \\
&= \sum_{i=1}^{n} \braket{u_A, x^{i}_{+} + x^{i}_{-}} \braket{u_B, y^{i}_{+} + y^{i}_{-}}\\
&= \sum_{i=1}^{n} \|x^{i}\|\, \|y^{i}\| \\
&= \|Z\|_{\pi}\, .
\end{align*}
To show the converse, notice first that $\|Z\|_{\pi}=\braket{u_{A}\otimes u_{B},\, Z}$ holds true for all separable $Z\in C_{A} \tmin C_{B}$. In fact, writing $Z=\sum_{i} x_{i}\otimes y_{i}$ for $x_{i},y_{i}\geq 0$ positive, we see that on the one hand $\|Z\|_{\pi}\leq \sum_{i} \|x_{i}\|\, \|y_{i}\|= \sum_{i} \braket{u_{A}, x_{i}} \braket{u_{B}, y_{i}} = \braket{u_{A}\otimes u_{B},\, Z}$, while on the other hand $\braket{u_{A}\otimes u_{B},\, Z} \leq \|Z\|_{\varepsilon}\leq\|X\|_{\pi}$ by inequality~\eqref{inj=<proj}. Now, for a generic $Z\in V_{A}\otimes V_{B}$ apply once again~\eqref{dual base eq} to construct $Z_{\pm}\in C_{A}\tmin C_{B}$ such that $Z=Z_{+}-Z_{-}$ and
\bbb
\|X\|_{A\tminfoot B} = \braket{u_{A}\otimes u_{B},\, Z_{+} + Z_{-}}\, .
\eee
By using the above observation and the triangle inequality, we find
\begin{align*}
\|Z\|_{\pi} - \braket{u_{A}\otimes u_{B},\, Z_{-}} &= \|Z\|_{\pi} - \|Z_{-}\|_{\pi} \\
&\leq \|Z+Z_{-}\|_{\pi} \\
&= \|Z_{+}\|_{\pi} \\
&=  \braket{u_{A}\otimes u_{B},\, Z_{+}}\, ,
\end{align*}
from which
\bbb
\|Z\|_{\pi} \leq \braket{u_{A}\otimes u_{B},\, Z_{-} + Z_+} = \|Z\|_{A \tminfoot B}\, .
\eee
This completes the proof of the equality $\|\cdot\|_{\pi} = \|\cdot\|_{A\tminfoot B}$.

Let us now show the second relation in~\eqref{d norms inj proj eq}. To find a lower bound on the separability norm, we will employ the expression~\eqref{d norm altern} with $\mathcal{M}=\text{LO}$. For arbitrary $Z\in V_{A}\otimes V_{B}$ and $\varphi\in V_{A}^{*},\, \lambda\in V_{B}^{*}$ such that $\|\varphi\|_{*}, \|\lambda\|_{*}\leq 1$, we show that $\varphi\otimes \lambda$ is a valid test functional to be plugged into~\eqref{d norm altern} since $\mu\coloneqq \left( \frac12 (u_{A}\otimes u_{B} + \varphi \otimes \lambda),\, \frac12 (u_{A}\otimes u_{B} - \varphi \otimes \lambda)\right) \in \braket{\text{LO}}$. Before we prove this latter claim, let us note in passing that $\mu$ is easily seen to be at least a separable measurement, as follows from Theorem~\ref{t-rk 2 sep}. Now, we show that $\mu\in \braket{\text{LO}}$. In fact, notice that
\begin{align}
\frac12 \left( u_{A}\otimes u_{B} + \varphi \otimes \lambda \right)\, &=\, \frac{u_{A}+\varphi}{2}\otimes \frac{u_{B}+\lambda}{2}\, +\, \frac{u_{A}-\varphi}{2}\otimes \frac{u_{B}-\lambda}{2}\, , \label{fg local 1} \\
\frac12 \left( u_{A}\otimes u_{B} - \varphi\otimes \lambda \right)\, &=\, \frac{u_{A}+\varphi}{2}\otimes \frac{u_{B}-\lambda}{2}\, +\, \frac{u_{A}-\varphi}{2}\otimes \frac{u_{B}+\lambda}{2}\, . \label{fg local 2}
\end{align}
Thanks to the fact that the unit balls of the dual to the local base norms have the form $B_{\|\cdot\|_{*}}=[-u,u]$ (Definition~\ref{def base}), we know that
\bbb
\left( \frac12 (u_{A}+ f),\, \frac12(u_{A}-f) \right) , \qquad \left( \frac12 (u_{B}+ g),\, \frac12(u_{B}-g) \right)
\eee
are valid measurements on $A$ and $B$, respectively. Using~\eqref{LO}, it is easy to see that $\mu$ is indeed obtainable from a product measurement via a coarse graining procedure as defined in~\eqref{coarse}. Then, equation~\eqref{d norm altern} yields $\|Z\|_{\text{LO}}\geq \braket{\varphi\otimes \lambda,\, Z}$, which becomes in turn $\|Z\|_{\text{LO}}\geq \|Z\|_{\varepsilon}$ once we maximise over the functionals $\varphi,\lambda$ satisfying $\|\varphi\|_{*},\|\lambda\|_{*}\leq 1$.
\end{proof}

\begin{cor} \label{cor d norms}
Any global base norm $\|\cdot\|$ in a bipartite GPT constructed according to~\eqref{CAB bound} must obey $\|x\otimes y\|=\|x\otimes y\|_{\text{\emph{LO}}}=\|x\|\, \|y\|$ for all $x\in V_A$ and $y\in V_B$, where $\|x\|,\,\|y\|$ stand for the local base norms. 
\end{cor}

\begin{proof}
On the one hand, since~\eqref{CAB bound} holds, the global base norms must be upper bounded by the one associated to the minimal tensor product, which coincides with $\|\cdot\|_\pi$ by Proposition~\ref{d norms inj proj}. On the other hand, the second identity in~\eqref{d norms inj proj eq} ensures that $\|\cdot\|_{\text{LO}}\geq \|\cdot\|_\varepsilon$. We already saw how injective and projective norm coincide on simple tensors. Then, putting all together we obtain
\begin{align*}
\|x\|\, \|y\| &= \| x \otimes y\|_\varepsilon \\
&\leq \|x\otimes y\|_{\text{LO}} \\
&\leq \|x\otimes y\| \\
&\leq \|x\otimes y\|_{A \tminfoot B} \\[-1.5ex]
&= \|x\otimes y\|_\pi \\
&= \|x\|, \|y\|\, ,
\end{align*}
concluding the proof.
\end{proof}

We are finally ready to prove one of our main results, that is, the optimality of the lower bound~\eqref{univ dh ratio lower} (up to an additive constant).

\begin{thm}[Upper bound on ultimate data hiding ratios] \label{thm univ} 
Let $A,B$ be two GPTs, and let their composite $AB$ obey~\eqref{CAB bound}. Then the data hiding ratios against locally constrained sets of measurements satisfy the bound
\bb
\begin{split}
R(\text{\emph{SEP}}) &\leq R(\text{\emph{LOCC}}) \\
&\leq R(\text{\emph{LOCC}}_\rightarrow) \\
&\leq R(\text{\emph{LO}}) \\[-1ex]
&\leq \max_{0\neq Z\in V_A\otimes V_B} \frac{\| Z \|_\pi}{\| Z \|_\varepsilon}\, . 
\end{split}
\label{thm univ eq1}
\ee
In particular, the corresponding ultimate data hiding ratios, as given in Definition~\ref{univ dh ratio}, are bounded as follows:
\bb
\begin{split}
\min\{d_A, d_B\} -1 &\leq R_{\text{\emph{SEP}}}(d_{A},d_{B}) \\
&\leq R_{\text{\emph{LOCC}}}(d_A,d_B) \\
&\leq R_{\text{\emph{LOCC}}_{\rightarrow}}(d_{A},d_{B}) \\
&\leq R_{\text{\emph{LO}}}(d_{A},d_{B}) \\
&\leq \min\{d_A,d_B\}\, .
\end{split}
\label{thm univ eq2}
\ee
\end{thm}

\begin{proof}
Since the inequalities~\eqref{chain dh} hold, we have to upper bound only the data hiding ratio against local operations. Using~\eqref{dh ratio} and Proposition~\ref{min is optimal}, we know that for a fixed pair of GPTs $A=(V_{A},C_{A},u_{A})$ and $B=(V_{B},C_{B},u_{B})$, this latter ratio satisfies
\bbb
R(\text{LO}) \leq \max_{Z\neq 0}\ \left\{ \|Z\|_{A\tminfoot B}\ \|Z\|_{\text{LO}}^{-1}\right\}\, .
\eee
Now, Proposition~\ref{d norms inj proj} states that $\|Z\|_{A\tminfoot B}=\|Z\|_{\pi}$ and $\|Z\|_{\text{LO}}^{-1}\leq \|Z\|_{\varepsilon}^{-1}$, from which we deduce~\eqref{thm univ eq1}. Using also Proposition~\ref{prop proj=<ninj}, we see that the right-hand side of~\eqref{thm univ eq1} is upper bounded by $\min\{d_A,d_B\}$ for all GPTs of fixed local dimensions $d_A,d_B$, proving also~\eqref{thm univ eq2}.
\end{proof}

The above Theorem~\ref{thm univ} is remarkable for several reasons. First, and most obviously, it solves the main problem that guided our investigation in the present chapter, by determining the exact scaling of the ultimate data hiding ratios against locally constrained sets of measurements in the pair of local dimensions. Second, it gives us a way to explicitly upper bound those ratios in any given model, via~\eqref{thm univ eq1}. Third, it shows a surprising similarity among all the classes of locally constrained measurements from the point of view of ultimate data hiding efficiency, which is not to be expected a priori.

Finally, by comparing Theorem~\ref{dh QM} and~\ref{thm univ} we can now tell how strong quantum mechanical data hiding is as compared to the strongest that is conceivable given the non-signalling constraint. When the strength is measured by the data hiding ratios, it turns out that quantum mechanics sits right in the middle between classical theories and the other extremal example provided by the spherical model. That is, the quantum mechanical data hiding ratio scales as the geometric mean between those pertaining to the two extremes.\footnote{If this rings a Bell, it should. Indeed, the same happens with the CHSH inequality, namely, the maximal quantum mechanical violation (`Tsirelson bound') is $2\sqrt{2}$, i.e. the geometric mean between the classical value, i.e. $2$, and the PR-box violation, which is $4$.}

\begin{rem}
Specialising Theorem~\ref{thm univ} to the modified version of quantum mechanics that we called $W$-theory (Definition~\ref{W theory}) yields an improvement of~\cite[Lemma 20]{Brandao-area-law}, in the form of the inequality
\bb
\|X\|_W\, \leq\, n^2 \|X\|_{\text{LO}}\, \leq\, n^2 \|X\|_\varepsilon\, ,
\label{improvement Brandao}
\ee
valid for all Hermitian operators $X$ on $\mathds{C}^{n_A}\otimes\mathds{C}^{n_B}$, where $n=\min\{n_A, n_B\}$, as usual. Observe that the rightmost side of~\eqref{improvement Brandao} coincides with the right-hand side of~\cite[Eq. (C1)]{Brandao-area-law}, while the leftmost side of~\eqref{improvement Brandao} is larger than the corresponding left-hand side of~\cite[Eq. (C1)]{Brandao-area-law}, as follows from~\eqref{W norm} and from Proposition~\ref{min is optimal} combined with the fact that the trace norm $\|\cdot\|_1$ is the base norm corresponding to the standard quantum mechanical composition rule.

Let us stress that neither of the two bounds $R_{\text{QM}}(\text{LO})\leq c\, \sqrt{n_A n_B}$~\cite{VV-dh} and $R_{\text{QM}}(\text{LO})\leq \min\{n_A^2, n_B^2\}$~\cite{Brandao-area-law} is tight. If $n_A$ and $n_B$ are very different from each other the latter bound will be more effective, while if they are of the same order the former will be preferable. In conclusion, as we already highlighted when discussing data hiding in quantum mechanics, determining the optimal scaling of $R_{\text{QM}}(\text{LO})$ remains an interesting open problem. At the same time, this example shows how consequences drawn from general results like Theorem~\ref{thm univ} can shed some light even on well-studied problems.
\end{rem}

\section{Data hiding in special classes of GPTs} \label{sec3 special}

Until now, we have been mainly interested in investigating the strongest cases of data hiding, in order to study the ultimate, intrinsic bounds characterising this non-classical phenomenon. For how well-motivated this inclination to universality can be, throughout this section we want to take a different approach and look into particular classes of models.

Subsection~\ref{subsec3 centr} is concerned with all the GPTs whose state space, just like in the spherical model, is centrally symmetric. It turns out that in this case all the data hiding ratios against locally restricted sets of measurements are equal up to an additive constant, and moreover can be computed as a maximal ratio between a projective and an injective norm. This has the merit of explaining the results of Subsection~\ref{subsec3 sph} in a more general context.

This is reminiscent of the proof of Theorem~\ref{thm univ}, where we upper bounded $R(\text{LO})$ with the maximal projective-injective ratio induced by the local base norms. However, while in that case the only piece of information we could extract was the existence of the upper bound, here computing such a quantity yields the exact data hiding ratio up to additive constants (Theorem~\ref{centr ratio}). That this is not just a alternative way of rephrasing the problem, but that on the contrary it can be instrumental in solving it, is apparent from our work in Subsection~\ref{subsec3 sph}, where we solved the spherical model thanks to this trick. We show how this approach can be pushed further, by using the machinery we develop to solve another `natural' model (Example~\ref{cubic}).

Throughout Subsection~\ref{subsec3 Werner}, we look into those GPTs whose vector space is endowed with a representation of a compact group that maps states to states but is otherwise irreducible on the section $\{x\in V:\, \braket{u,x}=0\}$. Although this assumption is quite strong, we show how it encompasses the main physically relevant examples of GPTs, like classical probability theory and quantum theory. With the group integral at hand, we are able to generalise the construction of Werner states and to use them for estimating data hiding in terms of simple geometrical parameters of the model (Theorem~\ref{dh Werner}).

\subsection{Centrally symmetric models} \label{subsec3 centr}

The solution of the spherical model we gave in Subsection~\ref{subsec3 sph}, namely through Lemma~\ref{norms sph} and Corollary~\ref{dh sph}, was based on two remarkable facts: on the one hand, we could achieve the data hiding ratio (up to an additive constant) by employing only tensors of the form $\myhat{M}$ for $M\in \mathds{R}^{(d_{A}-1)\times (d_{B}-1)}$, and on the other hand we were able to find simple expressions for base and separability norm of tensors of this simplified form. Here, we want to take the chance to generalise these intuitions a bit further, to encompass any centrally symmetric model (Definition~\ref{centr}).

For a thorough introduction to these models we refer the reader to Subsection~\ref{subsec2 centr symm}. Here, let us just recall that a centrally symmetric GPT is defined on a vector space $V = \mathds{R}^d = \mathds{R} u_* \oplus \mathds{R}^{d-1}$~\eqref{V decomp}, where $u_* = (1,0,\ldots, 0)^T$ is a distinguished state. For $x\in \mathds{R}^d$, we denote the second component with respect to the above decomposition with $\widebar{x}$. Conversely, for $v\in \mathds{R}^{d-1}$ one can construct $\myhat{v}\coloneqq (0,v)\in\mathds{R}^d$. The positive cone $C$ is defined by $C=\{(x_{0},\widebar{x})\in \mathds{R}\oplus\mathds{R}^{d-1}:\ x_{0}\geq |\widebar{x}|\}$, with $|\cdot|$ being an arbitrary norm on $\mathds{R}^{d-1}$. 
As can be easily verified, the dual cone to $C$ is given by $C ^* = \{(y_{0},\widebar{y})\in \mathds{R}\oplus\mathds{R}^{d-1}:\ y_{0}\geq |\widebar{y}|_*\}$~\eqref{centr dual eq}, where $|\cdot|_*$ is the dual to the norm $|\cdot|$. In particular, $C^*$ is centrally symmetric if so is $C$.
We remind the reader that the base norm of a centrally symmetric model is given by $\|(x_0,\widebar{x})\|=\max\{|x_0|, |\widebar{x}|\}$~\eqref{centr base norm}. For centrally symmetric models there exists a simple positive linear map, given by $T \coloneqq 1 \oplus (-I_{d-1})$~\eqref{T map}. 

When one combines two centrally symmetric models of dimensions $d_A, d_B$, the resulting vector space contains a `double' vector component $\mathds{R}^{(d_A-1)\times (d_B-1)}$ that is obtained by tensoring the vector components of the two spaces~\eqref{VAB decomp}. Analogously to the case of a single system, we can `lower' matrices $Z\in \mathds{R}^{d_A\times d_B}$ to $\widebar{Z}\in\mathds{R}^{(d_A-1)\times (d_B-1)}$ by retaining only this component, and conversely `lift' $M\in\mathds{R}^{(d_A -1)\times (d_B -1)}$ to $\myhat{M}\in\mathds{R}^{d_A\times d_B}$ by adding a row and a column of zeros.

As we proved in Proposition~\ref{sep centr prop}, a bipartite tensor of the form $U_* + \myhat{M}$, where $U_*=u_{A*}\otimes u_{B*}$, belongs to the maximal tensor product iff $|M|_\pi\leq 1$, and to the minimal tensor product iff $|M|_\varepsilon\leq 1$. Here, $|\cdot|_\varepsilon$ and $|\cdot|_\pi$ denote the injective and projective tensor norms constructed out of the local norms $|\cdot|$.
As a side remark, observe that the maps $T\otimes I,\, I\otimes T,\, T\otimes T$ preserve separability of states and of effects. Therefore, the norm $\|\cdot\|_{\text{SEP}}$ is left invariant by any of these maps. As is easy to verify, the same is true for all the locally constrained distinguishability norms, with the possible exception of $\|\cdot\|_{\text{LOCC}}$.

Concerning the analysis of the data hiding properties, what we did in the case of the spherical model was to consider only tensors of the form $\myhat{M}$, and to a posteriori justify this restriction by employing Theorem~\ref{thm univ} to show that the obtained result is tight up to an additive constant. As it turns out, this procedure can be always followed for centrally symmetric GPTs without loss of generality. This is the content of our first result.

\begin{prop} \label{prop centr R r}
Let $AB$ be a bipartite system formed by two centrally symmetric GPTs joined with any rule that respects~\eqref{CAB bound}. Then for all $Z\in \mathds{R}^{d_A\times d_B}$ we have
\bb
\frac{\|Z\|}{\|Z\|_{\text{\emph{LO}}}} \leq \frac{\big\|\widebar{Z}\big\|}{\big\|\widebar{Z}\big\|_{\text{\emph{LO}}}} + 2\, . \label{prop centr R r eq}
\ee
\end{prop}

\begin{proof}
Consider $Z=s\, u_*\otimes u_* + x \otimes u_* + u_*\otimes y + \widebar{Z}\in \mathds{R}^{d_A\times d_B}$, where we omitted the subscripts $A,B$ for the sake of simplicity. Applying the triangle inequality to the global base norm, we see that $\|Z\|\leq \| (s\,u_*+x)\otimes u_*\| + \|u_*\otimes y\| + \big\|\widebar{Z}\big\|$. Now, Corollary~\ref{cor d norms} guarantees that $\| (s\,u_*+x)\otimes u_*\| = \| (s\,u_*+x)\otimes u_*\|_{\text{LO}}=\|s\,u_*+x\|$ and $\|u_*\otimes y\|=\|u_*\otimes y\|_{\text{LO}}=\|y\|$. Moreover, it is also clear by discarding the system $B$ and performing an arbitrary operation on $A$, that $\|Z\|_{\text{LO}}\geq \|s\, u_*+x\|$. Proceeding in an analogous fashion with exchanged subsystems, remembering that $\Lambda$ defined in~\eqref{T map} leaves the base norm invariant, and exploiting the triangle inequality, yields
\begin{align*}
\|Z\|_{\text{LO}} &\geq \|s\, u_* + y\| \\
&= \|T (s\, u_*+y)\| \\
&= \|s\,u_* - y\| \\
&\geq \left\|\, \frac12 (s\, u_* + y) - \frac12 (s\, u_* - y) \, \right\| \\
&= \|y\|\, .
\end{align*}
Finally, using the readily verified invariance of $\|\cdot\|_{\text{LO}}$ under any of the maps $T\otimes I,\, I\otimes T,\, T\otimes T$, we obtain
\begin{align*}
\|X\|_{\text{LO}} &\geq \left\| \,\frac14 Z - \frac14 (T \otimes I)(Z) - \frac14 (I\otimes T)(Z) + \frac14 (T \otimes T)(Z)\, \right\|_{\text{LO}} \\
&= \big\|\widebar{Z}\big\|_{\text{LO}}\, .
\end{align*}
Putting all together, we find
\begin{align*}
\frac{\|Z\|}{\|Z\|_{\text{LO}}} &\leq \frac{\| s\,u_*+x\| + \|y\| + \big\|\widebar{Z}\big\|}{\|Z\|_{\text{LO}}} \\
&= \frac{\| s\,u_*+x\|}{\|X\|_{\text{LO}}} + \frac{\|y\|}{\|Z\|_{\text{LO}}} + \frac{\big\|\widebar{Z}\big\|}{\|Z\|_{\text{LO}}} \\
&\leq 2 + \frac{\big\|\widebar{Z}\big\|}{\big\|\widebar{Z}\big\|_{\text{LO}}}
\end{align*}
\end{proof}

What Proposition~\ref{prop centr R r} is telling us is that up to an additive constant we can restrict the search for data hiding against local operations to projected tensors of the form $\widebar{Z}$. From now on, we denote by $\widebar{R}(\mathcal{M})$ the \emph{restricted data hiding ratio} against a set of measurements $\mathcal{M}$ that is obtained by considering only those tensors. With this notation, Proposition~\ref{prop centr R r} can be cast into the form of the inequality $R(\text{LO}) \leq \widebar{R}(\text{LO}) + 2$. In order to carry out the analysis of restricted ratios, we need an analogue of Lemma~\ref{norms sph}.

\begin{prop} \label{norms centr}
Let $|\cdot|_\varepsilon,|\cdot|_\pi$ be the injective and projective norm constructed on $\mathds{R}^{(d_A-1)\times (d_B-1)}\simeq \mathds{R}^{d_A-1}\otimes \mathds{R}^{d_B-1}$ out of the local norms $|\cdot|$ on $\mathds{R}^{d_A-1}, \, \mathds{R}^{d_B-1}$. Then for $M\in\mathds{R}^{(d_A-1)\times (d_B-1)}$ the locally constrained distinguishability norms and the global base norm induced by the minimal tensor product are respectively given by 
\begin{align}
&\big\|\myhat{M}\big\|_{\text{\emph{LO}}} = \big\|\myhat{M}\big\|_{\text{\emph{LOCC}}_\rightarrow} = \big\|\myhat{M}\big\|_{\text{\emph{LOCC}}} = \big\|\myhat{M}\big\|_{\text{\emph{SEP}}} = |M|_\varepsilon\, , \label{norms centr eq1} \\[0.5ex]
&\big\|\myhat{M}\big\|_{A\tminitfoot B} = |M|_\pi\, .
\end{align}
\end{prop}

\begin{proof}
The argument follows the guidelines of the proof of Proposition~\ref{norms sph}, so we omit some of the details. First of all, thanks to Proposition~\ref{sep centr prop} tensor $W\in\mathds{R}^{d_{A}\times d_{B}}$ is separable then necessarily $|\widebar{W}|_{*\pi} \leq W_{00}$, and this condition is also sufficient when all cross terms $W_{i0}, W_{0j}$ ($i,j\geq 1$) vanish.

To compute the separability norm, on the one hand as in~\eqref{norms sph proof 4-5} we find
\begin{align*}
\big\|\myhat{M}\big\|_{\text{SEP}} &\leq
\max\left\{ \Tr \big[\widebar{W}^{T} M\big] :\ \big|\widebar{W}\big|_{*\pi} + |W_{00}| \leq 1 \right\} \\
&= |M|_{*\pi*} \\
&=|M|_\varepsilon \, ,
\end{align*}
where in the last step we used~\eqref{dual inj proj}. On the other hand, the complementary inequality $\big\| \myhat{M} \big\|_{\text{LO}}\geq |M|_\varepsilon$ is a simple consequence of the second relation in~\eqref{d norms inj proj eq}. In fact, since for all $v\in\mathds{R}^{d_A-1}$ the identity $|v|_*=\|\myhat{v}\|$ holds true, we obtain
\begin{align*}
\big\|\myhat{M}\big\|_{\text{LO}} &\geq \big\|\myhat{M}\big\|_\varepsilon \\
&= \max \left\{\braket{\varphi\otimes \lambda, \myhat{M} }:\ \|\varphi\|_*,\|\lambda\|_*\leq 1 \right\} \\
&\geq \max \left\{ v^T M w:\ |v|_*,|w|_*\leq 1 \right\} \\
&= |M|_\varepsilon\, .
\end{align*}
Putting all together, we obtain
\bbb
|M|_\varepsilon \leq \big\|\myhat{M}\big\|_{\text{LO}} \leq \big\|\myhat{M} \big\|_{\text{LOCC}_\rightarrow} \leq \big\| \myhat{M} \big\|_{\text{LOCC}} \leq \big\| \myhat{M} \big\|_{\text{SEP}} \leq |M|_\varepsilon\, ,
\eee
from which~\eqref{norms centr eq1} follows.
\end{proof}

\begin{thm} \label{centr ratio}
For a fixed pair of centrally symmetric GPTs whose composite is formed with the minimal tensor product rule according to Proposition~\ref{min is optimal}, all the four restricted data hiding ratios against locally constrained measurements coincide, and can be computed as
\begin{align}
\widebar{R} \coloneqq & \widebar{R}(\text{\emph{LO}}) = \widebar{R}(\text{\emph{LOCC}}_\rightarrow) = \widebar{R}(\text{\emph{LOCC}}) = \widebar{R}(\text{\emph{SEP}}) \label{centr ratio eq0}
\\
=& \max_{0\,\neq\, M\,\in\, \mathds{R}^{(d_A-1)\times (d_B-1)}} \frac{\,|M|_\pi}{\,|M|_\varepsilon}\, . \label{centr ratio eq1}
\end{align}
Moreover, the `true' data hiding ratios are equal to the restricted ones up to an additive constant. Namely,
\bb
\widebar{R} \leq R(\text{\emph{SEP}}) \leq R(\text{\emph{LOCC}}) \leq R(\text{\emph{LOCC}}_\rightarrow) \leq R(\text{\emph{LO}})\leq \widebar{R} + 2\, . \label{centr ratio eq2}
\ee
\end{thm}

\begin{proof}
First of all,~\eqref{centr ratio eq1} follows directly from Proposition~\ref{norms centr}. To show~\eqref{centr ratio eq2}, note that, by definition, restricted data hiding ratios are smaller than the true ones. In particular, $R(\text{SEP})\geq \widebar{R}(\text{SEP})=\widebar{R}$. Using~\eqref{chain dh} and the result of Proposition~\ref{prop centr R r} in the form of the inequality $R(\text{LO})\leq \widebar{R}(\text{LO})+2$, we obtain immediately~\eqref{centr ratio eq2}.
\end{proof}

\begin{rem}
Sometimes the quantity $\widebar{R}$ appearing in~\eqref{centr ratio eq1} can be better computed with the help of a simple trick. Namely, remember that for all norms $|\cdot|^{(1)},|\cdot|^{(2)}$ on any space $V$ we have $\max_{0\neq x\in V} \frac{|x|^{(1)}}{|x|^{(2)}} = \max_{0\neq \varphi\in V^*} \frac{|\varphi|^{(2)}_*}{|\varphi|^{(1)}_*}$. Thanks to this identity and to the duality relation~\eqref{dual inj proj}, we see that $\widebar{R}$ can alternatively be expressed as
\bb
\widebar{R} = \max_{0\,\neq\, M\,\in\, \mathds{R}^{(d_A-1)\times (d_B-1)}} \frac{\,|M|_{*\varepsilon}}{\,|M|_{*\pi}}\, . \label{Rbar dual}
\ee
\end{rem}

Theorem~\ref{centr ratio} suggests a precise recipe for computing data hiding ratios of centrally symmetric models. It is in fact enough to analyse the behaviours of injective and projective norms, compute $\widebar{R}$ defined in~\eqref{centr ratio eq1}, and finally use~\eqref{centr ratio eq2} to determine all the ratios against locally constrained measurements up to a universal additive constant. In fact, this is basically what we did in Subsection~\ref{subsec3 sph}, since trace norm and operator norm are exactly the injective and projective norms constructed out of local Euclidean spaces, as we saw in the proof of Proposition~\ref{prop proj=<ninj}. In the remaining part of this subsection, we are going to show how to apply all this machinery to solve another explicit example.

\begin{ex}[Cubic model] \label{cubic} 

Let us consider the class of cubic models $G_n$ defined in Subsection~\ref{subsec2 other ex}. We remind the reader that the state space of $G_n$ is an $n$-dimensional hypercube. Here, we are mainly interested in discussing the data hiding properties of GPTs of the form $G_n\tmin G_m$, for $n,m$ positive integers. Since the $n$-dimensional hypercube is centrally symmetric, this is a perfect playground for testing the machinery we developed throughout this subsection. Notice that in the case of $G_n$ the norm $|\cdot|$ on $\mathds{R}^n$ appearing in Definition~\ref{centr} is given by the $\infty$-norm $|v|_\infty\coloneqq \max_{1\, \leq\, i\, \leq\, n} |v_i|$, with dual $|w|_{\infty*}=|w|_1=\sum_{i=1}^n |w_i|$. It is worth noticing that the extreme points of the unit ball $B_{|\cdot|_1}$ coincide with the elements of the standard basis of $\mathds{R}^n$ up to a sign.

According to Theorem~\ref{centr ratio}, the first step in solving the GPT $G_n\tmin G_m$ is the determination of the restricted data hiding ratio $\widebar{R}$ in~\eqref{centr ratio eq1}, where in our case $|\cdot|_\varepsilon,|\cdot|_\pi$ are the injective and projective norm induced on $\mathds{R}^{n\times m}$ by the local $\infty$-norms. We already observed how this can be dually rephrased as~\eqref{Rbar dual}. In our case, this is helpful because the injective and projective dual norms $|\cdot|_{*\varepsilon}, |\cdot|_{*\pi}$ can be easily computed as follows. The former is given by
\bb
\begin{split}
|M|_{*\varepsilon} &= \max_{|v|_\infty,|w|_\infty\leq 1} v^T M w \\
&= \max_{s\in \{\pm 1\}^n,\, t\in \{\pm 1\}^m} s^T M t \\
&= \|M\|_{\infty\rightarrow 1}\, ,
\end{split}
\ee
while the latter can be found thanks to the chain of inequalities
\bb
|M|_{*\pi} = |M|_{\varepsilon*} = |M|_{\infty *} = |M|_1\, ,
\ee
where we used the fact that since
\bbb
|N|_{\varepsilon} = \max_{|v|_1, |w|_1\leq 1} v^T N w = \max_{i,j} |N_{ij}| = |N|_\infty\, ,
\eee
one has $|M|_{\varepsilon*}=|M|_1= \sum_{i,j} |M_{ij}|$, where we extended the definitions of $\infty$-norm and $1$-norm to matrices in an obvious way.

Now, we have to find the largest ratio $|M|_1/ \|M\|_{\infty\rightarrow 1}$ that is achievable by $M\in \mathds{R}^{n\times m}$. The argument comprises two parts: first, we have to exhibit an explicit example $M_0$ displaying a high ratio, and secondly, we have to show that this is optimal up to a constant factor.

Let us start with the first task. From now on, we will assume without loss of generality $n\leq m$. Suppose that there exists a matrix $H\in \{\pm 1\}^{n\times m}$ made of signs such that $HH^T=m\mathds{1}_n$ (such matrices are often called \emph{partial Hadamard matrices}). Then simple considerations very close in spirit to Lindsey's lemma~\cite{ERDOS, cut-norm-H} (see also~\cite{ALON-SPENCER, Alon-sparse}) show that for all $s\in \{\pm 1\}^n$ and $t\in \{\pm 1\}^m$ we have
\begin{align*}
s^T H t &\leq \sum_{j} \bigg| \sum_i s_i H_{ij}\bigg| \\
&\leq \sqrt{m}\, \sqrt{\sum_j \bigg| \sum_i s_i H_{ij}\bigg|^2} \\
&= \sqrt{m}\, \sqrt{\sum_{ijk} s_i s_k H_{ij} H_{kj}} \\
&= \sqrt{m}\, \sqrt{\sum_{ik} s_i s_k \, n\delta_{ik}} \\
& = n\sqrt{m}\, .
\end{align*}
Maximising over sign vectors $s,t$ yields $\|H\|_{\infty\rightarrow 1}\leq n\sqrt{m}$. Since $|H|_1=nm$, we obtain the lower bound $|H|_1 \big/ \|H\|_{\infty\rightarrow 1}\geq \sqrt{n}$. Now, the existence of a matrix $H$ with the properties required for the above construction is not guaranteed for arbitrary integers $n,m$. In fact, this is never the case when $n=m>2$ and $4\nmid n$, while when $n=m=4k$ it is the content of the so-called \emph{Hadamard conjecture}. However, observe that Hadamard matrices are elementarily guaranteed to exist for $n=m=2^k$, as the explicit example $\left( \begin{smallmatrix} 1 & 1 \\ -1 & 1 \end{smallmatrix} \right)^{\otimes k}$ shows. Since for all $n$ there is $n'$ power of $2$ such that $\frac{n}{2}\leq n'\leq n$, we get $|H|_1 \big/ \|H\|_{\infty\rightarrow 1}\geq \sqrt{n'}\geq \sqrt{\frac{n}{2}}$. This shows that
\bb
\max_{0\, \neq\, M\, \in\, \mathds{R}^{n\times m}} \frac{|M|_{*\pi}}{|M|_{*\varepsilon}}\, =\, \max_{0\, \neq\, M\, \in\, \mathds{R}^{n\times m}} \frac{|M|_1}{\|M\|_{\infty\rightarrow 1}}\, \geq\, \sqrt{\frac{n}{2}}\, .
\label{lower Hadamard}
\ee
We note in passing that the crude estimate~\eqref{lower Hadamard} can be improved to $\sqrt{n}$ when $n\leq \frac{m}{2}$ by the same tricks. Further asymptotic refinements can be obtained with the help of sophisticated results such as~\cite{half-Hadamard}, but this is beyond the scope of the present paper.

Now, we have to show that the above lower bound is optimal, i.e. that the scaling of the two sides of~\eqref{lower Hadamard} are the same. In order to do so, we resort to a celebrated inequality known as Khintchine inequality~\cite{bound-B, best-constant-1, best-constant-2}, which states that for all $c\in \mathds{R}^n$, if $s_1,\ldots, s_n$ is an i.i.d. sequence with Rademacher distribution, then
\bb
\mathds{E}_s \bigg|\sum_i s_i c_i \bigg|\, \geq\, \frac{|c|_2}{\sqrt{2}}\, ,
\label{K ineq}
\ee
where $|c|_2^2=\sum_i c_i^2$. Thanks to this result, we know that whenever $M\in\mathds{R}^{n\times m}$ we have~\cite[Corollary 2.4]{Alon-sparse}
\begin{align*}
\|M\|_{\infty\rightarrow 1} &= \max_{s\in \{\pm 1\}^n,\, t\in \{\pm 1\}^m} s^T M t \\[0.5ex]
&= \max_{s\in \{\pm 1\}^n} \sum_j \bigg| \sum_i s_i M_{ij} \bigg| \\[0.5ex]
&\geq \mathds{E}_s \sum_j \bigg|\sum_i s_i M_{ij}\bigg| \\[0.5ex]
&= \sum_j \mathds{E}_s \bigg|\sum_i s_i M_{ij}\bigg| \\[0.5ex]
&\geq \frac{1}{\sqrt{2}}\, \sum_j \sqrt{\sum_i M_{ij}^2} \\
&\geq \frac{1}{\sqrt{2n}}\,\sum_{ij} |M_{ij}| \\ 
&= \frac{1}{\sqrt{2n}}\,|M|_1\, .
\end{align*}
This is nothing but the complementary upper bound to~\eqref{lower Hadamard}, and reads
\bb
\max_{0\, \neq\, M\, \in\, \mathds{R}^{n\times m}} \frac{|M|_{*\pi}}{|M|_{*\varepsilon}}\, =\, \max_{0\, \neq\, M\, \in\, \mathds{R}^{n\times m}} \frac{|M|_1}{\|M\|_{\infty\rightarrow 1}}\, \leq\, \sqrt{2n}\, .
\label{upper Hadamard}
\ee
The two inequalities~\eqref{lower Hadamard} and~\eqref{upper Hadamard} together solve the problem of data hiding in the cubic model. We summarise this solution as follows.

\begin{prop} \label{dh cubic}
Let the composite of two cubic models $A=G_n$, $B=G_m$ be given by the minimal tensor product, $AB=G_n\tminit G_m$. Then the restricted data hiding ratio $\widebar{R}_{G_n\tminitfoot G_m}$ defined via Proposition~\ref{norms centr} satisfies
\bb
\sqrt{\frac{\min\{n,m\}}{2}} \leq \widebar{R}_{G_n\tminitfoot G_m}  \leq \sqrt{2\, \min\{n,m\}}\, .
\ee
Consequently, all the data hiding ratios against locally constrained measurements scale as 
\bb
R_{G_n \tminitfoot G_m}(\mathcal{M})\, =\, \Theta\left(\sqrt{\min\{n,m\}}\right)\, =\, \Theta\left(\sqrt{\min\{d_A,d_B\}}\right) ,
\ee
for $\mathcal{M}=\text{\emph{LO}},\, \text{\emph{LOCC}}_\rightarrow,\, \text{\emph{LOCC}},\, \text{\emph{SEP}}$.
\end{prop}
\end{ex}

\subsection{Werner construction for symmetric models} \label{subsec3 Werner}

A remarkable feature possessed by all the examples of GPTs we have seen so far, including the two that are undoubtedly the most physically relevant, i.e. classical probability theory (Subsection~\ref{subsec2 ex class}) and quantum theory (Subsection~\ref{subsec2 ex QM}), is the existence of a wide group of symmetries, i.e. linear transformations sending states to states. In the classical case, these are just permutations of the entries of the probability vector, while in the quantum case a symmetry is any conjugation by a unitary matrix. As for the spherical and cubic models (Subsection~\ref{subsec2 centr symm} and Example~\ref{cubic}), there are the natural actions of $SO(d-1)$ and of signed permutations on the last $d-1$ components of the state vector. We will show how these symmetries can be exploited in order to define a relevant class of bipartite states in a theory made of two copies of the same symmetric GPT. Let us start with the following definition.

\begin{Def}[Completely symmetric models] \label{compl symm}
A GPT $(V,C,u)$ is said to be \emph{completely symmetric} if there is a compact group $G$ and a representation $\zeta: G\rightarrow \mathcal{L}(V)$ that: (i) is such that $\zeta_g$ sends normalised states to normalised states, for all $g\in G$; and (ii) is irreducible and nontrivial on the preserved subspace $V_0\coloneqq \{x\in V:\, u(x)=0\}$.
\end{Def}

\begin{note}
Throughout this subsection, we will denote the action of a functional $\varphi$ on a vector $x$ with $\varphi(x)$ rather than $\braket{\varphi, x}$. This is to avoid confusion with other scalar products we will encounter along the way.
\end{note}

We will see at the end of this Subsection that the main examples of GPTs we have considered so far (namely, classical and quantum theory, and spherical and cubic model) are in fact completely symmetric. However, for now we are interested in keeping the reasoning as abstract and general as possible.
Thus, consider a completely symmetric GPT as in Definition~\ref{compl symm}. We start by listing some elementary consequences of the existence of a group symmetry. Let us remind the reader that if $\zeta:G\rightarrow \mathcal{L}(V)$ is a representation, then the dual space $V^*$ is naturally endowed with the dual representation $\zeta^*:G\rightarrow \mathcal{L}(V^*)$ given by $\zeta^*_g=\zeta_{g^{-1}}^T$. With this notation, the fact that $V_0$ is preserved by the action of $G$ can be written as $\zeta_g^*u=u$ for all $g\in G$. In other words, $\mathds{R}u$ is a trivial component of $V^*$ under the action of $G$ through $\zeta^*$. Since it is well-known that if $V$ is real and $G$ is compact then $\zeta$ and $\zeta^*$ are isomorphic, there must necessarily exist also a $G$-invariant vector $u_*\in V$. We deduce from the fact that $V_0$ is irreducible and nontrivial that $u_*\notin V_0$, so that we are free to rescale it in such a way that $u, (u_*)=1$. Note that this fixes the decomposition of $V$ into $G$-irreducible representations as
\bb
V = \mathds{R}u_*\oplus V_0\, , \label{decomp V}
\ee
where the two pieces are non-isomorphic. Now, we proceed to show that $u_*$ is in fact a state, as implied by property (i) in Definition~\ref{compl symm}. In order to do so, we use the Haar integral on $G$, denoted by $\int_G dg$ and whose existence is guaranteed by the compactness of $G$. For every state $\omega$, we have that $u_*=\int_G dg\, \zeta_g\, \omega$, because the right-hand side is $G$-invariant and normalised and therefore must coincide with $u_*$. This expression for $u_*$ as a positive combination of states $\zeta_g\, \omega$ reveals that $u_*$ is itself a state.

As is easy to see by referring to~\eqref{decomp V}, the dual vector space $V^*$ decomposes as
\bb
V^* = \mathds{R}u\oplus V_0^* \label{decomp V*}
\ee
under the action of $G$, where the first addend is a trivial representation and the second one is $G$-isomorphic to $V_0$. Therefore, any $G$-isomorphism $\chi:V^*\rightarrow V$ must map $u$ into a multiple of $u_*$ and $V_0^*$ into $V_0$.
If we think of $\chi$ as a tensor in $V\otimes V$, we can write $\chi = \alpha\, u_*\!\otimes u_* + \beta\, \mathcal{E} = \alpha U_* +\beta \mathcal{E}$, where we used the shorthand $U_*\coloneqq u_*\otimes u_*$, and $\mathcal{E} \in V_0\otimes V_0$ is (canonically identified with) a fixed $G$-isomorphism $V_0^*\rightarrow V_0$. Depending on the representation we choose, we can alternatively write $(\zeta_g\otimes \zeta_g)(\mathcal{E})=\mathcal{E}$ for all $g\in G$ or $\mathcal{E} \zeta_g^*=\zeta_g \mathcal{E} $ for all $g\in G$. From this latter expression we see that $\mathcal{E}_*\coloneqq (\mathcal{E}^{-1})^T:V\rightarrow V^*$ is also a $G$-isomorphism. As for the tensor $\mathcal{E}_*\in V_0^*\otimes V_0^*$, we will rephrase this as $(\zeta_g^*\otimes \zeta_g^*)(\mathcal{E}_*)=\mathcal{E}_*$ for all $g\in G$.

It is perhaps convenient to think of $\mathcal{E}$ and $\mathcal{E}_*$ also as scalar products $\braket{\cdot,\cdot}_\mathcal{E}$, $\braket{\cdot,\cdot}_{\mathcal{E}_*}$ on $V_0^*$ and $V_0$, respectively. This can be done via the definitions $\braket{f,g}_\mathcal{E} \coloneqq (f\otimes g)(\mathcal{E})$ and $\braket{v,w}_{\mathcal{E}_*}\coloneqq \mathcal{E}_* (v\otimes w)$. It is well known that it is possible to choose $\mathcal{E}$ such that both these scalar product are positive definite. We will always make this assumption throughout the rest of this Section. If $\{f_i\}_{i=1}^{d-1}$ is an orthonormal basis for $\braket{\cdot,\cdot}_\mathcal{E}$, we will have
\bb
\mathcal{E} = \sum_{i=1}^{d-1} f_i^*\otimes f_i^*\, \in\, V_0\otimes V_0\, ,\qquad \mathcal{E}_* = \sum_{i=1}^{d-1} f_i\otimes f_i\,\in\, V_0^*\otimes V_0^*\, ,
\label{varepsilon expl}
\ee
with $\{f_i^*\}_i$ being the dual basis to $\{f_i\}_i$. As a simple consequence, we see that $\mathcal{E}_*(\mathcal{E})=d-1$, and the two norms induced by the above scalar products are dual to each other. Moreover, observe that the completion $\{u,f_1,\ldots,f_d\}$ is an orthonormal basis for a global $G$-invariant scalar product. We are now ready to give the following definition.

\vspace{2ex}
\begin{Def}[Werner states] \label{Werner def}
For a composite $AB$ made of two copies $A,B$ of the same completely symmetric GPT $(V,C,u)$ and such that~\eqref{CAB bound} is obeyed, a \emph{Werner state} is a normalised state in $V\otimes V$ that corresponds to a $G$-isomorphism $V^*\rightarrow V$. 
\end{Def}

\vspace{2ex}
With the language developed throughout the above discussion, we can express a generic Werner state as
\bb
\chi_s = U_* + s\, \mathcal{E}\, ,
\label{Werner}
\ee
where $s$ is a real parameter whose range depends on specific features of the model. For any bipartite cone $C_{AB}$ satisfying~\eqref{CAB bound}, let us define
\bb
k_\pm \coloneqq \max\left\{ s:\, \chi_{\pm s}\in C_{AB} \right\}\, ,
\label{k+-}
\ee
so that the allowed range of $s$ in~\eqref{Werner} will be $[-k_-,k_+]$. Besides the obvious observation that $k_\pm\geq 0$, there seems to be nothing we can say a priori about these parameters. In order to proceed further, we need a little lemma.

\begin{lemma} \label{lemma G projector}
Let $(V,C,u)$ be a completely symmetric GPT with Werner states given by~\eqref{Werner}. If $\mathcal{E}, \mathcal{E}_*$ are given by~\eqref{varepsilon expl}, the identity
\bb
\int_G dg\ \zeta_g\otimes \zeta_g = U_* U + \frac{1}{d-1}\ \mathcal{E}\, \mathcal{E}_*
\label{G projector}
\ee
between linear operators on $V\otimes V$ holds true. Here, $U\coloneqq u\otimes u$, and $u_*u:V\rightarrow V$ acts as $(u_* u)(x)=\braket{u,x} u_*$ for all $x\in V$.
\end{lemma}

\begin{proof}
For an arbitrary $Z\in V\otimes V$, the properties of the Haar measure guarantee that $\int_G dg\, \zeta_g\otimes \zeta_g (Z)$ is a $G$-invariant tensor, and thus can be expressed as $\alpha\, U_* + \beta\, \mathcal{E}$ for some $\alpha,\beta\in \mathds{R}$. Applying $U$ on both sides we see that
\begin{align*}
\alpha &= \int_G dg\  U \left( \zeta_g\otimes \zeta_g\, Z \right) \\ 
&= \int_G dg\ \left( \zeta_g^* \otimes \zeta_g^*\, U \right) (Z) \\
&= \int_G dg\, U(Z) \\
&= U(Z)\, .
\end{align*}
Moreover, it is easy to verify that since $\mathcal{E}_* (\mathcal{E})=d-1$ and $\zeta_g^*\otimes \zeta_g^*\, \mathcal{E}_*=\mathcal{E}_*$ for all $g\in G$, the analogous equality $(d-1)\beta= \mathcal{E}_* (Z)$ holds, thus completing the proof of~\eqref{G projector}.
\end{proof}

With Lemma~\ref{lemma G projector} in our hands, we can say a bit more about the constants $k_\pm$ introduced in~\eqref{k+-}. Namely, we can determine lower bounds that correspond to the \emph{separability region} for the family of Werner states~\eqref{Werner}. As expected, these lower bounds will depend only on the local structure of the model, not on the particular choice of the composite cone, which will instead affect the values of $k_\pm$.

\vspace{2ex}
\begin{prop} \label{prop sep Werner}
For a completely symmetric GPT $(V,C,u)$ of dimension $d$ and a scalar product $\braket{\cdot,\cdot}_{\mathcal{E}_*}$ on the corresponding $V_0$, introduce the quantities
\begin{align}
m_+ &\coloneqq\, \max\left\{\braket{v,v}_{\mathcal{E}_*}:\ v\in V_0,\, u_*+v\in C\right\}, \label{m+}\\
m_-  &\coloneqq\, \max\left\{ \braket{v,w}_{\mathcal{E}_*}:\ v,w\in V_0,\, u_*+v, u_*-w\in C\right\}\, .
\label{m-}
\end{align}
Then the constants $k_\pm$ defined through~\eqref{k+-} satisfy $k_\pm\geq \frac{m_\pm}{d-1}$, and a Werner state $\chi_s$ as given in~\eqref{Werner} is separable if and only if $-\frac{m_-}{d-1}\leq s\leq \frac{m_+}{d-1}$.
\end{prop}

\begin{proof}
Take $v\in V_0$ such that $u_*+v\in C$ and $\braket{v,v}_{\mathcal{E}_*}=m_+$. Compute
\begin{align*}
&\int_G dg\ (\zeta_g\otimes \zeta_g)\left((u_*+v)\otimes (u_*+v)\right) \\
&\qquad = \left( U_* U + \frac{1}{d-1}\ \mathcal{E}\, \mathcal{E}_* \right)\left((u_*+v)\otimes (u_*+v)\right) \\
&\qquad = U_* + \frac{\mathcal{E}_*(v\otimes v)}{d-1}\ \mathcal{E} \\
&\qquad = U_* + \frac{\braket{v,v}_{\mathcal{E}_*}}{d-1}\ \mathcal{E} \\
&\qquad = U_* + \frac{m_+}{d-1}\ \mathcal{E}\, .
\end{align*}
From the leftmost side we see that this is an allowed separable state of the bipartite system. Looking at the rightmost side and comparing it to~\eqref{Werner}, we see that $k_+\geq \frac{m_+}{d-1}$ and that $\chi_{m_+/(d-1)}\in C_A\tmin C_B$. An analogous construction shows that we can find $v,w\in V_0$ such that $u_*+v, u_*-w\in C$ and
\bbb
\int_G dg\ \zeta_g\otimes \zeta_g \left((u_*+v)\otimes (u_*-w)\right) = U_* - \frac{m_-}{d-1}\ \mathcal{E}\, \in\, C_A\tmin C_B\, ,
\eee
so that $k_-\geq \frac{m_-}{d-1}$ and $\chi_{-k_-}\in C_A\tmin C_B$.

In order to show that $\chi_s\in C_A\tmin C_B$ if and only if $s\in \left[- \frac{m_-}{d-1},\, \frac{m_+}{d-1}\right]$, we start by proving that $\omega \in C_A\tmin C_B$ implies $-m_-\leq \mathcal{E}_*(\omega)\leq m_+$. In fact, if $\omega=\sum_i p_i\, (u_*+v_i)\otimes (u_*+w_i)$ we obtain $\mathcal{E}_*(\omega) = \sum_i p_i \braket{v_i,w_i}_{\mathcal{E}_*}$, and the claim follows from the inequalities
\bbb
\braket{v_i,w_i}_{\mathcal{E}_*} = - \braket{v_i,-w_i}_{\mathcal{E}_*} \geq -m_-\, ,
\eee
valid because $u_*+v_i,u_*-(-w_i)\in C$ and thus $\braket{v_i,-w_i}_{\mathcal{E}_*}\leq k_-$, and 
\bbb
\braket{v_i,w_i}_{\mathcal{E}_*} \leq \sqrt{\braket{v_i,v_i}_{\mathcal{E}_*} \braket{w_i,w_i}_{\mathcal{E}_*}} \leq m_+\, .
\eee
Applying this to a Werner state $\chi_s$ we obtain that a necessary condition for separability is $-m_-\leq \mathcal{E}_*(\chi_s)=(d-1) s\leq m_+$, concluding the proof.
\end{proof}

In order to study the data hiding properties of the family of Werner states, we need to understand the separability conditions at the dual level. Luckily enough, there is no need to repeat any calculation. This is because if $(V,C,u)$ is completely symmetric then $(V^*,C^*,u_*)$ is itself a completely symmetric GPT. Therefore, we start by giving the following `dual' definitions:
\begin{align}
k_\pm^* &\coloneqq\, \max\left\{ s:\, U\pm s \mathcal{E}_*\in C^*_{AB} \right\}\, ,
\label{k+-*} \\
m_+^* &\coloneqq\, \max\left\{\braket{f,f}_{\mathcal{E}}:\ f\in V_0^*,\, u+f\in C^*\right\}, \label{m+*} \\
m_-^* &\coloneqq\, \max\left\{ \braket{f,g}_{\mathcal{E}}:\ f,g\in V_0^*,\, u+f, u-g\in C^*\right\}\, .
\label{m-*}
\end{align}
Exactly as in Proposition~\ref{prop sep Werner}, we obtain
\bb
k_\pm^* \geq \frac{m_\pm^*}{d-1}\, .
\label{k+-* lower bound}
\ee
Moreover, one can prove the following.

\begin{lemma}
In a completely symmetric GPT $(V,C,u)$, the constants $k_\pm$, $k_\pm^*$, $m_\pm$, $m_\pm^*$ defined by~\eqref{k+-},~\eqref{k+-*},~\eqref{m+},~\eqref{m-},~\eqref{m+*},~\eqref{m-*}, satisfy $m_-\leq m_+$, $m_-^*\leq m_+^*$, and moreover
\bb
k_\pm^*\,k_\mp = \frac{1}{d-1}\, ,\qquad 1\,\leq\, m_\pm m_\mp^*\, \leq\, d-1\, .
\label{k m dual}
\ee
\end{lemma}

\begin{proof}
The inequalities $m_-\leq m_+$ and $m_-^*\leq m_+^*$ follow trivially from the definition by applying once Cauchy-Schwartz inequality. Thus, let us proceed to show the relation between $k_\pm$ and $k_\pm^*$. By definition of dual cone, $U\pm s \mathcal{E}_*\in C^*_{AB}$ if and only if $(U+ s \mathcal{E}_*)(\omega)\geq 0$ for all $\omega\in C_{AB}$, which can be assumed to be normalised without loss of generality. However, because of the group symmetry it is enough to test Werner states. In fact, since $(\zeta_g^*\otimes \zeta_g^*)(U+ s \mathcal{E}_*)=U+ s \mathcal{E}_*$ for all $g\in G$, we obtain
\begin{align*}
(U+ s \mathcal{E}_*)(\omega) &= \int_G dg\, \left((\zeta_g^*\otimes \zeta_g^*)(U+ s \mathcal{E}_*)\right) (\omega) \\
&= \int_G dg\ (U+ s \mathcal{E}_*) \left((\zeta_g\otimes \zeta_g) (\omega) \right)\\
&= (U+ s \mathcal{E}_*) (\chi_t) \\
&= 1 + (d-1) s t\, ,
\end{align*}
where $(d-1)t = \mathcal{E}_*(\omega)$. The rightmost side of the above equation is positive for all $\varphi\in [-k_-,k_+]$ if and only if $-\frac{1}{(d-1) k_+}\leq k\leq \frac{1}{(d-1) k_-}$.

Now, let us devote our attention to the bounds on the $m$ quantities in~\eqref{k m dual}. Clearly, it is enough to show that $1\leq m_+ m_-^*\leq d-1$, up to exchanging primal and dual GPT. Consider $f\in V_0^*$ such that $\braket{f,f}_{\mathcal{E}}=1$. Then it is easy to see that $u-\frac{f}{\sqrt{m_+}}\in C^*$, since for all normalised $\omega \in C$, which can be conveniently parametrised as $\omega=u_*+v$ with $v\in V_0$ satisfying $\braket{v,v}_{\mathcal{E}_*}\leq m_+$, one has
\begin{align*}
\Big(u \pm \frac{f}{\sqrt{m_+}}\Big)(\omega) &= 1 \pm \frac{1}{\sqrt{m_+}}\, f(v) \\
&\geq 1 - \frac{1}{\sqrt{m_+}}\, \sqrt{\braket{f,f}_{\mathcal{E}} \braket{v,v}_{\mathcal{E}_*}} \\
&\geq 1 - \sqrt{\braket{f,f}_{\mathcal{E}}}
&\geq 0\, .
\end{align*}
Since $u\, \pm \frac{f}{\sqrt{m_+}}\in C^*$, from the definition~\eqref{m-*} we infer that 
\bbb
m_-^*\geq \braket{\frac{f}{\sqrt{m_+}},\frac{f}{\sqrt{m_+}}}_{\mathcal{E}}=\frac{1}{m_+}\, .
\eee
In order to prove the upper bound $m_+m_-^*\leq d-1$, let us write
\bbb
m_+ \leq (d-1) k_+ = \frac{1}{k_-^*} \leq \frac{d-1}{m_-^*}\, ,
\eee
where we employed in order: the results of Proposition~\ref{prop sep Werner}, the first relation in~\eqref{k m dual} and finally~\eqref{k+-* lower bound}.
\end{proof}

With the tools we have developed so far, we are ready to discuss quantitatively the existence of allowed and separable measurements displaying Werner symmetry.

\begin{prop} \label{prop dual Werner}
In a completely symmetric GPT $(V,C,u)$, for $\alpha,\beta\in\mathds{R}$ the functionals $\left( \alpha U + \beta \mathcal{E}_*,\, (1-\alpha) U - \beta \mathcal{E}_* \right)$ form an allowed measurement if and only if
\bb
- k_-^*\alpha \leq \beta \leq k_+\alpha\, ,\qquad -k_+^*(1-\alpha) \leq \beta \leq k_-^*(1-\alpha)\, .
\label{dual Werner}
\ee
In the $(\alpha, \beta)$-plane, these two conditions identify a parallelogram with vertices
\bb
(0,0),\, (1,0),\ \left(\frac{k_-^*}{k_+^* + k_-^*},\ \frac{k_+^* k_-^*}{k_+^* + k_-^*} \right),\ \left(\frac{k_+^*}{k_+^* + k_-^*},\, -\frac{k_+^* k_-^*}{k_+^* + k_-^*} \right) .
\ee
The separability conditions for the measurement under examination can be deduced from~\eqref{dual Werner} by making the substitutions $k_\pm^*\mapsto \frac{m_\pm^*}{d-1}$.
\end{prop}

\begin{proof}
We omit the details, since the proof consists in a systematic application of the definitions~\eqref{k+-*}, together with the dual conditions to those already given in Proposition~\ref{prop sep Werner}.
\end{proof}

\vspace{2ex}
Equipped with Proposition~\ref{prop dual Werner}, we are ready to explore the data hiding properties of Werner states. Since we did not make any general claim about anything but separability, we will compute the highest data hiding efficiency against separable measurements that is obtainable by using Werner states in Definition~\ref{dh}.

\vspace{2ex}
\begin{thm}[Data hiding with Werner states] \label{dh Werner}
For a composite system made of two copies of the same completely symmetric GPT, the highest data hiding efficiency against separable measurements that is achievable with only Werner states is given by
\bb
\begin{split}
R_{\text{\emph{Werner}}}(\text{\emph{SEP}}) &= \max_{0\neq (a,b)\in\mathds{R}^2}\, \frac{\| a U_* + b \mathcal{E} \|}{\,\| a U_* + b \mathcal{E} \|_{\text{\emph{SEP}}}}\\
&= 1\, +\, \frac{2}{k_+ + k_-}\, \max\left\{\frac{1}{m_+^*} - k_-,\, \frac{1}{m_-^*} - k_+ \right\} .
\end{split}
\label{dh Werner eq}
\ee
\end{thm}

\begin{proof}
Let us start by computing the base norm $\|a U_*+ b \mathcal{E}\|$, for $a,b\in \mathds{R}$. Thanks to Proposition~\ref{prop dual Werner}, we have to test just one nontrivial measurement, namely
\bbb
\left( \frac{k_-^*}{k_+^* + k_-^*}\, U + \frac{k_+^* k_-^*}{k_+^* + k_-^*}\, \mathcal{E},\ \frac{k_+^*}{k_+^* + k_-^*}\, U - \frac{k_+^* k_-^*}{k_+^* + k_-^*}\, \mathcal{E} \right) .
\eee
Thus, using~\eqref{d norm} we find
\begin{align}
\| a U_*+ b \mathcal{E} \| &= \left| \left( \frac{k_-^*}{k_+^* + k_-^*}\, U + \frac{k_+^* k_-^*}{k_+^* + k_-^*}\, \mathcal{E}\right) \left( a U_* + b\mathcal{E}\right)\right| \nonumber \\
&\qquad + \left|\left( \frac{k_+^*}{k_+^* + k_-^*}\, U - \frac{k_+^* k_-^*}{k_+^* + k_-^*}\, \mathcal{E}\right) \left( a U_* + b \mathcal{E}\right)\right| \label{base Werner 1} \\[1ex]
&= \left| \frac{k_-^*}{k_+^* + k_-^*}\, a + \frac{k_+^* k_-^*}{k_+^* + k_-^*}\, (d-1) b\right| \nonumber \\
&\qquad + \left| \frac{k_+^*}{k_+^* + k_-^*}\, a - \frac{k_+^* k_-^*}{k_+^* + k_-^*}\, (d-1) b\right| \label{base Werner 2}\\[1ex]
&= \max\left\{ |a|,\ \frac{| 2(d-1)\,b\, k_+^* k_-^* + a (k_-^* - k_+^*) |}{k_+^* + k_-^*} \right\} , \label{base Werner 3}
\end{align}
where we used the elementary formula $|\alpha+\beta|+|\gamma-\beta|=\max\{ |\alpha+\gamma|,\, |2\beta+\alpha-\gamma| \}$ in the last step. As suggested by Proposition~\ref{prop dual Werner}, we can obtain the separability norm by replacing everywhere $k_\pm^*$ with $\frac{m_\pm^*}{d-1}$:
\bb
\| a U_*+ b \mathcal{E} \|_{\text{SEP}} = \max\left\{ |a|,\ \frac{| 2\,b\, m_+^* m_-^* + a (m_-^* - m_+^*) |}{m_+^* + m_-^*} \right\}\, . \label{sep Werner}
\ee
Observe that plugging~\eqref{k m dual} into~\eqref{base Werner 3} yields the somehow more handy expression
\bb
\| a U_*+ b \mathcal{E} \| = \max\left\{ |a|,\ \frac{| 2\,b\, + a (k_- - k_+) |}{k_+ + k_-} \right\} .
\label{base Werner}
\ee
Now, computing the ratio $\max_{0\neq (a,b)\in\mathds{R}^2}\, \frac{\| a U_* + b \mathcal{E} \|}{\,\| a U_* + b \mathcal{E} \|_{\text{SEP}}}$ amounts to minimising~\eqref{sep Werner} for a fixed value of~\eqref{base Werner}. It is very easy to see that such a minimum is always achieved for pairs $(a,b)$ such that the two expressions in the maximum appearing in~\eqref{sep Werner} coincide. Up to scalar multiples, there are exactly two such pairs, namely those satisfying $b=\pm \frac{a}{m_\mp^*}$. By substituting these values into~\eqref{base Werner} and dividing by~\eqref{sep Werner}, we find
\begin{align}
R_{\text{Werner}}(\text{SEP}) &= \max_{0\neq (a,b)\in\mathds{R}^2}\, \frac{\| a U_* + b \mathcal{E} \|}{\,\| a U_* + b \mathcal{E} \|_{\text{SEP}}} \label{dh Werner 0}\\
&= \max\left\{ 1,\ \frac{\big| \frac{2}{m_-^*}\, + k_- - k_+ \big|}{k_+ + k_-},\ \frac{\big| -\frac{2}{m_+^*}\, + k_- - k_+ \big|}{k_+ + k_-} \right\} . \label{dh Werner 1}
\end{align}
Now, using~\eqref{k m dual} it is easy to see that 
\begin{align*}
\frac{2}{m_\pm^*} \pm (k_+-k_-) &\geq \frac{2}{(d-1) k_\pm^*} \pm (k_+-k_-) \\
&= 2 k_\mp \pm (k_+-k_-) \\
&= k_+ + k_-\, .
\end{align*}
Thanks to the above inequalities, we can further simplify~\eqref{dh Werner 1} to
\begin{align*}
R_{\text{Werner}}(\text{SEP})\, &=\, \max\left\{ \frac{\big| \frac{2}{m_-^*}\, + k_- - k_+ \big|}{k_+ + k_-},\ \frac{\big| -\frac{2}{m_+^*}\, + k_- - k_+ \big|}{k_+ + k_-} \right\}\, =\\
&=\, \frac{1}{k_+ + k_-}\, \max\left\{ \frac{2}{m_+^*} + (k_+-k_-),\ \frac{2}{m_-^*} - (k_+-k_-) \right\}\, =\\
&=\, 1\, +\, \frac{2}{k_+ + k_-}\, \max \left\{ \frac{1}{m_+^*} - k_-,\ \frac{1}{m_-^*} - k_+ \right\} ,
\end{align*}
finally proving~\eqref{dh Werner eq}.
\end{proof}

\vspace{2ex}
The above result shows how the maximal data hiding ratio obtainable by employing only Werner states depends on just few geometric parameters characterising the model under examination. The usefulness of this theorem rests on its immediate applicability to several natural classes of highly symmetric GPTs. Since computing the relevant parameters is often a simple and intuitive task, as we shall see in a moment,~\eqref{dh Werner eq} gives a quick lower bound on all the data hiding ratios against locally constrained sets of measurements.
To demonstrate the power of Theorem~\ref{dh Werner}, we apply it to the symmetric models that we have examined so far: classical probability theory, quantum mechanics, spherical model, and cubic model.

\begin{itemize}

\item \emph{Classical probability theory} (Subsection~\ref{subsec2 ex class}). By looking at the definition~\eqref{classical}, it is easy to see that the relevant group here is the symmetric group $S_d$. We have $u=(1,\ldots,1)^T=d\, u_*$, and the action on the subspace $V_0=\{x\in\mathds{R}^d:\ \sum_i x_i=0\}$ is well-known to be irreducible. This can be either proved with elementary tools or shown in one line via character theory and Burnside's formula. In fact, denote with $\eta:S_d\rightarrow \mathds{R}^{d\times d}$ the standard representation of $S_d$ on $\mathds{R}^d$. Since we can exhibit an explicit one-dimensional irrep (vectors of constant entries), it is enough to show that $\frac{1}{d!} \sum_{\pi\in S_d} \left( \Tr \eta(\pi)\right)^2 = \langle \chi_\eta, \chi_\eta\rangle=2$. Observing that $\Tr \eta(\pi)$ is the number of elements fixed by $\pi$ and applying Burnside's formula to the natural action of $S_d$ on $\{1,\ldots, d\}^{\times 2}$, which has exactly $2$ orbits, we find $\frac{1}{d!} \sum_{\pi\in S_d} \left( \Tr \eta(\pi)\right)^2=2$.

Going back to the analysis of classical probability theory as a completely symmetric model, and exploiting the identification between elements in the tensor product and $d\times d$ real matrices we can also write $U=uu^T= d^2 U_*$ and $\mathcal{E}=\mathds{1}-\frac{1}{d}uu^T=\mathcal{E}_*$. Furthermore, observe that since the upper and lower bound in~\eqref{CAB bound} coincide, the bipartite cone is composed of all the entrywise positive matrices, collectively denoted by $\mathds{R}^{d\times d}_+$. Computing the relevant quantities is now elementary:
\begin{align*}
m_+^* \!&= \max\left\{v^Tv:\, u+v\in\mathds{R}^d_+,\ \sum\nolimits_i v_i=0 \right\} =  d(d-1) = d^2 m_+ ,\\
m_-^* \!&= \max\Big\{v^Tw:\, u\!+\!v, u\!-\!w\!\in\!\mathds{R}^d_+,\, \sum\nolimits_i v_i\!=\!\sum\nolimits_j w_j = 0 \Big\} = d = d^2 m_- ,\\
k_+ \!&= \max\left\{ s:\, u_*u_*^T+ s \left(\mathds{1} -d u_*u_*^T\right)\in \mathds{R}^{d\times d}_+\right\} = \frac{1}{d} ,\\
k_- \!&= \max\left\{ s:\, u_*u_*^T- s \left(\mathds{1} -d u_*u_*^T\right)\in \mathds{R}^{d\times d}_+\right\} = \frac{1}{d(d-1)} ,
\end{align*}
Applying~\eqref{dh Werner eq}, we find $R_{\text{Werner}}^{\text{Cl}}(\text{SEP})=1$, which is expected since the collapse of the hierarchy~\eqref{CAB bound} when either of the two cones is simplicial forbids the existence of data hiding altogether.

\item \emph{Quantum mechanics} (Subsection~\ref{subsec2 ex QM}). We refer to~\eqref{quantum} for the notation. The symmetry group in quantum mechanics is $U(n)$, the group of unitary $n\times n$ matrices. It is easy to see that we can choose $u=\mathds{1}=n\, u_*$ and that the orthogonal complement to this trivial representation (that is, the space of traceless Hermitian matrices) is irreducible. In fact, this follows already from the same result for the classical case, since permutation matrices are also unitaries.

For a bipartite system we see that $U=\mathds{1}=n^2\, U_*$ and $\mathcal{E}=F-\frac{\mathds{1}}{n}=\mathcal{E}_*$, where $F\ket{\alpha\beta}=\ket{\beta\alpha}$ denotes the flip operator. If the composite is assembled according to the standard quantum mechanical rule~\eqref{cone bipartite quantum}, we find
\begin{align*}
m_+^* \!&= \max\left\{\Tr X^2:\, \mathds{1}+X\geq 0,\, \Tr X\!=\!0 \right\}\, =\, n(n-1)\, =\, n^2 m_+\, ,\\
m_-^* \!&= \max\left\{\Tr XY:\, \mathds{1}\!+\! X,\, \mathds{1}\! -\! Y\!\geq\! 0,\, \Tr X\!=\!\Tr Y\!=\!0 \right\} = n = n^2 m_- ,\\
k_+ \!&= \max\left\{ s:\, \mathds{1}/ n^2+ s \left( F\!-\!\mathds{1}/n \right)\geq 0 \right\} = \frac{1}{n(n+1)} ,\\
k_- \!&= \max\left\{ s:\, \mathds{1}/ n^2 - s \left( F\!-\!\mathds{1}/n \right)\geq 0 \right\} = \frac{1}{n(n-1)} ,
\end{align*}
from which it follows easily $R_{\text{Werner}}^{\text{QM}}(\text{SEP})=n$. According to Theorem~\ref{dh QM}, this is even the optimal data hiding ratio against all separable measurements.

If the composite is formed with the minimal tensor product rule (what we called `$W$-theory' in Definition~\ref{W theory}, we write instead
\begin{align*}
k_+ &= \frac{m_+}{n^2-1} = \frac{1}{n(n+1)}\, ,\\
k_- &= \frac{m_-}{n^2-1} = \frac{1}{n(n^2-1)}\, ,
\end{align*}
and thus $R_{\text{Werner}}^{\text{W}}(\text{SEP})=2n-1$, in accordance with Proposition~\ref{dh W}.

\item \emph{Spherical model} (Subsection~\ref{subsec2 centr symm}). Using the same notation as in~\eqref{spherical}, we see that in this case the relevant group is $O(d-1)$, the set of $(d-1)\times (d-1)$ orthogonal matrices acting on the last $d-1$ components of $\mathds{R}^d$. We can choose $u=(1,0,\ldots,0)^T=u_*$, and again the orthogonal complement is irreducible as follows from the same result for classical probability theory. For a bipartite system $U=uu^T=U_*$ and $\mathcal{E}=\myhat{\mathds{1}}_{d-1}=\mathcal{E}_*$. If the composite is formed with the minimal tensor product rule, then it is very easy to verify that
\begin{align*}
m_\pm^* &=\, \max\{ v^Tv:\ u+v\in C_d\} = 1\, = m_\pm\, , \\
k_\pm &= \frac{m_\pm}{d-1} = \frac{1}{d-1}\, ,
\end{align*}
so that $R_{\text{Werner}}^{\text{Sph}}(\text{SEP})=d-1$, which coincides with the lower bound for the data hiding ratio against separable measurements given in Corollary~\ref{dh sph}.

\item \emph{Cubic model} (Example~\ref{cubic}). We follow the notation previously established. The symmetry group of the cubic model in $d$ dimensions is simply the symmetry group of the $(d-1)$-dimensional hypercube. As for the spherical model, we have $u=(1,0,\ldots,0)^T=u_*$, $U=uu^T=U_*$ and $u^\perp$ is irreducible. Furthermore, we can choose $\mathcal{E}=\myhat{\mathds{1}}_{d-1}=\mathcal{E}_*$. However, even for a minimal tensor product composite we have
\begin{align*}
m_\pm^*&=\, \max\{ v^Tv:\ v\in\mathds{R}^{d-1},\ |v|_1\leq 1 \} = 1\, ,\\
m_\pm &=\, \max\{ v^Tv:\ v\in\mathds{R}^{d-1},\ |v|_\infty\leq 1 \} = d-1\, ,\\
k_\pm &=\, \frac{m_\pm}{d-1} = 1\, ,
\end{align*}
and therefore $R_{\text{Werner}}^{\text{G}}(\text{SEP})=1$. This example shows how it might be the case that Werner states do not display any data hiding property, despite the fact that there is global data hiding in the cubic model, as shown by Proposition~\ref{dh cubic}.

\end{itemize}

\section{Conclusions} \label{sec3 conclusions}

Throughout this chapter, we have presented a general theory of data hiding in bipartite systems composed of general probabilistic theories. It significantly generalises ideas and results from quantum mechanics; in particular, we were able to determine the maximum so-called data hiding ratio in terms of the state space dimension, by making a connection between this problem and Grothendieck's tensor norms. This maximum is essentially attained for GPTs over spherical cones.

Inspired by the prominent role played by Werner states in quantum mechanical data hiding, we investigated Werner-like states in classes of theories with symmetric state spaces. By using these states as ansatzes, we could find a general lower bound on the data hiding ratio that depends on few geometrically meaningful parameters.

For quantum mechanics on finite-dimensional spaces we proved a new upper bound on the data hiding ratio against LOCC operations by exploiting the celebrated quantum teleportation protocol. We saw how this improves previous results and definitively settles the problem of determining the optimal data hiding ratio against LOCC for fixed local dimensions. However, the same problem for the smaller set of local operations remains open, as we showed that none of the bounds that we provide or that are available in the literature is generally optimal.

Perhaps surprisingly, data hiding ratios in quantum mechanics are not as large as the maximal ones, being of the order of the square root of the latter, which are exhibited by spherical cones.
This answers the general question we posed in Subsection~\ref{subsec2 fundamental}, point (iii), for a particular kind of non-classical phenomenon, i.e. data hiding.

If one were to summarise the results of our research in one single sentence, one could say that the laws of Nature, which move the sun and the other stars, are intrinsically non-classical, but not as non-classical as they could have been.



\part{Some aspects of quantum entanglement}

\chapter{Bipartite depolarising maps} \label{chapter4}


\section{Introduction} \label{sec4 introduction}

We now leave behind the technically fierce world of general probabilistic theories, and let our modest investigation set course for the more kindly and familiar waters of quantum information theory.
Namely, while most of Chapter~\ref{chapter2} has been devoted to discussing the concept of entanglement in the GPT framework, in this second part of the thesis we focus instead on some aspects of quantum entanglement. In this chapter, in particular, we address the problem of characterising the entanglement transformation properties of a certain class of quantum channels acting on a bipartite system. The maps we chose to analyse, which generalise the well-known depolarising channels~\cite{Horodecki1999}, are good candidates to model simple forms of noise acting on quantum devices.

The present chapter is organised as follows. The rest of this section is devoted to discussing the main problem and the motivation behind it (Subsection~\ref{subsec4 the problem}) and to highlighting our original contributions (Subsection~\ref{subsec4 our contributions}). After reviewing in Section~\ref{sec4 preliminaries} some basic concepts about notable classes of maps in quantum information, with special attention devoted to the depolarising channel and to its generalisation, we move on to the investigation of the properties of the maps we introduce. Section~\ref{sec P} is dedicated to positivity, the most basic requirement for maps that are of relevance for quantum information. Section~\ref{sec CP} is then devoted to complete positivity, while the main goal of Section~\ref{sec EB} is to find necessary and sufficient conditions under which the maps under examination are entanglement-breaking. The characterisation of entanglement annihilation is the subject of the following Section~\ref{sec EA}, where we present what is probably the main result of the chapter, Theorem~\ref{EA}, and discuss its applications to the solution of some open problems recently raised in~\cite{EA3, EA4}. Finally, in Section~\ref{sec4 conclusions} we draw the conclusions and discuss some future perspectives.

\subsection{The problem} \label{subsec4 the problem}

In quantum information science, the notion of entanglement plays a major role from both the foundational and the operational point of view~\cite{Horodecki-review, NC}, and thus its study constitutes one of the cornerstones of the field. Entanglement is however a very fragile resource and once distributed, the local participants in any kind of communication protocol can never increase it via local operations, even if assisted by unlimited classical communication (i.e. with so-called LOCC protocols, see Subsection~\ref{subsec2 measur bip}). Much effort has thus been invested into characterising the resource of entanglement and its manipulation under LOCC operations~\cite{Horodecki-review,siewert}. 

This programme can be carried out in several ways. For example, the problems of experimentally detecting~\cite{siewert} or quantifying entanglement from an operationally meaningful point of view~\cite{siewert} have been considered. Another related question concerns the transformation of entanglement. Naturally, the importance of this problem stems primarily from physics. In fact, on the one hand we want to manipulate our resource by implementing any given transformation that is physically accessible, in order to make it usable for running quantum information protocols. On the other hand, once we have some amount of entanglement stored in our quantum device, we should also be able to counter the deleterious effects of any spurious interaction with the external world. For an account of the progress that has been made in these directions, we point the reader to~\cite{Horodecki-review}.

The very basic question that we tackle in this chapter is the following: \emph{given a model of the noise acting on our entangled state, how long do we have to wait for the entanglement to be completely lost?} More loosely speaking, \emph{how strong does the noise have to be in order to destroy the initial entanglement?} Here we are taking a passive viewpoint, but of course equivalent questions can be asked in a dynamical setting, where one sends states through physical channels (such as optic fibres) instead of them being subjected to a harmful external noise.

Of course, we are going to answer this question in a very special case, i.e. for a very special model of noise. The reason behind this restriction will become clearer soon. However, the reason why we should impose \emph{some kind} of restriction, that is, why we can not solve the problem in a closed form in full generality, can already be appreciated.
In fact, as we mentioned in Section~\ref{sec2 sep problem}, the problem of deciding whether a given quantum state is entangled or not (called the \emph{separability problem}) is in general computationally NP-hard (under certain assumptions on the approximation error)~\cite{GurvitsNPhard}. This precludes a closed solution of the above problem in full generality, but nonetheless some paradigmatic cases can be addressed in a practically useful way. Complete solutions for the separability problem indeed exist for various special classes of states~\cite{Werner, HorodeckiPPT, Horodecki1999, Werner-symmetry, SPA, MaxAbelian, Marcus1, axi, Buchy, SepBos}.
A central tool for addressing these problems are positive maps, the canonical example being transposition, from which the probably most widely used entanglement test derives~\cite{PeresPPT}.

Let us now delve a bit more into the mathematical formulation of the aforementioned questions. In general, under reasonable assumptions~\cite{NC} any physical transformation of a quantum system can be represented by a completely positive (CP) map.
Studying which CP maps preserve and more importantly which maps adversely affect entanglement is therefore our main goal here. Two important examples of deleterious action are constituted by: (i) maps that act on a bipartite system as a whole and break any entanglement between the two parties, called \emph{entanglement-annihilating}~\cite{EA1}; and (ii) maps that act on a single quantum system and disentangle it from the rest of the world, called \emph{entanglement-breaking}~\cite{HSR}.
 
In this chapter, we tackle the above problems for a special class of maps acting on a bipartite system that are suitable generalisations of \emph{depolarising channels}~\cite{erasure, entanglement-assisted, Bruss-dep, King-dep-capacity}. The one-parameter class of depolarising channels provides undoubtedly some of the most physically significant examples of quantum channels, its importance ultimately stemming from the fact that it models the process of addition of white noise to the input system. At the same time, its mathematically simple structure allows to find a closed-form solution to many problems~\cite{Horodecki1999, King-dep-capacity}. Indeed, the depolarising channel $\Delta_\lambda$ acting on an $n$-level quantum system can be expressed as 
\bb
\Delta_\lambda \coloneqq \lambda I + (1-\lambda) \frac{\mathds{1}}{n} \tr\, ,
\label{dep}
\ee
meaning that its action on an input density matrix $\rho$ is given by $\Delta_\lambda (\rho) \coloneqq \lambda \rho + (1-\lambda)\frac{\mathds{1}}{n}$. More details will be provided in Subsection~\ref{subsec4 dep}.

\subsection{Our contributions} \label{subsec4 our contributions}

The material covered in this chapter is taken from the homonymous paper~\cite{LLMH}:
\begin{itemize}
\item L. Lami and M. Huber. Bipartite depolarizing maps. \emph{J. Math. Phys.}, 57(9):092201, 2016.
\end{itemize}

As anticipated, here we address the problems discussed in the preceding Subsection~\ref{subsec4 the problem} for certain maps, that are chosen as suitable generalisations of those in~\eqref{dep}. Since we want them to act on a bipartite system $AB$ and to contain the same terms as in~\eqref{dep}, they will be most naturally expressed as a linear combination of four elementary maps of the form $\mathcal{N}^i_A\otimes \mathcal{N}^j_B$ ($i,j=0,1$), where $\mathcal{N}^i$ can be either the identity channel $I$ ($i=0$) or the completely depolarising map $\mathds{1}\tr$ ($i=1$). Therefore, we will consider the family of maps
\bb
\Gamma[\alpha,\beta,\gamma] \coloneqq \mathds{1}_{AB} \tr_{AB} + \alpha\, \mathds{1}_{A} \tr_A \otimes I_B + \beta\, I_A\otimes \mathds{1}_{B} \tr_B + \gamma\, I_{AB}\, .
\label{Phi}
\ee

\begin{note}
In the original paper~\cite{LLMH}, we use the letter $\Phi$ instead of $\Gamma$ for the maps in~\eqref{Phi}. This naturally creates a potential conflict with the notation for maximally entangled states, which are denoted there by $\varepsilon$. Here, we chose to stick to the more customary convention of reserving the letter $\Phi$ for the maximally entangled state, and thus we chose $\Gamma$ for the bipartite depolarising maps in~\eqref{Phi}.
\end{note}

It is immediately obvious that maps of the form~\eqref{Phi} generalise those in~\eqref{dep} in the case of a bipartite system. The main motivation for studying this class comes from the fact that it can accommodate a variety of models for the noise acting on that system. As a first and most obvious example, consider the case of the two shares of an entangled state being stored in a single device, which is then subjected to white noise of the form~\eqref{dep}. In our case, such noise takes the form $(\Delta_\lambda )_{AB} \propto \Gamma \left[0,0,\frac{n_A n_B\lambda}{1-\lambda}\right]$. The second example we want to consider corresponds to the case of the two subsystems being sent through different depolarising channels, or -- equivalently -- being subjected to different local white noises in physically separated storing devices. The relevant channel becomes $(\Delta_\lambda)_A\otimes (\Delta_\mu)_B \propto \Gamma \left[ \frac{n_B \mu}{1-\mu},\, \frac{n_A \lambda}{1-\lambda},\, \frac{n_A n_B \lambda \mu}{(1-\lambda)(1-\mu)} \right]$, thus falling again within the class~\eqref{Phi}.

Despite the richer structure of the maps $\Gamma$ in~\eqref{Phi}, in this chapter we determine exactly the parameter regions for which these maps are: positive (Theorem~\ref{positivity}); completely positive (Theorem~\ref{complete positivity}); entanglement-breaking (Theorem~\ref{EB}) and entanglement-annihilating (Theorem~\ref{EA}). 
The results concerning the complete positivity and the entanglement-breaking regions are extensions of those in~\cite{Vollbrecht02, Piani}. The full characterisation we give here was not available before, but a modified version of the argument in~\cite{Vollbrecht02} can cover the case $n_A=n_B$. In~\cite{Piani}, a class of states is considered, that intersects the Choi states of the maps~\eqref{Phi} in a one-dimensional subspace, for all $n_A,n_B$.

On the contrary, the characterisation of the entanglement-annihilating region for the maps $\Gamma$ (Theorem~\ref{EA}) is fully original. Besides its physical significance, our initial motivation for tackling this question was to answer the main open problems in~\cite{EA3, EA4}. There, the problem of determining all pairs $\lambda, \mu$ for which the local depolarising noise $\Delta_\lambda\otimes \Delta_\mu$ is entanglement-annihilating for the system $AB$ was considered. However, the bounds established in~\cite{EA3, EA4} are not strong enough to solve the problem completely. Instead, an application of Theorem~\ref{EA} yields the desired characterisation (Corollary~\ref{EA cor}), showing in particular that a product of depolarising maps is entanglement-annihilating iff it is (positive and) PPT-inducing.

As by-products of the proofs, some minor results are found, and some interesting techniques are devised and applied to the quantum separability problem. An important tool in our analysis is in fact the observation that separable matrices are closed with respect to the so-called \emph{Hadamard} (or entry-wise) \emph{product}~\cite{HJ1, HJ2}, essentially because the latter can be stochastically implemented via local operations and classical communication (LOCC). We believe that this can be a useful technique in entanglement theory, so in Subsection~\ref{subsec4 Hadamard} we take a small detour to explore some of its properties.
Along the way to the main results of the chapter, we stumble upon some small results previously found by others. For instance, we retrieve a fairly simple example of a PPT entangled state~\eqref{PPT ent state} living in a system $AB:A'B'$ that was originally discovered in~\cite{Ishizaka1}, and make use of a positive indecomposable map~\eqref{indecomposable map} acting on a bipartite system $AB$, which is an instance of a more general construction presented in~\cite[\S IX]{Piani}.

\section{Preliminaries} \label{sec4 preliminaries}

\subsection{Classes of maps in quantum information} \label{subsec4 maps}

As discussed in Subsection~\ref{subsec2 ex QM}, states of an $n$-level quantum system are represented by $n\times n$ positive semidefinite matrices $\rho$ with normalised trace (i.e. subjected to the constraint $\Tr \rho=1$), generically called \textbf{density matrices}. The real span of the set of density matrices is the set of $n\times n$ Hermitian matrices $\mathcal{H}_n$. The extremal points of the set of quantum states are rank-one orthogonal projectors $\ket{\psi}\!\!\bra{\psi}$, called \textbf{pure states}.
Joining two quantum systems $A$ and $B$ of $n_A$ and $n_B$ levels, respectively, yields another quantum system of $n_A n_B$ levels. Therefore, bipartite density matrices are naturally thought of as living in $\mathcal{H}_{n_A} \otimes \mathcal{H}_{n_B}\simeq \mathcal{H}_{n_A n_B}$. Often, we will say for short that $AB$ is an $n_A\times n_B$ quantum system.

Bipartite pure states $\ket{\psi}_{AB}$ admit a remarkable decomposition in terms of local bases. Namely, using the notation $n\coloneqq \min\{n_A, n_B\}$, one can find orthonormal vectors $\ket{e_1}_A,\ldots, \ket{e_n}_A$ and $\ket{f_1}_B, \ldots, \ket{f_n}_B$ on the first and second subsystem, respectively, as well as non-negative scalars $\lambda_1, \ldots , \lambda_n \geq 0$, such that
\bb
\ket{\psi}_{AB} = \sum_{i=1}^n \sqrt{\lambda_i} \ket{e_i}_A \ket{f_i}_B\, ,
\label{Schmidt decomposition}
\ee
a representation known as \textbf{Schmidt decomposition}. Observe that if $\ket{\psi}$ is appropriately normalised then $\sum_{i=1}^n \lambda_i =1$.

In this chapter, we are mainly interested in linear maps acting on density matrices, or equivalently on $\mathcal{H}_n$. The reason why this is the case, remember, is that we care about transformations undergone by quantum states when some external noise is acting on them, either because they are stored in a non-isolated environment, or because they are sent through an imperfect channel to a distant agent. For a complete review of the topic, we refer the reader to the classic textbook~\cite[VIII]{NC}.
A very natural requirement to impose on a map that aims to represent a physical transformation is positivity. A map $\Lambda :\mathcal{H}_n\rightarrow \mathcal{H}_m$ is called \textbf{positive} if $\Lambda(\rho)\geq 0$ whenever $\rho\geq 0$ (where $X\geq 0$ stands for `$X$ is positive semidefinite'). This is a special instance of the definition we gave in Subsection~\ref{subsec2 Woronowicz} for general maps between ordered vector spaces.

As it turns out, this requirement is not strong enough to ensure that $\Lambda$ is physically implementable. In fact, we need to make sure that also the partial application of $\Lambda$ on one share of a bipartite system in a state $\rho$, that we represent as $(\Lambda \otimes I)(\rho)$, yields a physical state. This is the case iff $\Lambda \otimes I_k$ is a positive map for all integers $k$, where $I_k$ stands for the identity map acting on $\mathcal{H}_k$. A map satisfying the above requirement is usually called \textbf{completely positive} (CP), a concept we already defined in more general terms in Subsection~\ref{subsec2 measur bip}. This is a strictly stronger requirement than positivity, a classic example showing this being the transposition map $T: \mathcal{H}_n\rightarrow \mathcal{H}_n$ that acts as $T(X)\coloneqq X^T$. The transposition operation plays a relevant role in the theory, so relevant that it deserves some dedicated nomenclature: a map $\Lambda$ is called \textbf{completely copositive} (coCP) if $T\Lambda$ (equivalently, $\Lambda T$) is completely positive.

Despite sounding a bit less natural at first glance, complete positivity is actually much more easily checkable than positivity, since one can show that a map $\Lambda : \mathcal{H}_n\rightarrow \mathcal{H}_m$ is completely positive iff its \textbf{Choi state}
\bb
R_\Lambda \coloneqq \left( \Lambda \otimes I\right) (\Phi)
\label{Choi state}
\ee
is positive semidefinite, a statement known as \textbf{Choi's theorem}~\cite{Choi}. Here, $\Phi$ denotes the rank-one projector onto the \textbf{maximally entangled state}
\bb
\ket{\Phi} \coloneqq \frac{1}{\sqrt{n}} \sum_{i=1}^n \ket{ii}\, .
\label{max ent state}
\ee
As an easy corollary, one sees that $\Lambda \otimes I_k$ is positive for all integers $k$ iff it is positive for $k=n$ (the dimension of the input Hilbert space of $\Lambda$).
The correspondence between maps and states expressed by~\eqref{Choi state} is usually called \textbf{Choi-Jamiolkowski isomorphism}.

A completely positive map $\Lambda$ that is also \textbf{trace-preserving}, meaning that $\Tr \Lambda (X) = \Tr X$ for all Hermitian $X$, is called a \textbf{quantum channel}. Physically, quantum channels are exactly the state transformations that one obtains by letting the input system interact with an ancillary system prepared in an fixed state for some time (so that system and ancilla together undergo a unitary evolution), and by discarding the ancilla afterwards. This is known as \textbf{Stinespring's dilation theorem}~\cite{Stinespring}.

Despite not representing directly any physical transformation, positive maps constitute a valuable tool for entanglement detection. We remind the reader the definition given in Subsection~\ref{subsec2 bipartite} and (for the quantum case) in Subsection~\ref{subsec2 ex QM}: a state $\rho_{AB}$ is called \textbf{separable}~\cite{Werner} if there are collections of local states $\sigma^i_A$ and $\tau^i_B$ and probabilities $p_i$ such that
\bb
\rho_{AB} = \sum_i p_i\, \sigma^i_A \otimes \tau^i_B\, ,
\label{separable state}
\ee
and is called \textbf{entangled} otherwise. We already defined the cone of unnormalised quantum separable states in~\eqref{separable}. As we saw in Subsection~\ref{subsec2 Woronowicz}, already in~\cite{HorodeckiPPT} it was proved that a bipartite quantum state $\rho$ is separable iff $(\Lambda \otimes I)(\rho)\geq 0$ for all positive maps $\Lambda$, a statement that in Subsection~\ref{subsec2 Woronowicz} we called -- following the terminology of~\cite{GurvitsBarnum} -- \textbf{Woronowicz criterion}. As the reader may remember, an equivalent formulation of the criterion involves observables instead of maps: $\rho$ is separable iff $\Tr \rho W\geq 0$ for all $W\in\mathcal{H}_{n_A n_B}$ that are \textbf{entanglement witnesses}, i.e. are such that $\braket{\psi \varphi| W | \psi\varphi}\geq 0$ for all pure states $\ket{\psi}, \ket{\varphi}$ of the local subsystems (Subsection~\ref{subsec2 ex QM}). This reformulation is made possible by the remarkable fact that \emph{entanglement witnesses are precisely the Choi states of positive maps}.

A prominent example of a positive map used as a test for entanglement is the transposition. The resulting criterion, called \textbf{PPT criterion}~\cite{PeresPPT}, states that a separable state $\rho_{AB}$ is such that $\rho_{AB}^{T_A} = (T_A\otimes I_B)(\rho_{AB}) \geq 0$. This test is so strong that it turns out to be even \emph{sufficient} to ensure separability in low-dimensional systems, namely when $n_A n_B\leq 6$~\cite{HorodeckiPPT}. In higher dimensions this is no longer the case, and one can find examples of PPT entangled states.

Of particular relevance here are the properties of maps with respect to the entanglement of the input states. A map $\Lambda : \mathcal{H}_n\rightarrow \mathcal{H}_m$ is called \textbf{entanglement-breaking}~\cite{HSR} if it outputs a separable state whenever it acts on a share of a bipartite system prepared in \emph{any} initial state, i.e. if $(\Lambda \otimes I_k )(\rho)$ is separable for all states $\rho$ of all bipartite $n\times k$ quantum systems. Clearly, an entanglement-breaking map $\Lambda$ is in particular completely positive, and even more, has the property that $\Lambda'\Lambda$ is completely positive for all positive maps $\Lambda'$. As shown in~\cite{HSR}, a map $\Lambda$ is entanglement-breaking iff its Choi state~\eqref{Choi state} is separable. This allows for an operational characterisation of entanglement-breaking maps as those obtained by measuring and re-preparing the input state.

Since we learnt that PPT-ness is a mathematically convenient relaxation of the more elusive concept of separability, we can accordingly relax the notion of entanglement-breaking to that of \textbf{PPT-binding} map. This is defined as a map $\Lambda$ such that $(\Lambda\otimes I_k)(\rho)$ is PPT (instead of separable) for all bipartite input states $\rho$ of all bipartite $n\times k$ quantum systems. As is easy to check thanks to Choi's theorem, this happens iff the Choi state $R_\Lambda$ is PPT.

If the input system is already bipartite, one could look at the entanglement between the two subsystems that compose it, instead of focusing on that between the system and the rest of the world. This attitude leads to the definition of \textbf{entanglement-annihilating} maps~\cite{EA1}, which are maps acting on some bipartite system $AB$ and characterised by the property that $\Lambda_{AB}(\rho_{AB})$ is separable for all density matrices $\rho_{AB}$. Although it is sufficient to check pure input states $\rho_{AB}=\ket{\psi}\!\!\bra{\psi}_{AB}$, a more compact criterion on the same lines as that for entanglement-breaking maps is in general not available. By virtue of the above definitions, one sees that a map $\Lambda$ is entanglement-breaking iff $\Lambda \otimes I_k$ is entanglement-annihilating for all $k$.

We can use again the PPT test to obtain a convenient mathematically convenient relaxation of the entanglement-annihilation condition. Namely, if $\Lambda_{AB}$ is entanglement-annihilating, then $(T_A\otimes I_B)\, \Lambda_{AB}$ is a positive map, with $T_A$ denoting partial transposition on the first subsystem. Maps for which the latter condition holds are called \textbf{PPT-inducing} in~\cite{EA4}.

Observe that positive, completely positive, entanglement-breaking, PPT-binding, entanglement-annihilating and PPT-inducing maps all form convex sets.

\subsection{Depolarising maps} \label{subsec4 dep}

The \textbf{depolarising channel} constitutes one of the simplest yet most physically significant examples of a quantum channel, and has been the subject of extensive study~\cite{erasure, entanglement-assisted, Bruss-dep, King-dep-capacity}. For a real parameter $\lambda$, the channel $\Delta_\lambda$ acting on an $n$-dimensional system is defined by the formula~\eqref{dep}. We remind the reader~\cite{Horodecki1999} that the map $\Delta_\lambda$ is:
\begin{itemize}
\item positive iff $-\frac{1}{n-1}\leq \lambda\leq 1$;
\item completely positive iff $-\frac{1}{n^2-1}\leq \lambda\leq 1$;
\item completely copositive iff $-\frac{1}{n-1}\leq \lambda\leq \frac{1}{n+1}$;
\item entanglement-breaking iff it is at the same time CP and coCP, that is, iff $-\frac{1}{n^2-1}\leq\lambda\leq \frac{1}{n+1}$. 
\end{itemize}

In particular, it is useful to remember that whenever $\Delta_\lambda$ is positive it is also either completely positive or completely copositive (or both). Another notable fact is that $\mathds{1}\tr - I \propto \Delta_{- 1/(n-1)}$ is a completely copositive map, i.e. that $\mathds{1}\tr - T$ (where $T$ denotes transposition) is completely positive. Indeed, its Choi state is proportional to the projector onto the antisymmetric subspace of $\mathds{C}^n\otimes \mathds{C}^n$. These observations are used several times in the rest of the chapter.

\subsection{Bipartite depolarising maps} \label{subsec4 bip dep}

The main subject of study in the present chapter are the `bipartite depolarising maps' defined in~\eqref{Phi}. As a first remark, one could wish to consider a more general linear combination than~\eqref{Phi}, with another parameter for the coefficient of $\mathds{1}_{AB} \tr_{AB}$. However, already the positivity condition easily implies that such a parameter must be non-negative. Up to a normalisation constant, one can take it to be either $1$ or $0$. The latter case can be deduced from the former, because $\alpha\, \mathds{1}_{A} \tr_A \otimes I_B + \beta\, I_A\otimes \mathds{1}_{B} \tr_B + \gamma\, I_{AB}$ is positive iff $\frac{1}{M}\, \mathds{1}_{AB}\tr_{AB} + \alpha\, \mathds{1}_{A} \tr_A \otimes I_B + \beta\, I_A\otimes \mathds{1}_{B} \tr_B + \gamma\, I_{AB}$ is positive for all $M>0$, which is in turn equivalent to $\Gamma[M \alpha, M \beta, M \gamma]$ being positive for all $M>0$.

Observe that the maps $\Gamma$ we defined commute with all the local unitary conjugations, i.e. with all maps of the form $\rho_{AB}\mapsto U_A \otimes V_B\, \rho_{AB}\, U_A^\dag \otimes V_B^\dag$, for all local unitaries $U_A, V_B$. This is the same as saying that the corresponding Choi states are left invariant by conjugations by $U_A\otimes V_B \otimes U^*_{A'}\otimes V^*_{B'}$.
But there is more: up to including a coefficient for the first addend (which is an irrelevant modifications in the sense clarified above), these maps are \emph{all} the maps on $AB$ displaying this feature, as the following reasoning shows. Before delving into the details, we remind the reader that the states of a bipartite system $AA'$ commuting with all the conjugations by local unitaries of the form $U_A\otimes U^*_{A'}$ are exactly the linear combinations of the identity $\mathds{1}_{AA'}$ and the maximally entangled state $\Phi_{AA'}$~\cite{Horodecki1999}. Now, up the application of the Choi-Jamiolkowski isomorphism, we have to prove that the set of $ABA'B'$ states that are left invariant by conjugations by $U_A\otimes V_B \otimes U^*_{A'}\otimes V^*_{B'}$ coincides with the set of linear combinations of $\mathds{1}_{ABA'B'}$, $\mathds{1}_{AA'}\otimes \Phi_{BB'}$, $\Phi_{AA'} \otimes \mathds{1}_{BB'}$, and $ \Phi_{AA'} \otimes \Phi_{BB'}$, i.e. with the tensor product of the two locally invariant subspaces of $AA'$ and $BB$.
This is nothing but a particular instance of a more general phenomenon: given two representations of compact groups $\zeta_i: G_i\rightarrow GL(V_i)$ (with $i=1,2$), the invariant subspace of the \emph{external} tensor product $\zeta_1\boxtimes \zeta_2$ is the tensor product of the two $\zeta_i$-invariant subspaces; in fact, the latter is trivially contained in the former, while at the same time the dimensions are equal, thanks to character theory:
\begin{align*}
\braket{\chi_1,\, \chi_{\zeta_1\boxtimes\zeta_2}} &= \int dg_1 dg_2\ \Tr [(\zeta_1\otimes\zeta_2)\,(g_1,g_2)] \\
&= \int dg_1 dg_2\ \Tr \zeta_1(g_1)\, \Tr \zeta_2(g_2) \\
&= \braket{\chi_1, \chi_{\zeta_1}} \braket{\chi_1, \chi_{\zeta_2}} .
\end{align*}
In our case, the two groups are $G_1=SU(n_A)$ and $G_2=SU(n_B)$, the two spaces are the set of Hermitian operators on $AA'$ and $BB'$, while the representations are defined through $\zeta_1(U)(X_{AA'}) = U_A\otimes U^*_{A'}\, X_{AA'}\, U^\dag_A\otimes U^T_{A'}$ and analogously for $BB'$.

Note that for $n_A=n_B$ the separability properties of the Choi states corresponding to~\eqref{Phi} have been already considered in~\cite{ChrusKoss}. In that paper, the entanglement-breaking conditions for~\eqref{Phi} are found under the above simplifying assumption $n_A=n_B$. However, we will see that the general case $n_A\neq n_B$ is in a sense more interesting, because new phenomena such as the existence of PPT entangled states appear, as already observed in~\cite{Piani}.
A family similar to the Choi states of~\eqref{Phi} but possessing $U_A\otimes V_B \otimes U_{A'}\otimes V^*_{B'}$ instead of $U_A\otimes V_B \otimes U^*_{A'}\otimes V^*_{B'}$ symmetry was an important tool in~\cite{Vollbrecht02}. Translating the results of that paper onto the channel level would produce straightforwardly another $3$-parameter class of maps for which necessary and sufficient entanglement-breaking conditions are available.
As usual, we will denote by $n$ the minimum between the two local dimensions $n_A, n_B$, i.e. the maximum Schmidt rank of a global pure state. A trick that turns out to be useful in analysing the above $\Gamma[\alpha,\beta,\gamma]$ requires the construction of the following family of maps acting on an $n$--dimensional system:
\bb
\chi[a,c] \coloneqq \mathds{1}_{n} \tr + a\, D + c\, I\ ,  \label{chi}
\ee
where $D$ is the projection onto the diagonal, i.e. $D(X)=\mathds{1}\circ X$, with the symbol $\circ$ denoting Hadamard or entry-wise product, defined by $(X\circ Y)_{ij}\coloneqq X_{ij} Y_{ij}$.

\subsection{Entanglement properties of the Hadamard product} \label{subsec4 Hadamard}

Now that we mentioned it, let us take a small detour to explore some of the properties exhibited by the Hadamard product in relation to the separability problem. For an introduction to other basic properties, we refer the reader to~\cite[Chapter 5]{HJ2}.
The main feature of this seemingly bizarre operation is that the cone of positive matrices is closed with respect to it; in other words, if $M\geq 0$ and $N\geq 0$ then $M\circ N\geq 0$ (this is usually called \emph{Schur product theorem}). Actually, it is easily verified by means of Choi's theorem that for a given $M\geq 0$ the corresponding \emph{Hadamard channel} $\phi_M(\cdot)\coloneqq M\circ(\cdot)$ is not only positive but even completely positive.

Thanks to the Schur product theorem, given two quantum states $\rho,\sigma$, it makes sense to consider the (unnormalised) state $\rho\circ\sigma$. We stress that the definition of Hadamard product is explicitly dependent on the basis we choose to represent the operators $\rho,\sigma$ as matrices. As it turns out, if $\rho_{AB},\sigma_{AB}$ are bipartite states and we fix a product basis to represent them (as we will always do from now on), then the Hadamard product behaves well with respect to the entanglement properties of the input states. Namely, the following simple yet very useful lemma holds.

\begin{lemma} \label{Hadamard sep lemma}
Let $\rho_{AB}, \sigma_{AB}$ be two bipartite density matrices represented with respect to a product basis. Then:
\begin{enumerate}[(a)]
\item $\rho_{AB}\circ\sigma_{AB}$ can be obtained stochastically from $\rho_{AB}\otimes\sigma_{A'B'}$ through a local measurement on the bipartite system $AA':BB$;
\item in particular, if $\rho_{AB}, \sigma_{AB}$ are separable, the same is true for $\rho_{AB}\circ\sigma_{AB}$.
\end{enumerate}
\end{lemma}

\begin{proof}
Clearly, claim (b) follows straightforwardly once (a) has been established. In order to prove (a), it suffices to write
\begin{equation}
\rho_{AB}\circ\sigma_{AB} = \Pi^0_{AA'}\otimes\Pi^0_{BB'}\, \rho_{AB}\otimes\sigma_{A'B'}\, \Pi^0_{AA'}\otimes\Pi^0_{BB'}\, , \label{Had LOCC}
\end{equation}
here $\Pi^0_{AA'}=\sum_{i=1}^{n_A}\ket{i_A}\!\!\bra{i_A i_{A'}}$ and analogously for $\Pi^0_{BB'}$.
\end{proof}

Lemma~\ref{Hadamard sep lemma}(b) is a useful trick up our sleeve. This is all we will use in the rest of the chapter, particularly in the proof of Theorem~\ref{EA}. However, let us take some time to proceed a bit further.

As it turns out, there are examples of weakly entangled states that can be (stochastically) efficiently distilled by taking Hadamard powers. This holds also in the multipartite setting. For instance, the computation described in~\cite{GMESchur} (together with some later results from~\cite{Buchy}) shows the existence of $(n-2)$-separable $n$-qubit states having a genuinely multipartite entangled Hadamard square.

As a generalisation of the separability-preserving property of the Hadamard product, let us analyse the behaviour of the Schmidt rank function $SR(\cdot)$ defined in~\cite{SchmidtNumber}. Let us remind the reader that the Schmidt rank of a pure state $\ket{\psi}_{AB}$ of a bipartite $n_A\times n_B$ system $AB$ is the minimum $k$ such that one can write 
\bbb
\ket{\psi}_{AB} = \sum_{i=1}^k \ket{v_i} \ket{w_i}
\eee
for appropriate local vectors $\ket{v_i}, \ket{w_i}$. The Schmidt rank (or Schmidt number) of a bipartite state $\rho_{AB}$ is then defined as the minimum $k$ such that $\rho_{AB}$ can be written as a convex combination of bipartite pure states of Schmidt rank no larger than $k$. 

Clearly,~\eqref{Had LOCC} implies that 
\bb
SR(\rho_{AB}\circ \sigma_{AB})\leq SR_{AA':BB'}(\rho_{AB}\otimes\sigma_{A'B'})=SR(\rho_{AB}) SR(\sigma_{AB})\, .
\ee
Now, we proceed to show that this bound can always be saturated, as long as it does not conflict with the trivial requirements $SR(\omega_{AB})\leq\min\{n_A,n_B\}\eqqcolon n$.

Before coming to the details, we remind the reader of some well-known facts.
\begin{enumerate}[(i)]
\item The rank of the $n\times m$ Vandermonde matrix $M_{ij}=\lambda_j^i$ is always equal to $\min\{n,k\}$, where $k$ is the number of distinct elements among the complex numbers $\lambda_1,\ldots,\lambda_m$.
\item The Schmidt rank of a vector $\ket{Z}=\sum_i \ket{v_i}\ket{w_i}$ coincides with the rank of $v^T w$, where $v$ is a matrix defined via $\ket{v_i}=\sum_j v_{ij} \ket{j}$ for some fixed computational basis, and $w$ is constructed analogously. As a first consequence, we deduce that $SR(Z)=r$ whenever $\ket{Z}=\sum_{i=1}^r \ket{v_i}\ket{w_i}$ with $\{\ket{v_i}\}_{i=1}^r,\,\{\ket{w_i}\}_{i=1}^r$ linearly independent families in their own local spaces. As a second corollary, observe that $SR\left(\sum_i \ket{v_i}\ket{v_i^*}\right)=\rk(v)$.
\end{enumerate}

We are now in position to prove the following.

\begin{prop}
Let $r,s,n$ be positive integers such that $r,s\leq n$. Then there are (pure) states $\rho_{AB},\, \sigma_{AB}$ of an $n\times n$ bipartite quantum system such that $SR(\rho_{AB}\circ\sigma_{AB}) = \min\{ n, rs\}$.
\end{prop}

\begin{proof}
Consider a bipartite $n\times n$ quantum system. Choose a large $N\geq\max\{n,rs\}$, and define $\omega\coloneqq e^{2\pi i/N}$. Then, for all $0\leq i\leq r-1$ and $0\leq j\leq s-1$, introduce the local (unnormalised) states
\begin{equation*}
\ket{\alpha_i} \coloneqq \sum_{l=0}^{n-1} \omega^{il} \ket{l}\ ,\qquad \ket{\beta_j}\,\coloneqq\,\sum_{l=0}^{n-1} \omega^{jrl} \ket{l}\, . \end{equation*} 
Using these states as building blocks, we construct
\begin{equation*} \ket{\psi}\, \coloneqq\, \sum_{i=0}^{r-1} \ket{\alpha_i \alpha^*_i}\ ,\qquad \ket{\varphi}\, \coloneqq\, \sum_{j=0}^{s-1} \ket{\beta_j \beta^*_j}\, . \end{equation*}
It is easy to see that $\{ \ket{\alpha_i} \}_{0\leq i\leq r-1}$ and $\{ \ket{\beta_j} \}_{0\leq j\leq s-1}$ are two linearly independent families, since all the $\{\omega^i \}_{i=0}^{r-1}$ are distinct, as well as all the $\{\omega^{jr} \}_{j=0}^{s-1}$. Consequently, we see that $SR(\ket{\psi}\!\!\bra{\psi})=r$ and $SR(\ket{\varphi}\!\!\bra{\varphi})=s$. Taking the Hadamard product we obtain
\begin{equation*} \ket{\psi}\!\!\bra{\psi} \circ \ket{\varphi}\!\!\bra{\varphi}\ =\ \ket{\psi\circ\varphi}\!\!\bra{\psi\circ\varphi}\, , \end{equation*}
with
\begin{equation*} \ket{\psi\circ\varphi}\ =\ \sum_{i=0}^{r-1}\sum_{j=0}^{s-1} \ket{\alpha_i\circ\beta_j}\ket{\alpha_i^*\circ\beta_j^*}\, . \end{equation*}
Observe that $\ket{\alpha_i\circ\beta_j}=\sum_{l} \omega^{(i+rj)l} \ket{l}$; therefore, the Schmidt rank of the above state is equal to the rank of the $n\times rs$ matrix $M$ whose entries are given by $M_{l, i+rj}\coloneqq \omega^{(i+rj)l}$. Since this is a Vandermonde matrix with $rs$ distinct generating numbers, we conclude that $\rk M=\min\{rs,n\}$.
\end{proof}


\section{Positivity region} \label{sec P}

This section is devoted to answering the most basic question concerning the maps defined in~\eqref{Phi}: \emph{for what range of the parameters $\alpha,\beta,\gamma$ is $\Gamma[\alpha, \beta, \gamma]$ positive?}
The most direct way of solving the problem is reported in the following Subsection~\ref{subsec4 positivity direct}. However, as in many cases, a direct proof is not necessarily the most illuminating. Thus, in Subsection~\ref{subsec4 positivity} we choose to follow a longer path to prove the main result of this section (Theorem~\ref{positivity}), passing through Proposition~\ref{red pos prob} and Theorem~\ref{pos chi}. This is instructive because it shows the link between the map $\Gamma$ of~\eqref{Phi}, acting on a bipartite system, and the map $\chi$ of~\eqref{chi}, whose input is instead a single system.

\subsection{Direct approach} \label{subsec4 positivity direct}

Let us start with a preliminary result of elementary linear algebra.

\begin{lemma} \label{lemma idiota}
Let $M>0$ be a strictly positive definite matrix, and let $\psi = \ket{\psi}\!\!\bra{\psi}$ be a (non necessarily normalised) pure state. For $\kappa$ a real coefficient, $M + \kappa\, \psi\geq 0$ iff $1+ \kappa \braket{\psi | M^{-1} |\psi}\geq 0$.
\end{lemma}

\begin{proof}
Observe that
\bbb
M + \kappa \ket{\varphi}\!\! \bra{\varphi} = M^{1/2} (\mathds{1} + \kappa\, M^{-1/2} \ket{\varphi}\!\!\bra{\varphi} M^{-1/2}) M^{1/2}
\eee
is positive semidefinite iff so is $\mathds{1} + \kappa\, M^{-1/2} \ket{\varphi}\!\!\bra{\varphi} M^{-1/2}$, which happens iff $1+ \kappa \braket{\varphi | M^{-1} |\varphi}\geq 0$.
\end{proof}

Now, the main result of the present section is as follows.

\begin{thm} \label{positivity}
The map $\Gamma[\alpha,\beta,\gamma]$ defined by~\eqref{Phi} is positive iff:
\bb
\begin{split}
\alpha+1,\, \beta+1 &\geq 0\, ,\\
\frac{\alpha+\beta}{n} + \gamma +1 &\geq 0 \, ,\\
\alpha+\beta+\gamma+1 &\geq 0 \, .
\end{split}
\label{positivity eq}
\ee
Here, we used the notation $n\coloneqq \min\{n_A, n_B\}$.
\end{thm}

Before delving into the proof of the above result, let us try to represent the region determined by the inequalities~\eqref{positivity eq} on a $3$-dimensional plot. This will also help us to get some geometrical understanding of the various steps of the argument.

\begin{figure}[ht] 
\centering
\includegraphics[height=6cm, width=6cm, keepaspectratio]{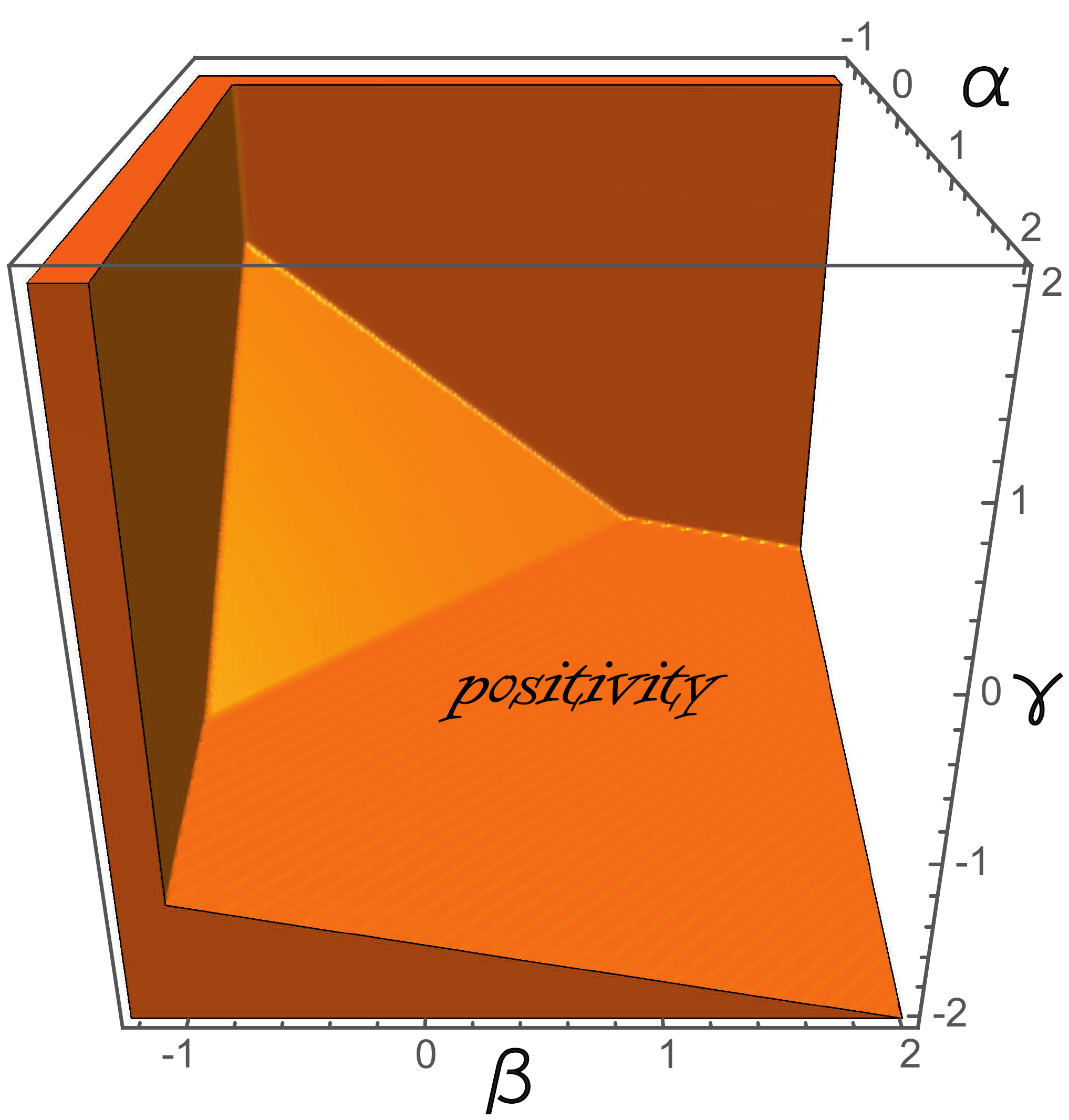}
\caption[]{The solid region represents the parameter region for which the map $\Gamma [ \alpha,\beta,\gamma ]$ of~\eqref{Phi} is not positive. Here the the $n=4$ case is considered.}
\label{Positivity region}
\end{figure}

\begin{proof}[Proof of Theorem~\ref{positivity}]
First of all, conditions~\eqref{positivity eq} are necessary, as can be seen by choosing suitable input states. In the following, we often write $\Gamma$ instead of $\Gamma[\alpha, \beta,\gamma]$.
\begin{itemize}
\item A maximally entangled state $\ket{\Phi}$ gives $\Gamma(\Phi) = \left(1+\frac{\alpha+\beta}{n}\right) \mathds{1} + \gamma \Phi \geq 0$, from which $1+\frac{\alpha+\beta}{n}+\gamma \geq 0$ follows.
\item A product state $\ket{11}$ gives $\Phi(\ket{11}\!\!\bra{11}) = \mathds{1} + \alpha\, \mathds{1}\otimes\ket{1}\!\!\bra{1} + \beta\, \ket{1}\!\!\bra{1}\otimes \mathds{1} + \gamma \ket{11}\!\!\bra{11} \geq 0$. Taking the matrix element on the same state produces $1+\alpha+\beta+\gamma\geq 0$.
\item As above, input $\ket{11}$ but take the matrix element on pure states $\ket{21}$ and $\ket{12}$, producing the condition $1+\alpha\geq 0$ and $1+\beta\geq 0$.
\end{itemize}
Observe that if $\alpha+\beta\geq 0$ the third condition in~\eqref{positivity eq} is subsumed under the second one, while on the contrary if $\alpha+\beta<0$ the second condition is subsumed under the third one. This completes the proof of the necessity of the inequalities~\eqref{positivity eq}.

In order to prove their sufficiency, observe the region that they determine in Figure~\ref{Positivity region}. It is apparent that every point belonging to that region is a convex combination of three points on the three half-lines that constitute the `edges' of the region. We proceed to list them.
\begin{enumerate}[(1)]
\item The vertical line that lies at the intersection between the $\alpha\gamma$ and $\beta\gamma$ coordinate planes:
\bb
\left\{ (-1,-1,\gamma):\, \gamma\geq 1\right\} .
\label{vertical line pos}
\ee
\item The lower line which runs parallel to the $\beta\gamma$ plane:
\bb
\left\{ \left(-1,\beta,-1-\frac{\beta-1}{n}\right):\, \beta\geq 1 \right\} .
\label{lower line 1 pos}
\ee
\item The lower line which runs parallel to the $\alpha\gamma$ plane:
\bb
\left\{ \left(\alpha,-1,-1-\frac{\alpha-1}{n} \right):\, \alpha\geq 1 \right\} .
\label{lower line 2 pos}
\ee
\end{enumerate}
If we prove that all of the above half-lines are entirely composed of positive maps, we are done. Let us proceed in the same order.
\begin{enumerate}[(1)]
\item $\Gamma[-1,-1,\gamma] = (\mathds{1}\tr-I)\otimes(\mathds{1}\tr-I) + (\gamma-1) I$ is positive if $\gamma\geq 1$, because the first addend is the tensor product of two completely copositive maps (Subsection~\ref{subsec4 dep}). In other words, up to composing it with a global transposition (that is positive and invertible), it is nothing but the product of two completely positive maps.
\item $\Gamma\big[ -1,\beta,-1-\frac{\beta-1}{n} \big] = (\mathds{1}\tr+I)\otimes (\mathds{1}\tr-I) + (\beta-1)\, I\otimes \big(\mathds{1}\tr-\frac{1}{n} I\big)$ is positive if $\beta\geq 1$, as we are set to show.
\begin{itemize}
\item The first addend is positive since it is a tensor product of an entanglement-breaking channel on the first subsystem and a positive map on the second subsystem (Subsection~\ref{subsec4 dep}).
\item The second addend is positive because by evaluating it directly on a pure state $\psi = \ket{\psi}\!\!\bra{\psi}$ one gets a positive result. In fact, writing $\ket{\psi}$ as in~\eqref{Schmidt decomposition}, one has
\begin{align*}
\bigg( I\otimes \Big(\mathds{1}\tr - \frac{I}{n} \Big) \bigg) (\psi_{AB}) &= (\tr_B \psi_{AB})\otimes \mathds{1}_B - \frac1n \psi_{AB} \\
&= \bigg( \sum_{i=1}^n \lambda_i \ket{e_i}\!\! \bra{e_i}\bigg) \otimes \mathds{1} - \frac1n \ket{\psi}\!\!\bra{\psi} \\
&\geq 0\, ,
\end{align*}
where the last step follows easily from Lemma~\ref{lemma idiota} with $\kappa=-1$, considering that
\bbb
\frac1n\, \Big\langle \psi\, \Big| \bigg( \sum_{i=1}^n \lambda_i \ket{e_i}\!\! \bra{e_i}\bigg)^{-1} \otimes \mathds{1}\ \Big|\, \psi \Big\rangle = \sum_{i=1}^n \lambda_i = 1\, .
\eee
\end{itemize}
The above argument works essentially because $\mathds{1}_B \tr_B - \frac1n I_B$ is completely positive if $n=n_B$ and at least $n$-positive if $n=n_A$.
\item The third case is completely analogous to the second one and can be treated in a symmetrically identical way.
\end{enumerate}
\end{proof}

\subsection{An alternative proof of Theorem~\ref{positivity}} \label{subsec4 positivity}

As we mentioned, there is a less direct pathway to arrive at Theorem~\ref{positivity} that is worth exploring. In fact, it involves a characterisation theorem for the maps $\chi$ presented in~\eqref{chi}, which might be of independent interest.
The first important observation we present is then the reduction of the problem of positivity for $\Gamma[\alpha,\beta,\gamma]$ to the analogous problem for $\chi[a,c]$. 

\begin{prop} \label{red pos prob}
The map $\Gamma[\alpha,\beta,\gamma]$ is positive iff $\alpha,\beta\geq -1$ and $\chi[\alpha+\beta, \gamma]$ is positive.
\end{prop}

\begin{proof}
Clearly, in order to ensure the positivity of $\Gamma$ we have to require that $\Gamma (\psi )\geq 0$ for all global pure states $\psi = \ket{\psi}\!\!\bra{\psi}$. Since $\Gamma$ commutes with the conjugation by local unitaries, we can suppose that the input state is in Schmidt normal form as in~\eqref{Schmidt decomposition}, with $\ket{e_i}=\ket{i}$ and $\ket{f_j}=\ket{j}$.
Then,
\begin{align*}
\Gamma (\psi) &= \mathds{1} + \sum_{i,j} (\alpha \lambda_j + \beta \lambda_i) \ket{ij}\!\!\bra{ij} + \gamma\, \psi\\
&= \sum_{i\neq j} (1+\alpha \lambda_j + \beta \lambda_i) \ket{ij}\!\!\bra{ij} \\
&\qquad + \sum_{ij} \left( \left(1+(\alpha+\beta)\lambda_i \right)\delta_{ij} + \gamma\, \sqrt{\lambda_i \lambda_j} \right) \ket{ii}\!\!\bra{jj}\ .
\end{align*}
From the above block decomposition it is apparent that $\Gamma (\psi)\geq 0$ iff $1+\alpha \lambda_j + \beta \lambda_i\geq 0$ and
\bbb \sum_{ij} \left( \left(1+(\alpha+\beta)\lambda_i \right)\delta_{ij}\ +\ \gamma\, \sqrt{\lambda_i \lambda_j} \right) \ket{i}\!\!\bra{j}\ \geq\ 0\ . \eee
Once we require the validity of these conditions for all the probability distributions $\lambda$, the first one reads $\alpha,\beta\geq -1$. At the same time, the above matrix is exactly $\chi[\alpha+\beta,\gamma]\left(\tilde{\psi}\right)$,
with $\ket{\tilde{\psi}}\coloneqq\sum_i \sqrt{\lambda_i}\ket{i}$, and therefore we are led to require the positivity of the map $\chi[\alpha+\beta,\gamma]$ as the second and last condition.
\end{proof}

The above proposition shows how the map $\chi[\alpha+\beta,\gamma]$ is in fact the restriction of $\Gamma[\alpha,\beta,\gamma]$ to the maximally coherent subspace. This, in turn, reduces our main problem to the question of determining what is the region of positivity of the map~\eqref{chi}. Now, we proceed to show how to solve this latter problem.

\begin{thm} \label{pos chi}
The map $\chi[a,c]$ defined by~\eqref{chi} is positive iff
\bb
\begin{split}
a +2 &\geq 0 \, ,\\
\frac{a}{n}+c+1 &\geq 0\, ,\\
a+c+1 &\geq 0\, .
\end{split}
\label{pos chi eq}
\ee
\end{thm}

\begin{proof}
Suppose first that $a\geq 0$. Denoting with $D_\lambda$ the diagonal matrix with diagonal $\lambda\in\mathds{R}^n$, the positivity of $\chi[a,c]$ amounts to require that
\bb
\mathds{1} + a\, D_\lambda + c\, \Ket{\psi}\!\!\Bra{\psi} \geq 0 
\label{pos chi pr eq1}
\ee
for all pure states $\Ket{\psi} = \sum_{i=1}^n \sqrt{\lambda_i}\,\Ket{i}$, where $\lambda$ is an arbitrary probability distribution over the alphabet $\{1, \ldots, n\}$ (phases on $\ket{\psi}$ are irrelevant, as the application of a diagonal unitary immediately reveals). Since $\mathds{1}+a\, D_\lambda>0$, one can apply Lemma~\ref{lemma idiota} with $\kappa=c$ and rewrite the above condition as
\bbb
1 + c\, \bra{\psi} \left(\mathds{1}+a D_\lambda\right)^{-1} \ket{\psi} \geq 0\, ,
\eee
i.e.
\bbb
c \geq -\, \frac{1}{\bra{\psi} \left(\mathds{1}+a D_\lambda\right)^{-1} \ket{\psi}} = -\, \frac{1}{\sum_i \frac{\lambda_i}{1+a \lambda_i}}\, .
\eee
We have to impose the above constraint for all probability vectors $\lambda\in\mathds{R}^n$. The strictest condition is that corresponding to the vector $\lambda$ that achieves the maximum of $f(\lambda) \coloneqq \sum_i \frac{\lambda_i}{1+a\lambda_i}$. From the strict concavity of $\frac{x}{1+ax}$ on $x\geq 0$ it follows that $f(\lambda)$ is strictly concave on the simplex of the allowed vectors $\lambda$. Therefore, any internal stationary point is automatically a global maximum. Applying the method of Lagrange multipliers gives us $\lambda_i\coloneqq \text{const} = \frac{1}{n}$ as the unique stationary point of $f$, and thus $f_{\max} = \frac{1}{1+\frac{a}{n}}$.
The final condition on $c$ becomes exactly
\bbb
c \geq -1-\frac{a}{n}\, ,
\eee
which is what~\eqref{pos chi eq} dictates for the present case $a\geq 0$.

Now, let us consider the opposite possibility, $a<0$. Since the diagonal entries of the operator in~\eqref{pos chi pr eq1} are $1+(a+c)\lambda_i$, imposing their positivity for all the allowed $\lambda$ amounts to requiring that $a+c+1\geq 0$. Moreover, observe that if $a<-2$ it is possible to force $\mathds{1}+a D_\lambda$ to have a $2$-dimensional negative eigenspace. In this case, an addition of pure state could not make the whole operator positive. Therefore, we have to demand $a\geq -2$. In order to show that these two conditions are sufficient, consider an arbitrary pure state $\ket{v}$ and take the matrix element of~\eqref{pos chi pr eq1} on it.
\bb \bra{v} \left( \mathds{1} + a D_\lambda + c \ket{\psi}\!\!\bra{\psi} \right) \ket{v}\ =\ 1\, +\, a\, \sum_i \lambda_i |v_i|^2\, +\, c\ \Big| \sum_i \sqrt{\lambda_i} v_i\Big|^2 \label{pos chi pr eq2} \ee
In order to prove that the above quantity is always positive if $a\geq -2$ and $a+c+1\geq 0$, we want to formalise the following intuition. On the one hand, it is not possible to make the coefficient of $c$ small without choosing at least two non-zero (and comparable) $v_i$, thus reducing the negative impact of the coefficient of $a$. On the other hand, if one wants to concentrate the weights $\sqrt{\lambda_i} v_i$ on a single element, then $a$ appears always summed with $c$. We can make the above reasoning rigorous by considering the following inequality, valid for arbitrary $z_1,\ldots,z_n\in\mathds{C}$:
\bb
\sum_i |z_i|^2 \leq \frac{1}{2}\,\bigg(\Big(\sum_i |z_i|\Big)^2 + \Big|\sum_i z_i\,\Big|^2\bigg)\, .
\label{geom ineq}
\ee
Its proof follows easily once one rewrites the difference of its two sides as
\bbb
\sum_{ij} \left( |z_i| |z_j| + \Re (z_i^* z_j) - 2 |z_i|^2\delta_{ij} \right) \geq 0\, ,
\eee
where the latter inequality holds trivially because the left-hand side is a sum of positive terms.\footnote{A nice geometrical interpretation arises when one considers a closed polygon, that is, numbers $z_1,\ldots, z_n$ such that $\sum_i z_i = 0$. In this case,~\eqref{geom ineq} states that the sum of the square sides is at most \emph{half} of the square perimeter.}
Applying~\ref{geom ineq} with $z_i=\sqrt{\lambda_i} v_i$ yields
\bbb
\sum_i \lambda_i |v_i|^2 \leq \frac{1+k}{2}\, ,
\eee
where we noticed that $\big|\sum_i \sqrt{\lambda_i} |v_i|\big| \leq \left( \sum_i \lambda_i\right)^{1/2} \left( \sum_i |v_i|^2 \right)^{1/2} \leq 1$ and wrote for brevity $k\coloneqq |\braket{\psi | v}|^2$. Then, taking into account that $a<0$,~\eqref{pos chi pr eq2} can be lower bounded by
\begin{align*}
\bra{v} \left( \mathds{1} + a D_\lambda + c \ket{\psi}\!\!\bra{\psi} \right) \ket{v} &\geq 1 + a \frac{1+k}{2} + c k\\
&= (1+a+c) k + \left(1+ \frac{a}{2}\right) (1-k)\\
&\geq 0 \, ,
\end{align*}
i.e. it is non-negative, as we claimed.
\end{proof}

\begin{rem}
One could also give a direct proof of the positivity of $\chi[a,c]$ when $-2\leq a\leq 0,\, 1+a+c\geq 0$, based on Hadamard product arguments. Clearly, up to taking convex combinations with known positive maps, one has to prove only the positivity of the extreme point $\chi[-2,1] = \mathds{1}\tr - 2 D + I$. Then, the crucial observation consists in writing the above map as a Hadamard square of a completely copositive map, i.e. $\chi[-2,1]=\left(\mathds{1}\tr -I\right)^{\circ 2}$, where the basis we choose from now on to take Hadamard products is the canonical basis of operators $\{\ket{i}\!\!\bra{j}\}_{i,j}$ (or its Choi--Jamiolkowski equivalent $\{\ket{ij}\}_{i,j}$). The following reasoning then shows that Hadamard multiplying two completely copositive maps yields another completely copositive map. Using the notation~\eqref{Choi state}, one has
\begin{equation*}
R_{\Lambda_1}^{T_A},\, R_{\Lambda_2}^{T_A} \geq 0\quad\Longrightarrow\quad R_{\Lambda_1\circ \Lambda_2}^{T_A} = \left( R_{\Lambda_1}\circ R_{\Lambda_2} \right)^{T_A} = R_{\Lambda_1}^{T_A}\circ R_{\Lambda_2}^{T_A} \geq 0\, .
\end{equation*}
\end{rem}

As an easy corollary of the above discussion, we deduce the following result.

\begin{proof}[Alternative proof of Theorem~\ref{positivity}]
Follows by combining Proposition~\ref{red pos prob} and Theorem~\ref{pos chi}. 
\end{proof}

\section{Complete Positivity region} \label{sec CP}

Determining the range of $\alpha,\beta,\gamma$ for which the map $\Gamma[\alpha,\beta,\gamma]$ given by~\eqref{Phi} is completely positive requires the construction of the Choi state associated to $\Phi$. Calling $A'B'$ the twin system of $AB$, a maximally entangled state $AB:A'B'$ reads
\begin{align*}
\ket{\Phi}_{AB:A'B'} &= \frac{1}{\sqrt{n_A n_B}} \sum_{i=1}^{n_A} \sum_{j=1}^{n_B}\, \ket{ij}_{AB}\ket{ij}_{A'B'} \\
&= \ket{\Phi}_{AB} \ket{\Phi}_{A'B'}\, .
\end{align*}
As a consequence, the Choi state $R_{\Gamma} = (\Gamma_{AB}\otimes I_{A'B'})(\Phi_{AB:A'B'})$ becomes
\bb
\begin{split}
R_{\Gamma} = \frac{\mathds{1}_{ABA'B'}}{n_A n_B} + \alpha\, \frac{\mathds{1}_{AA'}}{n_A}\otimes \Phi_{BB'} + \beta\, \Phi_{AA'}\otimes\frac{\mathds{1}_{BB'}}{n_B} + \gamma\, \Phi_{AA'}\otimes \Phi_{BB'}\, .
\end{split}
\label{Choi state Phi}
\ee
The following result follows easily.

\begin{thm} \label{complete positivity}
The map $\Gamma[\alpha,\beta,\gamma]$ defined by~\eqref{Phi} is completely positive iff
\bb
\begin{split}
1+ n_B \alpha &\geq 0\, ,\\
1+ n_A \beta &\geq 0\, ,\\
1+n_B \alpha + n_A \beta + n_A n_B \gamma &\geq 0\, .
\end{split}
\label{complete positivity eq}
\ee
\end{thm}

\begin{proof}
Since the four addends appearing in~\eqref{Choi state Phi} commute, the diagonalisation of their sum is straightforward. It is easily seen that the distinct eigenvalues of the operator in~\eqref{Choi state Phi} are
\bbb \frac{1}{n_A n_B} , \quad \frac{1}{n_A n_B} + \frac{\alpha}{n_A}, \quad \frac{1}{n_A n_B} + \frac{\beta}{n_B}, \quad \frac{1}{n_A n_B} + \frac{\alpha}{n_A} + \frac{\beta}{n_B} + \gamma\, . \eee
Imposing positivity of the last three expressions yields~\eqref{complete positivity eq}, as claimed.
\end{proof}

As can be easily verified, conditions~\eqref{complete positivity eq} imply that every completely positive map in the $\Gamma$ class can be written as a convex combination of three points lying on the three `edge' half-lines coming out from the vertex $\big(-\frac{1}{n_B},\, -\frac{1}{n_A},\, \frac{1}{n_A n_B} \big)$. These can be represented as follows:
\begin{enumerate}[(1)]
\item the vertical half-line is 
\bbb
\Big\{ \big(- 1 /n_B,\, - 1/n_A,\, \gamma \big) :\, \gamma\geq 1 /n_A n_B \Big\} ;
\eee
\item one of the other two is 
\bbb
\Big\{ \big(- 1 /n_B,\, \beta,\, - \beta / n_B \big):\, \beta\geq - 1 /n_A \Big\} ,\eee
\item while the last one is symmetrically described as
\bbb
\Big\{ \big(\alpha,\, - 1 /n_A,\, - \alpha /n_A \big):\, \alpha\geq - 1 /n_B \Big\} .
\eee
\end{enumerate}

\section{Entanglement-breaking region} \label{sec EB}

We now move on to the question of finding the region in the $\alpha,\beta,\gamma$ parameter space that defines an entanglement-breaking map through equation~\eqref{Phi}. Obviously, such a region must be contained in the complete positivity solid defined via~\eqref{complete positivity eq}. Reformulating the problem with the help of the Choi-Jamiolkowski isomorphism, we want to determine necessary and sufficient conditions for the separability of the state~\eqref{Choi state Phi}. This problem has already been solved in the special case $n_A=n_B$ in~\cite{ChrusKoss}, but we will see that the most interesting phenomena appear when one considers the asymmetric case $n_A\neq n_B$.

In Subsection~\ref{subsec4 EB} we deduce the main result of the section, Theorem~\ref{EB}, using some abstract arguments based on the symmetry properties of the problem. This is an instance of the general procedure detailed in Subsection~\ref{subsec2 sep symm}. In Subsection~\ref{subsec4 EB direct}, instead, we perform a double check and exhibit explicit separable decompositions showing that the maps identified by Theorem~\ref{EB} are indeed entanglement-breaking.

\subsection{Computation of the entanglement-breaking region} \label{subsec4 EB}

This subsection is devoted to the determination of the region in the parameter space $\alpha,\beta,\gamma$ for which the map $\Gamma[\alpha,\beta,\gamma]$ in~\eqref{Phi} is entanglement-breaking. As we already saw, this amounts to deciding the separability of the corresponding Choi state $(R_{\Gamma})_{ABA'B'}$ defined by~\eqref{Choi state Phi} with respect to the bipartition $AB:A'B'$.
An important preliminary observation is that $(R_{\Gamma})_{ABA'B'}$ belongs to the subspace $\mathcal{V}$ of linear operators that commute with unitaries of the form $U_A\otimes V_B \otimes U^*_{A'}\otimes V^*_{B'}$ (for all local unitaries $U,V$). This subspace is a central section in the sense of Subsection~\ref{subsec2 sep symm}. In fact, it constitutes a special case of Example~\ref{local groups ex}, with $V_1=V_2=\mathcal{H}_{n_A n_B}$, $G=SU(n_A) \times SU(n_B)$, and the local representations given by
\begin{align*}
\zeta_1 (U, V) (Z_{AB}) &\coloneqq U_A \otimes V_B\ Z_{AB}\ U_A^\dag \otimes V_B^\dag\, ,\\
\zeta_2 (U, V) (Z_{AB}) &\coloneqq U_A^* \otimes V_B^*\ Z_{AB}\ U_A^T \otimes V_B^T\, .
\end{align*}
We can therefore exploit Proposition~\ref{impr HhBh} in order to simplify the solution of the separability problem. To this purpose, observe that under the natural identification $V_{AB} = \mathcal{H}_{n_A n_B} = V_{AB}^*$ (made possible by the Hilbert-Schmidt product) one has $\mathcal{V}^*=\mathcal{V}$.

\begin{thm} \label{test EB}
If $n_A=n_B$, the map $\Gamma[\alpha,\beta,\gamma]$ in~\eqref{Phi} is entanglement-breaking iff it is completely positive and PPT-binding. If $n_A<n_B$, it is entanglement-breaking iff
\bb \Gamma\, ,\qquad T\,\Gamma \quad\text{and}\quad \bigg( I_A\otimes\Big( \mathds{1} \text{\emph{Tr}}\, -\frac{I}{n_A}\Big)_B \bigg)\, \Gamma \label{test EB} \ee
are completely positive. If $n_A>n_B$ a reversed but analogous condition holds (just exchange subscripts $A$ and $B$).
\end{thm}

\begin{proof}
As we already mentioned, $\Gamma$ is entanglement-breaking iff its Choi state $(R_\Gamma)_{ABA'B'}$~\eqref{Choi state Phi} is separable with respect to the bipartition $AB:A'B'$. Proposition~\ref{impr HhBh} implies that this happens iff $\tr W R_\Gamma\geq 0$ for all entanglement witnesses $W$ that share the same symmetry of $R_\Gamma$, i.e. that commute with unitaries of the form $U_A\otimes V_B \otimes U^*_{A'}\otimes V^*_{B'}$, for all $U\in SU(n_A)$, $V\in SU(n_B)$. In other words, $W$ can be assumed to be the Choi matrix of a positive map $\Gamma'\coloneqq \Gamma[\alpha',\beta',\gamma']$ belonging to the set defined by~\eqref{Positivity region} (or a limit point of the form $\lim_{\kappa \rightarrow\infty} \frac{1}{\kappa}\,\Gamma[\kappa \alpha',\kappa\beta',\kappa\gamma']$, with $(\kappa\alpha',\kappa\beta',\kappa\gamma')$ defining a positive map for all $\kappa>0$; this case will not be considered further because it does not introduce any new constraint).

We now show that instead of requiring the scalar condition $\Tr [R_{\Gamma'} R_\Gamma] \geq 0$ to hold for all positive $\Gamma'$ of the family~\eqref{Phi}, we can equivalently impose that $\Gamma'\Gamma$ is completely positive for all positive $\Gamma'$ as in~\eqref{Phi}. To show this, let us write
\begin{align*}
\Tr [R_{\Gamma'} R_\Gamma ] &= \Tr \left[ (\Gamma'\otimes I)(\Phi)\, (\Gamma\otimes I)(\Phi) \right] \\
&= \Tr \left[ \Phi\, \left((\Gamma')^\dag \Gamma\otimes I\right)(\Phi) \right] \\
&= \braket{\Phi | R_{(\Gamma')^\dag \Gamma} | \Phi} \\
&= \braket{\Phi | R_{\Gamma' \Gamma} | \Phi}\, ,
\end{align*}
where for the last step we observed that all the maps in~\eqref{Phi} are self-adjoint. Now, clearly $\Tr [R_{\Gamma'} R_\Gamma ]\geq 0$ follows from the complete positivity of $\Gamma'\Gamma$, which indeed implies the stronger condition $R_{\Gamma'\Gamma}\geq 0$. On the other hand, if $\Tr [R_{\Gamma'} R_\Gamma ]\geq 0$ holds for all positive maps $\Gamma'$ of the family~\eqref{Phi}, then we have seen that $\Gamma$ must be entanglement-breaking, which in turn implies that $\Gamma'\Gamma$ is completely positive whenever $\Gamma'$ is positive.

The above reasoning shows that the complete positivity of $\Gamma'\Gamma$ for all positive $\Gamma'$ as in~\eqref{Phi} is necessary and sufficient for $\Gamma$ to be entanglement-breaking.
From now on, we assume without loss of generality $n_A\leq n_B$. As detailed in the proof of Theorem~\ref{positivity}, the positivity region defined by~\ref{Positivity region} is the convex hull of the three half-lines~\eqref{vertical line pos},~\eqref{lower line 1 pos} and~\eqref{lower line 2 pos}.
Consequently, it suffices to test the three families of `witness maps' $\Gamma'$ corresponding to those three half-lines. For each family with parameter $\kappa$ ($\kappa=\gamma,\beta,\alpha$ in order), it suffices to test first the case $\kappa=1$ and then the limit $\kappa \rightarrow\infty$. This leads to six tests corresponding to just as many positive maps: 
\begin{enumerate}[(i)]
\item $\Gamma[-1,-1,1] = (\mathds{1}\tr-I)\otimes(\mathds{1}\tr-I)$;
\item $\lim_{\kappa\rightarrow \infty} \frac1\kappa \Gamma[-1,-1,\kappa] = I$;
\item $\Gamma[-1,1,-1] = (\mathds{1}\tr+I)\otimes (\mathds{1}\tr-I)$;
\item $\lim_{\kappa\rightarrow \infty} \frac1\kappa \Gamma[-1,\kappa ,-1- (\kappa-1)/n_A] = I\otimes \Big(\mathds{1}\tr-\frac{1}{n_A} I\Big)$;
\item $\Gamma[1, -1, -1] = (\mathds{1}\tr-I)\otimes (\mathds{1}\tr+I)$;
\item $\lim_{\kappa\rightarrow \infty} \frac1\kappa \Gamma[\kappa, -1 ,-1- (\kappa-1)/n_B] = \Big(\mathds{1}\tr-\frac{1}{n_B} I\Big) \otimes I$.
\end{enumerate}

Now, we want to understand which one of the above six tests can be subsumed under the PPT condition that $\Gamma$ and $T\Gamma$ are CP. This happens when the witness is \emph{decomposable}, i.e. it is a positive combination of a CP and a coCP map. Let us analyse the above maps one by one:
\begin{enumerate}[(i)]
\item Completely copositive, as is the tensor product of two completely copositive maps (Subsection~\ref{subsec4 dep}).
\item Obviously completely positive.
\item Completely copositive, as is the tensor product of two completely copositive maps (Subsection~\ref{subsec4 dep}).
\item Only positive, but not completely positive unless $n_A= n_B$; it will be clear later that it is actually indecomposable whenever $n_A\neq n_B$; for the time being, all we can conclude is that we can not a priori discard this test.
\item The same as (iii).
\item This is not the same as (iv), because the condition $n_A\leq n_B$ ensures that $\mathds{1}\tr-\frac{1}{n_B} I$, and thus the entire map, are completely positive.
\end{enumerate}

From the above discussion it should be clear that if $n_A=n_B$ then the PPT test is both necessary and sufficient to ensure that $\Gamma$ is entanglement-breaking, while if $n_A<n_B$ the only condition that can not be absorbed in the PPT test is the one in (iv). This is the same as saying that for $n_A<n_B$ a map $\Gamma$ of the class in~\eqref{Phi} is entanglement-breaking iff $T\Gamma$ and $\Big( I_A\otimes\big( \mathds{1} \tr\, -\frac{I}{n_A}\big)_B \Big)\, \Gamma$ are completely positive, which concludes the proof.
\end{proof}

Now that necessary and sufficient conditions for $\Gamma$ to be entanglement-breaking have been written down in the form~\eqref{test EB}, it is only a matter of finding out the shape of the corresponding solid. An interesting question, as usual, is whether we really need the second test or on the contrary the PPT condition is actually sufficient. We already saw that the suspected answer is that if $n_A<n_B$ we \emph{do} need the second test. In other words, in that case there are PPT entangled states of the form~\eqref{Choi state Phi}. The main result of this section reads as follows.

\begin{figure}[h] 
\centering
\includegraphics[height=5.5cm, width=5.5cm, keepaspectratio]{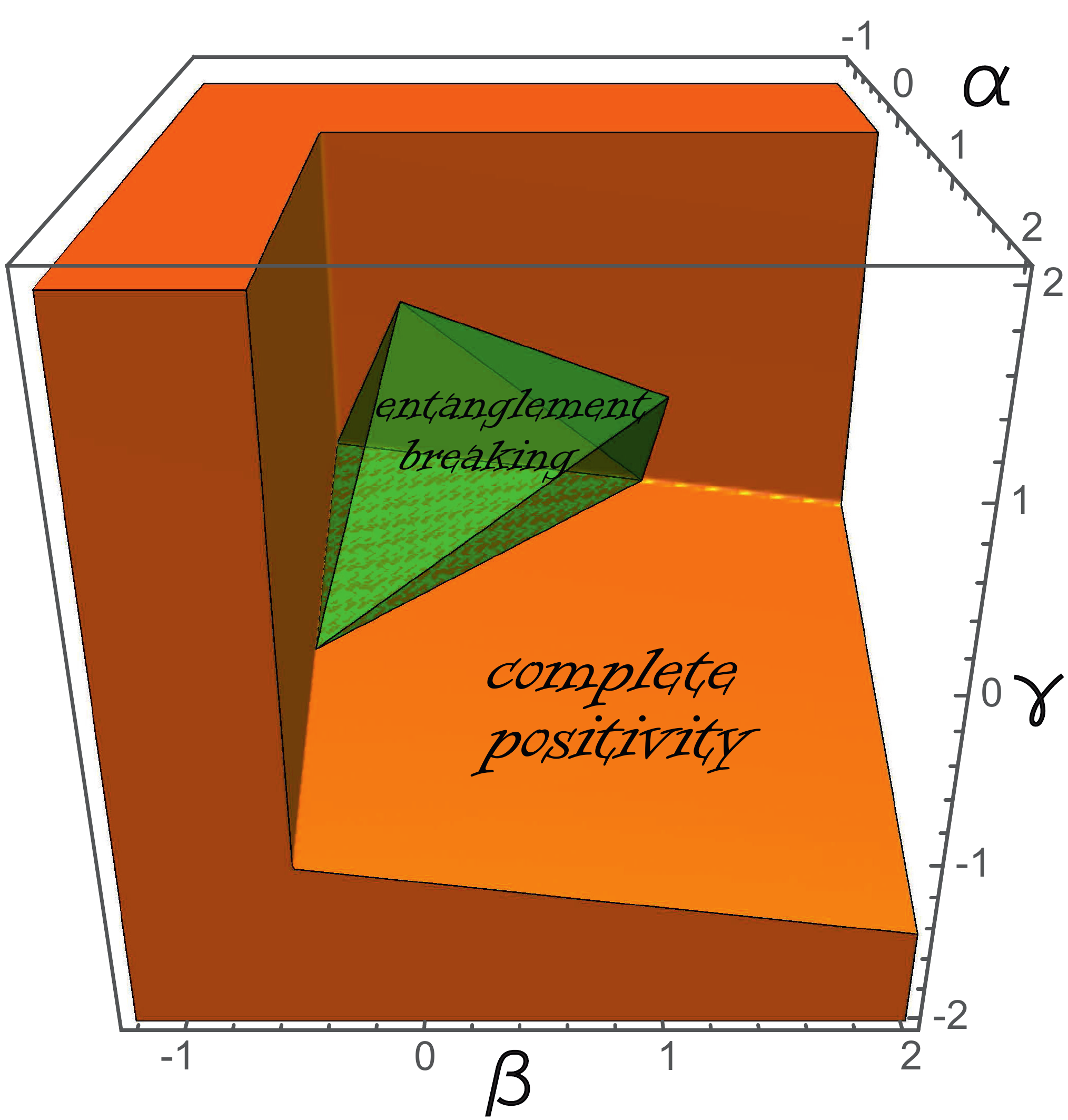}\qquad \includegraphics[height=5.5cm, width=5.5cm, keepaspectratio]{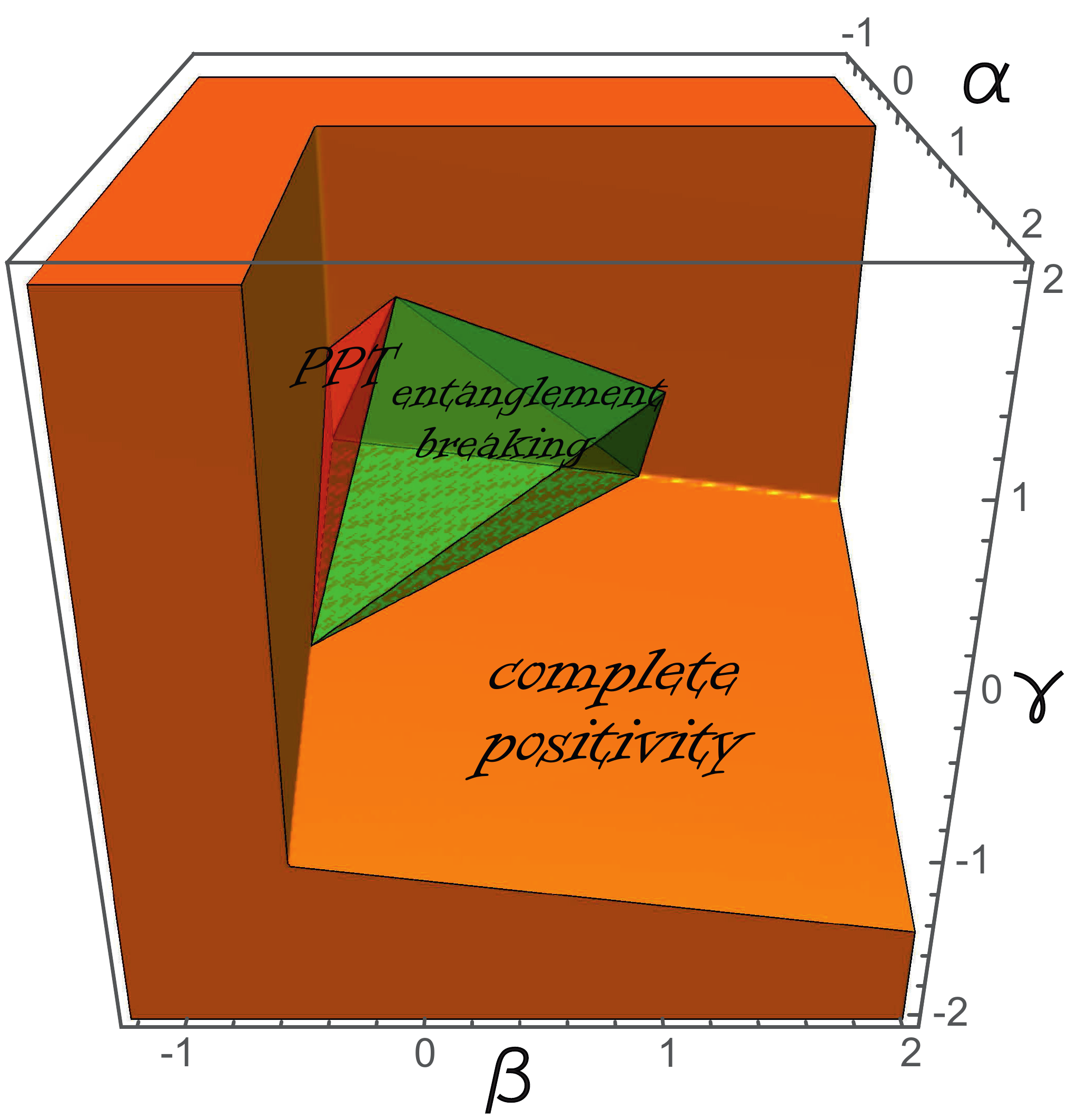}
\caption[]{The three planes identify the complete positivity region for the maps defined in~\eqref{Phi} in the $\alpha,\beta,\gamma$ space. Inside that, the entanglement-breaking solid (in green) is shown on the left. On the right, we added the pyramid for which this map is PPT but not entanglement-breaking (in red). Such a region exists iff $n_A\neq n_B$; here we chose the case $n_A=2,\ n_B=6$.}
\label{EB region}
\end{figure}

\begin{thm} \label{EB}
Assume that $n_A\leq n_B$. The map $\Gamma[\alpha,\beta,\gamma]$ defined by~\eqref{Phi} is entanglement-breaking iff the following conditions are met:
\bb
\begin{split}
1+ n_B \alpha &\geq 0\, ,\\
1+n_B \alpha+ n_A\beta +n_A n_B \gamma &\geq 0\, ,\\[1.5ex]
1-\alpha+\beta-\gamma &\geq 0\, , \\
1+\alpha-\beta-\gamma &\geq 0\, , \\
1-\alpha-\beta+\gamma &\geq 0\, ,\\[1.5ex]
(n_A n_B -1)(n_A\beta+1) - (n_B-n_A)(\alpha +n_A\gamma) &\geq 0\, .
\end{split}
\label{EB eq}
\ee
The last inequality can be omitted if $n_A=n_B$. The solid described by the above system is a double pyramid with triangular basis (see Figure~\ref{EB region}). The basis has vertices
\bbb
\left( -1 /n_B,\, -1/n_B,\, 1\right) , \quad \left( 1,\, -1/n_A,\, - 1/n_A\right) , \quad \left( - 1 /n_B,\, 1,\, -1 /n_B\right) , 
\eee
while the culminating vertices of the two pyramids are
\bbb
\left( -1/n_B,\, - 1/n_A,\, 1 /n_A n_B\right) , \quad \left( 1,\, 1,\, 1\right) .
\eee
\end{thm}

\begin{proof}
Thanks to Theorem~\ref{test EB}, we have just to impose the complete positivity of $\Gamma$, $T\Gamma$ and $\big(I\otimes\left(\mathds{1}\tr-I/n_A \right) \big)\, \Gamma$.
\begin{itemize}
\item $\Gamma$ is CP. This gives the first two conditions of~\eqref{EB eq} together with the requirement $\beta\geq -1/n_A$. However, the latter can be neglected because it follows from the other inequalities of the system~\eqref{EB eq}. Indeed, multiplying the second inequality of the first line by $(n_B-n_A)/n_B$ and summing the third line produces exactly $n_A\beta+1\geq 0$.
\item $T\Gamma$ is CP. Taking the partial transposition $T_{A'B'}$ of the Choi state~\eqref{Choi state Phi} yields
\begin{align*}
n_A n_B\, (\Gamma_{AB}\otimes T_{A'B'})\, \left( \Phi_{AB:A'B'} \right) &= \mathds{1}_{ABA'B'} + \alpha\,\mathds{1}_{AA'}\otimes F_{BB'} \\ 
&\quad + \beta\, F_{AA'}\otimes \mathds{1}_{BB'} + \gamma\, F_{AA'}\otimes F_{BB'} ,
\end{align*}
where $F$ is the operator that performs the swap between two subsystems. Since the four addends in the above equation commute, finding the eigenvalues of their sum is straightforward: they are given by $1+\alpha +\beta +\gamma$, $1+\alpha -\beta -\gamma$, $1 -\alpha +\beta -\gamma$, $1-\alpha -\beta +\gamma$. As is easy to see, $1+\alpha+\beta+\gamma\geq 0$ is already implied by the complete positivity conditions; in fact, using the second inequality in~\eqref{EB eq} to lower bound $\gamma$ gives
\begin{align*}
1+\alpha+\beta+\gamma &\geq 1-\frac{1}{n_A n_B} +\left(1-\frac{1}{n_A}\right)\alpha+\left(1-\frac{1}{n_B}\right)\beta \\
&\geq \left(1-\frac{1}{n_A}\right) \left(1-\frac{1}{n_B}\right) \\
&\geq 0\, ,
\end{align*}
where we used also $\alpha\geq -1/n_B$ and $\beta\geq -1/n_A$ in the second step. We found also the third, fourth and fifth line of~\eqref{EB eq}.
\item $\big(I\otimes\left(\mathds{1}\tr-I/n_A \right) \big)\, \Gamma$ is CP. Imposing this condition requires just mechanical calculations, because the composed map under examination belongs to the same parametric class defined by~\eqref{Phi}, and therefore Theorem~\ref{complete positivity} applies. Besides the last line of~\eqref{EB eq}, we obtain two additional inequalities:
\begin{align}
n_A n_B - 1 - (n_B - n_A) \alpha &\geq 0\, , \label{additional 1} \\
\alpha+(n_A n_B -1)\beta + n_A \gamma +n_B-\frac{1}{n_A} &\geq 0\, . \label{additional 2}
\end{align}
As is easy to see,~\eqref{additional 1} is redundant, because the upper bound $\alpha\leq \frac{n_A n_B -1}{n_B - n_A}$ is weaker than $\alpha\leq 1$ which follows from third, fourth and fifth line of~\eqref{EB eq} combined. As a matter of fact,~\eqref{additional 2} is also useless, because noting that $\alpha+n_A \gamma\geq -(1+n_A\beta)/n_B$ (thanks to the second inequality of~\eqref{EB eq}) gives us
\begin{align*}
\alpha+(n_A n_B -1)\beta + n_A \gamma +n_B-\frac{1}{n_A} &\geq \left(n_A n_B - 1 -\frac{n_A}{n_B}\right)\left(\beta+\frac{1}{n_A}\right) \\
&\geq 0\, ,
\end{align*}
where we used also $\beta\geq -1/n_A$.
\end{itemize}

The shape of the solid defined by~\eqref{EB eq} can be seen in Figure~\ref{EB region}. Finding the vertices is now an elementary exercise. 
\end{proof}

It is also possible to give a more direct proof of Theorem~\ref{EB}, consisting in finding explicitly separable expressions for the vertices of the solid defined by~\eqref{EB eq} and represented in Figure~\ref{EB region}. For details, we refer the reader to the forthcoming Subsection~\ref{subsec4 EB direct}.

From the above proof of Theorem~\ref{EB} we learnt that the PPT criterion is not sufficient for deciding separability as soon as $n_A\neq n_B$. Figure~\ref{EB region} shows the $\Gamma$ maps that are PPT but not entanglement-breaking, forming another pyramid with a face of the entanglement-breaking solid as a basis and the point $\big( -\frac{1}{n_B},\, -\frac{1}{n_A},\, 1-\frac{n_B-n_A}{n_A n_B}\big)$ as the culminating vertex.

\begin{rem}
Several facts of independent interest can be deduced from the above discussion. Let us recall some of them.
\begin{itemize}
\item The state
\bb
\begin{split}
R_{ABA'B'} &\coloneqq \mathds{1}_{ABA'B'} - \mathds{1}_{AA'}\otimes \Phi_{BB'} - \Phi_{AA'}\otimes \mathds{1}_{BB'} \\
&\qquad + (n_A n_B-n_B+n_A)\, \Phi_{AA'}\otimes \Phi_{BB'}\, ,
\end{split}
\label{PPT ent state}
\ee
corresponding to the vertex of the PPT entangled pyramid depicted in red on the right of Figure~\ref{EB region}, is a PPT entangled state of the bipartite system $AB:A'B'$. This construction can also be found in~\cite[Lemma 4]{Ishizaka1}.
\item If $n_A<n_B$, the map
\bb
I_A\otimes\bigg(\mathds{1}\tr-\frac{I}{n_A} \bigg)_B
\label{indecomposable map}
\ee
acting on a bipartite system $AB$ is positive but \emph{indecomposable}, as can be seen by noting that it detects the PPT entangled state $R_{ABA'B'}$ defined by~\eqref{PPT ent state}. As proved in~\cite[\S IX]{Piani} this remarkable fact is fully general: any $k$-positive but not completely positive map $\Lambda$ gives rise to an indecomposable map $\Lambda\otimes I_k$ by taking a tensor product with the identity channel.
\item The map
\bb
\Gamma\left[ - 1/n_B, - 1/n_B, 1 \right] = \mathds{1}_{AB}\tr - \frac{1}{n_B}\, \mathds{1}_A \tr \otimes I_B  - \frac{1}{n_B}\, I_A\otimes \mathds{1}_B \tr + I_{AB}
\ee
is entanglement-breaking when acting on a bipartite system $AB$ such that $n_A\leq n_B$. 
\end{itemize}
\end{rem}

\subsection{A more direct verification of Theorem~\ref{EB}} \label{subsec4 EB direct}

The proof of the above Theorem~\ref{EB} rests crucially on the argument showing that the necessary conditions we found by composing $\Gamma$ with some suitably chosen positive maps are also sufficient to ensure that $\Gamma$ is entanglement-breaking. In turn, that argument is an application of the theory of symmetries as developed in Subsection~\ref{subsec2 sep symm}. The curious reader might wonder, whether it is instead possible to show directly that the maps corresponding to the vertices of the solid~\eqref{EB eq} are entanglement-breaking. This is precisely what we are set to do now.

We start by reminding the reader that for a bipartite quantum system $AA'$ (with $n_A=n_{A'}=n$) one can specialise Example~\ref{local groups ex} by taking $V_1=V_2=\mathcal{H}_{n}$, $G=SU(n)$, and
\begin{align*}
\zeta_1 (U) (X) &\coloneqq UXU^\dag ,\\
\zeta_2 (U) (X) &\coloneqq U^* X U^T .
\end{align*}
The corresponding \emph{isotropic projection}, defined as in~\eqref{local groups projection}, acts as~\cite{Horodecki1999}
\begin{align}
\Pi^{\text{iso}} (\cdot) \coloneqq & \int dU\ U\otimes U^*\, (\cdot)\ U^\dag\otimes U^T \label{iso proj0} \\
= & \ket{\Phi}\!\!\bra{\Phi} (\cdot) \ket{\Phi}\!\!\bra{\Phi} + \frac{\mathds{1}-\Phi}{n^2-1}\, \Tr [(\mathds{1}-\Phi)\,(\cdot)]\, . \label{iso proj}
\end{align}
It follows directly from~\eqref{iso proj0} that $\Pi^{\text{iso}}$ is separability-preserving as a superoperator.

Now we are ready to prove that each of the vertices of the double pyramid of Figure~\ref{EB region} corresponds to an entanglement-breaking map (or, equivalently, to a separable Choi operator~\eqref{Choi state Phi}). Only one of the five vertices requires a special treatment, while the remaining four are easily seen to correspond to products of entanglement-breaking maps on $A$ and $B$, which are in turn entanglement-breaking maps on $AB$.

\begin{itemize}

\item Basis vertex $\left( -\frac{1}{n_B},\, -\frac{1}{n_B},\, 1\right)$.

The corresponding Choi state reads
\begin{align*}
R_{\Gamma \left[  -1/n_B,\, -1/n_B,\, 1 \right]} &= \frac{\mathds{1}_{ABA'B'}}{n_A n_B} - \frac{1}{n_A n_B}\, \mathds{1}_{AA'}\otimes \Phi_{BB'} \\
&\qquad - \frac{1}{n_B^2}\,\Phi_{AA'}\otimes \mathds{1}_{BB'} + \Phi_{AA'}\otimes \Phi_{BB'}\, .
\end{align*}
While being aware that in general $A$ and $B$ do not need to have the same dimension, we can nevertheless introduce a sort of maximally entangled state
\bbb
\ket{\tilde{\Phi}}_{AB} \coloneqq \frac{1}{\sqrt{n_A}}\, \sum_{i=1}^{n_A}\, \ket{i}_A\ket{i}_B .
\eee
With this notation, some calculations reveal that one can write
\bbb
R_{\Gamma\left[  -1/n_B,\, -1/n_B,\, 1 \right]} = \frac{n_A(n_B^2-1)}{n_B}\, \left(\Pi^{\text{iso}}_{AA'}\otimes \Pi^{\text{iso}}_{BB'}\right) \left( \tilde{\Phi}_{AB} \otimes \tilde{\Phi}_{A'B'} \right) ,
\eee
where $\tilde{\Phi} \coloneqq \ket{\tilde{\Phi}}\!\!\bra{\tilde{\Phi}}$. Since the right-hand side consists of the application of a map that preserves separability with respect to every cut to a state that is already $(AB:A'B')$- separable, we must conclude that the left-hand side is indeed $(AB:A'B')$-separable.

\item Basis vertex $\left( 1,\, -\frac{1}{n_A},\, -\frac{1}{n_A} \right)$.

The corresponding map reads
\begin{align*}
\Gamma \left[ 1,\, - 1/n_A,\, - 1/n_A \right] &= \mathds{1}\tr \otimes\mathds{1}\tr + \mathds{1}\tr\otimes I - \frac{1}{n_A}\, I\otimes \mathds{1}\tr - \frac{1}{n_A}\, I \\
&= \left( \mathds{1}\tr -\frac{I}{n_A} \right)\otimes \left(\mathds{1}\tr + I\right) ,
\end{align*}
and the rightmost side, being a tensor product of two entanglement-breaking maps on $A$ and $B$, is entanglement-breaking on the composite system $AB$. 

\item Basis vertex $\left( -\frac{1}{n_B},\, 1,\, -\frac{1}{n_B} \right)$.

This case is completely analogous to the previous one:
\begin{align*}
\Gamma \left[ -1/n_B,\, 1,\, -1/n_B \right] &= \mathds{1}\tr \otimes \mathds{1}\tr - \frac{1}{n_B} \mathds{1}\tr\otimes I + I\otimes \mathds{1}\tr - \frac{1}{n_B}\, I \\
&= \left( \mathds{1}\tr + I \right)\otimes \left(\mathds{1}\tr - \frac{I}{n_B}\right) .
\end{align*}

\item Culminating vertex $\left( -\frac{1}{n_B},\, -\frac{1}{n_A},\, \frac{1}{n_A n_B} \right)$.

We have
\begin{align*}
\Gamma \left[ -1 /n_B,\, - 1/n_A,\, 1/n_A n_B \right] &= \mathds{1}\tr \otimes \mathds{1}\tr -\frac{1}{n_B} \mathds{1}\tr\otimes I \\
&\qquad - \frac{1}{n_A} I\otimes \mathds{1}\tr + \frac{1}{n_A n_B}\, I \\
&= \left( \mathds{1}\tr -\frac{I}{n_A} \right)\otimes \left(\mathds{1}\tr - \frac{I}{n_B}\right) ,
\end{align*}
which is a tensor product of entanglement-breaking maps.

\item Culminating vertex $(1,\, 1,\, 1)$.

The last case is
\begin{align*}
\Gamma[1,\, 1,\, 1] &= \mathds{1}\tr\otimes \mathds{1}\tr + \mathds{1}\tr\otimes I + I\otimes\mathds{1}\tr + I \\
&= \left( \mathds{1}\tr +I \right)\otimes \left( \mathds{1}\tr +I \right) ,
\end{align*}
again a tensor product of entanglement-breaking maps. This concludes our sanity check of the correctness of Theorem~\ref{EB}.
\end{itemize}

\section{Entanglement-annihilating region} \label{sec EA}

We now move on to the problem of characterising the triple of parameters $\alpha,\beta,\gamma$ for which the map $\Gamma [\alpha,\beta,\gamma]$ is entanglement-annihilating. This is done in Subsection~\ref{subsec4 main}, where we prove the main result of the section and perhaps of the whole chapter, Theorem~\ref{EA}. In the following Subsection~\ref{subsec4 applications} we then show how to apply this result to the solution of some open problems that have been raised recently~\cite{EA3, EA4}.

\subsection{Main result} \label{subsec4 main}
 
We remind the reader that an entanglement-annihilating map $\Lambda_{AB}$ acting on a bipartite system is characterised by the property that $\Lambda_{AB}(\rho_{AB})$ is separable for all (pure) input states $\rho_{AB}$ (Subsection~\ref{subsec4 maps}).
There is a naive necessary criterion that must be satisfied in order for this to be the case: if $\Lambda_{AB}$ has to be entanglement-annihilating, then $(I_A\otimes T_B)\, \Lambda_{AB}$ must be positive (with $T_B$ denoting partial transposition). As we mentioned in Subsection~\ref{subsec4 maps}, maps for which the latter condition holds are called PPT-inducing in~\cite{EA4}. We are now in position to state and prove our main result.

\begin{thm} \label{EA}
Given a map $\Gamma[\alpha,\beta,\gamma]$ defined by~\eqref{Phi}, the following are equivalent:
\begin{enumerate}[(a)]
\item $\Gamma$ is is entanglement-annihilating ;
\item $\Gamma$ is positive and PPT-inducing;
\item $\Gamma$ is positive and in addition $\gamma\leq \alpha+\beta+2$;
\item the following conditions are met:
\bb
\begin{split}
\alpha + 1 &\geq 0\, ,\\
\beta + 1 &\geq 0\, ,\\ 
\frac{\alpha+\beta}{n}+\gamma+1 &\geq 0\, ,\\ 
\alpha+\beta+\gamma+1 &\geq 0\, ,\\ 
\alpha+\beta- \gamma +2 &\geq 0\, .
\end{split}
\label{EA eq}
\ee
\end{enumerate}
\end{thm}

\begin{figure}[h] 
\centering
\includegraphics[height=6cm, width=6cm, keepaspectratio]{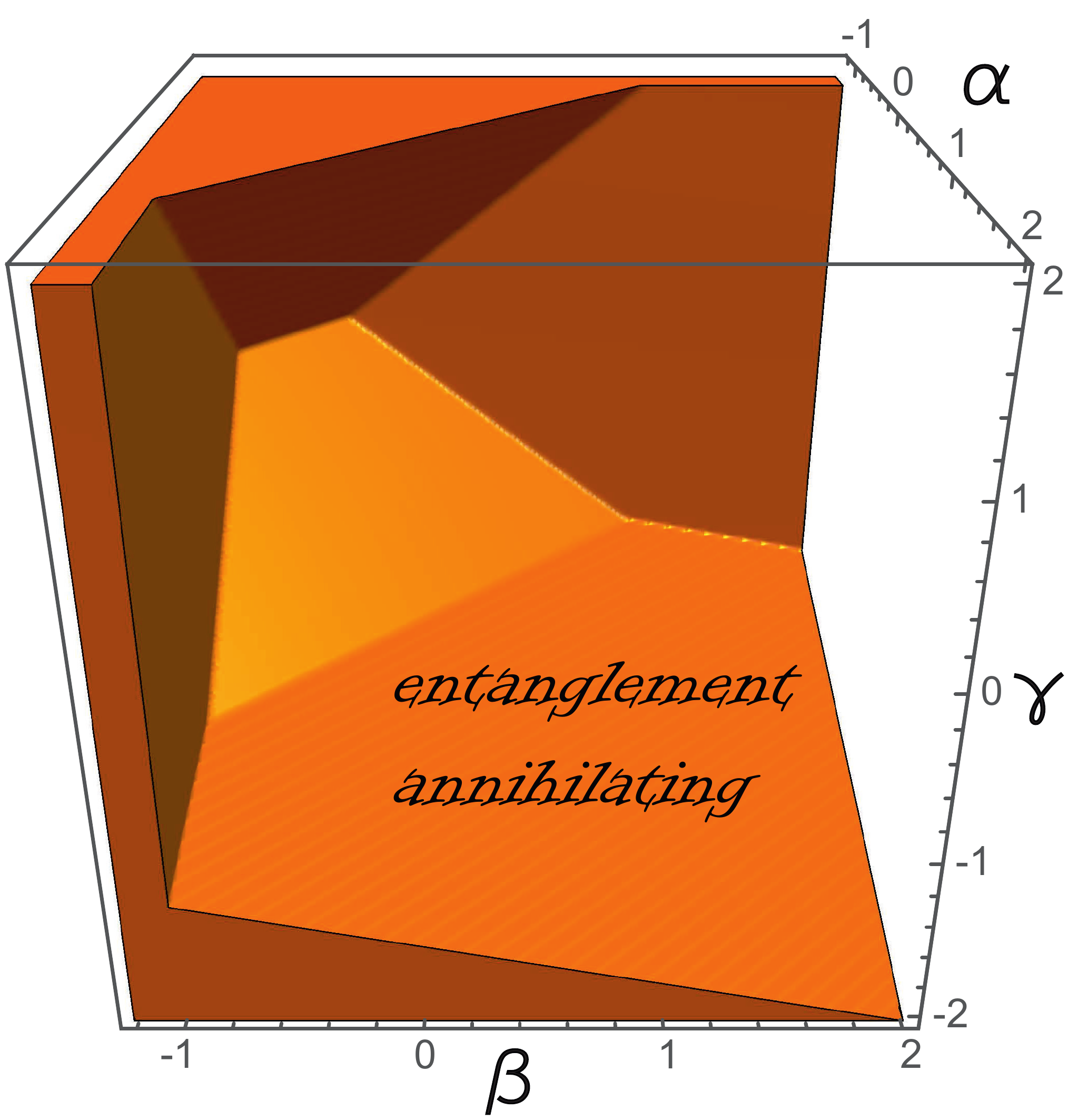}
\caption[]{The convex region outside of the solid represents the parameter range for which the channel~\eqref{Phi} is entanglement-annihilating. Here the case $n=4$ is shown. Compare with Figure~\ref{Positivity region} and note the extra plane on the top identifying the additional condition $\gamma\leq \alpha+\beta+2$.}
\label{EA region}
\end{figure}

\begin{proof} Let us break down the argument.
\begin{description}
\item[$(a)\Rightarrow (b).$] We already saw that an entanglement-annihilating map is necessarily PPT-inducing (and obviously positive).
\item[$(b)\Rightarrow (c).$] Input to $\Gamma$ the pure state $\ket{\psi}=\frac{\ket{11}+\ket{22}}{\sqrt{2}}$; the positivity of the partial transpose of the resulting state requires $\gamma\leq \alpha+\beta+2$. 
\item[$(c)\Rightarrow (d).$] Trivially obtained by using Theorem~\ref{positivity}. The region identified by these conditions is represented in Figure~\ref{EA region}.
\item[$(d)\Rightarrow (a).$] As one could expect, this is the only point that requires a bit of care. Observing Figure~\ref{EA region}, we note that every point of the positive and PPT-inducing region is a convex combination of four points on the four half-lines forming the edges of the set. If we prove that all these four half-lines are composed entirely of entanglement-annihilating maps, we are done.
\begin{itemize}

\item The two lower half-lines have already been studied in the proof of Theorem~\ref{positivity}. The one on the right of Figure~\ref{EA region} corresponds to~\eqref{lower line 1 pos}, and is composed of maps of the form
\begin{align*}
\Gamma\left[ -1,\,\beta,\, -1- (\beta-1)/n \right] &= (\mathds{1}\tr+I)\otimes (\mathds{1}\tr-I) \\
&\quad + (\beta-1)\, I\otimes \Big(\mathds{1}\tr-\frac{1}{n} I\Big)\, ,
\end{align*}
with $\beta\geq 1$. Both of the addends of the above equation are easily seen to be entanglement-annihilating, because they are tensor products of a positive and an entanglement-breaking map on the two subsystems. Consequently, their sum is entanglement-annihilating as well. The same reasoning applies to the other lower half-line~\eqref{lower line 2 pos}.

\item The two upper half-lines of Figure~\ref{EA region} are again symmetrically related and can be treated in the same way. The one on the right, for instance, is composed of maps of the form
\begin{align*}
\Gamma\left[ -1,\, \beta,\, \beta+1 \right] &= \Big(\mathds{1}\tr-\frac{I}{2}\Big)\otimes (\mathds{1}\tr-I) \\
&\quad + \Big(\beta+\frac{1}{2}\Big)\, I\otimes \left(\mathds{1}\tr+ I \right) ,
\end{align*}
with $\beta\geq -\frac{1}{2}$. The second addend is entanglement-annihilating because it is a tensor product of the identity and an entanglement-breaking map. Proving that also the first addend is entanglement-annihilating is not completely trivial. Consider an arbitrary pure state $\ket{\psi}_{AB}=\sum_{i=1}^n \sqrt{\lambda_i}\ket{i}_A\ket{i}_B$, that we are assuming to be Schmidt decomposed~\eqref{Schmidt decomposition} in the computational basis without loss of generality. We can also take the coefficients to be sorted in decreasing order, i.e. $\lambda_1\geq\ldots\geq \lambda_n>0$, and such that $\sum_i \lambda_i = 1$. Define $\psi_{AB} \coloneqq \ket{\psi}\!\!\bra{\psi}_{AB}$, and denote the reduced state by $\rho_{\psi}\coloneqq \tr_B \psi_{AB}$. We write
\begin{align*}
2\, \Big(\mathds{1}\tr-\frac{I}{2}\Big)\otimes (\mathds{1}\tr-I)\, ( \psi ) &= 2\, \mathds{1} - 2\, \mathds{1}\otimes \rho_{\psi} - \rho_\psi\otimes\mathds{1} + \psi =\\
&= \sum_{i\neq j} \sqrt{\lambda_i\lambda_j}\, \ket{ii}\!\!\bra{jj} \\
&\quad + \sum_{i,j} (2 - 2\lambda_j -\lambda_i + \lambda_i \delta_{ij})\, \ket{ij}\!\!\bra{ij}\, ,
\end{align*}
where the indexes are understood to run from $1$ to $n$. Basically, our strategy to prove that the above state is separable will consist in a comparison with a known separable state. Defining $Q\coloneqq\sum_{i\neq j} \ket{ij}\!\!\bra{ij}$, and denoting with $\Phi$ the projector onto the maximally entangled state~\eqref{max ent state}, the operator
\bb
Q + n\, \Phi = \sum_{i\neq j} \ket{ii}\!\!\bra{jj} + \sum_{i,j} \ket{ij}\!\!\bra{ij}
\label{notable state}
\ee
turns out to be separable.\footnote{Although this seems to be known to experts in the field~\cite[Eq. (18)]{MaxAbelian}, we were not able to find an explicit proof in the literature, so let us provide one. Define $\ket{+}\coloneqq \frac{1}{\sqrt{n}}\, \sum_{j=1}^n \ket{j}$, and for $\theta\in\mathds{R}^n$ let $D_\theta \coloneqq \sum_{j=1}^n e^{i\theta_j} \ket{j}\!\!\bra{j}$ be the corresponding diagonal unitary. Then, one sees that
\begin{align*}
Q + n\, \Phi = \int_{0}^{2\pi} \frac{d^n\theta}{(2\pi)^n}\ D_\theta \otimes D_\theta\ \ket{+}\!\!\bra{+} \otimes \ket{+}\!\!\bra{+}\ D_\theta^\dag \otimes D_\theta^\dag\, ,
\end{align*}
implying that the left-hand side is indeed separable, as claimed.}
A first strategy could be based on a conjugation by a local diagonal matrix $D_\lambda=\text{diag}(\lambda_1,\ldots,\lambda_n)$. One could write
\begin{align*}
D_\lambda \otimes \mathds{1}\, \left( Q + n\, \Phi \right)\, D_\lambda \otimes\mathds{1} &= \sum_{i\neq j} \sqrt{\lambda_i \lambda_j}\, \ket{ii}\!\!\bra{jj} + \sum_{i,j} \lambda_i \ket{ij}\!\!\bra{ij} =\\
&= 2\, \mathds{1} - 2\, \mathds{1}\otimes \rho_{\psi} - \rho_\psi \otimes\mathds{1} + \psi\\
&\quad - \sum_{i,j}\, (2 - 2\lambda_i - 2\lambda_j + \lambda_i \delta_{ij})\, \ket{ij}\!\!\bra{ij}\, .
\end{align*}
If $2 - 2\lambda_i - 2\lambda_j + \lambda_i \delta_{ij}\geq 0$ for all $i,j$ we would be done, because carrying the last addend on the left-hand side of the equation would yield a separable decomposition of the required state. However, the latter inequality fails to hold if $i=j=1$ and $\lambda_1>2/3$. To include also this case, we must think of something different. 

Construct $\ket{\tilde{\psi}}=\sum_i \sqrt{\lambda_i}\, \ket{i}$ and $\tilde{\psi}\coloneqq \ket{\tilde{\psi}}\!\!\bra{\tilde{\psi}}$, and use Theorem~\ref{pos chi} to claim that 
\begin{align*}
M &\coloneqq \mathds{1} - 2 D_\lambda + \tilde{\psi} \\
&= \sum_{i\neq j} \sqrt{\lambda_i\lambda_j}\, \ket{i}\!\!\bra{j} + \sum_i\, (1-\lambda_i)\, \ket{i}\!\!\bra{i} \\
&\geq 0\, .
\end{align*}
Then, define the completely positive map $\phi_M$ acting as $\phi_M(X)\coloneqq M\circ X$, where $\circ$ denotes Hadamard product. One has
\begin{align*}
(I\otimes \phi_M)(Q+n\,\Phi) &= \sum_{i\neq j} \sqrt{\lambda_i\lambda_j}\, \ket{ii}\!\!\bra{jj} + \sum_{i,j}\, (1-\lambda_j)\, \ket{ij}\!\!\bra{ij} \\
&= 2\, \mathds{1} - 2\, \mathds{1}\otimes \rho_{\psi} - \rho_\psi\otimes\mathds{1} + \psi \\
&\quad - \sum_{i,j}\, (1-\lambda_i-\lambda_j+\lambda_i \delta_{ij})\, \ket{ij}\!\!\bra{ij}\, .
\end{align*}
Since if $1-\lambda_i-\lambda_j+\lambda_i \delta_{ij}\geq 0$ for all $i,j$ (thanks to $\sum_i \lambda_i = 1$), we can conclude.
\end{itemize}
\end{description}
\end{proof}

We found particularly surprising that the state $ 2\, \mathds{1} - 2\, \mathds{1}\otimes \rho_{\psi} - \rho_\psi\otimes\mathds{1} + \psi$ is separable for all the global pure states $\ket{\psi}$. Furthermore, the techniques we employed to prove this fact are to some extent original and possibly prone to be used more extensively within the context of the quantum separability problem. We will see in a moment that besides being interesting in itself, Theorem~\ref{EA} is also useful in closing some open problems recently raised in the literature.
Recently, we have learnt that the result can be employed to study the relation between the two seemingly unrelated concepts of entanglement-annihilating and entanglement-breaking channel, following~\cite[\S IV.A]{Alex-tensor-stable}.\footnote{I am indebted to Alexander M\"{u}ller-Hermes for this comment.}

\subsection{Applications} \label{subsec4 applications}

Throughout this subsection, we apply the results in Section~\ref{sec EA} to solve some open problems recently posed in~\cite{EA3, EA4}. In those papers, the scenario in which a depolarising channel acts locally on two equal sides of a bipartite quantum system is considered. More specifically, using the notation in~\eqref{dep} with $-\frac{1}{n^2-1}\leq \lambda\leq 1$, it is asked what is the condition on $\lambda_1,\lambda_2$ under which $\Delta_{\lambda_1}\otimes \Delta_{\lambda_2}$ becomes entanglement-annihilating. In~\cite[Eq. (5)]{EA3}, a \emph{sufficient condition} is found, that reads
\begin{equation} (n^2-1) \lambda_1 \lambda_2 \leq 1 + \frac{(n-2)(n+1)}{n+2}\, (\lambda_1+\lambda_2)\, . \label{FZ eq1} \end{equation}
In particular, for the symmetric case $\lambda_1=\lambda_2=\lambda$, the explicit form
\begin{equation} \lambda \leq \frac{\,n-2+n\sqrt{\frac{2n}{n+1}}\,}{(n-1)(n+2)} \label{FZ eq1 sym} \end{equation}
is deduced from~\eqref{FZ eq1}. At the same time, the condition
\begin{equation}
\lambda \leq \frac{1+\sqrt{3}}{n+1+\sqrt{3}} \label{FZ eq2}
\end{equation}
is shown to be necessary in order for $\Delta_\lambda\otimes \Delta_\lambda$ to be PPT-inducing. In~\cite{EA4} it is also conjectured, that~\eqref{FZ eq2} is also \emph{sufficient} for enforcing the PPT-inducing behaviour (if the $\lambda$ range is restricted to the complete positivity interval). Observe that:
\begin{enumerate}[(i)]
\item there is a gap between the region~\eqref{FZ eq1 sym} in which the entanglement annihilation is guaranteed and the region~\eqref{FZ eq2} outside which the global map is not even PPT-inducing;
\item in~\cite{EA4} it is conjectured that inside that gap the map is still PPT-inducing, while no supposition is made about the entanglement annihilation.
\end{enumerate}

Applying Theorem~\ref{EA} straightforwardly solves all these problems, showing that~\eqref{FZ eq2} is indeed a \emph{necessary and sufficient} condition for the map $\Delta_\lambda\otimes \Delta_\lambda$ to be PPT-inducing or, equivalently, entanglement-annihilating. In particular, this proves the conjecture in~\cite{EA4}. Before discussing the details of the above solution, we stress that it is not necessary to assume the complete positivity of the local maps (as done in both~\cite{EA3} and~\cite{EA4}) in order for our problem to make sense. On the contrary, it is enough to demand that the product $\Delta_{\lambda_1}\otimes \Delta_{\lambda_2}$ is positive. A straightforward application of Theorem~\ref{positivity} shows that this is the case iff
\begin{equation} -\frac{1}{n-1} \leq \lambda_1,\lambda_2 \leq 1\quad \text{and}\quad \lambda_1 \lambda_2\geq -\frac{1}{n^2-1}\, . \label{pos delta delta} \end{equation}
By comparison, remind that the complete positivity conditions are
\begin{equation} -\frac{1}{n^2-1} \leq \lambda_1,\lambda_2 \leq 1\, . \label{cp delta delta} \end{equation}
Now, we are in position to show our solution to the aforementioned open questions.

\begin{cor} \label{EA cor}
The product $\Delta_{\lambda_1}\otimes \Delta_{\lambda_2}$ of two local depolarising maps~\eqref{dep} is entanglement-annihilating iff, besides the positivity conditions~\eqref{pos delta delta}, the inequality
\begin{equation} (n^2+2n-2) \lambda_1 \lambda_2 \leq 2 + (n-2) (\lambda_1 + \lambda_2) \label{improv 1} \end{equation}
holds. In the symmetric case $\lambda_1=\lambda_2=\lambda$,~\eqref{improv 1} becomes simply
\begin{equation} - \frac{\sqrt{3}-1}{n+1-\sqrt{3}} \leq \lambda \leq \frac{1+\sqrt{3}}{n+1+\sqrt{3}}\, . \label{improv 1 sym} \end{equation}
\end{cor}

\begin{proof}
We can apply Theorem~\ref{EA} in the form of condition 3. Writing
\begin{equation*}
\Delta_{\lambda_1}\otimes \Delta_{\lambda_2} = \lambda_1 \lambda_2\ \Gamma \left[\frac{n \lambda_2}{1-\lambda_2},\, \frac{n \lambda_1}{1-\lambda_1},\, \frac{n^2 \lambda_1 \lambda_2}{(1-\lambda_1)(1-\lambda_2)} \right]\, ,
\end{equation*}
we have only to impose the further constraint
\begin{equation*}
\frac{n^2 \lambda_1 \lambda_2}{(1-\lambda_1)(1-\lambda_2)} \leq \frac{n \lambda_2}{1-\lambda_2} + \frac{n \lambda_1}{1-\lambda_1} + 2\, ,
\end{equation*}
which becomes~\eqref{improv 1} after elementary algebraic manipulations. Deducing~\eqref{improv 1 sym} is now a simple exercise. Observe that the lower bound on $\lambda$ expressed by~\eqref{improv 1 sym} is superfluous if the complete positivity condition is imposed (and in that case we end up with~\eqref{FZ eq2}), but must be retained if only the positivity is imposed.
\end{proof}

We want to answer another question that is left open in~\cite{EA3}. Besides the local depolarising noise, in that paper a global depolarizing noise of the form $\left( \Delta_\lambda \right)_{AB} = \lambda I_{AB} + (1-\lambda)\frac{\mathds{1}_{AB}}{n_A n_B}\tr$ is also considered, in the simplest case $n_A=n_B=n$. It is observed that $\left( \Delta_\lambda\right)_{AB}$ is not PPT-inducing if $\lambda>\frac{2}{n^2+2}$, but an explicit entanglement-annihilating construction is provided only for $\lambda\leq\frac{n+2}{(n+1)(n^2-n+2)}$.

It is very simple to observe that $\left(\Delta_{2/(n^2+2)}\right)_{AB}=\frac{1}{n^2+2}(\mathds{1}\tr +2I)_{AB}$ is already entanglement-annihilating, indeed. This follows easily from Theorem~\ref{EA}, but is also a consequence of the well-known fact~\cite{VidalTarrach} that $\mathds{1}+2\rho$ is separable for all normalised density matrices $\rho$ on a bipartite system.

\section{Conclusion} \label{sec4 conclusions}

This concludes the characterisation of the extended depolarising channel. As we have shown, this natural generalisation to bipartite systems of the paradigmatic noisy channel displays a rich structure that nonetheless allows for analytical solutions of all of the basic problems. While explicitly working out the parameter regions for which the map is positive, completely positive, entanglement-breaking and entanglement-annihilating, we have developed some new techniques to deal with the separability problem in quantum information. In particular, we have made some detailed observations on the Hadamard product in relation to entanglement theory, and showed how these tricks can be useful in proving separability of special states or classes of states. This led us to the solution of some outstanding problems in entanglement annihilation.
Along the way, we also retrieved a simple example of a positive indecomposable map and constructed a notable entanglement-annihilating map. 

The maps we consider naturally emerge in situations where two separate systems are subject to white noise and thus provide a useful tool in predicting the physical impact of noise on entanglement. Furthermore they provide new ways to reveal bound entanglement and an analytical characterisation of entanglement for a natural class of states. We hope that the observations we point out elucidate certain facts in a way that is useful also for future applications. The solution of the depolarising channel now provides a promising basis for studying natural generalisations. First of all one should look at straightforward extensions to the multipartite case, which will involve many more parameters, but keep the local unitary symmetries (in fact the number of parameters scales exponentially in the number of parties only). Another potential path to pursue is to study the case of coloured noise. While this would in general lead to a number of parameters growing in the dimension, thermal noise in equidistant Hamiltonians would still give a physically important five parameter family that may yield tractable solutions.

\chapter{Gaussian entanglement revisited} \label{chapter5}

\section{Introduction} \label{sec5 introduction}

Quantum entanglement and the problem of its identification and characterisation has guided our investigation in Chapter~\ref{chapter2} (see especially Section~\ref{sec2 sep problem}), as well as in Chapter~\ref{chapter4}, where we had to analyse the entanglement transformation properties of certain physically relevant maps.
In the present chapter, instead, we give new insights into the separability problem for Gaussian states. These are special states that arise naturally, from both the theoretical and the experimental point of view, when one deals with continuous variable quantum systems, i.e. collections of quantum oscillators such as electromagnetic modes.

This chapter is organised as follows. As usual, the rest of this section is devoted to introducing the problem (Subsection~\ref{subsec5 G sep}) and to pointing the reader to our main original contributions (Subsection~\ref{subsec5 contributions}). In Section~\ref{sec5 methods} we recall some basic technical facts concerning Schur complements and matrix means, and fix our notation for the continuous variable formalism. Section~\ref{sec5 simplified} is then devoted to establishing a simplified separability condition for Gaussian states (Theorem~\ref{simp sep lemma}), something we will make extensive use of throughout the rest of the chapter. In Section~\ref{sec5 revisited} we revisit some known results in Gaussian entanglement theory by employing the tools introduced in the previous sections. In particular, we provide a new compact proof of the sufficiency of the PPT condition for separability of $1$ vs $n$-mode Gaussian states (Theorem~\ref{PPT thm}). Section~\ref{sec5 novel} contains most of our novel findings, including an extension of the PPT-separability equivalence to bipartite Gaussian states that are invariant under the exchange of any two modes pertaining to one of the two parties (Theorem~\ref{PPt sym}), and the solution to the problem~\cite{passive} of characterising those Gaussian states that can not be made entangled by means of any passive operation (Theorem~\ref{abs sep Gauss}). Finally, Section~\ref{outro} concludes the chapter with a brief summary and some future perspectives related to this work.

\subsection{Gaussian separability problem} \label{subsec5 G sep}

Continuous variable quantum systems~\cite{CERF, weedbrook12, adesso14, BUCCO} play a major role in quantum information science, which should not come as a surprise for at least two good reasons. First, they are somewhat privileged by Nature, since the electromagnetic field is essentially a collection of harmonic oscillators, one for each frequency. This is of outstanding practical other than theoretical importance, since sending light (via optical fibres, through the atmosphere or empty space) is right now, and perhaps for many decades to come, the only feasible way to engineer quantum communication among distant facilities. Second, harmonic oscillators are among the few quantum systems that we know how to solve analytically (essentially since the very early days of quantum mechanics), which makes them good candidates for accurate experimental control.

There is another lesson that we can learn from the choice Nature made for the laws that govern the electromagnetic field. Namely, the Hamiltonian of said field turns out to be \emph{quadratic in the field operators.} By employing relatively simple optical devices one can explore the whole set of such Hamiltonians, but it is not that easy to go out of this family and make the photons interact with each other directly. For this reason, it seems natural to expect that thermal states of quadratic Hamiltonians, also called \emph{Gaussian states}, will be among the most common quantum states one can reliably prepare in a laboratory. This intuition turns out to be correct, at least loosely speaking, and in fact Gaussian states are among the most widely employed quantum resources.

From the information theoretic standpoint, those states and the quantum channels that preserve their structure, called \emph{Gaussian channels}, have been used to successfully implement paradigmatic platforms for continuous variable quantum information processing. Among the others, let us mention unconditional quantum teleportation in optical and atomic domains~\cite{furuscience,spincoherent,fernnp13}, quantum cryptography with coherent states~\cite{Grosshans2003}, and sub-shot-noise interferometry in gravitational wave detectors~\cite{SGravi1,SGravi2}.

The epithet `Gaussian' comes from the fact that, exactly as for their classical counterparts, i.e. multivariate Gaussian random variables, also quantum Gaussian states are entirely described by a mean vector -- encoding the expected values of the field operators -- and the associated covariance matrix, called a \emph{quantum covariance matrix} (QCM). For a system made of $n$ electromagnetic modes, the QCM is a $2n\times 2n$ real matrix. The study of Gaussian states and of their entanglement properties is entirely ascribed to the characterisation of the associated QCMs. For this reason, the corresponding separability problem can be studied with methods of matrix analysis, linear algebra and symplectic geometry, which come directly from \emph{classical} mechanics~\cite{ARNOLD}.


The main question we address in this chapter is as follows. \emph{Given a QCM $V_{AB}$, can we determine whether the corresponding bipartite Gaussian state $\rho_{AB}^G$ is entangled or separable?} It is known~\cite{Werner01} that this problem can be rephrased as a semidefinite programming (SDP), and thus admits an efficient algorithmic solution. In fact, an explicit and provably efficient algorithm that decides the problem has been put forward~\cite{Giedke01}. This is in stark contrast to what happens for the separability problem in standard finite-dimensional quantum systems, which is known to be NP-hard under certain assumptions on the required accuracy~\cite{GurvitsNPhard, GharibianNPhard}.

When the two local systems have a limited number of modes, or when the state displays some further symmetry, there are even closed solution available, in the form of necessary and sufficient separability conditions. The most famous of such conditions is undoubtedly the already mentioned positive partial transpose (PPT) criterion~\cite{PeresPPT}, which turns out to be a powerful tool also in the analysis of continuous variable systems.
In fact, in 1999 it was proved~\cite{Simon00} that two-mode Gaussian states are separable iff they are PPT, while the following year the same result was shown to hold for Gaussian states of $1$ vs $n$ modes~\cite{Werner01}. Since PPT entangled Gaussian states exist already for $2$ vs $2$ modes~\cite{Werner01}, this closes the problem of determining all the pairs $(m,n)$ such that the PPT criterion is both necessary and sufficient in $m$ vs $n$ modes.

However, if further symmetries are imposed on the state, things can change. For instance, later it has been discovered that PPT-ness is again sufficient for `bi-symmetric' Gaussian states, characterised by their invariance under local permutations of any two modes within any of the two subsystems~\cite{Serafini05}, and for `isotropic' Gaussian states, whose QCM has a fully degenerate `symplectic spectrum'~\cite{holwer,botero03,giedkemode}.

Finally, let us mention that in the Gaussian world there are more compelling reason than just mathematical convenience to look at the PPT criterion. Namely, it is known that for bipartite Gaussian states, while entanglement can never be distilled by Gaussian operations alone~\cite{nogo1,nogo2,nogo3}, entanglement distillability~\cite{Bennett-distillation, HorodeckiBound} under general local operations assisted by classical communications is equivalent to violation of the PPT condition~\cite{Giedke01,GiedkeQIC}.

\subsection{Our contributions} \label{subsec5 contributions}

The material presented in this chapter is part of the homonymous paper~\cite{revisited}:
\begin{itemize}

\item L. Lami, A. Serafini, and G. Adesso. Gaussian entanglement revisited. \emph{New J. Phys.}, 20(2):023030, 2018.

\end{itemize}

The objective I had in mind when I started to think about the problems discussed in this chapter was to simplify the presentation of the beautiful theory of Gaussian entanglement, which developed rapidly at the beginning of this century thanks to the work of many authors~\cite{Simon00, Werner01, Giedke01, GiedkeQIC, 3-mode-sep, nogo1, nogo2, nogo3, giedkemode, botero03, passive, Serafini05, adesso07}.\footnote{Let me take the chance to thank Alessio Serafini for the incredibly stimulating course in quantum optics he taught at the Scuola Normale Superiore in Pisa in the academic year 2013-2014, which was of inspiration for this investigation.}
In particular, one of my immediate goals was to develop a mathematical formalism that would help shortening the somehow heterogeneous proof of the equivalence between separability and PPT-ness for $1$ vs $n$-mode Gaussian states. Indeed, the way it developed historically, the argument is composed of two separate parts: first, Simon~\cite{Simon00} tackled the case $n=1$ by means of a direct construction; later, Werner and Wolf~\cite{Werner01} were able to reduce the case of arbitrary $n$ to that of $n=1$ thanks to an ingenious application of symplectic geometry techniques.

This problem is our first testing ground. In Subsection~\ref{subsec5 ppt} we show how to unify the two parts of the proof in single compact reasoning that is technically much simpler and furthermore works directly for the case of generic $n$ (Theorem~\ref{PPT thm}). Key to our proof is  the intensive use of \emph{Schur complements}~\cite{ZHANG}, which have enjoyed applications in various areas of (Gaussian) quantum information theory~\cite{Giedke01, giedkemode, nogo1, nogo2, nogo3, eisemi, steerability, Simon16, Lami16, LL-log-det}, and -- as further reinforced by this work -- may be appreciated as a mathematical cornerstone for continuous variable quantum theory. These mathematical tools are instrumental in establishing a simplified separability condition (Theorem~\ref{simp sep lemma}) that is more easily handled than the original one in~\cite{Werner01}, at least in this context. In fact, the former requires convex optimisation over marginal covariance matrices on one subsystem only, yielding a significant simplification over the latter, which instead require optimisation on both.
Among the other things, this new condition allows us to retrieve immediately the recently proved result that Gaussian states are separable iff they are completely extendible with Gaussian extensions~\cite{Bhat16}.

Subsection~\ref{subsec5 mode-wise} contains instead an alternative proof of the sufficiency of the PPT criterion for `isotropic' Gaussian states, characterised by the property that their covariance matrix has a single distinct symplectic eigenvalue (Theorem~\ref{iso thm}). In the traditional approach, the sufficiency of PPT for their separability follows from a well known `mode-wise' decomposition of pure-state covariance matrices~\cite{holwer,botero03,giedkemode}, and from the fact that the covariance matrix of an isotropic state is just a multiple of that of a pure Gaussian state. On the contrary, main ingredients of this novel proof are advanced matrix analysis tools such as the operator geometric mean, already found to be useful in the context of quantum optics~\cite{Lami16, LL-log-det}. Along the way, we prove a curious result on matrix means (Lemma~\ref{lemma ha=g}) that to the best of our knowledge was not known before.


In Section~\ref{sec5 novel} we show how to apply some of the methods discussed in this chapter to deduce some new results, rather than just new proofs of already established results. In Subsection~\ref{subsec5 inva} we show how to apply our simplified separability condition (Theorem~\ref{simp sep lemma}) to prove that Gaussian states invariant under partial transposition are necessarily separable (Corollary~\ref{inva cor}), a result previously known only for the partial transposition of qubit subsystems~\cite{sep-2xN}.
We then show in Subsection~\ref{subsec5 symm}, Theorem~\ref{iso thm}, that the $1$ vs $n$-mode PPT-separability equivalence can be further extended to a class of arbitrary bipartite multimode Gaussian states that we call `mono-symmetric', i.e. invariant under local exchanges of any two modes on one of the two subsystems (see Figure~\ref{mononucleosi} for a pictorial representation of the proof idea). This result, which, to the best of our knowledge, is observed and proved here for the first time, generalises the case of bi-symmetric states studied in~\cite{Serafini05}, providing as a by-product a simplified proof for the latter as well.

Finally, in Subsection~\ref{subsec5 pass} we consider the well-known class of Gaussian passive operations (i.e. the ones that preserve the average number of excitations of the input state, such as beam splitters and phase shifters), which play a central role in quantum optics~\cite{introeisert,adesso14,BUCCO}, and we prove that a bipartite Gaussian state that always remains PPT under such a set of operations must also always stay separable.
This novel result complements the seminal study of~\cite{passive}, in that the latter only considered the possibility of turning a PPT state into a non-PPT one through passive operations -- essentially, the question of generating distillable entanglement -- which is not the same as the question of generating inseparability, because as we mentioned Gaussian PPT bound entangled states do exist~\cite{Werner01}. Here we settle the latter, more general and fundamental question.
All the previous results enable us to substantially extend the range of equivalence between Gaussian separability and PPT in contexts of strong practical relevance. Last but not least, we address the separability problem directly, and derive a novel simplified necessary and sufficient condition for Gaussian separability.

\section{The toolbox: Schur complements, matrix means and Gaussian states} \label{sec5 methods}

This section has the purpose of acquainting the reader with some of the technical tools to be employed later in the chapter, and in fact throughout the rest of the thesis. For this reason, we take the time to introduce some definitions and properties that will be useful later, even if they are not strictly needed for the derivation of this chapter's results. In Subsection~\ref{subsec5 Schur} we discuss Schur complements extensively, while Subsection~\ref{subsec5 matrix means} is dedicated to an essential introduction to the theory of matrix means. Subsection~\ref{subsec5 G states} is then devoted to Gaussian states and their fundamental properties.

\subsection{Schur complements} \label{subsec5 Schur}

One of the messages of the present paper is to lend further support to the fact that methods based on Schur complements can be successfully employed to derive fundamental results in continuous variable quantum information, following a streak of applications to various contexts including separability, distillability, steerability, entanglement monogamy, characterisation of Gaussian maps, and related problems~\cite{Giedke01, giedkemode, nogo1, nogo2, nogo3, eisemi, steerability, Simon16, Lami16, LL-log-det}. As a divertissement to set the stage, let us present a compact, essential compendium of such methods. Useful references on Schur complements are~\cite{HJ1} or the monograph~\cite{ZHANG}.

Given a square matrix $M$ partitioned into blocks as
\begin{equation}
M = \begin{pmatrix} A & X \\ Y & B \end{pmatrix} , \label{block}
\end{equation}
the \textbf{Schur complement} of its (square, invertible) principal submatrix $A$, denoted by $M/A$, is defined as
\begin{equation}
M/A \coloneqq B - YA^{-1} X\, .
\label{schur}
\end{equation}
We observe that the following factorisation formula holds:
\bb
M = \begin{pmatrix} \id & 0 \\ YA^{-1} & \id \end{pmatrix} \begin{pmatrix} A & 0 \\ 0 & B - YA^{-1} X \end{pmatrix} \begin{pmatrix} \id & A^{-1} X \\ 0 & \id \end{pmatrix} .
\label{Schur factor}
\ee
From this the determinant formula
\begin{equation}
  \det M = (\det A)(\det M/A)
  \label{det factor}
\end{equation}
and the additivity of ranks
\begin{equation}
\rk M = \rk A + \rk (M/A) .
\label{rank add}
\end{equation}
follow straightforwardly. Moreover, observe that if $M$ is symmetric then~\eqref{Schur factor} identifies a kind of normal form under congruence.

From a point of view of matrix analysis, Schur complements arise naturally when one wants to express the inverse of a block matrix in a compact form. Namely, for a matrix $M$ partitioned as in~\eqref{block} one can prove the useful formula~\cite{ZHANG}
\begin{equation}
  M^{-1} = \begin{pmatrix} A^{-1} + A^{-1} X (M/A)^{-1} Y  A^{-1} & -A^{-1} X (M/A)^{-1} \\[0.7ex]
                                            -(M/A)^{-1} Y  A^{-1} & (M/A)^{-1} \end{pmatrix} .
  \label{inv}
\end{equation}
Naturally, an analogous expression holds when one exchanges the roles of $A$ and $B$ and similarly those of $X$ and $Y$. Incidentally, from this latter fact many useful matrix identities can be easily derived. Furthermore, observe that $(M/A)^{-1}$ is a submatrix of $M^{-1}$. A direct computation shows that its Schur complement within $M^{-1}$ is $M^{-1} \big/ (M/A)^{-1}\ =\ A^{-1}$.

Another useful property is \emph{congruence covariance}. In fact, if $N=\left( \begin{smallmatrix} N_1 & 0 \\ 0 & N_2 \end{smallmatrix}\right)$ is decomposed conformally to the partition in~\eqref{block}, it holds
\bb
\big(N M N^T\big) \Big/ \big(N_1^T A N_1 \big) \geq N_2 \left( M/A \right) N_2^T
\label{congruence covariance}
\ee
for all square $N_1$, $N_2$ of appropriate size, with equality if $N_1$ is invertible.

The Schur complement notation is useful mainly because computations are more easily carried out when it is employed. One of the most useful rules that this formalism obeys is the \emph{quotient property}, that allows us to simplify concatenations of Schur complements as if they were ordinary fractions. Consider a matrix $M$ as in~\eqref{block}, and take a square invertible sub-block $A_1$ of its upper left block $A$. Then $A/A_1$ turns out to be a square, invertible submatrix of $M/A_1$, and moreover
\bb
M/A = (M/A_1) \big/ (A/A_1)\, .
\label{quotient property}
\ee

On a different line, Schur complements are the answer to a number of questions that arise pretty naturally in linear algebra. Many of these applications stem from the fact that the positivity conditions of $2\times 2$ Hermitian block matrices can be easily written out in terms of Schur complements.

\begin{lemma} \label{pos cond}
Consider a Hermitian matrix
\begin{equation} H = \begin{pmatrix}  A & X \\ X^\dag & B \end{pmatrix} . \label{H part} \end{equation}
Then $H$ is strictly positive definite ($H>0$) if and only if $A > 0$ and $H/A=B-X^\dag A^{-1}X > 0$. Then, by taking suitable limits, $H$ is positive semidefinite ($H\geq 0$) if and only if $A\geq 0$ and $B-X^{\dag}(A+\epsilon\mathds{1})^{-1} X\geq 0$ for all $\epsilon>0$.
\end{lemma}

A consequence of this result that will be relevant to us is the following.

\begin{cor} \label{Schur variational cor}
Let $H$ be a Hermitian matrix partitioned as in~\eqref{H part}. If $A>0$, then
\begin{equation}  H/A = \sup \Big\{ C=C^\dag:\ H > 0\oplus C \Big\}\, . \label{variational} \end{equation}
Here we mean that the matrix set on the right-hand side has a supremum (i.e.~a minimum upper bound) with respect to the positive semidefinite partial order, and that this supremum is given by the Schur complement on the left-hand side.
\end{cor}

We note in passing that from the above variational representation it follows immediately that $H/A$ is monotonically non-decreasing and concave in $H>0$. Consequently, $(H/A)^{-1}$ is monotonically non-increasing and convex in $H>0$.


\subsection{Matrix means} \label{subsec5 matrix means}

Somehow related to Schur complements are the so-called matrix means. As one might expect from their name, these are functions taking two positive matrices as inputs and yielding another positive matrix as output. For an excellent introduction to this topic, we refer the reader to~\cite[IV]{BHATIA}. 
Given two strictly positive matrices $A,B>0$, the simplest mean one can define is the \textbf{arithmetic mean} $(A+B)/{2}$, whose generalisation from scalars to matrices does not present difficulties. Another easily defined object is the \textbf{harmonic mean}~\cite{parallel-sum, ando79}, denoted by $A!B$ and given by
\begin{equation}
A!B \coloneqq \left(\frac{A^{-1}+B^{-1}}{2}\right)^{-1} . \label{harmonic}
\end{equation}
Incidentally, the harmonic mean can also be defined as a Schur complement, with the help of the identity 
\bb
A!B=A-A(A+B)^{-1}A=\begin{pmatrix} A & A \\ A & A+B \end{pmatrix} \Big/ (A+B)\, ,
\ee
which immediately implies that $A!B$ is monotone and jointly concave in $A$ and $B$, i.e.~concave in the pair $(A,B)$.

The least trivially defined among the elementary means is undoubtedly the \textbf{geometric mean} $A\# B$ between strictly positive matrices $A,B>0$~\cite{geometric-mean, ando79}, which can be constructed as
\begin{align}
A\# B \coloneqq&\, \max\{X=X^{\dag}:\ A\geq XB^{-1} X\} \label{geometric} \\
=& \, \max\Big\{ X=X^\dag:\ \begin{pmatrix} A & X \\ X & B \end{pmatrix}\geq 0\Big\}\, , \label{geometric block}
\end{align}
where the above maximisation is as usual with respect to the positive semidefinite partial order (the fact that the particular set of matrices we chose admits an absolute maximum is already nontrivial), and~\eqref{geometric block} is equivalent to~\eqref{geometric} thanks to Lemma~\ref{pos cond}. From the variational characterisation~\eqref{geometric block} we see that the geometric mean $A\# B$ is strictly monotonic in $A,B>0$.
Furthermore, with a bit of work one can show that $A\# B$ is explicitly given by~\cite{geometric-mean}
\begin{equation}
A\# B = A^{1/2} \left( A^{-1/2} B A^{-1/2} \right)^{1/2} A^{1/2}\, . \label{geom expl}
\end{equation}
Having multiple expressions for a single matrix mean is always useful, as some properties that are not easy to prove within one formulation may become apparent when a different approach is taken. For instance, the fact that $A\# B$ is covariant under congruences, i.e.~\cite[Corollary 2.1]{ando79}
\bb
(MAM^{\dag})\#(MBM^{\dag})=M(A\#B)M^{\dag} \qquad \forall\ \text{invertible $M$,}
\label{congruence covariance geometric mean}
\ee
is far from transparent if one looks at~\eqref{geom expl}, while it becomes almost obvious when~\eqref{geometric} is used.
On the contrary, the fact that $A\#B=(AB)^{1/2}$ when $[A,B]=0$ is not easily seen from~\eqref{geometric}, but it is readily verified employing~\eqref{geom expl}. Incidentally, these two latter requirements (congruence invariance and reduction to standard geometric mean for commuting matrices) identify uniquely the expression~\eqref{geom expl} as the `correct' matrix geometric mean.
The expression for the determinant of the geometric mean, given by
\bb
\det (A\# B) = \sqrt{(\det A) (\det B)}\, ,
\label{det geometric mean}
\ee
is another instance of a property that is easily readable from the explicit formula~\eqref{geom expl}. 

The geometric mean is jointly concave in $A,B>0$, which amounts to saying that the map $(A,B)\mapsto A\# B$ is concave. This is most easily seen via~\eqref{geometric block}, from which it also follows that
for all positive maps $\Lambda: \mathcal{H}_n\rightarrow\mathcal{H}_m$ (see Subsection~\ref{subsec4 maps}) it holds~\cite[Theorem 3]{ando79}
\bb
\Lambda (A\# B) \leq \Lambda (A) \# \Lambda (B)\, .
\label{geometric mean positive maps}
\ee

Now, let us discuss an interesting geometric interpretation of~\eqref{geom expl} that has drawn considerable attention in the mathematical community. For more details we refer the reader to~\cite[VI]{BHATIA}.
The set of positive definite matrices of fixed size $d$ can be seen as a manifold in the real space $\mathds{R}^{d \times d}$.
We can turn it into a Riemannian manifold by introducing on the tangent space the metric $ds^2\coloneqq \Tr[(A^{-1} dA)^2]$ (sometimes called \textbf{trace metric}). It turns out the geodesic connecting two positive matrices $A$ and $B$ in this metric, parametrised by $t\in [0,1]$, is given by
\begin{equation}
\gamma(t) = A^{1/2} \left( A^{-1/2} B A^{-1/2} \right)^{t} A^{1/2} \eqqcolon A\#_t B\, ,
\label{geom geod}
\end{equation}
sometimes called the \textbf{weighted geometric mean}.
Confronting~\eqref{geom expl} with the above expression~\eqref{geom geod}, we see in particular that $A\# B$ is nothing but the geodesic midpoint between $A$ and $B$.

A consequence of this observation is that the monotonicity of the geometric mean under positive maps~\eqref{geometric mean positive maps} extends to the weighted case as well, reading
\bb
\Lambda (A\#_t B) \leq \Lambda (A) \#_t \Lambda (B)
\label{weighted geometric mean positive maps}
\ee
for all matrices $A,B>0$, positive maps $\Lambda$, and $t\in [0,1]$.
To show the above inequality, just apply~\eqref{geometric mean positive maps} iteratively to conclude that~\eqref{weighted geometric mean positive maps} holds at least when $t$ is a dyadic rational. Then, continuity implies that it must hold indeed for all $t\in [0,1]$.
This standard reasoning is totally analogous to the one normally used to show that mid-point convexity and convexity are equivalent for continuous functions.
Also the determinantal identity~\eqref{det geometric mean} can be easily generalised to this more general case. In fact, employing~\eqref{geom geod} it is readily seen that
\begin{equation}
\det (A\#_t B) = (\det A)^{1-t} (\det B)^t .
\label{det geom}
\end{equation}

Of course, we can ask ourselves, how the geometric mean compares to the other means (arithmetic and harmonic) in the matrix setting. As it happens with scalars, the inequality
\begin{equation}
A!B \leq A\#B \leq \frac{A+B}{2} \label{hga}
\end{equation}
holds true for all $A,B>0$~\cite[Corollary 2.1(iv)]{ando79}. In view of the above inequality, it could be natural to wonder, whether the geometric mean between the leftmost and rightmost sides of~\eqref{hga} bears some relation with $A\#B$. That this could be a fruitful thought is readily seen by asking the same question for real numbers. In fact, when $0<a,b\in\mathds{R}$ it is elementary to verify that 
\bbb
\sqrt{ab}=\sqrt{\frac{a+b}{2}\, \left( \frac{1/a+1/b}{2}\right)^{-1}}\,.
\eee
Our first result is a little lemma extending this to the non-commutative case. We were not able to find a proof in the literature, so we provide one.

\begin{lemma} \label{lemma ha=g}
For $A,B>0$ strictly positive matrices, the identity
\begin{equation}
A\#B = \left(\frac{A+B}{2}\right)\#\left( A!B \right)
\label{ha=g}
\end{equation}
holds true.
\end{lemma}

\begin{proof}
We start by defining 
\begin{align*}
\tilde{A} &\coloneqq \left(A+B\right)^{-1/2}A\left(A+B\right)^{-1/2}\, ,\\
\tilde{B} &\coloneqq \left(A+B\right)^{-1/2}B\left(A+B\right)^{-1/2}\, .
\end{align*}
It is easy to see that $[\tilde{A},\tilde{B}]=0$, for instance because 
\bbb
\tilde{A}+ \tilde{B}=\left(A+B\right)^{-1/2}(A+B)\left(A+B\right)^{-1/2}=\mathds{1}\, .
\eee
Therefore, the identity $\tilde{A}\#\tilde{B}=(\tilde{A}\tilde{B})^{1/2}$ holds. Now, on the one hand the congruence covariance of the geometric mean implies that
\begin{align*}
\tilde{A}\#\tilde{B} &= \left( \left(A+B\right)^{-1/2}A\left(A+B\right)^{-1/2} \right) \# \left( \left(A+B\right)^{-1/2}B\left(A+B\right)^{-1/2} \right) \\[0.4ex]
&= \left(A+B\right)^{-1/2}(A\# B)\left(A+B\right)^{-1/2}\, .
\end{align*}
On the other hand,
\begin{align*}
\tilde{A}\tilde{B} &= \left(A+B\right)^{-1/2}A\left(A+B\right)^{-1}B\left(A+B\right)^{-1/2} \\[0.4ex]
&= \left(A+B\right)^{-1/2}\left(B^{-1}(A+B)A^{-1}\right)^{-1}\left(A+B\right)^{-1/2}  \\[0.4ex]
&= \frac12\, \left(A+B\right)^{-1/2}(A!B)\left(A+B\right)^{-1/2}\, .
\end{align*}
Putting all together, we see that
\begin{align*}
\left(A+B\right)^{-1/2}\!(A\# B)\left(A+B\right)^{-1/2} &= \tilde{A}\#\tilde{B} \\
&= (\tilde{A}\tilde{B})^{1/2} \\
&= \frac{1}{\sqrt{2}}\, \left(\left(A+B\right)^{-1/2}\!(A!B)\left(A+B\right)^{-1/2}\right)^{1/2}\! .
\end{align*}
Conjugating by $(A+B)^{1/2}$, we obtain
\begin{align*}
A\# B &= \frac{1}{\sqrt{2}}\, \left(A+B\right)^{1/2} \left(\left(A+B\right)^{-1/2}(A!B)\left(A+B\right)^{-1/2}\right)^{1/2} \left(A+B\right)^{1/2} \\[0.4ex]
&= \left(\frac{A+B}{2}\right)\#\left( A!B \right) ,
\end{align*}
where the last step is an application of~\eqref{geom expl}.
\end{proof}

\subsection{Gaussian formalism} \label{subsec5 G states}

In the remainder of this section, we provide a brief introduction to the main concepts of the Gaussian formalism. Not all of what we say here will be needed for the sake of the present chapter. However, we will refer to this introduction for the rest of the thesis when dealing with Gaussian states. For further details we refer the reader to the textbook~\cite{BUCCO}, whose notation and conventions we follow.

\subsubsection{Canonical commutation relations}

Quantum continuous variables describe quantum theory applied to an infinite-dimensional Hilbert space equipped with position and momentum operators $x_j,p_k$ ($j,k=1,\ldots, n$) satisfying the so-called canonical commutation relations $[x_j,p_k]=i\delta_{jk}$ (in natural units, $\hbar=1$). Such a Hilbert space describes, for instance, a collection of $n$ quantum harmonic oscillators, or modes of the electromagnetic radiation field, with $x_j, p_k$ being the non-commutative analogues of the classical electric and magnetic fields.

The operators $x_j,p_k$ are often grouped together to form a single vector of $2n$ operators $r \coloneqq  (x_1, \ldots, x_n, p_1,\ldots,p_n)$. The canonical commutation relations then take the form
\begin{equation}
[r,r^T] = i \Omega \coloneqq i \begin{pmatrix} 0 & \id \\ -\id & 0 \end{pmatrix} .
\label{CCR}
\end{equation}
We can also consider a different block decomposition of the system, corresponding to the ordering $r = (x_1, p_1, \ldots, x_n, p_n)$. In this new convention, one has
\begin{equation}
[r,r^T] = \omega^{\oplus n}\, , \qquad \omega \coloneqq \begin{pmatrix} 0 & {1} \\ - {1} & 0 \end{pmatrix} .
\label{CCR mode-wise}
\end{equation}
From now on, we will prefer the former to the latter convention, unless otherwise specified. We will refer to expressions written according to~\eqref{CCR mode-wise} as decomposed \textbf{mode-wise}.

An important object one can form is the displacement operator. For any $z\in\mathds{R}^{2n}$, we define
\begin{equation}
D_z \coloneqq e^{iz^T\Omega r}\, .
\label{displacement}
\end{equation}

\subsubsection{Gaussian states}

As we said, Nature has a special preference for quadratic Hamiltonians. A prominent example is the free-field Hamiltonian $\mathcal{H}_0=\frac{1}{2} r^T r$. For this reason, thermal states of quadratic Hamiltonians are extremely easily produced in the lab, in fact so easily that they deserve a special name, \textbf{Gaussian states}. As the name suggests, they can be fully described by a real displacement vector $w\in\mathds{R}^{2n}$ and a real, $2n\times 2n$ \textbf{quantum covariance matrix} (QCM) $V$, defined respectively as 
\begin{align}
w &\coloneqq \Tr [\rho\, r]\, , \label{displacement vector} \\ 
V &\coloneqq \Tr \big[\rho\, \{r-w,r^T\!-w^T\} \big]\, , \label{covariance matrix}
\end{align}
where the anticommutator $\{H,K\}\coloneqq HK+KH$ is needed in the quantum case in order to make the above expression real, and $w\coloneqq w\cdot \text{id}$ as operators on the Hilbert space.\footnote{It is customary not to divide by $2$ when defining the covariance matrix in the quantum case, the reason being that in this way the forthcoming~\eqref{Heisenberg} looks simpler.}
As it turns out, quantum covariance matrices that come out of \eqref{covariance matrix} are exactly those real positive definite matrices that moreover satisfies the Heisenberg uncertainty relation~\cite{simon94}
\begin{equation}
V+i\Omega \geq 0\, . \label{Heisenberg}
\end{equation}
The reason why some lower bound like~\eqref{Heisenberg} should hold is that, unlike in the classical case, in the quantum case not all the covariances can be small at the same time, and in particular not covariances that pertain to conjugate variables.
Note that~\eqref{Heisenberg} can equivalently be written as $V-i\Omega \geq 0$ upon applying transposition (as $V^T=V$, $\Omega^T=-\Omega$).
From now on we will often refer to real matrices satisfying \eqref{Heisenberg} as QCMs. 

The Gaussian state $\rho^G(V,w)$ with QCM $V$ and displacement vector $w$ admits the representation
\begin{equation}
\rho^G(V,w) = \int \frac{d^{2n} u}{(2\pi)^n} \ e^{-\frac{1}{4} u^T V u - iw^T u} D_{\Omega u}\,,
\label{Fourier-Weyl representation}
\end{equation}
which justifies the alternative definition of Gaussian states as the continuous variable states associated with a Gaussian characteristic function. For more on representations related to~\eqref{Fourier-Weyl representation}, see~\cite[IV]{BUCCO}.

\subsubsection{Bipartite systems}

Here we are mostly interested in the entanglement properties of Gaussian states, so we need to consider bipartite systems. The QCM $V_{AB}$ of a Gaussian state $\rho^G_{AB}$ pertaining to a $(m+n)$-mode bipartite system $AB$ can be naturally written in block form according to the splitting between  the subsystems $A$ and $B$:
\begin{equation}
V_{AB} = \begin{pmatrix} V_A & X \\ X^T & V_B \end{pmatrix}\, . 
\label{V explicit}
\end{equation}
According to the same splitting, the matrix $\Omega$ appearing in~\eqref{CCR} takes the form
\begin{equation}
\Omega_{AB} = \begin{pmatrix} \Omega_A & 0 \\ 0 & \Omega_B \end{pmatrix} = \Omega_A\oplus\Omega_B\, ,
\label{Omega bipartite}
\end{equation}
with $\Omega_A =  \omega^{\oplus m}$ and $\Omega_B=\omega^{\oplus n}$.

\subsubsection{Deterministic dynamics}

Clearly, linear transformations $r \rightarrow S r$ that preserve the commutation relations~\eqref{CCR} play a special role within this framework. Any such transformation is described by a \textbf{symplectic} matrix, i.e.~a matrix $S$ with the property that $S\Omega S^T=\Omega$. Symplectic matrices form a non-compact, connected Lie group that is additionally closed under transposition, and is typically denoted by $\mathrm{Sp}(2n,\mathds{R})$~\cite{pramana}. The importance of these operations arises from the fact that for any symplectic $S$ there is a unitary evolution $U_S$ on the Hilbert space -- that we call \textbf{symplectic unitary} -- such that $U_S^\dag r U_S = Sr$. Most importantly, such a unitary is the product of a finite number of factors\footnote{In fact, up to two such factors. I thank Uther P.F. Shackerley-Bennet and Alessio Serafini for bringing this to my attention.} of the form $e^{i\mathcal{H}_Q}$, where $\mathcal{H}_Q$ is a quadratic Hamiltonian, and as such it can be easily implemented in laboratory. Under conjugation by $U_{S}$, Gaussian states transform as
\begin{equation}
U_S^\dag\, \rho^G(V,w)\, U_S = \rho^G\left(SVS^T,\, Sw \right)\, .
\label{Gauss transform U_S}
\end{equation}

Unitary evolutions model the dynamics of closed quantum systems, but most systems one deals with in the laboratory interact with the environment, hence undergo some more general dynamics. A very general scenario is as follows: first, the input state is put in contact with an external system in a reference Gaussian state; second, the joint system evolves with a symplectic unitary; and third, the ancillary system, or part of it, is traced away, i.e. physically discarded. This transformation identifies a so-called \textbf{Gaussian channel}~\cite[\S 5.3]{BUCCO}. When acting on Gaussian states, the action of any such channel $\mathcal{N}$ can be cast as an action on covariance matrix and displacement vector as follows:
\begin{equation}
\mathcal{N}:\ \left\{\begin{array}{lcl}
V & \longmapsto & XVX^{T}+Y\\
w & \longmapsto & Xw+s
\end{array}
\right.  , \label{N cm}
\end{equation}
where $X,Y$ are matrices of appropriate size, and $s$ is a real vector. In order for the channel $\mathcal{N}$ to be completely positive, $X$ and $Y$ should satisfy the matrix inequality $Y+i\Omega\geq iX\Omega X^{T}$.

\subsubsection{Non-deterministic dynamics}

Of course, our actions on a quantum system are not limited to letting it evolve. We can also access it through a measurement, and in our case, not surprisingly, \textbf{Gaussian measurements} are the most relevant. A Gaussian measurement can be though of as a collection $\{E_t\}_{t\in\mathds{R}^{2n}}$ of positive Gaussian operators of the form $E_t \coloneqq \rho^G (\gamma, t)$, where $\gamma$ is a QCM obeying~\eqref{Heisenberg}, referred to as the \textbf{seed} of the measurement, and the notation is that of~\eqref{Fourier-Weyl representation}. As is easy to see, these operators form a valid quantum measurement, as they satisfy the normalisation condition $\int \frac{d^{2n}t}{(2\pi)^n} E_t =\id$.
We can then ask ourselves what happens when one performs said measurement on the second subsystem $B$ of a bipartite system $AB$ that is initially in a Gaussian state $\rho_{AB} = \rho^G(V_{AB}, w_{AB})$. One can show that~\cite[\S 5.4.5]{BUCCO} 
\bb
\tr_B \left[ \id \otimes E_t\, \rho_{AB}\right] = p(t)\, \rho^G\left( \tilde{V}_A, \tilde{w}_A(t) \right)\, ,
\label{G measurement}
\ee
where $p(t)$ is the probability of getting the outcome $t$. Using the notation of~\eqref{V explicit}, one has
\begin{align}
p(t) &= \frac{2^{n_B} e^{-(t-w_B)^T (V_B+\gamma)^{-1}(t-w_B)}}{\sqrt{\det (V_B+\gamma)}}\, , \label{G measurement probability} \\[0.4ex]
\tilde{V}_A &= V_A - X (V_B + \gamma)^{-1} X^T \, , \label{G measurement QCM} \\[0.4ex]
\tilde{w}_A(t) &= w_A + X(t-s_B)\, . \label{G measurement displacement}
\end{align}
Observe that~\eqref{G measurement QCM} can be equivalently written as
\bb
\tilde{V}_A  = (V_{AB} + 0\oplus\gamma) \big/ (V_B +\gamma)
\label{G measurement QCM Schur}
\ee
with the help of the Schur complement formalism. Remarkably, the final QCM $\tilde{V}_A$ does \emph{not} depend on the outcome of the measurement $t$, as long as this is known. On the contrary, forgetting it would correspond to a partial trace over the system $B$, and in this case we would have $\tilde{V}_A = V_A$. 

The most general Gaussian map we want to consider is an arbitrary concatenation of Gaussian channels and Gaussian partial measurements. Maps of this form act as follows on the covariance matrix and displacement vector of an input Gaussian state:
\bb
\Gamma_{1\rightarrow 2}:\ \left\{ \begin{array}{lcl} V_1 & \longmapsto & \gamma_2 - \delta_{12}^T \left( \gamma_1 + \Sigma V_1 \Sigma \right)^{-1} \delta_{12} \\ w_1 & \longmapsto & s_2 + \delta_{12}^T \left( \gamma_1 + \Sigma V_1 \Sigma \right)^{-1} \Sigma (w_1+s_1) \end{array} \right. ,
\label{CP Gauss}
\ee
where $\gamma_{12}=\left( \begin{smallmatrix} \gamma_1 & \delta_{12} \\ \delta_{12}^T & \gamma_2 \end{smallmatrix} \right)$ is a QCM of the joint system $12$, i.e. it satisfies~\eqref{Heisenberg} with $\Omega=\Omega_{12}$, $s_{12}$ is a real vector, and $\Sigma$ is the matrix that changes the sign of all the momenta, i.e.
\bb
\Sigma = \begin{pmatrix} \id & 0 \\ 0 & -\id \end{pmatrix} = \begin{pmatrix} 1 & 0 \\ 0 & -1 \end{pmatrix}^{\oplus n}\, ,
\label{Sigma}
\ee
with the rightmost expression being with respect to a mode-wise decomposition. The QCM $\gamma_{12}$ that appears in~\eqref{CP Gauss} turns out to be that of the Choi state corresponding to the map $\Gamma_{1\rightarrow 2}$. We will then refer to it as the \textbf{Choi covariance matrix} of $\Gamma_{1\rightarrow 2}$. Observe that the first line of~\eqref{CP Gauss} inherits the Schur complement structure from~\eqref{G measurement QCM Schur}, and can be rewritten accordingly as
\bb
\Gamma_{1\rightarrow 2}: V_1 \longmapsto (\gamma_{12} + \Sigma V_1 \Sigma) \big/ ( \gamma_1 + \Sigma V_1 \Sigma)\, .
\label{CP Gauss Schur}
\ee
Although it may not be apparent from the above expressions, the action of any Gaussian channel, specified by~\eqref{N cm}, can also be written as the limit of expressions of the form~\eqref{CP Gauss}, which makes~\eqref{CP Gauss} the most general physically allowed transformation we can think of within the Gaussian realm.
For details and a derivation of these formulae, see~\cite[Eq. (10a) and (10b)]{nogo3} or~\cite[\S 5.5]{BUCCO}.

\subsubsection{Williamson normal form}

It turns out that all Gaussian states can be brought into a remarkably simple normal form via unitary transformations induced by quadratic Hamiltonians. In fact, a theorem by Williamson~\cite{willy,willysim} implies that for all strictly positive matrices $V>0$ there is a symplectic transformation $S$ such that 
\begin{equation}
V = S\begin{pmatrix} \Lambda & 0 \\ 0 & \Lambda \end{pmatrix}S^T , \qquad
\Lambda = \text{diag}\, (\nu_1, \ldots, \nu_n)>0\, .
\label{Williamson}
\end{equation}
The diagonal elements $\nu_i>0$, each taken with multiplicity one, are called symplectic eigenvalues of $V$, and are uniquely determined by $V$ (up to their order, which can be assumed decreasing by convention with no loss of generality). Accordingly, we will refer to $\vec{\nu}=(\nu_1,\nu_2,\ldots,\nu_n)$ as the \textbf{symplectic spectrum} of $V$. Notably, Heisenberg's uncertainty relation~\eqref{Heisenberg} can be conveniently restated as $\Lambda\geq \mathds{1}$, or equivalently $\nu_{j}\geq 1$ for all $j=1,\ldots, n$.

A Gaussian state $\rho^G(V,w)$ can be shown to be pure if and only if all of its symplectic eigenvalues are equal to $1$, which corresponds to the matrix equality $V\Omega V\Omega=-\id$. Correspondingly, a QCM $V$ satisfying $\nu_{j}=1$ for all $j=1,\ldots,n$ will be called a \textbf{pure QCM}. Equivalent conditions for a QCM $V$ to be pure are: (i) $\det V=1$; or (ii) $\rk (V+i\Omega) = n$ (i.e. half the maximum). Note that pure QCMs $V$ are themselves symplectic matrices, $V=S S^T \in \mathrm{Sp}(2n,\mathds{R})$, and are the extremal elements in the convex set of QCMs. Mixed Gaussian states can be thought of as reduced states of a global Gaussian pure state by the addition of a (fictional) auxiliary system. The global pure state is called a \textbf{purification} of the original state. At the level of covariance matrices, this means that all QCMs can be seen as submatrices of symplectic matrices acting on a higher number of modes. For details, see Lemma~\ref{lemma pur}.

Finally, note that for the sake of our problems the displacement vector $w$ is often irrelevant since it can be made to vanish by local unitaries, resulting from the action of the displacement operator~\eqref{displacement} on each individual mode. Since non-local properties such as entanglement are invariant under local unitaries, all the results we are going to present will not depend on the first moments. Therefore, in what follows, we will completely specify any Gaussian state under our investigation as $\rho^G(V)$ in terms of its QCM $V$ alone.

\section{Simplified separability criterion for Gaussian states} \label{sec5 simplified}

One of the main motivations behind the work we present in this chapter was to revisit some standard results in the theory of Gaussian entanglement, possibly simplifying their proofs with the help of the methods of matrix analysis we discussed above. In order to do this, we first need to develop a new necessary and sufficient separability condition for Gaussian states, which is the purpose of this section.

As is apparent from the discussion above, the entanglement properties of a bipartite Gaussian state should admit a convenient translation at the level of QCMs. Recall that, in general, a bipartite quantum state $\rho_{AB}$ is separable if and only if it can be written as a convex mixture of product states as in~\eqref{separable state}.
For a Gaussian state $\rho^G_{AB}$ of a bipartite continuous variable system, we have then the following.

\begin{lemma} \emph{\cite[Proposition 1]{Werner01}.} \label{sep} A Gaussian state $\rho^G_{AB}(V_{AB})$ with $(m+n)$-mode QCM $V_{AB}$ is separable if and only if there exist an $m$-mode QCM $\gamma_A \geq i \Omega_A$ and an $n$-mode QCM $\gamma_B \geq i \Omega_B$  such that
\begin{equation} V_{AB} \geq \gamma_A\oplus\gamma_B\, . \label{sep eq} \end{equation}
\end{lemma}

In view of the above result, a QCM $V_{AB}$ satisfying~\eqref{sep eq} for some marginal QCMs $\gamma_{A}$, $\gamma_{B}$ will itself be called \emph{separable} from now on. The criterion in~\eqref{sep eq} is necessary and sufficient for separability of QCMs, and can be evaluated numerically via convex optimisation~\cite{Giedke01,eisemi}. However, such optimisation runs over both marginal QCMs, hence can be sometimes difficult to handle theoretically.

The first main result of this paper is to show that the necessary and sufficient separability condition~\eqref{sep eq}, for any $m$ and  $n$, can be further simplified. This result is quite neat and of importance in its own right. In particular, it allows us to recast the Gaussian separability problem as a convex optimisation over the marginal QCM of \emph{one} subsystem only (say $A$ without loss of generality), presumably resulting also in an appreciable reduction of computational resources, especially in case party $A$ comprises a much smaller number of modes than party $B$.\footnote{We do not engage in a thorough comparison of the computational resources needed to decide~\eqref{sep eq} and~\eqref{simp sep 1}, since our interest in~\eqref{simp sep 1} is mainly theoretical, as we shall see.}

\begin{thm}[Simplified separability condition for an arbitrary QCM] \label{simp sep lemma} $ \\ $
A  QCM $V_{AB}$ of $m+n$ modes is separable if and only if there exists an $m$-mode QCM $\gamma_A \geq i \Omega_A$ such that 
\begin{equation}
V_{AB} \geq \gamma_A\oplus i\Omega_B\, .
\label{simp sep 1}
\end{equation}
In terms of the block form~\eqref{V explicit} of $V_{AB}$, when $V_{B}>i\Omega_{B}$ the above condition is equivalent to the existence of a real matrix $\gamma_A$ satisfying
\begin{equation}
i\Omega_A \leq \gamma_A \leq V_A - X (V_B-i\Omega_B)^{-1} X^T\, .
\label{simp sep 2}
\end{equation}
If $V_{B}-i\Omega_{B}$ is not invertible, we require instead $i\Omega_A \leq \gamma_A \leq V_A - X (V_B+\epsilon\mathds{1}_{B}-i\Omega_B)^{-1} X^T$ for all $\epsilon>0$.
\end{thm}

\begin{proof}
Since both sets of QCMs $V_{AB}$ defined by~\eqref{sep eq} and~\eqref{simp sep 1} are clearly topologically closed, we can just show without loss of generality that their interiors coincide. This latter condition can be rephrased as an equivalence between the two following statements: (a) $V_{AB}>\gamma_{A}\oplus \gamma_{B}$ for some QCMs $\gamma_{A},\gamma_{B}$; and (b) $V_{AB} > \gamma_A\oplus i\Omega_B$ for some QCM $\gamma_{A}$.

Now, once $\gamma_A<V_{A}$ is fixed, the supremum of all the matrices $\gamma_B$ satisfying $V_{AB}>\gamma_{A}\oplus \gamma_{B}$ is given by the Schur complement $(V_{AB}-(\gamma_A\oplus 0_B))/(V_A-\gamma_A)$, as the variational characterisation~\eqref{variational} reveals. Therefore, statement (i) is equivalent to the existence of $i\Omega_{A}\leq\gamma_A<V_{A}$ such that $(V_{AB}-(\gamma_A\oplus 0_B))/(V_A-\gamma_A)> i\Omega_B$. This is the same as to require $V_{AB}> \gamma_A\oplus i\Omega_B$, as the positivity conditions of Lemma~\ref{pos cond} immediately show.

Until now, we have proved that the separability of $V_{AB}$ can be restated as $V_{AB} \geq \gamma_A\oplus i\Omega_B$ for some appropriate QCM $\gamma_{A}$. Employing Lemma~\ref{pos cond}, we see that this is turn equivalent to~\eqref{simp sep 2}, or to its $\epsilon$-modified version when $V_{B}-i\Omega_{B}$ is not invertible.
\end{proof}

It is worth noticing that both Lemma~\ref{sep} and Theorem~\ref{simp sep lemma} extend straightforwardly to encompass the case of full separability of multipartite Gaussian states. In the case of Lemma~\ref{sep}, this extension was already formulated in~\cite{Werner01,3-mode-sep}. As for Theorem~\ref{simp sep lemma}, the corresponding necessary and sufficient condition for the full separability of a $k$-partite QCM $V_{A_{1}\cdots A_{k}}$ would read  $V_{A_{1}\cdots A_{k}}\geq \gamma_{A_{1}}\oplus\ldots\oplus \gamma_{A_{k-1}}\oplus i\Omega_{A_{k}}$ for appropriate QCMs $\gamma_{A_1},\ldots, \gamma_{A_{k-1}}$.

\begin{rem}
It has been recently observed~\cite{Bhat16} that condition~\eqref{simp sep 2} is equivalent to the corresponding Gaussian state $\rho^G_{AB}(V_{AB})$ with QCM $V_{AB}$ being completely extendible with Gaussian extensions. We remind the reader that a bipartite state $\rho_{AB}$ is said to be `completely extendible' (Subsection~\ref{subsec2 Woronowicz}) if for all $k$ there exists a state $\rho_{AB_{1}\cdots B_{k}}$ that is: (i) symmetric under exchange of any two $B_{i}$ systems; and (ii) an extension of $\rho_{AB}$ in the sense that $\text{Tr}_{B_{2}\cdots B_{k}}\rho_{AB_{1}\cdots B_{k}}=\rho_{AB}$. When the original state $\rho_{AB}^G$ is Gaussian, it is natural to consider extensions $\rho^G_{AB_{1}\cdots B_{k}}$ of Gaussian form as well. Interestingly enough, the above Theorem~\ref{simp sep lemma} provides a simple alternative proof of the remarkable fact (also proved in~\cite{Bhat16}) that Gaussian states are separable if and only if completely extendible with Gaussian extensions.
\end{rem}

\section{Sufficiency of the PPT condition -- Revisited} \label{sec5 revisited}

In this section we revisit some of the classic results in Gaussian entanglement theory with the help of the mathematical tools we have developed so far, in particular Theorem~\ref{simp sep lemma}. Namely, in Subsection~\ref{subsec5 ppt} we present an alternative (and simpler) proof of the equivalence between separability and PPT-ness for $1$ vs $n$-mode Gaussian states, while in Subsection~\ref{subsec5 mode-wise} we tackle the same problem for isotropic Gaussian states.

\subsection{PPT implies separability for $1$ vs $n$-mode Gaussian states -- Revisited} \label{subsec5 ppt}

Most of the present chapter is devoted to the investigation of known and new conditions under which separability becomes equivalent to PPT for Gaussian states, so that the problem of deciding whether a given QCM is separable or not admits a handy formulation.
For any bipartite state $\rho_{AB}$, recall that the PPT criterion (Subsection~\ref{subsec4 maps}) provides a useful  necessary condition for separability~\cite{PeresPPT}: if a bipartite state $\rho_{AB}$ is separable, then $\rho_{AB}^{T_B}\geq 0$, where the suffix $T_B$ denotes transposition with respect to the degrees of freedom of subsystem $B$ only. As already mentioned, in finite-dimensional systems, PPT is also a sufficient condition for separability in $2\times 2$ and $2\times 3$ quantum systems~\cite{HorodeckiPPT}.
In continuous variable systems, the PPT criterion turns out to be also sufficient for separability of QCMs when either $A$ or $B$ is composed of one mode only~\cite{Simon00,Werner01}.

\begin{thm}[PPT is sufficient for Gaussian states of $1$ vs $n$ modes] \label{PPT thm} $ \\ $
Let $V_{AB}$ be a bipartite QCM such that either $A$ or $B$ are composed of one mode only. Then $V_{AB}$ is separable if and only if
\begin{equation}
V_{AB}\, \geq\, \begin{pmatrix} i\Omega_A & 0 \\ 0 & \pm i\Omega_B\end{pmatrix} = i\Omega_A \oplus (\pm i\Omega_B)\, ,
\label{PPT}
\end{equation}
which amounts to the corresponding Gaussian state being PPT, $\left(\rho^{G}_{AB}\right)^{T_B} \geq 0$.
\end{thm}

For completeness, we recall that the partial transpose of an $(m+n)$-mode QCM $V_{AB}$, i.e. the covariance matrix of the partially transposed density operator $\left( \rho^{G}_{AB}\right)^{T_B}$, is given by 
\bb
V_{AB}^{T_B} = \Theta V_{AB} \Theta\, ,
\label{partial transpose QCM}
\ee
where 
\bb
\Theta \coloneqq \id_A \oplus \Sigma_B
\label{Theta}
\ee
with the notation of~\eqref{Sigma}~\cite{Simon00}. Accordingly, we can say that the QCM $V_{AB}$ is PPT if and only if $V_{AB}^{T_B}$ is a valid QCM obeying~\eqref{Heisenberg}, which is equivalent to~\eqref{PPT} since $\Theta \Omega_{AB}\Theta = i\Omega_A \oplus (-i\Omega_B)$.

The original proof of Theorem~\ref{PPT thm} came in two steps. First, Simon~\cite{Simon00} proved it in the particular case when both $A$ and $B$ are made of one mode only by performing an explicit analysis of the symplectic invariants of $V_{AB}$; this seminal analysis is quite straightforward to follow and particularly instructive, but eventually a bit cumbersome, since it requires to distinguish between three cases, according to the sign of $\det X$, where $X$ is the off-diagonal block of the QCM $V_{AB}$ partitioned as in~\eqref{V explicit}. Later on, Werner and Wolf~\cite{Werner01} reduced the problem for the $1$ vs $n$-mode case with arbitrary $n$ to the $1$ vs $1$-mode case; the proof of this reduction is geometric in nature and rather elegant, but also relatively difficult.

Our purpose here is to use Schur complements to provide the reader with a simple,
direct proof of Theorem~\ref{PPT thm}. Before coming to that, there is a preliminary lemma we want to discuss.

\begin{lemma} \label{2x2 interval}
Let $M,N$ be $2\times 2$ Hermitian matrices. There is a real symmetric matrix $R$ satisfying $M\leq R\leq N$ if and only if $M\leq N, N^*$, where $*$ denotes complex conjugation.
\end{lemma}

\begin{proof}
The necessity of the condition $M\leq N, N^*$ is easily verified, so we just prove sufficiency.
The only complex entry in a $2\times 2$ Hermitian matrix is in the off-diagonal element. Suppose without loss of generality that $\Im M_{12}\ge0$ and $\Im N_{12}\le0$ (both conditions in the statement are in fact symmetric under complex conjugation of $M$ or $N$).
It is easy to verify that a $p$ such that $0 \leq p \leq 1$ and $\Im (pM+(1-p)N)_{12}=0$ always exists, and
we see that $R\coloneqq pM+(1-p)N$ is a real symmetric matrix. Moreover, since $R$ belongs to the segment joining $M$ and $N\geq M$ we conclude that $M\leq R\leq N$.
\end{proof}

\begin{rem}
Lemma~\ref{2x2 interval} admits an appealing physical interpretation which also leads to an intuitive proof. This interpretation is based on the fact that $2\times 2$ Hermitian matrices can be seen as events in $4$-dimensional Minkowski space-time through the correspondence $x_0 \mathds{1} + \vec{x}\cdot\vec{\sigma} \leftrightarrow (x_0, \vec{x})$. Furthermore, $M\leq N$ translates in Minkowski space-time to `$N$ is in the absolute future of $M$', since the remarkable determinantal identity $\det (x_0 \mathds{1} + \vec{x}\cdot\vec{\sigma}) = x_0^2 - \vec{x}^2$ holds true. Now, the complex conjugation at the matrix level becomes nothing but a spatial reflection with respect to a fixed spatial plane in Minkowski space-time. Thus, our original question is: is it true that whenever both an event $N$ and its spatial reflection $N^*$ are in the absolute future of a reference event $M$ then there is another event $R$ which is: (i) in the absolute future of $M$; (ii) in the absolute past of both $N$ and $N^*$; and (iii) lies right on the reflection plane? The answer is clearly yes, and there is a simple way to obtain it. Start from $M$ and shoot a photon to the location of that event between $N$ and $N^*$ that will happen on the other side of the reflection plane. After some time the photon hits the plane, and this event $R$ clearly satisfies all requirements.
\end{rem}

Now we are ready to give our direct proof of the equivalence between PPT and separability for $1$ vs $n$-mode Gaussian states, leveraging the simplified separability condition of Theorem~\ref{simp sep lemma}.

\begin{proof}[Proof of Theorem~\ref{PPT thm}]
Suppose without loss of generality that $A$ is composed of one mode only. As in the proof of Theorem~\ref{simp sep lemma}, since both sets of QCMs $V_{AB}$ defined by~\eqref{sep eq} and~\eqref{PPT} are topologically closed, we can assume that $V_{AB}$ is in the interior of the PPT set, i.e.~that $V_{AB}>i\Omega_{A}\oplus (\pm i\Omega_{B})$. Our goal will be to show that in this case $V_{AB}$ belongs to the separable set, as characterized by Theorem~\ref{simp sep lemma}. Since $V_{B}-i\Omega_{B}$ is taken to be invertible, the PPT condition reads
\begin{equation*}
V_A - X (V_B \mp i\Omega_B)^{-1} X^T\, \geq\, i\Omega_{A}\, .
\end{equation*}
Now, define $M=i\Omega_{A}$ and $N=V_A - X (V_B + i\Omega_B)^{-1} X^T$, and observe that $N^{*}=V_A - X (V_B - i\Omega_B)^{-1} X^T$. Thanks to Lemma~\ref{2x2 interval}, we can find a real matrix $\gamma_{A}$ such that
\begin{equation*}
V_A - X (V_B \mp i\Omega_B)^{-1} X^T\, \geq\, \gamma_{A}\, \geq\, i\Omega_{A}\, .
\end{equation*}
Choosing the negative sign in the above inequality, we see that the second condition~\eqref{simp sep 2} in Theorem~\ref{simp sep lemma} is met, and therefore $V_{AB}$ is separable.
\end{proof}

\subsection{PPT implies separability for multimode isotropic Gaussian states -- Revisited} \label{subsec5 mode-wise}

It is well known that the PPT criterion is in general sufficient, as well as obviously necessary, for pure bipartite states to be separable~\cite{PeresPPT}. This may be seen by a direct inspection of the Schmidt decomposition~\eqref{Schmidt decomposition} of a pure state.
Let us note, incidentally, that a stronger statement holds, namely that any bound entangled state (in any dimension) must have at least rank $4$~\cite{chen08}.

The Schmidt decomposition theorem is in fact so important that a Gaussian version of it, that is, the determination of a normal form of pure QCMs under local symplectic operations, is of central importance in continuous variable quantum information.
As can be shown at the covariance matrix level~\cite{holwer,giedkemode} or at the density operator level~\cite{botero03}, every pure bipartite Gaussian state $\rho^G_{AB}(V_{AB})$ can be brought into a tensor product of two-mode squeezed vacuum states and single-mode vacuum states by means of local unitaries with respect to the $A$ vs $B$ partition. In particular, by acting correspondingly with local symplectic transformations, any pure QCM $V_{AB}$ (where pure means $\det V_{AB} = 1$) can be transformed into a direct sum of (pure) two-mode squeezed vacuum QCMs and (pure) single-mode vacuum QCMs.
More precisely, at the level of QCMs, one can formulate this fundamental result as follows.

\begin{thm}[Mode-wise decomposition of pure Gaussian states~\cite{holwer,botero03,giedkemode}] \label{mode-wise thm}
Let $V_{AB}$ be a bipartite QCM of $m+n$ modes $A_1,\ldots,A_m,B_1,\ldots,B_n$, assuming $m \leq n$ (with no loss of generality). If $V_{AB}$ is a pure QCM, i.e.~all its symplectic eigenvalues are equal to $1$ (which amounts to $\det V_{AB}=1$), then there exist local symplectic transformations $S_A \in \mathrm{Sp}(2m,\mathds{R})$, $S_B \in \mathrm{Sp}(2n,\mathds{R})$ mapping $V_{AB}$ into the following normal form:
\begin{equation}\label{modewise}
(S_A \oplus S_B) V_{AB} (S_A^T \oplus S_B^T) = \bigoplus_{j=1}^m \widebar{V}_{A_jB_j}(r_j)\, \oplus \bigoplus_{k=m+1}^n \mathds{1}_{B_k}\,,
\end{equation}
where $\widebar{V}_{A_jB_j} = \begin{pmatrix} c_j \id & s_j {\zeta} \\  s_j {\zeta}  & c_j \id \end{pmatrix}$
with $c_{j}=\cosh(2r_j)$ and $s_j=\sinh(2r_j)$, for a real squeezing parameter $r_j$, is the pure QCM of a two-mode squeezed vacuum state of modes $A_j$ and $B_j$, and $\mathds{1}_{B_k}$ is the pure QCM of the single-mode vacuum state of mode $B_k$.
In particular, with respect to the block form~\eqref{V explicit}, for any pure QCM $V_{AB}$ the marginal QCMs $V_A$ and $V_B$ have matching symplectic spectra, given by
$\vec{\nu}_A=(c_1,\ldots,c_m)$ and $\vec{\nu}_B=(c_1,\ldots,c_m,\underbrace{1,\ldots,1}_{n-m})$.
\end{thm}

Leaving apart its far-reaching applications, in the context of the present paper this result is mainly instrumental for assessing the separability of so-called \emph{isotropic} multimode Gaussian states. The QCM of any such state of $m+n$ modes is characterised by the property of having a completely degenerate symplectic spectrum, i.e.~formed of only one distinct symplectic eigenvalue $\nu \geq 1$ (repeated $m+n$ times). This means that the QCM $V_{AB}$ of any isotropic state is proportional by a factor $\nu$ to a pure QCM. Hence, Theorem~\ref{mode-wise thm} tells us that $V_{AB}$ can be brought into a direct sum of two-mode QCMs via a local symplectic congruence (local with respect to any partition into groups of modes $A$ and $B$), as first observed in~\cite{holwer}. Thanks to Theorem~\ref{PPT thm}, this guarantees the following.

\begin{thm} \label{iso thm}
The PPT criterion is necessary and sufficient for separability of all isotropic Gaussian states of an arbitrary number of modes.
\end{thm}

However, notwithstanding the importance of Theorem~\ref{mode-wise thm} per se, one could strive to seek a more direct way to obtain Theorem~\ref{iso thm}. Our purpose in this subsection is in fact to provide an alternative proof of this result, which does not appeal to the mode-wise decomposition theorem at all, and uses directly Lemma~\ref{sep} instead, leveraging matrix analysis tools such as the notions of matrix means introduced in Section~\ref{subsec5 matrix means}.


Let us start with a preliminary result, equivalent to~\cite[Proposition 12]{manuceau} or to~\cite[Lemma 13]{Lami16}. We include a proof for the sake of completeness.

\begin{lemma} \label{QCM geom lemma}
Let $V>0$ be a positive matrix. Then
\bb
\gamma^{\#}_V \coloneqq V {\#} (\Omega V^{-1} \Omega^T)
\label{gamma sharp}
\ee
is a pure QCM. Furthermore, the following are equivalent:
\begin{enumerate}[(a)]
\item the matrix $V$ satisfies Heisenberg's uncertainty principle~\eqref{Heisenberg};
\item $V \geq \Omega V^{-1} \Omega^T$; and 
\item $V \geq \gamma_V^\#$.
\end{enumerate}
Moreover, $V$ is a pure QCM iff condition (b) (equivalently, (c)) is saturated with equality, and $V>i\Omega$ iff condition (b) (equivalently, (c)) is a strict inequality.
\end{lemma}

\begin{proof}
Let the Williamson form of $V$ be given by~\eqref{Williamson}. Then we can write
\begin{align*}
\Omega V^{-1} \Omega^T &= \Omega S^{-  T} (\Lambda^{-1} \oplus \Lambda^{-1}) S^{-1}\Omega^T \\
&\texteq{(1)} S \Omega (\Lambda^{-1} \oplus \Lambda^{-1})\Omega^T S^{T} \\
&\texteq{(2)} S (\Lambda^{-1} \oplus \Lambda^{-1}) \Omega \Omega^T S^{T} \\
&\texteq{(3)} S (\Lambda^{-1} \oplus \Lambda^{-1}) S^{T}\, ,
\end{align*}
where we used in order: (1) the identities $\Omega S^{-  T} = S\Omega$, $S^{-1}\Omega^T = \Omega^T S^{T}$, all consequences of the defining symplectic identity $S\Omega S^{T}=\Omega$; (2) the fact that $\Omega$ commutes with $\Lambda^{-1} \oplus \Lambda^{-1}$; and (3) the orthogonality relation $\Omega\Omega^T=\mathds{1}$. Now the first claim becomes obvious, since $\Lambda=\Lambda^{-1}=\mathds{1}$ if and only if $V$ is a pure QCM. In general, as it can be seen from the above expression, $V$ and $\Omega V^{-1} \Omega^T$ are brought in Williamson form by simultaneous congruences with the same symplectic matrix $S$.

Observe that Heisenberg's uncertainty principle~\eqref{Heisenberg}, which can be rephrased as $\Lambda\geq \id$, translates directly to $\Lambda\geq \Lambda^{-1}$ since $\Lambda>0$. Via the above calculation, this leads to $V=S(\Lambda \oplus \Lambda) S^T\geq S(\Lambda^{-1} \oplus \Lambda^{-1}) S^T= \Omega V^{-1} \Omega^T$. We have then shown the equivalence between claims (a) and (b).

As for (c), the covariance of the geometric mean under congruence ensures that
\begin{align*}
V\# (\Omega V^{-1}\Omega^{T}) &= \left( S \begin{pmatrix} \Lambda & 0 \\ 0 & \Lambda \end{pmatrix} S^{T} \right) \# \left( S \begin{pmatrix} \Lambda^{-1} & 0 \\ 0 & \Lambda^{-1} \end{pmatrix} S^{T} \right) \\[0.8ex]
&= S \left( \begin{pmatrix} \Lambda & 0 \\ 0 & \Lambda \end{pmatrix} \# \begin{pmatrix} \Lambda^{-1} & 0 \\ 0 & \Lambda^{-1} \end{pmatrix}\right) S^{T} \\[0.4ex]
&= SS^{T}\, ,
\end{align*}
where the last passage is an easy consequence of the fact that $A\# A^{-1}=\mathds{1}$ for all $A>0$. Again, upon congruence by $S$ Heisenberg's uncertainty relation $\Lambda\geq \id$ becomes $V=S (\Lambda \oplus \Lambda) S^{T}\geq SS^{T}=V\# (\Omega V^{-1}\Omega^{T})$, which proves the equivalence between (a) and (c). Finally, the last claims follow easily from the above reasoning.
\end{proof}

Now we are ready to explain our direct argument to show separability of PPT isotropic Gaussian states, alternative to the use of the mode-wise decomposition.

\begin{proof}[Proof of Theorem~\ref{iso thm}]
Thanks to~\eqref{partial transpose QCM} and~\eqref{Theta}, the PPT condition~\eqref{PPT} for a QCM $V_{AB}$ takes the form $\Theta V \Theta\geq i \Omega$, where $\Theta$ is defined in~\eqref{Theta}. Thanks to Lemma~\ref{QCM geom lemma}, this becomes in turn
\begin{equation*}
\Theta V \Theta \geq (\Theta V \Theta) \# (\Omega \Theta V^{-1} \Theta \Omega^{T})
\end{equation*}
and finally
\begin{equation*}
V \geq V \# (\Theta \Omega \Theta\, V^{-1}\, \Theta \Omega^{T}\Theta) = (g V) \# (Z \Omega\, (g V)^{-1}\, \Omega^{T} Z)
\end{equation*}
after conjugating by $\Theta$, applying once more the covariance of the geometric mean under congruences, introducing a real parameter $g>0$ (to be fixed later), and defining $Z\coloneqq \mathds{1}_{A}\oplus (-\mathds{1}_{B})=\Theta \Omega \Theta \Omega^T$. Now, we apply Lemma~\ref{lemma ha=g} to the above expression, obtaining
\begin{equation*}
V \geq \left(\frac{g V+Z\Omega\, (gV)^{-1}\, \Omega^{T} Z}{2}\right) \# \left( (gV)\, ! \left( Z\Omega\, (gV)^{-1}\, \Omega^{T} Z \right) \right)
\end{equation*}
Although it is not yet transparent, we are done, as the right-hand side of the above inequality is exactly of the form $\gamma_{A}\oplus\gamma_{B}$ when $V$ is the QCM of an isotropic Gaussian state. In fact, let $g>0$ be such that $gV$ is a pure QCM, satisfying $gV=\Omega\, (gV)^{-1}\, \Omega^{T}=\left( \begin{smallmatrix} P & Y \\ Y^{T} & Q \end{smallmatrix} \right)$, where the block decomposition is with respect to the $A$ vs $B$ splitting. Then on the one hand since $Z=\mathds{1}_{A}\oplus (-\mathds{1}_{B})$ we find
\begin{equation*}
\frac{g V+Z\Omega\, (gV)^{-1}\, \Omega^{T} Z}{2} = \begin{pmatrix} P & 0 \\ 0 & Q \end{pmatrix} ,
\end{equation*}
while on the other hand
\begin{align*}
(gV)\, ! \left( Z\Omega\, (gV)^{-1}\, \Omega^{T} Z \right) &= 2 \left( (gV)^{-1} + Z\Omega\, (gV)\, \Omega^{T} Z \right)^{-1} \\[0.4ex]
&= 2 \Omega \left( \Omega (gV)^{-1} \Omega^{T} + Z (gV) Z \right)^{-1} \Omega^{T} \\[0.4ex]
&= 2 \Omega \left( gV + Z (gV) Z \right)^{-1} \Omega^{T} \\[0.6ex]
&= \begin{pmatrix} \Omega P^{-1}\Omega^{T} & 0 \\ 0 & \Omega Q^{-1} \Omega^{T}\end{pmatrix} ,
\end{align*}
where we used the definition~\eqref{harmonic} of harmonic mean and the easily verified fact that $[Z,\Omega]=0$. Putting all together, we find
\begin{align*}
V &\geq \begin{pmatrix} P & 0 \\ 0 & Q \end{pmatrix}\#\begin{pmatrix} \Omega P^{-1}\Omega^{T} & 0 \\ 0 & \Omega Q^{-1} \Omega^{T}\end{pmatrix} \\[1ex]
&= \begin{pmatrix} P \# \Omega P^{-1} \Omega^{T} & 0 \\ 0 & Q \# \Omega Q^{-1} \Omega^{T} \end{pmatrix} \\[0.4ex]
&= \big(\gamma_P^\#\big)_{A}\oplus \big(\gamma_Q^\#\big)_{B}\, ,
\end{align*}
where the notation is as in~\eqref{gamma sharp}. Since Lemma~\ref{QCM geom lemma} ensures that $\gamma_P^ \# = P\# (\Omega P^{-1} \Omega^{T})$ is a QCM since $P>0$ (and analogously for $Q$), a direct invocation of Lemma~\ref{sep} allows us to conclude the proof.
\end{proof}

\section{Novel results on Gaussian entanglement} \label{sec5 novel}

While in the previous Section~\ref{sec5 revisited} we reviewed classic results and provided new proof of them, our purpose in this section is to apply some of our methods to gain some new insights into the nature of Gaussian entanglement. Namely, in Subsection~\ref{subsec5 inva} we focus on Gaussian states that are not only PPT but even invariant under partial transposition, and show that they are necessarily separable, which mimics an analogous result found in the finite-dimensional setting in~\cite{sep-2xN}. Next, Subsection~\ref{subsec5 mode-wise} is devoted to extending Theorem~\ref{PPT thm} to the more general case of a bipartite Gaussian state of $m$ vs $n$ modes which is invariant under the exchange of any two modes of one of the two subsystems. Finally, in Subsection~\ref{subsec5 pass} we close an open problem of~\cite{passive}, by showing that bipartite Gaussian states that remain PPT when an arbitrary passive operation is applied to them are necessarily separable.

\subsection{Gaussian states that are invariant under partial transpose are separable} \label{subsec5 inva}

As an example of straightforward application of Theorem~\ref{simp sep lemma}, we study here the separability of a special class of PPT Gaussian states, i.e.~those that are \emph{invariant} under partial transposition of one of the subsystems. This problem has an analogue in finite-dimensional quantum information, already studied in~\cite{sep-2xN}, where it was shown that bipartite states on $\mathds{C}^{2}\otimes \mathds{C}^{N}$ that are invariant under partial transpose on the first system are necessarily separable.\footnote{The proof reported in~\cite{sep-2xN} is rather long, so here we provide a shorter one, again based on Schur complements. A state on $\mathds{C}^{2}\otimes \mathds{C}^{N}$ that is invariant under partial transposition on the first subsystem can be represented in block form as $\rho=\left( \begin{smallmatrix} A & X \\ X & B \end{smallmatrix}\right)$. By a continuity argument, we can suppose without loss of generality that $A>0$. Rewrite $\rho = \left( \begin{smallmatrix} A & X \\ X & XA^{-1}X \end{smallmatrix}\right) + \ket{1}\!\!\bra{1}\otimes (B-XA^{-1}X)$. Both terms are positive by Lemma~\ref{pos cond}. Since the second one is separable, let us deal only with the first one, call it $\tilde{\rho}$. We have $\tilde{\rho} = \id_2 \otimes A^{1/2} \left( \begin{smallmatrix} \id & Y \\ Y & Y^{2} \end{smallmatrix}\right) \id_2 \otimes A^{1/2}$, where $Y\coloneqq A^{-1/2}X A^{-1/2}$ is Hermitian. Denoting by $Y = \sum_{i} y_{i} \ket{e_{i}}\!\!\bra{e_{i}}$ its spectral decomposition, we obtain the following manifestly separable representation of $\tilde{\rho}$:
\begin{equation*}
\tilde{\rho} = \id_2 \otimes A^{1/2} \bigg( \sum_{i} \left( \begin{smallmatrix} 1 & y_{i} \\ y_{i} & y_{i}^{2} \end{smallmatrix} \right) \otimes \ket{e_i}\!\!\bra{e_i}\bigg) \id_2 \otimes A^{1/2}\, . \end{equation*}}
Here we show that for Gaussian states an even stronger statement holds, in that invariance under partial transposition implies separability for any number of local modes.

\begin{cor} \label{inva cor}
A bipartite Gaussian state $\rho_{AB}^{G}$ that is invariant under partial transposition of one of the two subsystems is necessarily separable.
\end{cor}

\begin{proof}
Without loss of generality, we can assume that the partial transpose on the $B$ system leaves the state invariant. We now show that under the this assumption the separability condition~\eqref{simp sep 2} is immediately satisfied, since the rightmost side is already a real, symmetric matrix. In fact, equating the original QCM~\eqref{V explicit} with the one obtained after partial transpose on the $B$ system, according to~\eqref{partial transpose QCM} and~\eqref{Theta}, we get the identities $X=X \Sigma$ and $V_{B}=\Sigma V_{B}\Sigma$, with $\Sigma$ defined by~\eqref{Sigma}. As a consequence,
\begin{align*}
X(V_{B}-i\Omega_{B})^{-1} X^{T} &= X \Sigma (V_{B} - i\Omega_{B})^{-1} \Sigma X^{T} \\[0.4ex]
&= X \left(\Sigma (V_{B} - i\Omega_{B}) \Sigma \right)^{-1} X^{T} \\[0.4ex]
&= X \left(V_{B} + i\Omega_{B} \right)^{-1} X^{T}\, ,
\end{align*}
where we used also $\Sigma \Omega_B \Sigma = -\Omega_B$. This shows that $X(V_{B}-i\Omega_{B})^{-1} X^{T}$ is equal to its complex conjugate, and is therefore (despite appearances) a real symmetric matrix. Hence the separability condition~\eqref{simp sep 2} is satisfied with $\gamma_{A} = V_A - X(V_{B}-i\Omega_{B})^{-1} X^{T}$.
\end{proof}

\subsection{PPT implies separability for multimode mono-symmetric Gaussian states} \label{subsec5 symm}

\begin{figure}[tb]
\centering \includegraphics[width=12cm]{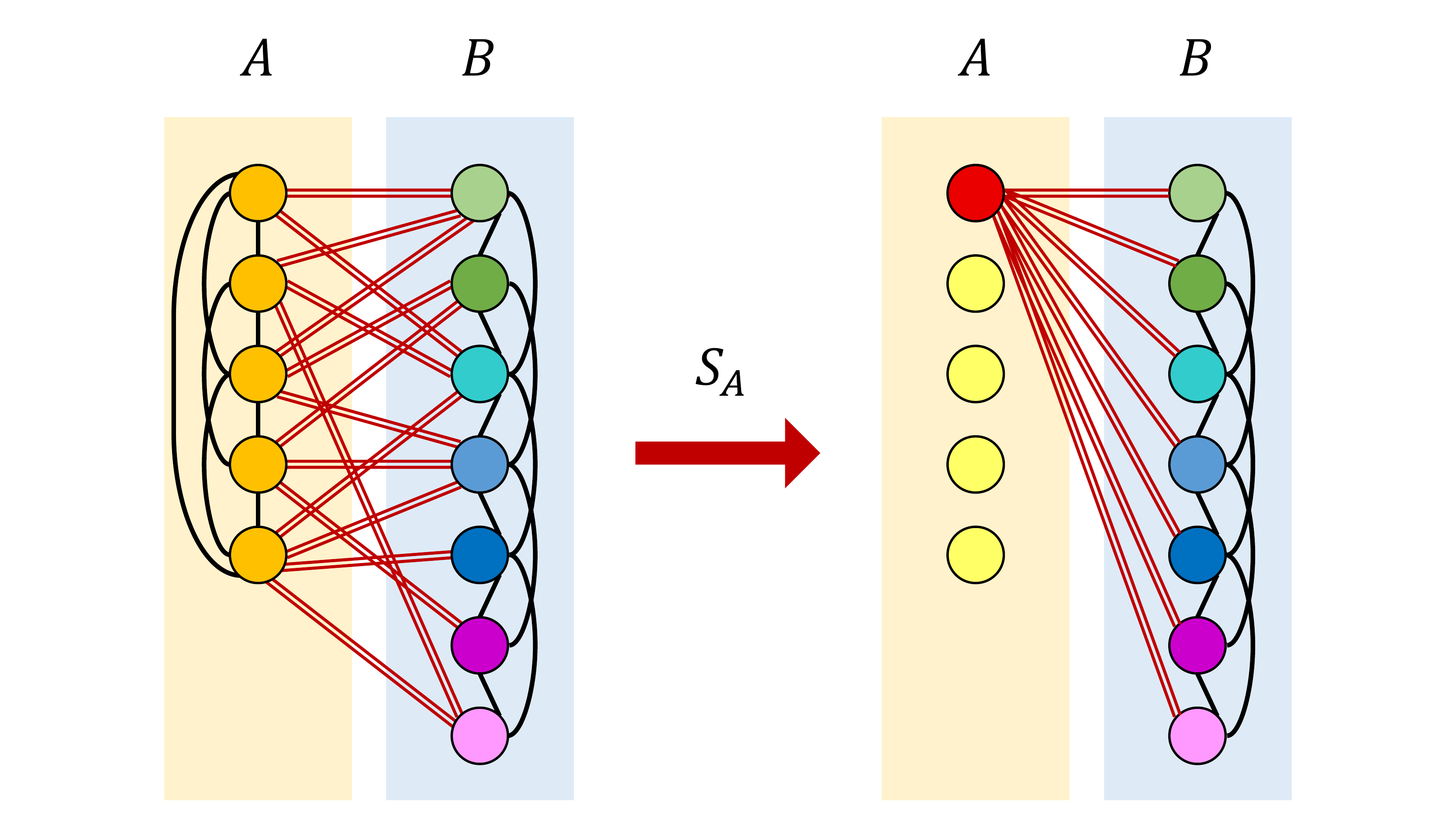}
\caption{Mono-symmetric Gaussian states of two parties $A$ (with $m$ modes) and $B$ (with $n$ modes) are invariant under exchange of any two modes within party $A$. By means of a suitable symplectic transformation on subsystem $A$, these states can be reduced to a $1$ vs $n$-mode Gaussian state and a collection of $m-1$ uncorrelated single-mode states on $A$'s side. Since PPT is equivalent to separability for $1$ vs $n$-mode Gaussian states, it follows that PPT is necessary and sufficient for separability of all $m$ vs $n$-mode mono-symmetric Gaussian states. In the schematics, entanglement between pairs of modes from the same party is depicted as a single solid (black) line, while entanglement across a mode from $A$ and a mode from $B$ is depicted as a double (dark red) line.
\label{mononucleosi}}
\end{figure}

Throughout this Section, we show how the PPT criterion is also necessary and sufficient for deciding the separability of bipartite Gaussian states of $m$ vs $n$ modes that are symmetric under the exchange of any two among the first $m$ modes. These states will be referred to as \textbf{mono-symmetric} (with respect to the first party $A$).  As can be easily seen, this novel result (see Figure~\ref{mononucleosi} for a graphical visualisation) is a generalisation of both Theorem~\ref{PPT thm} and of one of the main results in~\cite{Serafini05}, where the subclass of bi-symmetric states was considered instead, bi-symmetric meaning that they are invariant under swapping any two modes either within the first $m$ or within the last $n$ (that is, they are mono-symmetric in both $A$ and $B$).

\begin{thm} \label{PPt sym}
Let $\rho^G_{AB}(V_{AB})$ be a mono-symmetric Gaussian state of $m+n$ modes, i.e.~specified by a QCM $V_{AB}$ that is symmetric under the exchange of any two of the $m$ modes of subsystem $A$. Then there exists a local unitary operation on $A$ corresponding to a symplectic transformation $S_A \in \mathrm{Sp}(2m, \mathds{R})$ that transforms $\rho^G_{AB}$ into the tensor product of $m-1$ uncorrelated single-mode Gaussian states $\tilde{\rho}^G_{A_j}(\tilde{V}_{A_j})$ ($j=2,\ldots,m$) and a bipartite Gaussian state $\tilde{\rho}^G_{A_1B}(\tilde{V}_{A_1B})$ of $1$ vs $n$ modes. At the QCM level, this reads
\begin{equation}\label{monolocale}
(S_A \oplus \mathds{1}_B) V_{AB} (S_A^T \oplus \mathds{1}_B) = \bigg(\bigoplus_{j=2}^m \tilde{V}_{A_j}\bigg) \oplus \tilde{V}_{A_1B} \, ,
\end{equation}
the above direct sum being with respect to a mode-wise decomposition. The separability properties of $V_{AB}$ and $\tilde{V}_{A_1 B}$ are equivalent, in particular $\rho^G_{AB}(V_{AB})$ is separable if and only if it is PPT.
\end{thm}

\begin{proof}
Throughout this proof, we resort to a mode-wise decomposition, which is more apt to capture the symmetry of the problem. We will prove~\eqref{monolocale} directly at the QCM level, by constructing a suitable local symplectic $S_A$.
By virtue of the symmetry under the exchange of any two modes of subsystem $A$, if we decompose $V_{AB}$ as in~\eqref{V explicit}, the submatrices $V_A$ and $X$ have the following structure (with respect to a mode-wise decomposition):
\begin{equation}\label{strucaz}
V_{A} = \begin{pmatrix} \alpha & \varepsilon & \ldots & \varepsilon \\[-1ex] \varepsilon & \alpha & & \vdots \\[-1ex] \vdots & & \ddots & \varepsilon \\[-0.5ex] \varepsilon & \ldots & \varepsilon & \alpha \end{pmatrix}\, ,\qquad
X = \begin{pmatrix} \kappa_{1} & \kappa_{2} & \ldots & \kappa_{n} \\ \kappa_{1} & \kappa_{2} & \ldots & \kappa_{n} \\ \vdots & & & \vdots \\ \kappa_{1} & \kappa_{2} & \ldots & \kappa_{n} \end{pmatrix} \, ,
\end{equation}
where each one of the blocks $\alpha,\varepsilon,\kappa_{j}$ in~\eqref{strucaz} is a $2\times 2$ real matrix, with $\alpha$ and $\varepsilon$ symmetric~\cite{adescaling}.

We can now decompose the real space of the first $m$ modes as $\mathds{R}^{2m}=\mathds{R}^{m} \otimes \mathds{R}^{2}$. According to this decomposition, we may rewrite $V_A$ and $X$ as follows:
\begin{equation}
V_{A} = \mathds{1}_m\otimes (\alpha-\varepsilon) + m\ket{+}\!\!\bra{+}\otimes \varepsilon\, ,\qquad X = \sqrt{m}\,\sum_{j=1}^{n} \ket{+}\!\!\bra{j}\otimes \kappa_{j}\, ,
\end{equation}
where $\ket{+}=\frac{1}{\sqrt{m}}\sum_{i=1}^{m}\ket{i}$, with $\{\ket{i}\}_{i=1}^m$ denoting the standard basis for $\mathds{R}^m$. Observe that the symplectic form $\Omega_A$ on subsystem $A$ decomposes accordingly as $\Omega_A = \mathds{1}_m\otimes \omega$. If $O$ is an $m\times m$ orthogonal  matrix such that $O\ket{+}=\ket{1}$, we easily see that on the one hand $O\otimes\mathds{1}_2\ \Omega_A\ O^{T}\otimes\mathds{1}_2 = \Omega_A$, i.e.~$O\otimes\mathds{1}_2$ is symplectic, while on the other hand
\begin{align*}
O\otimes\mathds{1}_2\ V_{A}\ O^{T}\otimes\mathds{1}_2 &= \ket{1}\!\!\bra{1} \otimes (\alpha + (m-1)\varepsilon) + \sum_{i=2}^{m} \ket{i}\!\!\bra{i}\otimes (\alpha-\varepsilon) \\[0.4ex]
&= \begin{pmatrix} \alpha+(m-1)\varepsilon & 0 & \ldots & 0 \\[-1ex] 0 & \alpha-\varepsilon & & \vdots \\[-1ex] \vdots & & \ddots & 0 \\[-0.5ex] 0 & \ldots & 0 & \alpha-\varepsilon \end{pmatrix}
\end{align*}
and
\begin{align*}
O\otimes\mathds{1}_2\ X  &= \sqrt{m}\, \sum_{j=1}^{n} \ket{1}\!\!\bra{j}\otimes \kappa_{j} \\[0.4ex]
&= \begin{pmatrix} \sqrt{m}\,\kappa_{1} & \sqrt{m}\, \kappa_{2} & \ldots & \sqrt{m}\,\kappa_{n} \\ 0 & 0 & \ldots & 0 \\ \vdots & & & \vdots \\ 0 & 0 & \ldots & 0 \end{pmatrix} .
\end{align*}
Therefore, the initial QCM $V_{AB}$ has been decomposed as a direct sum of $m-1$ one-mode QCMs $\tilde{V}_{A_j}=\alpha-\varepsilon$, and of one $(1+m)$-mode QCM $\tilde{V}_{A_1B}$, via a local symplectic operation on subsystem $A$, i.e. $S_A = O \otimes \mathds{1}_2$. This proves~\eqref{monolocale} constructively. Applying Theorem~\ref{PPT thm}, one then gets immediately that the PPT condition is necessary and sufficient for separability in this case.
\end{proof}

This original result yields a substantial enlargement to the domain of validity of PPT as a necessary and sufficient criterion for separability of multimode Gaussian states, reaching beyond any existing literature. In practice, Theorem~\ref{PPt sym} tells us that, in any mono-symmetric Gaussian state, all the correlations (including and beyond entanglement) shared among the whole $m$ modes of $A$ and the whole $n$ modes of $B$ can be  localised onto correlations between a single mode $A_1$ of $A$ vs the whole $B$, by means of a local unitary (symplectic at the QCM level) operation at $A$'s side only. Being unitary, this operation is fully reversible, meaning that the correlations with $B$ can be redistributed back and forth between $A_1$ and the whole set of $A$ modes with no information loss. This also means that quantitative results on any measure of such correlations between $A$ and $B$ encoded in $V_{AB}$ can be conveniently evaluated in the much simpler  $1$ vs $n$-mode normal form $\tilde{V}_{A_1B}$ constructed in the proof Theorem~\ref{PPt sym}, ignoring the $m-1$ uncorrelated modes.

In the special case of $V_{AB}$ being the QCM of a bi-symmetric state, i.e.~with full permutation symmetry within both $A$ and $B$, it is immediate to observe that applying a similar construction by means of a local unitary at $B$'s side as well fully reduces $V_{AB}$ to a two-mode QCM $\tilde{V}_{A_1 B_1}$, with equivalent entanglement properties as the original $V_{AB}$, plus a collection of $m+n-2$ uncorrelated single modes. This reproduces the findings of~\cite{Serafini05}.

Similarly to what discussed in the Remark at the end of Subsection~\ref{sec5 simplified}, the results of Theorem~\ref{PPt sym} can also be straightforwardly extended to characterise full separability and, conversely, multipartite entanglement of arbitrary multimode Gaussian states which are partitioned into $k$ subsystems, with the requirement of local permutation invariance within some of these subsystems. It is clear that, by suitable local symplectic transformations, each of those locally symmetric parties can be localised onto a single mode correlated with the remaining parties, thus removing the redundancy in the QCM. Gaussian states of this sort generalise the so-called multi-symmetric states studied in~\cite{moleculo}, where local permutation invariance was enforced within all of the subsystems, resulting in a direct multipartite analogue of bi-symmetric states.

\subsection{Entangling Gaussian states via passive optical operations} \label{subsec5 pass}

Throughout this subsection, we finally complete the solution of a problem posed in~\cite{passive} and there addressed under some additional constraints. Let us start by recalling that symplectic operations can be divided into two main categories, namely those such as squeezers that require an exchange of energy between the system and the apparatus, called \textbf{active}, and those that can be implemented using only beam splitters and phase plates, called \textbf{passive}. A symplectic matrix $K$ represents a passive transformation if and only if it is also \emph{orthogonal}, meaning that $KK^{T}=\id$ (it may be worth adding that symplectic orthogonal transformations form the maximal compact subgroup of the symplectic group).
As it turns out, symplectic orthogonal matrices can be represented in an especially simple form if we resort to a position-momentum block decomposition. Namely, one has the parametrisation~\cite{BUCCO}
\begin{equation}
K = W^\dag \begin{pmatrix} U & \\ & U^* \end{pmatrix} W\, ,
\end{equation}
where
\begin{equation*}
W \coloneqq \frac{1}{\sqrt{2}} \begin{pmatrix} \id & i\id \\ \id & -i\id \end{pmatrix}
\end{equation*}
and $U$ is a generic, $n\times n$ unitary matrix, with $U^*$ denoting its complex conjugate (as usual, $n$ is the number of modes).

Since the implementation of passive operations is so inexpensive in quantum optics and entangled states so useful for quantum technologies, the question first posed in~\cite{passive} was a natural one: \emph{what bipartite Gaussian states are such that they can be entangled via a global, passive operation?} However, in this full generality the problem was left unanswered in~\cite{passive}. Instead, another related question was investigated and answered there, namely whether \emph{distillable} Gaussian entanglement can be produced in the same fashion. For Gaussian states, as mentioned in Subsection~\ref{subsec5 G sep}, distillability is well known to be equivalent to non-positivity of the partial transpose~\cite{Giedke01,GiedkeQIC}, so the authors of~\cite{passive} proceeded to identify the class of Gaussian states that can be made to violate the PPT condition by means of a passive transformation. However, it is important to realise that since PPT and separability are not the same for general multimode Gaussian states, the two questions are a priori different. Here we show that the answer to the original question above turns out to be yet another situation where the PPT condition is necessary and sufficient to ensure separability of Gaussian states. In other words, we will prove that a bipartite Gaussian state that can not be made distillable (i.e.~non-PPT) via passive operations is necessarily separable, and thus it stays separable under the application of said passive operations. Let us start with a technical lemma that we deduce from recent results obtained in~\cite{bhatia15}.

\begin{lemma} \label{lemma eig vs sp eig}
Let $A>0$ be a strictly positive $2n\times 2n$ matrix. Let $\nu_{i} (A)$ and $\lambda_i (A)$ denote its symplectic and ordinary (orthogonal) eigenvalues, respectively, arranged in nondecreasing order. Then
\begin{equation}
\nu_{1}(A)^2\geq \lambda_1(A) \lambda_2(A)\, .
\label{ineq sympl}
\end{equation}
In particular, every positive matrix whose two smallest eigenvalues obey the inequality $\lambda_1 \lambda_2\geq 1$ is automatically a legitimate QCM.
\end{lemma}

\begin{proof}
From \cite[Equation (71)]{bhatia15} we deduce $\prod_{j=1}^{k} \nu_{n-j+1}(A)^2\leq \prod_{j=1}^{2k} \lambda_{2n-j+1}(A)$ for all $k=1,\ldots,n$, 
with equality for $k=n$, when both terms equal the determinant of $A$. We can use this observation to deduce that $\prod_{j=1}^k \nu_j(A)^2\geq \prod_{j=1}^{2k} \lambda_j(A)$ for all $k=1,\ldots,n$. The special case $k=1$ yields the claim.
\end{proof}

Now, we are ready to present our strengthening of~\cite[Proposition 1]{passive}.

\begin{thm} \label{abs sep Gauss}
Let $V$ be a bipartite QCM of an $n$-mode system. Then the following are equivalent:
\begin{enumerate}[(a)]
\item $KVK^T$ is separable for all Gaussian passive transformations $K$;
\item $KVK^T$ is PPT for all Gaussian passive transformations $K$; and
\item the two smallest eigenvalues of $V$ satisfy $\lambda_1(V)\lambda_2(V)\geq 1$.
\end{enumerate}
\end{thm}

\begin{proof}
The implication $(a)\Rightarrow (b)$ is obvious, while $(c)\Rightarrow (b)$ already follows from Lemma~\ref{lemma eig vs sp eig} together with the fact that the partial transpose at the level of QCMs is a congruence by orthogonal transformation and thus does not change the ordinary spectrum. One of the main contributions of~\cite{passive} is the proof that $(b)$ and $(c)$ are in fact equivalent. In view of this discussion, we have just to show that $(c)\Rightarrow (a)$. To this end, we will assume that $V$ satisfies $\lambda_1(V)\lambda_2(V)\geq 1$ and construct two local QCMs $\gamma_A, \gamma_B$ that satisfy the hypothesis of the original separability criterion given by Lemma~\ref{sep}. Call $\lambda_1(V)= k$ and observe that if $k\geq 1$ then $V\geq \id=\id_A\oplus \id_B$ and we are done. Otherwise, assume $k<1$ and denote by $\ket{x}$ the normalised eigenvector corresponding to the minimal eigenvalue of $V$, i.e.~$V\ket{x}= k \ket{x}$ and $\braket{x|x}=1$. Since $\lambda_2(V)\geq \frac1k$ and a fortiori $\lambda_i(V)\geq \frac1k$ for all $i\geq 2$, we can write
\begin{equation*}
V\geq k \ket{x}\!\!\bra{x} + \frac1k \left( \id - \ket{x}\!\!\bra{x} \right) .
\end{equation*}
Now, decompose the vector $\ket{x}$ into its $A$ and $B$ components as $\ket{x}=\left(\begin{smallmatrix} \sqrt{p} \ket{y}_A \\ \sqrt{1-p} \ket{z}_B \end{smallmatrix}\right)$, where $0\leq p\leq 1$ and $\braket{y|y}=1=\braket{z|z}$. Then, Lemma~\ref{lemma eig vs sp eig} guarantees that the matrices
\begin{align*}
\gamma_A &\coloneqq k \ket{y}\!\!\bra{y} + \frac1k \left( \id - \ket{y}\!\!\bra{y} \right) \\
\gamma_B &\coloneqq k \ket{z}\!\!\bra{z} + \frac1k \left( \id - \ket{z}\!\!\bra{z} \right)
\end{align*}
are legitimate QCMs. For this reason, showing that $V_{AB} - \gamma_A\oplus \gamma_B\geq 0$ would complete our proof. By direct computation, we find
\begin{align*}
V_{AB} - \gamma_A\oplus \gamma_B &\geq -\left( \frac1k - k\right) \begin{pmatrix} p \ket{y}\!\!\bra{y} & \sqrt{p(1-p)} \ket{y}\!\!\bra{z} \\ \sqrt{p(1-p)} \ket{z}\!\!\bra{y} & (1-p) \ket{z}\!\!\bra{z} \end{pmatrix} \\[0.5ex]
&\quad + \frac1k \begin{pmatrix} \id & 0 \\ 0 & \id \end{pmatrix} \\[0.5ex]
&\quad - \begin{pmatrix}  -\left( \frac1k - k\right) \ket{y}\!\!\bra{y} + \frac1k \id & 0 \\ 0 & -\left( \frac1k - k\right) \ket{z}\!\!\bra{z} + \frac1k \id \end{pmatrix} \\[1ex]
&= \left( \frac1k - k\right) \begin{pmatrix} (1-p) \ket{y}\!\!\bra{y} & - \sqrt{p(1-p)} \ket{y}\!\!\bra{z} \\ -\sqrt{p(1-p)} \ket{z}\!\!\bra{y} & p \ket{z}\!\!\bra{z} \end{pmatrix} \\[1ex]
&= \left( \frac1k - k\right) \begin{pmatrix} \sqrt{1-p} \ket{y} & -\sqrt{p} \ket{z} \end{pmatrix}^T \begin{pmatrix} \sqrt{1-p} \ket{y} & -\sqrt{p} \ket{z} \end{pmatrix} \\[0.4ex]
&\geq 0\, ,
\end{align*}
and we are done.
\end{proof}

\begin{rem}
In some sense, one can think of the question posed in~\cite{passive} and answered here in Theorem~\ref{abs sep Gauss} as a continuous variable analogue of the \emph{absolute separability} problem in finite-dimensional quantum information, which asks for the characterisation of those spectra $\sigma=(\lambda_1,\ldots,\lambda_{mn})$ such that every bipartite quantum state on $\mathds{C}^m\otimes \mathds{C}^{n}$ with spectrum $\sigma$ is separable~\cite{kus01}. For a recent review of the state of the art, we refer the reader to~\cite{abs-sep-review}.
A suggestive argument concerning this analogy goes as follows. An arbitrary unitary transformation $\rho\mapsto U\rho U^\dag$ corresponds to an internal time evolution according to some unknown Hamiltonian. Then, the absolutely separable states are exactly those bipartite states whose correlations are so weak that they can not be made entangled by any internal evolution. In the case of continuous variable quantum systems, one may hold the free-field Hamiltonian $\mathcal{H}=\frac12 r^Tr$ as the privileged one, so that it makes sense to restrict oneself to those unitary evolutions that preserve this particular Hamiltonian. If the original state is Gaussian and the unitaries are generated by quadratic Hamiltonians, so that they are represented by symplectic matrices, preserving the free-field Hamiltonian is the defining feature of passive transformations, and one obtains exactly the problem we solved here.

As is often the case, the technical details and the nature of the solution are  simpler in the Gaussian realm. We found that the condition for being `absolutely separable' in the Gaussian sense is expressed by a simple inequality involving only the two smallest ordinary eigenvalues of the QCM, and that there are no `absolutely PPT' states that are not `absolutely separable' too. This latter equivalence has been conjectured to hold for the original problem in discrete-variable systems as well, but so far only partial answers are available. Namely, the conditions for absolute PPT-ness can be written explicitly~\cite{hildebrand07}, but whether or not they imply absolute separability is in general unknown. However, the answer to this latter question has been shown to be affirmative for the case of two qubits~\cite{verstraete01} and more recently for qubit-qudit systems~\cite{johnston13}.
\end{rem}

\section{Summary and outlook} \label{outro}

Throughout this chapter we presented some significant advances in the mathematical and physical study of separability and entanglement distillability in Gaussian states of continuous variable quantum systems. Based on the properties of Schur complements and other matrix analysis tools, we obtained a simplified necessary and sufficient condition for the separability of all multimode Gaussian states, requiring optimisation over the set of local covariance matrices of one subsystem only (Theorem~\ref{simp sep lemma}). Exploiting this result, we presented a compact proof
of the equivalence between PPT and separability for  $1$ vs $n$-mode Gaussian states (Theorem~\ref{PPT thm}), a seminal result in continuous variable quantum information theory~\cite{Simon00,Werner01}, as well as extended the criterion to multimode classes of so-called mono-symmetric (Theorem~\ref{PPT thm}) and isotropic Gaussian states (Theorem~\ref{iso thm}), through novel derivations.
Furthermore, we completed  the investigation of entanglement generation under passive operations by extending seminal results~\cite{passive} to consider the generation of any, possibly PPT, Gaussian entangled state: in this context we showed that, if passive operations cannot turn an initial Gaussian state into a non-PPT one, then no PPT entanglement can be generated through them either (Theorem~\ref{abs sep Gauss}). This can be interpreted as establishing the equivalence between absolute separability and absolute PPT-ness in the Gaussian world.
Side results of our analysis include a novel proof that Gaussian states invariant under partial transposition are separable (Corollary~\ref{inva cor}), as well as an independent proof of the equivalence between Gaussian separability and complete extendibility with Gaussian extensions~\cite{Bhat16}.

In the context of this study, and with the methods illustrated above, it would be interesting to research more general combinations of symmetries and conditions on the symplectic spectra of quantum covariance matrices whereby the sufficiency of the PPT separability criterion might be further extended. For instance, is it possible to obtain a Gaussian analogue of the results in~\cite{chen08}, whereby bound entangled Gaussian states can only exist given some simple condition on their symplectic spectrum? For example, both for mono-symmetric and isotropic states, large degeneracies in their symplectic spectra (for the marginal covariance matrix of one subsystem, and for the global covariance matrix of the bipartite system, respectively) were responsible for the sufficiency of the PPT condition for separability. It would be desirable to provide a full systematic characterisation of such requirements, possibly drawing inspiration from and/or shedding new insight on the Gaussian quantum marginal problem~\cite{tyc}.

Finally, let us stress how matrix analysis tools such as those heavily hammered in this chapter have already been proved useful for qualitative and quantitative analysis of entanglement and other correlations in general states of continuous variable systems~\cite{Giedke01, giedkemode, nogo1, nogo2, nogo3, eisemi, wise, AdessoSerafini, MonSteer, Simon16, Lami16, LL-log-det}. 

Equipped with our renewed mathematical tools, we are now prepared to climb up to the stars.

\part[ Quantum correlations in continuous variable systems]{Quantum correlations in continuous variable systems}

\chapter{Schur complement inequalities and monogamy of quantum correlations} \label{chapter6}


\section{Introduction} \label{sec6 introduction}

The sun is setting on our investigation of quantum entanglement. However, this does not relieve us from the challenge of studying more general kinds of correlations that bipartite quantum states exhibit, which we are now prepared to take up.
In fact, until now the expression `non-classical correlations' in the title of the present thesis has been interpreted mainly in two distinct senses. The first of them is connected with the notion of entanglement, a concept whose importance in quantum information is hardly overestimated, and to the study of which we devoted a great part of our effort. The second, stronger meaning has to do with violation of Bell inequalities, leading to the phenomenon dubbed `nonlocality' in Chapter~\ref{chapter2}. As it turns out, these are two extremes of a wider range of interpretations one can give to the above words. For instance, an intermediate scenario can lead to the notion of \emph{steering}~\cite{schr, wise}, which is equally important for some applications.
If one were to summarise the content of this chapter in a single sentence, one could say that it aims to establish properties of and relations among various forms of non-classical correlations in bipartite quantum systems, and more specifically in continuous variable quantum systems whose state is Gaussian. 

The chapter is organised as follows. The rest of this section is devoted to introducing and motivating our investigation (Subsection~\ref{subsec6 corr}) and to pointing the reader to our contributions (Subsection~\ref{subsec6 contributions}). In Section~\ref{sec6 steering} we discuss the concept of steering and the associated resource theory, with particular regard to the Gaussian setting.
The aim of Section~\ref{sec6 G steerability} is to prove some matrix inequalities that involve quantum covariance matrices and to use them to derive properties of Gaussian steerability measures.
Section~\ref{sec6 G corr hierarchy} is then dedicated to the proof of a new hierarchical relation among the quantifiers of Gaussian steerability, entanglement and total correlations based on the R\'enyi-2 entropies (Theorem~\ref{I2E}). Several consequences are drawn from this fact.
Finally, Section~\ref{sec6 conclusions} contains the conclusions and a discussion of the outlook of this work.

\subsection{Quantum correlations in Gaussian states} \label{subsec6 corr}


We spent most part of Chapter~\ref{chapter2} studying mainly two types of non-classicality exhibited by composite systems in general probabilistic theories. The first and most fundamental phenomenon connected with non-classicality is the emergence of entanglement, i.e. the existence of bipartite states that can not be expressed as a convex combination of products of local states. This notion is intrinsically algebraic and as such does not bear a direct operational interpretation. However, one can turn it into something more concrete by using local measurement to generate from a state of a physical system -- whatever its mathematical description is -- concrete correlations among the measurement outcomes. When the correlations can not be explained by a classical way of thinking, i.e. by a `hidden variable model', we say that they are Bell-nonlocal, or simply nonlocal. This second concept has nothing to do with the way we describe the system with a mathematical model, and on the contrary is fully intrinsic.

A useful way to think of nonlocality -- especially in an information theoretical sense -- is in terms of a task. Assume that two parties, Alice and Bob want to convince a third party, Charlie, that the state they share is entangled. Charlie does not know the details of the physical system under examination and can not access it, but he can nevertheless set up a test as follows. First, he forbids all communication between Alice and Bob. Then, he demands they provide him with a list of labels identifying possible measurements on their respective sides. In a multiple-round protocol, he picks labels from this list, asks them to perform the corresponding measurements (one for each side), and records the outcomes. For certain states (for this reason called nonlocal) and appropriately chosen measurements, Alice and Bob might be able to succeed and convince Charlie they do have shared entanglement, as otherwise he would not be able to explain the correlations he sees.

The above scenario features two \emph{untrusted} parties, since Charlie believes both Alice and Bob could try to cheat him. There is however an alternative scenario where only one of the two parties (say Alice) is untrusted, and Bob's good faith is instead acknowledged. In this case, we can identify Charlie with Bob, and the task becomes the following: first, Alice provides Bob with a list of labels identifying possible measurements on her side and with the states of Bob's share of the system that each outcome of those measurements will yield; then, over multiple rounds, Bob picks labels at random, asks Alice to perform the corresponding measurements and to tell him the outcomes, and tests whether his state is what Alice claimed to be beforehand. When the state and the measurements are carefully chosen, Bob will have the feeling that Alice is able to `steer' his system (which is isolated from the external environment) into very different states, hence the name \emph{steering} that this phenomenon bears~\cite{wise}. Incidentally, let us mention that this peculiar feature of the theory was already identified by Schr\"odinger in person, who felt somewhat uncomfortable with it~\cite{schr}. Anyway, if Bob is unable to find a classical explanation for this steering capability, he has to conclude that Alice possesses a system that is entangled with his. The global state is then said to be \emph{steerable} from $A$ to $B$ with respect to the chosen measurements. For a formal definition, we refer the reader to Subsection~\ref{sec6 steering}.

In the case of continuous variable bipartite quantum systems, arguments similar to that in Subsection~\ref{subsec5 G sep} suggest that a special case of the above scenario is of particular relevance, namely when the global state is Gaussian and the set of available measurements on Alice's side consists of all Gaussian measurements.\footnote{The reader could wonder what happens when one considers instead nonlocality in the Gaussian framework. It turns out that because of the representation~\eqref{Fourier-Weyl representation} Gaussian measurements applied to bipartite Gaussian states can never lead to a violation of a Bell inequality, a fact that was already clear to Bell himself~\cite[XXI]{SPEAKABLE}. For further details on ways around this problem, see~\cite[IV]{BUCCO}.} In this case, steerability can be characterised in closed form by means of a condition on the global covariance matrix analogous to that for separability,~\ref{sep eq}~\cite{wise}. Inspired by this characterisation, a measure of \emph{Gaussian steerability} that quantifies the amount by which this condition is violated was introduced in~\cite{steerability}. We will devote some time discussing this measure and its properties, see the forthcoming Subsection~\ref{subsec6 contributions}.

The Gaussian steerability measure of~\cite{steerability} is only one of the many correlation measures one can construct by looking at the covariance matrix of the state. Other important examples are provided by the Gaussian entanglement of formation~\cite{Wolf03, EoF-symmetric-G} and by its R\'enyi-2 version~\cite{AdessoSerafini}. The immediate advantage of using the R\'enyi-2 entropy instead of the more natural von Neumann entropy~\eqref{entropy} lies in the fact that it respects the quadratic nature of Gaussian states. As we will see, this enables us to prove a lot of properties that these quantifiers obey. 
However, we do not want to give the reader the impression that mathematical convenience is the only reason to prefer R\'enyi-2 quantifiers in the Gaussian framework. In fact, such quantifiers can be more operationally relevant when one deals with \emph{measured} quantities. The reason why this is the case will become clearer in the forthcoming Chapter~\ref{chapter7}.

\subsection{Our contributions} \label{subsec6 contributions}


This chapter is based on the (almost) homonymous paper~\cite{Lami16}:
\begin{itemize}

\item L. Lami, C. Hirche, G. Adesso, and A. Winter. Schur complement inequalities for covariance matrices and monogamy of quantum correlations. \emph{Phys. Rev. Lett.}, 117(22):220502, 2016.

\end{itemize}

Our first goal in this chapter will be to establish a collection of results for the Schur complement of a covariance matrix of a Gaussian state. As we said, covariance matrices bear a direct impact on the quantitative characterization of various forms of quantum correlations in continuous variable systems, and in turn on their usefulness for quantum technologies. Our analysis is inspired by recent works~\cite{AdessoSerafini, Gross, Kor, Simon16}, in which an inequality sharing the same formal structure as the strong subadditivity of entropy was obtained, by purely algebraic methods, for the log-determinant of positive semidefinite matrices $V_{ABC} \geq 0$:
\begin{equation}
\label{SSA 6}
\log\det V_{ABC} + \log\det V_{C} \leq \log\det V_{AC} +\log\det V_{BC}\,.
\end{equation}
In the rest of the thesis, we will call~\eqref{SSA 6} \textbf{strong subadditivity (SSA) of log-det entropy.} More reasons why the name is fitting will become apparent in Chapter~\ref{chapter7}.
If one identifies $V_{ABC}$ with the QCM of a $(n_A+n_B+n_C)$-mode tripartite \emph{quantum} system, which requires the extra condition corresponding to Heisenberg's uncertainty principle~\eqref{Heisenberg}, i.e. $V_{ABC} + i \Omega_{ABC} \geq 0$, to be obeyed,
then the scalar inequality~\eqref{SSA 6} has relevant implications, yielding alternative quantifiers of correlations~\cite{AdessoSerafini}, monogamy constraints for Gaussian entanglement~\cite{AdessoSerafini}, and limitations for joint steering of single-mode states in a multipartite scenario~\cite{Kor, Simon16, MonSteer}.

In Section~\ref{sec6 G steerability} we show, \emph{inter alia}, that such an inequality admits a powerful \emph{operator} strengthening directly at the level of covariance matrices (Theorems~\ref{pt incr sch} and~\ref{sch compl mon}). This paves the way to several applications. First, we take the Gaussian steerability measure introduced in~\cite{steerability} and prove it is a monotone under the relevant sets of free operations in the Gaussian subtheory of the recently established resource theory of steering~\cite{Gallego} (Theorem~\ref{G prop}). Second, we tackle monogamy questions related to this quantifier. Since steerability is intrinsically asymmetric, there is more than one possible monogamy inequality to be investigated. Our findings show that the measure is monogamous with respect to the \emph{steered} party, but not in general with respect to the \emph{steering} party, unless some further assumptions are made (Theorem~\ref{G mono}).
The theory of Schur complements (Subsection~\ref{subsec5 Schur}) plays a crucial role in the proofs of all the main results of Section~\ref{sec6 G steerability}, some of which are highly nontrivial, and require the use of other sophisticated matrix analysis techniques.

Next, in Section~\ref{sec6 G corr hierarchy} we study different forms of correlations in bipartite Gaussian states, especially when measured through quantifiers based on the R\'enyi-2 entropy. Our main result there is Theorem~\ref{I2E}, which tells us that a strong inequality characterising `well-behaved' theories of bipartite correlations~\cite{LiLuo} holds for Gaussian entanglement of formation and mutual information of Gaussian states when measured by means of the R\'enyi-2 entropy. The importance of this result lies also in the fact that an analogous inequality does \emph{not} hold for the same quantifiers based on the standard von Neumann entropy~\eqref{entropy}. Moreover, a straightforward corollary of Theorem~\ref{I2E} yields the monogamy of the R\'enyi-2 Gaussian entanglement of formation for arbitrary many modes per party (Corollary~\ref{E mono}), a fairly general statement which goes beyond all previous results.
Let us add that the proof of Theorem~\ref{I2E} we present, surprisingly short and direct, relies heavily on the properties of the matrix geometric mean we already discussed in Subsection~\ref{subsec5 matrix means}.

\section{Quantum steering} \label{sec6 steering}

The aim of this section is to provide a basic introduction to the quantum phenomenon known as steering. In Subsection~\ref{subsec6 generalities} we discuss steering in general, as presented in~\cite{wise}, and examine the Gaussian special case more closely. Subsection~\ref{subsec6 resource} is then devoted to reviewing the basics of the subsequently developed resource theory of steering~\cite{Gallego}. Finally, in Subsection~\ref{subsec6 G steerability} we introduce the Gaussian steerability measure of~\cite{steerability}.

\subsection{Generalities} \label{subsec6 generalities}

Among the various surprising phenomena that occur in the quantum world, steering was among the first ones to be recognised. In a famous paper~\cite{schr}, Schr\"odinger himself introduced the concept and term, and noted with disappointment that `it is rather discomforting that the theory should allow a system to be steered or piloted into one or the other type of state at the experimenter's mercy in spite of his having no access to it.' A more modern approach to the problem can be found in~\cite{wise}, where steering is seen from an information theoretical point of view as a \emph{resource}. The question then becomes: \emph{are there tasks than can be accomplished thanks to this capability of quantum systems to be steered by distant parties?}

As shown in~\cite{schr}, the answer turns out to be affirmative, and we already sketched the description of such a task in Subsection~\ref{subsec6 corr}. Let us repeat it here in a slightly more general -- and more dramatised -- form. The scenario features two distant parties, Alice and Bob, connected with a one-way quantum channel that goes from Alice to Bob. Alice can send as many copies of a quantum state as she wants. Bob has a quantum laboratory where he can perform arbitrary quantum measurements with his powerful equipment, so in particular he can test that the state Alice is sending corresponds to what she is claiming. Alice maintains that said state is actually one of the two shares of a bipartite entangled state.
How can she prove her claim to Bob, if he can not access her lab and does not trust her?

Alice maintains that her lab is equipped with some instruments with which she can implement some quantum measurements. In what follows, we will index such measurements with the label $x$. The possible values of $x$ are publicly declared by Alice. The outcomes of those measurements, also made public by Alice, are indexed by $a$. The protocol is made by several identical rounds, all starting with Bob choosing a random label $x$ and then asking Alice to output an (alleged) measurement outcome $a$.
The crux of the game consists in Alice's ability to foresee the state that Bob will have after she claims to have performed the measurement $x$ and recorded the outcome $a$. Let us denote such state by $\rho_{a,x}$. We assume that the experiment can be repeated as many times as Bob wants, so that he can then test his quantum system using his powerful equipment and make sure that this is indeed the case. Another thing Bob will learn over multiple rounds is the probability of Alice outputting $a$ when he asks for the measurement $x$.

It is useful to combine these two objects (states and probabilities) in a single operator $\rho(a|x)\coloneqq p(a|x)\rho_{a,x}$.
With this notation, the only thing Bob can access is the collection $\{\rho(a|x)\}_{a,x}$, called an \textbf{assemblage}~\cite{Gallego}. In what follows, the two variables specifying the outcome of the measurement and the label identifying what measurement we are making will be always added as subscripts, in this order, when specifying the assemblage we consider.
Any assemblage corresponding to a physically relevant situation must be \emph{non-signalling}, i.e. it must satisfy the condition
\bb
\sum_a \rho(a|x) = \sum_a \rho(a|x') \eqqcolon \rho_B\quad \forall\ x,x'\, .
\label{non-signalling assemblage}
\ee
Of course, very natural assemblages are those generated by a bipartite quantum state via measurements, i.e. such that
\bb
\rho(a|x) = \tr_A [E_a^x\otimes \id\, \rho_{AB}]
\label{quantum assemblage}
\ee
for some state $\rho_{AB}$ and some collection of measurements $\big\{ E_a^x\big\}_{a,x}$, whose elements will obey $E_a^x\geq 0$ for all $a,x$ and $\sum_a E^x_a = \id_A$ for all $x$. As is easy to realise, such assemblages correspond to situations where Alice is really trying to distribute to Bob a share of a bipartite quantum state.
However, it is known that not all non-signalling assemblages are compatible with the requirement of being generated by an original bipartite quantum state~\cite{postq-st}.

Let us come back to the main question here: given an assemblage $\{\rho(a|x)\}_{a,x}$, can Bob find an alternative classical explanation for what he is observing, or is any explanation going to involve entanglement? An explanation of the former kind will look like this: he will posit that Alice is trying to cheat him, as she is in fact sending not shares of entangled states but merely random quantum states that are generated according to a `hidden variable' $\lambda$ she has access to.
In order for this explanation to be consistent with the observations, it must be that
\bb
\rho(a|x) = \sum_\lambda p(\lambda) p(a|x \lambda)\, \xi_\lambda \qquad \forall\ a,x\, , \label{unst}
\ee
where $p(\lambda)$ is some a priori probability, $\xi_\lambda$ some collection of states, and $p(a|x\lambda)$ a classical decision rule to output a label $a$ based on $x$ and $\lambda$. Equation~\eqref{unst} defines the $A\rightarrow B$ \textbf{unsteerability} of the assemblage $\{\rho(a|x)\}_{a,x}$. Given a quantum state $\rho_{AB}$ and a set of measurements $\mathcal{M}$ on $A$, if the collection of measurements $\big\{ E_a^x\big\}_{a,x}$, where $x\in \mathcal{M}$, is such that the associated assemblage~\eqref{quantum assemblage} is $A\rightarrow B$ unsteerable, we say that $\rho_{AB}$ itself is $A\rightarrow B$ unsteerable with measurements in $\mathcal{M}$.

A scenario that is of particular relevance to us is when the state $\rho_{AB}=\rho_{AB}^G(V_{AB}, w_{AB})$ is a Gaussian state of a bipartite continuous variable system, and the set of measurements $\mathcal{M}$ comprises all Gaussian measurements on $A$. The question of what the steering properties of such state are in this setting boils down to a condition that involves only its covariance matrix $V_{AB}$, partitioned as in~\eqref{V explicit}. In fact, it can be shown~\cite{wise} that $\rho_{AB}$ is $A\rightarrow B$ unsteerable with Gaussian measurements iff
\bb
V_{AB} / V_A \geq i\Omega_B\, ,
\label{unsteerability Schur}
\ee
which can be equivalently written as 
\bb
V_{AB} \geq 0_A \oplus i\Omega_B
\label{unsteerability block}
\ee
by virtue of Corollary~\ref{Schur variational cor}.

\subsection{Resource theory of steering} \label{subsec6 resource}

If one wants to build a resource theory of steering, in which steerable states (or better, assemblages) are seen as resources that allow us to perform certain tasks, a prerequisite is the identification not only of free (unsteerable) states, but of free operations as well. A basic requirement free operations have to meet is that they can not create steerable states out of unsteerable ones, i.e. they have to preserve unsteerability.
A thorough discussion of the free operations that are operationally relevant for applications can be found in~\cite{Gallego}.
We will limit ourselves to introducing some natural candidates for the role of free operations, and to verifying that they do not change unsteerability of assemblages.

\begin{note}
In what follows, we will indicate all probability distributions with the same letter $p$, which is very convenient for limiting the amount of symbols in a single equation. What distinguishes one from the other is the labels they carry, that identifies not only the particular realisation, but also the relevant random variable (together with the associated probability distribution) uniquely. For instance, we will write Bayes' rule like $p(a|b)=p(b|a)\frac{p(a)}{p(b)}$, since it is understood that $b$ can not be another realisation of the random variable $A$ that produces $a$, otherwise we would have denoted it by $a'$.
\end{note}

There are at least three elementary operations Alice and Bob can perform on their assemblages without changing its unsteerability. Let us discuss them one by one. Then, the free operations of the resource theory of steering will be defined as arbitrary concatenations of those elementary transformations.

\begin{enumerate}[(1)]

\item Alice can measure Bob's share with a quantum instrument $\{\mathcal{E}_\omega\}_\omega$ before handing it over to Bob. The outcome $\omega$ is recorded by Alice, and now the game begins. What is the assemblage associated with the problem? Well, it is clearly $\left\{ \frac{\mathcal{E}_{\omega}(\rho(a|x))}{p(\omega)} \right\}_{a,x}$, where $p(\omega) = \Tr [ \mathcal{E}_{\omega}(\rho_{B})]$ is the prior probability of getting the outcome $\omega$. Let us stress that although this new assemblage does depend on $\omega$, it does not feature $\omega$ as a \emph{variable}.
It turns out that by doing all this Alice can never make steerable an assemblage that was previously unsteerable. In fact, if~\eqref{unst} holds then
\begin{align*}
\frac{\mathcal{E}_{\omega}(\rho(a|x))}{p(\omega)} &= \sum_{\lambda} p(\lambda) p(a|x\lambda) \frac{\mathcal{E}_{\omega}(\xi_{\lambda})}{p(\omega)} \\
&= \sum_{\lambda} p'(\lambda) p(a|x\lambda) \xi'_{\lambda}\, ,
\end{align*}
where
\begin{align*}
p'(\lambda) &\coloneqq \frac{p(\lambda)}{p(\omega)} \Tr [\mathcal{E}_{\omega}(\xi_{\lambda})] = \frac{p(\lambda)}{p(\omega)} p(\omega|\lambda) =  p(\lambda|\omega)\, ,\\
\xi'_{\lambda} &\coloneqq \frac{\mathcal{E}_{\omega}(\xi_{\lambda})}{ \Tr [\mathcal{E}_{\omega}(\xi_{\lambda})] }\, .
\end{align*}

\item Alice can pre-process classical information, that is, she can: (i) ask Bob to start the protocol by outputting a new classical label $y$ instead of $x$; (ii) use the former to produce a label $x$ according to some classical decision rule $p(x|y)$; and (iii) run the original protocol with this $x$, outputting $a$ and forgetting $x$. Bob's final state when the input is $y$ and the outcome is $a$ becomes
\begin{align*}
\rho_{a,y} &= \sum_x p(x|a,y) \rho_{a,x} \\
&= \sum_x \frac{p(a,x,y)}{p(a,y)}\, \rho_{a,x} \\
&= \sum_x \frac{p(a,x|y)}{p(a|y)\,}\, \rho_{a,x} \\
&= \sum_x \frac{p(x|y) p(a|x,y)}{p(a|y)}\, \rho_{a,x} \, .
\end{align*}
Using the fact that $p(a|x,y)=p(a|x)$, valid since Bob's input $y$ is fed into the original device $p(a|x)$ only through $x$, we obtain
\begin{align*}
\rho(a|y) &= p(a|y) \rho_{a,y} \\
&= \sum_x p(x|y) p(a|x)\, \rho_{a,x} \\
&= \sum_x p(x|y) \rho(a|x) \, .
\end{align*}
From this we see that the original assemblage is unsteerable whenever the old one was such, since
\begin{align*}
\rho(a|y) &= \sum_x p(x|y) \rho(a|x) \\
&= \sum_{x,\lambda} p(x|y) p(\lambda) p(a|x,\lambda) \xi_\lambda \\
&= \sum_\lambda p(\lambda) p(a|y,\lambda) \xi_\lambda\, ,
\end{align*}
where $p(a|y,\lambda)=\sum_x p(x|y) p(a|x,\lambda)$.

\item Alice can post-process all of her classical information before outputting the final label. Namely, instead of $a$ she can choose to output a new label $b$, which is produced with probability $p(b | a x)$, and later delete $a$. According to Bayes' rule, Bob's state corresponding to an outcome $b$ will be
\begin{align*}
\rho_{b,x} &= \sum_{a} p(a,x|b) \rho_{a,x} \\
&= \sum_{a} \frac{p(a,b,x)}{p(b,x)}\, \rho_{a,x} \\
&= \sum_{a} \frac{p(a,b|x)}{p(b|x)}\, \rho_{a,x} \\
&= \sum_{a} \frac{p(a|x) p(b|a,x)}{p(b|x)}\, \rho_{a,x}\, ,
\end{align*}
i.e. $\rho(b|x) = \sum_{a} p(b|a,x) \rho(a|x)$. If the original assemblage was unsteerable, then we have
\begin{align*}
\rho(b|x) &= \sum_a p(b|a,x) \rho(a|x) \\
&= \sum_{a,\lambda} p(b|a,x) p(\lambda) p(a|x,\lambda)\, \xi_\lambda \\
&= \sum_\lambda p(b|x,\lambda) \xi_\lambda\, ,
\end{align*}
where $p(b|x,\lambda)=\sum_a p(a|x,\lambda) p(b|a,x)$, and hence the new assemblage $\{\rho(b|x)\}_{b,x}$ is unsteerable, too.

\end{enumerate}

\begin{rem}
Although we did not say it explicitly, all of the above operations preserve the fundamental requirement of being non-signalling.
\end{rem}

The most general operation on assemblages that can be implemented using as building blocks the above three elementary steps is called in~\cite{Gallego} a \textbf{deterministic one-way LOCC}, or \textbf{1W-LOCC} for short. Of course, as we defined them these operations are specific to assemblages, while the set $\text{LOCC}_\rightarrow$ considered in Chapter~\ref{chapter3} comprised protocols acting on quantum states.

According to~\cite{Gallego}, any valid \emph{quantifier} of $A\rightarrow B$ steerability should be mandatorily: (i) vanishing on unsteerable assemblages and (ii) nonincreasing on average under 1W-LOCC operations, and optionally (iii) convex. A quantity satisfying (i), (ii), and (iii) is referred to as a \textbf{convex steering monotone}; examples of such monotones are discussed in~\cite{Gallego}.

What happens if we apply a 1W-LOCC to a quantum assemblage such as~\eqref{quantum assemblage}? Do we get another quantum assemblage? The answer to this question is readily seen to be affirmative: the quantum state that generates the new assemblage is of the form $\frac{(I\otimes \mathcal{E})(\rho_{AB})}{\Tr \mathcal{E}(\rho_B)}$, where $\mathcal{E}$ is a trace non-increasing completely positive map (i.e. an element of a quantum instrument), which corresponds to a specific outcome as selected by Alice. This is a useful observation to keep in mind.

Finally, let us consider the resource theory of steering in the Gaussian scenario. When the underlying states are Gaussian, it is very natural to restrict ourself to Gaussian free operations. In the case of 1W-LOCC, this means the following. First, the quantum instrument considered in point (1) of the above list has to be obtained by composing: (i) additions of Gaussian ancillas; (ii) symplectic unitaries; and (iii) Gaussian partial measurements (Subsection~\ref{subsec5 G states}). In particular, each of the maps $\mathcal{E}_\omega$ is a Gaussian completely positive map whose action is expressed by~\eqref{CP Gauss}. Second, the classical channels in points (2) and (3) of the above discussion have to be classical Gaussian channels. Assemblages deriving from Gaussian states together with \textbf{Gaussian 1W-LOCC operations} as we described them define a \textbf{resource theory of Gaussian steering} that is a sub-theory of the standard resource theory of steering we discussed in this subsection.

\subsection{Gaussian steerability measure} \label{subsec6 G steerability}

We now turn our attention to the problem of \emph{quantifying} steerability in a meaningful way. Mathematically, this amounts to identifying real measures on assemblages that are \emph{monotone} (say, monotonically non-increasing) under all free operations. We are mostly interested in the case of steerability of Gaussian states by Gaussian measurements\footnote{Some of the results presented in this chapter hold also when either the state or the measurements are Gaussian, but we focus here on the simplest case of all, i.e. when both the state and the measurement are Gaussian.}, so quantifiers can be equivalently thought of as being defined on the set of bipartite QCMs $V_{AB}$, which we assume partitioned with respect to the $A:B$ splitting as in~\eqref{V explicit}.

In~\cite{steerability}, a Gaussian steerability measure was introduced by leveraging the condition~\eqref{unsteerability Schur}. In order to discuss it, we need to introduce some definitions first.
In what follows, denote by $\nu_i(A)$ the $i$--th smallest symplectic eigenvalue of a strictly positive $2n\times 2n$ matrix $A>0$. Construct the two functions
\begin{equation}
g_\pm (A) =  {\sum}_{i=1}^n \max\left\{\pm \log \nu_i(A),\, 0 \right\}\, .
\label{func g}
\end{equation}
Clearly, by the Williamson decomposition theorem~\eqref{Williamson}, $g_-(A)=0$ is equivalent to $A$ satisfying~\eqref{Heisenberg}, i.e. being a legitimate QCM.
Let us mention in passing that the function $g_-$ finds many applications in continuous variable quantum information. For instance, the logarithmic negativity~\cite{negativity, plenioprl} of a bipartite state $\rho_{AB}$, defined as  $E_N(\rho_{AB}) =  \log \|\rho_{AB}^{T_B} \|_1$ (where $T_B$ denotes partial transposition), takes the form $E_N(\rho_{AB})=g_-\big(V_{AB}^{T_B}\big)=g_-(\Theta V_{AB} \Theta)$ if $\rho_{AB}=\rho_{AB}^G(V_{AB}, w_{AB})$ is a Gaussian state with QCM $V_{AB}$; here, the partial transpose of the QCM is expressed according to~\eqref{partial transpose QCM}. More details on the properties of the functions in~\eqref{func g} will be provided in Subsection~\ref{app prop g}.

Now, let us come back to our measure of Gaussian steerability, which on a Gaussian state $\rho^G_{AB}$ with QCM $V_{AB}$ takes the form~\cite{steerability}
\begin{equation}
\mathcal{G}(A\rangle B)_V = g_-(V_{AB}/V_A)\, ,
\label{G steer}
\end{equation}
in the case of party $A$ steering party $B$. As we mentioned, we will consider the function $\mathcal{G}(A\rangle B)$, from now on dubbed \textbf{Gaussian steerability}, as defined on bipartite covariance matrices instead of states, but this is no loss of generality. Observe that by virtue of the properties of $g_-$ discussed above,~\eqref{G steer} measures the amount by which the unsteerability condition~\eqref{unsteerability Schur} is violated, and is therefore a good candidate for a valid quantifier.

A point to stress is that the definition~\eqref{G steer} can be extended to every state that possesses a covariance matrix, even if it is not Gaussian. While $\mathcal{G}(A\rangle B)_V>0$ is necessary and sufficient for steerability from $A$ to $B$ of a Gaussian state with QCM $V_{AB}$ by means of Gaussian measurements on $A$~\cite{steerability, wise}, it can be shown that it is only sufficient if either the state~\cite{JOSAB} or the measurements~\cite{NoGauss1, NoGauss2} are non-Gaussian. This extends significantly the range of applicability of many of our results.

\section{Schur complement inequalities and Gaussian steerability} \label{sec6 G steerability}

This aim of this section is to show that the Gaussian steerability of~\eqref{G steer} is a valid monotone of the Gaussian resource theory of steering. In order to achieve our goal, in Subsection~\ref{subsec6 Schur} we first derive some fundamental inequality that Schur complements, and in particular the Schur complements of quantum covariance matrices, must obey. After a brief technical detour to study some mathematical properties of Subsection~\ref{app prop g}, we apply these inequalities to prove the main results of the section, Theorems~\ref{G prop} and~\ref{G mono}.

\subsection{Schur complement inequalities} \label{subsec6 Schur}

For an introduction to Schur complements, we refer the reader to the dedicated monograph~\cite{ZHANG}. Here, we follow the notation and conventions of Subsection~\ref{subsec5 Schur}, where a succinct list of properties of Schur complements is also provided.

Our starting point is the inequality~\eqref{SSA 6}, from which the authors of~\cite{Simon16} deduce that
\begin{equation}
\label{log det ineq}
\log\det V_{AC} +\log\det V_{BC} - \log\det V_{A} - \log\det V_{B} \geq 0
\end{equation}
for all QCMs $V_{ABC}$.\footnote{We stress that while~\eqref{SSA 6} holds for all strictly positive matrices,~\eqref{log det ineq} requires $V_{ABC}$ to be a QCM, i.e. to satisfy~\eqref{Heisenberg}.} This is the form of~\eqref{SSA 6} that is particularly relevant in Gaussian quantum information.

For the sake of the presentation, let us provide a one-line proof of~\eqref{log det ineq}. We already observed that Corollary~\ref{Schur variational cor} implies that the function $H\mapsto H/A$ is concave in $H>0$. Applying the logarithm (which is a matrix monotone and concave) and taking the trace one obtains the concavity of $\Tr \log (H/A) = \log \det (H/A)$ in $H>0$.
Then, rewrite the left-hand side of~\eqref{log det ineq} as
\bbb
\log\det (V_{AC}/V_A) + \log\det (V_{BC}/V_B)\, .
\eee
Since this is a concave function of $V_{ABC}$, one can evaluate it on pure QCMs, which are easily seen to saturate it.
Inequalities~\eqref{SSA 6} and~\eqref{log det ineq} are equivalent expressions of strong subadditivity for the log-determinant, and can be converted into each other by `purifying' $V_{ABC}$~\cite{Simon16}.

This suggests that the Schur complement of QCMs can define a natural notion of quantum conditional covariance, as previously noted for classical Gaussian variables~\cite{simoprl}.
Hence we will study the Schur complement $V_{AB}/V_B$, thereby proving that many well-known properties of the standard conditional entropy $H(A|B)=H(AB)-H(B)$, where $H$ denotes respectively Shannon or von Neumann entropy for a classical or quantum system,
have a straightforward equivalent within this framework.


We start by recalling that a canonical formulation of strong subadditivity in classical and quantum information theory is $H(A|BC) \leq H(A|C)$, i.e.~partial trace on the conditioning system increases the conditional entropy~\cite{lieb73a, lieb73b, NC}. Guided by our formal analogy, our first result is thus a generalization of~\eqref{SSA 6}.

\begin{thm}[Partial trace in the denominator increases Schur complement]
\label{pt incr sch}
If $V_{ABC} \geq 0$ is any tripartite positive semidefinite matrix, then
\begin{equation} V_{ABC} / V_{BC} \leq V_{AC} / V_C\, . \label{INEQ 1} \end{equation}
\end{thm}

\begin{proof}
Since $V_{ABC}\geq W_A \oplus 0_{BC}$ implies $V_{AC}\geq W_A\oplus 0_C$, employing Corollary~\ref{Schur variational cor} we find
\begin{align*}
V_{ABC} / V_{BC} &= \max\big\{ W_A:\, V_{ABC}\geq W_A\oplus 0_{BC} \big\} \\
&\leq \max\big\{ W_A:\, V_{AC}\geq W_A\oplus 0_C \big\} \\
&= V_{AC} / V_C\, ,
\end{align*}
which proves the claim.
\end{proof}

Clearly, taking the determinant of~\eqref{INEQ 1} and applying the factorization property~\eqref{det factor} of the Schur complement under determinants, equation~\eqref{det factor}, yields~\eqref{SSA 6} immediately.
Notice further that the invariance of $V_{AB}/V_B$ under symplectic operations on $B$, implied by the congruence covariance property of the Schur complement, equation~\eqref{congruence covariance}, and its monotonicity under partial trace, suffice to guarantee its monotonicity under general Gaussian channels $\mathcal{N}_B$ on $B$. 

But there is more: perhaps surprisingly, the Schur complement is also monotonically increasing under general non-deterministic \emph{classical} Gaussian operations on $B$. These maps act at the level of covariance matrices as in~\eqref{CP Gauss}, but the `Choi covariance matrix' of the map, i.e. $\gamma =\left( \begin{smallmatrix} \gamma_1 & \delta_{12} \\ \delta_{12}^T & \gamma_2 \end{smallmatrix} \right)$, is only required to be a strictly positive definite matrix, instead of a legitimate QCM, which would yield a completely positive Gaussian map. We formulate this statement as follows.

\begin{thm}[Classical Gaussian maps in the denominator increase Schur complement]
\label{CP incr sch}
If $\Gamma_{B\rightarrow B'}$ is a non-deterministic classical Gaussian map acting as in~\eqref{CP Gauss}, with the corresponding `Choi covariance matrix' $\gamma_{BB'}$ only required to be strictly positive definite, then
\begin{equation*}
\Gamma_{B\rightarrow B'}(V_{AB}) \big/ \Gamma_{B\rightarrow B'}(V_B) \geq V_{AB} / V_B\, .
\end{equation*}
\end{thm}

\begin{proof}
We employ the formula~\eqref{CP Gauss Schur}, which in our case reads $\Gamma_{B\rightarrow B'}:\ V_B\longmapsto (\gamma_{BB'}+V_B) \big/ (\gamma_B+V_B)$. Then,
\begin{align*}
&\Gamma_{B\rightarrow B'}(V_{AB}) \big/ \Gamma_{B\rightarrow B'}(V_B) \\
&\quad= \big((\gamma_{BB'}+V_{AB}) / (\gamma_B+V_B)\big) \Big/ \big((\gamma_{BB'}+V_B) / (\gamma_B+V_B) \big) \\
&\quad\texteq{(1)} (\gamma_{BB'}+V_{AB})  \big/ (\gamma_{BB'}+V_B) \\
&\quad\textgeq{(2)} V_{AB} / V_B\, .
\end{align*}
The justification of these steps is as follows: (1) is an instance of the quotient property~\eqref{quotient property}; (2) follows from the elementary bound
\bbb
\begin{pmatrix} A & X \\ X^T & B+\sigma \end{pmatrix} \bigg/ (B+\sigma) \geq \begin{pmatrix} A & X \\ X^T & B \end{pmatrix} \bigg/ B\, ,
\eee
in turn an immediate consequence of the monotonicity of the matrix inversion on strictly positive definite matrices.
\end{proof}

Next, we would like to obtain from~\eqref{INEQ 1} an operator generalization of~\eqref{log det ineq} by applying the symplectic purification trick. This requires a certain amount of work, so we state a couple of preliminary lemmas. From now on, we will always assume that the $V$ matrices are not only positive semidefinite but actual QCMs, which obey~\eqref{Heisenberg}.

\begin{lemma} \label{pure Schur lemma 1}
Let $V_{AB}$ be a bipartite pure QCM. Then
\bb
V_{AB}/V_A=\Omega_B^T V_B^{-1} \Omega_B\, .
\label{pure Schur 1 eq}
\ee
\end{lemma}

\begin{proof}
Since $V_{AB}$ is symplectic, it satisfies $V_{AB}^{-1} = \Omega_{A}^T V_{AB}^T \Omega_{AB} = \Omega_{AB}^T V_{AB} \Omega_{AB}$. Comparing the $B$ blocks thanks to~\eqref{inv} yields~\eqref{pure Schur 1 eq}.
\end{proof}

\begin{lemma} \label{pure Schur lemma 2}
Let $V_{ABC}$ be a tripartite pure QCM. Then
\bb
V_{AB}/V_B = \Omega_A^T (V_{AC}/V_C)^{-1} \Omega_A\, .
\label{pure Schur 2 eq}
\ee
\end{lemma}

\begin{proof}
Applying Lemma~\ref{pure Schur lemma 1} to the bipartite pure QCM $V_{AC:B}$ (i.e. the same matrix $V_{ABC}$ but seen as representing a pure QCM pertaining to the bipartite system $AC:B$), one obtains
\bbb
V_{ABC} / V_B = \Omega_{AC}^T V_{AC}^{-1} \Omega_{AC}\, .
\eee
Projecting onto the $A$ component thanks to the identities 
\begin{align*}
V_{AB}/V_{B} &= \Pi_A (V_{ABC}/V_B) \Pi_A^T\, ,\\
\Pi_A \Omega_{AC}^T V_{AC}^{-1} \Omega_{AC} \Pi_A^T &= \Omega_A^T (V_{AC}/V_C)^{-1} \Omega_A\, ,
\end{align*}
the latter following from the block inversion formula~\eqref{inv}, one deduces~\eqref{pure Schur 2 eq}.
\end{proof}

We then get the following for any tripartite quantum covariance matrix.

\begin{thm}[Schur complement of QCMs is monogamous]
\label{sch compl mon}
If $V_{ABC}\geq i \Omega_{ABC}$ is any tripartite QCM, then
\begin{equation}
V_{AC} / V_A \geq \Omega_C^T (V_{BC} / V_B)^{-1} \Omega_C\, . \label{INEQ 2}
\end{equation}
\end{thm}

\begin{proof}
Consider a symplectic purification $V_{ABCD}$ of the system $ABC$. Applying first~\eqref{INEQ 1} and then~\eqref{pure Schur 2 eq} yields~\eqref{INEQ 2}: $V_{AC}/V_A \geq V_{ACD}/V_{AD} = \Omega_C^T (V_{BC}/V_B)^{-1} \Omega_C$.
Alternatively, observe that the difference between right- and left-hand side of~\eqref{INEQ 2} is concave in $V_{ABC}$ (as $V_{AC}/V_A$ is concave and $(V_{BC}/V_B)^{-1}$ is convex), and it vanishes on pure QCMs by Lemma~\ref{pure Schur lemma 2}.
\end{proof}

The following special case of the above theorem will be useful later.

\begin{cor} \label{sch compl mon cor}
If $V_{AB}\geq i \Omega_{AB}$ is any bipartite QCM, then
\bb
V_{AB} / V_B \geq \Omega_A^T V_A^{-1} \Omega_A\, .
\label{sch compl mon cor eq}
\ee
\end{cor}

\begin{proof}
When the $B$ subsystem is uncorrelated with the rest, \eqref{INEQ 2} becomes $V_{AC}/V_A \geq \Omega_C^T V_{C}^{-1} \Omega_C$, which yields the claim after relabelling.
\end{proof}

We remark that the operator inequalities~\eqref{INEQ 1} and~\eqref{INEQ 2} are significantly stronger than the scalar ones~\eqref{SSA 6} and~\eqref{log det ineq} reported in~\cite{AdessoSerafini, Gross, Kor, Simon16}, as the former establish algebraic limitations directly at the level of covariance matrices, in a similar spirit to the marginal problem~\cite{tyc}, for arbitrary multipartite states.
These powerful tools will be instrumental in the investigation of quantum correlations, especially steering and entanglement, conducted in the forthcoming Subsection~\ref{subsec6 G steerability}.

\subsection{Mathematical detour: properties of the functions $\boldsymbol{g_\pm}$} \label{app prop g}

Before we delve into the study of the properties of the Gaussian steerability~\eqref{G steer}, we need to gain some knowledge of the basic properties of the functions $g_\pm$ defined in~\eqref{func g}.
A decisive ingredient of our analysis is a version of the Courant-Fischer-Weyl variational principle for symplectic eigenvalues proved recently in~\cite{bhatia15}.

First of all, observe that 
\bb
g_\pm(A)=g_\pm(SAS^T)\qquad \forall\ \text{symplectic $S$,}
\label{symplectic invariance g+-}
\ee
since those functions are defined only in terms of symplectic eigenvalues. Moreover, for all $A,B>0$ it holds that
\begin{align}
g_\pm (A^{-1}) &= g_\mp (A) \, , \label{elem prop g 1}\\
g_+(A)-g_-(A) &= \frac{1}{2}\,\log\det A\, , \label{elem prop g 2}\\
g_\pm (A\oplus B) &= g_\pm (A) + g_\pm (B) \, .  \label{elem prop g 3}
\end{align}
The next lemma will prove helpful many times.

\begin{lemma} \label{g- convex}
The function $g_-$ is monotonically non-increasing and convex, while $g_+$ is monotonically non-decreasing but neither convex nor concave.
\end{lemma}

\begin{proof}
The fact that $g_+,g_-$ are monotone in their inputs can be seen as an easy consequence of the symplectic equivalent of Weyl's monotonicity theorem first proved in~\cite[Lemma 2]{giedkemode} (see also~\cite[Theorem 8.15]{GOSSON}). This states that if $A,B>0$ are $2n\times 2n$ real matrices that satisfy $A\geq B$, then their ordered symplectic eigenvalues are such that $\nu_i (A)\geq \nu_i(B)$ for all $i=1,\ldots, n$. The claim follows by performing elementary algebraic manipulations.

Now, let us prove the convexity of $g_-(A)$. Our proof employs the recently found variational expression
\begin{equation}
\prod_{i=1}^k\, \nu_i(A)\ = \min \left\{ \sqrt{\det(S^T A S)}:\ S^T\Omega_{2n} S=\Omega_{2k} \right\}  \label{var expr}
\end{equation}
for the product of the $k$ smallest symplectic eigenvalues~\cite[Theorem 5]{bhatia15}. In the above expression, $\Omega_{2k}$ denotes the standard symplectic form on $k$ modes. We find
\begin{align}
g_-(A) &= \max_{1\leq k\leq n}\, \sum_{i=1}^k (-\log \nu_i(A)) \label{var expr g 0} \\
&= - \frac{1}{2}\, \min\left\{ \log\det(S^T A S):\ S^T\Omega_{2n} S=\Omega_{2k} ,\ 1\leq k\leq n \right\} . \label{var expr g}
\end{align}
Since $\log \det$ is well-known to be concave on positive matrices~\cite{logconcave}, and $F(x)\coloneqq \min_{y\in Y} f(x,y)$ is always concave in $x$ if $f(x,y)$ was concave in $x$ for all fixed $y\in Y$, we infer that $g_-$ is indeed convex. Finally, in order to see that $g_+$ is neither convex nor concave it suffices to test it on positive multiples of the identity.
\end{proof}

\begin{rem} \emph{Why Lemma~\ref{g- convex} does not imply that the logarithmic negativity is convex}. The formula $E_N = g_-(\Theta V_{AB} \Theta)$ for the logarithmic negativity, discussed in Subsection~\ref{subsec6 G steerability}, together with the convexity of $g_-$, could lead us to think that the logarithmic negativity is convex in the input state, which is known to be false~\cite{negativity, plenioprl}. The reason why this reasoning does not work is that $E_N$ is expressible in terms of the covariance matrix only for Gaussian states, that do not constitute a convex set. However, it is true that if $\{\rho_i\}_i$ is a family of Gaussian states such that their convex combination $\sum_i p_i \rho_i$ is again Gaussian, then
\begin{equation} E_N \bigg(\sum_i p_i \rho_i \bigg) \leq \sum_i p_i E_N(\rho_i)\, . \end{equation}
One could dub this property \textbf{Gaussian-convexity}. The logarithmic negativity is thus an example of a Gaussian-convex function which is in general non-convex.
\end{rem}

Lemma~\ref{g- convex} can be used to prove that $g_-(V_{AB})$ decreases if the coherences between subsystems $A$ and $B$ are erased.

\begin{lemma}
\label{dec red g-}
Let $V_{AB}>0$ be a bipartite positive definite matrix (not necessarily a QCM). Then
\begin{equation}
g_-(V_{AB}) \geq g_-(V_A) + g_-(V_B)\, . \label{dec red g- eq}
\end{equation}
\end{lemma}

\begin{proof}
We will give a straightforward proof based on the convexity of $g_-$, but an alternative argument can be deduced directly from the variational expression~\eqref{var expr g}. Observing that $\mathds{1}_A\oplus (-\mathds{1}_B)$ is a symplectic operation, and using the symplectic invariance of $g_-$~\eqref{symplectic invariance g+-}, one finds
\begin{equation}
g_-(V_{AB})\ =\ g_-\Big( \big(\mathds{1}_A\oplus (-\mathds{1}_B) \big) \, V_{AB}\, \big(\mathds{1}_A\oplus (-\mathds{1}_B) \big) \Big)\, ,
\end{equation}
from which we infer
\begin{align*}
g_-(V_{AB}) &= \frac{1}{2} \left( g_-(V_{AB}) + g_-\Big( \big(\mathds{1}_A\oplus (-\mathds{1}_B) \big) \, V_{AB}\, \big(\mathds{1}_A\oplus (-\mathds{1}_B) \big) \Big) \right) \\
&\geq g_-\left(\, \frac{1}{2}\, V_{AB} + \frac{1}{2}\, \big(\mathds{1}_A\oplus (-\mathds{1}_B) \big) \, V_{AB}\, \big(\mathds{1}_A\oplus (-\mathds{1}_B) \big)\, \right) \\
&= g_-(V_A\oplus V_B) \\
&= g_-(V_A) + g_-(V_B)\, .
\end{align*}
thanks to Lemma~\ref{g- convex}.
\end{proof}

\begin{rem}
One could be tempted to conjecture inequalities linking $g_-(V_{AB})$ with $g_-(V_A)$ and $g_-(V_{AB}/V_A)$. However, in general on the one hand $g_-(V_{AB})\ngeq g_-(V_A) + g_-(V_{AB}/V_A)$ (counterexample: bipartite QCM that is steerable from $A$ to $B$) and on the other hand $g_-(V_{AB})\nleq g_-(V_A) + g_-(V_{AB}/V_A)$ (numerical counterexamples are easily found).
\end{rem}

We have seen that the function $g_-$ admits a variational expression~\eqref{var expr g} in terms of a maximum (or the negative of a minimum). Now, we explore an alternative variational principle for $g_-$ that yields it directly as a minimum instead. The following lemma completes our toolbox.

\begin{lemma} \label{second var g+-}
The functions $g_\pm$ admit the following representations:
\begin{align}
g_+(W) &= \min_{W\leq\, Z\,\geq i\Omega} \ \frac{1}{2} \log\det Z\, , \label{second var g+} \\
g_-(W) &= \min_{\Omega^T W^{-1}\Omega \leq \, Z\, \geq i\Omega} \ \frac{1}{2} \log\det Z\, . \label{second var g-}
\end{align}
\end{lemma}

\begin{proof}
Since~\eqref{elem prop g 1} holds, we can limit ourselves to showing the first
The monotonicity of $g_+$ implies that 
\bbb
g_+(W)\leq g_+(Z)=\frac{1}{2}\log\det Z\, ,
\eee
whenever $W\leq Z \geq i\Omega$, where the last equality holds because $Z$ is a QCM. This shows that $g_+(W) \leq \min_{W\leq\, Z\,\geq i\Omega} \ \frac{1}{2} \log\det Z$. On the other hand, we can perform a Williamson decomposition~\eqref{Williamson} on $W$ and write it as $W=S\left(\begin{smallmatrix} \nu & 0 \\ 0 & \nu \end{smallmatrix}\right)S^T$, with $S$ symplectic and $\nu\geq \id$ diagonal. This allows us to construct the ansatz $\tilde{Z}\coloneqq S\left(\begin{smallmatrix} \tilde{\nu} & 0 \\ 0 & \tilde{\nu} \end{smallmatrix}\right) S^T$, where $\tilde{\nu}_i \coloneqq \max\{ \nu_i, 1 \}$, which satisfies $W \leq \tilde{Z}\geq i\Omega$ and
\begin{equation}
\frac{1}{2}\log\det \bar{Z} = \sum_i \max\{ 0,\, \log\nu_i(W)\} = g_+(W)\, ,
\end{equation}
hence we are done.
\end{proof}

\begin{rem}
We remind the reader that for any $W>0$ the condition $Z \geq \Omega^T W^{-1} \Omega$ is equivalent to
\begin{equation}
\begin{pmatrix} W & \Omega \\ \Omega^T & Z \end{pmatrix} \geq 0\, ,
\end{equation}
by virtue of Lemma~\ref{pos cond}.
\end{rem}

\subsection{Monotonicity of Gaussian steerability} \label{subsec6 G steerability}

We are now in position to employ the mathematical results obtained in Subsections~\ref{subsec6 Schur} and~\ref{app prop g} to our problem, that is, showing that the Gaussian steerability in~\eqref{G steer} is a valid monotone within the resource theory of Gaussian steering. The requirements that such a monotone should satisfy, as put forward in~\cite{Gallego}, are reported in the discussion at the end of Subsection~\ref{subsec6 resource}. Let us translate them into the Gaussian framework.
\begin{enumerate}[(i)]

\item In order for $\mathcal{G}(A\rangle B)$ to vanish on Gaussian unsteerable states, it must be that $\mathcal{G}(A\rangle B)_V =0$ whenever $V_{AB}/V_A \geq i\Omega_B$~\eqref{unsteerability Schur}. This is automatically guaranteed by construction, thanks to~\eqref{func g}.

\item The monotonicity under 1W-LOCC is more difficult. As we said at the end of Subsection~\ref{subsec6 resource}, the measure needs to be monotonically non-increasing under the transformation 
\bbb
\rho_{AB} \longmapsto \frac{(I\otimes \mathcal{E})(\rho_{AB})}{\Tr [\mathcal{E}(\rho_B)]}\, ,
\eee
where $\mathcal{E}$ is an arbitrary completely positive map on the $B$ system. Since here we care only about the Gaussian resource theory, we can assume $\mathcal{E}$ to be a Gaussian CP map, whose action is given by~\eqref{CP Gauss}.

\item We move on to the third, optional requirement, namely the convexity, or in our case Gaussian-convexity (see the Remark after Lemma~\ref{g- convex}). A good starting point is the observation that the QCM $\widebar{V}$ of a convex combination $\sum_i p_i \rho_i$ of states $\rho_i$ with covariance matrices $V_i$ satisfies~\cite{Werner01}
\bbb
\widebar{V}_i \geq \sum_i p_i V_i\, .
\eee
In particular, convexity in the quantum state is automatically guaranteed by: (iii.a) anti-monotonicity in the QCM, and (iii.b) convexity in the QCM.

\end{enumerate}

The following result shows that the Gaussian steerability measures meets all requirements (i)-(iii), hence fully validating it within the Gaussian resource theory of steering. This settles a question that had been left open in~\cite{steerability, Gallego}. As a by-product, we also establish further properties such as the monotonicity under non-deterministic Gaussian maps on the \emph{steering} party (Theorem~\ref{G prop}(c)). Our findings extend those of~\cite{steerability} considerably, for in~\cite{steerability} (some of) the stated properties were only proved in the special case of one-mode steered subsystem, i.e. $n_B=1$.

\begin{thm}[Properties of Gaussian steerability]
\label{G prop}
The Gaussian steerability measure, defined on Gaussian states by~\eqref{G steer}, satisfies the following properties:
\begin{enumerate}[(a)]
\item $\mathcal{G}(A\rangle B)_V$ is convex and non-increasing in the QCM $V_{AB}$, hence Gaussian-convex in the state $\rho_{AB}^G$, which meets condition (iii) above;
\item $\mathcal{G}(A\rangle B)$ is additive under tensor products, i.e.~under direct sums of QCMs:
\bb
\mathcal{G}(A_1 A_2\rangle B_1 B_2)_{V_{A_1B_1}\oplus W_{A_2 B_2}} =  \mathcal{G}(A_1\rangle B_1)_{V_{A_1B_1}} +  \mathcal{G}(A_2\rangle B_2)_{W_{A_2 B_2}}\, ;
\label{G additive}
\ee
\item $\mathcal{G}(A\rangle B)$ is non-increasing under general, non-deterministic Gaussian maps on the steering party $A$;
\item $\mathcal{G}(A\rangle B)$ is non-increasing under general, non-deterministic Gaussian maps on the steered party $B$, meeting requirement (ii) above;
\item for any QCM $V_{ABC}$, it holds $\mathcal{G}(A\rangle C)_V  \leq  g_+(V_{BC}/V_B)$.
\end{enumerate}
\end{thm}


Before delving into the proof of the above theorem, let us comment a bit on the various claims. As mentioned before, (a) and (d) serve the main purpose of the section, i.e. showing that the Gaussian steerability measure~\eqref{G steer} is a valid monotone of the Gaussian resource theory of steering. Identity~\eqref{G additive} in (b) is desirable -- even if not mandatory -- for a well-behaved measure of quantum correlations. As a matter of fact, most standard measures of entanglement do not satisfy it (a notable exception is the squashed entanglement~\eqref{squashed eq}, whose additivity was shown in~\cite{squashed}).

As for claim (c), one could expect monotonicity under \emph{Gaussian channels} on the steering party, because of the following reasoning. Taking a quantum assemblage as in~\eqref{quantum assemblage} and applying before the measurement a quantum channel $\mathcal{N}$ on $A$ one gets 
\bbb
\tr_A \left[ E^x_a \otimes \id\, (\mathcal{N}\otimes I)(\rho_{AB}) \right] = \tr_A \left[ \mathcal{N}^\dag (E^x_a) \otimes \id\, \rho_{AB} \right] ,
\eee
where the adjoint map $\mathcal{N}^\dag$ is defined by $\Tr [X \mathcal{N}(Y)] = \Tr [\mathcal{N}^\dag(X) Y]$, exactly as in~\eqref{adjoint}. The operators $\mathcal{N}^\dag(E^x_a)$ form a valid measurement because $\mathcal{N}$ is trace-preserving. If $\{E^x_a\}_{a}$ covers over all Gaussian measurements when $x$ runs over its range, and $\mathcal{N}$ is a Gaussian channel, then $\big\{\mathcal{N}^\dag(E^x_a)\big\}_a$ constitutes a subset of all Gaussian measurements. Naturally, we do not want the steerability to increase when the set of measurements that are available to Alice is \emph{smaller}, which leads to (c).

Claim (e) establishes an upper bound that can be useful in practical situations. An example of such scenario would feature a malicious agent $A$ who is trying to steer system $C$. Thanks to inequality (e), $B$ and $C$ can estimate the maximum steerability $A$ can achieve by measuring properties of their covariance matrix $V_{BC}$ alone.

\begin{proof}[Proof of Theorem~\ref{G prop}]
We prove the claims one by one.
\begin{enumerate}[(a)]
\item \emph{$\mathcal{G}(A\rangle B)_V$ is convex and non-increasing as a function of the QCM $V_{AB}>0$.} \\[0.5ex]
Both properties follow straightforwardly by combining concavity and monotonicity of the Schur complement (deduced from Corollary~\ref{Schur variational cor}) with the convexity of $g_-$ (Lemma~\ref{g- convex}). Let us prove convexity for instance. Since the Schur complement is concave, for any $V_{AB},W_{AB}>0$ and $0\leq p\leq 1$ we obtain
\begin{equation*}
(pV_{AB}+(1-p)W_{AB}) \big/ (pV_A+(1-p) W_A) \geq p\, V_{AB}/V_A + (1-p)\, W_{AB}/W_A\, .
\end{equation*}
Applying the fact that $g_-$ is monotonically non-increasing and convex gives
\begin{align*}
\mathcal{G}(A\rangle B)_{pV_{AB}+(1-p)W_{AB}} &= g_- \big( (pV_{AB}+(1-p)W_{AB}) \big/ (pV_A+(1-p) W_A) \big) \\ 
&\leq g_- \big( p V_{AB}/V_A + (1-p) W_{AB}/W_A\big) \\
&\leq\ p\, g_-(V_{AB}/V_A) + (1-p)\, g_- (W_{AB}/W_A) \\ 
&= p\, \mathcal{G}(A\rangle B)_V + (1-p)\, \mathcal{G}(A\rangle B)_W \, . 
\end{align*}

\item \emph{$\mathcal{G}(A\rangle B)$ is additive under tensor products.} \\[0.5ex]
Elementary, since
\begin{align*}
\mathcal{G}(A_1A_2\rangle B_1B_2)_{V_{A_1B_1}\oplus W_{A_2 B_2}} &= g_-\left( (V_{A_1B_1}\oplus W_{A_2 B_2}) \big/ (V_{A_1}\oplus W_{A_2}) \right) \\
&= g_-\left( V_{A_1B_1} / V_{A_1} \oplus W_{A_2 B_2}/ W_{A_2} \right)\, \\
&= g_-\left( V_{A_1B_1} / V_{A_1}\right) + g_-\left(W_{A_2 B_2} / W_{A_2} \right) \\ 
&= \mathcal{G}(A_1\rangle B_1)_{V_{A_1B_1}} + \mathcal{G}(A_2\rangle B_2)_{W_{A_2 B_2}}\, ,
\end{align*}
where the third equality follows from~\eqref{elem prop g 3}.

\item \emph{$\mathcal{G}(A\rangle B)$ is monotonically non-increasing under general, non-deterministic Gaussian maps on the steering party $A$.} \\[0.5ex]
Using the monotonicity of the Schur complement under general Gaussian maps, as expressed by Theorem~\ref{CP incr sch}, one gets
\begin{equation*}
\Gamma_{A\rightarrow A'} (V_{AB}) \big/ \Gamma_{A\rightarrow A'}(V_A) \geq V_{AB} / V_A\, .
\end{equation*}
Applying $g_-$ to both sides and reversing the inequality as prescribed by Lemma~\ref{g- convex}, we obtain
\begin{equation}
\mathcal{G}(A'\rangle B)_{\Gamma_{A\rightarrow A'}(V_{AB})} \leq \mathcal{G}(A\rangle B)_{V_{AB}} \, ,
\end{equation}
which is the claim.

\item \emph{$\mathcal{G}(A\rangle B)$ is monotonically non-increasing under general, non-deterministic Gaussian maps on the steered party $B$.} \\[0.5ex]
This is the most difficult claim to prove. First of all, we recall that any general, non-deterministic Gaussian map can always be obtained by: (i) adding an uncorrelated ancillary system in a Gaussian state; (ii) performing a global symplectic operation; and (iii) measuring some of the modes by means of a Gaussian measurement. 
Clearly, $\mathcal{G}(A\rangle B)$ is invariant under the addition of an ancillary steered system in an uncorrelated state because of the additivity proved in point (b). Furthermore, the invariance under symplectic operations on $B$ is guaranteed by~\eqref{symplectic invariance g+-}. Thus, we are only left to prove that the Gaussian steerability decreases when a partial Gaussian measurement is performed on the steered system.

According to~\eqref{G measurement QCM Schur}, given a composite system $ABC$ in a Gaussian state with QCM $V_{ABC}$, when one makes a Gaussian measurement on the subsystem $C$ that has seed $\gamma_C$, the reduced post-measurement state of subsystem $AB$ is Gaussian and with a QCM given by $\tilde{V}_{AB}=(V_{ABC}+\gamma_C)/(V_C+\gamma_C)$ (independently of the outcome), where we adopted the shorthand notation $\gamma_C = 0_{AB} \oplus \gamma_C$. Bearing that in mind, we are claiming that for all QCMs $V_{ABC}$ one has
\begin{equation}
g_-\big( \tilde{V}_{AB} /\tilde{V}_A \big) \leq g_-(V_{ABC}/V_A)\, . \label{steerability decreases measurement}
\end{equation}
Call $W_{BC} \coloneqq V_{ABC}/V_A$. Then, a simple calculation that uses the quotient property of the Schur complement~\eqref{quotient property} shows that
\begin{align*}
\tilde{V}_{AB} /\tilde{V}_A &= \big( (V_{ABC}+\gamma_C) / (V_C+\gamma_C) \big)\, \big/\, \big( (V_{AC}+\gamma_C) / (V_C+\gamma_C) \big) \\[0.4ex]
&= (V_{ABC}+\gamma_C) / (V_{AC}+\gamma_C)\, \\[0.4ex]
&= \big( (V_{ABC}+\gamma_C)/V_A \big)\, \big/\, \big( (V_{AC}+\gamma_C)/V_A \big) \\[0.4ex] 
&= \big( V_{ABC}/V_A + \gamma_C \big)\, \big/\, \big( V_{AC}/V_A+\gamma_C \big) \\[0.4ex]
&= (W_{BC}+\gamma_C) / (W_C+\gamma_C)\, .
\end{align*}
Thus,~\eqref{steerability decreases measurement} takes the form
\begin{equation}
g_-\big((W_{BC}+\gamma_C)/(W_C+\gamma_C)\big) \leq g_-\left( W_{BC} \right)\, , \label{steerability decreases measurement 2}
\end{equation}
to be proved for all $W_{BC}> 0$. Now, since the measured matrix $(W_{BC}+\gamma_C)/(W_C+\gamma_C)$ is concave in $\gamma_C$, and $g_-$ is decreasing and convex by Lemma~\ref{g- convex}, we can restrict ourselves to prove inequality~\eqref{steerability decreases measurement 2} only in the case in which $\gamma_C$ is a pure QCM, i.e. when it is symplectic.

Now we apply the above Lemma~\ref{second var g+-} (together with the remark immediately below it). Suppose we found a matrix $Z_{BC}\geq i\Omega_{BC}$ such that
\begin{equation}
\begin{pmatrix} W_{BC} & \Omega_{BC} \\ \Omega_{BC}^T & Z_{BC} \end{pmatrix} \geq 0\, ,\qquad \frac{1}{2}\log\det Z_{BC} = g_-(W_{BC})\, . \label{global matrix 1}
\end{equation}
Then, consider the matrix
\begin{equation}
\begin{pmatrix} 0_B\oplus \gamma_C & 0_B \oplus \Omega_C^T \\[1ex] 0_B\oplus \Omega_C & 0_B \oplus \gamma_C \end{pmatrix} \geq 0\, , \label{global matrix 2}
\end{equation}
where the last inequality holds because $\gamma\geq \Omega^T\gamma^{-1}\Omega$ for all $\gamma\geq i\Omega$, as follows from Lemma~\ref{QCM geom lemma}(b). Adding~\eqref{global matrix 2} to~\eqref{global matrix 1} we get
\begin{equation}
\begin{pmatrix} W_{BC}+\gamma_C & \Omega_B\oplus 0_C \\[1ex] \Omega_B^T\oplus 0_C & Z_{BC}+\gamma_C \end{pmatrix} \geq 0\, . \label{global matrix 3}
\end{equation}
Taking the Schur complement with respect to the two $C$ components, thanks to the two crucial zero blocks we have just formed, we obtain
\begin{equation}
\begin{pmatrix} (W_{BC}+\gamma_C) / (W_C+\gamma_C) & \Omega_B \\[1ex] \Omega_B^T & (Z_{BC}+\gamma_C) / (Z_C+\gamma_C) \end{pmatrix} \geq 0\, .
\label{global matrix 4}
\end{equation}
Remarkably, since $Z_{BC}\geq i\Omega_{BC}$ one finds easily $(Z_{BC}+\gamma_C) / (Z_C+\gamma_C)\geq i\Omega_B$. Therefore, the same Lemma~\ref{second var g+-} gives us
\begin{equation}
g_-\left( (W_{BC}+\gamma_C) / (W_C+\gamma_C) \right) \leq \frac{1}{2} \log \det (Z_{BC}+\gamma_C) / (Z_C+\gamma_C)\, .
\end{equation}
The proof is ended once we show that
\begin{equation}
\det (Z_{BC}+\gamma_C) / (Z_C+\gamma_C) \leq \det Z_{BC}\, ,
\label{determinant decreases pure POVM}
\end{equation}
for all QCMs $Z_{BC}\geq i\Omega_{BC}$ and for all pure QCMs $\gamma_C$. This rather surprising fact is hard to prove at the level of QCMs, but it becomes more tractable once we go back to the Hilbert space picture behind. This can be done thanks to the identity $\Tr [\rho^2 ]=1\big/\!\sqrt{\det V}$, relating the \emph{purity} $\Tr[\rho^2]$ of a Gaussian state $\rho$ to the determinant of its QCM $V$~\cite[\S 3.5]{BUCCO}. Such an identity allows us to restate~\eqref{determinant decreases pure POVM} as the claim that \emph{purity of Gaussian states increases when pure Gaussian measurements are applied.}

Let us now translate also the measurement into the Hilbert space picture. Since $\gamma_C$ is pure, the Gaussian measurement will be represented by a collection of rank-one (unnormalized) operators $\{ \psi(t)_C\}_t$ such that $\int \frac{d^{2n_C} t}{(2\pi)^{n_C}} \psi(t)_C = \id_C$ (Subsection~\ref{subsec5 G states}). Furthermore, it follows from~\eqref{G measurement}, \eqref{G measurement probability}, \eqref{G measurement QCM}, and~\eqref{G measurement displacement} that the outcomes of this measurement on a Gaussian state $\rho_{BC}$ with covariance matrix $Z_{BC}$ are Gaussian states of the form $\tilde{\rho}(t)_B=U(t)^\dag \tilde{\rho}_B U(t)$, where $U(t)$ are displacement unitaries depending on $t$ and $\tilde{\rho}_B$ is a Gaussian state independent of $t$ with QCM $(Z_{BC}+\gamma_C)/(Z_C+\gamma_C)$. With these hypotheses, we now see that Lemma~\ref{rk-1 POVM lemma} below allows us to conclude that $\Tr [\tilde{\rho}_B^2] \geq \Tr [\rho_{BC}^2]$, which immediately yields~\eqref{determinant decreases pure POVM} since both $\tilde{\rho}_B$ and $\rho_{BC}$ are Gaussian states.

\item \emph{The upper bound $\mathcal{G}(A\rangle C)_V \leq g_+(V_{BC}/V_B)$ holds for any QCM $V_{ABC}$.} \\[0.5ex]
Taking~\eqref{INEQ 2}, applying $g_-$ thanks to Lemma~\ref{g- convex}, and using the symplectic invariance of $g_-$~\eqref{symplectic invariance g+-} yields exactly
\begin{equation}
\mathcal{G}(A\rangle C) = g_-(V_{AC}/V_A) \leq g_+(V_{BC}/V_B)\, .
\end{equation}
\end{enumerate}
\end{proof}

In dealing with point (d) of Theorem~\ref{G prop} above, we used some unproven property of Gaussian measurements with pure seeds. We now clarify this issue by stating two lemmas which complete the above argument.

\begin{lemma} \label{x x_dag lemma}
Let
\begin{equation}
\begin{pmatrix} A & X \\ X^\dag & B \end{pmatrix} \geq 0
\end{equation}
be a positive definite block matrix. Then
\begin{equation}
\|X\|_2^2 \leq \|A\|_2 \|B\|_2\, ,
\label{x x_dag eq}
\end{equation}
where $\|M\|_2\coloneqq \sqrt{\text{\emph{Tr}}\, M^\dag M}$ denotes the Hilbert-Schmidt norm.
\end{lemma}

\begin{proof}
One writes 
\begin{align*}
\|X\|_2^2 &= \Tr XX^\dag \\
&= \Tr A\, A^{-1/2} XX^\dag A^{-1/2} \\
&\textleq{(1)} \|A\|_2\, \| A^{-1/2} XX^\dag A^{-1/2}\|_2 \\
&= \|A\|_2\, \sqrt{\Tr A^{-1/2} XX^\dag A^{-1} XX^\dag A^{-1/2}} \\
&= \|A\|_2\, \sqrt{\Tr \left(X^\dag A^{-1} X \right)^2} \\
&= \|A\|_2\, \|X^\dag A^{-1} X\|_2 \\
&\textleq{(2)} \|A\|_2\, \|B\|_2\, .
\end{align*}
The justification of the steps is as follows: (1) is just the Cauchy-Schwartz inequality for the Hilbert-Schmidt product; (2) follows from the inequality $X^\dag A^{-1}X\leq B$, deduced from Lemma~\ref{pos cond}, and from the fact that the Hilbert-Schmidt norm is an increasing function on positive matrices.
\end{proof}

\begin{lemma} \label{rk-1 POVM lemma}
Let $\{ \psi^i_C \}_i$ be a measurement composed of multiples of rank-one projectors. Assume that the outcomes of this measurement when performed on a bipartite system $BC$ in a state $\rho_{BC}$ are always unitarily equivalent to a fixed density operator on the remaining system $B$, i.e.
\begin{equation}
\text{\emph{Tr}}_C \left[ \id_B \otimes \psi^i_C\, \rho_{BC}\right] = p_i\, U_i^\dag \tilde{\rho}_B U_i \qquad \forall\ i\, .
\end{equation}
Then
\begin{equation}
\text{\emph{Tr}}\, [\tilde{\rho}_B^2 ]\geq \text{\emph{Tr}}\, [\rho_{BC}^2]\, . \label{rk-1 POVM eq}
\end{equation}
\end{lemma}

\begin{proof}
For the sake of brevity, in what follows we suppose that $i$ is an index running over a finite alphabet, but the argument below extends straightforwardly to the more general case in which it belongs to a measurable space.
Since the identity
\begin{equation}
\tilde{\rho}_B =\frac{1}{p_i} U_i \tr_C \left[ \id_B \otimes \psi^i_C\, \rho_{BC}\right] U_i^\dag\, ,
\end{equation}
is valid for all indices $i$, we obtain
\begin{align}
\Tr [\tilde{\rho}_B^2] &= \|\tilde{\rho}_B\|_2^2 \nonumber \\
&= \bigg( \sum_i p_i \,\|\tilde{\rho}_B\|_2 \bigg)^2 \nonumber \\
&= \bigg(\sum_i \, \Big\| \tr_C \left[ \id_B \otimes \psi^i_C\, \rho_{BC}\right] \Big\|_2\ \bigg)^2 \nonumber \\
&= \sum_{ij} \,\Big\| \tr_C \big[ \id_B \otimes \psi^i_C\, \rho_{BC}\big] \Big\|_2\, \Big\| \tr_C \big[ \id_B \otimes \psi^j_C\, \rho_{BC}\big] \Big\|_2\, , \label{rk-1 POVM eq1}
\end{align}
where we omitted the subscripts $C$ for the sake of brevity. Now, using the notation $\psi^i = \ket{\psi^i}\!\!\bra{\psi^i}$ (with $\ket{\psi^i}$ not necessarily normalised) consider the map $\mathcal{N}_{C \rightarrow C'}$ from $C$ to a new system $C'$ as defined by
\begin{equation}
\mathcal{N}(X) \coloneqq \bigg( \sum_i \ket{i}\!\!\bra{\psi^i} \bigg)\, X\, \bigg( \sum_j \ket{j}\!\!\bra{\psi^j} \bigg)^\dag = \sum_{ij} \ket{i}\!\!\bra{\psi^i} X \ket{\psi^j}\!\!\bra{j}\, .
\end{equation}
Obviously, $\mathcal{N}$ is completely positive and trace-preserving. In particular, we deduce that $\left(I_B\otimes \mathcal{N}_{C\rightarrow C'}\right)(\rho_{BC})\geq 0$ is a legitimate quantum state. This latter density matrix has blocks indexed by $i,j$ and given by 
\bbb
\tr_C \left[ \id_B \otimes \ket{\psi^j}\!\!\bra{\psi^i}_C\, \rho_{BC} \right] .
\eee
Thanks to Lemma~\ref{x x_dag lemma}, we know that for all $i\neq j$,
\begin{align*}
&\Big\| \tr_C \left[ \id_B \otimes \ket{\psi^j}\!\!\bra{\psi^j}_C\, \rho_{BC} \right] \Big\|_2^2 \\
&\qquad \leq \Big\| \tr_C \big[ \id_B \otimes \psi^i_C\, \rho_{BC} \big] \Big\|_2 \, \Big\| \tr_C \big[ \id_B \otimes \psi^j_C\, \rho_{BC} \big] \Big\|_2\, .
\end{align*}
This is also trivially true when $i=j$. Therefore,
\begin{align*}
&\sum_{ij} \,\Big\| \tr_C \big[ \id_B \otimes \psi^i_C\, \rho_{BC}\big] \Big\|_2\, \Big\| \tr_C \big[ \id_B \otimes \psi^j_C\, \rho_{BC}\big] \Big\|_2 \\
&\qquad \geq \sum_{ij} \,\Big\| \tr_C \left[ \id_B \otimes \ket{\psi^j}\!\!\bra{\psi^i}_C\, \rho_{BC} \right] \Big\|_2^2 \\
&\qquad = \sum_{ij} \tr_{BC} \left[ (\id_B \otimes \psi^i_C)\, \rho_{BC} \, (\id_B \otimes \psi^j_C) \, \rho_{BC} \right] \\
&\qquad = \Tr [\rho_{BC}^2]\, , 
\end{align*}
where we used the normalization condition $\sum_i \psi^i_C = \id_C$. Inserting the above inequality into~\eqref{rk-1 POVM eq1} yields the claim~\eqref{rk-1 POVM eq}.
\end{proof}

\begin{rem}
By following the above argument, we see that all claims of Theorem~\ref{G prop} with the exception of (d) hold for arbitrary continuous variable states, not necessarily Gaussian. On the other hand, our current proof of the monotonicity of $\mathcal{G}(A \rangle B)$ under general, non-deterministic Gaussian maps on $B$ relies on the specification to Gaussian states of $AB$. We leave it as an open problem whether a more general proof of part (d) could be obtained, valid even for non-Gaussian states.
\end{rem}

\subsection{Monogamy of Gaussian steerability} \label{subsec6 G steerability monogamy}

Our framework allows us to address the general problem of the monogamy of $\mathcal{G}(A \rangle B)$. For a state with QCM $V_{AB_1\ldots B_k}$, consider the following inequalities
\begin{align}
\mathcal{G}(A\rangle B_1\ldots B_k) &\geq {\sum}_{j=1}^k \mathcal{G}(A\rangle B_j) \label{mon steer 1}\,,\\
\mathcal{G}(B_1\ldots B_k \rangle A) &\geq {\sum}_{j=1}^k \mathcal{G}(B_j\rangle A) \label{mon steer 2}\,.
\end{align}
In a very recent study~\cite{MonSteer}, both inequalities were proved in the special case of a $(k+1)$-mode system with one single mode per party, i.e. $n_A=n_{B_j}=1$ ($j=1,\ldots,k$). We now show that  only one of these constraints holds in full generality.

\begin{thm}[Monogamy of Gaussian steerability] \label{G mono}
(a) Inequality~\eqref{mon steer 1} holds for any multimode QCM $V_{A B_1 \ldots B_k}$. (b) Inequality~\eqref{mon steer 2} holds for any multimode QCM $V_{A B_1 \ldots B_k}$ such that either $A$ comprises a single mode ($n_A=1$), or $V_{A B_1 \ldots B_k}$ is pure, but can be violated otherwise.
\end{thm}
\begin{proof} (a) It suffices to prove the inequality $\mathcal{G}(A\rangle BC)\, \geq\, \mathcal{G}(A\rangle B) + \mathcal{G}(A\rangle C)$ for a tripartite QCM $V_{ABC}$, as~\eqref{mon steer 1} would follow by iteration. Observe that $V_{AB}/V_A$ and $V_{AC}/V_A$ form the diagonal blocks of the bipartite matrix $V_{ABC}/V_A$. Then, applying~\eqref{dec red g- eq} yields
\begin{align*}
\mathcal{G}(A\rangle BC)_V &= g_-(V_{ABC}/V_A) \\
&\geq g_-(V_{AB}/V_A) + g_-(V_{AC}/V_A) \\
&= \mathcal{G}(A\rangle B)_V + \mathcal{G}(A\rangle BC)_V\, ,
\end{align*}
which concludes the proof. (b) For the case $n_A=1$ with $n_{B_j}$ arbitrary, one exploits the fact that only one term $\mathcal{G}(B_j\rangle A)$ in the right-hand side of~\eqref{mon steer 2} can be nonzero, due to the impossibility of jointly steering a single mode by Gaussian measurements as implied by~\eqref{log det ineq} (see the original papers~\cite{Kor, Simon16}), combined with the monotonicity of $\mathcal{G}(B_1\ldots B_k \rangle A)$ under partial traces on the steering party as implied by Theorem~\ref{G prop}. Finally, the case when $V_{A B_1 \ldots B_k}$ is pure follows from the forthcoming Corollary~\ref{E mono}, namely by combining the last claim of Theorem~\ref{I2E} with~\eqref{mon E2}:
\begin{align*}
\mathcal{G}(B_1\ldots B_k \rangle A)_V &= E^G_{F,2} (B_1\ldots B_k:A)_V \\
&\geq \sum_{j=1}^k E^G_{F,2} (B_j:A)_V \\
&\geq \sum_{j=1}^k \mathcal{G}(B_j\rangle A)_V\, ,
\end{align*}
where the first equality holds specifically for pure states.
\end{proof}

The Gaussian steerability is thus not monogamous with respect to a common steered party $A$ when the latter is made of two or more modes, with violations of~\eqref{mon steer 2} existing already in a tripartite setting ($k=2$) with $n_{B_1}=n_{B_2}=1$ and $n_A=2$; a counterexample is reported in Appendix~\ref{app Schur}. 
What is truly monogamous is the log-determinant of the Schur complement, which only happens to be directly linked to the function $g_-$ when $n_A=1$.

\section{Correlations hierarchy in Gaussian states} \label{sec6 G corr hierarchy}

The previous section was concerned with the problem of quantifying steering in an appropriate way. For bipartite Gaussian states, we found that a suitable quantifier based directly on the covariance matrix does the job. In this section, we look at the more general problem of measuring other kinds of correlations (both classical and quantum) by looking at the QCM. In Subsection~\ref{subsec6 corr} we provide a brief introduction to an axiomatic framework to deal with the problem that was put forward in~\cite{LiLuo}. In the most favourable case, such framework requires a strong constraint to be met, namely that the quantum part of the correlations should not exceed \emph{half} of the total. After having introduced R\'enyi-2 Gaussian quantifiers in Subsection~\ref{subsec6 R2 GEoF}, in the following Subsection~\ref{subsec6 G ent hierarchy} we proceed to show that bipartite Gaussian states do satisfy this strong inequality when their correlations are measured with said R\'enyi-2 measures. This observation is the content of Theorem~\ref{I2E}, which constitutes the main result of the section. Finally, Subsection~\ref{subsec6 applications} presents some applications of this result, most importantly concerning the monogamy of the R\'enyi-2 Gaussian entanglement of formation (Corollary~\ref{E mono}).

\subsection{Correlations in bipartite quantum systems} \label{subsec6 corr}

Given a bipartite quantum state $\rho_{AB}$, can we draw a line between its classical and its quantum correlations? How to measure the two, and what requirements to impose on candidate quantifiers? These questions have been debated extensively in the literature~\cite{Vedral-1, Vedral-2, VV-correlations, LiLuo}, and we do not attempt to give a full account of the discussion here. However, for the sake of the presentation, let us summarise briefly the point of view of~\cite{LiLuo}. 

The \textbf{quantum mutual information} is given by
\bb
I(A:B)_\rho \coloneqq S(\rho_{A}) + S(\rho_B) - S(\rho_{AB}) \, ,
\label{mutual info}
\ee
where $S(\rho)\coloneqq -\Tr [\rho \log \rho]$ is the von Neumann entropy given by~\eqref{entropy}.\footnote{This is nothing but a special case of the already mentioned conditional mutual information~\eqref{cond mutual info}, which reduces to~\eqref{mutual info} for a decorrelated $C$.} As it turns out,~\eqref{mutual info} is an excellent candidate for a measure of total correlations, because of its operational significance. In fact, it has been shown that the mutual information quantifies the maximal amount of information that can be sent securely using $\rho_{AB}$ in a one-time-pad cryptographic scheme~\cite{Schumacher-correlations}, and also the minimal amount of noise that is needed to erase the correlations of $\rho_{AB}$~\cite{VV-correlations}. In view of all these considerations, we will take~\eqref{mutual info} as our measure of total correlations.

Now, assume that we can always divide the correlations in $I(A:B)$ between classical and quantum. That is, assume that we can come up with two functions $C(\rho)$ and $Q(\rho)$, which measure respectively the classical and quantum correlations, and such that $C+Q=I(A:B)$. As argued in~\cite{LiLuo}, a reasonable assumption to make is that the inequality 
\bb
Q(\rho) \leq C(\rho)
\label{Q<C}
\ee
holds for all states $\rho$. Since $Q+C=I(A:B)$,~\eqref{Q<C} is equivalent to
\bb
Q(\rho) \leq \frac12 I(A:B)_\rho\, .
\label{Q<I/2}
\ee
An intuitive argument in favour of taking~\eqref{Q<C} (or~\eqref{Q<I/2}) as a postulate goes as follows~\cite{LiLuo}.

First, examine its status for pure states. On the one hand, the quantum correlations contained in a pure state $\ket{\psi}_{AB}$ should be measured by its entanglement, and virtually any measure of entanglement is commonly required to reduce to the \textbf{entanglement entropy} $S\big( \tr_B \,\psi_{AB} \big)$ on pure states~\cite[\S 15.6]{GeometryQuantum}. Therefore, we can take as granted that $Q(\psi) = S\big( \tr_B \,\psi_{AB} \big)$ holds on pure states. On the other hand, the classical correlations as measured for instance by the \emph{classical mutual information}, i.e. by the maximal mutual information that is obtainable by local measurements, are also known to coincide with the entanglement entropy for pure states~\cite{locking}. This leads us to the assumption that $Q(\psi)=C(\psi)$ holds for pure states.

Mixed states are naturally obtained as classical mixtures of pure states. Classical mixing can increase $C$ but it should never increase the quantum correlations $Q$. Given that the two are equal on pure states, inequality~\eqref{Q<C} seems like a reasonable expectation for general mixed states. This matches also the intuition of the authors of~\cite{LiLuo, VV-correlations}. And in fact, many entanglement measures $Q$ do satisfy~\eqref{Q<I/2}, most notably the squashed entanglement~\eqref{squashed eq} we already encountered in Chapter~\ref{chapter3}. However, some other, equally important entanglement monotones, such as the entanglement of formation, do not obey~\eqref{Q<I/2}~\cite{LiLuo}. Even worse, the entanglement of formation can even exceed the quantum mutual information itself, as observed in~\cite{aspects-generic-entanglement, LiLuo}.

We do not take a firm position on the status of~\eqref{Q<I/2} as a fundamental assumption. Indeed, we do not see any a priori clear reason why there should be a way to divide the total correlations into a classical and a quantum part. On the contrary, it could well be that they are inextricably linked and that any attempt of separating them will lead eventually, \emph{in the most general case}, to a contradiction. Instead, here we regard~\eqref{Q<I/2} more as a \emph{desirable} property that we can hope to find to hold in some special cases, i.e. for some restricted sets of states and for suitable measures of correlations. This is exactly what we will establish in the next subsection: \emph{for Gaussian states, quantifiers based on the R\'enyi-2 entropy do satisfy the inequality~\eqref{Q<I/2}}, and can thus be regarded -- from this point of view -- as defining a desirable framework for the problem.



\subsection{R\'enyi-2 Gaussian entanglement of formation} \label{subsec6 R2 GEoF}

The aim of this subsection is to provide the reader with an introduction to R\'enyi-2 quantifiers for Gaussian states, extensively discussed in~\cite{AdessoSerafini, adesso14}.

We start by recalling the notion of \textbf{quantum R\'enyi-$\boldsymbol{\alpha}$ entropy} of a state $\rho$, given by
\begin{equation}
S_\alpha(\rho) \coloneqq \frac{1}{1-\alpha} \log \Tr[ \rho^{\alpha}]\, .
\label{Renyi ent}
\end{equation}
It is not hard to show that~\eqref{Renyi ent} reproduces the conventional von Neumann entropy in the limit $\alpha\rightarrow 1$.
As expected, the R\'enyi-$\alpha$ entropy becomes a function of the covariance matrix alone for Gaussian states. Adopting the shorthand notation $S_\alpha\big(\rho^G(V,w)\big)=S_\alpha(V)$, we have 
\begin{equation}
S_\alpha (V) = -\frac{1}{\alpha-1}\, \sum_{i=1}^n \log \frac{2^\alpha}{\big(\nu_i(V)+1\big)^\alpha-\big(\nu_i(V)-1\big)^\alpha}
\label{S_alpha}
\end{equation}
for $\alpha>1$, and
\begin{equation}
S_1(V) = \sum_{i=1}^n\left(\frac{\nu_i(V)+1}{2}\,\log\frac{\nu_i(V)+1}{2} - \frac{\nu_i(V)-1}{2}\,\log\frac{\nu_i(V)-1}{2}\right) \label{S_1}
\end{equation}
for the von Neumann case $\alpha=1$.
For details, see~\cite[Eq. (108) and (109)]{adesso14} or~\cite[Eq. (3.96)]{BUCCO}. Here we are mostly interested in the case $\alpha=2$, in which case we have
\begin{equation}
S_2(\rho) = \frac12 \log \det V
\label{Renyi-2 Gauss}
\end{equation}
for an arbitrary Gaussian state $\rho$ with QCM $V$.

R\'enyi entropies can be used to define a corresponding measure of entanglement. For a bipartite quantum state $\rho_{AB}$, the \textbf{R\'enyi-$\boldsymbol{\alpha}$ entanglement of formation} is constructed as the convex hull of the R\'enyi-$\alpha$ entropy of entanglement defined on pure states
\cite{Horodecki-review}, i.e.
\begin{equation}
\begin{split}
  E_{F,\alpha}(A:B)_{\rho}
    &\coloneqq \inf \sum\nolimits_i p_i \, S_{\alpha}\bigl(\psi^i_A\bigr) \\
           &\quad\text{ s.t. } \rho_{AB} = \sum\nolimits_i p_i\, \psi^i_{AB} ,
\end{split}
  \label{EoF}
\end{equation}
where $\psi^{i}_{AB}$ are density matrices of pure states, $\psi^i_A = \tr_B\, \psi^i_{AB}$ is the reduced state (marginal), and $S_\alpha$ is defined in~\eqref{Renyi ent}.

For quantum Gaussian states, an upper bound to this quantity can be derived by restricting the decompositions appearing in the above infimum
to be comprised of pure Gaussian states only. One obtains what is called \textbf{R\'enyi-$\boldsymbol{\alpha}$ Gaussian entanglement of formation}, a monotone under Gaussian local operations and classical communication, that in terms of
the QCM $V_{AB}$ of $\rho_{AB}$ is given by the simpler formula~\cite{Wolf03}
\begin{equation}
\begin{split}
  E^{G}_{F,\alpha}(A:B)_{V}
    &= \inf S_{\alpha}(\gamma_{A}) \\
    &\quad \text{ s.t. } \gamma_{AB} \text{ pure QCM and } \gamma_{AB}\leq V_{AB},
\end{split}
  \label{GEoF}
\end{equation}
where with a slight abuse of notation we denoted with $S_{\alpha}(W)$ the R\'enyi-$\alpha$ entropy of a Gaussian state with QCM $W$ as given by~\eqref{S_alpha}, and $\gamma_{AB}$ stands for the QCM of a pure Gaussian state. Observe that $E_{F,\alpha}(A:B)$ and also $E^G_{F, \alpha}$ are symmetric under the exchange of $A$ and $B$, since \emph{the entropies of the two local shares of a pure state are the same.}
Incidentally, it has been proved~\cite{EoF-symmetric-G, Giovadd} that for some two-mode Gaussian states the formula~\eqref{GEoF} reproduces exactly~\eqref{EoF}, i.e.~Gaussian decompositions in~\eqref{EoF} are globally optimal.

\begin{rem}
We wrote an infimum in~\eqref{GEoF}, but with little work one can show that this is in fact a minimum, because the set over which we are optimising is compact, and the objective function is clearly uniformly continuous (remember that $\det \gamma_A\geq 1$.
\end{rem}

The most commonly used $E_{F,\alpha}$ is the one corresponding to the von Neumann entropy, $\alpha=1$. However, we already saw that in the Gaussian setting the choice $\alpha=2$ is also natural, an intuitive reason being that it respects the quadratic nature of the states. More operationally motivated reasons will be discussed in the forthcoming Chapter~\ref{chapter7}. Thus, from now on we will focus on the case $\alpha=2$. Under this assumption, thanks to~\eqref{Renyi-2 Gauss} we see that~\eqref{GEoF} becomes
\begin{equation}
\begin{split}
  E^{G}_{F,2}(A:B)_{V}
    &= \inf \frac12 \log \det \gamma_{A} \\
    &\quad \text{ s.t. } \gamma_{AB} \text{ pure QCM and } \gamma_{AB}\leq V_{AB} .
\end{split}
  \label{G R2 EoF}
\end{equation}

We will find it convenient to rewrite the above equation in a slightly different form, that employs the \textbf{R\'enyi-2 mutual information}
\bb
I_M(A:B)_V \coloneqq \frac{1}{2}\, \log \frac{\det V_A \det V_B}{\det V_{AB}}\, ,
\label{I2}
\ee
a quantity that will be the subject of more thorough investigation in Chapter~\ref{chapter7}. Using the readily verified fact that $I_M(A:B)_\gamma = \log \det \gamma_A = \log \det \gamma_B$ when $\gamma_{AB}$ is the QCM of a pure state~\cite{adesso14} (essentially because $\det\gamma_{AB}=1$ and the two marginals have equal entropies), we obtain
\begin{equation}
\begin{split}
  E^{G}_{F,2}(A:B)_{V}
    &= \inf \frac12 I_M(A:B)_{\gamma} \\
    &\quad \text{ s.t. } \gamma_{AB} \text{ pure QCM and } \gamma_{AB}\leq V_{AB} .
\end{split}
  \label{G R2 EoF alt}
\end{equation}
Among the other things,~\eqref{G R2 EoF alt} has the advantage of being evidently faithful on Gaussian states~\cite{AdessoSerafini}. In other words, it becomes zero iff the condition~\eqref{sep eq} in Lemma~\ref{sep} is met, i.e. iff the Gaussian state with QCM $V_{AB}$ is separable.

\subsection{Gaussian entanglement and correlations hierarchy} \label{subsec6 G ent hierarchy}


In this subsection we prove that the inequality~\eqref{Q<I/2} holds for Gaussian states when $Q$ is chosen to be the R\'enyi-2 Gaussian entanglement of formation of~\eqref{G R2 EoF}. In fact, in Appendix~\ref{app Schur} we show that one could rather take $Q$ equal to \emph{any} R\'enyi-$\alpha$ Gaussian entanglement of formation as long as $\alpha \geq 2$. Our main result reads as follows.

\begin{thm}[Gaussian R\'enyi-2 correlations hierarchy]\label{I2E}
Let $V_{AB}$ be an arbitrary bipartite QCM. Then the R\'enyi-2 mutual information~\eqref{I2}, the R\'enyi-2 Gaussian entanglement of formation~\eqref{G R2 EoF}, and the Gaussian steerability~\eqref{G steer} satisfy
\begin{equation}
\frac12 I_M(A:B)_V \geq E^G_{F,2}(A:B)_V \geq \mathcal{G}(A \rangle B)_V\, . \label{I>2E}
\end{equation}
If $V_{AB}$ is pure, all the above three quantities coincide with the reduced R\'enyi-2 entropy $\frac{1}{2}\log\det V_A$.
\end{thm}

\begin{proof}
First of all, let us prove the rightmost inequality. Consider the optimal pure QCM $\gamma_{AB}\leq V_{AB}$ that saturates the infimum (actually, the minimum) in~\eqref{G R2 EoF}. Then write
\begin{align*}
E^G_{F,2}(A:B)_V &= \frac{1}{2}\log\det \gamma_B \\
&\texteq{(1)} g_+(\gamma_B) \\
&\texteq{(2)} g_-(\gamma_B^{-1}) \\
&\texteq{(3)} g_-(\Omega_B^T\gamma_B^{-1} \Omega_B) \\
&\texteq{(4)} g_-(\gamma_{AB}/\gamma_A) \\[0.6ex]
&= \mathcal{G}(A\rangle B)_\gamma \\
&\textgeq{(5)} \mathcal{G}(A\rangle B)_V\, .
\end{align*}
These steps can be justified as follows: (1) follows straight from the definition of $g_+$,~\eqref{func g}, together with the fact that $\gamma_B$ is a legitimate $QCM$ and hence all its simplectic eigenvalues are no smaller than $1$; (2) is an application of~\eqref{elem prop g 1}; (3) uses the symplectic invariance of $g_\pm$,~\eqref{symplectic invariance g+-}; (4) follows from Lemma~\ref{pure Schur lemma 1}; and (5) is a consequence of the anti-monotonicity of $\mathcal{G}(A\rangle B)_V$ in $V$, Theorem~\ref{G prop}(a).

We now move on to the proof of the leftmost inequality in~\eqref{I>2E}. The key step leverages the construction of a special pure state to use as an ansatz in~\eqref{G R2 EoF alt}. Consider $\gamma_V^\# = V\# (\Omega V^{-1}\Omega^T)$, which is a pure QCM by Lemma~\ref{QCM geom lemma}. Plugging $\gamma_V^\#$ into~\eqref{G R2 EoF alt} is possible again thanks to Lemma~\ref{QCM geom lemma}(c), which guarantees that the inequality $\gamma_V^\#\leq V$ holds. Doing this we obtain 
\begin{align*}
E^G_{F,2}(A:B)_V &\leq \frac{1}{2}\, \log\det (\gamma_{V}^\#)_A \\
&= \frac{1}{2}\, \log\det \left( \Pi_A \left(V_{AB}\#(\Omega_{AB} V_{AB}^{-1} \Omega_{AB}^T)\right) \Pi_A^T\right)\, .
\end{align*}
where we denoted by $\Pi_A$ the projector onto the $A$ component.
To proceed further, we employ the inequality~\eqref{geometric mean positive maps}~\cite[Theorem 3]{ando79}, with $\Lambda(\cdot)\mapsto \Pi_A(\cdot)\Pi_A^T$, $A\mapsto V_{AB}$ and $B\mapsto \Omega_{AB} V_{AB}^{-1} \Omega_{AB}^T$, finding
\begin{align*}
\Pi_A \left(V_{AB}\#(\Omega_{AB} V_{AB}^{-1} \Omega_{AB}^T)\right) \Pi_A^T &\leq V_A \# \left( \Pi_A (\Omega_{AB} V_{AB}^{-1} \Omega_{AB}^T) \Pi_A^T \right) \\[0.4ex]
&= V_A\# \left( \Omega_A (V_{AB}/V_B)^{-1} \Omega_A^T \right)\, ,
\end{align*}
where for the last step we used the formula~\eqref{inv} for the inverse of a $2\times 2$ block matrix. Inserting this operator inequality into the above upper bound for $E^G_{F,2}(A:B)_V$ we obtain
\begin{align*}
E^G_{F,2}(A:B)_V &\leq \frac{1}{2}\, \log\det \left( V_A\# \left( \Omega_A (V_{AB}/V_B)^{-1} \Omega_A^T \right)\right) \\
&= \frac{1}{4}\,\log\frac{\det V_A}{\det V_{AB}/V_B} \\[0.4ex]
&= \frac{1}{4}\, \log\frac{\det V_A \det V_B}{\det V_{AB}} \\
&= \frac{1}{2}\, I_M(A:B)_V\, ,
\end{align*}
and we are done.
\end{proof}

Intuitively, this proves that the involved measures quantitatively capture the general hierarchy of correlations~\cite{ABC} in arbitrary Gaussian states~\cite{adesso14}: the Gaussian steerability is generally smaller than the entanglement degree, which accounts for a portion of quantum correlations up to half the total ones.

Theorem~\ref{I2E} is a significant improvement over previous results. For instance, using~\cite[Eq.~(14) and~(17)]{AdessoSerafini} we deduce the weaker inequality $E^G_{F,2}(A:B) \leq I_M(A:B)$ (proved there only for two-mode Gaussian states). We will generalise the inequality~\eqref{I>2E} in the next chapter, where we will show that on the leftmost side one can put instead the R\'enyi-2 \emph{conditional} mutual information computed on \emph{any} tripartite Gaussian extension of the state (Theorem~\ref{thm I cond G R2 EoF}).

\subsection{Applications} \label{subsec6 applications}

An outstanding consequence of Theorem~\ref{I2E} is that the R\'enyi-2 measure of entanglement can now be proved monogamous for \emph{arbitrary} Gaussian states with any number of modes per party.

\begin{cor}[Monogamy of Gaussian R\'enyi-2 entanglement] \label{E mono}
The Gaussian R\'enyi-2 entanglement of formation~\eqref{G R2 EoF} is monogamous for any multipartite Gaussian state, i.e.
\begin{equation}
E^G_{F,2}(A: B_1\ldots B_k)_V \geq {\sum}_{j=1}^k E^G_{F,2}(A:B_j)_V \label{mon E2}\,.
\end{equation}
for all QCMs $V_{AB_1\ldots B_k}$.
\end{cor}

\begin{proof}  It suffices again to prove that 
\bbb
E^G_{F,2}(A: BC)_V \geq E^G_{F,2}(A: B)_V + E^G_{F,2}(A: C)_V
\eee
holds for any tripartite QCM $V_{ABC}$. In order to do this, take the pure QCM $\gamma_{ABC}\leq V_{ABC}$ that saturates the infimum in the definition of $E^G_{F,2}(A: BC)$, and notice that 
\begin{align*}
E^G_{F,2}(A: BC) &= \frac{1}{2} \log\det \gamma_A \\
&= \frac{1}{2} I_M(A:BC)_\gamma \\
&= \frac{1}{2} I_M(A:B)_\gamma  + \frac{1}{2} I_M(A:C)_\gamma\, ,
\end{align*}
where the last equality holds specifically for pure states, being ultimately a consequence of the fact that the two marginals of any pure state have equal R\'enyi entropies. Applying~\eqref{I>2E} to each of the two rightmost addends yields 
\begin{align*}
E^G_{F,2}(A: BC)_V &\geq E^G_{F,2}(A: B)_\gamma + E^G_{F,2}(A: C)_\gamma \\
&\geq E^G_{F,2}(A: B)_V + E^G_{F,2}(A: C)_V\, ,
\end{align*}
where the last step follows as $E^G_{F,2}$ is a non-increasing function of the QCM, which is clear already from the definition~\eqref{G R2 EoF}.
\end{proof}

Corollary~\ref{E mono} yields the \emph{most general} result to date regarding quantitative monogamy of continuous variable entanglement~\cite{adesso14}, as all previous proofs (for the R\'enyi-2 measure~\cite{AdessoSerafini} or other quantifiers~\cite{hiroshima2007,strongmono}) were restricted to the special case of \emph{one} mode per party. 
A monogamy inequality is a powerful tool in dealing with entanglement measures. For instance, when combined with monotonicity under local operations, it leads to the additivity of the measure under examination.

\begin{cor} \label{additivity G R2 EoF cor}
The Gaussian R\'enyi-2 entanglement of formation~\eqref{G R2 EoF} is additive under tensor products (equivalently, direct sum of covariance matrices). In formula,
\begin{equation}
E_{F,2}^{G} (A_1 A_2: B_1 B_2)_{V_{A_1 B_1} \oplus W_{A_2 B_2}} = E_{F,2}^{G} (A_1: B_1)_{V} + E_{F,2}^{G} (A_2: B_2)_{W}\, .
\label{additivity GEoF}
\end{equation}
\end{cor}

\begin{proof}
Applying first~\eqref{mon E2} and then the monotonicity of $E_{F,2}^{G}$ under the operation of discarding some local subsystems, we obtain
\begin{align*}
&E_{F,2}^{G} (A_1 A_2: B_1 B_2)_{V_{A_1 B_1} \oplus W_{A_2 B_2}} \\[0.8ex]
&\quad \geq E_{F,2}^{G} (A_1 A_2 : B_1)_{V_{A_1 B_1}\oplus W_{A_2}} + E_{F,2}^{G} (A_1 A_2: B_2)_{V_{A_1} \oplus W_{A_2 B_2}} \\[0.8ex]
&\quad\geq E_{F,2}^{G} (A_1: B_1)_{V} + E_{F,2}^{G} (A_2: B_2)_{W}\, .
\end{align*}
The opposite inequality follows by inserting factorised ansatzes $\gamma_{A_1 B_1}\oplus \tau_{A_2 B_2}$ into~\eqref{G R2 EoF}.
\end{proof}

As established in this section, the R\'enyi-2 Gaussian entanglement of formation alias R\'enyi-2 Gaussian squashed entanglement also emerges as a rare example of an additive entanglement monotone (within the Gaussian framework) which satisfies the general monogamy inequality \eqref{mon E2}. By contrast, the conventional (R\'enyi-$1$) entanglement of formation can not fundamentally be monogamous~\cite{lancien2016}.

\section{Conclusions} \label{sec6 conclusions}

We have derived fundamental inequalities for the Schur complement of positive semidefinite matrices and explored their far-reaching applications to quantum information theory. This enabled us to recover seemingly unrelated findings from recent literature, like the strong subadditivity for log-determinant of covariance matrices~\cite{AdessoSerafini, Kor, Simon16} and basic properties of measures of continuous variable entanglement~\cite{AdessoSerafini} and steering~\cite{steerability, MonSteer}, and to reach substantially beyond.  In particular, we proved that the Gaussian steerability~\cite{steerability, JOSAB} for Gaussian states is a convex monotone under Gaussian local operations and classical communication, i.e. it is a fully fledged steering measure~\cite{Gallego} within the Gaussianity restriction; we further proved it is monogamous with respect to the steering party but not with respect to the steered party if the latter has more than one mode and the overall state is mixed. We also proved that the Gaussian R\'enyi-2 measure of entanglement~\cite{AdessoSerafini} is monogamous for any Gaussian state with an arbitrary number of modes per party. This key result is a simple corollary of a general hierarchical relation here established for measures of correlations based on log-determinant of covariance matrices.

This work further reveals how pursuing \emph{prima facie} technical advances in classical information theory and linear algebra can significantly impact on the identification of possibilities and limitations for quantum technologies, which had eluded a general quantitative analysis so far.
It will be worth investigating adaptations of our results to the study of quantum correlations in discrete variable stabiliser states, useful resources for quantum computing~\cite{rauss} which share deep mathematical analogies with continuous variable Gaussian states~\cite{Gross06, Gross}.


\chapter{From log-det inequalities to Gaussian entanglement via recoverability theory} \label{chapter7}

\section{Introduction} \label{sec7 intro}

In the preceding Chapter~\ref{chapter6} we stumbled upon a matrix inequality dubbed \emph{strong subadditivity of log-det entropy} (log-det SSA), which is reported here as~\eqref{SSA}.
We showed that this relation has a major impact on Gaussian quantum information, either directly or through suitable generalisations.
In this chapter we develop further insights into the properties of the above inequality, and explore some of its applications to classical and quantum information theory.

The material is structured as follows. The rest of this section is devoted to introducing the framework where our work belongs (Subsection~\ref{subsec7 a bridge}) and to pointing the reader to our main contributions (Subsection~\ref{subsec7 original}).
In Section~\ref{sec7 exact recov} we derive various conditions that characterise the case of saturation of log-det SSA~\eqref{SSA} with equality. Then, in Section~\ref{sec7 improvements} we turn to the case of near-saturation, which leads us on the one hand to the theory of recovery maps, and on the other hand to establishing simple and faithful lower bounds on the log-det conditional mutual information. Up to that point, all results
hold for general positive definite matrices.
After that, in Section~\ref{sec7 strength} we turn our attention to quantum Gaussian states and their quantum covariance
matrices. There, we introduce a measure of entanglement for quantum Gaussian states based on the log-det conditional mutual information and prove its faithfulness and additivity. Quite remarkably, we show that the measure coincides with the R\'enyi-2 Gaussian entanglement of formation introduced in~\cite{AdessoSerafini}, equipping the latter with an interesting operational interpretation in the context of recoverability. We conclude in Section~\ref{sec7 conclusions} with a number of open questions.

\subsection{Building bridges between matrix analysis, probability theory and Gaussian quantum information} \label{subsec7 a bridge}

The concept of Gaussian random variable, which is of central importance in this last part of the present thesis, has been long known to constitute a bridge between the two seemingly unrelated fields of matrix analysis and probability theory. The key of this correspondence is to construct, given a positive definite $n\times n$ matrix $A>0$, an $n$-dimensional Gaussian random variable $X$ with covariance matrix $A$. 
One can show that all the R\'enyi entropies of the random variable $X$ are equal (up to a constant that depends only on $n$) to the log-det entropy we defined in Chapter~\ref{chapter6}, given by $\frac12 \ln \det A$. This provides us with a systematic way to turn information theoretical inequalities into determinantal inequalities for positive matrices, a technique which has been the subject of growing interest in the last decades (see e.g.~the reviews given in~\cite{logconcave, Dembo}). In fact, the strong subadditivity of log-determinant entropies~\eqref{SSA} can be immediately obtained in this way.

Analogously to their classical counterparts, also quantum Gaussian states are completely described by their covariance matrix (up to an often irrelevant displacement operation). Moreover, the outcomes of Gaussian measurements performed on Gaussian states are (unsurprisingly) classical Gaussian random variables themselves, which allows us to connect also this framework to the one described above. In particular, when one looks at Shannon-entropic quantities constructed out of \emph{measured} correlations of quantum Gaussian states, the relevant quantity is again the log-det entropy. 

One could wonder whether there is a more mathematically direct way to obtain the log-det entropy starting from the quantum state itself. As already mentioned in the previous chapter, one such way is to look at R\'enyi-2 quantifiers. In fact, the R\'enyi-2 entropy, given in general by the formula $S_2(\rho) \coloneqq - \ln \Tr[ \rho^{2}]$ (see~\eqref{Renyi ent}), reduces to $S_2 (\rho)= \frac12 \log \det V$~\eqref{Renyi-2 Gauss} when evaluated on Gaussian states. This paves the way to direct operational interpretations for the R\'enyi-2 entropy in the Gaussian setting.

To summarise, when one looks at measured correlation quantifiers in the quantum Gaussian setting, the natural quantities that appears is the Shannon entropy of the measurement outcome, which in turn reduces to the log-det entropy of the (quantum) covariance matrix and thus coincides (up to additive constants) with the R\'enyi-2 entropy of the quantum state itself. 

\begin{note}
All entropies that appear in this chapter are measured with the natural logarithm $\ln$ instead of the binary logarithm $\log$. This is in agreement with the most common conventions in continuous variable information theory.
\end{note}

\begin{note}
Throughout this chapter, positive definite matrices $V>0$ will be occasionally referred to as \emph{covariance matrices}, a nomenclature that is very much in the spirit of highlighting the connections with probability theory. As usual, we instead call \emph{quantum covariance matrix} (QCM) any matrix $V$ of even size that satisfies the stronger condition $V+i\Omega\geq 0$, corresponding to the Heisenberg principle for Gaussian states~\eqref{Heisenberg}. Here, $\Omega$ is the standard symplectic form defined in~\eqref{CCR}.
\end{note}

\subsection{Our contributions} \label{subsec7 original}

This chapter is based on the homonymous paper~\cite{LL-log-det}:
\begin{itemize}

\item L. Lami, C. Hirche, G. Adesso, and A. Winter. From log-determinant inequalities to Gaussian entanglement via recoverability theory. \emph{IEEE Trans. Inf. Theory}, 63(11):7553--7568, 2017.

\end{itemize}
Our goal here is to conduct a deeper investigation of the log-det SSA~\eqref{SSA}, and to explore some of its far-reaching implications in quantum information theory. Our analysis rests crucially on the connection between Gaussian random variables and positive definite matrices we have outlined here, which allows us to use tools taken from matrix analysis~\cite{BHATIA} to explore properties of the log-det conditional mutual information~\eqref{I_2-matrix}, an approach we already followed in the previous Chapter~\ref{chapter6}.  

More in detail, in Section~\ref{sec7 exact recov} we will show how to find known and new necessary and sufficient conditions under which saturation with equality occurs (Theorem~\ref{thm satur}). Along the way, we develop new insights into the properties of information theoretic quantities based on the log-det entropy. Most notably, in Theorem~\ref{thm I cond geom} we show that: (a) log-det mutual information and \emph{conditional} mutual information can be obtained from one another by taking the \emph{inverse} of the covariance matrix; and (b) the log-det mutual information is convex on the geodesics of the \emph{trace metric}, defined on the sets of positive definite matrices $A$ by the formula $ds^2= \Tr [(A^{-1} dA)^2]$.
This is surprising, since in general the mutual information is known to be neither convex nor concave in the probability distribution.
Moreover, since this metric is also used to define the matrix geometric mean (see~\eqref{geom geod}), the above result establishes further connections between matrix analysis and information theory, in the very spirit of the present work.

Subsequently, in Section~\ref{sec7 improvements} we move on to the problem of improving the log-det SSA inequality, with special emphasis on the case of near saturation. In this context, we discuss the role of the classical transpose channel, also known as Petz recovery map, and find its action explicitly (Proposition~\ref{prop Petz}). We then use this result to prove some extensions of the saturation theorem, in the form of faithful lower bounds on the log-det conditional mutual information. Exploiting the aforementioned connection with the theory of matrix means, we further establish a somewhat different lower bound -- of a less information theoretical and more matrix analytic nature -- from which the saturation conditions are easily readable (Theorem~\ref{thm lower b}).

Section~\ref{sec7 strength} is devoted to exploring some of the applications of log-det SSA in the field of Gaussian quantum information. As we said, for Gaussian states the log-det entropy is equivalent to the R\'enyi entropy of order $2$. Our first result, Theorem~\ref{thm I cond G R2 EoF}, provides a strengthening of log-det SSA for quantum covariance matrices that involves the R\'enyi-2 Gaussian entanglement of formation, an entanglement measure we already encountered in Chapter~\ref{chapter6} (see~\eqref{G R2 EoF}). We then employ this result to define a log-det entropy equivalent of the squashed entanglement~\eqref{squashed eq}, which is then shown to coincide with the R\'enyi-2 Gaussian entanglement of formation (Theorem~\ref{thm GSq=GEoF}).
In particular, we prove that in the simpler Gaussian setting there is no need for extensions of unbounded size, which makes the measure efficiently computable. This is remarkable, since no attempt to establish a similar result for the standard squashed entanglement based on the von Neumann entropy has succeeded yet.
Finally, simpler proofs of useful properties like monogamy, faithfulness, and additivity can be found thanks to the identity between these two Gaussian entanglement measures.



\section{Mathematical preliminaries} \label{sec7 preliminaries}

The goal of this introductory section is to acquaint the reader with some of the tools we will use in the rest of the chapter. Namely, in Subsection~\ref{subsec7 info theory G} we discuss the basics of information theory with Gaussian random variables, explaining in detail the nature of the connection we mentioned in Subsection~\ref{subsec7 a bridge}. Next, Subsection~\ref{subsec7 log-det} is devoted to a more systematic introduction of log-det entropies, which we already encountered at the end of Chapter~\ref{chapter6}.

\subsection{Information theory with Gaussian random variables} \label{subsec7 info theory G}

As we already mentioned, the idea of using information theoretical reasoning to prove determinantal inequalities for positive definite matrices has received much attention recently~\cite{logconcave, Dembo}. The key of the above correspondence is to associate, to each positive matrix\footnote{In this chapter we consider only real matrices since they are more relevant for the applications we are interested in, but all the results we find apply also to the Hermitian case with minor modifications.} $A\in\mathcal{M}_n(\mathds{R})$, an $n$-dimensional Gaussian random variable
$X\in\mathds{R}^n$ with mean $0$ and variance (aka covariance matrix)
$\operatorname{Var} X = \mathds{E}\  X X^T = A$.
The density of $X$ is given by
\begin{equation}
  p_A(x) = \frac{e^{-\frac12 x^T A^{-1} x}}{\sqrt{(2\pi)^n\det A}}\, .
  \label{gauss}
\end{equation}
This has the nice feature that for two independent Gaussian random
variables $X$ and $Y$ with mean $0$ and covariance matrices $A$ and $B$, respectively,
the sum $A+B$ is the covariance matrix of $X+Y$.

Under the density~\eqref{gauss}, the differential entropy
\bb
h(X) \coloneqq -\int d^n x\, p_A(x) \ln p_A(x)
\label{differential entropy}
\ee
of~\eqref{gauss} takes the form
\begin{equation}
  h(X) = \frac{1}{2}\ln\det A + \frac{n}{2}\left(\ln 2\pi + 1 \right) ,
  \label{ent}
\end{equation}
while the relative entropy
$D(p_A\|p_B) \coloneqq \int d^n x\, p_A(x) \ln \frac{p_A(x)}{p_B(x)}$ is given by
\begin{equation}
  D(p_A\|p_B) = \frac{1}{2} \ln\frac{\det B}{\det A} + \frac{1}{2} \Tr(B^{-1}\! A) - \frac{n}{2}\, .
  \label{rel ent}
\end{equation}
Here and in the remainder of the chapter we denote by $\ln$ the natural logarithm. The positivity of~\eqref{rel ent} as a function of the matrices $A$ and $B$ can be seen as an instance of Klein's inequality applied to the natural logarithm~\cite{Wehrl}.

In this picture, general inequalities involving entropies can be turned into
inequalities involving determinants thanks to~\eqref{ent} and~\eqref{rel ent}.
A prominent example of the usefulness of this approach is constituted by
\emph{strong subadditivity} (SSA), the basic `Shannon-type' entropy
inequality~\cite{Yeung}.
Consider a Gaussian distributed vector
$X_{ABC} = (X_A,X_B,X_C)^T \in \mathds{R}^{n_A+n_B+n_C}$
with covariance matrix $V_{ABC}$:
\begin{equation}
  V_{ABC} = \begin{pmatrix} A & X & Y \\
                            X^{T} & B & Z \\
                            Y^{T} & Z^{T} & C \end{pmatrix}
    \geq 0\, .
  \label{global CM}
\end{equation}
The positivity of the conditional mutual information $I(X_{A}:X_{B}|X_{C})$ then yields the already encountered strong subadditivity of log-det entropy~\eqref{SSA 6}. Since this inequality is our main subject of study in this chapter, we report it here: 
\begin{equation}
  \ln\det V_{AC} + \ln\det V_{BC} - \ln\det V_{ABC} - \ln\det V_C \geq 0\, .
  \label{SSA}
\end{equation}
Observe how the above combination of log-det entropies mimics that appearing in the celebrated SSA of the quantum von Neumann entropy~\cite{Robinson67,Lanford68,lieb73a,lieb73b}, which is nowadays widely regarded as one of the cornerstones upon which quantum information theory is built~\cite{NC}.
In~\eqref{SSA}, the local reductions $V_{AC}$, $V_{BC}$ and $V_{C}$ are the principal
submatrices of $V_{ABC}$ corresponding to the components $AC$, $BC$ and $C$,
respectively:
\begin{equation}
  V_{AC} = \begin{pmatrix} A & Y \\ Y^{T} & C \end{pmatrix}, \quad
  V_{BC} = \begin{pmatrix} B & Z \\ Z^{T} & C \end{pmatrix}, \quad
  V_C = C\, .
  \label{eq:reductions}
\end{equation}
Let us observe that since~\eqref{SSA} is balanced, the contribution of the inhomogeneous second terms of~\eqref{ent} cancel out.

\begin{note}
Throughout the present chapter, we denote with $A$, $B$ and $C$ both the three groups of components of the random vector $X_{ABC}$ and -- occasionally -- the corresponding blocks in the covariance matrix~\eqref{global CM}. Since the former are always employed as subscripts, this should not lead to any ambiguity. 
\end{note}

Inequality~\eqref{SSA} was considered in~\cite{Ando09} (see also~\cite[\S 4.5]{PETZ}), although it has been known long before under the name of Hadamard-Fisher inequality (see for instance~\cite[Exercise 14, p.485]{HJ1}).\footnote{I thank Minghua Lin for bringing this to my attention.} From the point of view of matrix analysis,~\eqref{SSA} lends itself to straightforward generalisations. In fact,
inequalities of the same form have recently been investigated. In
particular, the problem of determining all the continuous functions
$f:\mathds{R}_{+}\rightarrow\mathds{R}$ such that
for all block matrices $V_{ABC} \geq 0$,
\begin{equation}
  \Tr f(V_{AC}) + \Tr f(V_{BC}) - \Tr f(V_{ABC}) - \Tr f(V_C) \geq 0\, ,
  \label{SSA gen}
\end{equation}
was considered in full generality in~\cite{Audenaert10}, where a sufficient
condition was found:~\eqref{SSA gen} holds as soon as $-f'$ is matrix monotone.
Later on, it was shown that this condition is also necessary~\cite{Lewin14}.
By virtue of L\"owner's theorems characterising matrix monotone
functions~\cite[V]{BHATIA-MATRIX}, this yields an explicit
characterisation of all the functions $f$ obeying~\eqref{SSA gen}. Here we
are mainly concerned with the particular choice $f(x)=\ln x$, that
turns~\eqref{SSA gen} into~\eqref{SSA}.

\subsection{Log-det entropy} \label{subsec7 log-det}

The differential R\'enyi-$\alpha$ entropy of a Gaussian random variable $X$ with density
$p_A(x)$, i.e.~$H_\alpha(X) \coloneqq \frac{1}{1-\alpha}\ln \int d^n x\, p_A(x)^\alpha$,
is given by
\begin{equation*}
  h_\alpha(X) = \frac{1}{2} \ln\det A + \frac{n}{2}\left( \ln 2\pi + \frac{1}{\alpha-1}\ln\alpha \right),
\end{equation*}
showing that all the differential R\'enyi entropies of Gaussian random vectors are
essentially equivalent to the differential Shannon entropy~\eqref{ent}, up to a characteristic
universal additive offset.
In view of this and of the above remarks, we are motivated,
given a vector-valued random variable $X$ with covariance
matrix $V$, to introduce the notation
\begin{equation}
  \label{logdetent}
  M(X) \coloneqq M(V) \coloneqq \frac12 \ln \det V\, .
\end{equation}
We will refer to the quantity defined in~\eqref{logdetent}, already encountered in Subsection~\ref{subsec6 R2 GEoF}, as the \textbf{log-det entropy} of $V$. We borrow from Subsection~\ref{subsec6 R2 GEoF} also the definition of \emph{log-det mutual information}, defined for a bipartite covariance matrix $V_{AB} > 0$ by
\begin{equation}
  I_{M}(A:B)_{V} \coloneqq \frac12 \ln\frac{\det V_{A}\det V_{B}}{\det V_{AB}} = M(V_{A})+M(V_{B})-M(V_{AB})\, .
\label{logdetMI}
\end{equation}
For tripartite covariance matrices $V_{ABC}>0$ we can also construct a \textbf{log-det conditional mutual information}:
\begin{equation}\begin{split}
  I_{M}(A:B|C)_{V} \coloneqq&\ \frac12 \ln\frac{\det V_{AC}\det V_{BC}}{\det V_{C}\det V_{ABC}} \\
                   =&\ M(V_{AC})+M(V_{BC})  -M(V_{ABC})-M(V_{C})\, .
  \label{I_2-matrix}
\end{split}\end{equation}

Every (balanced) entropic inequality thus yields a corresponding inequality for the log-determinants of positive block matrices~\cite{Chan-balanced}. Thanks to the work of Zhang and Yeung~\cite{ZhangYeung}
and followers~\cite{Dougherty,Matus}, infinitely many
independent such inequalities, so-called `non-Shannon-type inequalities',
are known by now.

The question of what are the precise constraints on the determinants of the $2^n$ principal
submatrices of a positive matrix of size $n\times n$ has been raised much earlier,
either directly in a matrix setting~\cite{JohnsonBarrett}
or more recently in the guise of the balanced entropy inequalities of Gaussian
random variables (both real valued or vector valued)~\cite{Hassibi,Shadbakht}.
Remarkably, the latter papers show that while the entropy region of three Gaussian
real random variables is convex but not a cone, the entropy region of three Gaussian
random vectors is a convex cone and that the \emph{linear} log-det inequalities
for three Gaussian random variables (and equivalently Gaussian random vectors)
are the same as the inequalities for the differential entropy of any three
variables -- which in turn coincide with the Shannon inequalities,
cf.~\cite{Yeung,Chan-balanced}.
It is conjectured that the same identity between Gaussian vector inequalities
and general differential inequalities holds for any number parties.


Our concrete interest in~\eqref{SSA} is partly motivated by its applications in quantum information theory with continuous variables~\cite{adesso14}, as first explored in~\cite{AdessoSerafini, Gross}. 
Every continuous variable quantum state $\rho$ of $n$ modes (Gaussian or non-Gaussian), subject to mild regularity conditions, has a corresponding $2n\times 2n$ quantum covariance matrix $V$ of the phase space variables defined as in~\eqref{covariance matrix}. 
Thus, quantities like log-det entropy and (conditional) mutual information can be associated with any sufficiently regular quantum state via its QCM.
By construction, such entropic quantifiers captures the correlations encoded in the second moments of the state. These are exactly the correlations that one can access through Gaussian measurements, whose outputs are in fact Gaussian multivariate random variables (Subsection~\ref{subsec5 G states}), as quantified by combinations of Shannon entropies.


As we saw in Subsection~\ref{subsec6 R2 GEoF}, for Gaussian states the log-det entropy reduces to the quantum R\'enyi-2 entropy, according to~\eqref{Renyi-2 Gauss}. We report that identity here since it is of central importance in the rest of the chapter:
\begin{equation}
S_{2}(\rho) \coloneqq -\ln \Tr [\rho^{2}] = \frac12 \ln \det V = M(V)\, .
\label{Renyi-2 entropy Gaussian 7}
\end{equation}
In particular, in the relevant case of tripartite quantum Gaussian states, the general inequality~\eqref{SSA} for log-det entropy takes the form of a
SSA inequality for the R\'enyi-2 entropy~\cite{AdessoSerafini, Gross, Simon16}, holding in addition to the standard
one for R\'enyi-1 entropy aka von Neumann entropy, which is valid for arbitrary (Gaussian or not) tripartite quantum states.
Note that in general it is not advisable to form entropy expressions from R\'enyi entropies, since they do not obey any nontrivial constraints in a general multi-partite system~\cite{LMW}. In information theory, this is addressed by defining directly well-behaved notions of conditional R\'enyi entropy and R\'enyi mutual information~\cite{TOMAMICHEL}. Here, we evade those issues as we are restricting to Gaussian states. In fact, thanks to the correspondence discussed here, the following important result is readily established.

\begin{lemma} \label{correspondence lemma}
The balanced inequalities~\cite{Chan-balanced} obeyed by the R\'enyi-2 entropies of Gaussian states are precisely those obeyed by the Shannon entropies of Gaussian multivariate random variables.
\end{lemma}

\begin{proof}
Follows from the correspondence law~\eqref{Renyi-2 entropy Gaussian 7} and from the observation that every positive definite matrix $A>0$ is a positive multiple of a QCM $A/\nu_{\min} (A)$, where $\nu_{\min}(A)$ is the minimum symplectic eigenvalue of $A$~\eqref{Williamson}. If the inequality under examination is balanced, this scaling does not affect it.
\end{proof}

Let us stress that the claim of Lemma~\ref{correspondence lemma} does not hold for non-balanced inequalities. An example illustrating this fact is provided by the inequality~\eqref{log det ineq}, proved in~\cite{Simon16}, which can not hold for all positive definite $V$ (that is, for all classical covariance matrices), as it can be easily seen by rescaling it via $V\mapsto kV$, for $k>0$. Observe that the new matrix $V$ becomes unphysical for sufficiently small $k$, as it violates the uncertainty principle~\eqref{Heisenberg}.

\section{SSA saturation and exact recovery} \label{sec7 exact recov}

In this section we proceed to study the conditions under which~\eqref{SSA}
is saturated with equality. A necessary and sufficient condition was already found
in~\cite{Ando09} (for a comprehensive discussion, see~\cite{PETZ}),
but here we present new proofs as well as alternative formulations,
which may provide new insights.

Let us start by fixing our notation concerning classical Gaussian channels,
whose action can be described as follows. Denote the input random variable
by $X$, and consider an independent Gaussian variable $Z \sim P_K$, where
$P_{K}$ is a normal distribution with covariance matrix $K$ and zero mean.
Then the output variable $Y$ of the Gaussian channel $N$ is given by
$N(X) \coloneqq Y \coloneqq H X + Z$ for some matrix $H$ of appropriate size. At the level of covariance matrices this
translates to
\begin{equation}
  N : V \longmapsto V' = H V H^T + K ,
  \label{N}
\end{equation}
where the only constraint to be obeyed is $K\geq 0$.

The following theorem gathers some notable facts concerning log-det conditional mutual information,
and provides a neat example of how useful the interplay between matrix analysis
and information theory with Gaussian random variables can be. In particular, the theory of Schur complements and matrix means we presented in Subsections~\ref{subsec5 Schur} and~\ref{subsec5 matrix means}, respectively, will play a relevant role in what follows.
The results we prove now, in particular, are used extensively throughout the chapter, and some of them play an important role already in the proof of the main theorem of this section, Theorem~\ref{thm satur}.

\begin{thm}
  \label{thm I cond geom}
  For all positive, tripartite matrices $V=V_{ABC}>0$, the following identities hold true:
  \begin{align}
    I_{M}(A:B|C)_{V} &= I_{M}(A:B)_{V_{ABC}/V_C}\, ,     \label{I cond Schur}\\
    I_{M}(A:B|C)_{V} &= I_{M}(A:B)_{V^{-1}}\, .                \label{I cond inv}
  \end{align}
Furthermore, for all pairs of positive definite matrices $V_{AB},W_{AB}>0$, the log-det mutual information is convex on the trace metric geodesic connecting
  them as in~\eqref{geom geod}, i.e.
  \begin{equation}
  I_M(A:B)_{V\#_t W} \leq (1-t) I_M(A:B)_V + t I_M(A:B)_W \label{I conv geod}
  \end{equation}
for all $t\in [0,1]$.
\end{thm}

\begin{proof}
Let us start by showing~\eqref{I cond Schur}. Using repeatedly the determinant factorisation property~\eqref{det factor}, we find
\begin{align*}
    I_{M}(A:B)_{V_{ABC}/V_C} &= \frac12 \ln \frac{\det (V_{AB}/V_C) \det (V_{BC}/V_C)}{\det (V_{ABC}/V_C)} \\[0.8ex]
    &= \frac12 \ln \frac{(\det V_{AB})(\det V_C)^{-1} (\det V_{BC})(\det V_C)^{-1}}{(\det V_{ABC})(\det V_C)^{-1}} \\[0.8ex]
    &= \frac12 \ln \frac{(\det V_{AB}) (\det V_{BC} )}{(\det V_{ABC})(\det V_C)} \\[0.8ex]
    &= I_M(A:B|C)_V\, .
\end{align*}
We now move to~\eqref{I cond inv}. The block inverse formula~\eqref{inv} give us
\begin{align*}
  (V^{-1})_{AB} &= (V_{ABC}/V_{C})^{-1} ,  \\
  (V^{-1})_{A}  &= (V_{ABC}/V_{BC})^{-1} , \\
  (V^{-1})_{B}  &= (V_{ABC}/V_{AC})^{-1} .
\end{align*}
Putting all together we find
\begin{align*}
    I_{M}(A:B)_{V^{-1}} &= \frac12 \ln\frac{\det (V^{-1})_{A}\det (V^{-1})_{B}}{\det (V^{-1})_{AB}} \\[0.8ex]
    &= \frac12 \ln\frac{\det (V_{ABC}/V_{BC})^{-1} \det (V_{ABC}/V_{AC})^{-1}}{\det (V_{ABC}/V_{C})^{-1}} \\[0.8ex]
    &= \frac12 \ln\frac{\det (V_{ABC}/V_{C})}{\det (V_{ABC}/V_{BC}) \det (V_{ABC}/V_{AC})} \\[0.8ex]
    &= \frac12 \ln\frac{(\det V_{ABC})(\det V_{C})^{-1}}
                        {(\det V_{ABC})(\det V_{BC})^{-1} (\det V_{ABC})(\det V_{AC})^{-1}} \\[0.8ex]
    &= \frac12 \ln \frac{\det V_{AC}\det V_{BC}}{\det V_{ABC}\det V_{C}} \\[0.8ex]
    &= I_{M}(A:B|C)_V\, ,
\end{align*}
which is what we wanted to show.

Finally, let us consider~\eqref{I conv geod}. We start by applying the monotonicity of the weighted geometric mean under positive maps~\eqref{weighted geometric mean positive maps}, to the positive map $\Phi(\cdot)\coloneqq \Pi_A (\cdot) \Pi_A^{T}$, where $\Pi_A$ is the projector onto the $A$ components. This yields 
\bbb
(V\#_t W)_A = \Pi_A (V\#_t W) \Pi_A^T \leq V_A \#_t W_A\, .
\eee
Taking the determinant of both sides of this latter inequality and using for the right hand side the explicit formula~\eqref{det geom} we obtain 
\bbb
\det \left(V\#_t W \right)_A \leq \det\left(V_A \#_t W_A\right) = (\det V_A)^{1-t} (\det W_A)^t .
\eee
Together with the analogous inequality for the $B$ system, this gives
\begin{align*}
   I_M(A:B)_{V\#_t W} &= \frac12 \ln \frac{\left(\det (V\#_t W)_A \right) \left( \det (V \#_t W)_B \right)}{\det (V\#_t W)_{AB}} \\[0.8ex]
   &\leq \frac12 \ln \frac{(\det V_A)^{1-t} (\det W_A)^t (\det  V_B)^{1-t}  (\det W_B)^t}{(\det V_{AB})^{1-t} (\det W_{AB})^t} \\[0.8ex]
   &= (1-t) I_M(A:B)_V + t I_M(A:B)_W ,
\end{align*}
concluding the proof.
\end{proof}

\begin{rem}
Inequality~\eqref{I conv geod} is especially notable because in general the log-det mutual information is not convex over the set of positive matrices. However, it is convex when restricted to geodesics in the trace metric, as we have just shown. Moreover, we note in passing that an analogous inequality to~\eqref{I conv geod} does not seem to hold for the log-det conditional mutual information.
\end{rem}

We are now ready to provide necessary and sufficient conditions for the saturation of the log-det SSA inequality~\eqref{SSA}.

\begin{thm}
\label{thm satur}
  For an arbitrary $V_{ABC}>0$ written in block form as in~\eqref{global CM}, the following are equivalent:
  \begin{enumerate}[(a)]
    \item $I_{M}(A:B|C)_{V}=0$, i.e.~\eqref{SSA} is saturated;
    \item $V_{ABC}/V_{BC}=V_{AC}/V_{C}$, i.e.~\eqref{INEQ 1} is saturated;
    \item $(V^{-1})_{AB}=(V^{-1})_{A}\oplus (V^{-1})_{B}$;
    \item $X=YC^{-1}Z^{T}$ (see~\cite{Ando09} or~\cite[Theorem~4.49]{PETZ});
    \item there is a classical Gaussian channel $N_{C\rightarrow BC}$ such
          that $(I_A\oplus N_{C\rightarrow BC})(V_{AC})=V_{ABC}$.
  \end{enumerate}
\end{thm}

\begin{proof} We divide the proof as follows. 
\begin{description}
\item[$(a)\! \Leftrightarrow\! (b).$] Saturation of~\eqref{SSA} and~\eqref{INEQ 1} are equivalent concepts, since it is very easy to verify that if $M\geq N>0$ then $M=N$ if and only if $\det M=\det N$.

\item[$(a)\!\Leftrightarrow\! (c).$] It is well-known that $W_{AB}>0$ satisfies $\det W_{AB} = \det W_{A} \det W_{B}$ iff its off-diagonal block is zero, i.e. iff $W_{AB}=W_{A}\oplus W_{B}$. For instance, this can be easily seen as a consequence of~\eqref{det factor}. Thanks to Theorem~\ref{thm I cond geom}, identity~\eqref{I cond inv}, applying this observation with $W=V^{-1}$ yields the claim.

\item[$(b)\!\Rightarrow\! (d).$] This is known in linear algebra~\cite{Ando09}, but for
the sake of completeness we provide a different proof that fits more with the
spirit of the present work. Namely, we see that the variational representation of Schur complements~\eqref{variational} guarantees that~\eqref{INEQ 1} is saturated if and only if
\begin{equation}
\begin{split}
  V_{ABC} - (V_{AC}/V_C) \oplus 0_{BC}\
    &= \lmatrix A-V_{AC}/V_C & X & Y \\
                      X^T  & B & Z \\
                      Y^T  & Z^T  & C \rmatrix \\[0.8ex]
    &= \lmatrix YC^{-1}Y^T  & X & Y \\
                      X^T  & B & Z \\
                      Y^T  & Z^T  & C \rmatrix \\[0.8ex]
    &\geq 0\, .
\end{split}
  \label{eq cond eq1}
\end{equation}
A necessary condition for~\eqref{eq cond eq1} to hold is obtained by taking suitable matrix elements:
\begin{equation*}
\begin{split}
  0 &\leq \lmatrix v \\ w \\ -C^{-1}Y^T  v \rmatrix^{T}
\lmatrix YC^{-1}Y^T  & X & Y \\
                           X^T  & B & Z \\
                           Y^T  & Z^T  & C \rmatrix
\lmatrix v \\ w \\ -C^{-1}Y^T  v \rmatrix \\[0.8ex]
    &=    2 v^T  (X-YC^{-1}Z^T ) w + w^T  B w .
\end{split}
\end{equation*}
This can only be true for all $v$ and $w$ if $X=YC^{-1}Z^T $.
Moreover, this latter condition (together with the positivity of $V_{ABC}$)
is enough to guarantee that~\eqref{eq cond eq1} is satisfied. Indeed, we can write
\begin{align*}
  \lmatrix YC^{-1}Y^T  & YC^{-1}Z^T  & Y \\
                  ZC^{-1}Y^T  & B & Z \\
                  Y^T  & Z^T  & C \rmatrix &= \lmatrix 0 & & \\ & B-ZC^{-1}Z^T  & \\ & & 0 \rmatrix  + \lmatrix YC^{-\frac12} \\ ZC^{-\frac12} \\ C^{\frac12} \rmatrix \lmatrix YC^{-\frac12} \\ ZC^{-\frac12} \\ C^{\frac12} \rmatrix^T \\[0.4ex]
                  &\geq 0\, ,
\end{align*}
where $B-ZC^{-1}Z^T \geq 0$ follows from
$\left(\begin{smallmatrix} B & Z \\ Z^T & C \end{smallmatrix}\right) \geq 0$.

\item[$(d)\!\Rightarrow\! (e).$] If in~\eqref{N} we define
\begin{equation}\begin{split}
  H &= H_R \coloneqq \begin{pmatrix} \mathds{1} & 0 \\ 0 & ZC^{-1} \\ 0 & \mathds{1} \end{pmatrix} , \\[0.8ex]
  K &= K_R \coloneqq \begin{pmatrix} 0 & & \\ & B-ZC^{-1}Z^T  & \\ & & 0 \end{pmatrix} ,
  \label{gaus Petz eq2}
\end{split}\end{equation}
we obtain straightforwardly
\begin{align*}
  (I_A\oplus N_{C\rightarrow BC}) (V_{AC}) &= H_R \lmatrix A & X \\ X^T  & C \rmatrix H_R^T + K_R  \\[0.8ex]
    &= \lmatrix A & X & Y \\ X^T  & B & Z \\ Y^T  & Z^T  & C \rmatrix  \\[0.8ex]
    &= V_{ABC} ,
\end{align*}
provided that $X=YC^{-1}Z^{T}$. We will see in the next section
that this map is nothing but a specialisation to the Gaussian case of a
general construction known as `transpose channel' or `Petz recovery map'.

\item[$(e)\!\Rightarrow\! (b).$] Since in the previous chapter we have shown that classical Gaussian channels acting on
$C$ never decrease the Schur complement $V_{AC}/V_C$ (Theorem~\ref{CP incr sch}), it is clear that the equality in~\eqref{INEQ 1}
is a necessary condition for the existence of a Gaussian recovery map $N_{C\rightarrow BC}$.
\end{description}
\end{proof}

\section{Improvements of log-det SSA} \label{sec7 improvements}

This section is devoted to presenting some improvements of the inequality~\eqref{SSA}. In Subsection~\ref{subsec7 G recov} we employ techniques inspired from the recently developed \emph{recoverability theory} to found an equivalent expression for $I_M(A:B|C)_M$. We leverage standard information theoretical results to obtain faithful lower bounds for this expression, leading to our first strengthening of the inequality. In Subsection~\ref{subsec7 I_M-lower-bound} we take instead a matrix analytic point of view and to improve the saturation conditions of Theorem~\ref{thm satur} more directly.

\subsection{Gaussian recoverability} \label{subsec7 G recov}

We start by discussing the role of some well-known remainder terms for inequalities
of the form~\eqref{SSA}. These terms have been introduced recently in the
context of sufficient statistics~\cite{Petz-old} and its approximate
variants~\cite{VV-Markov}, or so-called `recoverability'.
In~\cite{Fawzi-Renner}, a form involving recovery maps was proposed for such
a term in the fully quantum case (i.e., considering the SSA for von Neumann entropy)
based on the fidelity of recovery, and subsequently strengthened to a bound involving the
measured relative entropy~\cite{BHOS}; in both cases the given bounds
turn out to be operationally meaningful quantities~\cite{CHMOSWW}.
The much simpler classical reasoning (with a better bound) was presented in~\cite{VV-Markov}.
We will translate these results into the Gaussian setting
in order to find an explicit expression for a remainder term to be added to~\eqref{SSA}.

For classical probability distributions $p$ and $q$ over a discrete alphabet,
in~\cite{VV-Markov} the following inequality was shown, which improves on the
monotonicity of the relative entropy under channels:
\begin{equation}
  D(p\|q) - D(Np\|Nq) \geq D\left( p \| R N p \right) ,
  \label{recov}
\end{equation}
where $N = (N_{ji})$ is any stochastic map (channel)
and the action of the transpose channel
$R=R_{q,N}$ on an input distribution $r$ is uniquely
defined via the requirement that $N_{ji}q_i = R_{ij} (Nq)_j$ for all $i$ and $j$.
Explicitly,
\begin{equation}
  (R_{q,N}\, r)_i \coloneqq \sum_j \frac{q_i N_{ji}}{(Nq)_j} r_j .
  \label{Petz}
\end{equation}
Observe that $R_{q,N}$ is a bona fide channel, since
\begin{equation*}
  \sum_i (R_{q,N})_{ij} = \sum_i \frac{q_i N_{ji}}{(Nq)_j}
                        = \frac{(Nq)_j}{(Nq)_j}
                        = 1\, .
\end{equation*}
For obvious reasons, we will call the right hand side of~\eqref{recov} the
\textbf{relative entropy of recovery}. The proof of~\eqref{recov} is a simple
application of the concavity of the logarithm, and we
reproduce it here for the benefit of the reader.
\begin{align}
  D\left(p\|R_{q,N}Np\right)
     &= \sum_i p_i \Big(\ln p_i - \ln (R_{q,N}Np)_i \Big) \nonumber\\[0.8ex]
     &= \sum_i p_i \Big(\ln p_i - \ln \sum_j \frac{q_i\,N_{ji}}{(Nq)_j}\,(Np)_j \Big) \label{petz eq1} \\
     &\leq \sum_i p_i \Big(\ln p_i - \sum_j N_{ji} \ln \frac{q_i}{(Nq)_j}\,(Np)_j \Big) \label{petz eq2} \\
     &= D(p\|q) - D\left(Np\|Nq\right) . \nonumber
  \label{petz eq2}
\end{align}

Although we wrote out the proof only for random variables taking values in a discrete
alphabet, all of the above expressions make perfect sense also in more general cases,
e.g. when $i$ and $j$ are multivariate real variables.
If $N$ is a classical Gaussian channel acting as in~\eqref{N}, it
can easily be verified that the `transition probabilities' $N(x,y)$ satisfying
\begin{equation}
  (Np)(x) = \int dy\, N(x,y) p(y)
\end{equation}
take the form
\begin{equation}
  N(x,y) = \frac{e^{-\frac{1}{2} (x-Hy)^T K^{-1} (x-Hy)}}{\sqrt{(2\pi)^n\det K}}\, .
  \label{N Gauss}
\end{equation}

Following again~\cite{VV-Markov}, we observe that if the output of the random channel
$N$ is a deterministic function of the input, then~\eqref{recov} is always
saturated with equality.
This can be seen by noticing that in that case for all $i$ there is only one
index $j$ such that $N_{ji}\neq 0$ (and so $N_{ji}=1$). Therefore, the step
from~\eqref{petz eq1} to~\eqref{petz eq2} is an equality. This remark is useful for a very important special case. Consider a triple of random variables $XYZ$
distributed according to $p(xyz)$, a second probability distribution $q(xyz)=p(x)p(yz)$,
and the channel $N$ consisting of discarding $Y$. Obviously, in this case the
output is a deterministic function of the input. It is easily seen that the
reconstructed global probability distribution $\tilde{p}=R_{q,N}Np$ is
\begin{equation}
  \tilde{p}(xyz) = p(xz) p(y|z) .
\end{equation}
Then the saturation of~\eqref{recov} allows us to write
\begin{equation}
  I(X:Y|Z) = D(p\|q) - D(Np\|Nq) = D(p\|\tilde{p}) .
  \label{I rel ent}
\end{equation}

From now on, we will consider the case in which $N$ is a classical Gaussian channel transforming covariance matrices according to the rule~\eqref{N}. As can be easily verified, if also $q$ is a multivariate Gaussian distribution, then $R_{q,N}$ becomes a classical Gaussian channel as well. We compute its action in the case we are mainly interested in, that is, when the left-hand side of~\eqref{recov} corresponds to the difference of the two sides of~\eqref{SSA}, and verify that it coincides with the recovery map introduced in Section~\ref{sec7 exact recov} (via the general action~\eqref{N} with the substitutions~\eqref{gaus Petz eq2}).

\begin{rem}
Before we present our result, a quick side note. While here we are interested in the \emph{classical} Gaussian Petz recovery map, a very recent study of mine~\cite{LL17} has investigated the quantum Petz map for Gaussian states, showing it to be a Gaussian channel. The results and techniques in~\cite{LL17} are however of somewhat different nature than those presented here, since they involve quantum states instead of classical random variables. 
\end{rem}

\begin{prop}
  \label{prop Petz}
  Let $q$ be a tripartite Gaussian probability density with zero mean and covariance matrix
  \begin{equation*}
    V_A\oplus V_{BC} = \begin{pmatrix} A & 0 & 0 \\ 0 & B & Z \\ 0 & Z^T & C \end{pmatrix} ,
  \end{equation*}
  and let the channel $N$ correspond to the action of discarding the $B$ components, i.e.
  $H = \Pi_{AC}
   = \left( \begin{smallmatrix} \mathds{1} & 0 & 0 \\ 0 & 0 & \mathds{1} \end{smallmatrix}\right)$
  and $K=0$ in~\eqref{N}. Then, the action $C\rightarrow BC$ of the Petz recovery
  map~\eqref{Petz} on Gaussian variables with zero mean can be written at the level
  of covariance matrices as~\eqref{N}, where $H_R$ and $K_R$ are given by~\eqref{gaus Petz eq2}.
\end{prop}

\begin{proof}
The Petz recovery map~\eqref{Petz} is a composition of three operations: first the pointwise division by a Gaussian distribution, then the transpose of a deterministic channel, and eventually another pointwise Gaussian multiplication. It should be obvious from~\eqref{gauss} that a pointwise multiplication by a Gaussian distribution with covariance matrix $A$ is a Gaussian (non--deterministic) channel that leaves the mean vector invariant and acts on covariance matrices as $V\mapsto V'=(V^{-1}+A^{-1})^{-1}$. Furthermore, it can be proved that the transpose $N^T$ of the channel $N$ in~\eqref{N} sends Gaussian variables with zero mean to other Gaussian variables with zero mean, while on the inverses of the covariance matrices it acts as
\begin{equation}
  N^T : V^{-1} \longmapsto\ (V')^{-1} = H^T (V+K)^{-1} H\, .
  \label{N^T}
\end{equation}
A way to prove the above equation is by using~\eqref{N Gauss} to compute directly
the action of $N^T$ on a Gaussian input distribution.

After the preceding discussion, it should be clear that under our hypotheses
the action of the Petz recovery map can be written as
\begin{equation}
\sigma_{AC} \longmapsto \sigma'_{ABC} = \Big( V_A^{-1}\oplus V_{BC}^{-1} + (\sigma_{AC}^{-1}-V_A^{-1}\oplus V_C^{-1})\oplus 0_B \Big)^{-1} .
\label{gaus Petz eq3}
\end{equation}
The Woodbury matrix identity (see~\cite{Woodbury}, or~\cite[Eq. (6.0.10)]{ZHANG}),
\begin{equation}
  (S+UTV)^{-1} = S^{-1} - S^{-1}U\left(VS^{-1}U+T^{-1}\right)^{-1} V S^{-1},
  \label{Wood}
\end{equation}
can be used to bring~\eqref{gaus Petz eq3} into the canonical form~\eqref{N}:
\begin{align*}
  \sigma'_{ABC} &= \left( V_A^{-1}\oplus V_{BC}^{-1} + (\sigma_{AC}^{-1}-V_A^{-1}\oplus V_C^{-1})\oplus 0_B \right)^{-1} \\[0.8ex]
                &= \big( V_A^{-1}\oplus V_{BC}^{-1} + \Pi_{AC}^T  (\sigma_{AC}^{-1}-V_A^{-1}\oplus V_C^{-1}) \Pi_{AC} \big)^{-1} \\[0.8ex]
                &= V_A\oplus V_{BC} - (V_A\oplus V_{BC}) \Pi_{AC}^T \\
                &\quad \cdot \Big( (\sigma_{AC}^{-1}\! - V_A^{-1}\!\oplus V_C^{-1})^{-1}\! + \Pi_{AC} (V_A \oplus V_{BC}) \Pi_{AC}^T\Big)^{-1} \Pi_{AC} (V_A \oplus V_{BC}) \\[0.8ex]
                &= V_A \oplus V_{BC} - (V_A \oplus V_{BC}) \Pi_{AC}^T \\
                &\quad \cdot \Big(\! -V_A\! \oplus\! V_C  - (V_A\! \oplus\! V_C) (\sigma_{AC}-V_A\! \oplus\! V_C)^{-1} (V_A\! \oplus\!  V_C) + V_A\! \oplus \!V_C \Big)^{-1} \\
                &\quad \cdot \Pi_{AC} (V_A \oplus V_{BC}) \\[0.8ex]
                &= V_A\oplus V_{BC} \\
                &\quad+ (V_A\oplus V_{BC}) \Pi_{AC}^T (V_A^{-1} \oplus V_C^{-1}) (\sigma_{AC}  -  V_A\oplus V_C) (V_A^{-1}\!\oplus V_C^{-1}) \\
                &\quad\quad \cdot \Pi_{AC} (V_A\oplus V_{BC}) \\[0.8ex]
                &= H_R \sigma_{AC} H_R^T + K_R\, ,
\end{align*}
where $H_R = (V_A\oplus V_{BC}) \Pi_{AC}^T (V_A^{-1}\oplus V_C^{-1}$ and $K_R$ are defined according to~\eqref{gaus Petz eq2}.
This concludes the proof.
\end{proof}

We are ready to employ the classical theory of recoverability in order
to find the expression of the relative entropy of recovery in the Gaussian case.

\begin{prop}
For all tripartite covariance matrices $V_{ABC}>0$ written in block
form as in~\eqref{global CM}, we have
\begin{equation}
I_{M}(A:B|C)_{V} = \frac12 \ln\frac{\det V_{AC}\det V_{BC}}{\det V_{ABC}\det V_C} = D\!\left( V_{ABC} \| \tilde{V}_{ABC}\right) ,
\label{SSA+}
\end{equation}
where
\begin{equation}
\tilde{V}_{ABC} \coloneqq \begin{pmatrix} A & YC^{-1} Z^T  & Y \\ ZC^{-1}Y^T  & B & Z \\ Y^T  & Z^T  & C \end{pmatrix}
\label{V tilde}
\end{equation}
and the relative entropy function $D(\cdot\|\cdot)$ is given by~\eqref{rel ent}.
\end{prop}

\begin{proof}
This is just an instance of~\eqref{I rel ent} applied to the continuous
Gaussian variable $(X_{A},X_{B},X_{C})$.
\end{proof}

The identity~\eqref{SSA+} is useful in deducing new constraints that will
be much less obvious coming from a purely matrix analysis perspective.
For instance, it is well known that $D(p\|q)\geq -\ln \mathcal{F}^2(p,q)$ {(see e.g.~\cite{newRenyi,KMRA})},
where the fidelity is given by $\mathcal{F}(p,q)=\sum_i \sqrt{p_i q_i}$
in the discrete case.
In case of Gaussian variables with the same mean, it holds
\begin{equation}
  \mathcal{F}^2(p_A,p_B) = \frac{\det (A!B)}{\sqrt{\det A\det B}}\, ,
  \label{fid gaus}
\end{equation}
where $(A!B)\coloneqq 2\left(A^{-1}+B^{-1}\right)^{-1}$ is the \emph{harmonic mean}
of $A$ and $B$.
Inserting this standard lower bound into~\eqref{SSA+} we obtain
\begin{equation}
  \frac{\det V_{AC}\det V_{BC}}{\det V_{ABC}\det V_C}
    \geq \frac{\det V_{ABC}\det \tilde{V}_{ABC}}{\left( \det (V_{ABC}!\tilde{V}_{ABC}) \right)^2}\, ,
  \label{SSA+ fid}
\end{equation}
leading to
\begin{equation}
   I_{M}(A:B|C)_{V}
    \geq \frac12 \ln\frac{\det V_{ABC}\det \tilde{V}_{ABC}}{\left( \det (V_{ABC}!\tilde{V}_{ABC}) \right)^2}\, .
  \label{SSA+ fid2}
\end{equation}
Using furthermore
\begin{align*}
  \det \tilde{V}_{ABC} &= \det\tilde{V}_{BC} \det (\tilde{V}_{ABC}/\tilde{V}_{BC}) \\
                       &= \det V_{BC} \det (\tilde{V}_{AC}/\tilde{V}_{C}) \\
                       &= \det V_{BC} \det (V_{AC}/V_{C})\, ,
\end{align*}
we also arrive at the inequality
\begin{equation}
  \det V_{ABC} \leq \det (V_{ABC}!\tilde{V}_{ABC})\, .
\end{equation}
To illustrate the power of this relation,
we note that inserting the harmonic-geometric mean inequality
for matrices~\eqref{hga}
\begin{equation*}
A!B \leq A\# B
\end{equation*}
yields again SSA~\eqref{SSA} in the form
$\det \tilde{V}_{ABC} \geq \det V_{ABC}$.

\subsection{A lower bound on $I_{M}(A:B|C)_{V}$} \label{subsec7 I_M-lower-bound}

Throughout this subsection, we explore some alternative ways of strengthening Theorem~\ref{thm satur}, i.e. of finding suitable lower bounds on the log-det conditional mutual information $I_{M}(A:B|C)_{V}$. The expression we are seeking should have two main features: (i) it should be easily computable in terms of the blocks of $V_{ABC}$; and (ii) the explicit saturation condition in Theorem~\ref{thm satur}(d) should be easily readable from it. This latter requirement can be accommodated, for example, if the lower bound involves some kind of distance between the off-diagonal block $X$ and its `saturation value' $YC^{-1}Z^{T}$. We start with a preliminary result.

\begin{prop}
  \label{prop I(A:B)}
  For all matrices
  \begin{equation*}
    V_{AB} = \begin{pmatrix} A & X \\ X^{T} & B \end{pmatrix} \geq 0\, ,
  \end{equation*}
  we have
  \begin{equation}
    I_{M}(A:B)_{V} \geq \frac12 \text{\emph{Tr}}\, [A^{-1} X B^{-1} X^{T}] = \frac12 \big\| A^{-1/2} X B^{-1/2} \big\|^{2}_{2}\, .
  \label{prop I(A:B) eq}
  \end{equation}
\end{prop}

\begin{proof}
Using, in this order, the standard factorisation of the determinant in terms
of the Schur complement, the identity $\ln \det V = \Tr \ln V$ (valid when $V>0$),
and the inequality $\ln(\mathds{1}+\Delta)\leq \Delta$ (for Hermitian
$\Delta > -\mathds{1}$), we find
\begin{align*}
  I_{M}(A:B)_{V} &=    \frac12 \ln \frac{\det V_{A}\det V_{B}}{\det V_{AB}} \\[0.8ex]
                 &=   -\frac12 \ln \det V_{A}^{-1/2}(V_{AB}/V_{B})V_{A}^{-1/2} \\[0.8ex]
                 &=   -\frac12 \ln \det (\mathds{1} - A^{-1/2}XB^{-1}X^T A^{-1/2}) \\[0.8ex]
                 &=   -\frac12 \Tr \ln (\mathds{1} - A^{-1/2}XB^{-1}X^T A^{-1/2})  \\[0.8ex]
                 &\geq \frac12 \Tr A^{-1/2}XB^{-1}X^T A^{-1/2} \\[0.8ex]
                 &=    \frac12 \Tr A^{-1}XB^{-1}X^T \\[0.8ex]
                 &=    \frac12  \big\| A^{-1/2} X B^{-1/2} \big\|^{2}_{2}\, ,
\end{align*}
and we are done.
\end{proof}

\begin{thm}
  \label{thm lower b}
  For all $V_{ABC}>0$ written in block form as in~\eqref{global CM}, we have
  the following chain of inequalities:
\begin{align*}
  &I_{M}(A:B|C)_{V} \\
  &\quad\geq \frac12 \text{\emph{Tr}}\, \Big[ (V_{AC}/V_{C})^{-1} (X\! -\! YC^{-1}Z^{T}) (V_{BC}/V_{C})^{-1} (X\! - \! YC^{-1}Z^{T})^{T} \Big] \\[0.8ex]
  &\quad\geq \frac12 \text{\emph{Tr}}\,\Big[ A^{-1} (X-YC^{-1}Z^{T}) B^{-1} (X-YC^{-1}Z^{T})^{T} \Big] \\[0.8ex]
  &\quad= \frac12 \left\| A^{-1/2}(X-YC^{-1}Z^{T})B^{-1/2} \right\|^{2}_{2}\, .
\end{align*}
\end{thm}

\begin{proof}
We want to use the identity~\eqref{I cond inv} to lower bound $I_{M}(A:B|C)_{V}$ via Proposition~\ref{prop I(A:B)}.
In order to do so, we need to write out the $A$-$B$ off-diagonal block of the
inverse $(V_{ABC})^{-1}$. Introducing the projectors onto the $A$ and $B$
components, denoted by $\Pi_{A}$ and $\Pi_{B}$ respectively, this amounts to saying that we are seeking an
explicit expression for $\Pi_{A} (V_{ABC})^{-1} \Pi_{B}^{T}$. Remember
that for an arbitrary bipartite block matrix $W_{12}$ the block-inversion formula~\eqref{inv} gives
\begin{align}
  \Pi_{1} (W_{12})^{-1} \Pi_{1}^{T} &= (W_{12}/W_{2})^{-1} , \label{inv1} \\[0.8ex]
  \Pi_{1} (W_{12})^{-1} \Pi_{2}^{T} &= - W_{1}^{-1} (\Pi_{1}W_{12}\Pi_{2}^{T})
                                                            (W_{12}/W_{1})^{-1} . \label{inv2}
\end{align}
This allows us to write
\begin{align*}
  &\Pi_{A} (V_{ABC})^{-1} \Pi_{B}^{T} \\
     &\quad= \Pi_{A} \Pi_{AB} (V_{ABC})^{-1} \Pi_{AB}^{T} \Pi_{B}^{T} \\[0.8ex]
     &\quad= \Pi_{A} (V_{ABC}/V_{C})^{-1} \Pi_{B}^{T} \\[0.8ex]
     &\quad= -(V_{AC}/V_{C})^{-1} \bigl( \Pi_{A} V_{ABC}/V_{C} \Pi_{B}^{T} \bigr) \bigl( (V_{ABC}/V_{C}) \big/ (V_{AC}/V_{C}) \bigr)^{-1} \\[0.8ex]
     &\quad= -(V_{AC}/V_{C})^{-1} \bigl( X - YC^{-1} Z^{T} \bigr) (V_{ABC}/V_{AC})^{-1} .
\end{align*}
By exchanging $A$ and $B$ in this latter expression and subsequently taking the transpose, we arrive also at
\begin{equation*}
  \Pi_{A} (V_{ABC})^{-1} \Pi_{B}^{T} = -(V_{ABC}/V_{BC})^{-1} \big( X - YC^{-1} Z^{T} \big) \cdot (V_{BC}/V_{C})^{-1} .
\end{equation*}
Now we are ready to invoke Proposition~\ref{prop I(A:B)} to write
\begin{align*}
  &I_{M}(A:B|C)_{V} \\
  &\quad= I_{M}(A:B)_{V^{-1}} \\[0.8ex]
           &\quad\geq \frac12 \Tr \Big[(V^{-1})_{A}^{-1} (\Pi_{A} V^{-1} \Pi_{B}^{T}) (V^{-1})_{B}^{-1} (\Pi_{B}^{T} V^{-1} \Pi_{A}) \Big] \\[0.8ex]
           &\quad= \frac12 \Tr \Bigl[ (V_{ABC}/V_{BC}) \bigl( (V_{ABC}/V_{BC})^{-1} (X-YC^{-1} Z^{T}) (V_{BC}/V_{C})^{-1} \bigr) \\
           &\qquad\qquad \cdot (V_{ABC}/V_{AC}) \bigl( (V_{AC}/V_{C})^{-1} (X-YC^{-1} Z^{T}) (V_{ABC}/V_{AC})^{-1} \big)^{T} \Bigr] \\[0.8ex]
           &\quad= \frac12 \Tr \Big[ (V_{AC}/V_{C})^{-1} ( X - YC^{-1} Z^{T}) (V_{BC}/V_{C})^{-1} ( X - YC^{-1} Z^{T})^{T} \Big]\, .
\end{align*}
Since on the one hand $V_{AC}/V_{C}\leq V_{A}=A$, and on the other hand the
expression $\Tr R K S K^{T}$ is clearly monotonic in $R,S\geq 0$,
we finally obtain
\begin{align*}
  I_{M}(A:B|C)_{V} &\geq \frac12 \Tr \bigl[ A^{-1} ( X - YC^{-1} Z^{T}) B^{-1} ( X - YC^{-1} Z^{T})^{T} \bigr] \\[0.8ex]
                   &= \frac12 \bigl\| A^{-1/2}(X-YC^{-1}Z^{T})B^{-1/2} \bigr\|^{2}_{2}\, .
\end{align*}
and we are done.
\end{proof}

It can easily be seen that the above result satisfies the requirements stated in the beginning of the section. In fact, the above lower bound for $I_M(A:B|C)_V$ is (i) easily computable in terms of the blocks of $V_{ABC}$, and (ii) zero iff the condition in Theorem~\ref{thm satur}(d) is met.

\section{SSA for quantum covariance matrices and R\'enyi-2 Gaussian squashed entanglement}
\label{sec7 strength}

In this final section we show how to apply results on log-det conditional mutual information to infer properties of Gaussian states in quantum optics.
First, in Subsection~\ref{subsec7 more G} we have a closer look at purifications of Gaussian states, stating some elementary results that complement those presented in Subsection~\ref{subsec5 G states}. The, Subsection~\ref{subsec7 SSA QCMs} presents a lower bound on the log-det conditional mutual information that holds specifically for quantum covariance matrices (Theorem~\ref{thm I cond G R2 EoF}). Using this result, in the following Subsection~\ref{subsec7 Gauss sq} we define a R\'enyi-2 Gaussian version of the squashed entanglement we already encountered in Subsection~\ref{subsec3 telep arg}. Our main result there is a proof of the remarkable fact that in the Gaussian setting this quantity coincides with the R\'enyi-2 Gaussian entanglement of formation (Theorem~\ref{thm GSq=GEoF}), which in particular provides an efficient way to compute it.

\subsection{More about Gaussian states} \label{subsec7 more G}

We start by proving some results on Gaussian purifications that will be instrumental in the rest of the section.
Remember that the commutation relations of the canonical operators $x_i, p_j$ are encoded in the $2n\times 2n$ matrix $\Omega$ defined in~\eqref{CCR}.
In the rest of the chapter, the antisymmetric, non-degenerate quadratic form identified by $\Omega$ will be called \textbf{standard symplectic product}, and the linear space $\mathds{R}^{2n}$ endowed with this product will be referred to as a \textbf{symplectic space}. In what follows, the symplectic space associated with a quantum optical system $A$ will be denoted with $\Sigma_A$.\footnote{Not to be confused with the matrix that changes the signs of the momenta~\eqref{Sigma}, denoted with the same symbol.} For an introduction to symplectic geometry, we refer the reader to the excellent monograph~\cite{GOSSON}.

When the system under examination is made of several parties (each comprising a certain number of modes), the global QCM will have a block structure as in~\eqref{global CM}. The symplectic form in this case is simply given by the direct sum of the local symplectic forms~\eqref{Omega bipartite}. This can be rephrased by saying that the symplectic space associated with the system $AB$ is the direct sum of the symplectic spaces associated with $A$ and $B$, in formula $\Sigma_{AB}=\Sigma_{A}\oplus \Sigma_{B}$~\cite[Eq. (1.4)]{GOSSON}. Conversely, discarding a subsystem corresponds to performing an orthogonal projection of the QCM onto the corresponding symplectic subspace~\cite[\S 1.2.1]{GOSSON}, in formula $V_{A}=\Pi_{A} V_{AB} \Pi_{A}^T$.

Pure Gaussian states enjoy many useful properties that we will exploit multiple times throughout this section. To explore them, it is essential to make a clever use of the complementarity between the two pictures, at the Hilbert space level on the one hand and at the QCM level on the other hand. Let us illustrate this point by presenting three lemmas.

\begin{lemma} \label{lemma pure reduction}
Let $V_{AB}$ be a QCM of bipartite system $AB$. Denote by $V_{A}=\Pi_{A} V_{AB} \Pi_{A}^T$ the reduced QCM corresponding to the subsystem $A$, and analogously for $V_{B}$. If $V_{A}$ is pure, then $V_{AB} = V_{A} \oplus V_{B}$.
\end{lemma}

\begin{proof}
The statement becomes obvious at the Hilbert space level. In fact, the reduced state on $A$ of a bipartite state $\rho_{A B}$ is given by $\rho_{A} = \tr_{B}\, \rho_{A B}$, where $\tr_{B}$ denotes partial trace~\cite{NC}. Evaluating the ranks of both sides of this equation shows that if $\rho_{A}$ is pure then the global state must be factorised, i.e. the QCM must have vanishing off-diagonal blocks.
\end{proof}

Extending the system as to include auxiliary degrees of freedom is a standard technique in quantum information, popularly referred to as going to the `Church of the larger Hilbert space'\footnote{This expression was coined by John A. Smolin, see also~\cite{Church}.}. Such a technique can be most notably employed in order to \emph{purify} the system under examination, as detailed in the following lemma~\cite{holwer}.

\begin{lemma} \label{lemma pur}
For all QCMs $V_A$ pertaining to a system $A$ there exists an extension $AE$ of $A$ and a pure QCM $\gamma_{AE}$ such that $\Pi_A \gamma_{AE} \Pi_A^T = V_A$, where $\Pi_A$ is the projector onto the symplectic subspace $\Sigma_A\subset \Sigma_{AE}$.
\end{lemma}

\begin{proof}
See~\cite[\S III.D]{holwer}.
\end{proof}

Finally, let us present here another useful observation.

\begin{lemma} \label{lemma fact out}
For all QCMs $V_A\geq i\Omega_A$ of a system $A$, there is a decomposition $\Sigma_A=\Sigma_{A_1} \oplus \Sigma_{A_2}$ of the global symplectic space into a direct sum of two symplectic subspaces such that
\begin{equation}
V_A = V_{A_1} \oplus \eta_{A_2}\, ,
\end{equation}
where $V_{A_1} > i\Omega_{A_1}$ and $\eta_{A_2}$ is a pure QCM. Furthermore, for every purification $\gamma_{AE}$ of $V_A$ (see Lemma~\ref{lemma pur}) there is a symplectic decomposition of $E$ as $\Sigma_{E}=\Sigma_{E_1} \oplus \Sigma_{E_2}$ such that: (a) $\gamma_{AE}=\gamma_{A_1E_1} \oplus \eta_{A_2} \oplus \tau_{E_2}$, with $\eta_{A_2}, \tau_{E_2}$ pure QCMs; (b) $n_{A_1} = n_{E_1}$; and (c) $\gamma_{E_1}>i\Omega_{E_1}$.
\end{lemma}

\begin{proof}
The first claim follows from Williamson's decomposition~\eqref{Williamson}. The subspace $\Sigma_{A_2}$ corresponds to those symplectic eigenvalues of $V_A$ that are equal to $1$. 

Now, let us prove the second claim. Consider an arbitrary pure QCM $\gamma_{AE}$ that satisfies $\gamma_A=V_A=V_{A_1}\oplus \eta_{A_2}$. Since in particular $\gamma_{A_2}=\eta_{A_2}$, we can apply Lemma~\ref{lemma pure reduction} and conclude that $\gamma_{AE}=\gamma_{A_1 E} \oplus \eta_{A_2}$. Claim (a) then tells us that $\gamma_{E}=\gamma_{E_1}\oplus \tau_{E_2}$, with $\gamma_{E_1}> i\Omega_{E_1}$ and $\tau_{E_2}$ pure. Again, Lemma~\ref{lemma pure reduction} yields $\gamma_{AE}=\gamma_{A_1E_1}\oplus \eta_{A_2}\oplus \tau_{E_2}$, corresponding to statement (b). Hence, we have only to show that $n_{A_1}=n_{E_1}$. In order to show this, let us write
\begin{equation*}
\gamma_{A_1 E_1} = \begin{pmatrix} V_{A_1} & L \\ L^T & \gamma_{E_1} \end{pmatrix} .
\end{equation*}
We can invoke Lemma~\ref{pure Schur lemma 1} to deduce the identity $V_{A_1} - L\gamma_{E_1}^{-1} L^T = \Omega V_{A_1}^{-1} \Omega^T$, that is, $L \gamma_{E_1}^{-1} L^T = V_{A_1} - \Omega V_{A_1}^{-1} \Omega^T$. Since the right hand side has maximum rank $2n_{A_1}$ thanks to the strict inequality $V_{A_1}>i\Omega$ (last claim of Lemma~\ref{QCM geom lemma}), we conclude that $2n_{E_1}\geq \rk \left( L \gamma_{E_1}^{-1} L^T \right) = 2 n_{A_1}$, and hence $n_{E_1}\geq n_{A_1}$. But the same reasoning can be applied with $A_1$ and $E_1$ exchanged, thus giving $n_{A_1}\geq n_{E_1}$, which concludes the proof.
\end{proof}

\subsection{SSA for quantum covariance matrices} \label{subsec7 SSA QCMs}

We are now ready to apply our results to strengthen the SSA inequality~\eqref{SSA} in the quantum case.
This subsection is thus devoted to finding
a sensible lower bound on the log-det conditional mutual information for
all QCMs. This bound will be given by the \emph{R\'enyi-2 Gaussian entanglement of formation}, already encountered in Subsection~\ref{subsec6 R2 GEoF} (see~\eqref{G R2 EoF} and~\eqref{G R2 EoF alt}). We report its expression here:
\begin{equation}
\begin{split}
  E^{G}_{F,2}(A:B)_{V}
    &= \inf \frac12 \log \det \gamma_{A} \\
    &\quad \text{ s.t. } \gamma_{AB} \text{ pure QCM and } \gamma_{AB}\leq V_{AB} .
\end{split}
  \label{G R2 EoF 7}
\end{equation}
As we saw, the above equation can be rephrased in an equivalent form using the identity between local entropies of a pure state, which in our case reads $M(\gamma_A)=M(\gamma_B)=\frac12 I_M(A:B)_\gamma$ for all pure QCMs $\gamma_{AB}$. We obtain
\begin{equation}
\begin{split}
  E^{G}_{F,2}(A:B)_{V}
    &= \inf \frac12 I_M(A:B)_{\gamma} \\
    &\quad \text{ s.t. } \gamma_{AB} \text{ pure QCM and } \gamma_{AB}\leq V_{AB} .
\end{split}
  \label{G R2 EoF alt 7}
\end{equation}
Remember that in the preceding Chapter~\ref{chapter6} we used~\eqref{G R2 EoF alt 7}, together with the notable inequality $E^G_{F,2}(A:B)_V \leq \frac12 I_M(A:B)_V$~\eqref{I>2E}, to show that the R\'enyi-2 Gaussian entanglement of formation is monogamous~\eqref{mon E2}.
We are now in position to apply some of the tools we have been developing so far to prove a generalisation of said inequality~\eqref{I>2E} that is of interest to us since it constitutes also a strengthening of~\eqref{SSA}.

\begin{thm}
  \label{thm I cond G R2 EoF}
  For all tripartite QCMs $V_{ABC}\geq i\Omega_{ABC}$,
  it holds that
  \begin{equation}
    \frac12 I_{M}(A:B|C)_{V} \geq E^{G}_{F,2}(A:B)_{V}\, .
    \label{ext I con}
  \end{equation}
\end{thm}

\begin{proof}
We employ a similar trick to the one used in the proof of Theorem~\ref{I2E}: for any QCM $V_{ABC}$, using the notation of~\eqref{gamma sharp}, define
\begin{equation*}
  \gamma_{AB} \coloneqq \gamma^\#_{V_{ABC}/V_{C}}\, .
\end{equation*}
By Lemma~\ref{QCM geom lemma}, we see that $\gamma_{AB}$ is a pure QCM. Now we proceed to show that $\gamma_{AB}\leq V_{AB}$. On the one hand, the very definition of Schur complement implies that $V_{ABC}/V_{C}\leq V_{AB}$,
while on the other hand~\eqref{sch compl mon cor eq} gives us 
\bbb
V_{ABC}/V_{C} \geq \Omega_{AB}^T V_{AB}^{-1}\Omega_{AB}\, ,
\eee
which upon matrix inversion becomes
\bbb
\Omega_{AB} (V_{ABC}/V_{C})^{-1} \Omega_{AB}^{T}\leq V_{AB}\, .
\eee
Since the geometric mean is well-known to be monotonic (see Subsection~\ref{subsec5 matrix means} as well as~\cite{ando79}),
we obtain immediately 
\begin{align*}
\gamma_{AB} &= (V_{ABC}/V_C)\# \big(\Omega_{AB} (V_{ABC}/V_C) \Omega_{AB}^T\big) \\[0.4ex]
&\leq V_{AB} \# V_{AB} \\[0.4ex]
&= V_{AB}\, .
\end{align*}
This shows that $\gamma_{AB}$ can be used as an ansatz in~\eqref{G R2 EoF alt 7}.
We can then write
\begin{align*}
     E^{G}_{F,2}(A:B)_{V} &\leq \frac12 I_M(A:B)_\gamma \\[0.4ex]
     &= \frac12 I_M(A:B)_{(V_{ABC}/V_{C})\# (\Omega_{AB} (V_{ABC}/V_{C})^{-1}\Omega_{AB}^T ) } \\[0.4ex]
     &\textleq{(1)} \frac14 I_M(A:B)_{V_{ABC}/V_{C}} + \frac14 I_M(A:B)_{\Omega_{AB} (V_{ABC}/V_{C})^{-1}\Omega_{AB}^T} \\[0.8ex]
     &\texteq{(2)} \frac14 I_M(A:B)_{V_{ABC}/V_{C}} + \frac14 I_M(A:B)_{(V_{ABC}/V_{C})^{-1}} \\[0.8ex]
     &\texteq{(3)} \frac14 I_M(A:B|C)_V + \frac14 I_M(A:B|C)_V \\[0.8ex]
     &= \frac12 I_M(A:B|C)_V\, ,
\end{align*}
where we employed, in order: (1) the convexity of log-det mutual information on the trace metric geodesics~\eqref{I conv geod}, (2) the obvious fact that since $\Omega_{AB}=\Omega_A \oplus \Omega_B$ is a local symplectic operation, the equality $I_M(A:B)_{\Omega W \Omega^T }=I_M(A:B)_W$ holds true; and (3) the identity~\eqref{I cond Schur} for the first term and~\eqref{I cond inv} followed again by~\eqref{I cond Schur} for the second.
\end{proof}

\subsection{R\'enyi-2 Gaussian squashed entanglement} \label{subsec7 Gauss sq}

As we mentioned in Subsection~\ref{subsec3 telep arg}, in finite-dimensional quantum systems the positivity of conditional
mutual information allows to construct a powerful entanglement measure
called \emph{squashed entanglement}~\eqref{squashed eq} (see also the original paper~\cite{squashed}). Let us report the definition here for convenience. For a bipartite state $\rho_{AB}$, one constructs
\begin{equation*}
  E_{\text{sq}}(A:B)_{\rho} \coloneqq \inf_{\rho_{ABC}} \frac12 I(A:B|C)_{\rho}\, ,
\end{equation*}
where the infimum ranges over all possible ancillary quantum systems $C$
and over all the possible states $\rho_{ABC}$ that have $\rho_{AB}$ as their marginal.
We are now in position to discuss a similar quantity tailored to Gaussian states.
First, we can restrict the infimum by considering only Gaussian extensions,
which corresponds to the step leading from~\eqref{EoF} to~\eqref{GEoF} in the construction of the Gaussian entanglement of formation.
Secondly, as it was done to arrive at the definition~\eqref{G R2 EoF} of the R\'enyi-2 Gaussian entanglement of formation (here reported as~\eqref{G R2 EoF 7}), we can substitute von Neumann entropies with R\'enyi-2 entropies. The result is
\begin{equation}
  E^{G}_{\text{sq},2}(A:B)_{V} \coloneqq \inf_{V_{ABC}} \frac12 I_{M}(A:B|C)_{V}\, ,
  \label{Gauss sq}
\end{equation}
where the infimum is on all extended QCMs $V_{ABC}$ satisfying the condition
$\Pi_{AB} V_{ABC}\Pi_{AB}^T = V_{AB}$ on the $AB$ marginal (and~\eqref{Heisenberg}).
We name the quantity in~\eqref{Gauss sq} \textbf{R\'enyi-2 Gaussian squashed
entanglement}, stressing that it is a quantifier specifically tailored to Gaussian states
and different from the R\'enyi squashed entanglement defined
in~\cite{Wilde} for general states, where an alternative expression
for the conditional R\'enyi-$\alpha$ mutual information is adopted instead.

Despite the complicated appearance of the expression~\eqref{Gauss sq}, it turns out that \emph{the R\'enyi-2 Gaussian squashed entanglement coincides with the R\'enyi-2 Gaussian entanglement of formation for all bipartite QCMs}. This unexpected fact shows once more that R\'enyi-2 quantifiers are particularly well behaved when employed to analyse Gaussian states, while at the same time it provides us with a novel, alternative expression of $E^{G}_{F,2}$ that can be used to understand its basic properties in a different, and sometimes more intuitive, way. Before stating the main result of this subsection, we need some preliminary results.

\begin{lemma} \label{lemma follia 0}
Let $\gamma_{AB}$ be a pure QCM of a bipartite system $AB$ such that $n_A=n_B=n$ and $\gamma_A>i\Omega_A$. Then
\begin{equation*}
\left(\gamma_{AB}+ i \Omega_{AB}\right) \big/ \left( \gamma_A + i\Omega_A \right) = 0_B\, .
\end{equation*}
\end{lemma}

\begin{proof}
From Williamson's decomposition~\eqref{Williamson}, we see that whenever $\gamma_{AB}$ is pure one has $\rk(\gamma_{AB}+i\Omega_{AB}) = n_A + n_B = 2n$ (i.e. half the maximum). Since already $\rk(\gamma_A + i\Omega_A) = 2n$, the additivity of ranks under Schur complements~\eqref{rank add} tells us that $\rk \left( \left(\gamma_{AB}+ i \Omega_{AB}\right) \big/ \left( \gamma_A + i\Omega_A \right) \right) = 0$, concluding the proof.
\end{proof}

\begin{prop} \label{follia prop}
Let $V_{AB}$ be a QCM of a bipartite system, and let $\gamma_{ABC}$ be a fixed purification of $V_{AB}$ (see Lemma~\ref{lemma pur}). Then, for all pure QCMs $\tau_{AB} \leq V_{AB}$ there exists a one-parameter family of pure QCMs $\sigma_C (t)$ (where $0<t\leq 1$) on $C$ such that
\begin{equation}
\gamma'_{AB}(t)\coloneqq \left(\gamma_{ABC} + 0_{AB} \oplus \sigma_C(t) \right) \big/ \left( \gamma_C +\sigma_C(t) \right)
\label{follia prop eq}
\end{equation}
is a pure QCM for all $t>0$, and $\lim_{t\rightarrow 0^+} \gamma'_{AB}(t) = \tau_{AB}$. Equivalently, there is a sequence of Gaussian measurements on $C$, identified by pure seeds $\sigma_C(t)$, such that the QCM of the post-measurement state on $AB$~\eqref{G measurement QCM Schur} is pure and tends to $\tau_{AB}$. 
\end{prop}

\begin{proof}
Let us start by applying Lemma~\ref{lemma fact out} to decompose the symplectic space of $AB$ as $\Sigma_{AB}=\Sigma_R \oplus \Sigma_S$ in such a way that $V_{AB}=V_R \oplus \eta_S$, where $V_R > i\Omega_R$ and $\eta_S$ is a pure QCM. According to Lemma~\ref{lemma fact out}, the purification $\gamma_{ABC}$ can be taken to be of the form $\gamma_{ABC}=\gamma_{RC_{1}} \oplus \eta_S\oplus \delta_{C_{2}}$, with $\gamma_{C_{1}}>i\Omega_{C_{1}}$, $n_{C_{1}}=n_R$, and $\delta_{C_{2}}$ pure. If $\tau\leq V$ is a pure QCM, a projection onto $\Sigma_S$ reveals that $\tau_S = \Pi_S \tau \Pi_S^T \leq \eta_S$. Since $\tau_S$ must be a legitimate QCM, and pure states are minimal within the set of QCMs, we deduce that $\tau_S=\eta_S$. Then, an application of Lemma~\ref{lemma pure reduction} allows us to conclude that $\tau = \tau_R \oplus \eta_S$, and accordingly $\tau_R \leq V_R$.

We claim that for all pure $\tau_R < V_R$ there is a pure QCM $\sigma_{C_{1}}$ such that
\begin{equation}
\left(\gamma_{RC_{1}} + 0_{R} \oplus \sigma_{C_{1}} \right) \big/ \left( \gamma_{C_{1}} +\sigma_{C_{1}} \right) = \tau_R\, .
\label{follia prop eq2}
\end{equation}
Constructing the extension $\sigma_{C}\coloneqq \sigma_{C_{1}}\oplus \tilde{\sigma}_{C_{2}}$, where $\tilde{\sigma}_{C_{2}}$ is an arbitrary pure QCM, we see that~\eqref{follia prop eq2} can be rewritten as 
\begin{equation}
\left(\gamma_{ABC} + 0_{AB} \oplus \sigma_{C} \right) \big/ \left( \gamma_{C} +\sigma_{C} \right) = \tau_R \oplus \eta_{S}\, .
\label{follia prop eq2bis}
\end{equation}
In fact, adding the ancillary system $C_{2}$ does not produce any effect on the Schur complement, since there are no off-diagonal block linking $C_{2}$ with any other subsystem. Analogously, the $S$ component of the $AB$ system can be brought out of the Schur complement because it is in direct sum with the rest.

In light of~\eqref{follia prop eq2bis}, we know that once~\eqref{follia prop eq2} has been established, in~\eqref{follia prop eq} we can achieve all QCMs $\gamma'$ that can be written as $\tau_{R}\oplus \eta_{S}$, with $\tau_{R}<V_{R}$. It is not difficult to see that this would allow us to conclude. Before proving~\eqref{follia prop eq2}, let us see why. The main point here is that every pure QCM $\tau_R\leq V_R$ can be thought of as the limit of a sequence of pure QCMs $\tau_R(t)< V_R$. An explicit formula for such a sequence reads $\tau_R(t) = \tau_R \#_t \gamma_{V_R}^\#$, where $\gamma_{V_R}^\#$ is the pure QCM defined in~\eqref{gamma sharp}, and $\#_t$ denotes the weighted geometric mean~\eqref{geom geod}. Observe that: (1) $\tau_R(t)$ is a QCM since it is known that the set of QCMs is closed under weighted geometric mean~\cite[Corollary 8]{bhatia15}; (2) $\tau_R(t)$ is in fact a pure QCM, because according to~\eqref{det geom} its determinant satisfies $\det \tau_R(t) = \left(\det \tau_R\right)^{1-t} \big(\det \gamma_{V_R}^\# \big)^t = 1$; (3) $\lim_{t\rightarrow 0^+} \tau_R(t)=\tau_R$ as can be seen easily from~\eqref{geom geod}; and (4) $\tau_R(t)< V_R$ for all $t>0$. This latter fact can be justified as follows. Since $V_R > i\Omega_R$, from the last claim of Lemma~\ref{QCM geom lemma} we deduce $\gamma_{V_R}^\# < V_R$. Taking into account that $\tau_R \leq V_R$, the claim follows from the strict monotonicity of the weighted geometric mean, in turn an easy consequence of~\eqref{geom geod}.

Now, let us prove~\eqref{follia prop eq2}. We start by writing
\begin{equation*}
\gamma_{RC_{1}} = \begin{pmatrix} V_R & L \\ L^T & \gamma_{C_{1}} \end{pmatrix} ,
\end{equation*}
where $V_R>i\Omega_R$, $\gamma_{C_{1}}> i\Omega_{C_{1}}$, and the off-diagonal block $L$ is square. As a matter of fact, more is true, namely that $L$ is also invertible. The simplest way to see this involves two ingredients: (1) the identity $\Omega V_{R}^{-1} \Omega^T = \gamma_{RC_{1}} / \gamma_{C_{1}} = V_{R} - L \gamma_{C_{1}}^{-1} L^T$, easily seen to be a special case of~\cite[Eq. (8)]{Lami16}; and (2) the fact that $V_R> \Omega V_R^{-1} \Omega^T$ because of Lemma~\ref{QCM geom lemma}. Combining these two ingredients we see that
\begin{equation*}
V_{R} > \Omega V_{R}^{-1} \Omega^T = V_{R} - L \gamma_{C_{1}}^{-1} L^T ,
\end{equation*}
which implies $L \gamma_{C_{1}}^{-1} L^T>0$ and in turn the invertibility of $L$. Now, for a pure QCM $\tau_R< V_R$, take $\sigma_{C_{1}} = L^T (V_R-\tau_R)^{-1} L - \gamma_{C_{1}}$. On the one hand, 
\bbb
\big(\gamma_{RC_{1}} \!+ 0_R\!\oplus\! \sigma_{C_{1}}\big) \big/ \big(\gamma_{C_{1}} \!+ \sigma_{C_{1}}\big) = V_R - L \left(\gamma_{C_{1}}\! + \sigma_{C_{1}}\right)^{-1}\! L^T = \tau_R
\eee
by construction. On the other hand, write
\begin{align*}
\sigma_{C_{1}} \!- i\Omega_{C_{1}} &= L^T (V_R - \tau_R)^{-1} L - (\gamma_{C_{1}} + i\Omega_{C_{1}}) \\[0.8ex]
&= L^T (V_R - \tau_R)^{-1} L - L^T (V_R + i \Omega_R)^{-1} L \\[0.8ex]
&= L^T \left( (V_R - \tau_R)^{-1} - (V_R + i \Omega_R)^{-1} \right) L \\[0.8ex]
&= L^T (V_R - \tau_R)^{-1} \left( (V_R + i \Omega_R) - (V_R - \tau_R) \right) (V_R + i \Omega_R)^{-1} L \\[0.8ex]
&= L^T (V_R - \tau_R)^{-1} \left( \tau_R + i \Omega_R \right) (V_R + i \Omega_R)^{-1} L\, ,
\end{align*}
where we employed Lemma~\ref{lemma follia 0} in the form $\gamma_{C_{1}} + i \Omega_{C_{1}} = L^T (V_R + i \Omega_R)^{-1} L$ and performed some elementary algebraic manipulations. Now, from the third line of the above calculation it is clear that $\sigma_{C_{1}} - i \Omega_{C_{1}}\geq 0$, since from $V_R - i \Omega_R \geq V_R - \tau_R > 0$ we immediately deduce $(V_R - \tau_R)^{-1} \geq (V_R + i \Omega_R)^{-1}$. This shows that $\sigma_{C_{1}}$ is a valid QCM. Moreover, observe that
\begin{align*}
\rk \left(\sigma_{C_{1}} - i \Omega_{C_{1}}\right) &= \rk \left( L^T (V_R - \tau_R)^{-1} \left( \tau_R + i \Omega_R \right) (V_R + i \Omega_R)^{-1} L \right) \\[0.8ex]
&\quad= \rk \left( \tau_R + i \Omega_R \right) \\
&\quad= n_R \\[0.8ex]
&\quad= n_{C_{1}}\, ,
\end{align*}
which tells us that $\sigma_{C_{1}}$ is also a pure QCM.
\end{proof}

Now, we are ready to state the conclusive result of the present chapter.

\begin{thm}
  \label{thm GSq=GEoF}
  For all bipartite QCMs $V_{AB}\geq i\Omega_{AB}$, the R\'enyi-2 Gaussian squashed
  entanglement coincides with the R\'enyi-2 Gaussian entanglement of formation, i.e.
  \begin{equation}
  E^{G}_{\text{\emph{sq}},2}(A:B)_{V} = E_{F,2}^{G}(A:B)_V\, .
  \label{GSq=GEoF}
  \end{equation}
\end{thm}

\begin{proof}
The inequality $E^{G}_{\text{sq},2}(A:B)_V \geq E_{F,2}^{G}(A:B)_V$ is an easy consequence
of~\eqref{ext I con} together with~\eqref{Gauss sq}. To show the converse, we employ the expression~\eqref{G R2 EoF alt 7} for the R\'enyi-2 Gaussian entanglement of formation. Consider an arbitrary purification $\gamma_{ABC}$ of $V_{AB}$, and pick a pure state $\tau_{AB}\leq V_{AB}$. By construction, we have $\gamma_{AB}=V_{AB}$. Now, thanks to Proposition~\ref{follia prop} one can construct a sequence of measurements identified by $\sigma_C(t)$ such that~\eqref{follia prop eq2} holds. Then, we have
\begin{align*}
\frac12 I_M(A:B)_\tau &= \frac12 I_M(A:B)_{\lim_{t\rightarrow 0^+} \left(\gamma_{ABC} + 0_{AB} \oplus \sigma_C(t) \right) / \left( \gamma_C +\sigma_C(t) \right) } \\
&\texteq{(1)} \lim_{t\rightarrow 0^+} \frac12 I_M(A:B)_{\left(\gamma_{ABC} + 0_{AB} \oplus \sigma_C(t) \right) / \left( \gamma_C +\sigma_C(t) \right) } \\
&\texteq{(2)} \lim_{t\rightarrow 0^+} \frac12 I_M(A:B|C)_{\gamma_{ABC} + 0_{AB} \oplus \sigma_C(t)} \\
&\textgeq{(3)} E^{G}_{\text{sq},2}(A:B)_V\, ,
\end{align*}
where we used, in order: (1) the continuity of the log-det mutual information; (2) the identity~\eqref{I cond Schur}; and (3) the fact that the QCMs $\gamma_{ABC} + 0_{AB} \oplus \sigma_C(t)$ constitute valid extensions of $V_{AB}$ and hence legitimate ansatzes in~\eqref{Gauss sq}.
\end{proof}

\begin{rem}
A by-product of the above proof of Theorem~\ref{thm GSq=GEoF} is that in~\eqref{Gauss sq} we can restrict ourselves to systems of bounded size $n_C \leq n_{AB} = n_A + n_B$. Moreover, up to limits the extension can be taken of the form $\gamma_{ABC} + 0_{AB} \oplus \sigma_C$, where $\gamma_{ABC}$ is a fixed purification of $V_{AB}$ and $\sigma_C$ is a pure QCM.
\end{rem}

This surprising identity between two seemingly very different entanglement measures, even though tailored to Gaussian states, is remarkable. On the one hand, it provides an interesting operational interpretation for the R\'enyi-2 Gaussian entanglement of formation in terms of log-det conditional mutual information, via the recoverability framework. On the other hand, it simplifies the notoriously difficult evaluation of the squashed entanglement, in this case restricted to Gaussian extensions and log-det entropy, because it recasts it as an optimisation of the form~\eqref{G R2 EoF 7}, which thus involves matrices of bounded instead of unbounded size (more precisely, of the same size as the mixed QCM whose entanglement is being computed).
In general, Theorem~\ref{thm GSq=GEoF} allows us to export useful properties between the two frameworks it connects. For instance, it follows from the identity~\eqref{GSq=GEoF} that the R\'enyi-2 Gaussian squashed entanglement is faithful on Gaussian states and a monotone under Gaussian local operations and classical communication; in contrast, proving the property of faithfulness for the standard squashed entanglement was a very difficult step to perform~\cite{faithful}. On the other hand, the arguments establishing many basic properties of the standard squashed entanglement can be imported from~\cite{squashed} and applied to~\eqref{Gauss sq}, providing new proofs of the same properties for the R\'enyi-2 Gaussian entanglement of formation. Let us give an example of how effective is the interplay between the two frameworks by providing an alternative, one-line proof of Corollary~\ref{E mono}, i.e. of the inequality
\bbb
E^G_{F,2}(A: B_1\ldots B_k)_V \geq {\sum}_{j=1}^k E^G_{F,2}(A:B_j)_V\, .
\eee

\begin{proof}[Alternative proof of Corollary~\ref{E mono}]
As usual, it suffices to show the above inequality in the case of three parties:
\bbb
E_{F,2}^{G}(A:BC) \geq E_{F,2}^{G}(A:B) + E_{F,2}^{G}(A:C)\, .
\eee
Thanks to Theorem~\ref{thm GSq=GEoF}, we can prove this relation for the R\'enyi-2 Gaussian squashed entanglement instead. We use basically the same argument as in~\cite[Proposition 4]{squashed}. Namely, call $V_{ABC}$ the QCM of the system $ABC$. Then for all extensions $V_{ABCE}$ of $V_{ABC}$ one has
\begin{align*}
I_M(A:BC|E)_V &= I_M(A:B|E)_V + I_M (A:C|BE) \\[0.8ex]
&\geq 2\, E^{G}_{\text{sq},2}(A:B)_V + 2\, E^{G}_{\text{sq},2}(A:C)_V\, ,
\end{align*}
where we applied the chain rule for the conditional mutual information together with the obvious facts that $V_{ABE}$ is a valid extension of $V_{AB}$ and $V_{ABCE}$ a valid extension of $V_{AC}$.
\end{proof}

This proof is substantially identical to the one provided in~\cite{squashed} to show the monogamy of the standard squashed entanglement. Then, in a way, the identity~\eqref{GSq=GEoF} explains why the R\'enyi-2 Gaussian entanglement of formation is monogamous and additive, something that looked a bit like an accident in Chapter~\ref{chapter6}. Now, we can say that it is not an accident, and on the contrary follows from the fact that the R\'enyi-2 Gaussian entanglement of formation coincides with its corresponding squashed entanglement, and the latter is monogamous and additive almost by construction.

\section{Conclusions} \label{sec7 conclusions}

In this chapter, we analysed the SSA inequality for the log-det entropy from a matrix analysis viewpoint and explored some of its applications. We first derived new necessary and sufficient conditions for saturation of said inequality. In the context of classical recoverability, we then provided an explicit form for the  Gaussian Petz recovery map and further obtained a strengthening of SSA by constructing a faithful lower bound to a log-det entropy based conditional mutual information. We finally specialised to quantum Gaussian states, for which the log-det entropy reduces to the R\'enyi-2 entropy, and defined a corresponding Gaussian version of the squashed entanglement measure. Surprisingly, we showed that the latter measure coincides with the R\'enyi-2 entanglement of formation defined via a Gaussian convex roof construction~\cite{AdessoSerafini}.
In turn, this has allowed us to build a bridge connecting the two frameworks, that can be used to establish new properties of a measure by looking at the other, or to provide simpler and more instructive proofs of known properties.
This work, together with the preceding Chapter~\ref{chapter6}, casts further light on the connections between matrix analysis (in particular determinantal inequalities) and information theory in both classical and quantum settings. 

Despite the progress, many of the quantifiers we talked about in this chapter still present some thorny open questions.
A major problem concerns the entanglement of formation of Gaussian states, since to the best of our knowledge it is still unknown whether in this case Gaussian decompositions always attain the global infimum in the convex roof optimisation~\eqref{EoF}.
Analogously, one could wonder whether for Gaussian states also the optimisation that defines the standard squashed entanglement~\cite{squashed} can be restricted to Gaussian extensions. This would considerably simplify the computation of the squashed entanglement on states very relevant for applications in quantum optics.
Finally, it could be interesting to establish whether the equivalence between the R\'enyi-2 Gaussian squashed entanglement defined here and the R\'enyi-2 Gaussian entanglement of formation defined in~\cite{AdessoSerafini} further extends to a third measure of entanglement, namely the recently introduced Gaussian intrinsic entanglement~\cite{GIE}. This latter measure is operationally relevant, since it constitutes -- by construction -- an upper bound on the secret key rate that is achievable by the application of local Gaussian measurements.

In future work we will focus on these problem, thus following the guiding star of quantum information science, operational relevance.

\appendix

\chapter{Computation rules for polars} \label{app polars}

This appendix contains the proofs of the computation rules \eqref{double polar 1}, \eqref{double polar 2}, \eqref{computation rule 1} and \eqref{computation rule 2}, which were used extensively in the main text. We will mainly adapt material from~\cite[IV \S 1.5]{SCHAEFER} and~\cite{computation}. In the following, let $E$ be a Banach space with Banach dual $E^*$. For the definition of the polars $M^\circ, N^\circ$ of two subsets $M\subseteq E$ and $N\subseteq E^*$, see \eqref{polar M} and~\eqref{polar N}. Some observations are in order:
\begin{enumerate}[(a)]
\item for an arbitrary family $\{M_i\}_i$ of subsets either of $E$ or of $E^*$, one has $\left( \bigcup_i M_i \right)^\circ = \bigcap_i M_i^\circ$;
\item polar sets are always convex, and moreover norm-closed (if in $E$) or weak*-closed (if in $E^*$);
\item for $M\subseteq E$ and $N\subseteq E^*$, one has 
\begin{align}
M^\circ = \left( \co(M) \right)^\circ = \left( \cl(M) \right)^\circ = \left( \cl_{\mathcal{w}} (M)\right)^\circ\, , \label{polar triviality 1} \\
N^\circ = \left( \co(N) \right)^\circ = \left( \cl(N) \right)^\circ = \left( \cl_{\mathcal{w}*} (N)\right)^\circ\, ; \label{polar triviality 2}
\end{align}
\item if $C\subseteq E$ is a cone, then for all $M\subseteq E$ one has $(M\pm C)^\circ = M^\circ \cap (\mp C^*)$ .
\end{enumerate}

The following lemma contains as special cases the two computation rules for double polars~\eqref{double polar 1} and~\eqref{double polar 2}.

\begin{lemma} \label{polar 1 lemma}
For $M\subseteq E$ and $N\subseteq E^*$, one has 
\begin{align}
M^{\circ\circ} &= \clit \left(\coit \left(M\cup \{0\}\right)\right) \label{double polar eq1} \\
N^{\circ\circ} &=\clit_{w*} \left(\coit \left(M\cup \{0\}\right)\right) . \label{double polar eq2}
\end{align}
\end{lemma}

\begin{proof}
Let us start by proving~\eqref{double polar eq1}. Since $\{0\},M\subseteq M^{\circ\circ}$ and the latter set is convex and closed, one has $\cl \left( \co \left( M \cup \{0\} \right)\right)\subseteq M^{\circ \circ}$. Conversely, consider $x\notin \cl \left( \co \left( M \cup\{0\} \right)\right)$. We have to show that $x\notin M^{\circ\circ}$, i.e. that there is $\varphi\in M^\circ$ such that $\braket{\varphi,x}>1$. Applying Corollary~\ref{Hahn-Banach cor} (a) with $M=\{x\}$ and $N=\cl \left( \co \left( M \cup\{0\} \right)\right)$, we can find $0<\epsilon<1$ and a continuous functional $\varphi'\in E^*$ such that $\braket{\varphi',y} \geq  -1+\epsilon$ for all $y\in \cl \left( \co \left( M\cup \{0\} \right)\right)$ and $\braket{\varphi',x}\leq -1$. By restricting $y$ to belong to $M$, we see immediately that $\varphi\coloneqq -\frac{\varphi'}{1-\epsilon}\in M^\circ$ but $\braket{\varphi,x}\geq \frac{1}{1-\epsilon}>1$, yielding the claim.

The proof of~\eqref{double polar eq2} proceeds in an analogous fashion. On the one hand, it is relatively easy to verify that $\cl_{\mathcal{w}*} \left(\co \left(N\cup \{0\}\right)\right)\subseteq N^{\circ\circ}$. On the other hand, if $\varphi\notin \cl_{\mathcal{w}*} \left(\co \left(N\cup \{0\}\right)\right)$ by definition we can find $y\in E$ that satisfies $\left( \varphi + y^{-1}\left( [-1,1]\right)\right) \cap \co \left(N\cup \{0\}\right)=\emptyset$, where we are seeing $y$ as a functional on $E^*$. This is the same as saying that $|\braket{\varphi - \psi,y}| > 1$ for all $\psi\in \co \left(N\cup \{0\}\right)$. Since the set of values taken by $\braket{\varphi - \psi, y}$ for different choices of $\psi$ is clearly convex, we can assume that $\braket{\varphi - \psi, y} > 1$ holds for all $\psi\in \co \left(N\cup \{0\}\right)$. Picking $\psi=0$ we find $\braket{\varphi, y} > 1$, while restricting to $\psi\in N$ and defining $x\coloneqq \frac{y}{\braket{\varphi, y}-1}$ yields $x\in N^\circ$ after a quick computation. Since $\braket{\varphi, x} = \frac{\braket{\varphi, y}}{\braket{\varphi, y}-1} > 1$, we see that $\varphi\notin N^{\circ\circ}$, implying that $N^{\circ\circ}\subseteq \cl_{\mathcal{w}*} \left(\co \left(N\cup \{0\}\right)\right)$ and concluding the proof. 
\end{proof}

\begin{rem}
The reader might have noticed some sort of asymmetry between~\eqref{double polar eq1}, where the norm closure appears, and~\eqref{double polar eq2}, where we have instead the weak* closure. This asymmetry in fact artificial, in the sense that $\cl(N)=\cl_{\mathcal{w}}(N)$ for all convex subsets $N\subseteq E$ thanks to Corollary~\ref{norm cl = weak cl conv cor}, and thus in~\eqref{double polar eq2} we can replace the norm closure with the weak closure, if needed.
\end{rem}

\begin{lemma} \label{polar 2 lemma}
Let $M,N\subseteq E$ be convex, closed, and such that $0\in M\cap N$. Then
\bb
(M\cap N)^\circ = \clit_{w*} \left( \coit \left(M^\circ \cup N^\circ \right) \right) .
\ee
\end{lemma}

\begin{proof}
We start by observing that
\begin{align*}
\left( \cl_{\mathcal{w}*} \left( \co \left(M^\circ \cup N^\circ \right) \right)\right)^\circ &\texteq{(1)} \left( M^\circ \cup N^\circ \right)^\circ \\
&\texteq{(2)} M^{\circ\circ} \cap N^{\circ \circ} \\
&\texteq{(3)} M \cap N \, .
\end{align*}
The justification of the above derivation is as follows: (1) is an application of the identities~\eqref{polar triviality 2}; (2) is deduced from observation (a) above; and (3) follows from Lemma~\ref{polar 1 lemma}. Now, employing again Lemma~\ref{polar 1 lemma} we find
\bbb
\cl_{\mathcal{w}*} \left( \co \left(M^\circ \cup N^\circ \right) \right) = \left( \cl_{\mathcal{w}*} \left( \co \left(M^\circ \cup N^\circ \right) \right)\right)^{\circ \circ} = (M\cap N)^\circ\, ,
\eee
which concludes the proof.
\end{proof}

The above rule for computing polars of intersections is so useful, that it makes sense to extend it further. In particular, we are interested in the following strengthening.

\begin{lemma} \emph{\cite[Lemma 1]{computation}.} \label{kung-fu lemma}
Let $M,N\subseteq E$ be convex and such that $0\in M\cap N$.
\begin{enumerate}[(a)]
\item If $\clit(M\cap N) = \clit(M) \cap \clit(N)$ then
\bbb
(M\cap N)^\circ = \clit_{w*} \left( \coit \left(M^\circ \cup N^\circ \right) \right) .
\eee
\item The identity $\clit(M\cap N) = \clit(M) \cap \clit(N)$ holds if $M,N$ are closed or if $0\in \interit(M) \cap \interit (N)$.
\end{enumerate}
\end{lemma}

\begin{proof}
We start with claim (a). Using~\eqref{polar triviality 1} together with Lemma~\ref{polar 2 lemma}, we find
\begin{align*}
(M\cap N)^\circ &= \left( \cl (M\cap N)\right)^\circ \\
&= \left( \cl (M) \cap \cl(N) \right)^\circ \\
&= \cl_{\mathcal{w}*} \left( \co \left(\cl(M)^\circ \cup \cl(N)^\circ \right) \right) \\
&= \cl_{\mathcal{w}*} \left( \co \left( M^\circ \cup N^\circ \right) \right) .
\end{align*}
Now, let us come to claim (b). If $M,N$ are closed then so is $M\cap N$, and hence trivially $\cl(M\cap N) = M\cap N = \cl(M) \cap \cl(N)$. In general, it is easy to verify that $\cl(M\cap N)\subseteq \cl(M)\cap \cl(N)$ holds always. Thus, it remains to show that there is equality when $0\in \inter(M)\cap \inter(N)$. Take $\delta>0$ such that $\|y\|\leq \delta\Rightarrow y\in M\cap N$. Consider $x\in \cl(M) \cap \cl(N)$, and let us show that $t x \in M\cap N$ for all $0\leq t<1$, so that $x=\lim_{t\rightarrow 1} tx \in \cl(M\cap N)$. One can pick $x'\in M$ such that $\left\| x - x' \right\|\leq \frac{(1-t)\delta}{t}$, which allows us to write $t x = t x' + (1-t) y$, with $y\coloneqq \frac{t}{1-t} (x-x')$ satisfying $\| y \|\leq \delta$ and thus $y\in M$. Since we have written $tx$ as a convex combination of two points $x',y\in M$, and $M$ is convex, we deduce that $tx\in M$. Analogously, one can show that $tx\in N$, so that $tx\in M\cap N$ for all $t\in [0,1)$, hence implying that $x \in \cl(M\cap N)$.
\end{proof}

Now, we are ready to prove~\eqref{computation rule 1} and~\eqref{computation rule 2} from the main text.

\begin{cor} \label{cor polar 3}
Let $E$ be an ordered Banach space with closed positive cone $E_+$, and let the Banach dual $E^*$ be ordered by the cone $E_+^*$ of positive continuous functionals. Denote by $B$ and $B^\circ$ the unit balls of $E$ and $E^*$, respectively. Then $[B]^\circ =\, ] B^\circ [$ and $[B^\circ]^\circ = \clit\, ]B[$. 
\end{cor}

\begin{proof}
Let us start with the first claim. We have:
\begin{align*}
[B]^\circ &= \left((B+E_+) \cap (B-E_+) \right)^\circ \\
&\texteq{(1)} \cl_{\mathcal{w}*} \left( \co \left( (B+E_+)^\circ \cup (B-E_+)^\circ \right) \right) \\
&\texteq{(2)} \cl_{\mathcal{w}*} \left( \co \left( (B^\circ \cap - E_+^*) \cup (B^\circ \cap E_+^*) \right) \right) \\
&\texteq{(3)} \co \left( (B^\circ \cap - E_+^*) \cup (B^\circ \cap E_+^*) \right) \\
&=\ ] B^\circ [ \, .
\end{align*}
The justification of these steps is as follows: (1) follows from Lemma~\ref{kung-fu lemma}, which is applicable since the sets $B\pm E_+$ are clearly convex and have the origin as an interior point; (2) see observation (d) above; (3) since $(B^\circ \cap - E_+^*) \cup (B^\circ \cap E_+^*)$ is a weak*-closed subset of the unit ball $B^\circ$, which is weak*-compact by the Banach-Alaoglu theorem (Theorem~\ref{Banach-Alaoglu}), it is itself weak*-compact; therefore, it is not difficult to realise that $\co \left( (B^\circ - E_+^*) \cup (B^\circ + E_+^*) \right)$ is already weak*-closed.

The second claim can be shown in an analogous fashion, but we have first to convince ourselves that $B^\circ + E_+^*$ is weak*-closed. This can be done once more thanks to the Banach-Alaoglu theorem (Theorem~\ref{Banach-Alaoglu}). In fact, if $(\lambda_\alpha +\mu_\alpha)_\alpha\subseteq B^\circ + E_+^*$ is such that $\wstarlim_\alpha\, (\lambda_\alpha + \mu_\alpha) = \varphi\in E^*$, the weak*-compactness of $B^\circ$ ensures that we can extract a subnet $(\lambda_\beta)_\beta$ of $(\lambda_\alpha)_\alpha\subseteq B^\circ$ that converges to some $\lambda\in B^\circ$ in the weak* topology. It is then easy to realise that on this subnet $\wstarlim_\beta\, \mu_\beta = \varphi-\lambda$ necessarily. Moreover, since $E_+^*$ is weak*-closed, we get $\varphi-\lambda\in E_+^*$ and in turn $\varphi = \lambda + (\varphi-\lambda)\in B^\circ + E_+^*$. Naturally, the same reasoning also shows that $B^\circ - E_+^*$ is weak*-closed. Now we can employ Lemma~\ref{polar 2 lemma} and conclude that
\begin{align*}
[B^\circ]^\circ &= \left((B^\circ+E_+^*) \cap (B^\circ-E_+^*) \right)^\circ \\
&= \cl_{\mathcal{w}} \left( \co \left((B^\circ+E_+^*)^\circ \cup (B^\circ-E_+^*)^\circ \right)\right) \\
&= \cl_{\mathcal{w}} \left( \co \left((B-E_+) \cup (B+E_+)\right)\right) \\
&= \cl_{\mathcal{w}}\, ]B[ \\
&= \cl\, ]B[\, ,
\end{align*}
where in the last step we used once more the fact that norm closure and weak closure coincide for convex sets.
\end{proof}

\chapter{Complements on the proof of Ludwig's embedding theorem} \label{app Ludwig}

Throughout this appendix, we will show directly that any two Banach spaces $E_1,E_2$ satisfying the hypotheses of Theorem~\ref{Ludwig emb thm} must be connected by an isometric order isomorphism, in the sense discussed at the beginning of Subsection~\ref{subsec uniqueness}. Let us start with a preliminary lemma that will allow us to characterise isometric order isomorphisms between base norm spaces in a computationally convenient way.

\begin{lemma} \label{lemma transf bases}
Let $E_1,E_2$ be two base norm spaces with corresponding bases $K_1,K_2$, and let $\Phi:E_1\rightarrow E_2$ be an injective linear map. Then $\Phi$ is an isometric order isomorphism iff $\Phi(K_1)=K_2$.
\end{lemma}

\begin{proof}
On the one hand, if $\Phi$ is an isometric order isomorphism, then by applying Lemmas~\ref{strictly pos base lemma} and~\ref{base norm on positive} we see that
\begin{align*}
K_2 &= \left\{ y \in E_{2+}:\ \|y\|=1 \right\} \\
&= \left\{ \Phi(x):\ x \in E_{1+},\, \|\Phi(x)\|=1 \right\} \\
&= \left\{ \Phi(x):\ x \in E_{1+},\, \|x\|=1 \right\} \\
&= \Phi \left( \left\{ x \in E_{1+}:\ \|x\|=1 \right\} \right) \\
&= \Phi(K_1)\, .
\end{align*}
On the other hand, let us assume that $\Phi(K_1)=K_2$ and that $\Phi$ is injective. Then $x\in \lambda K_1\Leftrightarrow \Phi(x)\in \lambda K_2$, for all $x\in E_1$ and $\lambda\geq 0$. Applying this observation repeatedly, we see that for all $x,y\in E_1$ one has
\begin{align*}
x\geq 0 \ &\Longleftrightarrow\ \exists\, \lambda\geq 0:\ x\in \lambda K_1 \\
&\Longleftrightarrow \ \exists\, \lambda\geq 0:\ \Phi(x)\in \lambda K_2 \\
&\Longleftrightarrow \ \Phi(x)\geq 0\, ,
\end{align*}
hence $\Phi$ is an order isomorphism. Verifying that $\Phi$ is also an isometry is again straightforward using~\eqref{base norm K}:
\begin{align*}
\|x\| &= \inf \left\{ |\alpha|+|\beta|:\ x=\alpha a - \beta b,\, \alpha,\beta\in\mathds{R},\, a,b\in K_1 \right\} \\
&= \inf \left\{ |\alpha|+|\beta|:\ \Phi(x)=\alpha \Phi(a) - \beta \Phi(b),\, \alpha,\beta\in\mathds{R},\, \Phi(a),\Phi(b)\in K_2 \right\} \\
&= \inf \left\{ |\alpha|+|\beta|:\ \Phi(x)=\alpha a' - \beta b',\, \alpha,\beta\in\mathds{R},\, a',b'\in K_2 \right\} \\
&= \|\Phi(x)\|\, .
\end{align*}
The last thing to check is that $\Phi$ is also surjective. This follows immediately from the fact that $K_2$ spans the positive cone $E_{2+}$, which is in turn spanning since $E_2$ is a base norm space.
\end{proof}

Let us recall here also the definition of \textbf{adjoint} of a linear map between Banach spaces. For details, we refer the reader to~\cite[\S 4.10]{RUDIN}. The adjoint $\Phi^*:E_2^*\rightarrow E_1^*$ of a map $\Phi:E_1\rightarrow E_2$ between two Banach spaces $E_1,E_2$ is constructed as the unique linear operator such that
\bb
\braket{\varphi_2, \Phi(x_1)} = \braket{\Phi^*(\varphi_2), x_1}\qquad \forall\ \varphi_2\in E_2^*,\ x_1\in E_1\, .
\label{adjoint}
\ee
It is fairly easy to verify that the following holds.

\begin{lemma} \label{lemma iso *-1}
If a linear map $\Phi:E_1\rightarrow E_2$ between two ordered Banach spaces is an isometric order isomorphism, then the same is true for $\Phi^{-1}:E_2\rightarrow E_1$ and for $\Phi^*:E_2^*\rightarrow E_1^*$. 
\end{lemma}

We are going to need two elementary results in linear algebra, which we state now without proof.

\begin{lemma} \label{lemma aff ext}
Let $V,W$ be real vector spaces, and let $K\subseteq V$ be a convex set, whose affine hull we denote by $\affit(K)=\{\sum_{i=1}^n \lambda_i x_i:\, x_i\in K,\, \sum_{i=1}^n \lambda_i=1\}$. Then every convex-linear map $f:K\rightarrow W$ admits a unique extension to an affine map $f:\affit (K)\rightarrow W$. 
\end{lemma}

\begin{lemma} \label{lemma lin ext}
Let $A\subseteq V$ be an affine subspace of a real vector space $V$. If $0\notin A$, then every affine map $f:A\rightarrow W$ (where $W$ is another real vector space) can be extended to a linear map $f:V\rightarrow W$. If $\Spanit(A)=V$, then this extension is unique.
\end{lemma}

Another trivial yet useful observation is a slight generalisation of the dual formula~\eqref{dual formula norm Banach eq} for the norm in a Banach space.

\begin{lemma} \label{lemma weak* dense}
Let $E$ be a Banach space with dual $E^*$. If $W$ is weak*-dense in $E^*$, then
\bbb
\|x\| = \sup_{\varphi \in W,\, \|\varphi\|_*\leq 1} |\braket{\varphi,x}|
\eee
for all $x\in E$.
\end{lemma}

Let us come now to the main result of this appendix, i.e. the direct proof of the uniqueness claim of Theorem~\ref{Ludwig emb thm}.

\begin{prop}
Let $\Omega\subseteq E_1$, $\Lambda \subseteq E_1^*$ convex subsets of a a Banach space $E_1$ and of its dual $E_1^*$. Let $E_1$ be a base norm space with base $\clit (\Omega)$, and assume $\Spanit (\Lambda)$ to be dense in $E_1^*$ in the weak*-topology. Let $\phi:\Omega \rightarrow E_2$ and $\psi:\Lambda\rightarrow E_2^*$ be embeddings of $\Omega$ and $\Lambda$ into another dual pair of Banach spaces $E_2$ and $E_2^*$ such that:
\begin{enumerate}[(i)]
\item $\braket{\psi(\lambda),\phi(\omega)}=\braket{\lambda,\omega}$ for all $\omega\in \Omega$ and $e\in \Lambda$;
\item $E_2$ is a base norm space with base $\clit \left(\phi(\Omega)\right)$;
\item $\Spanit \left( \psi(\Lambda)\right)$ is dense in $E_2^*$ with the weak*-topology.
\end{enumerate}
Then $\phi$ and $\psi$ can be extended to isometric order isomorphisms $\Phi:E_1\rightarrow E_2$ and $\Psi: E_1^*\rightarrow E_2^*$ such that $\Psi=(\Phi^*)^{-1}$
\end{prop}

\begin{proof} $\\[-5ex]$
\begin{itemize}

\item First of all, $\psi:\Lambda\rightarrow E_2^*$ is linear, in the sense that whenever one considers $\lambda,\eta\in \Lambda$ and $\alpha,\beta\in\mathds{R}$ such that also $\alpha \lambda + \beta \eta\in \Lambda$, one has $\psi(\alpha \lambda + \beta \eta) = \alpha \psi(\lambda) + \beta \psi(\eta)$. This can be verified as follows.
\begin{align*}
&\left\|\psi\left(\alpha \lambda + \beta \eta\right) - \alpha \psi(\lambda) - \beta \psi(\eta) \right\|_* \\
&\qquad \texteq{(1)} \sup_{\omega'\in \cl\left(\phi(\Omega)\right)} \left| \braket{\psi\left( \alpha \lambda + \beta \eta\right) - \alpha \psi(\lambda) - \beta \psi(\eta), \omega'} \right| \\
&\qquad \texteq{(2)} \sup_{\omega'\in \phi(\Omega)} \left| \braket{\psi\left( \alpha \lambda+ \beta \eta\right) - \alpha \psi(\lambda) - \beta \psi(\eta), \omega'} \right| \\
&\qquad= \sup_{\omega \in \Omega} \left| \braket{\psi\left(\alpha \lambda + \beta \eta\right) - \alpha \psi(\lambda) - \beta \psi(\eta), \phi(\omega)} \right| \\
&\qquad= \sup_{\omega \in \Omega} \left| \braket{\psi\left(\alpha \lambda + \beta \eta\right), \phi(\omega)} - \alpha \braket{\psi(\lambda),\phi(\omega)} - \beta \braket{\psi(\eta), \phi(\omega)} \right| \\
&\qquad\texteq{(3)} \sup_{\omega \in \Omega} \left| \braket{\alpha \lambda + \beta \eta, \omega} - \alpha \braket{\lambda,\omega} - \beta \braket{\eta, \omega} \right| \\
&\qquad= 0\, . 
\end{align*}
The justification of these steps is the following: (1) we applied Corollary~\ref{dual base norm cor}; (2) we restricted the supremum from $\cl\left( \phi(\Omega)\right)$ to $\phi(\Omega)$, which is possible because the function we are taking the supremum of is clearly continuous; (3) we used hypothesis (i) to drop $\phi$ and $\psi$ inside the scalar product.

\item Thanks to the aforementioned observations, it is clear that $\psi$ can be extended uniquely to a linear map $\psi:\Span (\Lambda) \rightarrow E_2^*$. Few manipulations similar to those we carried out above show also that $\psi$ is norm-preserving. In fact, employing hypothesis (ii) together with Corollary~\ref{dual base norm cor} yields
\begin{align*}
\|\psi(y)\|_* &= \sup_{\omega'\in \cl(\phi(\Omega))} |\braket{\psi(y), \omega'}| \\
&= \sup_{\omega'\in \phi(\Omega)} |\braket{\psi(y), \omega'}| \\
&= \sup_{\omega\in \Omega} |\braket{\psi(y), \phi(\omega)}| \\
&= \sup_{\omega\in \Omega} |\braket{y, \omega}| \\
&= \|y\|_*
\end{align*}
for all $y\in \Span(\Lambda)$. For the second last step we made use of the equality in hypothesis (i) extended from $\Lambda\times \Omega$ to $\Span(\Lambda)\times \Omega$.

\item $\phi:\Omega\rightarrow E_2$ is convex-linear. In fact, for all $\omega_1,\omega_2\in \Omega$, $\lambda\in \Lambda$, and $p\in [0,1]$, we have 
\begin{align*}
\braket{\phi(p\omega_1+(1-p)\omega_2), \psi(\lambda)} &= \braket{p\omega_1+(1-p)\omega_2 , \lambda} \\
&= p \braket{\omega_1, \lambda} + (1-p) \braket{\omega_2, \lambda} \\
&= p \braket{\phi(\omega_1), \psi(\lambda) } + (1-p) \braket{\phi(\omega_2), \psi(\lambda)} \\
&= \braket{p \phi(\omega_1) + (1-p) \phi(\omega_2), \psi(\lambda)}\, .
\end{align*}
Since $\Span \left( \psi(\Lambda)\right)$ is weak*-dense in $E_2$, the above equality implies that
\bbb
\phi(p\omega_1+(1-p)\sigma) = p \phi(\omega) + (1-p) \phi(\omega_2)\, ,
\eee
showing that $\phi:\Omega\rightarrow E_2$ is indeed convex-linear, as claimed.

\item Now, we can use the above observation together with Lemma~\ref{lemma aff ext} to extend $\phi$ to an affine map defined on the whole $\aff(\Omega)$. Since $\aff(\Omega)\subseteq \aff \left( \cl(\Omega)\right)$ and the latter (being the linear span of a base of a base norm space) does not contain the origin, we can immediately apply Lemma~\ref{lemma lin ext} and extend further $\phi$ to a linear map $\phi:\Span(\Omega)\rightarrow E_2$. Let us show that also $\phi$ preserves the norm, i.e. that $\|\phi(x)\|=\|x\|$ for all $x\in E_1$.
\begin{align*}
\|\phi(x)\| &= \sup \left\{ |\braket{y',\phi(x)}|:\ y'\in E_2,\ \|y'\|_*\leq 1\right\} \\
&\texteq{(1)} \sup \left\{ |\braket{y',\phi(x)}|:\ y'\in\Span(\psi(\Lambda)),\ \|y'\|_*\leq 1\right\} \\
&= \sup \left\{ |\braket{\psi(y),\phi(x)}|:\ y\in\Span(\Lambda),\ \|\psi(y)\|_*\leq 1\right\} \\
&\texteq{(2)} \sup \left\{ |\braket{y,x}|:\ y\in\Span(\Lambda),\ \|\psi(y)\|_*\leq 1\right\} \\
&\texteq{(3)} \sup \left\{ |\braket{y,x}|:\ y\in\Span(\Lambda),\ \|y\|_*\leq 1\right\} \\
&\texteq{(4)} \|x\|\, .
\end{align*}
In the above derivation, we used: (1) Lemma~\ref{lemma weak* dense}; (2) hypothesis (i), extended from $\Omega\times \Lambda$ to $\Span(\Omega)\times \Span(\Lambda)$; (3) the already shown fact that $\psi$ preserves the norm; (4) again Lemma~\ref{lemma weak* dense}.

\item Since $\phi: \Span(\Omega)\rightarrow E_2$ is bounded, there is a continuous (linear) extension $\Phi:\cl\left(\Span(\Omega)\right)\rightarrow E_2$. Now, it is easy to verify that $\cl\left(\Span\, M \right) \supseteq \Span \left( \cl\, \Omega\right)$ for all subsets $M$ of any Banach space. Applying this with $M=\Omega$ yields $\cl\left(\Span(\Omega)\right) \supseteq \Span \left(\cl\, \Omega\right) = E_1$, where the last equality holds because $\cl\, \Omega$ is a base of the positive cone of a base norm space, and as such is spanning.
Hence $\cl\left(\Span(\Omega)\right) = E_1$, and we have constructed an extension $\Phi: E_1\rightarrow E_2$. Few useful properties of $\phi$ that are retained by $\Phi$ are as follows.
\begin{itemize}
\item Hypothesis (i) can be extended to $\braket{\psi(y), \Phi(x)} = \braket{y,x}$ for all $x\in E_1$ and $y\in \Span(\Lambda)$.
\item Performing a continuous linear extension preserves the norm, hence the identity $\|\Phi(x)\|=\|x\|$ is obeyed by all $x\in E_1$. This implies that $\Phi$ is an isometry, and in particular that it is injective.
\item Consequently, $\Phi\left( \cl(\Omega) \right) = \cl\left( \phi(\Omega) \right)$ holds. To show this, we first take $\omega \in \cl (\Omega)$ and construct a sequence $(\omega_n)_n\subset \Omega$ such that $\lim_n \| \omega - \omega_n\|=0$. Then, we verify that 
\bbb
\lim_n \|\Phi(\omega)-\Phi(\omega_n)\| = \lim_n \|\Phi(\omega-\omega_n)\|=\lim_n \|\omega -\omega_n\|=0\, ,
\eee
which implies that $\Phi(\omega) \in \cl\left( \phi(\Omega) \right)$ because $\Phi(\omega_n) = \phi (\omega_n)\in \phi(\Omega)$. Since $\omega\in \cl(\Omega)$ was generic, we obtain $\Phi\left( \cl(\Omega) \right) \subseteq \cl\left( \phi(\Omega) \right)$.

Conversely, pick $\omega'\in \cl\left( \phi(\Omega) \right)$, and let $(\omega_n)_n\subset \Omega$ be a sequence such that $\lim_n \|\omega'- \phi(\omega_n)\|=0$. Since $\phi$ is an isometry and $\left(\phi(\omega_n)\right)_n$ is a Cauchy sequence, the same is true for $(\omega_n)_n$. Hence, the sequence $(\omega_n)_n$ converges in norm to some $\omega \in \cl(\Omega)$, which in turn satisfies $\omega'=\Phi(\omega) \in \Phi\left( \cl(\Omega) \right)$ because $\Phi$ is a bounded (equivalently, continuous) map. Since $\omega'\in \cl\left( \phi(\Omega) \right)$ was generic, we have shown that $\cl\left( \phi(\Omega) \right) \subseteq \Phi\left( \cl(\Omega) \right)$ and thus completed the proof of the claim. 

Since $\cl(\Omega)$ and $\cl\left( \phi(\Omega) \right)$ are the bases of the base norm spaces $E_1$ and $E_2$, from Lemma~\ref{lemma transf bases} we immediately deduce that $\Phi$ is an isometric order isomorphism.
\end{itemize}

\item We are now in position to extend also the action of $\psi$ from $\Span(\Lambda)$ to the whole $E_1^*$. To do this, we can exploit the already proved surjectivity of $\Phi$. Namely, for a generic $y\in E_1^*$, we can simply \emph{define} the action of the transformed functional $\Psi(y)\in E_2^*$ to be
\bbb
\braket{\Psi(y), \Phi(x)} \coloneqq \braket{y,x}\, .
\eee
Observe that $\Psi$ is clearly an extension of $\psi$, since for $y\in\Span(\Lambda)$ by the known properties of $\Phi$ one has $\braket{\psi(y),\Phi(x)}=\braket{y,x}=\braket{\Psi(y),\Phi(x)}$ for all $x\in E_1$, which implies $\braket{\Psi(y)-\psi(y),\Phi(x)}=0$ and in turn $\Psi(y)=\psi(y)$, since $\Phi(x)$ runs over the whole space $E_2$. Clearly, since $\braket{y,x}=\braket{\Psi(y),\Phi(x)}=\braket{y,(\Psi^*\Phi)(x)}$ holds for all $x\in E_1$ and $y\in E_1^*$, we have $\Psi^*\Phi=\text{id}$ and therefore $\Psi=(\Phi^*)^{-1}$, as claimed. In particular, Lemma~\ref{lemma iso *-1} tells us that $\Psi$ is an isometric isomorphism because so is $\Phi$. This concludes the proof.

\end{itemize}
\end{proof}

\chapter{Tensor product rule for GPTs} \label{app tensor}

The purpose of this appendix is to give an elementary proof of  the tensor product rule for finite dimensional GPTs (Lemma~\ref{tensor product lemma}). Let us first restate it.

\begin{lemma*}
Let the finite-dimensional GPTs $A,B$ and $AB$ satisfy Axioms~\ref{ax composition maps},~\ref{ax factorisation probabilities},~\ref{ax local tomography}. Then there are linear isomorphisms $\mathcal{J}:V_A\otimes V_B\rightarrow V_{AB}$ and $\mathcal{J}_*:V_A^*\otimes V_B^*\rightarrow V_{AB}^*$ such that:
\begin{enumerate}[(a)]
\item $\mathcal{J}(\omega_A\otimes \tau_B) = \mathcal{j}(\omega_A, \tau_B)$ for all states $\omega_A\in \Omega_A$, $\tau_B\in \Omega_B$;
\item $\mathcal{J}_*(e_A\otimes f_B) = \mathcal{j}_*(e_A, f_B)$ for all effects $e_A\in [0,u_A]$, $f_B\in [0,u_B]$;
\item $\mathcal{J}_*(u_A\otimes u_B) = u_{AB}$; and
\item $\mathcal{J}_*^{-1} = \mathcal{J}^*$, where $\mathcal{J}^*$ is the adjoint of $\mathcal{J}$.
\end{enumerate}
\end{lemma*}

\begin{proof}
We first show that the composition map $\mathcal{j}:\Omega_A\times \Omega_B\rightarrow \Omega_{AB}\subseteq V_{AB}$ can be extended to a bilinear function acting on the whole $V_A\times V_B$.
Since $\mathcal{j}$ is convex-bilinear, for all fixed $\omega_A\in \Omega_A$ the resulting function $\mathcal{j}(\omega_A,\cdot):\Omega_B\rightarrow V_{AB}$ is convex-linear. We can then apply Lemmas~\ref{lemma aff ext} and~\ref{lemma lin ext} to extend it to a linear map $\mathcal{j}(\omega_A,\cdot): V_B\rightarrow V_{AB}$. This way we have sketched of obtaining a linear map starting from a state $\omega_A\in \Omega_A$ amounts to a function $\Omega_A\rightarrow \mathcal{L}(V_B, V_{AB})$. This latter function is easily seen to be convex-linear, and therefore (again by Lemmas~\ref{lemma aff ext} and~\ref{lemma lin ext}) can be extended to a linear map $V_A \rightarrow \mathcal{L}(V_B, V_{AB})$. A little thought shows that one can equivalently think of such a linear map as a bilinear function $V_A\times V_B\rightarrow V_{AB}$, which is immediately seen to extend $\mathcal{j}$ and therefore will be denoted again with $\mathcal{j}$.

Now, the universal property of the tensor product allows us to `lift' $\mathcal{j}$ to a \emph{linear} map $\mathcal{J} :V_A\otimes V_B\rightarrow V_{AB}$, which satisfies $\mathcal{J}(x_A\otimes x_B)=\mathcal{J}(x_A,x_B)$ for all $x_A\in V_A$, $x_B\in V_B$, meeting requirement (a). Clearly, the same reasoning can be applied to the effects, and one obtains instead a linear map $\mathcal{J}_*:V_A^*\otimes V_B^*\rightarrow V_{AB}^*$ that meets (b). 
The factorisation rule for the probabilities now reads
\bbb
\braket{\mathcal{J}_* (\varphi_A\otimes \varphi_B),\, \mathcal{J}(x_A \otimes x_B)} = \braket{\varphi_A, x_A} \braket{\varphi_B, x_B}\, .
\eee
The above identity implies that $\mathcal{J},\mathcal{J}_*$ are injective. Let us show this latter claim for $\mathcal{J}$, as the reasoning for $\mathcal{J}_*$ is totally analogous. If one has $\mathcal{J}(x_{AB})=0$ with $x_{AB}=\sum_i x_A^{(i)}\otimes x_B^{(i)} \in V_A\otimes V_B$, one can write
\begin{align*}
\braket{\varphi_A\otimes \varphi_B, x_{AB}} &= \sum_i \braket{\varphi_A, x_A^{(i)}} \braket{\varphi_B, x_B^{(i)}} \\
&= \sum_i \braket{\mathcal{J}_*(\varphi_A\otimes \varphi_B), \mathcal{J}( x_A^{(i)} \otimes x_B^{(i)})} \\
&= \braket{\mathcal{J}_*(\varphi_A\otimes \varphi_B), \mathcal{J}( x_{AB} ) } \\
&= 0\, .
\end{align*}
Since product functionals generate $V_A^*\otimes V_B^*$, this is possible only if $x_{AB}=0$.

Next, let us show that $\mathcal{J},\mathcal{J}_*$ are in fact surjective, so that they become linear isomorphisms. To this end, we first use Axiom~\ref{ax local tomography} at the level of effects to see that $\mathcal{J}_*(u_A\otimes u_B)=u_{AB}$, thus verifying (c). Since both sides of the equation are legitimate effects, we just need to check that they produce the same statistics on all local states:
\begin{align*}
&\braket{\mathcal{J}_*(u_A\otimes u_B) - u_{AB},\, \mathcal{J}(\omega_A\otimes \omega_B)} \\
&\qquad = \braket{\mathcal{J}_*(u_A\otimes u_B),\, \mathcal{J}(\omega_A\otimes \omega_B)}  - \braket{u_{AB},\, \mathcal{J}(\omega_A\otimes \omega_B)} \\
&\qquad = \braket{u_A, \omega_A} \braket{u_B, \omega_B} - \braket{u_{AB},\, \mathcal{J}(\omega_A\otimes \omega_B)} \\
&\qquad = 1-1\\
&\qquad =0\, .
\end{align*}

Then, we proceed to show that the image $\mathcal{J}_*(V_A^*\otimes V_B^*)$ of $\mathcal{J}_*$ has trivial annihilator, i.e. is such that
\bbb
\braket{\varphi_{AB}, x_{AB}}=0\quad \forall\ \varphi_{AB}\in \mathcal{J}_*(V_A^*\otimes V_B^*)\quad \Longrightarrow\quad x_{AB}=0\, ,
\eee
for all $x_{AB}\in V_{AB}$. In fact, if this were not the case, one could pick a state $\omega_{AB}$ that belongs to the relative interior of $\Omega_{AB}$ (equivalently, to the interior of the positive cone $C_{AB}$), a vector $x_{AB}\in V_{AB}$ enjoying the above property, and construct $\omega'_{AB}\coloneqq \omega_{AB} + \epsilon x_{AB}$. This $\omega'_{AB}$ is normalised, because
\bbb
\braket{u_{AB}, x_{AB}} = \braket{\mathcal{J}_*(u_A\otimes u_B), x_{AB}} = 0 
\eee
and hence $\braket{u_{AB}, \omega'_{AB}}=1$, and moreover is a legitimate state for sufficiently small $\epsilon$, as $\omega_{AB}$ was chosen to be in the interior of the positive cone. It is also easy to see that $\omega'_{AB}$ and $\omega_{AB}$ have the same local statistics, as $x_{AB}$ annihilates all local effects $\mathcal{J}_*(e_A\otimes e_B)$. This shows that the annihilator of $\mathcal{J}_*(V_A^*\otimes V_B^*)$ is the trivial subspace, and since the only subspace of the dual with trivial annihilator is the whole space, we conclude that $\mathcal{J}_*(V_A^*\otimes V_B^*)=V_{AB}^*$ and $\mathcal{J}_*$ is surjective, thus an isomorphism. A simple dimension count then implies that also $\mathcal{J}$ is an isomorphism, as
\bbb
\dim V_{AB} = \dim V_{AB}^* = \dim (V_A^* \otimes V_B^*) = \dim (V_A^*) \dim (V_B^*) = \dim V_A \dim V_B\, .
\eee
Finally, Axiom~\ref{ax local tomography} together with the definition of adjoint~\eqref{adjoint} shows that
\begin{align*}
\braket{\varphi_A, x_A} \braket{\varphi_B, x_B} &= \braket{\mathcal{J}_* (\varphi_A\otimes \varphi_B),\, \mathcal{J}(x_A \otimes x_B)} \\
&= \braket{(\mathcal{J}^*\mathcal{J}_*) (\varphi_A\otimes \varphi_B),\, x_A \otimes x_B}\, , 
\end{align*}
implying that $\mathcal{J}^*\mathcal{J}_*(\varphi_A\otimes \varphi_B) = \varphi_A\otimes \varphi_B$ for all $\varphi_A,\varphi_B$ (because product vectors span $V_A\otimes V_B$) and in turn that $\mathcal{J}^*\mathcal{J}_*=\text{id}$, because product functionals span $V_A^*\otimes V_B^*$. This shows that also condition (d) is met and thus concludes the proof.
\end{proof}

\begin{rem}
It is not enough to assume that the local tomography principle holds just at the level of states. In fact, consider $V_{AB} = (V_A\otimes V_B) \oplus \mathds{R}$, with $u_{AB}=(u_A\otimes u_B, 1)$. With this definition, states in $V_{AB}$ will have the form $\omega_{AB} = (k x_{AB}, 1-k)$, where $x_{AB}\in V_A\otimes V_B$ is such that $\braket{u_A\otimes u_B, x_{AB}}=1$, and $k\in \mathds{R}$ is for now generic. The injection map $\mathcal{J}:\Omega_A\times \Omega_{AB}\rightarrow \Omega_{AB}$ acts as $\mathcal{J}(\omega_A, \omega_B) = (\omega_A\otimes \omega_B, 0)$, thus it is clearly convex-bilinear, and analogously for $J_*$. The rule of factorisation of probabilities is satisfied, and moreover every state is uniquely determined by the statistics of local effects, since from the numbers
\bbb
\braket{(e_A\otimes e_B, 0),\, \omega_{AB}} = \braket{(e_A\otimes e_B, 0 ),\, (k x_{AB}, 1-k)} = k \braket{e_A\otimes e_B, x_{AB}}
\eee
one can reconstruct $k$ (by taking $e_A=u_A$, $e_B=u_B$) and consequently also $x_{AB}$, as product effects generate $V_A^*\otimes V_B^*$.
\end{rem}
\chapter{Upper bound on the \\ inseparability ratio $S(2)$} \label{app S(2)}

The problem of determining the largest $c$ satisfying Theorem~\ref{thm c ratio} has been left open. Here we show at least that such an optimal $c$ must be \emph{strictly} less than 2, by providing the explicit upper bound $\sqrt{3}$.

\begin{thm} \label{thm upper R(2)}
The real number $S(2)$ defined in~\eqref{S} satisfies $S(2)\leq\sqrt{3}$. In other words, any $c$ satisfying Theorem~\ref{thm c ratio} must not exceed $\sqrt{3}$.
\end{thm}

\begin{proof}
We consider a set of norms $N_{t}$ on $\mathds{R}^{2}$ parametrised by the real number $t\in [0,1]$. We take as the unit ball of $N_{t}$ the octagon
\begin{equation}
B_{N_{t}} \coloneqq \co \left\{(\pm 1,0),\ (0,\pm 1),\ \frac{1+t}{2}\, (1,\pm1),\ \frac{1+t}{2}\, (\pm1,1) \right\} .
\label{BN_epsilon}
\end{equation}
In formulae, one can see that
\begin{equation}
N_{t}(x,y)\, =\, |x|+|y|-\frac{2t}{1+t}\min\{|x|,|y|\}\, .\label{N_epsilon}
\end{equation}
The unit ball of the dual norm is another octagon:
\begin{equation}
\begin{split}
B_{N_{t *}}  \coloneqq \co\, \bigg\{ &\bigg(\pm 1,\, \frac{1-t}{1+t}\bigg),\ \bigg(\frac{1-t}{1+t},\, \pm1\bigg),\\
&\bigg(\pm1,\, -\frac{1-t}{1+t}\bigg),\ \bigg( -\frac{1-t}{1+t},\, \pm1\bigg) \bigg\}\, .
\end{split}
\label{BN_epsilon'}
\end{equation}
In the following, we will denote the injective and projective norms on $\mathds{R}^{2}\otimes \mathds{R}^{2}\simeq \mathcal{M}_{2}(\mathds{R})$ corresponding to $N_{t}$ by $\|\cdot\|_{\varepsilon(t)}$ and $\|\cdot\|_{\pi(t)}$, respectively. From the comparison of the unit balls it follows easily that
\begin{equation}
N_{t} \leq N_{0} \leq (1+t)^{2} N_{t}\, . \label{N_epsilon vs N_0}
\end{equation}
For
\begin{equation}
Z = \begin{pmatrix} a & b \\ c & d \end{pmatrix} \in \mathcal{M}_{2}(\mathds{R})\, , \label{Z not}
\end{equation}
it can be seen that
\begin{equation}
\|Z\|_{\pi(0)} = |Z|_{1} = |a|+|b|+|c|+|d|\, . \label{proj0}
\end{equation}
A way to show this is by computing the dual norm $\|H\|_{\pi(0)*}=\|H\|_{*\varepsilon(0)}$, which is the injective norm induced by $|\cdot|_{\infty}$ and therefore coincides with the maximum entry of the matrix $H$. The injective norm is slightly less transparent, but one can prove that
\begin{equation}
\|Z\|_{\varepsilon(0)} = \left\{ \begin{array}{lr} |Z|_{1} & \text{if } abcd\geq 0, \\[1ex] |Z|_{1} - 2\, \min\{|a|,|b|,|c|,|d|\} & \text{if } abcd<0. \end{array}\right. \label{inj0}
\end{equation}
For what follows it will be useful to define
\begin{equation}
p(Z) = \left\{ \begin{array}{lr} 0 & \text{if } abcd\geq 0, \\[1ex] 4\,\frac{\min\{|a|,|b|,|c|,|d|\}}{|a|+|b|+|c|+|d|} & \text{if } abcd<0 . \end{array}\right. \label{pZ}
\end{equation}
We can then write
\begin{equation}
r_{0}(Z) = \frac{\|Z\|_{\pi(0)}}{\|Z\|_{\varepsilon(0)}} = \frac{1}{1-p(Z)/2}
\end{equation}
for $t=0$. Incidentally, it is now clear that in this case there are matrices $Z$ such that $r_{0}(Z)=2$, an example of which is the Hadamard matrix $\left( \begin{smallmatrix} 1 & 1 \\ 1 & -1 \end{smallmatrix} \right)$. We have to perturb the model to establish tight upper bounds on $r_{t}(Z)$. In general, the comparison with~\eqref{N_epsilon vs N_0} yields
\begin{equation}
r_{t}(Z) \leq (1+t)^{2} r_{0}(Z) = \frac{(1+t)^{2}}{1-p(Z)/2} \label{ratio bound 1}
\end{equation}
for all $t\in [0,1]$. This is our first bound on the projective-injective ratio. Unfortunately, the information we obtain from it is again useless when $Z$ is sufficiently close to the Hadamard matrix we saw before (or to some equivalent version). Therefore, we will now look for another bound specifically targeted on this matrix. Clearly, such a bound will work only for $t>0$.

Consider a matrix $Z$ as in~\eqref{Z not}. Up to operations that leave the norm $N_{t}$ invariant, we can assume that $a,b,c\geq 0$ and that $|d|\leq a,b,c$. Consider first the case $d<0$. Then~\eqref{pZ} becomes $p(Z)=4|d|/|Z|_{1}$. Taking as an ansatz in the supremum appearing in~\eqref{inj} two times the first functional in the list~\eqref{BN_epsilon'}, we see that
\begin{align*}
\|Z\|_{\varepsilon(t)} &\geq \begin{pmatrix} 1 & \frac{1-t}{1+t} \end{pmatrix} \begin{pmatrix} a & b \\ c & d \end{pmatrix} \begin{pmatrix} 1 \\ \frac{1-t}{1+t} \end{pmatrix}\\
&= a+b+c+d - \frac{2t}{1+t}\, (b+c+2d) + \frac{4t^{2}}{(1+t)^{2}}\, d \\
&\geq a+b+c+d - \frac{2 t}{1+ t}\, (a+b+c+3d) + \frac{4t^{2}}{(1+t)^{2}}\, d \\
&= |a|+|b|+|c|+|d| -2|d| - \frac{2t}{1+t}\, (|a|+|b|+|c|+|d|-4|d|) \\
&\qquad - \frac{4t^{2}}{(1+t)^{2}}\, |d|\, .
\end{align*}
Dividing by $|Z|_{1}$ yields
\begin{equation}
\frac{\|Z\|_{\varepsilon(t)}}{|Z|_{1}} \geq 1 - \frac{p(Z)}{2}  - \frac{2t}{1+t}\, (1-p(Z)) - \frac{t^{2}}{(1+t)^{2}}\, p(Z)\, . \label{inj Z estimate}
\end{equation}
A posteriori, it is easily seen that~\eqref{inj Z estimate} is a fortiori true when $d\geq 0$ and hence $p(Z)=0$. In fact, in this case one sees that
\begin{align*}
\frac{\|Z\|_{\varepsilon(t)}}{|Z|_{1}} &= \frac{a+b+c -\frac{2t}{1+t}(b+c)+\left( \frac{1-t}{1+t}\right)^{2} d}{|Z|_{1}} \\
&= 1 - \frac{2t}{1+t} + \frac{d}{|Z|_{1}}\left( -1 + \frac{4t}{1+t} + \left( \frac{1-t}{1+t}\right)^{2} \right) \\
&= 1 - \frac{2t}{1+t} + \frac{4t^{2}}{(1+t)^{2}}\, \frac{d}{|Z|_{1}} \\
&\geq 1 - \frac{2t}{1+t}\, .
\end{align*}
Now, in order to find an upper bound on $\|Z\|_{\pi(t)}$, we employ the definition~\eqref{proj} together with the expression~\eqref{N_epsilon} for the norm $N_{t}$. Write
\begin{align*}
\|Z\|_{\pi(t)} &= \| (1,0)\otimes (a,b) + (0,1)\otimes (c,d)\|_{\pi(t)} \\[0.5ex]
&\leq\ N_{t}(1,0) N_{t}(a,b) + N_{t}(0,1) N_{t}(c,d) \\[0.5ex]
&= N_{t}(a,b) + N_{t}(c,d) \\
&= |a|+|b|+|c|+|d|-\frac{2t}{1+t} \left( \min\{ |a|,|b| \} + \min\{ |c|,|d| \} \right) \\
&\leq\ |Z|_{1} - \frac{4t}{1+t} \min\{|a|,|b|,|c|,|d|\}\, .
\end{align*}
Finally, using also~\eqref{pZ} we obtain
\begin{equation}
\frac{\|Z\|_{\pi(t)}}{|Z|_{1}} \leq 1 - \frac{t}{1+t}\, p(Z)\, . \label{proj Z estimate}
\end{equation}
Putting together~\eqref{inj Z estimate} and~\eqref{proj Z estimate}, we deduce
\begin{equation}
\begin{split}
r_{t}(Z) &\leq \frac{1 - \frac{t}{1+t}\, p(Z)}{1 - \frac{p(Z)}{2}  - \frac{2t}{1+t}\, (1-p(Z)) - \frac{t^{2}}{(1+t)^{2}}\, p(Z)} \\
&= \frac{1+t(1-p(Z))}{1-t -\frac{1-2 t-t^{2}}{2(1+t)}\, p(Z)}\, .
\end{split}
\label{ratio bound 2}
\end{equation}
This together with~\eqref{ratio bound 1} yields
\begin{equation}
r_{t}(Z) \leq \min\left\{\, \frac{(1+t)^{2}}{1-\frac{p(Z)}{2}},\, \frac{1+t(1-p(Z))}{1-t -\frac{1-2t-t^{2}}{2(1+t)}\, p(Z)}\,\right\} . \label{ratio bound}
\end{equation}
By definition of $S(2)$ we see that
\begin{equation}
S(2) \leq \min_{0\leq t \leq 1}\, \max_{0\leq p\leq 1} \min\left\{\, \frac{(1+t)^{2}}{1-\frac{p}{2}},\, \frac{1+t(1-p)}{1-t -\frac{1-2t-t^{2}}{2(1+t)}\, p}\,\right\} = \sqrt{3}\, ,
\end{equation}
where the last step follows from some elementary considerations that we do not report. The $t$ achieving the above minimum is $2-\sqrt{3}$.
\end{proof}

\begin{rem}
The above upper bound $\sqrt{3}$ is actually achieved by this model when $t=2-\sqrt{3}$. Even more is true, that is, for all $t\in [0,2-\sqrt{3}]$ the function on the right-hand side of~\eqref{ratio bound} is maximal for $p(Z)=1$, and the subsequent upper bound 
\begin{equation}
r_{t}(Z) \leq \frac{2(1+t)}{1+2t-t^{2}} \label{ratio bound tight}
\end{equation}
is tight and achieved by $Z=\left( \begin{smallmatrix} 1 & 1 \\ 1 & -1 \end{smallmatrix} \right)$. In fact, on the one hand it is easy to convince ourselves that
\begin{align*}
\left\| \begin{pmatrix} 1 & 1 \\ 1 & -1 \end{pmatrix} \right\|_{\varepsilon(t)} &= \begin{pmatrix} 1 & \frac{1-t}{1+t} \end{pmatrix} \begin{pmatrix} 1 & 1 \\ 1 & -1 \end{pmatrix} \begin{pmatrix} 1 \\ \frac{1-t}{1+t} \end{pmatrix} \\
&= \frac{2(1+2t-t^{2})}{(1+t)^{2}}\, .
\end{align*}
On the other hand,
\begin{align*}
\left\| \begin{pmatrix} 1 & 1 \\ 1 & -1 \end{pmatrix} \right\|_{\pi(t)} &= \left\| \begin{pmatrix} 1 & 1 \\ 1 & -1 \end{pmatrix} \right\|_{*\varepsilon(t)} \\
&\geq \begin{pmatrix} \frac{1+t}{2} & \frac{1+t}{2} \end{pmatrix} \begin{pmatrix} 1 & 1 \\ 1 & -1 \end{pmatrix} \begin{pmatrix} 1 \\ 0 \end{pmatrix} \\
&= 1+t, ,
\end{align*}
and this is actually an equality. Putting all together we see that
\begin{equation*}
r_{t} \begin{pmatrix} 1 & 1 \\ 1 & -1 \end{pmatrix} = \frac{2(1+t)}{1+2t-t^{2}}\, ,
\end{equation*}
which in turn for $t\in [0,2-\sqrt{3}]$ corresponds to the maximum of the right-hand side of~\eqref{ratio bound} over $0\leq p(Z)\leq 1$ and reduces to $\sqrt{3}$ when $t=2-\sqrt{3}$.
\end{rem}

\chapter{Equiprobable data hiding pairs} \label{app equi}

In this appendix we look into the consequences of restricting the definition of data hiding to equiprobable pairs of states. Let us start by stating the corresponding modified version of Definition~\ref{dh}.

\begin{Def} \label{equi dh}
Let $(V,C,u)$ be a GPT. For an informationally complete set of measurements $\mathcal{M} \subseteq \mathbf{M}$, we say that there is \emph{balanced data hiding against $\mathcal{M}$ with efficiency $\widetilde{R}\geq 1$} if there are two normalised states $\rho,\sigma\in\Omega$ such that the probability of error defined in~\eqref{pr error} satisfies
\begin{equation}
P_{e}^{\mathbf{M}}(\rho,\sigma; 1/2) = 0\, ,\qquad P_{e}^{\mathcal{M}}(\rho,\sigma; 1/2) = \frac12 \left( 1-\frac{1}{\widetilde{R}}\right) .
\label{equi dh eq}
\end{equation}
The highest balanced data hiding efficiency against $\mathcal{M}$ is called \emph{balanced data hiding ratio against $\mathcal{M}$} and will be denoted by $\widetilde{R}(\mathcal{M})$.
\end{Def}

\begin{rem}
Clearly, standard data hiding ratios are always larger or equal to the corresponding balanced versions, i.e. $R(\mathcal{M})\geq \widetilde{R}(\mathcal{M})$.
\end{rem}

Throughout the rest of this appendix, we will consider (balanced) data hiding against a fixed set of measurements $\mathcal{M}$ that we do not specify further. 
There are at least two reasons why the above Definition~\ref{equi dh} does not add much to the theory developed in the main text, at least from a fundamental perspective. The first such reason is that it turns out that highly efficient standard data hiding pairs of states are automatically almost equiprobable, as the following result establishes.

\vspace{2ex}
\begin{lemma}
Let $\rho,\sigma$ be states that obey~\eqref{dh eq} for some a priori probability $p\neq 1/2$. Then the standard data hiding efficiency $R$ satisfies 
\begin{equation}
R\, \leq\, \frac{1}{|2p-1|}\, .
\label{if high R then p almost 1/2}
\end{equation}
\end{lemma}

\begin{proof}
An obvious upper bound to the probability of error $P_e^\mathcal{M}(\rho,\sigma ; p)$ can be obtained by considering the strategy of guessing according to the maximum likelihood rule.
Since such strategy fails with probability $\min\{p,1-p\}$, we get
\bbb
P_e^\mathcal{M}(\rho,\sigma; p)\, \leq\, \min\{p,1-p\}\, =\, \frac{1-|2p-1|}{2}\, .
\eee
Using the definition of efficiency~\eqref{dh eq}, one finds immediately
\bbb
R\, =\, \frac{1}{1-2P_e^{\mathcal{M}}(\rho,\sigma ;p)}\, \leq\, \frac{1}{|2p-1|}\, ,
\eee
as claimed.
\end{proof}

While the above result establishes the `approximate equiprobability' of highly efficient data hiding pairs, balanced data hiding ratios still stand as a different concept, as in their definition we required the \emph{exact} equality $p=1/2$ from the start. 
In fact, it could indeed happen that $\widetilde{R}(\mathcal{M})<R(\mathcal{M})$ in some cases. However, we will show in a moment that anyway $R$ and $\widetilde{R}$ have to be of the same order. To this purpose, we need a modified version of Proposition~\ref{dh ratio}.

\begin{lemma} \label{equi dh ratio}
For an informationally complete set of measurements $\mathcal{M}\subseteq\mathbf{M}$ in an arbitrary GPT, the balanced data hiding ratio $\widetilde{R}(\mathcal{M})$ can be computed as
\begin{equation}
\widetilde{R}(\mathcal{M})\, =\, \max \left\{ \frac{\|y\|}{\|y\|_{\mathcal{M}}}:\ y\in V,\ y\neq 0,\ \braket{u,y}=0 \right\}\, .
\label{equi dh ratio eq}
\end{equation}
\end{lemma}

\begin{proof}
The argument follows closely that we gave in the proof of Proposition~\ref{dh ratio}. The only difference is that now one needs to characterise the set $\widetilde{K}$ of vectors $y\in V$ that can be expressed as $y=\frac12 \rho- \frac12 \sigma$ for some normalised states $\rho,\sigma\in \Omega$. It is not difficult to realise that $\widetilde{K} = \left\{y\in V:\ \|y\|\leq 1,\, \braket{u,y}=0 \right\}$, ultimately yielding~\eqref{equi dh ratio eq}.

The above characterisation can be justified as follows. On the one hand, if $y=\frac12\rho-\frac12 \sigma$ for $\rho,\sigma\in\Omega$ then $\|y\|\leq \frac12 \|\rho\|+\frac12 \|\sigma\| = \frac12+\frac12=1$, and moreover $\braket{u,y}=\frac12 (1-1)=0$. On the other hand, applying~\eqref{dual base eq} as usual we can write any $y\in V$ such that $\|y\| = 1$ as a difference $y=y_+-y_-$, where $y_\pm \geq 0$ and $\braket{u,y_++y_-}=1$. Requiring also $\braket{u,y}=0$ leads us to the conditions $\braket{u,y_+}=\braket{u,y_-}=\frac12$, so that $y_+=\frac12 \rho$ and $y_-=\frac12 \sigma$ for some normalised states $\rho,\sigma\in\Omega$. This shows that all vectors $y\in V$ such that $\|y\| = 1$ and $\braket{u,y}=0$ belong to $\widetilde{K}$. Since the latter set is convex and contains $0$, we deduce that indeed $\left\{y\in V:\ \|y\|\leq 1,\, \braket{u,y}=0 \right\}\subseteq \widetilde{K}$, which is what was left to show.
\end{proof}

We are now ready to prove the following result, which concludes our analysis.

\begin{prop} \label{equi vs standard}
The balanced and standard data hiding ratios, as given by Definition~\ref{dh} and~\ref{equi dh}, respectively, satisfy the inequalities
\bb
\widetilde{R}(\mathcal{M})\, \leq\, R(\mathcal{M})\, \leq\, 2 \widetilde{R}(\mathcal{M}) + 1\, .
\label{equi vs standard eq}
\ee
\end{prop}

\begin{proof}
We already saw that balanced data hiding ratios are always upper bounded by their standard versions, so let us focus on the complementary bound. We start by picking a vector $x\in V$ that saturates the maximum in~\eqref{dh ratio eq}, i.e. that is such that $\|x\|_\mathcal{M}=1$ and $\|x\|=R$. Choose a normalised state $\omega\in \Omega$, and construct $y\coloneqq x-\braket{u,x} \omega$, so that $\braket{u,y}=0$. Observe that $|\braket{u,x}|\leq \|x\|_\mathcal{M}=1$. To estimate the norms $\|y\|$ and $\|y\|_{\mathcal{M}}$, one can apply repeatedly the triangle inequality:
\begin{align*}
\|y\|\, &\geq\, \|x\| - |\braket{u,x}|\, \geq\, R(\mathcal{M})-1\, ,\\
\|y\|_\mathcal{M}\, &\leq\, \|x\|_\mathcal{M} + |\braket{u,x}|\, \leq\, 2\, .
\end{align*}
Therefore, we can use Lemma~\ref{equi dh ratio} to write
\bbb
\widetilde{R}(\mathcal{M})\, \geq\, \frac{\|y\|}{\|y\|_\mathcal{M}}\, \geq\, \frac{R(\mathcal{M})-1}{2}\, ,
\eee
which gives~\eqref{equi vs standard eq} upon elementary algebraic manipulations.
\end{proof}

\chapter{Quantum data hiding with Werner states} \label{app Werner}

Throughout this appendix, we briefly review the well-known techniques used to find lower bounds on data hiding ratios via Werner states, and apply them to prove \eqref{bound Werner} and \eqref{bound Werner W}. More precisely, we will show that those described in \eqref{bound Werner} and~\eqref{bound Werner W} are the optimal data hiding pairs within the Werner class, defined as the real span of symmetric and antisymmetric projector~\eqref{symm antisymm proj}. Formally, this amounts to showing that
\begin{align}
&\max_{(\alpha,\beta)\neq (0,0)} \frac{\|\alpha\rho_{S}+\beta\rho_{A}\|_{1}}{\|\alpha\rho_{S}+\beta\rho_{A}\|_{\text{SEP}}} = n\, ,\label{app Werner eq1} \\
&\max_{(\alpha,\beta)\neq (0,0)} \frac{\|\alpha\rho_{S}+\beta\rho_{A}\|_{W}}{\|\alpha\rho_{S}+\beta\rho_{A}\|_{\text{SEP}}} = 2n-1\, .
\label{app Werner eq2}
\end{align}
In other words, both maxima are achieved when $\alpha=n+1,\, \beta=-(n+1)$, which -- interestingly enough -- makes $\alpha \rho_S + \beta \rho_A \propto F$. As a side remark, let us notice here that the above separability norms could be easily replaced with $\text{LOCC}$ norms, since the extremal measurements in the Werner class are well-known to be $\text{LOCC}$~\cite[Propositions 1 and 3]{VV-dh-Chernoff}.

Consider a binary measurement $(E,\mathds{1}-E)$, which we will choose later to be either standard, or a witness, or else an $\text{LOCC}$. Thanks to the fact that to estimate distinguishability norms we are going to compute quantities of the form $\Tr E Z$ where $Z$ is in the Werner class and thus invariant under the `twirling' operation $\mathcal{T}(\cdot)\coloneqq \int dU\, U\otimes U (\cdot) U^{\dag}\otimes U^{\dag}$, a very standard trick going back to the first paper on data hiding~\cite{dh-original-1} allows us to assume that also the measurement elements $E,\mathds{1}-E$ belong to the Werner class. For $E = a\mathds{1}+b F$, it is well-known that the measurement $(E,\mathds{1}-E)$ is:
\begin{itemize}
\item a standard quantum mechanical measurement iff $0\leq E\leq \mathds{1}$, i.e. iff $0\leq a\pm b\leq 1$;
\item separable iff $E,\mathds{1}-E$ are both separable operators~\eqref{separable}, iff $0\leq a - b\leq 1$ and $0\leq a + nb\leq 1$;
\item a $W$-theory measurement iff $E,\mathds{1}-E$ are both witnesses~\eqref{witnesses}, iff $0\leq a\leq 1$ and $0\leq a+b\leq 1$.
\end{itemize}
While the first two points are well-known results~\cite{Werner-symmetry}, the latter deserves a quick comment. An operator $E$ is a witness iff $\braket{\alpha \beta| E|\alpha\beta}\geq 0$ for all states $\ket{\alpha},\ket{\beta}\in\mathds{C}^{n}$. We can thus use the properties of the flip operator to conclude that $E=a\mathds{1}+bF$ is a witness iff $a + pb\geq 0$ for all $p\in [0,1]$, which together with the analogous condition for $\mathds{1}-E$ reproduces the above constraints. Optimising over all measurements in the Werner class we conclude that
\begin{align}
\|\alpha\rho_{S}+\beta\rho_{A}\|_{1}\, &=\, |\alpha|+|\beta|\, , \label{app Werner eq2} \\
\|\alpha\rho_{S}+\beta\rho_{A}\|_{\text{SEP}}\, &=\, \frac{2}{n+1}\, |\alpha| + \left| \frac{n-1}{n+1}\alpha +\beta\right| , \label{app Werner eq3} \\
\|\alpha\rho_{S}+\beta\rho_{A}\|_{W}\, &=\, |\alpha-\beta|+2|\beta|\, . \label{app Werner eq4}
\end{align}
Computing the maxima in~\eqref{app Werner eq1} and showing that they are achieved for $\alpha=n+1,\, \beta=-(n+1)$ is now an elementary exercise.

\chapter{Comparing $\|\cdot\|_W$ and $\|\cdot\|_2$} \label{app W norm}

Here, we argue that the methods in~\cite{VV-dh} can not be applied directly to determine the data hiding ratio in $W$-theory against any locally constrained set of measurements. As detailed in Subsection~\ref{subsec3 dh W}, to this purpose it is enough to show the following.

\begin{lemma}
For all $\delta>0$, there exists a sequence of operators $Z_n$ acting on $\mathds{C}^n\otimes\mathds{C}^n$, such that $\|Z_n\|_2=1$ but $\|Z_n\|_W\geq \Omega(n^{3/2-\delta})$, where $\|\cdot\|_W$ is defined in~\eqref{W norm}.
\end{lemma}

\begin{proof}
From the results of~\cite{aspects-generic-entanglement} it is known that a random orthogonal projector $\Pi$ onto a subspace of $\mathds{C}^n\otimes \mathds{C}^n$ of dimension $k\gg n$ will satisfy
\bb
\braket{\alpha\beta | \Pi |\alpha \beta }\, \leq\, \frac{2k}{n^2} \qquad \forall\ \ket{\alpha}, \ket{\beta}\in \mathds{C}^n:\ \braket{\alpha|\alpha}=\braket{\beta|\beta}=1
\label{random proj}
\ee
with high probability as $n$ tends to infinity. Fixing $k=n^{1+2\delta}$ and picking one such $\Pi_n$ for all $n\in\mathds{N}$, we can construct $Z_n\coloneqq n^{-1/2-\delta}\, \Pi_n$. By definition, $\|Z_n\|_2=1$ for all $n\in\mathds{N}$. Moreover, defining $H\coloneqq \mathds{1} - \frac{n^2}{k}\, \Pi_n$, by~\eqref{random proj} we have $-1\leq \braket{\alpha\beta| H | \alpha\beta}\leq 1$ for all normalised $\ket{\alpha},\ket{\beta}$, i.e. $H$ belongs to the unit ball of the dual base norm of the GPT $\text{QM}_n\tmin\text{QM}_n$. Hence,
\bbb
\|Z_n\|_W \geq \Tr Z_n H = \frac{n^2-k}{\sqrt{k}} = \Omega(n^{3/2-\delta})\, ,
\eee
which yields the claim.
\end{proof}

\chapter{Additional remarks on some results of Chapter 6} \label{app Schur}

This appendix complements Chapter~\ref{chapter6}. Here we complete the proof of Theorem~\ref{G mono}(b) and extend Theorem~\ref{I2E} to include the case of the underlying entropy being a R\'enyi-$\alpha$ entropy with $\alpha\geq 2$.

\section{Gaussian steerability is not monogamous with respect to the steering party}

We present a counterexample showing that already in the simplest case $k=2$, $n_A=2,\, n_{B_1}=n_{B_2}=1$, there exist mixed states violating inequality~\eqref{mon steer 2}. This completes the proof of Theorem~\ref{G mono}(b).
A QCM that does the job, as found by numerical search, is as follows:
\begin{equation} V_{AB_1B_2} = \begin{pmatrix}
 1.2 & -0.3 & 0.4 & -2.7 & 1.8 & -1.9 & 0.4 & -0.1 \\
 -0.3 & 0.9 & -1.2 & 0.4 & -1.2 & 0.5 & -0.4 & 0.1 \\
 0.4 & -1.2 & 4.5 & 1.6 & -1.4 & 1.8 & -0.1 & -0.3 \\
 -2.7 & 0.4 & 1.6 & 12. & -9.5 & 10.1 & -1.4 & -0.3 \\
 1.8 & -1.2 & -1.4 & -9.5 & 11.9 & -11.5 & 1.6 & 0.8 \\
 -1.9 & 0.5 & 1.8 & 10.1 & -11.5 & 11.9 & -1. & -1.4 \\
 0.4 & -0.4 & -0.1 & -1.4 & 1.6 & -1. & 2.4 & -2. \\
 -0.1 & 0.1 & -0.3 & -0.3 & 0.8 & -1.4 & -2. & 2.8
\end{pmatrix} .
\end{equation}
Here, the first four rows and columns pertain to $A$, the fifth and sixth to $B_1$, the last two to $B_2$. It can be easily verified that the minimum symplectic eigenvalue of the above matrix with respect to the symplectic form $\Omega_A\oplus \Omega_{B_1}\oplus\Omega_{B_2}$ is $\nu_{\min}(V_{B_1B_2A})=1.01359$, so that $V_{B_1B_2 A}$ is a legitimate QCM (obeying~\eqref{Heisenberg}). However,
\begin{equation}
\mathcal{G}(B_1B_2\rangle A)_V - \mathcal{G}(B_1\rangle A)_V - \mathcal{G}(B_2\rangle A)_V = -0.816863 < 0\, ,
\end{equation}
violating~\eqref{mon steer 2} as claimed.

\section{R\'enyi extension of Theorem~\ref{I2E}}

This last part of the present appendix is devoted to discussing possible R\'enyi generalisations of~\eqref{I>2E}. In fact, by looking at the proof Theorem~\ref{I2E}, one could wonder what makes the R\'enyi-2 entropy special in this context. It turns out that for all R\'enyi-$\alpha$ entropies with $\alpha\geq 1$ (included the von Neumann one) we can always provide the upper bound
\begin{equation}
2\, E^G_{F,\alpha}(A:B)_V \leq S_\alpha \left( V_A\# \left( \Omega_A (V_{AB}/V_B)^{-1} \Omega_A^T \right)\right) ,
\label{cool generalisation}
\end{equation}
where the function $S_\alpha(V)$, defined in~\eqref{S_alpha}, gives the R\'enyi-$\alpha$ entropy of a Gaussian state with QCM $V$. However, the crucial inequality
\begin{equation}
S_\alpha(M\# N) \leq \frac{1}{2} S_\alpha(M)+\frac{1}{2} S_\alpha (N)\, , \label{ineq geom mean}
\end{equation}
which we used in the proof of Theorem~\ref{I2E} to simplify further the right-hand side of~\eqref{cool generalisation}, breaks down for $\alpha<2$. In particular, it can be violated for $\alpha=1$. On the contrary, $I_{\alpha}(A:B) \geq 2\, E^G_{F,\alpha}(A:B)$ holds true as long as $\alpha\geq 2$, as the next Lemma clarifies.

\begin{lemma} \label{lemma alpha}
Fix an integer $n\geq 1$. The inequality $S_\alpha(M\# N)\leq \frac{1}{2} S_\alpha(M)+\frac{1}{2} S_\alpha (N)$ holds for all $2n\times 2n$ real matrices $M,N>0$ if and only if $\alpha\geq 2$.
\end{lemma}

\begin{proof}
We claim that inequality~\eqref{ineq geom mean} is equivalent to the convexity of the function
\begin{equation}
f_\alpha (x) \coloneqq -\frac{1}{\alpha-1}\,\log\frac{2^\alpha}{(e^x+1)^\alpha - (e^x-1)^\alpha} \label{f_alpha}
\end{equation}
defined on $\mathds{R}_+$, where conformally to~\eqref{S_1} one defines
\begin{equation}
f_1 (x) \coloneqq \frac{e^x+1}{2}\,\log\frac{e^x+1}{2} - \frac{e^x-1}{2}\,\log\frac{e^x-1}{2}\, . \label{f_1}
\end{equation}
In fact, on the one hand choosing $M=e^x\mathds{1}$ and $N=e^y \mathds{1}$ yields
\begin{align*}
S_\alpha(M\# N) &= n\, f_\alpha\left(\frac{x+y}{2}\right)\, , \\
\frac{1}{2} S_\alpha(M)+\frac{1}{2} S_\alpha (N) &= \frac{n}{2} f_\alpha (x) + \frac{n}{2} f_\alpha (y)\, ,
\end{align*}
so that $f_\alpha$ is necessarily convex when~\eqref{ineq geom mean} holds. On the other hand, suppose that $f_\alpha$ is convex. From~\cite[Theorem 3]{bhatia15} we learn that $\log\hat{\nu}(M\# N)\prec \frac{1}{2}\log\hat{\nu}(M) +\frac{1}{2}\log\hat{\nu}(N)$, where $\hat{\nu}(M)\in\mathds{R}^{2n}_+$ is obtained by listing the symplectic eigenvalues of $M$ each repeated twice and sorting the entries of the resulting vector in descending order, the logarithm of vectors is intended entrywise, and the symbol $\prec$ denotes \emph{majorization}~\cite[II]{BHATIA-MATRIX}. What the above relation tells us is that the symplectic spectrum of the geometric mean is in a precise sense more disordered than the geometric mean of the two spectra. It is elementary to verify that whenever $x\prec y$ and $f$ is convex, $\sum_{i=1}^n f(x_i)\leq \sum_{i=1}^n f(y_i)$ holds true~\cite[Corollary II.3.4]{BHATIA-MATRIX}. Choosing as $f$ the function in~\eqref{f_alpha} and observing that $f_\alpha$ is always monotonically increasing, we obtain
\begin{align*}
S_\alpha(M\# N) &= \sum_{i=1}^n f_\alpha\left( \log \nu_i(M\# N) \right) \\
&= \frac{1}{2}\sum_{i=1}^{2n} f_\alpha\left( \log \hat{\nu}_i(M\# N) \right) \\
&\leq \frac{1}{2}\sum_{i=1}^{2n} f_\alpha\left( \frac{1}{2}\log\hat{\nu}_i(M) +\frac{1}{2}\log\hat{\nu}_i(N) \right) \\
&\leq \frac{1}{2}\sum_{i=1}^{2n} \frac{1}{2} \big( f_\alpha(\log\hat{\nu}_i(M)) + f_\alpha\left(\log\hat{\nu}_i(N)\right) \big) \\
&= \frac{1}{2} S_\alpha(M)+\frac{1}{2} S_\alpha (N)\, .
\end{align*}

Now, the main claim will follow once we show that $f_\alpha$ defined via~\eqref{f_alpha} is convex if and only if $\alpha\geq 2$. We can restrict our analysis to the case $\alpha>1$ since the function in~\eqref{f_1} is elementarily seen to be non-convex (actually, concave). Some tedious algebra leads us to the following expression for the second derivative of that function:
\begin{equation}
\begin{split}
f_\alpha''(x) &= \frac{\alpha}{\alpha-1}\, \frac{\cosh^\alpha(x/2) \sinh^\alpha(x/2)}{\sinh^2 x\, (\cosh^\alpha(x/2) - \sinh^\alpha(x/2))} \\
&\qquad \cdot \left( \cosh^\alpha(x/2) \sinh^{2-\alpha}(x/2) - \sinh^\alpha (x/2) \cosh^{2-\alpha}(x/2) - \alpha +1 \right) .
\end{split}
\label{f_alpha''}
\end{equation}
Since everything else in the above expression is positive, we have only to prove that $\cosh^\alpha (x/2) \sinh^{2-\alpha} (x/2)\, -\, \sinh^\alpha (x/2) \cosh^{2-\alpha} (x/2)\, \geq \, \alpha -1$ for all $x\geq 0$ if and only if $\alpha\geq 2$. That $\alpha\geq 2$ is necessary can be seen by taking the limit $x\rightarrow 0^+$. Conversely, if $\alpha=2+\delta$ with $\delta\geq 0$ one gets
\begin{align*}
&\cosh^\alpha (x/2) \sinh^{2-\alpha} (x/2) - \sinh^\alpha (x/2)\cosh^{2-\alpha} (x/2) \\
&\qquad= \cosh^2 (x/2)\, \tanh^{-\delta} (x/2) - \sinh^2 (x/2)\, \tanh^\delta (x/2) \\
&\qquad= \frac{1}{1-t^2}\, t^{-\delta} - \frac{t^2}{1-t^2}\, t^{\delta} \\
&\qquad\eqqcolon \varphi_t(\delta)\, ,
\end{align*}
where we defined $t\coloneqq \tanh(x/2)$. It is not difficult to see that the function $\varphi_t(\delta)$ is convex in $\delta$ since
\begin{equation*}
\varphi''_t(\delta) = \frac{t^{-\delta } \left(1-t^{2 \delta +2}\right) \log ^2(t)}{1-t^2} \geq 0\, .
\end{equation*}
From this fact we deduce that
\begin{equation*}
\varphi_t(\delta) \geq \varphi_t(0)+\delta\, \varphi'_t(0) = 1 + \delta = \alpha-1\, ,
\end{equation*}
implying that~\eqref{f_alpha''} is positive and hence that $f_\alpha$ defined by~\eqref{f_alpha} is convex for all $\alpha\geq 2$, as claimed.
\end{proof}

\bibliography{biblio}

\begin{thebibliography}{100}
\bibitem{Plavala-2016}
M.~Pl{\'a}vala.
\newblock All measurements in a probabilistic theory are compatible if and only
  if the state space is a simplex.
\newblock {\em Phys. Rev. A}, 94:042108, 2016.
\end{thebibliography}

\end{document}